\documentclass[11pt]{article}
\usepackage{eurosym}
\usepackage{amsfonts}
\usepackage{amssymb}
\usepackage{graphicx}
\usepackage{amsmath}
\usepackage{makeidx}
\usepackage{indentfirst}
\usepackage[T1]{fontenc}
\usepackage[utf8]{inputenc}

\newcounter{resultnum}[section]
\newcounter{conclusionnum}[section]
\newcounter{conditionnum}[section]
\newcounter{conjecturenum}[section]
\newcounter{examplenum}[section]
\newcounter{exercisenum}[section]
\newcounter{lemmanum}[section]
\newcounter{notationnum}[section]
\newcounter{theoremnum}[section]
\newcounter{definitionnum}[section]
\newcounter{corollarynum}[section]
\newcounter{remarknum}[section]
\newcounter{propositionnum}[section]
\newcounter{acknowledgementnum}[section]
\newcounter{algorithmnum}[section]
\newcounter{axiomnum}[section]
\newcounter{casenum}[section]
\newcounter{claimnum}[section]
\newcounter{summarynum}[section]
\newcounter{problemnum}[section]
\begin{document}

\title{General off-diagonal integrability of metric and nonmetric geometric
flow and Finsler-Lagrange-Hamilton modified Einstein equations }
\date{November 24, 2025}
\author{ \textbf{Sergiu I. Vacaru} \thanks{%
emails: sergiu.vacaru@fulbrightmail.org ; sergiu.vacaru@gmail.com } \\
{\small \textit{Astronomical Observatory, Taras Shevchenko National
University of Kyiv, Kyiv 01601, Ukraine;}} \\
{\small \textit{Department of Physics, California State University at
Fresno, Fresno, CA 93740, USA;}} \\
{\small \textit{\ Department of Physics, Kocaeli University, Kocaeli, 41001,
T\"{u}rkiye}} 
\vspace{.1 in} }

\maketitle


\begin{abstract}
Over the last seventy years, many Finsler-type geometric and modified gravity theories (MGTs) have been elaborated. They have been formulated in terms of different classes of Finsler generating functions, metric and nonmetric structures, nonlinear and linear connections, and various sets of postulated fundamental geometric objects with corresponding nonholonomic dynamical or evolution equations. In several approaches, the resulting gravitational and matter field equations were not completely defined geometrically, or were developed only for restricted models. We present a progress report with historical remarks and a summary of new results on Finsler–Lagrange–Hamilton (FLH) geometric flow and gravity theories. Such theories can be constructed in an axiomatic form on (co)tangent Lorentz bundles as nontrivial modifications of Einstein gravity. They are characterized by nonlinear dispersion relations and may encode nonassociative and noncommutative corrections from string theory, quantum effects, or other MGTs. To generate exact and physically relevant solutions of the FLH–modified Einstein equations, we have developed the anholonomic frame and connection deformation method. We provide a proof of the general integrability of such FLH geometric flows and MGTs, and we analyze new classes of generic off-diagonal solutions determined by generating functions and effective sources that, in general, depend on all spacetime and (co)fiber coordinates.  In general, such off-diagonal configurations do not exhibit horizon/hypersurface duality or holographic structures and thus lie outside the Bekenstein–Hawking thermodynamic paradigm. Instead, by extending G. Perelman’s entropy concept to relativistic FLH geometric flows, we introduce and compute new classes of geometric thermodynamic variables that characterize different FLH theories and their associated solution spaces.

\end{abstract}

\tableofcontents


\section{Introduction:\ Metric and Nonmetric \newline
Finsler-Lagrange-Hamilton geometric flows, GR and MGTs}

\label{sec1}The general relativity (GR) theory, i.e. Einstein's gravity, is
considered by the bulk of researchers as the almost standard theory of
gravity beginning the end of 1915, when A. Einstein and D. Hilbert completed
the formulation of gravitational field equations in certain heuristic and,
respectively, variational forms. The monographs \cite%
{hawking73,misner73,wald82,kramer03} contain summaries of results and
methods, and the most important cosmological and astrophysical applications
of GR before the "accelerating cosmology era". The Einstein gravity theory
was formulated axiomatically on Lorentz manifolds defined as relativistic
versions of the (pseudo) Riemannian spaces, when various types of metric,
tetradic (vierbeind) and spinor variables and 3+1 or 2+2 nonholonomic
decompositions were used. The discovery of late-time cosmic acceleration 
\cite{riess98,perlmutter99} resulted in extensive research on modified
gravity theories (MGTs) \cite{sotiriou10,nojiri11,capo11,clifton12,harko14}
and dark energy (DE) and dark matter (DM) physics \cite{copeland06}; and on
geometric and quantum information flows \cite{partner06}, see references
therein. In this work, we do not cite many works on non-Riemannian
geometries and related MGTs and do not discuss certain models if they do not
consider nonholonomic or Finsler geometry methods in explicit forms. For
reviews of results and geometric methods, readers are recommended to study 
\cite{hehl95,blum12,lheis23} and references therein.

\vskip4pt Finsler-like generalizations of the Einstein gravity form a
special class of MGTs when the geometric and physical objects depend not
only on some base manifold coordinates but also on so-called
velocity/momenta type coordinates (for different types of models with local
anisotropies). We studied such relativistic Finsler-Lagrange-Hamilton (FLH)
geometries and MGTs modelled on (co) tangent Lorentz bundles. We also
elaborated on higher order tangent bundle, supersymmetric, nonassociative
and noncommutative and other type extensions as we reviewed in chronological
form in \cite{partner06,vacaru18,bsssvv25}. Here, we provide and review an
extended list of our contributions to both metric-compatible and
noncompatible FLH theories and elaborate on new methods of constructing
exact solutions for nonmetric MGTs. Many such works were published during
1988-2008 in Eastern Europe \cite{vmon3}, being less known and not
correspondingly cited in Western countries.

\vskip4pt This paper can be considered as a progress report and a transfer of
knowledge on geometric methods for constructing exact and parametric
solutions in FLH theories. It also consists a review and a pedagogical
introduction into the anholonomic frame and connection deformation method,
AFCDM, formulated for constructing off-diagonal solutions in relativistic
Finsler geometric flow and gravity models. This work is different from the
nonassociative and noncommutative FLH theories (on eight dimensions, 8-d,
tangent and cotangent Lorentz bundles) studied in Part II of \cite{bsssvv25}%
. Here we consider real associative and commutative and nonmetric theories
on nonholonomic phase spaces involving conventional velocity or momentum
like coordinates. It redefines for Finsler theories the 4-d constructions
from Part I (and respective Tables) of that review article and of recent
published papers on off-diagonal solutions and G. Perelman's thermodynamics 
\cite{vv25,vv25a}. Here we note that a series of early results and geometric
methods on constructing off-diagonal solutions in GR and Finsler and
non-Finsler MGTs were reviewed in \cite%
{vacaru03,vacaru07a,sv11,vbubuianu17,partner02}.


\subsection{Historical remarks on Finsler-like relativistic geometric flow
and gravity theories}

We list and discuss seven most important steps to formulating and further
developments of respective directions in FLH geometry and physics, which are
important for the objectives of this work:

\begin{description}
\item[{1] Prehistory and beginning of History (1854-1934):}] B. Riemann
considered in his famous habilitation thesis (\cite{riemann1854}, 1854) the
first example of nonlinear quadratic element $ds^{2}=F^{2}(x,y),$ where $%
x=\{x^{i}\}$ are coordinated on a base manifold $V$ and $y=\{y^{a}\}$ are
certain fibler/velocity like coordinates. Such an $F(x,y)$ involving
homogeneity conditions on $y$ (when $F(x,\lambda y)=\lambda F(x,y),$ for $%
\lambda >0$), was later called a Finsler metric, or (equivalently) a Finsler
generating function. The Riemannian geometry was formulated and studied by
using quadratic elements, $ds^{2}\simeq g_{ij}(x)y^{i}y^{j},$ for $%
y^{a}\simeq dx^{a},$ and a symmetric metric tensor $g_{ij}(x).$ That was the
"prehistory" of Finsler geometry. \newline
\vskip0.1pt $-$ The "Finsler history" began in 1918 due to P. Finsler's
thesis \cite{finsler18}, when the term "Finsler geometry" was introduced in
E. Cartan's first monograph on Finsler geometry from 1935 \cite{cartan35}.
In that monograph (in coordinate form), the first examples of Finsler
nonlinear, $\widetilde{\mathbf{N}}=\{\widetilde{N}_{i}^{a}(x,y)\},$
connection (N-connection) and linear distinguished connection (d-connection) 
$\widetilde{\mathbf{D}}=\{\widetilde{\mathbf{D}}_{\alpha }(x,y)\}$ were
considered. Such values consist of two other cornerstone geometric objects
(the first one is $F(x,y)$). Mathematical details on so-called
Riemann-Finsler geometry can be found in monographs \cite%
{rund59,matsumoto86,bcs2000}. We consider that E. Cartan's contributions are
also fundamental because without his first use of N- and d-connections and
the concept of bundle space, Finsler geometric and physical models are
incomplete for elaborating applications in physics. For our purposes to
formulate and study relativistic Finsler generalizations of the Einstein
gravity theory and to construct exact and parametric solutions in such
theories, we follow the system of notations and conventions stated in \cite%
{vacaru18,partner06,vmon3}.\footnote{%
In our approach $V$ is a four dimensional (4-d) Lorentz manifold as in GR;
and $TV$ and $T^{\ast }V$ are respective tangent and cotangent Lorentz
bundles with respective local coordinates $u=\{u^{\alpha }=(x^{i},v^{a})$
and $\ ^{\shortmid }u=\{\ ^{\shortmid }u^{\alpha }=(x^{i},p_{a})\}.$ Indices
run values $i,j,...=1,2,3,4,$ where $x^{4}=ct$ is the time like coordinate
(typically we use systems of unities when the speed of light $c=1,$ and
consider necessary auxiliary constants to work with dimensionless geometric
and physical objects); and, for velocity and momentum type coordinates, $%
a,b,...=5,6,7,8,$ where $\alpha ,\beta ,...=1,2,...8$ are considered as
cumulative (total space) indices. In this work, we shall give priority to
the abstract geometric formulation of GR and MGTs as in \cite%
{misner73,vacaru18,partner06} (with various abstract left-up and low
labels), but certain coefficient formulas will be used for constructing
exact and parametric solutions. Boldface symbols are used for denoting
N-adapted geometric objects, and respective geometric constructions and "$\
^{\shortmid }$" emphasizes that the formulas involve momentum-like variables
on $T^{\ast }V.$ Tilde on geometric objects is used to emphasize that we
work in the framework of the Finsler-Cartan geometry. All necessary
definitions and notations will be provided and explained in the next
sections and appendices to this paper. In the introduction section, we
follow standard and abstract notations on geometric objects in FLH theories
elaborated in \cite{vacaru18,vmon3}.} 

\item[{2] Classical Geometric Ages (1935 - 1955):}] A series of fundamental
geometric papers were published on so-called metric noncompatible Finsler
geometry theories due to L. Berwald \cite{berwald26,berwald41} and S. -S.
Chern \cite{chern48} and others. Such Finsler geometric models are different
from the Finsler-Cartan theories and involve many conceptual and technical
difficulties in elaborating Finsler modifications of GR and standard
particle physics because of nontrivial nonmetricity fields, as we criticised
in \cite{vacaru10,vmon3,vacaru18}. For instance, it is a problem how to
define Finsler-spinors and Finsler-Dirac equations, conservation laws, and
finding exact and parametric solutions of respective locally anisotropic
modifications of the Einstein equations. In principle, a cure exists as in
the case of metric-affine gravity theories formulated on (co) tangent
Lorentz bundles following advanced geometric methods for constructing
generic off-diagonal solutions as in \cite{vacaru25b}. For FLH MGTs, this is
a task for a series of future works. \newline
\vskip0.1pt $-$ We also mention certain important works on the geometry of
nonholonomic manifolds due to G. Vr\v{a}nceanu and Z. Horak (1926-1955) \cite%
{vranceanu31,vranceanu57,horak27}; and on the global theory of N-connections
and Finsler geometry due to A. Kawaguchi (1937-1952) \cite%
{kawaguchi37,kawaguchi52} and C. Ehresmann (1955) \cite{ehresmann55}.
Summaries of results and detailed bibliography on the classical period of
Finsler geometry and first applications can be found in \cite%
{rund59,matsumoto86,vmon3,bcs2000,vacaru18}.\footnote{%
Here we note that a nonholonomic manifold is a standard manifold $M$ of
necessary smooth class endowed with a nonholonomic (equivalently,
anholonomic, or non-integrable) distribution. A nonlinear connection,
N-connection structure $\mathbf{N}=\{N_{i}^{a}\}$, used in Finsler geometry
models defined on tangent bundle $TM$ can be defined equivalently as a
Whitney sum: $\mathbf{N:}TTM=hTM\oplus vTM,$ where $h$ and $v$ are used,
respectively, for conventional horizontal and vertical splitting. All
necessary definitions and details will be provided in the next section and
Appendix.} We can consider that all FLH models consist of particular
examples of nonholonomic manifolds or (co) tangent bundle geometries defined
by respective N-connection structures.

\item[{3] Early Middle Ages (1950 - 1974):}] That was the beginning of
research on applications of Finsler geometry methods in MGTs and geometric
mechanics. For our purposes, we mention the first Finsler modification of
the Einstein equations, when instead of the Levi-Civita (LC) connection $%
\nabla $ the Cartan distinguished (d) connection $\widetilde{\mathbf{D}}$
was used (J. I. Horvath, 1950) \cite{horvath50}. We also cite the monograph 
\cite{yano73}, which contains a rigorous formulation of Finsler geometry as
an example of (co) tangent bundle geometry. It is defined by Sasaki lifts of
Hessians (vertical quadratic forms $\sim \partial ^{2}F^{2}/\partial
y^{a}\partial y^{b}$), using N-connections, to total metrics $g_{\alpha\beta
}$ on $TV.$ In our opinion, such constructions are very important because
they result in well-defined geometrically complete d-metrics on (co) tangent
Lorentz bundles and allow self-consistent extensions of metrics in GR. 
\newline
\vskip0.1pt -$3.1]$ \textit{Relativistic Lagrange mechanics as a generalized
Finsler geometry without homogeneity conditions, on tangent Lorentz bundle. }
Another very important work was that due to J. Kern (1974) \cite{kern74} who
introduced the concept of Lagrange geometry with nonlinear quadratic element 
$ds^{2}=L(x,v)$ on a tangent bundle $TM.$ This provides an alternative
geometrization of classical mechanics which is different from the standard
approach to geometric mechanics (see, for instance, \cite{deleon85}).
Finsler geometry is a particular example of homogeneous mechanics, when $%
L=F^{2}.$ Such generalizations are very important because Finsler metrics
with homogeneity conditions consists of a very special case, which is not
motivated for general nonlinear interactions and MDRs in MGTs or GR. \newline
\vskip0.1pt -$3.2]$ \textit{Relativistic Hamilton mechanics as a cotangent
Lorentz bundle model of generalized Finsler geometry and equivalent almost
(symplectic) K\"{a}hler-Lagrange/ Hamilton. }Using Legendre transforms, $%
L(x,v)\rightarrow H(x,p),$ the concept of Hamilton geometry, with nonlinear
quadratic elements $d\ ^{\shortmid }s^{2}=H(x,p)$, on cotangent bundle $%
T^{\ast }M,$ can be introduced. Such geometries can be constructed on
Lorentz manifolds with nontrivial N-connection structures and described
equivalently as almost (symplectic) K\"{a}hler-Lagrange/ Finsler (or K\"{a}%
hler-Hamilton/geometries, see \cite%
{matsumoto66,matsumoto86,brandt92,mhss2000}). Respectively, classical and
quantum geometric techniques are important for the deformation quantization
for such theories \cite{vacaru07,vacaru10a,vacaru13,biv16,vacaru18} (in
chronological form, such developments are relevant to the research described
for directions 6-7], see below).

\item[{4] Middle Ages (1959 - 1995): }] The period is characterized by many
works on Finsler and other types of locally anisotropic field theories and
gravity. It begins with the publication of the monograph \cite{rund59} (H.
Rund, 1959). It was translated into Russian by G. Asanov, who developed
(together with his postgraduate students, during 1980-1990) in the former
USSR some new directions with applications in physics of Finsler geometry 
\cite{asanov85,aa88,ap88}. Fortschritte der Physik published a series of review and original papers on the Finsler gravity theories and applications elaborated authors from the former USSR and India, \cite{apr88,asa90,asa91,bog94}. 
M. Masumoto's monograph \cite{matsumoto86} (1986)
played a very important role in the elaboration of various variants of
Finsler gravity theories and postulating certain versions of Finsler
gravitational field equations \cite%
{ikeda78,ikeda79,ikeda81,rutz93,vmon1,stavr99,vmon2,stavr04}. At that time,
certain methods of constructing solutions of locally anisotropic
gravitational field equations were not formulated, but only certain
post-Newtonian computations of possible Finsler anisotropy effects. 

\item[{5] Dark non-standard Ages (1975 - 2011): }] It was an "Orthodox"
period of GR and standard particle physics, when influential authors in
gravity and QFT \cite{bekenstein93,will01} concluded that Finsler theories
with generic anisotropies had substantial restrictions by experimental and
observational data. Their theoretical analysis had not involved nontrivial
N-connection structures, which made those conclusions quite uncertain and
ambiguous. Unfortunately, because of the mentioned works, tents of Finsler
papers (including manuscripts by this author) were rejected from Phys. Rev.
Lett. and Phys. Rev. D as "as unphysical". That situation was described in
Appendix B of the preprint variant of \cite{vacaru18}, see also references
therein. Nevertheless, many authors in the USA, Japan, Germany, UK, Canada,
Greece, Russia, Poland, Romania, R. Moldova, etc., published in other
mathematical and theoretical physics journals of number of papers on
generalized Finsler gravity theories \cite{beil93,beil03,brandt03,vacaru18}.
Here we note that important works (using the FLH geometric methods) were
performed using nonholonomic fibered structures on Lorentz manifolds, in
extra-dimension gravity, string theory, noncommutative Finsler gravity, etc. 
\cite{vacaru96a,vacaru96b,vcv00,kou08,vacaru09a,vacaru09b}. A series of important
monographs on nonholonomic manifolds and generalized Finsler gravity and
applications were published \cite%
{bejancu90,bejancu03,mhss2000,vmon1,vmon2,vmon3}.

\item[{6] Renaissance (1996 - 2011):}] The period is characterized by a new
series of works on modified dispersion relations (MDRs) and local Lorentz
invariance violations (LIVs) as attempts to solve certain important problems
in QG, string theory and other MGTs \cite{amelino96,amelino97,girelli06}.
For additional assumptions on nonholonomic geometric structures, various
classes of such theories can be formulated as (generalized) Finsler
geometries. This induced in the literature a non-correct opinion that
Finsler gravity models are theories with local anisotropies determined by
MDRs and LIVs \cite{kostelecky11,kostelecky12,kouretsis10,mavromatos11}. 
\newline
\vskip0.1pt -$6.1]$ \textit{Elaborating general methods for constructing
exact and parametric solutions in FLH MGTs.} As we mentioned above, FLH
geometric and physical models can be formulated in general self-consistent
forms on (co) tangent Lorentz bundles when the postulates of GR theory can
be extended from similar ones with Lorentz manifolds \cite%
{vacaru18,vacaru10b,vmon3}. Using nonholonomic methods as in Finsler
geometry, redefined in canonical 2+2 variables on Lorentz spacetimes, we can
construct new classes of off-diagonal exact and parametric solutions in GR
and MGTs. Extending the constructions to higher dimensions involving
nonholonomic dyadic decompositions and distortions of affine (linear)
connections, the anholonomic frame and connection deformation method (AFCDM)
was formulated \cite{vacaru00,vacaru01,vacaru02,vacaru07a,sv11,vbubuianu17}.
Perhaps, this is the most general geometric and analytic method which allows
us to decouple and integrate in certain general forms the gravitational and
matter field equations in FLH and other types of MGTs, and GR, see recent
reviews and results \cite{partner02,partner06,vacaru18}. The ansatz for
constructing such solutions is chosen as some generic off-diagonal metrics
that depend on the type of distorting relations for connections, and on
prescribed generating sources. \newline
\vskip0.1pt -$6.2]$\textit{\ Problems with the definition of Finsler-spinors
and Finsler-Dirac equations.} Another fundamental problem for elaborating
physically viable FLH is that of formulating Finsler generalizations of the
concept of Clifford structures/ spinors adapted to N-connection structures.
A related issue is that on how to formulate self-consistent variants of
Finsler-Dirac equations, and (in general) to construct FLH modifications of
the Einstein-Yang-Mills-Higgs-Dirac (EYMHD) equations. Such research
programs were performed for metric compatible Finsler connections (during
1996 - 2012) in our works \cite%
{vacaru96,vacaru96f,vacaru98,vmon1,vmon2,vv04,vmon3,vg05,vacaru05,vacaru05a,vacaru08,vacaru12}%
. We criticized \cite{vacaru10,vmon3,vacaru18} the approaches for
elaborating nonmetric Finsler gravity theories (involving Chern, Berwald and
other type metric noncompatible Finsler connections). That was because of
ambiguities with the definition of nonmetric versions of the Finsler-Dirac
equations and with the formulation of conservation laws for nonmetric FLH
deformed EYMHD systems. Such difficulties do not exist for the case of
Finsler-Cartan geometry, and for FLH generalizations with metric-compatible
Finsler connections. Recently, we discussed how to find a cure for nonmetric
MGTs using physically important solutions as in \cite{bsvv24,vacaru25b}. One
of the main purposes of this work is to show how the AFCDM can be applied
both to metric and nonmetric FLHs modified Einstein equations with general
generating sources. Proofs of the integrability of nonmetric FLH deformed
EYMHD equations will be provided in our future works. \newline
\vskip0.1pt -$6.3]$ \textit{FLH MGTs and string gravity, locally anisotropic
gauge gravity, supersymmetric and noncommutative Finsler models.} The works 
\cite%
{vacaru96a,vacaru96b,vcv00,vmon1,vacaru09a,vacaru08a,gvv15,biv16,partner02,partner06}
contain a series of original results on Finsler like (super) string and
supergravity theories formulated for extra dimension coordinates being of
velocity/ momentum type. The approach was developed for noncommutative
Finsler gravity models, Finsler-Ho\v{r}ava-Lifshitz theories, almost
Kaehler-Finsler models, deformation and gauge-like quantization of Finsler
gauge gravity theories etc. \cite%
{vacaru12a,vg95,vacaru96c,vd00,vacaru03,vacaru16a,vacaru18}. The AFCDM was
correspondingly generalized and applied to constructing physically important
solutions in such FLH\ theories. \newline
\vskip0.1pt -$6.4]$ \textit{Metric and nonmetric nonholonomic and
generalized Finsler geometric flows.} This direction of our research (see
reviews \cite{vacaru11,partner06,bnsvv24,vacaru25b}) was inspired by G.
Perelman's preprint \cite{perelman1} on the entropy of Ricci flows and
related proof of the Poincar\'{e}-Thurston conjecture, see mathematical
methods in \cite{hamilton82,monogrrf1,monogrrf2,monogrrf3}. Formulatig
Finsler generalizations (in a certain unified form of that conjecture) is
not possible because of the various types of Finsler geometries. This is
different from the case of Riemannian or K\"{a}hler geometry. Nevertheless,
we can construct abstract geometric and N-adapted variational forms of FLH
geometric flow models by using distortions of connections. The most
important motivation to formulate such Finsler-like generalizations is that
the concepts of G. Perelman W-entropy, and respective statistical and
geometric thermodynamic variables, can be computed for any type of FLH
geometries and MGTs defined by certain (generalized) Ricci scalar and Ricci
tensors. This is enough for analyzing the thermodynamic properties of large
classes of exact and parametric solutions in FLH gravity and geometric flow
theories. In general, the off-diagonal solutions are not characterized by
certain hypersurface, holographic, or duality properties. For such
solutions, the Bekenstein-Hawking paradigm \cite{bek2,haw2} is not
applicable and we have to elaborate on theories of nonholonomic and Finsler
geometric flows. Here, we cite our and co-authors' works beginning 2006, 
\cite{vacaru06dd,vv06,svnonh08,vacaru07ee,vacaru13,gheorghiuap16}. Recent
results and methods on nonholonomic or nonmetric geometric and quantum
information flows and FLH, and other types of MGTs can be found in \cite%
{bsvv24,bnsvv24,partner06}. Such works are also related to a series of
papers on Finsler-like diffusion, locally anisotropic kinetic processes,
locally anisotropic thermodynamics, etc. \cite%
{mrugala90,mrugala92,vacaru95a,vacaru00b,bv21,vacaru18}. \newline
\vskip0.1pt -$6.5]$ \textit{Certain different Renaissance directions} were
developed by other authors (we reviewed and discussed the most important
contributions till 2018 in Appendix B of the preprint variant of \cite%
{vacaru18}). Here we note that those results and methods do not allow for
the construction in general forms of certain off-diagonal solutions,
depending on the type of Finsler generating functions and prescribed
connections. They are different from our approach to FLH and can be
incomplete for applying the AFCDM. The fiber (tangent or vector ....) bundle
formalism was not applied in elaborating such theories, which put many
questions on mathematical rigorous formulation. In many cases, those models
and a few found diagonalizable solutions of Finsler-like gravitational
equations are with a trivial N-connection structure. Typically, they consist
of arbitrary lifts and deformations on $y$-variables that are not derived as
certain exact or parametric solutions involving Sasaki lifts and an
axiomatic formulation and self-consistent physical formulation.

\item[{7] Nowadays and Future (2012 - ...)}] After many observational
evidences on accelerating cosmology and validating papers of various MGTs
and DM and DE physics \cite%
{riess98,perlmutter99,sotiriou10,nojiri11,capo11,clifton12,harko14,copeland06}%
, Phys. Rev. D and other influential physical journals began to publish
papers on Finsler geometry and applications in modern physics, cosmology and
astrophysics \cite%
{kostelecky12,pfeifer11a,pfeifer11b,hohmann13,basilakos13,xi14,russell15,barcaroli15, schreck15,fuster15,fuster18}%
. New international and multi-disciplinary teams and research groups
(originating from Greece, Germany, China, Romania, R. Moldova, Spain, Italy,
India, Iran, etc.) have been organized during the last 15 years. We
discussed the main developments (till 2018) in Appendix B of the preprint
part of \cite{vacaru18}. The discovery of late-time cosmic acceleration \cite%
{riess98,perlmutter99} resulted in extensive research on modified gravity
theories (MGTs) \cite{sotiriou10,nojiri11,capo11,clifton12,harko14} and dark
energy (DE) and dark matter (DM) physics \cite{copeland06} Here we analyze
in brief some recent results on FLH MGTs and applications, which are
relevant to the purposes of this work: \newline
\vskip0.1pt -$7.1]$ \textit{Undetermined concepts of pseudo-Finsler
spacetime, causality, and problems with nonmetric Finsler gravitational
equations.} Such geometric models and possible applications were elaborated
in a series of works \cite%
{pfeifer11a,pfeifer11b,hohmann13,fuster15,fuster18,voicu23,heefer24}. This
group of authors try to impose a Finsler spacetime standard (publishing in
Phys. Rev. D and other influential journals). Former important and more
general results and methods (from the periods when Finsler gravity was
considered "unphysical") typically are not cited and not discussed, as we
described above in 5], and reviewed in \cite{vacaru18}). The main idea of
the mentioned team of authors was to extend the concept of pseudo-Riemannian
geometry in certain causal, with a well-defined theory of observations, and
for a concept of pseudo-Finsler spacetime, by postulating a version of
locally anisotropic gravitational equations. For instance, such a
pseudo-Finsler spacetime can be defined by some data $(V,L=F^{2}),$ where $V$
is a pseudo-Riemannian manifold and $F(x,v)$ is a Finsler generating
function. Additional assumptions are necessary to define certain causality
and postulates and to derive in geometric or variational forms some
Finsler-like gravitational equations, for instance, of type \cite%
{pfeifer11a,pfeifer11b}, etc. This is not enough for formulating
generalizations of Einstein gravity if we do not prescribe some classes of
nonlinear and distinguished Finsler d-connections on (co) tangent Lorentz
bundles. Well-motivated physical arguments (general theoretic/
phenomenological or observational/experimental one) are necessary for
prescribing an $F$ (why it should be homogeneous?) and certain types of N-
and d-connections, or to motivate certain very special relativity models. In 
\cite{hohmann13}, a procedure on how to make extensions of the Lorentzian
spacetime geometry "from Finsler to Cartan can vice versa" is analyzed.
Then, in \cite{voicu23}, a Birkhoff theorem was proved for Berwald-Finsler
spacetimes. Nevertheless, Berwald, Chern or other type of nonmetric
d-connections (reviewed in \cite{heefer24}) result in many difficulties in
definition of Finsler like versions of the Einstein-Dirac equations, as we
explained above in paragraph 6.2] (see also criticisms in \cite%
{vacaru10,vmon3,vacaru18}). In principle, a cure can be provided by
formulating Finsler methods involving the canonical d-connection $\widehat{%
\mathbf{D}}$ and distortions of linear connection structures as in \cite%
{vacaru25b} (necessary definitions and formulas will be provided in the next
section). \newline
{\qquad -} The variant of Finsler geometric extension of GR \cite%
{pfeifer11a,pfeifer11b} (and related models with metrizability conditions)
does not allow construction of exact or parametric solutions \cite{heefer24}
in certain general forms with nontrivial N-connection structure and using
Sasaki lifts \cite{yano73}. Such variants of Finsler-like gravitational
equations consist just of an example reflecting priorities of a group of
authors when other more general approaches were elaborated (as we motivated
in \cite{vmon3,vacaru18}). Any axiomatic approach for the Finsler gravity
theories \cite{pfeifer11a,pfeifer11b} depends on the type of geometric
structures, conditions of causality, (homogeneity conditions on $F,$ the
types of N- and d-connections) and other assumptions on deriving Finsler
modified gravitational equations. Mentioned authors do not cite and do not
apply the AFCDM described above in 6.1]. Our main idea was to elaborate on
FLH theories in general forms on (co) tangent Lorentz bundles, which allow a
straightforward generalization of the axiomatic approach to GR \cite%
{vacaru18}. Then we can apply the AFCDM for respective classes of
distortions of connections, Sasaki-type d-metrics with respective
gravitational polarizations, and effective sources, and construct very
general classes of off-diagonal solutions. After certain classes of
physically important solutions have been constructed in some general forms,
we can analyze if additional homogeneity conditions on (effective)
Finsler-Lagrange, or Cartan-Hamilton, generating functions and generating
sources can be imposed. This way, we can extract, for instance,
Fisler-Cartan, Finsler-Berwald, or other types of configurations. Certain
conditions for generating off-diagonal solutions for a base Lorentz
spacetime manifold (as in GR but, for instance, with anisotropic
polarizations of physical constants) can also be formulated. We shall
provide important examples in section \ref{sec5}. Such results can't be
obtained if we follow only the geometric approach and methods elaborated in 
\cite{pfeifer11a,pfeifer11b,hohmann13,fuster15,fuster18,voicu23,heefer24}
(only certain particular examples of vacuum solutions were found by these
authors). More than that, it is not clear how to define self-consistently
and solve the respective Finsler deformed EYMHD equations. \newline
\vskip0.1pt -$7.2]$ \textit{Barthel-Randers/-Kropina-type Finsler geometries
and cosmological implications. } A series of such applications in modern
cosmology and DM and DE physics was elaborated recently in \cite%
{hama21,boulai23,hama23}. N-adapted Sasaki-type metrics are used for
defining d-metrics on total tangent bundles. In \cite{hama21}, the
gravitational field equations are derived using Akbar-Zadeh geometric lifts
(Ricci type) tensors \cite{akbar88,akbar95}. Further developments \cite%
{boulai23,hama23} involve so-called $(\alpha ,\beta )$ metrics, Kropina
metrics, $F=\alpha ^{2}/\beta ,$ the Barthel connection etc. Such models can
be subjected to cosmological tests and may explain certain DE properties.
Nevertheless, the simplifications for respective classes of d-metrics and
d-connections (and respective osculator constructions) do not have
theoretical motivations in the conditions when the AFCDM allows us to
construct more general classes of off-diagonal locally anisotropic
cosmological solutions. Even if we begin with a variant of Barthel-Randers/
- Kropina-type configuration, the geometric evolution and off-diagonal
dynamics transform such Finsler geometric data into another type ones. The
issues on constructing Barthel-Randers/-Kropina-type deformations of EYMHD
systems and constructing respective classes of solutions have not been
analyzed in \cite{hama21,boulai23,hama23} and further developments. We
analyze how to solve such problems in the next sections. \newline
\vskip0.1pt -$7.3]$ \textit{Fibered structures in (pseudo) Finsler geometry,
causality, and variational formalism.} A series of such works was published
by research groups in mathematics and mathematical physics \cite%
{caponio20,javaloyes21,garcia22} (see also references therein). A class of
static vacuum solutions of the Einstein equations were extended for a model
of Finsler-Berwald spacetimes with vanishing Ricci scalar. For such
constructions, a concept of Finsler spacetime was considered for a smooth
finite dimensional manifold endowed with a Lorentz-Finsler metric. More
general models with the Chern connection and Akbar-Zadeh were also
considered, and the conditions for embedding such solutions to solve Rutz's
equations \cite{rutz93} were stated. Then, a metric-affine version of the
Finslerian Einstein equations was derived using the Palatini formalism \cite%
{javaloyes21}. Such equations depend on the type of N-connection and
d-connection structures. Nevertheless, it is not clear how to solve such
locally anisotropic gravitational and matter field equations in certain
general forms and to find physically important solutions. Physically viable
models of anisotropic gravity with a respective action integral for a
Finsler gravity theory can be obtained by pulling back and
Einstein-Cartan-like Lagrangian from the tangent bundle to the base manifold 
\cite{garcia22}. To find solutions (vacuum ones and for nontrivial matter
field sources) in such an approach is a difficult technical task. \newline
\vskip0.1pt -$7.4]$ \textit{Physically important black hole (BH), wormhole
(WH), and cosmological solutions.} Such new numeric and graphic methods were
elaborated in \cite{nekouee23,nekouee24,praveen25}. Typically, the
Finsler-Runders geometry with the Barthel connection is used for finding
solutions of the Finsler-like gravitational equations from \cite{xi14}
and/or using the Akbar-Zadeh definition of Ricci tensors for Finsler spaces 
\cite{akbar88,akbar95}. Such solutions are not derived in a unique and
complete form on tangent Lorentz bundles and consist of arbitrary lifts
because certain variants of Sasaki d-metrics are not defined \cite%
{vacaru00,vacaru09b,vacaru12a,vacaru18,v13,v14}. In principle, to
reconstruct such BH, WH, and locally anisotropic configurations is possible
in general and self-consistent forms as in \cite%
{v01t,v01q,vacaruplb16,partner06,bnsvv24} (we explain this in section \ref%
{sec5}). In our works, we applied the AFCDM in its
nonassociative/noncommutative, supersymmetric and other type
generalizations. For FLH theories (in general form, not only for certain
Finsler-Randers generating functions and Berwald or Barthel connections), we
consider different types of nonholonomic structures. This allows us to
construct exact and parametric off-diagonal solutions for FLH theories
defined on (co) tangent Lorentz bundles. \newline
\vskip0.1pt -$7.5]$ \textit{Finsler-Randers-Sasaki gravity and cosmology. }
A series of recent works \cite%
{triantafy20,kapsabelis21,kapsabelis23,savvopoulos23,miliaresis25}, see also 
\cite{basilakos13}, is devoted to constructing of exact and parametric
solutions (and applications in modern cosmology and astrophysics). This
direction of research is close to our research programs on FLH MGTs and
applications. Our collaboration with Prof. P. Stavrinos and other co-authors
in 2000 \cite{vmon3,vmon2} is reviewed \cite{vacaru18}. The most common
points are those that we elaborate our geometric and physical models on (co)
tangent Lorentz bundles, using Sasaki-type d-metrics, metric-compatible
Cartan or canonical d-connections, etc. The main difference is that in our
works we construct off-diagonal solutions encoding FLH deformations in most
general forms (not for any special classes of Finsler generating functions,
like Randers types encoding a covector field as a linear $y$-contributions).
Even if we begin with certain particular Finsler configurations (Randers or
another type) the geometric evolutions/ dynamics of corresponding Sasaki
type d-metrics and distortion of N- and d-connections transform
substantially the target d-metrics. To prove this is possible if we apply
the AFCDM as in \cite{vbubuianu17,bubuianu18,vacaru18,partner02} for FLH
geometric flow and MGTs. Here we also emphasize that using respective
nonholonomic frame transforms, and distortions of connections, we can
reproduce as particular cases any class of solutions in Finsler-like
theories 7.2-7.5].
\end{description}


For the directions 1-7], we cited certain series of works which resulted in
new directions and research programs involving nonholonomic and Finsler
geometry methods. The priority was given to the papers which are important
for formulating relativistic versions of FLH gravity theories on (co)
tangent Lorentz bundles involving modifications of the Einstein equations in
GR. We also provided the main references related to geometric and analytic
methods for constructing exact and parametric solutions in such theories and
performing certain quantum, supersymmetric, nonassocitive and noncommutative
generalizations of classical and QG models of FLH gravity models and
geometric and quantum information flows. Respective historical remarks and
detailed bibliography can be found in \cite{partner06,vacaru18,vmon3,vmon2}.
We emphasize that in our approach, we can extend in a natural causal form on
(co) tangent Lorentz bundles the general principle of relativity and
respective axioms of GR. This allows us to consider any types of frame and
coordinate transforms in total (and base or typical fibers and co-fibers)
spaces. Arbitrary (nonholonomic) distributions can also be introduced, and
they define respective N-connection structures, matter fields,
Finsler-Lagrange and/or Finsler-Hamilton generating functions, effective
sources, etc. Adapting the geometric constructions to respective classes of
nonholonomic distributions, we can model different types of Finsler
geometries and MGTs. Then, using nonholonomic dyadic decompositions and
distortions of d-connection structures, we can apply the AFCDM and integrate
physically important systems of nonlinear partial differential equations,
PDEs, in certain general off-diagonal forms. If necessary, we can impose
additional nonholonomic constraints, restrict the classes of generating
functions and generating sources, and extract, for instance,
LC-configurations, define metric or nonmetric Finsler-Cartan structures, or
model in equivalent forms certain Finsler-Randers models, etc. 

\subsection{The hypotheses, objectives and structure of the review}

In many mathematical works, a general S. S. Chern's definition \cite{chern96}
that the "Finsler geometry is just Riemannian geometry without the quadratic
restriction" is used. Such a definition also includes the Lagrange geometry 
\cite{kern74} if we drop the condition of homogeneity and the $y$-variables
are considered as velocity-type coordinates on a $TV,$ where $V$ is a
Lorentz 4-d manifold of signature $(+++-),$ see above sub-paragraph 3.1].
The Chern definition of d-connection and approach to Finsler geometry are
not enough for elaborating self-consistent and physically viable extensions
of GR and standard particle physics theories. We need additional assumptions
and geometric constructions. For instance, for an $L(x,v)$ or a homogeneous $%
F(x,v)$, with a nondegenerate Hessian (which allows us to construct a
so-called vertical metric, v-metrics), we can construct a geometry of
semi-sparays and use them to define certain canonical N-connection and
d-connection structures.\footnote{%
Necessary formulas \ (\ref{hessls}) and explanations are provided in next
section. In this subsection, we use only the definitions and notations which
are important for stating the hypotheses and objectives of this work.} We
can define Sasaki lifts \cite{yano73} of Hessian v-metric and N-connections
into d-metrics $\widetilde{\mathbf{g}}=\{\widetilde{\mathbf{g}}_{\alpha\beta
}(x,v)\}$ and elaborate on metric-compatible Lagrange/ Finsler - Cartan
geometry models. Such nonholonomic tangent bundle geometries are determined
by a triple of fundamental geometric objects $\ _{1}^{3}\widetilde{L}%
=(L(x,v):$ $\widetilde{\mathbf{N}}(x,v),\widetilde{\mathbf{g}}(x,v),%
\widetilde{\mathbf{D}}(x,v)\}.$ Here, the left labels "$\ _{1}^{3}$" state
the signature of base spacetime.

\vskip4pt In dual form, using conventional Legendre transforms, $%
L(x,v)\longleftrightarrow H(x,p),$ we can work on $T^{\ast }V,$ considering
a non-degenerate Hamiltonian $H(x,p)$ (in general, without homogeneity
conditions) as a Finsler-like generating function\footnote{%
see respective formulas (\ref{legendre1}) and (\ref{hesshs}) in next section}%
. This results in Hamilton-Cartan geometries determined by corresponding
triples of fundamental geometric objects, $\ _{1}^{3}\widetilde{H}=(H(x,p):\
^{\shortmid }\widetilde{\mathbf{N}}(x,p),\ ^{\shortmid }\widetilde{\mathbf{g}%
}(x,p),\ ^{\shortmid}\widetilde{\mathbf{D}}(x,p)\}$ (as described in the
above sub-paragraph 3.2], see details in \cite%
{cartan35,rund59,vmon3,mhss2000,vacaru18}). The relativistic spaces $\
_{1}^{3}\widetilde{L}$ and $\ _{1}^{3}\widetilde{H}$ are metric compatible
but involve nontrivial nonholonomically induced torsion structures.
Following another geometric principles (see details in \cite%
{berwald26,berwald41,chern48,bcs2000,vmon3,vacaru18}), we can introduce, for
instance, the Berwald (or Chern) Finsler-like d-connections, $^{B}\mathbf{D}%
(x,v)$ (or $^{C}\mathbf{D}(x,v)$). Such d-metrics which are characterized by
nontrivial nonmetricity fields $^{B}\mathbf{Q}:=\ ^{B}\mathbf{Dg\neq 0}$ (or 
$^{C}\mathbf{Q}:=\ ^{C}\mathbf{Dg\neq 0}$) and used to define respective
relativistic models for FLH geometries $\ _{1}^{3B}L$ and $\ _{1}^{3B}H,$ or 
$\ _{1}^{3C}L$ and $\ _{1}^{3C}H.$ Different types of Finsler geometries can
be mutually related by noholonomic frame transforms and distortion of
d-connection structures, for instance, in the form $^{B}\mathbf{D}=%
\widetilde{\mathbf{D}}+\ ^{B}\widetilde{\mathbf{Z}},$ where $\ ^{B}%
\widetilde{\mathbf{Z}}(x,v)$ is a respective distortion d-tensor from $%
\widetilde{\mathbf{D}}$ to $^{B}\mathbf{D}$. Corresponding distortions can
be defined in abstract geometric forms on $T^{\ast }V,$ when the geometric
d-objects are written with left labels $"\ ^{\shortmid }"\,\ $ (for
instance, $_{\shortmid }^{B}\mathbf{D}=\ ^{\shortmid }\widetilde{\mathbf{D}}%
+\ _{\shortmid }^{B}\widetilde{\mathbf{Z}}).$ We do not provide details on
definitions and explicit abstract or coordinate indexes formulas for the
Berwald and Chern models of Finsler geometry because we do not use such
particular cases of d-connections in this work. In the next sections, we
study FLH MGTs for general classes of metric and nonmetric d-connections
adapted to general N-connection structures. 

\vskip4pt Summarizing above ideas, we conclude that FHL modifications on
respective relativistic phase spaces, $TV$ and $T^{\ast }V$, of the
pseudo-Riemannian geometric objects $(hg(x),h\nabla (x))$ on a 4-d
Lorentzian manifold $V$ are defined by additional assumptions and N-adapted
data for fundamental geometric objects on (co) tangent Lorentz bundles which
are adapted to respective N-connection structures: 
\begin{equation}
\begin{array}{cccccc}
& \nearrow & \left\{ 
\begin{array}{c}
\ _{1}^{3}\widetilde{L}=(L:\widetilde{\mathbf{N}},\widetilde{\mathbf{g}},%
\widetilde{\mathbf{D}}\},\mbox{ see(\ref{hessls})}; \\ 
\ _{1}^{3B}L=(L:\ ^{B}\mathbf{N},\ ^{B}\mathbf{g},\ ^{B}\mathbf{D}\}; \\ 
\ _{1}^{3C}L=(L:\ ^{C}\mathbf{N},\ ^{C}\mathbf{g},\ ^{C}\mathbf{D}\};%
\end{array}%
\right. &  &  & \left[ \mathbf{N}(x,v),\mathbf{g}(x,v),\mathbf{D}(x,v)\right]
\\ 
(hg,h\nabla ) &  & \updownarrow \left[ 
\begin{array}{c}
\mbox{ Legendre transforms (\ref{legendre1})}; \\ 
\mbox{ almost symplectic,}%
\end{array}%
\right] & \left[ 
\begin{array}{c}
\mbox{frame} \\ 
\mbox{transforms}; \\ 
\mbox{distortions of} \\ 
\mbox{connections};%
\end{array}%
\right] & \rightarrow & \updownarrow \\ 
& \searrow & \left\{ 
\begin{array}{c}
\ _{1}^{3}\widetilde{H}=(H:\ ^{\shortmid }\widetilde{\mathbf{N}},\
^{\shortmid }\widetilde{\mathbf{g}},\ ^{\shortmid }\widetilde{\mathbf{D}}\},%
\mbox{ see(\ref{hesshs})}; \\ 
\ _{1}^{3B}H=(H:\ _{\shortmid }^{B}\mathbf{N}),\ _{\shortmid }^{B}\mathbf{g}%
,\ _{\shortmid }^{B}\mathbf{D}\}; \\ 
\ _{1}^{3C}H=(H:\ _{\shortmid }^{C}\mathbf{N},\ _{\shortmid }^{C}\mathbf{g}%
,\ _{\shortmid }^{C}\mathbf{D}\};%
\end{array}%
\right. &  &  & \left[ \ ^{\shortmid }\mathbf{N}(x,p),\ ^{\shortmid }\mathbf{%
g}(x,p),\ ^{\shortmid }\mathbf{D}(x,p)\right]%
\end{array}
\label{flh}
\end{equation}%
We argue that Chern's definition of Finsler geometry is not enough for
elaborating self-consistent models of FLH modifications of GR and other
types of MGTs defined on Lorentz manifolds. We need the concept of
nonholonomic metric-affine manifold with extensions to nonholonomic (co)
tangent bundles \cite{vmon3}, when the geometric objects are adapted to an
N-connection structure. This is different from the "not N-adapted" geometric
constructions used in \cite{hehl95,lheis23}, see critical and non-critical
remarks in \cite{vacaru10,vacaru18}. We can work respectively with the
geometric data $\left[ \mathbf{N},\mathbf{g},\mathbf{D}\right] $ or $\left[
\ ^{\shortmid }\mathbf{N},\ ^{\shortmid }\mathbf{g},\ ^{\shortmid }\mathbf{D}%
\right] $ as in (\ref{flh}) and such data can be respectively defined even
on nonholonomic (pseudo) Riemannian manifolds, Lorentz manifolds, their
tangent and cotangent bundles, or on (co) vector bundles of higher order, on
Lie algebroids etc. \cite{vmon3,vacaru05a}. 

\vskip4pt \textit{In our approach, any FLH geometry can be considered as an
example of nonholonomic geometry constructed on a (co) tangent bundle/
manifold when geometric objects are determined by prescribing respective
(Finsler-Lagrange, or Hamilton-Cartan) generating functions as in (\ref{flh})%
}. Using general nonholonomic frame transforms and distortions of linear
connections, all fundamental geometric objects (d-metrics, d-connections and
respective curvature, torsion and nonmetric d-tensors) can be defined in
abstract geometric forms, or in arbitrary frame and coordinate forms (in
particular, we can use N-adapted frames and compute respective ). Such
geometric and physical FLH models can be elaborated respectively on $TV$ and 
$T^{\ast }V,$ when the Einstein gravity theory is included for certain
nonholonomic constraints, or in some small parametric limits, in such MGTs 
\cite{hehl95,vmon1,vmon2,vmon3,lheis23,vacaru18,partner06}. The priorities
of such an approach are that: 1) The axiomatic formulation of GR is
naturally extended for $\left[ \mathbf{N},\mathbf{g},\nabla \lbrack \mathbf{%
g]}\right]$ or $\left[ \ ^{\shortmid }\mathbf{N},\ ^{\shortmid }\mathbf{g},\
^{\shortmid }\nabla \lbrack \ ^{\shortmid }\mathbf{g]}\right] ,$ where $%
\mathbf{D}=\nabla \lbrack \mathbf{g}]+\mathbf{Z}$ and $\ ^{\shortmid }%
\mathbf{D}=\ ^{\shortmid }\nabla \lbrack \ ^{\shortmid }\mathbf{g}]+\
^{\shortmid }\mathbf{Z;}$ and 2) there are metric compatible canonical
geometric objects $\widehat{\mathbf{D}}[\mathbf{g}]$ and $\widehat{\mathbf{Z}%
}[\mathbf{g}],$ respectively, $\ ^{\shortmid }\widehat{\mathbf{D}}[\mathbf{g}%
]$ and $\ ^{\shortmid }\widehat{\mathbf{Z}}[\mathbf{g}],$ which allow us to
apply the AFCDM and decouple and integrate in certain general off-diagonal
forms FLH modified relativistic geometric flow and generalized Einstein
equations; 3) all nonholonomic geometric and physical models can be
formulated in a causal and self-consistent physical forms using Lorentz
configurations determined by $(\mathbf{g},\nabla \lbrack \mathbf{g]),}$ or $%
(\ ^{\shortmid }\mathbf{g},\ ^{\shortmid }\nabla \lbrack \ ^{\shortmid }%
\mathbf{g]);}$ nonholonomic frames, distortions of connections and
generating functions and generating sources can be prescribed for metric or
nonmetric configurations as for FLH modified $f(R,T,Q,...)$ gravity theories
but extended on respective (co) tangent bundles. 

\vskip4pt One of the main tasks for any physical theory is to construct and
study physically important solutions of respective systems of fundamental
nonlinear PDEs. Such equations and their solutions typically describe
certain geometric evolution, dynamical field interactions, kinetic or
dynamical classical/ quantum processes. The AFCDM was elaborated in our
works as a geometric and analytic method for constructing generic
off-diagonal solutions in GR and various types of MGTs, see recent reviews
of results in \cite{vacaruplb16,vbubuianu17,vacaru18,partner02}. It uses
geometric methods elaborated from Finsler-like geometry theories for
conventional nonholonomic 2+2 and (2+2)+(2+2) etc., splitting, respectively,
for 4-d GR and MGTs and for 8-d FLH gravity theories. We note that a generic
off-diagonal metric can't be diagonalized by coordinate transforms in a
finite spacetime region, and the coefficients of metrics and (non) linear
connections may depend, in general, on all spacetime and phase space (fiber
or co-fiber) coordinates. 

\vskip4pt\ The main reason to study off-diagonal configurations in GR and
2+2 toy models of FLH theories (they can be defined by a formal N-connection
splitting) is that, in general, the exact or parametric solutions are
described by 6 independent coefficients (from 10 ones for a symmetric metric
tensor on a Lorentz manifold). Such configurations are different from the
cases of quasi-stationary and cosmological solutions described by a diagonal
ansatz with a maximum of 4 independent coefficients of metrics. We argue
that two additional degrees of freedom allow us to describe new models of
relativistic physics and locally anisotropic cosmology involving nonlinear
off-diagonal gravitational and (effective) matter field interactions. This
can be used, for instance, for elaborating various models of nonlinear
classical and quantum theories, locally anisotropic thermodynamics,
diffusion and kinetics, and effective DE and DM physics. We constructed
classes of generic off-diagonal solutions involving nontrivial gravitational
vacuum and pattern-creating structures (for instance, time
quasi-crystal-like), with locally anisotropic polarizations of physical
constants; or describing moving BHs, black ellipsoid/ torus (BE and BT)
configurations and nonholonomic WHs. Such results are discussed in details
in \cite%
{partner06,vacaruplb16,vbubuianu17,vacaru18,partner02,sv11,vacaru25,bsssvv25}%
. 

\vskip5pt The main goal of this paper is to formulate and analyze general
geometric methods of constructing exact and parametric solutions, physical
properties and some applications of generic off-diagonal solutions in 8-d
relativistic FLH MGTs. The AFCDM is applied for respective nonholonomic
dyadic (2+2)+(2+2) decompositions and distortions of d-connections. We shall
omit tedious proofs and cumbersome formulas presented in \cite%
{vbubuianu17,partner02,bsssvv25} (those works contain details and various
constructions with nonassociative and noncommutative variables, and other
examples of MGTs). A generalized abstract geometric and abstract index
formalism (which is similar to that in \cite{misner73}) is used to simplify
the formulas and proofs; when necessary, technical results will be
summarized in the Appendix. We provide explicit coefficient formulas
describing how physically important BH, WH, BT, and cosmological solutions
can be constructed in FLH MGTs. Then, we analyze how such solutions are
nonholonomically deformed into new classes of off-diagonal solutions in GR
and MGTs. Certain models with nontrivial gravitational vacuum, polarization
of physical constants, and deformation of horizons (in general, depending
also on velocity/ momentum variables) will be elaborated. Because the
general off-diagonal solutions in MGTs and GR do not involve, in general,
certain horizons, duality, or holographic configurations, the
Bekenstein-Hawking thermodynamic paradigm \cite{bek2,haw2} is not applicable
to characterise such solutions. We argue that other types of thermodynamic
variables defined by G. Perelman \cite{perelman1} can be used for
characterizing the fundamental properties of off-diagonal solutions
corresponding to conventional 2+2+2+... splitting \cite%
{gheorghiuap16,partner06}. Our models with nonholonomic and FLH
generalizations of the Ricci flow theory are formulated in certain forms
which allow us to apply the AFCDM for constructing new classes of
off-diagonal solutions for nonholonomic and FLH Ricci solitons. Such
nonholonomic geometric configurations are equivalent to Finsler modified
Einstein equations with nontrivial cosmological constants if certain
additional conditions are imposed. For respective classes of solutions, we
show how to define and compute the G. Perelman thermodynamic variables for
respective classes of physically important solutions. 

\vskip5pt We generalize for metric and nonmetric FLH theories the main 
\textbf{Hypotheses} (structured below as sub-paragraphs H1, H2,...)
formulated in certain particular cases, or in different forms and for other
types of nonholonomic structures, in our partner works \cite%
{vacaru18,vbubuianu17,partner06,bsssvv25,vmon3,bvvz24,bnsvv24}:

\begin{enumerate}
\item[H1.] \textsf{The axiomatic formulation of the Einstein gravity theory
on Lorentz manifolds can be extended for Finsler-like gravity theories in
physically motivated, self-consistent, and mathematically rigorous forms on
(co) tangent Lorentz bundles. In general forms, the nonholonomic geometric
formulation of such FLH MGTs is similar to metric-affine geometry, when the
fundamental geometric objects and physically important systems of nonlinear
PDEs can be adapted to a prescribed N-connection structure. The nonmetric
generalizations are performed by using distortions of Finsler-like
connections used in metric-compatible theories (such constructions may
involve, or not, nontrivial torsion structures). Explicit examples of FLH
MGTs can be derived by postulating respective classes of Lagrangian or
Hamiltonians, or certain almost symplectic structures, and by prescribing
corresponding classes of FLH generating functions. Such generating functions
and effective sources can be with homogeneity conditions, or with other
types of prescribed symmetries, N-connections, and d-connections. General
frame/coordinate transforms and distortions of connections on (co) tangent
Lorentz bundles (in particular, on base spacetime manifolds), and respective
classes of off-diagonal solutions of fundamental systems of nonlinear PDEs,
mix the types of FLH theories and may transform them into general
nonholonomic MAG theories.} 

\item[H2.] \textsf{Considering N-adapted geometric flows on a
(temperature-like) }$\tau $\textsf{-parameter involving respective
distortions of connections and Sasaky-type d-metrics, we can generalize the
theory of Ricci flows for respective nonolonomic geometric flows of FHL
theories. In relativistic forms, such FLH-flows are modelled on (co) tangent
Lorentz bundles. Considering N-adapted (3+1)+(3+1) splitting, the G.
Perelman's statistical and geometric thermodynamics can be generalized for
FLH and various types of MGTs. For such modifications, explicit formulations
and proofs of certain generalized Poincar\'{e}-Thurston conjectures are not
possible because the topological and geometric analysis constructions depend
on the types of N- and d-connections and respective distortions.
Nevertheless, formal generalizations and respective general decoupling and
off-diagonal integration properties of the modified R. Hamilton and D.
Friedan geometric flow equations can be proven for general classes of
distortion relations, generating functions and sources. This allows us to
compute the modified G. Perelman thermodynamic variables for respective
classes of FLH theories, and important classes of solutions, and speculate
on the physical importance and possible application of such theories or
classes of solutions. The dynamical FLH gravitational and (effective) matter
field equations stated by H1 can be considered as specific examples of FLH
Ricci solutions, which are derived as self-similar nonholonomic geometric
flow configurations for a fixed $\tau =\tau _{0}.$} 

\item[H3.] \textsf{The fundamental geometric and physically important
systems of nonlinear PDEs considered for H1 and H2 (FLH modified geometric
flow or gravitational and matter field equations) can be decoupled and
integrated in general off-diagonal forms using the AFCDM. Such geometric and
analytic methods allow us to generate physically important exact and
parametric solutions using respective dyadic variables with conventional
nonholonomic (2+2)+(2+2) decompositions. For respective subclasses of
generating and integration functions, we can define FLH modifications/
versions of BH, WH and BT solutions. In dual forms, with respective
nonlinear symmetries, we can elaborate on locally anisotropic cosmological
models with implications in modern DE and DM physics. Typically, the
Bekenstein-Hawking paradigm does not apply to characterizing physical
properties of such FLH theories and respective classes of solutions.
Changing the statistical thermodynamic paradigm to that for G. Perelman's
thermodynamics of Ricci flows, we can investigate in a more general form the
physical properties of such nonlinear systems under nonholonomic geometric
evolution and/or with dynamical off-diagonal interactions.} 

\item[H4.] \textsf{Certain Finsler-like MGTs are used to study new (non)
metric and nonlinear physical effects and observational features of
accelerating cosmology, DE and DM theories. This involves both conceptual
and technical problems if we consider FLH non-standard models of particle
physics with nonmetricity fields. In such cases, to define in a unique form
Finsler variants of the Dirac operator and respective FLH modifications of
the Einstein-Dirac, ED, equations are not possible. For the canonical
d-connection, which is metric compatible and can be related via frame
transforms to the Cartan-Finsler d-connection, the constructions for
Clifford structures are similar to those in GR, but result in ambiguities,
for instance, for the Chern-Finsler d-connection (because of nonmetricity).
In our approach, we can define FLH-modifications of certain well-defined
EYMHD systems possessing physically motivated solutions in nonholonomic
dyadic variables and canonically adapted (non) linear connection structures.
For such configurations, the AFCDM allows us to decouple and integrate in a
certain general off-diagonal form and generate both metric-compatible and
nonmetric solutions. Various classes of solutions in GR and MGTs defined on
Lorentz spacetime manifolds can be extracted for certain particular
parameterizations of general generation and integration data. Using the
formalism of N-adapted distortion of linear connections nonholonomic dyadic
variables, and canonically adapted (non) linear connection structures, the
AFCDM allows decoupling and integrating in a certain general off-diagonal
form the nonmetric FLH geometric evolution flow equations and related
dynamical (non) metric EYMHD equations. All new classes of off-diagonal
solutions for nonmetric FLH geometric flow equations, with nonholonomic
Ricci soliton configurations for EYMHD systems, etc., are characterized by
generalized G. Perelman thermodynamic variables.} 

\item[H5.] \textsf{FLH geometric flow and MGTs can be quantized by using
nonholonomic generalized methods of deformation quantization, generalized
Batalin-Fradkin-Vilkovisky quantization methods, etc. Generalized QG
theories with local anisotropy and modified dispersion relations, for FLH
models, can be elaborated in certain compatible forms with the (super)
string/ M-theories. In general, they involve nonassociative and
noncommutative R-flux modifications and nontrivial Lie-algebroid structures
and gerbes. Such constructions result in new types of nonmetric FLH
geometric and quantum flow information theories, and locally anisotropic
kinetic, diffusion and geometric thermodynamic models.}
\end{enumerate}


\vskip5pt The goals and structure of the article motivate the hypotheses
H1-H3 (further partner works will be devoted to H4 and H5; when \cite%
{vacaru96,vacaru96c,vacaru01,vd00,vacaru01,vacaru07,vacaru08,vacaru12,gvv15,bsvv24,vacaru25,vacaru25b}
contain necessary geometric methods and a number of preliminary results) as
follow:

In section \ref{sec2}, we outline the nonholonomic geometry of metric and
nonmetric FLH MGTs on (co) tangent Lorentz bundles by formulating an
abstract geometric approach with metrics and affine connections adapted to
N-connection structures (the \textbf{first aim} of this paper, i.e.
objective Obj1). This allows us to generalize the axiomatic principles of GR
for FLH theories. The \textbf{second aim} (Obj2) of this work is to
formulate such MGTs in a generalized form, when using distortions of
connections and respective effective sources we can construct nonholonomic
geometric and physical models with nontrivial torsion and nonmetricity
fields. Such theories include Finsler-like generalizations of $f(R,T,Q,...)$
gravity to physical d-objects depending on both spacetime and
velocity/momentum type variables. We also discuss how certain examples of
Finsler gravity theories and physically important (in many cases
undetermined or incomplete) solutions studied recently in the physical
literature can be included as particular cases in our approach. 

Two important goals of this article are stated in section \ref{sec3}. We
formulate the theory of nonmetric geometric flows of FLH systems in the
first subsection (consisting of the \textbf{third aim, }Obj3). Respective
nonmetric deformed G. Perelman functionals and thermodynamic variables are
postulated, which allows us to derive both in abstract geometric or
N-adapted variational forms, the metric and nonmetric geometric flow
equations. This consists of the \textbf{forth aim} (Obj4) of the paper.
Here, we emphasize that Perelman's thermodynamics is different and more
general than the constructions based on the Bekenstein-Hawking paradigm. The
(non) metric geometric flow approach is important for characterizing
off-diagonal solutions in GR and various types of MGTs, including also the
FLH models. 

In section \ref{sec4}, we provide details and proofs that nonmetric
geometric flow equations of the FLH modified Einstein equations can be
decoupled and integrated in general off-diagonal forms. This is possible for
various classes of distortions of Finsler-like connections and FLH theories,
consisting of particular examples of nonholonomic Finsler-Ricci solitons
(the \textbf{fifth aim, }Obj5). We consider examples for toy 2+2 FL and FH
models, then extend the constructions for nonholonomic (2+2)+(2+2) splitting
using velocity/ momentum-like variables and speculate how the LC
configurations or other type Finsler-like structures can be extracted in
general forms. As the \textbf{sixth aim} (Obj6), we formulate a method of
geometric and analytic computations of G. Perelman thermodynamic variables
for nonmetric geometric flows of FLH systems, using new types of nonlinear
symmetries relating generating functions and generating sources to certain
families of effective running cosmological constants. We discuss some
examples of how solutions for Finsler gravity theories constructed by other
authors can be introduced in our nonholonomic MAG approach. This is related
to the \textbf{seventh aim }(Obj7) to study how the nonholonomic geometric
flows and off-diagonal interactions can transform a class of FLH theories
into another class, and how this is described by respective sub-classes of
general solutions. 

The section \ref{sec5} is devoted to constructing explicit classes of
physically important exact and parametric generic off-diagonal solutions for
nonholonomic geometric running of FLH theories. The \textbf{eighth aim}
(Obj8) is to apply the AFCDM to generate quasi-stationary solutions
describing off-diagonal nonmetric FLH of regular BH solutions in GR to
ellipsoidal configurations, in particular, for BE solutions. We state the
conditions for generating BH, WH, BT and solitonic wave configurations in
FLH gravity. The second class of solutions (with different variables and
nonlinear symmetries which are defined in certain time dual forms to
quasi-stationary configurations; i.e. the \textbf{ninth aim,} Obj9) is
constructed for nonmetric cosmological configurations encoding FLH
deformations of the spherical symmetric cosmological systems in GR. As
explicit examples, we analyze accelerating models with spheroidal symmetry
and voids, with solitonic wave evolution, and small parametric deformations
of the FRW metrics to (co) tangent Lorentz bundle configurations. In brief,
we speculate on possible applications of such results and methods in DE and
DM physics. 

In section \ref{sec6}, we conclude the results of this work and discuss
further perspectives. Appendix A contains the technical results and the
necessary proofs for the AFCDM generalized for FLH geometric flow and MGTs.
We summarize this method in Tables 1-13 in Appendix B, which can be used for
constructing quasi-stationary and locally anisotropic cosmological solutions
in Finsler-like gravity theories. Finally, we note that this review also has
a pedagogical character for introducing and abstract geometric formalism of
FLH theories. So, we present necessary details for formulations and proofs
of formulas on nonholonomic phase spaces modelled as (co) tangent Lorentz
bundles (in general, endowed with dyadic splitting, distortions of
connections) to familiarize the reader with the formalism. Then, when we
consider the provided calculations and examples with conventional indices,
coordinate dependencies, etc., are enough for understanding the fundamental
ideas and geometric constructions, we omit details and provide the formulas,
for instance, only for FL structures and respective nonholonomic Ricci
soliton configurations. 

\section{Nonholonomic metric or nonmetric geometry and FLH-modified Einstein
gravity}

\label{sec2}In this section, we show how the Einstein gravity and MGTs
formulated (effectively) on Lorentz manifolds can be extended on (co)
tangent Lorentz bundles (equivalently, phase spaces) as relativistic FLH
theories. In general, such Finsler-like gravity theories involve nontrivial
torsion and nonmetricity fields. The gravitational field equations in metric
and nonmetric FLH MGTs are introduced in the most general forms with
effective sources determined by distortions of linear connection structures
and matter field configurations lifted on total phase spaces involving both
spacetime and velocity/momentum type coordinates.

\vskip4pt We provide six important motivations to elaborate and study
physical implications of FLH MGTs theories:

\begin{enumerate}
\item MDRs and models of locally anisotropic media can be described by
nonholonomic or generalized Finsler structures on phase spaces \cite%
{amelino96,amelino97,girelli06,vacaru12a,barcaroli15,bubuianu18,vacaru18}.

\item A subclass of Finsler geometries and physical models involves LIVs 
\cite{kostelecky11,kostelecky12,mavromatos11,russell15}, which could define
a new physics if such effects were be discovered.

\item In MGTs, we can define N-connection structures determined by
semi-spray equations, i.e. nonlinear geodesic equations. Such equations are
equivalent to the Euler-Lagrange and/or Hamilton equations \cite%
{kern74,bcs2000,deleon85}. This results into alternative formulations of
geometric mechanics, which may involve supersymmetric/ nonassociative and
other types nonholonomic variables \cite{vmon1,vmon3}.

\item Non-geometric star product R-flux deformations in string theory \cite%
{blum12,blum16,asch17} can be geometrized in nonassociative and nonholonomic
forms on 8-d phase spaces involving complex or real momentum variables \cite%
{partner06,partner02,bsssvv25}. Star product R-flux deformations can be also
characterized by MDRs encoding nonassociative and noncommutative data. We
also can elaborate on (super) string and supergravity theories when extra
dimension coordinates are velocity/momentum type \cite%
{vacaru96a,vacaru96b,vcv00,vmon3,vacaru18}. To prove general decoupling and
integration properties of physically important systems of nonlinear PDEs in
such theories, we have to consider nonholonomic dyadic decompositions and
certain classes of generalized metrics and linear connections adapted to
N-connection structures. Such geometric methods were elaborated in Finsler
geometry.

\item New classes of generic off-diagonal solutions with effective sources
in MGTs on (nonassociative, noncommutative, supersymetric, etc.) phase
spaces, or nonholonomic Lorentz manifolds, are characterized by G. Perelman
statistical and geometric thermodynamic models \cite{perelman1} which are
generalized for nonassociative Finsler-Lagrange-Hamilton geometric flow and
nonholonomic Ricci soliton theories \cite%
{vacaru06dd,vv06,svnonh08,vacaru07ee,vacaru13,gheorghiuap16}. We cite \cite%
{bsvv24,bnsvv24,partner06} for recent results and methods (which were
originally formulated in Finsler geometry) on nonholonomic or nonmetric
geometric and quantum information flows and FLH, and other types of MGTs.

\item We elaborated the AFCDM as a geometric and analytic method for
constructing generic off-diagonal solutions in GR and various types of MGTs
by using nonholonomic dyadic frames and certain canonical d-connections
defined as in Finsler geometry, see recent reviews of results in \cite%
{vacaruplb16,vbubuianu17,vacaru18,partner02,bsssvv25}. Conventional
nonholonomic 2+2 and (2+2)+(2+2) etc. splitting (respectively, for 4-d GR
and MGTs and for 8-d FLH gravity theories) were used. New classes of
Finsler-like solutions were constructed for elaborating various models of
nonlinear classical and quantum theories, locally anisotropic
thermodynamics, diffusion and kinetics, and effective DE and DM physics.
\end{enumerate}


\subsection{E. Cartan's approach, canonical d-connections, and relativistic
FLH geometry}

\subsubsection{Finsler-Cartan geometry}

The first example of Finsler metrics was considered by B. Riemann in 1954
his famous habilitation thesis \cite{riemann1854}, when the Riemannian
geometry was constructed as a particular example of curved space geometry
defined by a quadratic line element $ds^{2}=g_{ij}(x)v^{i}v^{j},$ for $%
v^{i}\sim dx^{i}.$ More general cases with 
\begin{equation}
ds^{2}=F^{2}(x,v),  \label{finslqlel}
\end{equation}%
where $F(x,v)$ is a general nonlinear function (or a functional on some
other tensor, etc. functions, depending on $(x,v))$ had not been studied in
that work. In modern literature, a \textit{generating function} $F(x,v)$ is
also called a \textit{Finsler metric }which results in certain ambiguities
because a metric tensor is not just a line element and we need additional
assumptions to define some (symmetric or nonsymmetric) tensors.\footnote{%
We show below how to define so-called vertical metric structures (Hessians)
and Sasaki type lifts d-metrics, see respective formulas (\ref{hessls}) and (%
\ref{cdms8}).} In standard Finsler models, certain homogeneity conditions, $%
F(x,\lambda v)=|\lambda |F(x,v),$ with the module taken for a real constant $%
\lambda $ (and other geometric or physical conditions) are considered.
Contrary to the Riemannian geometry, which is completely defined by a metric
tensor $g_{ij}(x),$ to prescribe a $F(x,v)$ is not enough for constructing
geometric or physical models in a rigorous and complete form. Additional
assumptions on the properties of geometric objects were used for
constructing different types of Finsler geometries. 

The monograph \cite{cartan35} contains in coordinate form the definition and
first examples of some N-connection and d-connection structures which define
in a complete mathematical form the so-called Finsler-Cartan geometry. Here
we note that E. Cartan introduced the term Finsler geometry for a class of
non-Riemannian spaces defined and studied originally in \cite{finsler18},
see historical remarks and details in \cite%
{rund59,matsumoto86,vmon3,vacaru18,bubuianu18,bsssvv25}. The monographs \cite%
{cartan38,cartan63} (see references therein) contain in local coordinate
forms all necessary concepts and formulas which are necessary for
constructing geometric models of gravity and particle field theories on
curved pseudo-Riemannian spacetimes and Finsler geometries. Here we note
certain definitions of fiber bundles, N- and d-connections, spinors, moving
frames, etc. The first example of Finsler-Cartan modified Einstein equations
was provided in \cite{horvath50} using the Cartan d-connection (which is
metric compatible) for Finsler geometry. That was not yet a generalization
of the Einstein gravity theory because $v=\{$ $v^{i}\}$ where treated as
velocity-type variables on a Riemannian manifold $M$ with local Euclidean
signature, and the concept of tangent Lorentz bundle was not involved. Here
we note that certain influential authors \cite{bcs2000} use the term
Riemann-Finsler geometry because B. Riemann considered Finsler-like
nonlinear quadratic elements many years before Finsler's thesis. In our
opinion, this is not quite correct because self-consistent and physically
important Finsler-like geometries can be formulated only by introducing
different types of Finsler linear connection structures adapted to
N-connections, for instance, due to Chern, Berwald, Hashiguchi, etc. For
reviews of definitions and necessary coefficient formulas, we cite \cite%
{rund59,matsumoto86,heefer24}) and distortions of connections from (\ref{flh}%
). Such d-connections may be metric compatible or not compatible, which has
different implications for constructing physical theories. 

In geometric mechanics, a Finsler generating function $F(x,v)$ can be used
as an effective Lagrangian $L(x.v):=F^{2}(x,v)$ and applied, for instance,
for constructing certain models of reonomic mechanics. In such theories,
corresponding homogeneity and nonholonomic constraints are introduced. In
equivalent form, the terms nonholonomic and anholonomic, i.e. non-integrable
constraints, are used in literature on Finsler geometry (by analogy to
nonholonomic mechanics \cite{deleon85}). In a more general context, Finsler
geometries can be considered as certain examples of nonholonomic geometries
originally formulated and studied in detail in \cite%
{horvath50,vranceanu31,vranceanu57,bejancu90,bejancu03,vmon1,vmon2}. The
idea to drop the conditions of homogeneity and consider a Finsler-like
generalization of Lagrange mechanics on tangent bundle $TM$, by introducing
the concept of \textit{Lagrange geometry, also called Lagrange-Finsler
geometry}, is due to J. Kern \cite{kern74}. In a rigorous form, such
geometries can be defined on the tangent bundle $TM$ \ and extended on
higher order tangent bundles like $TTM,TTTM$ etc., see \cite{mhss2000,vmon3}%
. We cite R. Miron and his co-authors' works with many reserves and critical
remarks on hidden plagiarism, intellectual slavery, and other ethical,
political and human rights problems during N. Ceau\c{s}escu's dictatorial
regime provided in appendix B of \cite{vacaru18}. 

We emphasize that the geometrization of mechanics and classical field
theories using generalized Finsler d-objects is very different from another
approaches \cite{deleon85}. Certain authors \cite%
{beem70,asanov85,aa88,ap88,pfeifer11a,pfeifer11b} constructed Finsler-like
theories by dropping the conditions of Euclidean signature for certain
effective or background metrics on base manifolds, typical fibers and
respective total bundles, but had not elaborated a complete geometric
formalism for relativistic Finsler spaces. In our works \cite%
{vacaru96a,vacaru96b,vmon1,vmon3,vacaru18}, we concluded that we have to
consider certain base Lorentz manifolds $V$ of necessary dimension and
smooth class, and respective (co) tangent bundles on such manifolds, if we
wont to study Finsler-Lagrange, FL, modifications of the GR and modern MGTs
in a causal form on $TV$, $TTV,\,\ $ etc. This allows us to elaborate in
covariant form on physically viable theories with local anisotropy but in
certain relativistic forms and causality structure which are similar to
those in special relativity and GR theories and in standard particle
physics. The Finsler-like generalizations distort such geometric
constructions (for instance, for different connection structures), but
certain effective metrics of signature $(+,+,...,+,-)$ are supposed to be
defined on typical base and fiber spaces. This way we avoid the problem of
constructing analyzing properties of a plethora of Finsler-generalized
causal structures, which depend on the type of Finsler geometry and
fundamental physical equations are postulated for respective theories. In
our approach, the main assumption is to begin our research with certain
well-defined geometric and physical models (of relativistic mechanics,
gravity, standard particle physics, etc.) on a (co) tangent Lorentz bundle,
or Lorentz manifold. Then, we can consider additional nonholonomic
transforms, nonholonomic distributions and distortions of connections which
result in certain generalized metric or nonmetric FL theories, or their dual
FH models. 

\subsubsection{N-connection structures for FL and FH spaces}

The GR theory is formulated on a Lorentz manifold $V[g,\nabla ]$ of
dimension 4, $\dim V=4$ (for the higher dimension theories, we can consider $%
\dim V\geq 4,\,\ $ when the manifolds are of necessary smooth class and
signature). We follow the conventions from \cite%
{vacaru18,vmon3,partner06,partner02,bsssvv25,bubuianu18} as (co) tangent
Lorentz bundle generalizations of the geometric methods from \cite%
{misner73,hawking73,wald82,kramer03}. A $g=\{g_{ij}(x)\}$ a four
dimensional, 4-d, pseudo-Riemannian manifold $V=\ _{1}^{3}V$ of signature $%
(+++-).$ Naturally and minimally, we can elaborate on Finsler-like
modifications of GR and MGTs and of the standard particle physics theories
on conventional phase spaces modelled as tangent and cotangent Lorentz
bundles, $TV$ and $T^{\ast }V.$\footnote{%
We note that, the geometric constructions can be performed similarly for
higher dimensions (with extra dimension pseudo-Riemannian coordinates as in
Kaluza-Klein or string gravity theories) and other types of MGTs. We can
elaborate also on $f(R)$ or $f(Q)$ gravity \cite%
{sotiriou10,nojiri11,capo11,clifton12,harko14,hehl95,lheis23} effectively
modelled on 4-d Lorentz manifolds. In this work, we assume that the GR can
be modeled only on a 4-d base spacetime manifold $V$ when the fundamental
geometric structures are subjected to Finsler-like nonholonomic
generalizations for MGTs formulated on certain total (co) tangent Lorentz
bundles.} 

A relativistic 4-d model of \textit{Lagrange geometry} \cite{kern74}, i.e. a 
\textit{Lagrange space} $\ _{1}^{3}L=(TV,L(x,v))$, consists a generalization
of the concept of Finsler space by dropping the homogeneity conditions for $%
F $ $(x,v).$ We call it as a Finsler-Lagrange, FL, geometry if it is
completed (see below) with additional Finsler-like N-connection and
d-connection structures. This is a 8-d phase space $\mathcal{\tilde{M}}$
with velocity type conventional coordinates $y\approx v$ is defined by a
fundamental function (equivalently, generating function) $L(x,v)$ subjected
to the conditions:1) $TV\ni (x,v)\rightarrow L(x,v)\in \mathbb{R},$ which is
a real valued function, differentiable on $\widetilde{TV}:=TV/\{0\},$ for $%
\{0\}$ being the null section of $TV,$ and continuous on the null section of 
$\pi :TV\rightarrow V;$ 2) Such a Lagrangian model is regular if the Hessian
(v-metric) 
\begin{equation}
\widetilde{g}_{ab}(x,v):=\frac{1}{2}\frac{\partial ^{2}L}{\partial
v^{a}\partial v^{b}}  \label{hessls}
\end{equation}%
is non-degenerate, i.e. $\det |\widetilde{g}_{ab}|\neq 0,$ and of constant
signature. In our works, we distinguish the so-called horizontal, h, space
coordinate indices ($x^{i},$ for $i,j,..=1,2,3,4$) from the vertical, v,
ones $(v^{a},$ for $a,b,...=5,6,7,8),$ when additional conventions will be
considered for certain lifts of geometric objects from the base to the total
spaces. 

In a similar form, a 4-d relativistic model of \textit{Hamilton geometry
(space, it includes generalizations of dual Finsler geometries)} $\
_{1}^{3}H=(T^{\ast }V,H(x,p))$ can be constructed for a fundamental function
(equivalently, generating Hamilton function) on a Lorentz manifold $V.$ We
call it a Finsler-Hamilton, FH, geometry if it is completed (see below) with
additional Finsler-like N-connection and d-connection structures. Such a 8-d
phase space$\ ^{\shortmid }\mathcal{\tilde{M}}$ with conventional
momentum-like coordinates $p=\{p_{a}\}$ is defined by a Hamiltonian $H(x,p)$
subjected to the conditions: 1) $T^{\ast }V\ni (x,p)\rightarrow H(x,p)\in 
\mathbb{R}$ is defined by a real valued function being differentiable on $%
\widetilde{T^{\ast }V}:=T^{\ast }V/\{0^{\ast }\},$ for $\{0^{\ast }\}$ being
the null section of $T^{\ast }V,$ and continuous on the null section of $\pi
^{\ast }:\ T^{\ast }V\rightarrow V;$ 2) such a model is regular if the
Hessian (cv-metric) 
\begin{equation}
\ \ ^{\shortmid }\widetilde{g}^{ab}(x,p):=\frac{1}{2}\frac{\partial ^{2}H}{%
\partial p_{a}\partial p_{b}}  \label{hesshs}
\end{equation}%
is non-degenerate, i.e. $\det |\ ^{\shortmid }\widetilde{g}^{ab}|\neq 0,$
and of constant signature.\footnote{%
We follow such additional conventions: A v-metric $\widetilde{g}_{ab}$ and a 
$\ $c-metric$\ ^{\shortmid }\widetilde{g}^{ab}$ are labeled by a tilde "%
\symbol{126}" to emphasize that such conventional v--metrics and c-metrics
are defined canonically by respective Lagrange and Hamilton generating
functions. General frame/ coordinate transforms on $TV$ and/or $T^{\ast }V$
allow us to express any "tilde" Hessian in a general form, respectively as a
vertical metric (v-metric), $g_{ab}(x,y),$ and/or co-vertical metric
(cv-metric), $\ ^{\shortmid }g^{ab}(x,p).$ We can also works with inverse
frame/ coordinate transforms by prescribing any v-metric (cv-metric). In
general, a $g_{ab}$ is different from the inverse of $\ ^{\shortmid }g^{ab},$
i.e. from $\ ^{\shortmid }g_{ab}$. In explicit form, certain Lagrange and/or
Hamilton models on corresponding $\mathcal{M}$ and/or $\ ^{\shortmid }%
\mathcal{M}$ can be always constructed by prescribing certain generating
functions $L(x,y)$ and/or $H(x,p).$ We shall omit tildes on geometrical/
physical objects if certain formulas hold true in general forms and not only
for some canonical structures and if that will not result in ambiguities.} 

We can elaborate on generalized Finsler theories with
Finsler-Lagrange-Hamilton, FLH, geometric objects defined on relativistic
phase spaces defined by certain general generating functions $L(x,v)$ on $TV$
or $H(x,p)$ on $T^{\ast }V.$ For simplicity, we can consider only regular
configurations for nonzero Hessians, even though such geometries can be
formulated for singular ones as in geometric mechanics involving
nonholonomic constraints on generalized coordinates \cite%
{vranceanu57,bejancu03}. Here, we note that there are Legendre transforms $%
L\rightarrow H,$ with $H(x,p):=p_{a}y^{a}-L(x,y)$ and $y^{a}$ determining
solutions of the equations $p_{a}=\partial L(x,y)/\partial y^{a}.$ In a
similar manner, the inverse Legendre transforms can be introduced, $%
H\rightarrow L,$ for 
\begin{equation}
L(x,y):=p_{a}y^{a}-H(x,p)  \label{legendre1}
\end{equation}%
and $p_{a}$ determining solutions of the equations $y^{a}=\partial
H(x,p)/\partial p_{a}.$ For regular configurations, we can work equivalently
both with Largange and/or Hamilton spaces. Here we emphasize that, in
general, the Lagrange mechanics is not equivalent to the Hamilton mechanics,
see details in \cite{deleon85,mhss2000,vmon3}. So, corresponding
Finsler-like (more exactly, Finsler-Cartan) extensions of GR are different,
for instance, for different types of formulated almost symplectic structures
and used N- and d-connection structures as we explain in \cite{vacaru18}. 

Using the antisymmetric product $\wedge $ on a Hamiltonian phase space $%
\widetilde{H}$, we can define a canonical symplectic structure $\theta
:=dp_{i}\wedge dx^{i}$ and a unique vector filed $\ \widetilde{X}_{H}:=\frac{%
\partial \widetilde{H}}{\partial p_{i}}\frac{\partial }{\partial x^{i}}-%
\frac{\partial \widetilde{H}}{\partial x^{i}}\frac{\partial }{\partial p_{i}}
$ determined by the equation $i_{\widetilde{X}_{H}}\theta =-d\widetilde{H},$
where $i_{\widetilde{X}_{H}}$ denotes the interior produce defined by $%
\widetilde{X}_{H}.$ This allows to perform a Hamilton calculus for any
functions $\ ^{1}f(x,p)$ and $\ ^{2}f(x,p)$ and respective canonical Poisson
structure $\{\ ^{1}f,\ ^{2}f\}:=\theta (\widetilde{X}_{^{1}f},\widetilde{X}%
_{^{2}f}).$ For instance, we can construct a structure which is related to
respective so-called Hamilton-Jacobi configurations: Let us consider a
regular curve $c(\zeta ),$ when $c:\zeta \in \lbrack 0,1]\rightarrow
x^{i}(\zeta )\subset U\subset V,$ for a real parameter $\zeta .$ It can be
lifted to $\pi ^{-1}(U)\subset \widetilde{TV}$ defining a curve in the total
space, when $\widetilde{c}(\zeta ):\zeta \in \lbrack 0,1]\rightarrow
\left(x^{i}(\zeta ),y^{i}(\zeta )=dx^{i}/d\zeta \right) $ with a
non-vanishing v-vector field $dx^{i}/d\zeta .$ The canonical Hamilton-Jacobi
equations for relativistic FH spaces are defined: 
\begin{equation*}
\frac{dx^{i}}{d\zeta }=\{\widetilde{H},x^{i}\}\mbox{ and }\frac{dp_{a}}{%
d\zeta }=\{\widetilde{H},p_{a}\}.
\end{equation*}

We can elaborate equivalent Lagrange and Hamilton models of relativistic
phase spaces formulated as $L$-dual effective phase spaces $\widetilde{H}%
^{3,1}$ and $\widetilde{L}^{3,1}$ described by fundamental generating
functions $\widetilde{H}$ and $\widetilde{L}$ which satisfy respectively
such important conditions: The Hamilton-Jacobi equations can be written as 
\begin{equation*}
\frac{dx^{i}}{d\zeta }=\frac{\partial \widetilde{H}}{\partial p_{i}}%
\mbox{
and }\frac{dp_{i}}{d\zeta }=-\frac{\partial \widetilde{H}}{\partial x^{i}},
\end{equation*}%
being equivalent to the Euler-Lagrange equations, 
\begin{equation}
\frac{d}{d\zeta }\frac{\partial \widetilde{L}}{\partial y^{i}}-\frac{%
\partial \widetilde{L}}{\partial x^{i}}=0.  \label{lagreiler}
\end{equation}%
The equations (\ref{lagreiler}), in their turn, are equivalent to the
nonlinear geodesic (semi-spray) equations 
\begin{equation}
\frac{d^{2}x^{i}}{d\zeta ^{2}}+2\widetilde{G}^{i}(x,p)=0,\mbox{ for }%
\widetilde{G}^{i}=\frac{1}{2}\widetilde{g}^{ij}(\frac{\partial ^{2}%
\widetilde{L}}{\partial y^{i}}y^{k}-\frac{\partial \widetilde{L}}{\partial
x^{i}}),  \label{ngeqf}
\end{equation}%
with $\widetilde{g}^{ij}$ being inverse to $\widetilde{g}_{ij}$ (\ref{hessls}%
). These equations state that point-like probing particles move in
corresponding phase spaces not along usual geodesics as on Lorentz
manifolds, but follow some nonlinear geodesic equations. 

Using $\widetilde{G}^{i}$ from (\ref{ngeqf}), we can define a canonical
N--connection in $L$--dual form following formulas 
\begin{equation}
\ \ \ \ ^{\shortmid }\widetilde{\mathbf{N}}=\left\{ \ ^{\shortmid }%
\widetilde{N}_{ij}:=\frac{1}{2}\left[ \{\ \ ^{\shortmid }\widetilde{g}_{ij},%
\widetilde{H}\}-\frac{\partial ^{2}\widetilde{H}}{\partial p_{k}\partial
x^{i}}\ ^{\shortmid }\widetilde{g}_{jk}-\frac{\partial ^{2}\widetilde{H}}{%
\partial p_{k}\partial x^{j}}\ ^{\shortmid }\widetilde{g}_{ik}\right]
\right\} \mbox{ and }\widetilde{\mathbf{N}}=\left\{ \widetilde{N}_{i}^{a}:=%
\frac{\partial \widetilde{G}}{\partial y^{i}}\right\} .  \label{cartnc}
\end{equation}%
This is a corn-stone geometric object for defining Finsler-like models on $%
TV $ or $TV.$ For the Finsler geometry with $L=F^{2},$ the N-connections
were introduced by E. Cartan in \cite{cartan35}. In rigorous mathematical
form, N-connections can be defined as certain C. Ehressmann connection \cite%
{ehresmann55}. For our purposes, we can consider them as nonholonomic
distributions defining conventional h and v, or c, distributions: 
\begin{equation}
\mathbf{\tilde{N}}(u):TTV=hTV\oplus vTV\mbox{ or }\ ^{\shortmid }\mathbf{%
\tilde{N}}(\ ^{\shortmid }u):TT^{\ast }V=hT^{\ast }V\oplus cT^{\ast }V,
\label{ncondistr}
\end{equation}%
where $\oplus $ is a direct (Whitney) sum. They define corresponding
nonholonomic (4+4) splitting of (co) tangent bundles. In local coordinates $%
u=$ $(x,v)$ or $\ ^{\shortmid }u=(x,p),$ N-connection are defined in
respective coefficient forms as $\mathbf{\tilde{N}}(u)=\{\tilde{N}%
_{i}^{a}(x,v)\}$ or $\ ^{\shortmid }\mathbf{\tilde{N}}(\ ^{\shortmid }u)=\{\
^{\shortmid }\tilde{N}_{i}^{a}(x,p)\}.$ 

\subsubsection{Canonical d-metrics in relativistic FL and FH geometry}

Any "tilde" N-connection $\ ^{\shortmid }\mathbf{\tilde{N}}$ allows us to
define respective systems of canonical N--adapted (co) frames, for instance,
when 
\begin{eqnarray}
\ ^{\shortmid }\widetilde{\mathbf{e}}_{\alpha } &=&(\ ^{\shortmid }%
\widetilde{\mathbf{e}}_{i}=\frac{\partial }{\partial x^{i}}-\ ^{\shortmid }%
\widetilde{N}_{ia}(x,p)\frac{\partial }{\partial p_{a}},\ ^{\shortmid }e^{b}=%
\frac{\partial }{\partial p_{b}}),\mbox{ on }T^{\ast }V;  \label{ccnadaph} \\
\ \ ^{\shortmid }\widetilde{\mathbf{e}}^{\alpha } &=&(\ ^{\shortmid
}e^{i}=dx^{i},\ ^{\shortmid }\mathbf{e}_{a}=dp_{a}+\ ^{\shortmid }\widetilde{%
N}_{ia}(x,p)dx^{i})\mbox{ on }(T^{\ast }V)^{\ast }.  \notag
\end{eqnarray}%
Generalizing the abstract geometric formalism from \cite{misner73}, see
details for phase spaces in \cite{vacaru18,vmon1,partner06,bsssvv25}, the
formulas (\ref{cartnc}) and (\ref{ccnadaph}) are defined respectively in
dual form by $L$ and $\mathbf{\tilde{N},}$ when 
\begin{eqnarray}
\ \widetilde{\mathbf{e}}_{\alpha } &=&(\widetilde{\mathbf{e}}_{i}=\frac{%
\partial }{\partial x^{i}}-\ \widetilde{N}_{i}^{a}(x,v)\frac{\partial }{%
\partial v^{a}},e_{b}=\frac{\partial }{\partial v^{b}}),\mbox{ on }TV;
\label{ccnadapl} \\
\ \widetilde{\mathbf{e}}^{\alpha } &=&(e^{i}=dx^{i},\mathbf{e}^{a}=dv^{a}+%
\widetilde{N}_{i}^{a}(x,v)dx^{i})\mbox{ on }(TV)^{\ast }.  \notag
\end{eqnarray}%
Above N-adapted frames are, in general, nonholonomic (equivalently,
anholonomic, i.e. non-integrable) because they satisfy certain anholonomy
conditions. For instance, 
\begin{equation}
\ \ ^{\shortmid }\widetilde{\mathbf{e}}_{\beta }\ ^{\shortmid }\widetilde{%
\mathbf{e}}_{\gamma }-\ ^{\shortmid }\widetilde{\mathbf{e}}_{\gamma }\
^{\shortmid }\widetilde{\mathbf{e}}_{\beta }=\ ^{\shortmid }w_{\beta \gamma
}^{\tau }[\ ^{\shortmid }\widetilde{N}_{ia}]\ \ ^{\shortmid }\widetilde{%
\mathbf{e}}_{\tau },  \label{anholcond8}
\end{equation}%
where the the formulas for the anholonomy coefficients $\
^{\shortmid}w_{\beta \gamma }^{\tau }[\ ^{\shortmid }\widetilde{N}_{ia}]$ on 
$\ ^{\shortmid }\mathcal{M}$ can be found by computing the commutators in
explicit form. A basis $\ ^{\shortmid }\widetilde{\mathbf{e}}_{\alpha
}\simeq \ ^{\shortmid }\mathbf{\partial }_{\alpha }$ is holonomic and can be
transformed into a coordinate one for trivial N-connection structures
satisfying the conditions $\ ^{\shortmid }w_{\beta \gamma }^{\tau }[\
^{\shortmid }\widetilde{N}_{ia}].$ Similar formulas can be defined on $%
\mathcal{M}$ using geometric objects without label "$^{\shortmid "}$. 

The Hessians (\ref{hessls}) and (\ref{hesshs}) define certain v- and
c-metric structures but not total metrics. The Sasaki \cite{yano73} lifts
and respective N-adapted frames (\ref{ccnadapl}) and (\ref{ccnadaph}) allow
us to define on total bundles certain canonical distinguished metric,
d-metric, structures 
\begin{eqnarray}
\widetilde{\mathbf{g}} &=&\widetilde{\mathbf{g}}_{\alpha \beta }(x,y)%
\widetilde{\mathbf{e}}^{\alpha }\mathbf{\otimes }\widetilde{\mathbf{e}}%
^{\beta }=\widetilde{g}_{ij}(x,y)e^{i}\otimes e^{j}+\widetilde{g}_{ab}(x,y)%
\widetilde{\mathbf{e}}^{a}\otimes \widetilde{\mathbf{e}}^{a}\mbox{
and/or }  \label{cdms8} \\
\ ^{\shortmid }\widetilde{\mathbf{g}} &=&\ ^{\shortmid }\widetilde{\mathbf{g}%
}_{\alpha \beta }(x,p)\ ^{\shortmid }\widetilde{\mathbf{e}}^{\alpha }\mathbf{%
\otimes \ ^{\shortmid }}\widetilde{\mathbf{e}}^{\beta }=\ \ ^{\shortmid }%
\widetilde{g}_{ij}(x,p)e^{i}\otimes e^{j}+\ ^{\shortmid }\widetilde{g}%
^{ab}(x,p)\ ^{\shortmid }\widetilde{\mathbf{e}}_{a}\otimes \ ^{\shortmid }%
\widetilde{\mathbf{e}}_{b}.  \label{cdmds8}
\end{eqnarray}%
Here we not that the terms d-metric, d-tensor, d-connection, d-spinor,
d-object etc. are used for all geometric objects which are such way adapted
to N-connection structures. The d-tensors $\widetilde{\mathbf{g}}$ and $\
^{\shortmid }\widetilde{\mathbf{g}}$ are completely determined by respective
geometric data $(\widetilde{L},\ \ \widetilde{\mathbf{N}};\widetilde{\mathbf{%
e}}_{\alpha },\widetilde{\mathbf{e}}^{\alpha };\widetilde{g}_{jk},\widetilde{%
g}^{jk})$ or $(\widetilde{H},\ ^{\shortmid }\widetilde{\mathbf{N}}; \
^{\shortmid }\widetilde{\mathbf{e}}_{\alpha },\ ^{\shortmid }\widetilde{%
\mathbf{e}}^{\alpha };\ \ ^{\shortmid }\widetilde{g}^{ab},\ \ ^{\shortmid }%
\widetilde{g}_{ab}).$ They consist cornerstone geometric structures for
defining relativistic FL and FH spaces. 

Using frame transforms, the d-metric structures (\ref{cdms8}) and (\ref%
{cdmds8}), with tildes, can be written, respectively, in general d-metric
forms without tildes. Such vierbein transforms can be parameterized
respectively as $e_{\alpha }=e_{\ \alpha }^{\underline{\alpha }}(u)\partial
/\partial u^{\underline{\alpha }}$ and $e^{\beta }=e_{\ \underline{\beta }%
}^{\beta }(u)du^{\underline{\beta }}$, where the local coordinate indices
are underlined in order to distinguish them from arbitrary abstract ones. In
respective formulas, the matrix $e_{\ \underline{\beta }}^{\beta }$ is
inverse to $e_{\ \alpha }^{\underline{\alpha }}$ for orthonormalized bases.
For Hamilton like configurations, one writes $\ ^{\shortmid }e_{\alpha }=\
^{\shortmid}e_{\ \alpha }^{\underline{\alpha }}(\ ^{\shortmid }u)\partial
/\partial \ ^{\shortmid }u^{\underline{\alpha }}$ and $\ ^{\shortmid
}e^{\beta }=\ ^{\shortmid }e_{\ \underline{\beta }}^{\beta }(\ ^{\shortmid
}u)d\ ^{\shortmid }u^{\underline{\beta }}.$ If such transforms are adapted
to certain N-connection structures, we may use $\mathbf{e}_{\alpha }=\{%
\mathbf{e}_{\ \alpha }^{\underline{\alpha }}\}$ and $\ ^{\shortmid }\mathbf{e%
}_{\alpha }=\{\ ^{\shortmid }\mathbf{e}_{\ \alpha }^{\underline{\alpha }}\},$
and impose additional conditions to generate orthonormal frames, or to
preserve certain h-v, or h-c, nonholonomic splitting for general d-metric
structures $\mathbf{g}$ and $\ ^{\shortmid }\mathbf{g.}$ Tilde labels are
appropriate to emphasize that certain d-metrics encode FL or FH data,
respectively, $\mathbf{\tilde{e}}_{\alpha }=\{\mathbf{\tilde{e}}_{\ \alpha}^{%
\underline{\alpha }}\}$ and $\ ^{\shortmid }\mathbf{\tilde{e}}_{\alpha }=\{\
^{\shortmid }\mathbf{\tilde{e}}_{\ \alpha }^{\underline{\alpha }}\}.$ We can
prescribe respective generating functions $L$ or $H$ to transform an
arbitrary phase space $\mathcal{M}$ or $\ ^{\shortmid }\mathcal{M}$ into a
relativistic FL or FH space, $\mathcal{\tilde{M}}$ or $\ ^{\shortmid }%
\mathcal{\tilde{M}}.$ Inversely, arbitrary frame transforms define
modifications of the nonholonomic geometric structures, $\mathcal{\tilde{M}%
\rightarrow M}$ or $\ ^{\shortmid }\mathcal{\tilde{M}\rightarrow }\
^{\shortmid }\mathcal{M}$. 

With respect to local coordinate frames, any general d--metric structures on 
$\mathcal{M}$ or $\ ^{\shortmid }\mathcal{M}$ can be written as 
\begin{eqnarray}
\mathbf{g} &=&\mathbf{g}_{\alpha \beta }(x,y)\mathbf{e}^{\alpha }\mathbf{%
\otimes e}^{\beta }=g_{\underline{\alpha }\underline{\beta }}(x,y)du^{%
\underline{\alpha }}\mathbf{\otimes }du^{\underline{\beta }}\mbox{
and/or }  \label{dmgener} \\
\ ^{\shortmid }\mathbf{g} &=&\ ^{\shortmid }\mathbf{g}_{\alpha \beta }(x,p)\
^{\shortmid }\mathbf{e}^{\alpha }\mathbf{\otimes \ ^{\shortmid }e}^{\beta
}=\ ^{\shortmid }g_{\underline{\alpha }\underline{\beta }}(x,p)d\
^{\shortmid }u^{\underline{\alpha }}\mathbf{\otimes }d\ ^{\shortmid }u^{%
\underline{\beta }}.  \notag
\end{eqnarray}%
Using frame transforms, $\mathbf{g}_{\alpha \beta }=e_{\ \alpha }^{%
\underline{\alpha }}e_{\ \beta }^{\underline{\beta }}g_{\underline{\alpha } 
\underline{\beta }}$ and $\ ^{\shortmid }\mathbf{g}_{\alpha \beta }=\
^{\shortmid }e_{\ \alpha }^{\underline{\alpha }}\ ^{\shortmid }e_{\ \beta }^{%
\underline{\beta }}\ ^{\shortmid }g_{\underline{\alpha }\underline{\beta }},$
corresponding off-diagonal coefficients of these d-metrics are parameterized
in the form: 
\begin{eqnarray}
g_{\underline{\alpha }\underline{\beta }} &=&\left[ 
\begin{array}{cc}
g_{ij}(x)+g_{ab}(x,y)N_{i}^{a}(x,y)N_{j}^{b}(x,y) & g_{ae}(x,y)N_{j}^{e}(x,y)
\\ 
g_{be}(x,y)N_{i}^{e}(x,y) & g_{ab}(x,y)%
\end{array}%
\right] \mbox{
and }  \notag \\
\ ^{\shortmid }g_{\underline{\alpha }\underline{\beta }} &=&\left[ 
\begin{array}{cc}
\ ^{\shortmid }g_{ij}(x)+\ ^{\shortmid }g^{ab}(x,p)\ ^{\shortmid
}N_{ia}(x,p)\ ^{\shortmid }N_{jb}(x,p) & \ ^{\shortmid }g^{ae}\ ^{\shortmid
}N_{je}(x,p) \\ 
\ ^{\shortmid }g^{be}\ ^{\shortmid }N_{ie}(x,p) & \ ^{\shortmid
}g^{ab}(x,p)\ 
\end{array}%
\right] .  \label{offd}
\end{eqnarray}%
Phase space metrics of type (\ref{offd}) are considered, for instance, in
the Kaluza--Klein theory and various string theories with extra dimension
coordinates \cite%
{beil93,beil03,asanov85,aa88,ap88,bejancu90,vg95,vacaru96c,vacaru03,vd00,vcv00,vmon1}%
. Such metrics are generic off-diagonal if the corresponding N-adapted
structures are nonholonomic, see (\ref{anholcond8}). 

To develop and apply the AFCDM in sections \ref{sec3} and \ref{sec4} and
construct exact and parametric solutions in FLH\ theories, we have to
consider conventional (2+2)+(2+2) splitting on respective phase spaces. Such
nonholonomic structures can be stated as dyadic (i.e. 2-d) decompositions
into four oriented shells $s=1,2,3,4.$ In brief, we write this as
s-decompositions and use respective $s$-labels in abstract form, or we shall
use indices and coordinates with additional s-label. Nonholonomic
s-splitting is defined by respective N-connection (equivalently,
s-connection), structures: 
\begin{eqnarray}
\ _{s}^{\shortmid }\mathbf{N}:\ \ _{s}T\mathbf{T}^{\ast }\mathbf{V} &=&\
^{1}hT^{\ast }V\oplus \ ^{2}vT^{\ast }V\oplus \ ^{3}cT^{\ast }V\oplus \
^{4}cT^{\ast }V,\mbox{ which is dual to }  \notag \\
\ _{s}\mathbf{N}:\ \ _{s}T\mathbf{TV} &=&\ ^{1}hTV\oplus \ ^{2}vTV\oplus \
^{3}vTV\oplus \ ^{4}vTV,\mbox{  for }s=1,2,3,4.  \label{ncon2222}
\end{eqnarray}%
In (\ref{ncon2222}), $\ ^{1}h$ is for a conventional 2-d shell (dyadic)
splitting on (co) tangent bundle (for local coordinates $x^{i_{1}})$ and $\
^{2}v$ is for a 2-d vertical like splitting with $x^{a_{2}}=y^{a_{2}}$
coordinates on shell $s=2.$ Then, on (co) fiber shell $s=3,$ the splitting
is conventional (co) vertical; we write $\ ^{3}v$ (or $\ ^{3}c$) and use
local coordinates $v^{a_{3}}$ $($or $p_{a_{3}}).$ Similarly, on the 4th
shell $s=4,$ we use respective symbols $\ ^{4}v$ and $v^{a_{4}}$ (or $\
^{4}c $ and $p_{a_{4}}).$ Hereafter, we consider that we can always write
necessary formulas of s-geometric objects on s-labelled phase spaces, $\ _{s}%
\mathcal{M}=TV$ and $\ _{s}^{\shortmid }\mathcal{M}=T^{\ast }V,$ when the
formulas for velocity and momentum type coordinates can be enabled with
necessary shell indices. 

Using a set of s-connection coefficients, we can construct N-/ s-adapted
bases as linear N-operators: 
\begin{eqnarray}
\ ^{\shortmid }\mathbf{e}_{\alpha _{s}}[\ ^{\shortmid }N_{\ i_{s}a_{s}}]
&=&(\ ^{\shortmid }\mathbf{e}_{i_{s}}=\ \frac{\partial }{\partial x^{i_{s}}}%
-\ ^{\shortmid }N_{\ i_{s}a_{s}}\frac{\partial }{\partial p_{a_{s}}},\ \
^{\shortmid }e^{b_{s}}=\frac{\partial }{\partial p_{b_{s}}})\mbox{ on }\
_{s}T\mathbf{T}_{\shortmid }^{\ast }\mathbf{V,}  \notag \\
\ ^{\shortmid }\mathbf{e}_{\alpha }[\ ^{\shortmid }N_{\ ia}] &=&(\
^{\shortmid }\mathbf{e}_{i}=\ \frac{\partial }{\partial x^{i_{s}}}-\
^{\shortmid }N_{\ ia}\frac{\partial }{\partial p_{a}},\ \ ^{\shortmid }e^{b}=%
\frac{\partial }{\partial p_{b}})\mbox{ on }\ T\mathbf{T}_{\shortmid }^{\ast
}\mathbf{V,}  \label{ccnadapc}
\end{eqnarray}%
and, dual s-adapted bases, s-cobases,%
\begin{eqnarray}
\ ^{\shortmid }\mathbf{e}^{\alpha _{s}}[\ ^{\shortmid }N_{\ i_{s}a_{s}}]
&=&(\ ^{\shortmid }\mathbf{e}^{i_{s}}=dx^{i_{s}},\ ^{\shortmid }\mathbf{e}%
_{a_{s}}=d\ p_{a_{s}}+\ ^{\shortmid }N_{\ i_{s}a_{s}}dx^{i_{s}})\mbox{ on }\
\ _{s}T^{\ast }\mathbf{T}_{\shortmid }^{\ast }\mathbf{V,}  \notag \\
\ ^{\shortmid }\mathbf{e}^{\alpha }[\ ^{\shortmid }N_{\ ia}] &=&(\
^{\shortmid }\mathbf{e}^{i}=dx^{i},\ ^{\shortmid }\mathbf{e}_{a}=d\ p_{a}+\
^{\shortmid }N_{\ ia}dx^{i})\mbox{ on }\ \ T^{\ast }\mathbf{T}_{\shortmid
}^{\ast }\mathbf{V.}  \label{ccnadapv}
\end{eqnarray}%
Such s-frames are not integrable because, in general, they satisfy certain
anholonomy conditions (\ref{anholcond8}) (in this case, with shell indices). 

\subsubsection{General covariance, modified dispersion relations and
nonholonomic FLH variables}

We suppose that MGTs derived in the framework of the M-theory, or string
gravity, and for quasi-classical limits of QG, can be characterized by MDRs (%
\ref{mdrg}) can be modelled with (small) values of and an indicator $\varpi $
are described by basic Lorentzian and non-Riemannian total phase space.

Dropping the conditions of homogeneity of the generating function, the
formula for the relativistic Finsler nonlinear quadratic line elements (\ref%
{finslqlel}) transform can be written for respective Lagrange and Hamilton
spaces: 
\begin{eqnarray}
ds_{L}^{2} &=&L(x,v),\mbox{ for models on  }TV;\mbox{ and }  \label{nqe} \\
d\ ^{\shortmid }s_{H}^{2} &=&H(x,p),\mbox{ for models on  }T^{\ast }V.
\label{nqed}
\end{eqnarray}%
Such quadratic elements can be positive or negative. We must consider
additional assumptions if our goal is to work with real analysis and
geometric models, such as those with fixed local pseudo-Riemannian
signatures. As an example, let us explain how such geometries can model
modified dispersion relations, MDRs, on a nonholonomic phase space $\
^{\shortmid }\mathcal{M}$. In various semi-classical MGTs (in general, they
can be nonassociative and noncommutative) and QGs, MDRs, can be
parameterized locally in the form 
\begin{equation}
c^{2}\overrightarrow{\mathbf{p}}^{2}-E^{2}+c^{4}m^{2}=\varpi (E,%
\overrightarrow{\mathbf{p}},m;\ell _{P},\kappa ,...).  \label{mdrg}
\end{equation}%
An indicator function $\varpi $ in (\ref{mdrg}) \ involves dependencies on a
conventional energy-momentum $p_{a}=(p_{\acute{\imath}},p_{4}=E),%
\overrightarrow{\mathbf{p}}=\{p_{\acute{\imath}}\},$ (for $a=1,2,3,4$). In
string gravity MGT, there are dependencies on the Planck length scale $\ell
_{p}:=\sqrt{\hbar G/c^{3}}\sim 10^{-33}cm$ and $\kappa :=\ell
_{s}^{3}/6\hbar $ being a string constant, were $\ell _{s}$ is a length
parameter (we can fix the light velocity $c=1$ for a respective system of
physical units). An indicator $\varpi (...)$ encodes in a functional form
possible contributions of MGTs which, in general, can be with LIVs,
generalized Finsler and/or string type contributions, etc. MDRs can be
extended to dependencies on 4-d spacetime coordinates $%
x^{i}=(x^{1},x^{2},x^{3},x^{4}=ct),$ or to include higher dimensions and for
various phase space models. Indicators $\varpi (...)$ are prescribed to
construct certain phenomenological models, or determined experimentally.
Such values can be computed in the framework of certain classical or quantum
theories of gravity and matter field interactions. For $\varpi =0,$ (\ref%
{mdrg}) transforms into a standard quadratic dispersion relation for a
relativistic point particle with mass $m,$ energy $E,$ and momentum $p_{%
\acute{\imath}}$ (for $\acute{\imath}=1,2,3$); such a particle propagates in
a 4-d, flat Minkowski spacetime. We also note that MGTs with MDRs were
studied as candidates for explaining acceleration cosmology and applications
in DE and DM, physics, see \cite%
{amelino96,amelino97,girelli06,kostelecky12,barcaroli15,vacaru18,bubuianu18}
and references therein. 

Any MDR (\ref{mdrg}) can be modeled on a Hamilton space $\ _{1}^{3}H$
determined by an Hamilton function 
\begin{equation*}
H(p):=E=\pm (c^{2}\overrightarrow{\mathbf{p}}^{2}+c^{4}m^{2}-\varpi (E,%
\overrightarrow{\mathbf{p}},m;\ell _{P}))^{1/2}.
\end{equation*}%
Changing the system of frames/ coordinates on total phase space $\
^{\shortmid }\mathcal{M}$, we obtain generating functions $H(x,p)$ depending
also on all spacetime and momentum coordinates on $T^{\ast }V.$ This way, we
can define and compute certain Hessian and FH d-metric structures (\ref%
{hesshs}) and (\ref{cdmds8}). Applying Legendre transforms (\ref{legendre1}%
), $H(x,p)\rightarrow L(x,v),$ for a nonlinear quadratic element (\ref{nqe}%
), we can construct an associated FL geometry defined by formulas (\ref%
{hessls}) and (\ref{cdms8}). So, FLH geometries are characterized by MDRs,
which can be used for experimental/ observational verifications of certain
models or classes of solutions in the framework of a MGT. 

Let us explain how metrics in GR can be extended as d-metrics on phase
spaces $\mathcal{M}$ and $\ ^{\shortmid }\mathcal{M}$. We can follow
Assumption 2.1 from \cite{vacaru18} that the standard gravity and particle
physics theories based on the special relativity and Einstein gravity
principles and axioms can be generalized from a 4-d Lorentz spacetime
manifold $V$ to (co) tangent bundles $TV$ or $T^{\ast }V.$ For flat typical
(co) fiber spaces, the total phase space metrics can be parameterized in the
form: 
\begin{eqnarray}
ds^{2} &=&g_{\alpha ^{\prime }\beta ^{\prime }}(x^{k^{\prime }})du^{\alpha
^{\prime }}du^{\beta ^{\prime }}=g_{i^{\prime }j^{\prime }}(x^{k^{\prime
}})dx^{i^{\prime }}dx^{j^{\prime }}+\eta _{a^{\prime }b^{\prime
}}dy^{a^{\prime }}dy^{b^{\prime }},\mbox{ for }v^{a^{\prime }}\sim
dx^{a^{\prime }}/d\zeta ;\mbox{ and/ or }  \label{lqe} \\
d\ ^{\shortmid }s^{2} &=&\ ^{\shortmid }g_{\alpha ^{\prime }\beta ^{\prime
}}(x^{k^{\prime }})d\ ^{\shortmid }u^{\alpha ^{\prime }}d\ ^{\shortmid
}u^{\beta ^{\prime }}=g_{i^{\prime }j^{\prime }}(x^{k^{\prime
}})dx^{i^{\prime }}dx^{j^{\prime }}+\eta ^{a^{\prime }b^{\prime
}}dp_{a^{\prime }}dp_{b^{\prime }},\mbox{ for }p_{a^{\prime }}\sim
dx_{a^{\prime }}/d\zeta .  \label{lqed}
\end{eqnarray}%
In these formulas, curves $x^{a^{\prime }}(\zeta )$ on $V$ are parameterized
by a positive parameter $\zeta .$ A pseudo--Riemannian spacetime metric $%
g=\{g_{i^{\prime }j^{\prime }}(x)\}$ can be chosen as a solution of the
Einstein equations for the Levi-Civita connection $\nabla .$ In diagonal
form, the vertical metric $\eta _{a^{\prime }b^{\prime }}$ and its dual $%
\eta ^{a^{\prime }b^{\prime }}$ are standard Minkowski metrics, $\eta
_{a^{\prime }b^{\prime }}=diag[1,1,1,-1].$ The geometric and physical phase
space models are elaborated for general frame/ coordinate transforms on the
base spacetime and in total spaces when the metric structures can be
parameterized equivalently by the same h-components of $g_{\alpha
^{\prime}\beta ^{\prime }}(x^{k^{\prime }})$ and $\ ^{\shortmid }g_{\alpha
^{\prime}\beta ^{\prime }}(x^{k^{\prime }})=g_{\alpha ^{\prime }\beta
^{\prime}}(x^{k^{\prime }})$, respectively, in quadratic elements (\ref{lqe}%
) and (\ref{lqed}). FLH gravitational interactions can be modelled by 8-d
frame transforms when 
\begin{equation}
\mathbf{g}_{\alpha \beta }(x,v)=e_{\ \alpha }^{\alpha ^{\prime }}\ (x,v)e_{\
\beta }^{\beta ^{\prime }}(x,v)\ g_{\alpha ^{\prime }\beta ^{\prime
}}(x^{k^{\prime }})\text{\mbox{ and }}\ ^{\shortmid }\mathbf{g}_{\alpha
\beta }(x,p)=\ ^{\shortmid }e_{\ \alpha }^{\alpha ^{\prime }}(x,p)\
^{\shortmid }e_{\ \beta }^{\beta ^{\prime }}(x,p)\ ^{\shortmid }g_{\alpha
^{\prime }\beta ^{\prime }}(x^{k^{\prime }}),  \label{frametransf}
\end{equation}%
for $x=\{x^{k}(x^{k^{\prime }})\},$ where $e_{\ \alpha }^{\alpha ^{\prime}}\
(x,v)$ or $\ ^{\shortmid }e_{\ \alpha }^{\alpha ^{\prime }}(x,p)$ can be
determined by respective generalized Einstein equations on nonholonomic
phase spaces. The frame transforms can be parameterized in a certain
N-adapted form if we work with d-objects on respective phase spaces. 

The formulas (\ref{offd}), (\ref{lqe}) and (\ref{lqed}) can be written in
"tilde" nonholonomic variables if we prescribe some nonholonomic
distributions as $L(x,v)$ or $H(x,p)$ and use them for constructing
N-adapted bases (\ref{ccnadapl}) \ or (\ref{ccnadaph}). Up to general frame
transforms, we model an effective FL phase space $\widetilde{\mathcal{M}}$
or an effective FH phase space $\ ^{\shortmid }\widetilde{\mathcal{M}}.$ We
can consider also arbitrary frame transforms, $\ ^{\shortmid }\widetilde{%
\mathbf{g}}_{\alpha \beta }=\ ^{\shortmid }e_{\ \alpha }^{\underline{\alpha }%
}\ ^{\shortmid }e_{\ \beta }^{\underline{\beta }}\ ^{\shortmid }\widetilde{g}%
_{\underline{\alpha }\underline{\beta }}$ or s-adapted ones, $\ ^{\shortmid }%
\widetilde{\mathbf{g}}_{\alpha _{s}\beta _{s}}=\ ^{\shortmid }e_{\ \alpha
_{s}}^{\underline{\alpha }}\ ^{\shortmid}e_{\ \beta _{s}}^{\underline{\beta }%
}\ ^{\shortmid }\widetilde{g}_{\underline{\alpha }\underline{\beta }}.$ For
instance, we elaborate a FH model of phase space with equivalent geometric
data $(\ ^{\shortmid }\widetilde{\mathcal{M}}:\ ^{\shortmid }\widetilde{%
\mathbf{g}},\ \ ^{\shortmid }\widetilde{\mathbf{N}})\simeq (\
_{s}^{\shortmid }\widetilde{\mathcal{M}}:\ _{s}^{\shortmid }\widetilde{%
\mathbf{g}},\ _{s}^{\shortmid }\widetilde{\mathbf{N}}).$ In general, we can
omit tilde labels and write $(\ _{s}^{\shortmid }\mathcal{M}:\
_{s}^{\shortmid }\mathbf{g},\ _{s}^{\shortmid}\mathbf{N}),$ which can be
used for applying the AFCDM for constructing off-diagonal physically
important solutions in sections \ref{sec4} and \ref{sec5}. In explicit form,
such solutions are derived in certain forms encoding also data for
generalized affine d-connections, generating and integration functions, and
effective sources. 

\subsubsection{Almost symplectic variables in FLH geometry}

\label{ssalmsymplectic}Original ideas and constructions on almost K\"{a}hler
modeling of Finsler geometry were proposed in \cite%
{matsumoto66,matsumoto86,oproiu85}. We cite \cite{mhss2000,vmon3,vacaru18}
for further developments of (higher order) Finsler-Lagrange and Hamilton
geometries and related almost symplectic approaches to modern geometric
mechanics as generalizations of \cite{kern74}. Respective nonholonomic
geometric methods were applied for performing deformation quantization (in
an almost K\"{a}hler - Fedosov or Gukov - Witten sense) of FLH theories, GR
and MGTs on metric-affine spaces, see a series of works \cite%
{vacaru07,vacaru10a,vacaru13,biv16,vacaru12a,vacaru16a}. 

Let us explain how certain MDRs (\ref{mdrg}), or FLH generating functions,
and related canonical N--connections $\widetilde{\mathbf{N}}$ and $\
^{\shortmid }\widetilde{\mathbf{N}},$ define canonical almost complex
structures $\widetilde{\mathbf{J}},$ on $\mathbf{TV},$ and $\ ^{\shortmid }%
\widetilde{\mathbf{J}},$ on $\mathbf{T}^{\ast }\mathbf{V}.$ We introduce the
linear operator $\widetilde{\mathbf{J}}$ acting as $\widetilde{\mathbf{J}}(%
\mathbf{\tilde{e}}_{i})=-\widetilde{\mathbf{e}}_{n+i}$ and $\widetilde{%
\mathbf{J}}(e_{n+i})=\widetilde{\mathbf{e}}_{i}$ for $\widetilde{\mathbf{e}}%
_{\alpha }=(\widetilde{\mathbf{e}}_{i},e_{b})$ (\ref{ccnadapl}). This
defines on $\mathbf{TV}$ an almost complex structure, when $\widetilde{%
\mathbf{J}}\mathbf{\circ \widetilde{\mathbf{J}}=-I}$ for and unity matrix $%
\mathbf{I}$ determined by a generating function $L(x,v).$ Similar structures
can be defined on $\mathbf{T}^{\ast }\mathbf{V}$ by considering a linear
operator $\ ^{\shortmid }\widetilde{\mathbf{J}}$ acting on $\ ^{\shortmid }%
\mathbf{\tilde{e}}_{\alpha }=(\ ^{\shortmid }\mathbf{\tilde{e}}_{i},\
^{\shortmid }e^{b})$ (\ref{ccnadaph}), when $\ ^{\shortmid }\widetilde{%
\mathbf{J}}(\ ^{\shortmid }\mathbf{\tilde{e}}_{i})=-\ ^{\shortmid }e^{n+i}$
and $\ ^{\shortmid }\widetilde{\mathbf{J}}(\ ^{\shortmid }e^{n+i})=\
^{\shortmid }\mathbf{\tilde{e}}_{i}$. So, $\ ^{\shortmid }\widetilde{\mathbf{%
J}}$ also defines an almost complex structure, when $\ ^{\shortmid }%
\widetilde{\mathbf{J}}\ \circ \ ^{\shortmid }\widetilde{\mathbf{J}}=-\
^{\shortmid }\ ^{\shortmid }\mathbf{I}$ for the unity matrix $\ ^{\shortmid }%
\mathbf{I}$ on $\mathbf{T}^{\ast }\mathbf{V}$ completely determined by a $%
H(x,p).$ We note that $\widetilde{\mathbf{J}}$ and $\ ^{\shortmid }%
\widetilde{\mathbf{J}}$ are standard almost complex structures only for the
Euclidean signatures (for pseudo-Euclidean signatures, we define such
operators in abstract geometric forms). Considering arbitrary
frame/coordinate transforms, we can write $\mathbf{J}$ and $\ ^{\shortmid }%
\mathbf{J}$ but we have to consider that the constructions for general
nonholonomic manifolds/ bundles involve, in general, not compatible almost
symplectic/ complex/ product structures being different from those on FLH
phase spaces. For physical applications, we can prescribe certain
well-defined almost symplectic data $(\widetilde{\mathbf{N}},\widetilde{%
\mathbf{g}}, \widetilde{\mathbf{J}}),$ or $(\ ^{\shortmid }\widetilde{%
\mathbf{N}},\ ^{\shortmid }\widetilde{\mathbf{g}},\ ^{\shortmid }\widetilde{%
\mathbf{J }})$, and then to consider general frame transforms and
distortions of d-connections.

Respective canonical Neijenhuis tensor fields on Lagrange and Hamilton phase
space can be considered as curvatures of respective N--connections: 
\begin{eqnarray}
\widetilde{\mathbf{\Omega }}(\widetilde{\mathbf{X}},\widetilde{\mathbf{Y}})
&:=&\mathbf{\ -[\widetilde{\mathbf{X}}\mathbf{,}\widetilde{\mathbf{Y}}]+[%
\widetilde{\mathbf{J}}\widetilde{\mathbf{X}},\widetilde{\mathbf{J}}%
\widetilde{\mathbf{Y}}]-\widetilde{\mathbf{J}}[\widetilde{\mathbf{J}}%
\widetilde{\mathbf{X}},\widetilde{\mathbf{Y}}]-\widetilde{\mathbf{J}}[%
\widetilde{\mathbf{X}},\widetilde{\mathbf{J}}\widetilde{\mathbf{Y}}]}%
\mbox{
and/or }  \notag \\
\ ^{\shortmid }\widetilde{\mathbf{\Omega }}(\ ^{\shortmid }\widetilde{%
\mathbf{X}},\ ^{\shortmid }\widetilde{\mathbf{Y}}) &:=&\ \mathbf{-[\
^{\shortmid }\widetilde{\mathbf{X}}\mathbf{,}\ ^{\shortmid }\widetilde{%
\mathbf{Y}}]+[\ ^{\shortmid }\widetilde{\mathbf{J}}\ ^{\shortmid }\widetilde{%
\mathbf{X}},\ ^{\shortmid }\widetilde{\mathbf{J}}\ ^{\shortmid }\widetilde{%
\mathbf{Y}}]-\ ^{\shortmid }\widetilde{\mathbf{J}}[\ ^{\shortmid }\widetilde{%
\mathbf{J}}\ ^{\shortmid }\widetilde{\mathbf{X}},\ ^{\shortmid }\widetilde{%
\mathbf{Y}}]-\ ^{\shortmid }\widetilde{\mathbf{J}}[\ ^{\shortmid }\widetilde{%
\mathbf{X}},\ ^{\shortmid }\widetilde{\mathbf{J}}\ ^{\shortmid }\widetilde{%
\mathbf{Y}}].}  \label{neijt}
\end{eqnarray}%
Hereafter, for simplicity, we shall omit tildes or hats for d--vectors and
write $\mathbf{X,}\mathbf{Y}$ and $\ ^{\shortmid }\mathbf{X,\ ^{\shortmid }Y}
$ if that does not result in ambiguities. Using general frame/coordinates,
the curvatures (\ref{neijt}) can be written in general form without tildes
or in index form: 
\begin{equation*}
\Omega _{ij}^{a}=\frac{\partial N_{i}^{a}}{\partial x^{j}}-\frac{\partial
N_{j}^{a}}{\partial x^{i}}+N_{i}^{b}\frac{\partial N_{j}^{a}}{\partial y^{b}}%
-N_{j}^{b}\frac{\partial N_{i}^{a}}{\partial y^{b}},\mbox{\ or\ }\ \ \mathbf{%
\ ^{\shortmid }}\Omega _{ija}=\frac{\partial \mathbf{\ ^{\shortmid }}N_{ia}}{%
\partial x^{j}}-\frac{\partial \mathbf{\ ^{\shortmid }}N_{ja}}{\partial x^{i}%
}+\ \mathbf{^{\shortmid }}N_{ib}\frac{\partial \mathbf{\ ^{\shortmid }}N_{ja}%
}{\partial p_{b}}-\mathbf{\ ^{\shortmid }}N_{jb}\frac{\partial \mathbf{\
^{\shortmid }}N_{ia}}{\partial p_{b}}.
\end{equation*}%
We have the conditions that certain almost complex structures $\mathbf{J}$
and $\ ^{\shortmid }\mathbf{J}$ transform into standard complex structures
if $\mathbf{\Omega }=0$ and/or $\ ^{\shortmid }\mathbf{\Omega }=0.$

Almost symplectic structures on $\mathbf{TV}$ and $\mathbf{T}^{\ast }\mathbf{%
V}$ can be defined by respective nondegenerate N-adapted 2--forms 
\begin{equation*}
\ \theta =\frac{1}{2}\ \theta _{\alpha \beta }(u)\ \mathbf{e}^{\alpha
}\wedge \mathbf{e}^{\beta }\mbox{ and }\ \ ^{\shortmid }\theta =\frac{1}{2}\
\ ^{\shortmid }\theta _{\alpha \beta }(\ ^{\shortmid }u)\ \ ^{\shortmid }%
\mathbf{e}^{\alpha }\wedge \ ^{\shortmid }\mathbf{e}^{\beta },
\end{equation*}%
when (using h-c components) 
\begin{equation}
\mathbf{\ }\ ^{\shortmid }\theta =\frac{1}{2}\mathbf{\ }\ ^{\shortmid
}\theta _{ij}(\ ^{\shortmid }u)e^{i}\wedge e^{j}+\frac{1}{2}\mathbf{\ }\
^{\shortmid }\theta ^{ab}(\ ^{\shortmid }u)\mathbf{\ ^{\shortmid }e}%
_{a}\wedge \mathbf{\ }\ ^{\shortmid }\mathbf{e}_{b}.  \label{aux03}
\end{equation}

Then, we state that a N--connection $\ ^{\shortmid }\mathbf{N}$ defines a
unique decomposition of a d--vector $\ ^{\shortmid }\mathbf{X=\ }X^{h}+\
^{\shortmid }X^{cv}$ on $T^{\ast }\mathbf{V},$ for $X^{h}=h\ ^{\shortmid }%
\mathbf{X}$ and $\ ^{\shortmid }X^{cv}=cv\ \ ^{\shortmid }\mathbf{X.}$
Respective projectors $h$ and $cv$ can be related to a dual distribution $\
^{\shortmid }\mathbf{N}$ on $\mathbf{V;}$ the properties $h+cv=\mathbf{I}%
,h^{2}=h,(cv)^{2}=cv,h\circ cv=cv\circ h=0$ are satisfied. We can introduce
the almost product operator $\ ^{\shortmid }\mathbf{P:=}I-2cv=2h-I$ acting
on $\ ^{\shortmid }\mathbf{e}_{\alpha }=(\ ^{\shortmid }\mathbf{e}_{i},\
^{\shortmid }e^{b})$ following formulas 
\begin{equation}
\ \mathbf{\ }^{\shortmid }\mathbf{P}(\ ^{\shortmid }\mathbf{e}_{i})=\
^{\shortmid }\mathbf{e}_{i}\mbox{\ and  }\ ^{\shortmid }\mathbf{P}(\
^{\shortmid }e^{b})=-\ \ ^{\shortmid }e^{b}.  \label{almps}
\end{equation}%
Similar formulas can be defined by a N--connection $\ \mathbf{N}$ inducing
an almost product structure $\mathbf{P}$ on $T\mathbf{V}.$ 

In almost symplectic models of FLH geometry, other important geometric
d-operators are used. For instance, the almost tangent (co) ones satisfy the
conditions 
\begin{eqnarray*}
\mathbb{J(}\mathbf{e}_{i}\mathbb{)} &=&e_{4+i}\mbox{\ and
\ }\ \mathbb{J}\left( e_{a}\right) =0,\mbox{ \ or \ }\mathbb{J=}\frac{%
\partial }{\partial y^{i}}\otimes dx^{i}; \\
\mathbf{\ }^{\shortmid }\mathbb{J}(\mathbf{\ }^{\shortmid }\mathbf{e}_{i}%
\mathbb{)} &=&\mathbf{\ }^{\shortmid }g_{ib}\mathbf{\ }^{\shortmid }e^{b}%
\mbox{\ and
\ }\ \mathbf{\ }^{\shortmid }\mathbb{J}\left( \mathbf{\ }^{\shortmid
}e^{b}\right) =0,\mbox{ \ or \ }\mathbf{\ }^{\shortmid }\mathbb{J=}\mathbf{\ 
}^{\shortmid }g_{ia}\frac{\partial }{\partial p_{a}}\otimes dx^{i}.
\end{eqnarray*}%
We can verify by straightforward computations that there are satisfied for
pairs of so-called $\mathcal{L}$--dual N--connections $\left( \mathbf{N,\ \
^{\shortmid }N}\right) ,$ see details in \cite{mhss2000,vmon3,vacaru18}, the
properties: 
\begin{equation*}
\mathbf{J}=-\delta _{i}^{a}e_{a}\otimes e^{i}+\delta _{a}^{i}\mathbf{e}%
_{i}\otimes \mathbf{e}^{a}\mbox{ and }\mathbf{\ }^{\shortmid }\mathbf{J}=-\
^{\shortmid }g_{ia}\ ^{\shortmid }e^{a}\otimes \ ^{\shortmid }e^{i}+\
^{\shortmid }g^{ia}\ ^{\shortmid }\mathbf{e}_{i}\otimes \ ^{\shortmid }%
\mathbf{e}_{a}
\end{equation*}%
hold for a $\mathcal{L}$--dual pair of almost complex structures $\left(%
\mathbf{J},\ ^{\shortmid }\mathbf{J}\right) .$ For such configurations, 
\begin{equation*}
\mathbf{P}=\mathbf{e}_{i}\otimes e^{i}-e_{a}\otimes \mathbf{e}^{a}%
\mbox{ and
}\ ^{\shortmid }\mathbf{P}=\ ^{\shortmid }\mathbf{e}_{i}\otimes \
^{\shortmid }e^{i}-\ ^{\shortmid }e^{a}\otimes \ ^{\shortmid }\mathbf{e}_{a}
\end{equation*}%
correspond to a $\mathcal{L}$--dual pair of almost product structures $%
\left( \mathbf{P},\ \ ^{\shortmid }\mathbf{P}\right) .$ This allows us to
define respective almost symplectic structures 
\begin{equation}
\theta =g_{aj}(x,v)\mathbf{e}^{a}\wedge e^{i}\mbox{ and }\ ^{\shortmid
}\theta =\delta _{i}^{a}\ ^{\shortmid }\mathbf{e}_{a}(x,p)\wedge \
^{\shortmid }e^{i}  \label{sympf}
\end{equation}

Above defined d-operators can be re-written in canonical forms by
considering N-adapted bases with tilde. For instance, we can write (\ref%
{sympf}) (using frame transforms) as $\widetilde{\theta }=\widetilde{g}%
_{aj}(x,y)\widetilde{\mathbf{e}}^{a}\wedge e^{i}$ and $\ ^{\shortmid }%
\widetilde{\theta }=\delta _{i}^{a}\ ^{\shortmid }\widetilde{\mathbf{e}}%
_{a}\wedge \ ^{\shortmid }e^{i}$ and consider tilde almost symplected data $%
\left(\ ^{\shortmid }\widetilde{\mathbf{J}},\mathbf{\ }^{\shortmid}%
\widetilde{\mathbb{J}}, \ ^{\shortmid}\widetilde{\mathbf{P}}, \ ^{\shortmid }%
\widetilde{\theta }\right) .$ The constructions can encode nonholonomic
dyadic splitting of type $\left( \ _{s}^{\shortmid }\widetilde{\mathbf{J}},\
_{s}^{\shortmid }\widetilde{\mathbb{J}}, \ _{s}^{\shortmid}\widetilde{%
\mathbf{P}}, \ _{s}^{\shortmid }\widetilde{\theta }\right) .$ 

It should be noted that the phase space nonholonomic geometry can be
formulated as an almost Hermitian model of a tangent Lorentz bundle $T%
\mathbf{V}$ equipped with a N--connection structure $\mathbf{N.}$ For this,
we consider a triple $\mathbf{H}^{8}=(T\mathbf{V},\theta ,\mathbf{J}),$
where $\theta \mathbf{(X,Y)}:=\mathbf{g}\left( \mathbf{JX,Y}\right) .$
Respectively, on a cotangent Lorentz bundle $T^{\ast }\mathbf{V}$ with a (or 
$\ ^{\shortmid }\mathbf{N}),$ we can define a triple $\ ^{\shortmid }\mathbf{%
H}^{8}=(T^{\ast }\mathbf{V}, \ ^{\shortmid }\theta ,\ ^{\shortmid }\mathbf{J}%
),$ where $\ ^{\shortmid }\theta (\ ^{\shortmid }\mathbf{X}, \ ^{\shortmid }%
\mathbf{Y}):= \ ^{\shortmid }\mathbf{g(}\ ^{\shortmid }\mathbf{J} \
^{\shortmid }\mathbf{X},\ ^{\shortmid }\mathbf{Y}).$ A space $\mathbf{H}^{8}$
(or $\ ^{\shortmid }\mathbf{H}^{8})$ is almost K\"{a}hler and denoted $%
\mathbf{K}^{8}$ if $d\ \theta =0$ (or $\ ^{\shortmid }\mathbf{K}^{8}$ if $d\
^{\shortmid }\theta =0).$ This property holds true in tilde variables with
1--forms, respectively, defined by a regular Lagrangian $L$ and Hamiltonian $%
H$ (related by a Legendre transform) when $\widetilde{\omega }=\frac{%
\partial L}{\partial y^{i}}e^{i}$ and $\ ^{\shortmid }\widetilde{\omega }%
=p_{i}dx^{i},$ for which $\widetilde{\theta }=d\widetilde{\omega }$ and $\
^{\shortmid }\widetilde{\theta }=d\ ^{\shortmid }\widetilde{\omega }$. So,
we have that $d\widetilde{\theta }=0$ and $d\ ^{\shortmid }\widetilde{\theta}%
=0.$ If such conditions are satisfied, for instance, for $\ ^{\shortmid }%
\widetilde{\mathbf{N}},$ we can consider arbitrary or nonholonomic dyadic
structures with $\ _{s}^{\shortmid }\widetilde{\mathbf{N}}$ and $d\ _{s}%
\widetilde{\theta }=0$ and $d\ _{s}^{\shortmid }\widetilde{\theta }=0.$ We
emphasize that such properties do not hold for arbitrary $\ ^{\shortmid }%
\mathbf{N}$ and $\ ^{\shortmid }\theta ,$ when, in general, $d\ ^{\shortmid
}\theta \neq 0.$ We have to introduce a special nonholonomic distribution $\
^{\shortmid }\widetilde{\mathbf{N}}$ determined by a $H$ with N-elongated
frames $\ ^{\shortmid }\widetilde{\mathbf{e}}_{\alpha }= (\ ^{\shortmid }%
\widetilde{\mathbf{e}}_{i},\ ^{\shortmid }e^{b}),$ see reviews \cite%
{vacaru18,vmon3}. 

\subsection{General and canonical FLH d-connections and distortion d-tensors}

Many Finsler-like gravity theories were formulated for different types of
nonlinear quadratic elements, N-connection and d-connection structures.
Self-consistent generalizations of GR as relativistic FLH theories are
possible on (co) tangent Lorentz bundles $\mathbf{TV}$ and $\mathbf{T}^{\ast
}\mathbf{V.}$ Technically, it is almost impossible to integrate and generate
physically important solutions of corresponding systems of nonlinear PDE if
we work only with d-metrics determined by nonlinear quadratic forms $L(x,y)$
(\ref{nqe}) or $H(x,p)$ (\ref{nqed}) (or with arbitrary nonholonomic fibered
4+4 structures (\ref{dmgener}) and (\ref{offd})). We have to consider
additional nonholonomic dyadic decompositions (\ref{ncon2222}) and
corresponding N-adapted distortions of Finsler-like d-connections to be able
to decouple and integrate certain general form corresponding geometric flow
and gravitational field equations. The goal of this subsection is to define
such general and canonical FLH d-connections and distortion d-tensors. 

\subsubsection{Affine connections, d- and s-connections in FLH geometry}

A distinguished connection (d--connection) can be defined as a linear
connection $\mathbf{D,}$ or $\ ^{\shortmid }\mathbf{D},$ which is compatible
with the almost product structure $\mathbf{DP}=0,$ or $\ ^{\shortmid}\mathbf{%
D\ ^{\shortmid }P}=0,$ see (\ref{almps}). Such a d--connection can be
defined to preserve under parallelism a respective N--connection splitting (%
\ref{ncondistr}), which can be prescribed to be a more special N-connection $%
\ ^{\shortmid }\mathbf{\tilde{N}}$ and then related to a nonholonomic dyadic
decomposition (\ref{ncon2222}). 

On a $\ ^{\shortmid }\mathcal{M}$, the coefficients of a d--connection $\
^{\shortmid }\mathbf{D}$ can be defined with respect to N--adapted frames (%
\ref{ccnadapc}) and (\ref{ccnadapv}) as 
\begin{equation*}
\ \ ^{\shortmid }\mathbf{D}_{\ ^{\shortmid }\mathbf{e}_{k}}\ ^{\shortmid }%
\mathbf{e}_{j}:=\ ^{\shortmid }L_{\ jk}^{i}\ ^{\shortmid }\mathbf{e}_{i},\
^{\shortmid }\mathbf{D}_{\mathbf{e}_{k}}\ ^{\shortmid }e^{b}:=-\ ^{\shortmid
}\acute{L}_{a\ k}^{\ b}\ ^{\shortmid }e^{a},\ ^{\shortmid }\mathbf{D}_{\
^{\shortmid }e^{c}}\ ^{\shortmid }\mathbf{e}_{j}:=\ ^{\shortmid }\acute{C}%
_{\ j}^{i\ c}\ ^{\shortmid }\mathbf{e}_{i},\ ^{\shortmid }\mathbf{D}_{\
^{\shortmid }e^{c}}\ ^{\shortmid }e^{b}:=-\ ^{\shortmid }C_{a}^{\ bc}\
^{\shortmid }e^{a}.
\end{equation*}%
Using respective labeling of h- and v-indices, such equations can be
considered for a $\mathbf{D}$ on $\mathcal{M}.$ We parameterize the
N-adapted coefficients of respective d-connections in the form 
\begin{equation}
\mathbf{D=\{\Gamma }_{\ \beta \gamma }^{\alpha }\}=\{L_{\ jk}^{i},\acute{L}%
_{\ bk}^{a},\acute{C}_{\ jc}^{i},C_{\ bc}^{a}\}\mbox{ or }\ ^{\shortmid }%
\mathbf{D=}\{\ ^{\shortmid }\mathbf{\Gamma }_{\ \beta \gamma }^{\alpha
}\}=\{\ ^{\shortmid }L_{\ jk}^{i},\ ^{\shortmid }\acute{L}_{a\ k}^{\ b},\
^{\shortmid }\acute{C}_{\ j}^{i\ c},\ ^{\shortmid }C_{a}^{\ bc}\}.
\label{dconc}
\end{equation}%
For explicit abstract or index computations, we can consider corresponding
h-- and c-splitting of covariant derivatives $\ ^{\shortmid }\mathbf{D}%
=\left( \ _{h}^{\shortmid }\mathbf{D,\ }_{v}^{\shortmid }\mathbf{D}\right) ,$
where $\ _{h}\mathbf{D}=\{L_{\ jk}^{i},\acute{L}_{\ bk}^{a}\},\ $ and $\
_{c}^{\shortmid }\mathbf{D}=\{\ ^{\shortmid }\acute{C}_{\ j}^{i\ c},\
^{\shortmid }C_{a}^{\ bc}\}.$

A d--connection $\mathbf{D}$ (\ref{dconc}) is characterized by three
fundamental geometric d-objects, which (by definition in abstract forms)
are: 
\begin{eqnarray}
\mathcal{T}(\mathbf{X,Y}) &:=&\mathbf{D}_{\mathbf{X}}\mathbf{Y}-\mathbf{D}_{%
\mathbf{Y}}\mathbf{X}-[\mathbf{X,Y}],\mbox{ torsion d-tensor,  d-torsion};
\label{fundgeom} \\
\mathcal{R}(\mathbf{X,Y}) &:=&\mathbf{D}_{\mathbf{X}}\mathbf{D}_{\mathbf{Y}}-%
\mathbf{D}_{\mathbf{Y}}\mathbf{D}_{\mathbf{X}}-\mathbf{D}_{\mathbf{[X,Y]}},%
\mbox{ curvature d-tensor, d-curvature};  \notag \\
\mathcal{Q}(\mathbf{X}) &:= &\mathbf{D}_{\mathbf{X}}\mathbf{g,}%
\mbox{nonmetricity d-fiels, d-nonmetricity}.  \notag
\end{eqnarray}%
Similar d-objects and formulas can be written for $\ ^{\shortmid }\mathbf{D,}
$ for instance, as $\ ^{\shortmid }\mathbf{X,}\ ^{\shortmid }\mathcal{T},%
\mathcal{R}$ $\ ^{\shortmid }$and $\ ^{\shortmid }\mathcal{Q}$. For further
considerations, we shall omit details on such d-tensors on $\ ^{\shortmid }%
\mathcal{M}$ if respective definitions and formulas consist of certain
abstract labeling of their analogues on $\mathcal{M}$ and the abstract
computations do not involve ambiguities. The N--adapted coefficients of the
fundamental geometric d-objects (\ref{fundgeom}) are computed by introducing
d-vectors $\mathbf{X}=\mathbf{e}_{\alpha }$ and $\mathbf{Y}=\mathbf{e}%
_{\beta },$ defined by (\ref{ccnadapc}) and (\ref{ccnadapv}), and
considering a h-v-splitting for $\mathbf{D}=\{\mathbf{\Gamma }_{\ \alpha
\beta }^{\gamma}\}$ into above formulas, see details in \cite%
{vacaru18,vmon2,partner02,bsssvv25}, 
\begin{eqnarray}
\mathcal{T} &=&\{\mathbf{T}_{\ \alpha \beta }^{\gamma }=\left( T_{\
jk}^{i},T_{\ ja}^{i},T_{\ ji}^{a},T_{\ bi}^{a},T_{\ bc}^{a}\right) \};
\label{fundgeomdcoef} \\
\mathcal{R} &\mathbf{=}&\mathbf{\{R}_{\ \beta \gamma \delta }^{\alpha }%
\mathbf{=}\left( R_{\ hjk}^{i}\mathbf{,}R_{\ bjk}^{a}\mathbf{,}R_{\ hja}^{i}%
\mathbf{,}R_{\ bja}^{c}\mathbf{,}R_{\ hba}^{i},R_{\ bea}^{c}\right) \mathbf{%
\};}  \notag \\
\ \mathcal{Q} &=&\mathbf{\{Q}_{\ \alpha \beta }^{\gamma }=\mathbf{D}^{\gamma
}\mathbf{g}_{\alpha \beta }=(Q_{\ ij}^{k},Q_{\ ij}^{c},Q_{\ ab}^{k},Q_{\
ab}^{c})\}.  \notag
\end{eqnarray}%
We say that any geometric data $\left( T\mathbf{V},\mathbf{N},\mathbf{g,D}%
\right) $ define a \ \textit{nonholonomic}, i.e. N-adapted, $\mathbf{N}$, 
\textit{metric-affine structure} (equivalently, metric-affine d-structure)
determined by a d-metric, $\mathbf{g,}$ see (\ref{dmgener}) and (\ref{offd}%
), and a d-connection $\mathbf{D}$ (\ref{dconc}) stated independently, but
both in N-adapted form on $\mathbf{V}$. In dual form, we write $%
\left(T^{\ast }\mathbf{V},\ ^{\shortmid }\mathbf{N},\ ^{\shortmid }\mathbf{%
g, \ ^{\shortmid }D}\right) ,$ when the v-indices are changed into
c-indices, for instance, in the form 
\begin{equation*}
\ ^{\shortmid }\mathcal{Q}=\{\ ^{\shortmid }\mathbf{Q}_{\ \alpha \beta
}^{\gamma }= \ ^{\shortmid }\mathbf{D}^{\gamma }\ ^{\shortmid }\mathbf{g}%
_{\alpha \beta }=(\ ^{\shortmid }\mathbf{Q}_{\ ij}^{k}, \ ^{\shortmid } 
\mathbf{Q}_{cij}, \ ^{\shortmid}\mathbf{Q}_{k}\ ^{ab}, \ ^{\shortmid }%
\mathbf{Q}_{c\ }^{\ ab})\}.
\end{equation*}

For dyadic decompositions, the symbols of geometric objects and/or indices
of such objects are labelled additionally with a shell label, for instance, $%
\ _{cs}^{\shortmid }\mathbf{D}=\{\ ^{\shortmid }\acute{C}_{\ j_{s}}^{i_{s}\
c_{s}},\ ^{\shortmid }C_{a_{s}}^{\ b_{s}c_{s}}\},$ when, for instance, $%
j_{2}=1,2,3,4$ and $a_{3}=5,6.$ In such cases, we use the terms s-connection
instead of d-connection (respectively, s-tensor instead of d-tensor) and
write in abstract form $\ _{s}^{\shortmid }\mathcal{M}$. Similar
decompositions can be performed for a $\ _{s}\mathbf{D}$ on $\ _{s}\mathcal{M%
}$. All formulas on $\mathcal{M}$ and $\ ^{\shortmid }\mathcal{M}$ can be
proven in abstract geometric and s-adapted forms. We omit such details in
this work; see \cite{vacaru18,vmon3,bsssvv25} and references therein. The
fundamental geometric s-objects can be labeled in the form 
\begin{eqnarray*}
\ _{s}\mathcal{T} &=&\{\mathbf{T}_{\ \alpha _{s}\beta _{s}}^{\gamma
_{s}}=\left( T_{\ j_{s}k_{s}}^{i_{s}},T_{\ j_{s}a_{s}}^{i_{s}},T_{\
j_{s}i_{s}}^{a_{s}},T_{\ b_{s}i_{s}}^{a_{s}},T_{\ b_{s}c_{s}}^{a_{s}}\right)
\}; \\
\ _{s}\mathcal{R} &\mathbf{=}&\mathbf{\{R}_{\ \beta _{s}\gamma _{s}\delta
_{s}}^{\alpha _{s}}\mathbf{=}\left( R_{\ h_{s}j_{s}k_{s}}^{i_{s}}\mathbf{,}%
R_{\ b_{s}j_{s}k_{s}}^{a_{s}}\mathbf{,}R_{\ h_{s}j_{s}a_{s}}^{i_{s}}\mathbf{,%
}R_{\ b_{s}j_{s}a_{s}}^{c_{s}}\mathbf{,}R_{\ h_{s}b_{s}a_{s}}^{i_{s}},R_{\
b_{s}e_{s}a_{s}}^{c_{s}}\right) \mathbf{\};} \\
\ \ _{s}\mathcal{Q} &=&\mathbf{\{Q}_{\ \alpha _{s}\beta _{s}}^{\gamma _{s}}=%
\mathbf{D}^{\gamma _{s}}\mathbf{g}_{\alpha _{s}\beta _{s}}=(Q_{\
i_{s}j_{s}}^{k_{s}},Q_{\ i_{s}j_{s}}^{c_{s}},Q_{\ a_{s}b_{s}}^{k_{s}},Q_{\
a_{s}b_{s}}^{c_{s}})\}.
\end{eqnarray*}%
Similar s-adapted formulas for $\ _{s}^{\shortmid }\mathcal{T}$, $\
_{s}^{\shortmid }\mathcal{R}$, and$\ _{s}^{\shortmid }\mathcal{Q}$ involve
transforming v-indices into c-indices on cotangent Lorentz bundles, for
instance, in the form 
\begin{equation}
\ _{s}^{\shortmid }\mathcal{T}=\{\ ^{\shortmid }\mathbf{T}_{\ \alpha
_{s}\beta _{s}}^{\gamma _{s}}=\left( \ ^{\shortmid }T_{\
j_{s}k_{s}}^{i_{s}},\ ^{\shortmid }T_{\ j_{s}}^{i_{s}\ a_{s}},\ ^{\shortmid
}T_{a_{s}\ j_{s}i_{s}},\ ^{\shortmid }T_{a_{s}\ i_{s}}^{\ b_{s}},\
^{\shortmid }T_{a_{s}\ }^{\ b_{s}c_{s}}\right) \}.  \label{stors}
\end{equation}

We con consider on $\mathcal{M}$ and$\ ^{\shortmid }\mathcal{M}$ arbitrary
affine connections, denoted in "non-boldface" forms as $D=\{\Gamma _{\ \beta
\gamma }^{\alpha }\}$ and $\ ^{\shortmid }D=\{\ ^{\shortmid }\Gamma _{\
\beta \gamma }^{\alpha }\}.$ If we introduce N-adapted frames and respective
d-connection structures, we can consider respective distortion d-tensors (or
s-tensors), $\mathbf{Z}=\{\mathbf{Z}_{\ \beta \gamma }^{\alpha}\}$ (or $\
_{s}\mathbf{Z=}\{\mathbf{Z}_{\ \beta _{s}\gamma _{s}}^{\alpha _{s}}\}$) and $%
\ ^{\shortmid }\mathbf{Z=}\{\ ^{\shortmid }\mathbf{Z}_{\ \beta \gamma
}^{\alpha }\}$ (or $\ _{s}^{\shortmid }\mathbf{Z=}\{\ ^{\shortmid }\mathbf{Z}%
_{\ \beta _{s}\gamma _{s}}^{\alpha _{s}}\})\mathbf{,}$ when 
\begin{equation}
D=\mathbf{D+Z}\mbox{ and }\ ^{\shortmid }D=\ ^{\shortmid }\mathbf{D+\
^{\shortmid }Z.}  \label{affinedist}
\end{equation}%
Any tensor can be transformed into a respective N- or s-tensor and inversely
if we define respective adapted frames. But general affine connections are
different from some general (or special Finsler-type) d-connections because
different geometric principles define them. Nevertheless, all geometric and
analytic constructions and respective computations can be related by
respective distortions (\ref{affinedist}).

Certain geometric data $\left( TV,g,D\right) $ define a general \textit{%
metric-affine structure }on a tangent Lorentz bundle $TV.$ We use
not-boldface symbols because, in general, it is not N-adapted. Such data $%
\left( T^{\ast }V,\ ^{\shortmid }g,\ ^{\shortmid }D\right) $ can be
considered also for cotangent Lorentz bundle $T^{\ast }V$. For such
metric-affine phase spaces, we can also introduce formal h-v, h-c, or diadic
splitting but the linear connections $D$ and $\ ^{\shortmid }D$ are not d-or
s-connections. The corresponding formulas for fundamental geometric objects
are written with "non-boldface" symbols, for instance, as $\mathcal{T}%
[D]=\{T_{\ \alpha \beta }^{\gamma }[D]\},\ ^{\shortmid }\mathcal{R}[\
^{\shortmid }D]\mathbf{=}\mathbf{\{\ ^{\shortmid }}R_{\ \beta \gamma
\delta}^{\alpha }[\ ^{\shortmid }D]\}$ etc. To avoid ambiguities, we can
emphasize functional dependencies $[D]$ or $[\ ^{\shortmid }D]$, stating
that we work with not N-adapted geometric structures. This does not allow us
to apply the AFCDM for general decoupling and integrating of fundamental
physical systems of nonlinear PDEs. But we can always consider s-adapted
frames and distortion of general affine connections to certain classes of
s-connections, 
\begin{equation}
D=\ _{s}\mathbf{D+}\ _{s}\mathbf{Z}\mbox{ and }\ ^{\shortmid }D=\
_{s}^{\shortmid }\mathbf{D+}\ _{s}^{\shortmid }\mathbf{Z.}
\label{affinesdist}
\end{equation}%
This allows us to define and computer distortion of fundamental geometric
objects in certain canonical forms then to construct generic off-diagonal
solutions (see section \ref{sec4}). 

\subsubsection{Physically important and canonical and dyadic FLH
d-connections}

Various classes of Finsler-like linear connections and d-connections were
considered for elaborating classical and quantum FLH and MGTs or for an
alternative geometrization of mechanics and nonholonomic geometric flow
theories. We reviewed them in chronological form in paragraphs 2-7] of the
previous section. In this subsection, we define eight type geometric and
physically important linear connections defining LC-configurations, almost K%
\"{a}hler-Lagrange and almost K\"{a}hler-Hamilton structures, nonholonomic
dyadic decompositions, etc., for relativistic FLH spaces. 

On a relativistic phase space $\mathcal{M}$, we can define in abstract and
N- or s-adapted forms such eight important linear connection structures: 
\begin{eqnarray}
\lbrack \mathbf{g,N]} &\mathbf{\simeq }&\mathbf{[}\widetilde{\mathbf{g}},%
\widetilde{\mathbf{N}}]\mathbf{\simeq \lbrack }\widetilde{\theta }:=%
\widetilde{\mathbf{g}}(\widetilde{\mathbf{J}}\cdot ,\cdot ),\widetilde{%
\mathbf{P}}\mathbf{,}\widetilde{\mathbf{J}}\mathbf{,}\widetilde{\mathbb{J}}]%
\mathbf{\simeq }[\ _{s}\mathbf{g,}\ _{s}\mathbf{N]}  \label{canonicalscon} \\
&\Longrightarrow &\left\{ 
\begin{array}{ccccc}
\nabla : &  & \nabla \mathbf{g}=0;\ \mathcal{T}\mathbf{[\nabla ]}=0, &  & %
\mbox{ LC--connection}; \\ 
\widetilde{\mathbf{D}}: &  & \widetilde{\mathbf{D}}\widetilde{\theta }=0,%
\widetilde{\mathbf{D}}\widetilde{g}=0 &  & 
\mbox{almost symplectic
Lagrange d-connection}; \\ 
\widehat{\mathbf{D}}: &  & \widehat{\mathbf{D}}\ \mathbf{g}=0;\ h\widehat{%
\mathcal{T}}=0,\ v\widehat{\mathcal{T}}=0, &  & 
\mbox{canonical Lagrange
d-connection}; \\ 
\ _{s}\widehat{\mathbf{D}}: &  & 
\begin{array}{c}
\ _{s}\widehat{\mathbf{D}}\ _{s}\mathbf{g}=0;\ ^{s}h\widehat{\mathcal{T}}%
=0,\ \ ^{s}v\widehat{\mathcal{T}}=0, \\ 
\ ^{s^{\prime }}h\ ^{s}v\widehat{\mathcal{T}}\neq 0\ ^{s^{\prime }}v\ ^{s}v%
\widehat{\mathcal{T}}\neq 0,s^{\prime }\neq s,%
\end{array}
&  & \mbox{canonical s-connection}; \\ 
\begin{array}{c}
\mathbf{D:} \\ 
_{Q}\mathbf{D:}%
\end{array}
&  & 
\begin{array}{c}
\mathcal{Q}\mathbf{:=Dg}\neq 0,\mathcal{T}\neq 0; \\ 
\mathcal{Q}\mathbf{:=}\ _{Q}\mathbf{Dg}\neq 0,\ _{Q}\mathcal{T}=0;%
\end{array}
&  & \mbox{ nonmetric N-adapted phase spaces }; \\ 
\begin{array}{c}
D\mathbf{:} \\ 
_{Q}D\mathbf{:}%
\end{array}
&  & 
\begin{array}{c}
\mathcal{Q}:=Dg\neq 0,\mathcal{T}\neq 0; \\ 
\mathcal{Q}:=\ _{Q}Dg\neq 0,\ _{Q}\mathcal{T}=0;%
\end{array}
&  & \mbox{  not-N-adapted  nonmetric phase spaces }.%
\end{array}%
\right.  \notag
\end{eqnarray}

For $\ ^{\shortmid }\mathcal{M},$ we can define important (dual) linear
connection structures by using respective similar abstract formulas. We
emphasize that the geometric constructions can be performed in a dual form
to those on $\ \mathcal{M}.$ It should be noted that, in general, the
Lagrange mechanics is not equivalent to Hamilton mechanics. The almost
symplectic models are with different types of almost K\"{a}hler N-adapted
connections if we try to elaborate on models with symplectomorphisms etc,
see details and coefficient formulas in \cite{mhss2000,vacaru18,vmon3}. We
outline eight dual d-connections which are not adaptation to general
symplectic transforms (such an adapting requests more sophisticated
definitions and cumbersome formulas): 
\begin{eqnarray}
\lbrack \ ^{\shortmid }\mathbf{g,\ ^{\shortmid }N]} &\mathbf{\simeq }&%
\mathbf{[}\ ^{\shortmid }\widetilde{\mathbf{g}},\ ^{\shortmid }\widetilde{%
\mathbf{N}}]\mathbf{\simeq \lbrack }\ ^{\shortmid }\widetilde{\theta }:=\
^{\shortmid }\widetilde{\mathbf{g}}(\ ^{\shortmid }\widetilde{\mathbf{J}}%
\cdot ,\cdot ),\ ^{\shortmid }\widetilde{\mathbf{P}}\mathbf{,}\ ^{\shortmid }%
\widetilde{\mathbf{J}}\mathbf{,}\ ^{\shortmid }\widetilde{\mathbb{J}}]%
\mathbf{\simeq }[\ _{s}^{\shortmid }\mathbf{g,}\ _{s}^{\shortmid }\mathbf{N]}
\label{canonicalsdcon} \\
&\Longrightarrow &\left\{ 
\begin{array}{ccccc}
\ ^{\shortmid }\nabla : &  & \ ^{\shortmid }\nabla \ ^{\shortmid }\mathbf{g}%
=0;\ \ ^{\shortmid }\mathcal{T}\mathbf{[\ ^{\shortmid }\nabla ]}=0, &  & %
\mbox{ LC--connection}; \\ 
\ ^{\shortmid }\widetilde{\mathbf{D}}: &  & \ ^{\shortmid }\widetilde{%
\mathbf{D}}\ ^{\shortmid }\widetilde{\theta }=0,\ ^{\shortmid }\widetilde{%
\mathbf{D}}\ ^{\shortmid }\widetilde{g}=0 &  & 
\mbox{alm. symple.
Hamilton  d-connect.}; \\ 
\ ^{\shortmid }\widehat{\mathbf{D}}: &  & \ ^{\shortmid }\widehat{\mathbf{D}}%
\ \mathbf{g}=0;\ h\ ^{\shortmid }\widehat{\mathcal{T}}=0,\ c\ ^{\shortmid }%
\widehat{\mathcal{T}}=0, &  & \mbox{canonical Hamilton d-connection}; \\ 
\ _{s}^{\shortmid }\widehat{\mathbf{D}}: &  & 
\begin{array}{c}
\ _{s}^{\shortmid }\widehat{\mathbf{D}}\ _{s}^{\shortmid }\mathbf{g}=0;\
^{s}h\ ^{\shortmid }\widehat{\mathcal{T}}=0,\ \ ^{s}c\ ^{\shortmid }\widehat{%
\mathcal{T}}=0, \\ 
\ ^{s^{\prime }}h\ ^{s}v\ ^{\shortmid }\widehat{\mathcal{T}}\neq 0\
^{s^{\prime }}v\ ^{s}v\ ^{\shortmid }\widehat{\mathcal{T}}\neq 0,s^{\prime
}\neq s,%
\end{array}
&  & \mbox{canonical dual s-connection}; \\ 
\begin{array}{c}
\ ^{\shortmid }\mathbf{D:} \\ 
\ _{Q}^{\shortmid }\mathbf{D:}%
\end{array}
&  & 
\begin{array}{c}
\ ^{\shortmid }\mathcal{Q}\mathbf{:=\ ^{\shortmid }D\ ^{\shortmid }g}\neq
0,\ ^{\shortmid }\mathcal{T}\neq 0; \\ 
\ ^{\shortmid }\mathcal{Q}\mathbf{:=}\ _{Q}^{\shortmid }\mathbf{D\
^{\shortmid }g}\neq 0,\ _{Q}^{\shortmid }\mathcal{T}=0;%
\end{array}
&  & \mbox{ nonmetric N-adapted  phase space}; \\ 
\begin{array}{c}
\ ^{\shortmid }D\mathbf{:} \\ 
_{Q}^{\shortmid }D\mathbf{:}%
\end{array}
&  & 
\begin{array}{c}
\ ^{\shortmid }\mathcal{Q}:=\ ^{\shortmid }D\ ^{\shortmid }g\neq 0,\
^{\shortmid }\mathcal{T}\neq 0; \\ 
\ ^{\shortmid }\mathcal{Q}:=\ _{Q}^{\shortmid }D\ ^{\shortmid }g\neq 0,\
_{Q}^{\shortmid }\mathcal{T}=0;%
\end{array}
&  & \mbox{ not-N-adapted  phase spaces}.%
\end{array}%
\right.  \notag
\end{eqnarray}

Let us explain some very important properties of the linear connections (\ref%
{canonicalscon}) and (\ref{canonicalsdcon}):

\begin{itemize}
\item[{[a]}] The LC-connections $\ \nabla $ and $\ ^{\shortmid }\nabla $ can
be defined in standard forms using corresponding d-metrics (\ref{cdms8}) and
(\ref{cdmds8}), or (\ref{lqe}) and (\ref{lqed}), or (\ref{dmgener}) and (\ref%
{offd}), or their off-diagonal representations for coordinate bases on phase
spaces. Such linear connections can be used for elaborating FLH models by
analogy to higher dimension extensions of the Einstein gravity, when
extra-dimension coordinates are velocity or momentum type. We can construct
diagonal configurations, for instance, certain BH solutions as in higher
dimension gravity, in string gravity theories, etc., see discussions in \cite%
{vacaru18,vmon3,bsssvv25}. To construct generic off-diagonal solutions using
only $\nabla $ or $\ ^{\shortmid}\nabla $ is a very difficult task because
we are not able to prove any general decoupling properties. We can encode in
such LC-configurations certain FLH data, but in general such theories are
not Finsler-like because the LC-connections are not adapted to certain
N-connection structures. Here we note that any metric-affine phase space
geometry defined by some data $(\mathcal{M},g,D)$ involves bi-connection, $%
(\nabla \lbrack g],D),$ and distortion configurations, $(\mathcal{M}%
,g,D=\nabla +Z).$ If we introduce a N-connection structure $\mathbf{N}$ on $%
\mathcal{M}$, we can perform N-adapted geometric constructions with $(%
\mathcal{M},\mathbf{N},\mathbf{g},\mathbf{D})$ as on nonholonomic manifolds
and (co) tangent bundles. Corresponding bi-connection, $(\ \nabla \lbrack 
\mathbf{g}],\mathbf{D}),$ and distortion N-adapted configurations, $(%
\mathcal{M},\mathbf{N},\mathbf{g},\mathbf{D}=\nabla \lbrack \mathbf{g}]+%
\mathbf{Z})$ can be also defined. To prescribe/ or define an N-connection
structure is crucial for constructing FLH theories even, in general, a
d-connection $\mathbf{D}$ can be an arbitrary one. A distortion d-tensor $%
\mathbf{Z}$ can be determined from certain fundamental geometric of modified
gravitational field equations for a postulated FLH or other type MGT. Here
we note that N- and d-connections can also be introduced in GR and
"non-Finsler" gravity theories if we prescribe a N-connection as
nonholonomic distribution on a $V$ and, respectively, $TV$. For non-FLH
theories, the N-connection structure is not obligatory of type (\ref{cartnc}%
) but can be a general one (\ref{ncon2222}). Nevertheless, nonholonomic
dyadic splitting and distortion of connection formalism are important for
all MAG theories because they allow applications of the AFCDM for
constructing off-diagonal solutions. For dual phase spaces, the above
formulas are determined by a distortion relation of type $(\ ^{\shortmid }%
\mathcal{M}, \ ^{\shortmid }\mathbf{N}, \ ^{\shortmid }\mathbf{g},\
^{\shortmid }\mathbf{D}=\ ^{\shortmid }\nabla \lbrack \ ^{\shortmid }\mathbf{%
g}]+ \ ^{\shortmid }\mathbf{Z}).$

\item[{[b]}] The almost symplectic d-connections $\ \widetilde{\mathbf{D}}$
and $\ ^{\shortmid }\widetilde{\mathbf{D}}$ (respectively on $\widetilde{%
\mathcal{M}}$ and $\ ^{\shortmid }\widetilde{\mathcal{M}}$) are very
important because they are also equivalent to the Cartan d-connection in
Finsler geometry, see details and index formulas in \cite%
{cartan35,rund59,vmon3,vacaru18,hohmann13}. In abstract geometric form, such
nonholonomic geometries are determined, respectively, by $(\widetilde{%
\mathcal{M}},\widetilde{\mathbf{N}},\widetilde{\mathbf{g}},\widetilde{%
\mathbf{D}}\mathbf{=}\nabla \lbrack \widetilde{\mathbf{g}}]+\widetilde{%
\mathbf{Z}})$ and $(\ ^{\shortmid }\widetilde{\mathcal{M}},\ ^{\shortmid }%
\widetilde{\mathbf{N}}, \ ^{\shortmid }\widetilde{\mathbf{g}},\ ^{\shortmid }%
\widetilde{\mathbf{D}}= \ ^{\shortmid }\nabla \lbrack \ ^{\shortmid }%
\widetilde{\mathbf{g}}]+\ ^{\shortmid }\widetilde{\mathbf{Z}}).$ The
corresponding phase space gravitational field equations possess very special
integration properties \cite{vacaru03,vacaru09b,vacaru12a} but not general
ones. The main priority of such d-connections defined for Finsler-like
variables is that we can perform DQ of FLH MGTs and GR \cite%
{vacaru07,vacaru10a,vacaru13,biv16,bsssvv25,vacaru16a}. Nevertheless, to
prove certain general off-diagonal decoupling properties of corresponding
dynamical or geometric flow equations is not possible if we work only with
the LC-connection. 

\item[{[c]}] The almost symplectic d-connections $\ \widetilde{\mathbf{D}}$
and $\ ^{\shortmid }\widetilde{\mathbf{D}}$ (respectively on $\widetilde{%
\mathcal{M}}$ and $\ ^{\shortmid }\widetilde{\mathcal{M}}$) are very
important because they are also equivalent to the Cartan d-connection in
Finsler geometry, see details and index formulas in \cite%
{cartan35,rund59,vmon3,vacaru18,hohmann13}. In abstract geometric form, such
nonholonomic geometries are determined, respectively, by $(\widetilde{%
\mathcal{M}},\widetilde{\mathbf{N}},\widetilde{\mathbf{g}},\widetilde{%
\mathbf{D}}=\nabla \lbrack \widetilde{\mathbf{g}}]+\widetilde{\mathbf{Z}})$
and $(\ ^{\shortmid }\widetilde{\mathcal{M}},\ ^{\shortmid }\widetilde{%
\mathbf{N}}, \ ^{\shortmid }\widetilde{\mathbf{g}},\ ^{\shortmid }\widetilde{%
\mathbf{D}}= \ ^{\shortmid }\nabla \lbrack \ ^{\shortmid }\widetilde{\mathbf{%
g}}]+\ ^{\shortmid }\widetilde{\mathbf{Z}}).$ The corresponding phase space
gravitational field equations possess very special integration properties 
\cite{vacaru03,vacaru09b,vacaru12a} but not general ones. The main priority
of such d-connections defined for Finsler-like variables is that we can
perform DQ of FLH MGTs and GR \cite%
{vacaru07,vacaru10a,vacaru13,biv16,vacaru16a}. Nevertheless, to prove
certain general off-diagonal decoupling properties of corresponding
dynamical or geometric flow equations is not possible if we work only with
the LC-connection. 

\item[{[d]}] The main goal of this work is to study nonholonomic
metric-affine FLH theories adapted to N-adapted structures determined by
respective geometric data $(\mathcal{M},\mathbf{g,N,D)}$ and $(\ ^{\shortmid}%
\mathcal{M},\ ^{\shortmid }\mathbf{g,\ ^{\shortmid }N,} \ ^{\shortmid }%
\mathbf{D)}$ and show how the AFCDM can be applied in such a case. We can
consider theories when, for instance, $_{Q}\mathbf{D}$ (see definition in (%
\ref{canonicalscon})) is a d-connection with nontrivial nonmetricity but
with zero torsion. This provides a tangent Lorentz bundle generalization of
MAG theories \cite{hehl95,vmon3}, in particular, of $f(Q)$ gravity \cite%
{lheis23,vacaru25b}. The AFCDM can be generalized for such 4-d and 8-d
theories (or other dimensions), which allows us to construct generic
off-diagonal solutions for nonmetric FLH theories (see sections \ref{sec4}
and \ref{sec5}). Finsler-like theories with nonmetricity were criticised in 
\cite{vacaru10,vmon3,vacaru18} because of the problems with definitions of
general nonmetric spinors and the Dirac equation. Nevertheless, we can
elaborate on nonmetric FLH modifications of ED systems using the same
nonholonomic methods as in \cite{vacaru25b} (using velocity/ momentum
variables, we shall study this problem in our further partner works). 

\item[{[e]}] We can consider metric-affine structures $(g,D)$ or $(\
^{\shortmid }g,\ ^{\shortmid }D)$ on respective (co) tangent Lorentz bundles
and postulate certain types of generalized gravitational field equations
with nontrivial torsion and nonmetricity fields. Such phase space MGTs can't
be integrated in general forms if certain special diagonal ansatz are not
considered. It is not clear how to define a self-consistent metric-affine
geometric flow models and respective nonmetric generalizations of EYMHD
systems etc. Introducing formal s-connection structures (\ref{ncon2222}),
with respective distortions and s-adapted frames, we can generate solutions
for physically important systems of nonlinear PDEs. We can speculate when
such MGTs can be related to certain FLH configurations if certain effective
backgrounds are determined by velocity/ momentum - like variables. This can
be performed for respective nonholonomic dyadic splitting when, for
instance, $(g,D)\rightarrow (\ _{s}\mathbf{g,\ _{s}N,}D=\ _{s}\mathbf{%
\widehat{\mathbf{D}}+}\ _{s}\mathbf{\mathbf{\widehat{\mathbf{Z}}}),}$ see
also distortions (\ref{affinesdist}).
\end{itemize}


FLH theories on (co) tangent Lorentz bundles (or other types of nonholonomic
bundle/ manifolds) have been modeled by different authors using different
types of d-metric and d-connection structures as we outlined in paragraphs
1-7] of section \ref{sec1}. We provided explicit examples and discussions
related to formulas (\ref{flh}), (\ref{affinedist}) and (\ref{affinesdist}).
Above defined affine connections and d- or s-connections (\ref{canonicalscon}%
) and (\ref{canonicalsdcon}) can be re-defined into each other, or related
to other types of ones using distortion formulas. We can fix $\ ^{F}\mathbf{D%
}=\ ^{B}\mathbf{D,}$ or $^{F}\mathbf{D}=\ ^{C}\mathbf{D,}$ or $^{F}\mathbf{D}%
=\ \widetilde{\mathbf{D}},$ and any other type of (generalized)
Finsler-Lagrange d-connection encoding velocity type variables. On dual
phase spaces with momentum like variables $(\ ^{\shortmid}\mathcal{M},\
^{\shortmid }\mathbf{g,\ ^{\shortmid }N),}$ the respective linear
connection/ d-connection structures are labeled, for instance, as $\
^{\shortmid }\widetilde{\mathbf{D}},\ _{H}^{\shortmid }\mathbf{D}$, etc.,
encoding some effective Hamilton structures. The geometric objects $(\mathbf{%
g,N)}$ or $(\ ^{\shortmid }\mathbf{g,\ ^{\shortmid }N)}$ can be arbitrary
ones, or related via frame transforms to other ones with (or not) adapted
N-or s-adapted structures, for instance, 
\begin{eqnarray*}
(\mathbf{g} &\mathbf{\simeq }&\ ^{F}\mathbf{g\simeq }\ ^{L}\mathbf{g\simeq }%
\ _{s}\mathbf{\mathbf{g\simeq }\ }\{g_{\alpha \beta }\}\mathbf{\mathbf{%
\simeq }\ }\{g_{\alpha _{s}\beta _{s}}\}\mathbf{,N\mathbf{\simeq }}\ ^{F}%
\mathbf{N\mathbf{\simeq }}\ ^{L}\mathbf{\mathbf{\mathbf{N}\simeq }}\ _{s}%
\mathbf{N\mathbf{\mathbf{\simeq }\ }\{}N_{i}^{a}\mathbf{\}\mathbf{\mathbf{%
\simeq }\ }}\{N_{i_{s-1}}^{a_{s}}\}),\mbox{ or } \\
(\ ^{\shortmid }\mathbf{g} &\mathbf{\simeq }&\ \ _{\shortmid }^{H}\mathbf{%
g\simeq }\ _{s}^{\shortmid }\mathbf{\mathbf{g\simeq }\ }\{\ ^{\shortmid
}g_{\alpha \beta }\}\mathbf{\mathbf{\simeq }\ }\{\ ^{\shortmid }g_{\alpha
_{s}\beta _{s}}\}\mathbf{,\ ^{\shortmid }N\mathbf{\simeq }}\ \ _{\shortmid
}^{H}\mathbf{\mathbf{\mathbf{N}\simeq }}\ _{s}^{\shortmid }\mathbf{N\mathbf{%
\mathbf{\simeq }\ }\{}\ ^{\shortmid }N_{ia}\mathbf{\}\mathbf{\mathbf{\simeq }%
\ }}\{\ ^{\shortmid }N_{i_{s-1}a_{s}}\}).
\end{eqnarray*}%
For such geometric data, we can postulate different types of such
FLH-modified Einstein equations, but there are both conceptual and technical
difficulties and constructing physically important solutions of
corresponding systems of nonlinear PDEs. 

To apply the AFCDM we can use necessary types of distortion relations:%
\begin{eqnarray}
\widehat{\mathbf{D}} &=&\nabla +\widehat{\mathbf{Z}},\widetilde{\mathbf{D}}%
=\nabla +\widetilde{\mathbf{Z}},\mbox{ and }\widehat{\mathbf{D}}=\widetilde{%
\mathbf{D}}+\mathbf{Z,}\mbox{  determined by }(\mathbf{g,N)};
\label{distortcan} \\
\ ^{L}\widehat{\mathbf{D}} &=&\ ^{L}\nabla +\ ^{L}\widehat{\mathbf{Z}},\ ^{L}%
\widetilde{\mathbf{D}}=\ ^{L}\nabla +\ ^{L}\widetilde{\mathbf{Z}},%
\mbox{ and
}\ ^{L}\widehat{\mathbf{D}}=\ ^{L}\mathbf{D}+\ ^{L}\mathbf{Z,}%
\mbox{
determined by }(\ ^{L}\mathbf{g,}\ ^{L}\mathbf{N)};  \notag \\
\ _{s}\widehat{\mathbf{D}} &=&\widetilde{\mathbf{D}}+\ _{s}\mathbf{Z,}\
_{s}^{L}\widehat{\mathbf{D}}=\ ^{L}\mathbf{D}+\ _{s}^{L}\mathbf{Z,}%
\mbox{
determined by }(\ _{s}\mathbf{g}\simeq \ ^{L}\mathbf{\mathbf{g},\ _{s}N}%
\simeq \ ^{L}\mathbf{N);}  \notag \\
\ _{s}\widehat{\mathbf{D}} &=&\ \mathbf{D}+\ _{s}\mathbf{Z}\mbox{ or }\
_{s}^{Q}\widehat{\mathbf{D}}=\ _{Q}\mathbf{D}+\ _{s}^{Q}\mathbf{Z}%
\mbox{ for
nonmetric affine connections },  \notag \\
\ _{s}\widehat{\mathbf{D}} &=&\ D+\ _{s}\mathbf{Z}\mbox{ or }\ _{s}^{Q}%
\widehat{\mathbf{D}}=\ _{Q}D+\ _{s}^{Q}\mathbf{Z}%
\mbox{ for nonmetric affine
connections }.  \notag
\end{eqnarray}%
Similar formulas can be defined for distortions on dual phase space $\
^{\shortmid }\mathcal{M}$ enabled if necessary with nonholonomic dyadic
structures. For instance, we can write:%
\begin{eqnarray}
&&...  \notag \\
\ _{s}^{\shortmid }\widehat{\mathbf{D}} &=&\ ^{\shortmid }\widetilde{\mathbf{%
D}}+\ _{s}^{\shortmid }\mathbf{Z,}\ _{s}^{H}\widehat{\mathbf{D}}=\ ^{H}%
\mathbf{D}+\ _{s}^{H}\mathbf{Z,}\mbox{determined by }(\ _{s}^{\shortmid }%
\mathbf{g}\simeq \ ^{H}\mathbf{\mathbf{g},\ _{s}^{\shortmid }N}\simeq \ ^{H}%
\mathbf{N);}  \label{distortcand} \\
&&... .  \notag
\end{eqnarray}

Additionally to (\ref{distortcan}), we can consider various types of known
FLH and other types linear connection structures and relate them via
corresponding distortion relations. The priority of "hat" connections is
that $\widehat{\mathbf{D}}$ can be used for general decoupling of
gravitational field equations in 2+2 dimensional MGTs; and $\ _{s}\widehat{%
\mathbf{D}}$ allows a general decoupling of conventional 2(3)+2+2+2 FLH
theories. If certain MGTs are formulated in non-hat geometric variables, we
can consider respective nonholonomic frame transforms and distortion
relations to certain canonical data which allows a general decoupling of
necessary systems of nonlinear PDEs. Using hat variables, various terms
defined by distortion d- and s-tensors (for instance, $\widehat{\mathbf{Z}},%
\widetilde{\mathbf{Z}}$ or $\mathbf{Z}$) are conventionally encoded into
(effective) generating sources, which are determined both by energy-momentum
tensors of matter field but also by distortions of connections (resulting in
effective sources). This allow us to find very general classes of
off-diagonal solutions which may have, or not, certain physical importance.
To extract LC-configurations for $\nabla $ or $\ ^{\shortmid }\nabla $ is
possible if additional nonholonomic constraints are imposed on integrating
and generating functions defining phase space configurations with vanishing
distortions. In a similar form but for another types of nonholonomic
distortions constraints, we can extract Finsler-like configurations with $\
^{B}\mathbf{D,}$ or $\ ^{C}\mathbf{D.}$ 

\subsubsection{Distorting curvature, torsion, nonmetricity and Ricci
s-tensors}

Introducing distortions of linear (d/s-) connections (\ref{distortcan}) or (%
\ref{distortcand}) into formulas (\ref{fundgeom}), we can compute in
abstract geometric form the respective curvature, torsion and nonmetricity
d/s-tensors and their distortions. For instance, we can compute the
canonical curvature d-tensors and distortion d-tensors on $\mathcal{M}$ and $%
\ ^{\shortmid }\mathcal{M}$, 
\begin{eqnarray*}
\widehat{\mathcal{R}}[\mathbf{g},\widehat{\mathbf{D}} &=&\nabla +\widehat{%
\mathbf{Z}}]=\mathcal{R}[\mathbf{g},\nabla ]+\widehat{\mathcal{Z}}[\mathbf{g}%
,\widehat{\mathbf{Z}}]\mbox{ and } \\
\ ^{\shortmid }\widehat{\mathcal{R}}[\ ^{\shortmid }\mathbf{g},\ ^{\shortmid
}\widehat{\mathbf{D}} &=&\ ^{\shortmid }\nabla +\ ^{\shortmid }\widehat{%
\mathbf{Z}}]=\ ^{\shortmid }\mathcal{R}[\ ^{\shortmid }\mathbf{g},\
^{\shortmid }\nabla ]+\ ^{\shortmid }\widehat{\mathcal{Z}}[\ ^{\shortmid }%
\mathbf{g},\ ^{\shortmid }\widehat{\mathbf{Z}}].
\end{eqnarray*}%
Contracting the first and third indices (we can consider distortions of
coefficient formulas (\ref{fundgeomdcoef})), we obtain such formulas for the
canonical Ricci d-tensors,%
\begin{eqnarray*}
\widehat{R}ic[\mathbf{g},\widehat{\mathbf{D}} &=&\nabla +\widehat{\mathbf{Z}}%
]=Ric[\mathbf{g},\nabla ]+\widehat{Z}ic[\mathbf{g},\widehat{\mathbf{Z}}]%
\mbox{ and } \\
\ ^{\shortmid }\widehat{R}ic[\ ^{\shortmid }\mathbf{g},\ ^{\shortmid }%
\widehat{\mathbf{D}} &=&\ ^{\shortmid }\nabla +\ ^{\shortmid }\widehat{%
\mathbf{Z}}]=\ ^{\shortmid }Ric[\ ^{\shortmid }\mathbf{g},\ ^{\shortmid
}\nabla ]+\ ^{\shortmid }\widehat{Z}ic[\ ^{\shortmid }\mathbf{g},\
^{\shortmid }\widehat{\mathbf{Z}}].
\end{eqnarray*}%
In dyadic form, above formulas can be re-written, for instance, for the
curvature s-tensor:%
\begin{eqnarray*}
\ _{s}^{\shortmid }\widehat{R}[\mathbf{g},\widehat{\mathbf{D}} &=&\nabla +%
\widehat{\mathbf{Z}}]=\mathcal{R}[\mathbf{g},\nabla ]+\ _{s}\widehat{Z}[%
\mathbf{g},\widehat{\mathbf{Z}}]\mbox{ and } \\
\ _{s}^{\shortmid }\widehat{R}[\ ^{\shortmid }\mathbf{g},\ ^{\shortmid }%
\widehat{\mathbf{D}} &=&\ ^{\shortmid }\nabla +\ ^{\shortmid }\widehat{%
\mathbf{Z}}]=\ _{s}^{\shortmid }R[\ ^{\shortmid }\mathbf{g},\ ^{\shortmid
}\nabla ]+\ _{s}^{\shortmid }\widehat{Z}[\ ^{\shortmid }\mathbf{g},\
^{\shortmid }\widehat{\mathbf{Z}}].
\end{eqnarray*}%
Using the almost symplectic (i.e. the Finsler-Cartan) d-connection, the
curvature and distortion d-tensors can be computed for the almost symplectic
Lagrange-Hamilton spaces as we \ defined in subsection \ref{ssalmsymplectic}%
, 
\begin{eqnarray*}
\widetilde{\mathcal{R}}[\widetilde{\mathbf{g}} &\simeq &\widetilde{\theta },%
\widetilde{\mathbf{D}}=\nabla +\widetilde{\mathbf{Z}}]=\mathcal{R}[%
\widetilde{\mathbf{g}}\simeq \widetilde{\theta },\nabla ]+\widetilde{%
\mathcal{Z}}[\widetilde{\mathbf{g}}\simeq \widetilde{\theta },\widetilde{%
\mathbf{Z}}]\mbox{ and } \\
\ ^{\shortmid }\widetilde{\mathcal{R}}[\ ^{\shortmid }\widetilde{\mathbf{g}}
&\simeq &\ ^{\shortmid }\widetilde{\theta },\ ^{\shortmid }\widetilde{%
\mathbf{D}}=\ ^{\shortmid }\nabla +\ ^{\shortmid }\widetilde{\mathbf{Z}}]=\
^{\shortmid }\mathcal{R}[\ ^{\shortmid }\widetilde{\mathbf{g}}\simeq \
^{\shortmid }\widetilde{\theta },\ ^{\shortmid }\nabla ]+\ ^{\shortmid }%
\widetilde{\mathcal{Z}}[\ ^{\shortmid }\mathbf{g}\simeq \ ^{\shortmid }%
\widetilde{\theta },\ ^{\shortmid }\widetilde{\mathbf{Z}}].
\end{eqnarray*}

To elaborate on the AFCDM for generating solutions of FLH gravitational
equations, we have to consider N-- and s-adapted formulas of fundamental
geometric objects, see (\ref{fundgeom}). The d-tensors (\ref{fundgeomdcoef})
of a general metric-affine d-connection$\ \mathbf{D}$ (\ref{dconc}) on $%
\mathcal{M}$ are defined by such coefficient formulas: 
\begin{eqnarray}
\mathcal{R} &=&\mathbf{\{R}_{\ \beta \gamma \delta }^{\alpha }=(R_{\
hjk}^{i},R_{\ bjk}^{a},P_{\ hja}^{i},P_{\ bja}^{c},S_{\ hba}^{i},S_{\
bea}^{c})\},\mbox{ d-curvature},  \notag \\
&\mbox{ for }& R_{\ hjk}^{i} =\mathbf{e}_{k}L_{\ hj}^{i}-\mathbf{e}_{j}L_{\
hk}^{i}+L_{\ hj}^{m}L_{\ mk}^{i}-L_{\ hk}^{m}L_{\ mj}^{i}-C_{\ ha}^{i}\Omega
_{\ kj}^{a},  \notag \\
&&R_{\ bjk}^{a} =\mathbf{e}_{k}\acute{L}_{\ bj}^{a}-\mathbf{e}_{j}\acute{L}%
_{\ bk}^{a}+\acute{L}_{\ bj}^{c}\acute{L}_{\ ck}^{a}-\acute{L}_{\ bk}^{c}%
\acute{L}_{\ cj}^{a}-C_{\ bc}^{a}\Omega _{\ kj}^{c},  \label{dcurv} \\
&& P_{\ jka}^{i} = e_{a}L_{\ jk}^{i}-D_{k}\acute{C}_{\ ja}^{i}+\acute{C}_{\
jb}^{i}T_{\ ka}^{b},\ P_{\ bka}^{c}=e_{a}\acute{L}_{\ bk}^{c}-D_{k}C_{\
ba}^{c}+C_{\ bd}^{c}T_{\ ka}^{c},  \notag \\
&& S_{\ jbc}^{i} = e_{c}\acute{C}_{\ jb}^{i}-e_{b}\acute{C}_{\ jc}^{i}+%
\acute{C}_{\ jb}^{h}\acute{C}_{\ hc}^{i}-\acute{C}_{\ jc}^{h}\acute{C}_{\
hb}^{i},\hspace{0in}\ S_{\ bcd}^{a}=e_{d}C_{\ bc}^{a}-e_{c}C_{\ bd}^{a}+C_{\
bc}^{e}C_{\ ed}^{a}-C_{\ bd}^{e}C_{\ ec}^{a};  \notag \\
&& \mathcal{T} = \{\mathbf{T}_{\ \alpha \beta }^{\gamma }=(T_{\ jk}^{i},T_{\
ja}^{i},T_{\ ji}^{a},T_{\ bi}^{a},T_{\ bc}^{a})\},\mbox{ d-torsion},  \notag
\\
&\mbox{ for }& T_{\ jk}^{i} = L_{jk}^{i}-L_{kj}^{i},T_{\ jb}^{i}=\acute{C}%
_{jb}^{i},T_{\ ji}^{a}=-\Omega _{\ ji}^{a},\ T_{aj}^{c}=\acute{L}%
_{aj}^{c}-e_{a}(N_{j}^{c}),T_{\ bc}^{a}=C_{bc}^{a}-C_{cb}^{a};  \label{dtors}
\\
&& \mathcal{Q} =\mathbf{\{Q}_{\gamma \alpha \beta }=\left(
Q_{kij},Q_{kab},Q_{cij},Q_{cab}\right) \},\mbox{d-nonmetricity},  \notag \\
&\mbox{ for }& Q_{kij}
=D_{k}g_{ij},Q_{kab}=D_{k}g_{ab},Q_{cij}=D_{c}g_{ij},Q_{cab}=D_{c}g_{ab}.
\label{dnonm}
\end{eqnarray}

Contracting the first and forth d-indices in (\ref{dcurv}), the
h-v-coefficients of the Ricci d-tensor on $\mathcal{M}$, $\mathbf{R}ic=\{%
\mathbf{R}_{\ \beta \gamma }:=\mathbf{R}_{\ \beta \gamma \alpha
}^{\alpha}\}, $ split into four groups of coefficients, 
\begin{equation}
\mathbf{R}_{\alpha \beta }=\{R_{ij}:=R_{\ ijk}^{k},\ R_{ia}:=-R_{\
ika}^{k},\ R_{ai}:=R_{\ aib}^{b},\ R_{ab}:=R_{\ abc}^{c}\}.  \label{driccic}
\end{equation}%
Using the inverse d-tensor of a d-metric (\ref{dmgener}), we can compute the
scalar curvature $\ ^{sc}R$ of $\ \mathbf{D}$ is by definition 
\begin{equation}
Rsc:=\mathbf{g}^{\alpha \beta }\mathbf{R}_{\ \alpha
\beta}=g^{ij}R_{ij}+g^{ab}R_{ab}.  \label{sdcurv}
\end{equation}

On a phase space $\ ^{\shortmid }\mathcal{M}$ with momentum variables, the
formulas (\ref{dcurv}), (\ref{dtors}) and (\ref{dnonm}) can be written for
respective labels " $^{\shortmid }$". For instance, the analogs of formulas (%
\ref{driccic}) and (\ref{sdcurv}) can be written in the form:%
\begin{eqnarray}
\ ^{\shortmid }\mathbf{R}_{\alpha \beta } &=&\{\ ^{\shortmid }R_{ij}:=\
^{\shortmid }R_{\ ijk}^{k},\ ^{\shortmid }R_{i}^{\ a}:=-\ ^{\shortmid }R_{\
ik}^{k\ a},\ ^{\shortmid }R_{\ i}^{a}:=\ ^{\shortmid }R_{b\ i}^{\ a\ b},\
^{\shortmid }R^{ab}:=\ ^{\shortmid }R_{\ c\ \ c}^{\ ab}\}\mbox{ and }  \notag
\\
\ ^{\shortmid }Rsc &=&\ ^{\shortmid }\mathbf{g}^{\alpha \beta }\ ^{\shortmid
}\mathbf{R}_{\ \alpha \beta }=\ ^{\shortmid }g^{ij}\ ^{\shortmid }R_{ij}+\
^{\shortmid }g_{ab}\ ^{\shortmid }R^{ab}.  \label{sdcurvd}
\end{eqnarray}

The formulas (\ref{dcurv})--(\ref{sdcurvd}) can be written also with dyadic
coefficients for any type of linear connections and d-/ s-connections (\ref%
{distortcan}) or (\ref{distortcand}). For canonical s-variables, such
details are provided in \cite{vacaru18,partner02,bsssvv25}.

Let us consider an important example for the canonical d-connection $%
\widehat{\mathbf{D}}=\{\widehat{\mathbf{\Gamma }}_{\ \alpha \beta }^{\gamma
}=(\widehat{L}_{jk}^{i},\widehat{L}_{bk}^{a},\widehat{C}_{jc}^{i},\widehat{C}%
_{bc}^{a})\}$ defined by N-adapted coefficients 
\begin{eqnarray}
\widehat{L}_{jk}^{i} &=&\frac{1}{2}g^{ir}(\mathbf{e}_{k}g_{jr}+\mathbf{e}%
_{j}g_{kr}-\mathbf{e}_{r}g_{jk}),  \label{cdc} \\
\widehat{L}_{bk}^{a} &=&e_{b}(N_{k}^{a})+\frac{1}{2}g^{ac}(\mathbf{e}%
_{k}g_{bc}-g_{dc}\ e_{b}N_{k}^{d}-g_{db}\ e_{c}N_{k}^{d}),  \notag \\
\widehat{C}_{jc}^{i} &=&\frac{1}{2}g^{ik}e_{c}g_{jk},\ \widehat{C}_{bc}^{a}=%
\frac{1}{2}g^{ad}(e_{c}g_{bd}+e_{b}g_{cd}-e_{d}g_{bc}).  \notag
\end{eqnarray}%
Such coefficients are computed for a general d--metric $\mathbf{g}%
=[g_{ij},g_{ab}]$ (\ref{dmgener}) with respect to N--adapted frames. Such
coefficients are different from those of the LC-connection $\nabla =\{\Gamma
_{\ \alpha \beta }^{\gamma }\}$ computed with respect to the same system of
reference. The N-adapted coefficients of the canonical distortion d-tensor
in (\ref{distortcan}) can be computed as $\widehat{\mathbf{Z}}=\{\widehat{%
\mathbf{Z}}_{\ \alpha \beta }^{\gamma }=\widehat{\mathbf{\Gamma }}_{\ \alpha
\beta }^{\gamma }-\Gamma _{\ \alpha \beta }^{\gamma }\}.$ Introducing $%
\widehat{\mathbf{\Gamma }}_{\ \alpha \beta }^{\gamma }$ \ (\ref{cdc}) into (%
\ref{dcurv})--(\ref{sdcurv}) (instead of the coefficients of a general
d-connection $\mathbf{\Gamma }_{\ \alpha \beta }^{\gamma })$, we can compute
the N-adapted coefficients of canonical fundamental d--objects. Such
coefficients are labelled, for instance, as $\widehat{\mathcal{R}}=\{%
\widehat{\mathbf{R}}_{\ \beta \gamma \delta }^{\alpha }=(\widehat{R}_{\
hjk}^{i},\widehat{R}_{\ bjk}^{a},...)\},\ \widehat{\mathcal{T}}=\{\widehat{%
\mathbf{T}}_{\ \alpha \beta }^{\gamma }=(\widehat{T}_{\ jk}^{i},\widehat{T}%
_{\ ja}^{i},...)\},$ for $\widehat{\mathcal{Q}}=\{\widehat{\mathbf{Q}}%
_{\gamma \alpha \beta }=(\widehat{Q}_{kij}=0,\widehat{Q}_{kab}=0)=0,$ and
similarly for $\widehat{\mathbf{R}}_{\alpha \beta }=\{\widehat{R}_{ij}:=%
\widehat{R}_{\ ijk}^{k},\ ...\}$ and $\widehat{R}sc:=\mathbf{g}^{\alpha
\beta }\widehat{\mathbf{R}}_{\ \alpha \beta }=g^{ij}\widehat{R}_{ij}+g^{ab}%
\widehat{R}_{ab}.$ Such formulas (in general, with additional nonholonomic
dyadic splitting) will be used in section \ref{sec4} to prove important
general decoupling and integration properties of FLH and geometric flow
modified Einstein equations. 

\subsection{Metric and nonmetric FLH-deformed Einstein equations}

Finsler-like gravity theories can be formulated for two classes of
d-connection structures: metric-compatible d-connections (for instance,
using the Cartan or the canonical ones) or noncompatible, for instance,
using the Chern or Berwald d-connections. Many variants of modified
gravitational field equations were postulated by different authors as we
discussed in paragraphs 3-7] of section \ref{sec1}. In our approach, we
argue that all FLH theories can be elaborated in a unified form as
nonholonomic metric-affine theories constructed on but on relativistic phase
spaces $\mathcal{M}$ and $\ ^{\shortmid }\mathcal{M}$ as in \cite%
{hehl95,vmon3,lheis23,vacaru18,partner02,partner06,bsssvv25}. Explicit cases
of physically important systems of nonlinear PDEs for FLH MGTs and certain
classes of solutions can be distinguished by respective types of
distortions, generating functions and effective sources, as we motivated in
detail in \cite{bvvz24,bnsvv24,vacaru25b} (see proofs in the next section). 

\subsubsection{Canonical nonholonomic metric affine structures in FLH gravity%
}

Any geometric and gravity model for nonholonomic metric-affine
generalizations of FLH theories can be formulated in N-adapted form using
the geometric objects (\ref{canonicalscon}) and (\ref{canonicalsdcon}) and
respective fundamental geometric d-objects (\ref{dcurv}) - (\ref{sdcurv}).
The formulations in canonical nonholonomic variables (with hat-symbols) are
important for decoupling and integrating corresponding modified
gravitational and geometric flow equations using the AFCDM.

In metric compatible form, we shall work with the canonical d-connection $%
\widehat{\mathbf{D}}=\{\widehat{\mathbf{\Gamma }}_{\beta \gamma }^{\sigma}\} 
$ \ (\ref{cdc}) with the nonholonomically induced d-torsion (it can be
considered as an auxiliary one to other types of torsion and nonmetricity
structures on $\mathcal{M}$):%
\begin{equation*}
\widehat{\mathcal{T}}=\{\widehat{\mathbf{T}}_{\beta \gamma }^{\alpha }=%
\widehat{\mathbf{\Gamma }}_{\beta \gamma }^{\sigma }-\widehat{\mathbf{\Gamma 
}}_{\gamma \beta }^{\sigma }+w_{\beta \gamma }^{\sigma }\}.
\end{equation*}%
This canonical d-torsion which is completely determined by the coefficients
of $\ \mathbf{g}$ and $\mathbf{N}$ and anholonomy coefficients $%
w_{\beta\gamma }^{\sigma }$ (for Cartan-Finsler configurations, see (\ref%
{anholcond8})). The contortion s-tensor is defined 
\begin{equation}
\ \widehat{\mathbf{K}}_{\beta \gamma \sigma }=\widehat{\mathbf{T}}_{\beta
\gamma \sigma }+\widehat{\mathbf{T}}_{\gamma \sigma \beta }+\widehat{\mathbf{%
T}}_{\sigma \gamma \beta }.  \label{cotors}
\end{equation}%
Such values with "hats" are induced by an N-connection structure and written
in N-adapted forms.\footnote{%
They include nonholonomic torsion components but in a form which is
different from (for instance) the Riemann-Cartan theory or string gravity
with torsion. In those gravity models, there are considered algebraic
equations for motivating torsion fields as generated by certain spin-like
fluids with nontrivial sources or certain other completely anti-symmetric
torsion fields.}

Using the canonical distortion relations (\ref{distortcan}) can be redefined
in the form 
\begin{equation}
\mathbf{D}=\nabla +\mathbf{L}=\widehat{\mathbf{D}}+\widehat{\mathbf{L}},%
\mbox{ where }\widehat{\mathbf{L}}=\mathbf{L}-\widehat{\mathbf{Z}},
\label{disf}
\end{equation}%
for an additional disformation d-tensor $\mathbf{L}=\{\mathbf{L}_{\ \beta
\lambda }^{\alpha }= \frac{1}{2}(\mathbf{Q}_{\ \beta \lambda }^{\alpha }- 
\mathbf{Q}_{\ \beta \ \lambda }^{\ \alpha } - \mathbf{Q}_{\ \lambda \
\beta}^{\ \alpha })\},$ with $\mathbf{Q}_{\alpha \beta \lambda }:= \mathbf{D}%
_{\alpha }\mathbf{g}_{\beta \lambda },$ for $\widehat{\mathbf{Q}}_{\alpha
\beta \lambda }= \widehat{\mathbf{D}}_{\alpha }\mathbf{g}_{\beta \lambda
}=0. $ For nonmetric geometric constructions with disformations (\ref{disf}%
), we can use respective nonmetricity d-vectors and vectors:%
\begin{equation*}
\mathbf{Q}_{\alpha }=\mathbf{g}^{\beta \lambda }\mathbf{Q}_{\alpha \beta
\lambda }=\mathbf{Q}_{\alpha \ \lambda }^{\ \lambda },\ ^{\intercal }\mathbf{%
Q}_{\beta }=\mathbf{g}^{\alpha \lambda }\mathbf{Q}_{\alpha \beta \lambda }=%
\mathbf{Q}_{\alpha \beta }^{\quad \alpha }.
\end{equation*}%
In our works, we shall use also such geometric d-objects: the nonmetricity
conjugate d-tensor and tensor,%
\begin{equation}
\widehat{\mathbf{P}}_{\ \ \alpha \beta }^{\gamma }=\frac{1}{4}(-2\widehat{%
\mathbf{L}}_{\ \alpha \beta }^{\gamma }+\mathbf{Q}^{\gamma }\mathbf{g}%
_{\alpha \beta }-\ ^{\intercal }\mathbf{Q}^{\gamma }\mathbf{g}_{\alpha \beta
}-\frac{1}{2}\delta _{\alpha }^{\gamma }\mathbf{Q}_{\beta }-\frac{1}{2}%
\delta _{\beta }^{\gamma }\mathbf{Q}_{\alpha }),  \label{nmcjdt}
\end{equation}%
and the nonmetricity scalar for respective d-connections and LC-connection,%
\begin{equation}
\widehat{\mathbf{Q}}=-Q_{\alpha \beta \lambda }\widehat{\mathbf{P}}^{\alpha
\beta \lambda },Q=-Q_{\alpha \beta \lambda }\mathbf{P}^{\alpha \beta \lambda
}\mbox{ and }Q=-Q_{\alpha \beta \lambda }P^{\alpha \beta \lambda }.
\label{nmsc}
\end{equation}

Considering distortion relations (\ref{distortcan}) and (\ref{disf})
relating certain $\nabla =\{\breve{\Gamma}_{\ \beta \gamma }^{\alpha }(u)\},$
$\mathbf{D}=\{\Gamma _{\ \alpha \beta }^{\gamma }\},\widehat{\mathbf{D}}%
=\{\Gamma _{\ \alpha \beta }^{\gamma }\}$ and $D=\{\Gamma _{\ \alpha
\beta}^{\gamma }\},$ we can compute corresponding distortion relations for
fundamental d-tensors (\ref{dcurv})--(\ref{sdcurvd}). For convenience, we
provide here respective parameterizations for distortions of the Ricci
d-tensor and respective Ricci scalars: 
\begin{align}
\mathbf{R}ic& =\breve{R}ic+\breve{Z}ic=\widehat{\mathbf{R}}ic+\widehat{%
\mathbf{Z}}ic,\mbox{ for respective coefficients }  \label{driccidist} \\
& \mathbf{R}ic=\{\mathbf{R}_{\beta \gamma }=\mathbf{R}_{\ \beta \gamma
\alpha }^{\alpha }\};\breve{R}ic=\{\breve{R}_{\beta \gamma }=\breve{R}_{\
\beta \gamma \alpha }^{\alpha }\},\breve{Z}ic=\{\breve{Z}_{\beta \gamma }=%
\breve{Z}_{\ \beta \gamma \alpha }^{\alpha }\};  \notag \\
& \widehat{\mathbf{R}}ic=\{\widehat{\mathbf{R}}_{\beta \gamma }=\widehat{%
\mathbf{R}}_{\ \beta \gamma \alpha }^{\alpha }\},\widehat{\mathbf{Z}}ic=\{%
\widehat{\mathbf{Z}}_{\beta \gamma }:=\widehat{\mathbf{Z}}_{\ \beta \gamma
\alpha }^{\alpha }\};\mbox{ and }  \notag \\
\mathbf{R}sc& =\mathbf{g}^{\alpha \beta }\mathbf{R}_{\ \alpha \beta }=\breve{%
R}sc+\breve{Z}sc=\widehat{\mathbf{R}}sc+\widehat{\mathbf{Z}}sc,\mbox{where }
\notag \\
& \breve{R}sc=g^{\beta \gamma }\breve{R}_{\beta \gamma },\breve{Z}%
sc=g^{\beta \gamma }\breve{Z}_{\beta \gamma };\widehat{\mathbf{R}}sc=\mathbf{%
g}^{\alpha \beta }\widehat{\mathbf{R}}_{\alpha \beta },\widehat{\mathbf{Z}}%
sc=\mathbf{g}^{\alpha \beta }\widehat{\mathbf{Z}}_{\alpha \beta }.  \notag
\end{align}%
Above geometric d-objects and objects can be used for elaborating and
analyzing physical properties of various models of nonmetric geometric flows
and MGTs. The motivation and priority of the canonical "hat" variables is
that they allow to decouple and integrate in certain general off-diagonal
forms corresponding physically important systems of nonlinear PDEs. The
Greeck indices in (\ref{driccidist}) run values $\alpha ,\beta
,...=1,2,....,8.$ For dyadic decompositions, $\alpha ,\beta ,...\rightarrow
\alpha _{s},\beta _{s},...\,$for $s=1,2,3,4,$ etc. 

\subsubsection{FL deformed nonmetric f(Q) gravity and modified Einstein
equations}

We can formulate on $\mathcal{M}$ a model of nonmetric Finsler-Lagrange
gravity using a gravitational Lagrange density $\ ^{g}\widehat{\mathcal{L}}=%
\frac{1}{2\kappa }\widehat{f}(\widehat{\mathbf{Q}}),$ with $\widehat{\mathbf{%
Q}}$ defined by formulas (\ref{nmsc}), and for the matter fields $^{m}%
\widehat{\mathcal{L}}$ defined in "hat" variables. The total relativistic
phase space action is postulated in the form:%
\begin{equation}
\widehat{\mathcal{S}}=\int \sqrt{|\mathbf{g}_{\alpha \beta }|}\delta ^{8}u(\
^{g}\widehat{\mathcal{L}}+\ ^{m}\widehat{\mathcal{L}}).  \label{actnmc}
\end{equation}%
In this formula, hat-labels state that all Lagrange densities and geometric
objects are written in nonholonomic dyadic form, with boldface indices;
using $\widehat{\mathbf{D}}$ and disformations (\ref{disf}) and the measure $%
\sqrt{|\mathbf{g}_{\alpha \beta }|}\delta ^{8}u,$ for $\delta
^{8}u=du^{1}du^{2}\delta u^{3}\delta u^{4}\delta u^{5}\delta u^{6}\delta
u^{7}\delta u^{8}$ with $\delta u^{a_{s}}= \mathbf{e}^{a_{s}}$ (for $s=2,3,4$%
) as in formulas (\ref{ccnadapc}) and (\ref{ccnadapv}) but considered with
velocity type variables.

In N-adapted variational form (see similar 4-d and 8-d constructions in \cite%
{bvvz24,bnsvv24,vacaru25b}), or applying "pure" nonholonomic geometric
methods as in \cite{misner73}, we can derive such nonmetric gravitational
field equations: 
\begin{align}
\frac{2}{\sqrt{|\mathbf{g}|}}\widehat{\mathbf{D}}_{\gamma }(\sqrt{|\mathbf{g}%
|}\widehat{f_{Q}}\widehat{\mathbf{P}}_{\ \ \alpha \beta }^{\gamma })+\frac{1%
}{2}\widehat{f}\mathbf{g}_{\alpha \beta }+\widehat{f_{Q}}(\widehat{\mathbf{P}%
}_{\beta \mu \nu }\mathbf{Q}_{\alpha }^{\ \mu \nu }-2\widehat{\mathbf{P}}%
_{\alpha \mu \nu }\mathbf{Q}_{\quad \beta }^{\mu \nu })& =\kappa \widehat{%
\mathbf{T}}_{\alpha \beta }  \label{cfeq3a} \\
\mbox{ and }\widehat{\mathbf{D}}_{\alpha }\widehat{\mathbf{D}}_{\beta }(%
\sqrt{|\mathbf{g}|}\widehat{f_{Q}}\widehat{\mathbf{P}}_{\ \ \gamma }^{\alpha
\beta })& =0.  \label{cfeq3b}
\end{align}%
In these formulas, for $\widehat{f_{Q}}:=\partial \widehat{f}/\partial 
\widehat{Q};$ with $\widehat{\mathbf{P}}_{\ \ \alpha \beta }^{\gamma }$ and $%
\mathbf{Q}_{\ \ \alpha \beta }^{\gamma }$ defined as in formulas (\ref%
{nmcjdt}) and (\ref{nmsc}); the energy-momentum d-tensor $\widehat{\mathbf{T}%
}_{\alpha \beta }$ is defined by N-adapted variations for $\ ^{m}\widehat{%
\mathcal{L}}$ on the d-metric; and all equations being written for
N-elongated frames. 

Using distortion relations (\ref{driccidist}), we can write the system (\ref%
{cfeq3a}) in a form provided in \cite{jhao22} for the Einstein tensor, $%
\breve{E}:=\breve{R}ic-\frac{1}{2}g\breve{R}sc$, computed for $\nabla $ on
4-d pseudo-Riemannian manifolds, and nonmetric generalizations, with
applications in DE physics \cite{jhao22,koussour23}.\footnote{%
In this work, we follow a different system of notations and chose an
opposite sign before $\ ^{m}T_{\alpha \beta }.$} Such effective
gravitational field equations can be generalized to 8-d phase spaces $%
\mathcal{M}$ in the form%
\begin{align}
\breve{E}_{\alpha \beta }& =\frac{\kappa }{f_{Q}}\ ^{m}T_{\alpha \beta }+\
^{Q}T_{\alpha \beta }+\ ^{z}T_{\alpha \beta }=\kappa \breve{T}_{\alpha \beta
},\mbox{ or }  \label{gfeq2a} \\
\breve{R}_{\alpha \beta }& =\breve{\Upsilon}_{\alpha \beta },\mbox{ where }%
\breve{\Upsilon}_{\alpha \beta }=\kappa (\breve{T}_{\alpha \beta }-\frac{1}{2%
}g_{\alpha \beta }\breve{T}),\mbox{ for }\breve{T}=g^{\alpha \beta }\breve{T}%
_{\alpha \beta },  \label{gfeq2b} \\
\ ^{Q}T_{\alpha \beta }& =\frac{1}{2}g_{\alpha \beta }(\frac{f}{f_{Q}}-Q)+2%
\frac{f_{QQ}}{f_{Q}}\nabla _{\gamma }(QP_{\ \ \alpha \beta }^{\gamma }),
\label{deemt} \\
\ ^{m}T_{\alpha \beta }& =-\frac{2}{\sqrt{|g|}}\frac{\delta (\sqrt{|g|}\ ^{m}%
\mathcal{L})}{\delta g^{\alpha \beta }}=\ ^{m}\mathcal{L}g_{\alpha \beta }+2%
\frac{\delta (\ ^{m}\mathcal{L})}{\delta g^{\alpha \beta }}  \label{emtlc} \\
\ ^{z}T_{\alpha \beta }& =%
\mbox{ [computed by respective distortion
relations] }(\ref{driccidist}).  \label{emtdist}
\end{align}%
The formula (\ref{emtlc}) \ holds if $\ ^{m}\mathcal{L}$ does not depend in
explicit form on $\Gamma _{\ \alpha \beta }^{\gamma }$ (it has to be
modified, for instance, for spinor fields). The nonmetric modifications of
GR (with trivial extensions on velocity variables) are encoded into the
effective energy-momentum tensor $\ ^{Q}T_{\alpha \beta }$ (\ref{deemt}).

FL configurations can be modeled by systems of nonlinear PDEs (\ref{cfeq3a}%
), or (\ref{gfeq2a}), or (\ref{gfeq2b}) if we prescribe "tilde" N-connection
and d-metric structures of type (\ref{cartnc}) and (\ref{cdms8}) for some
geometric data $(\widetilde{L},\ \ \widetilde{\mathbf{N}};\widetilde{\mathbf{%
e}}_{\alpha },\widetilde{\mathbf{e}}^{\alpha };\widetilde{g}_{jk},\widetilde{%
g}^{jk}).$ Such geometric data can be subjected to general frame transforms
with re-definition of frame structure and distortions of d-connections (\ref%
{driccidist}), for instance, choosing the variants 
\begin{eqnarray}
\ ^{L}\widehat{\mathbf{D}} &=&\ ^{L}\nabla +\ ^{L}\widehat{\mathbf{Z}},\ ^{L}%
\widetilde{\mathbf{D}}=\ ^{L}\nabla +\ ^{L}\widetilde{\mathbf{Z}},%
\mbox{ and
}\ ^{L}\widehat{\mathbf{D}}=\ ^{L}\mathbf{D}+\ ^{L}\mathbf{Z,}%
\mbox{ 
determined by }(\ ^{L}\mathbf{g,}\ ^{L}\mathbf{N)};  \notag \\
\ _{s}\widehat{\mathbf{D}} &=&\widetilde{\mathbf{D}}+\ _{s}\mathbf{Z,}\
_{s}^{L}\widehat{\mathbf{D}}=\ ^{L}\mathbf{D}+\ _{s}^{L}\mathbf{Z,}%
\mbox{ 
determined by }(\ _{s}\mathbf{g}\simeq \ ^{L}\mathbf{\mathbf{g},\ _{s}N}%
\simeq \ ^{L}\mathbf{N),}  \notag
\end{eqnarray}%
The nonholonomic distorted data $(\ ^{L}\mathbf{N};\ ^{L}\mathbf{e}_{\alpha
_{s}},\ ^{L}\mathbf{e}^{\alpha _{s}};\ ^{L}\mathbf{g}_{\alpha _{s}\beta
_{s}},\ _{s}^{L}\widehat{\mathbf{D}}),$ when a $L$-label is kept if $%
\widetilde{L}\rightarrow L$ for some general frame/coordinate frame
transforms on $\mathcal{M}$, can be used for applying the AFCDM.

\subsubsection{FH deformed nonmetric f($\ ^{\shortmid }$Q) gravity and
modified Einstein equations}

In similar forms, we can define nonmetric gravity models on dual phase space 
$\ ^{\shortmid }\mathcal{M}$ enabled with Hamilton generating functions.
Using a gravitational Lagrange density $\ _{\shortmid }^{g}\widehat{\mathcal{%
L}}=\frac{1}{2\kappa }\ ^{\shortmid }\widehat{f}(\ ^{\shortmid }\widehat{%
\mathbf{Q}}),$ with $\ ^{\shortmid }\widehat{\mathbf{Q}}$ defined by
formulas which are conventionally dual to (\ref{nmsc}) and for the matter
fields $\ _{\shortmid }^{m}\widehat{\mathcal{L}}$ defined in "hat" variables
involving Legendre transforms and respective co-fiber coordinates. The total
relativistic phase space action is postulated (if necessary, related to (\ref%
{actnmc})) as%
\begin{equation*}
\ ^{\shortmid }\widehat{\mathcal{S}}=\int \sqrt{|\ ^{\shortmid }\mathbf{g}%
_{\alpha \beta }|}\delta ^{8}\ ^{\shortmid }u(\ _{\ ^{\shortmid }}^{g}%
\widehat{\mathcal{L}}+\ _{\ ^{\shortmid }}^{m}\widehat{\mathcal{L}}).
\end{equation*}%
In this formula, hat-labels state that all Lagrange densities and geometric
objects are written in nonholonomic dyadic form, with boldface indices;
using $\widehat{\mathbf{D}}$ and disformations (\ref{disf}) and the measure $%
\sqrt{|\ ^{\shortmid }\mathbf{g}_{\alpha \beta }|}\delta ^{8}\
^{\shortmid}u, $ for $\delta ^{8}\ ^{\shortmid }u=du^{1}du^{2}\delta
u^{3}\delta u^{4}\delta \ ^{\shortmid }u^{5}\delta \ ^{\shortmid
}u^{6}\delta \ ^{\shortmid }u^{7}\delta \ ^{\shortmid }u^{8}$ with $\delta \
^{\shortmid }u^{a_{s}}=\ ^{\shortmid }\mathbf{e}^{a_{s}},$ (for $s=2,3,4$)
as in (\ref{ccnadapc}) and (\ref{ccnadapv}) using momentum type variables $\
^{\shortmid }u^{a_{s}}=(p_{a_{3}},p_{a_{4}}),$ for $s=3,4.$ We can consider
three variants to derive respective nonmetric phase space gravitational
equations: 1) to perform a N-adapted variational form for $\ ^{\shortmid }%
\widehat{\mathcal{S}};$ 2) or to apply geometric methods as in \cite%
{misner73}; or 3) to take dual formulas for (\ref{cfeq3a}) and (\ref{cfeq3b}%
). We obtain such nonmetric gravitational field equations on $\ ^{\shortmid }%
\mathcal{M}$: 
\begin{align}
\frac{2}{\sqrt{|\ ^{\shortmid }\mathbf{g}|}}\ ^{\shortmid }\widehat{\mathbf{D%
}}_{\gamma }(\sqrt{|\ ^{\shortmid }\mathbf{g}|}\ ^{\shortmid }\widehat{f_{Q}}%
\ ^{\shortmid }\widehat{\mathbf{P}}_{\ \ \alpha \beta }^{\gamma })+\frac{1}{2%
}\ ^{\shortmid }\widehat{f}\ ^{\shortmid }\mathbf{g}_{\alpha \beta }+\
^{\shortmid }\widehat{f_{Q}}(\ ^{\shortmid }\widehat{\mathbf{P}}_{\beta \mu
\nu }\ ^{\shortmid }\mathbf{Q}_{\alpha }^{\ \mu \nu }-2\ ^{\shortmid }%
\widehat{\mathbf{P}}_{\alpha \mu \nu }\ ^{\shortmid }\mathbf{Q}_{\quad \beta
}^{\mu \nu })& =\kappa \ ^{\shortmid }\widehat{\mathbf{T}}_{\alpha \beta }
\label{cfeq3ad} \\
\mbox{ and }\ ^{\shortmid }\widehat{\mathbf{D}}_{\alpha }\ ^{\shortmid }%
\widehat{\mathbf{D}}_{\beta }(\sqrt{|\ ^{\shortmid }\mathbf{g}|}\
^{\shortmid }\widehat{f_{Q}}\ ^{\shortmid }\widehat{\mathbf{P}}_{\ \ \gamma
}^{\alpha \beta })& =0.  \label{cfeq3bd}
\end{align}%
In these formulas, for $\ ^{\shortmid }\widehat{f_{Q}}:=\partial \
^{\shortmid }\widehat{f}/\partial \ ^{\shortmid }\widehat{Q};$ with $\
^{\shortmid }\widehat{\mathbf{P}}_{\ \ \alpha \beta }^{\gamma }$ and $\
^{\shortmid }\mathbf{Q}_{\ \ \alpha \beta }^{\gamma }$ defined in dual form
to formulas (\ref{nmcjdt}) and (\ref{nmsc}); the energy-momentum d-tensor $\
^{\shortmid }\widehat{\mathbf{T}}_{\alpha \beta }$ is defined by N-adapted
variations for $\ _{\shortmid}^{m}\widehat{\mathcal{L}}$ on the d-metric;
and all equations involve N-elongated frames. 

FH configurations are modeled in canonical form by systems of nonlinear PDEs
(\ref{cfeq3ad}) and (\ref{cfeq3bd}) if we prescribe "tilde" N-connection and
d-metric structures of type (\ref{cartnc}), (\ref{canonicalsdcon}) and (\ref%
{cdmds8}) for the nonholonomic geometric data $(\widetilde{H},\ ^{\shortmid }%
\widetilde{\mathbf{N}};\ ^{\shortmid }\widetilde{\mathbf{e}}_{\alpha },\
^{\shortmid }\widetilde{\mathbf{e}}^{\alpha };\ \ ^{\shortmid }\widetilde{g}%
^{ab},\ \ ^{\shortmid }\widetilde{g}_{ab}).$ Such geometric data can be
subjected to general frame transforms with re-definition of frame structure
and distortions of d-connections (\ref{distortcand}), for instance, choosing
the variants 
\begin{eqnarray}
\ _{\shortmid }^{H}\widehat{\mathbf{D}} &=&\ _{\shortmid }^{H}\nabla +\
_{\shortmid }^{H}\widehat{\mathbf{Z}},\ _{\shortmid }^{H}\widetilde{\mathbf{D%
}}=\ _{\shortmid }^{H}\nabla +\ _{\shortmid }^{H}\widetilde{\mathbf{Z}},%
\mbox{ and
}\ _{\shortmid }^{H}\widehat{\mathbf{D}}=\ _{\shortmid }^{H}\mathbf{D}+\
_{\shortmid }^{H}\mathbf{Z,}\mbox{ 
determined by }(\ _{\shortmid }^{H}\mathbf{g,}\ _{\shortmid }^{H}\mathbf{N)};
\notag \\
\ _{s}^{\shortmid }\widehat{\mathbf{D}} &=&\ ^{\shortmid }\widetilde{\mathbf{%
D}}+\ _{s}^{\shortmid }\mathbf{Z,}\ _{s}^{H}\widehat{\mathbf{D}}=\
_{\shortmid }^{H}\mathbf{D}+\ _{s}^{H}\mathbf{Z,}\mbox{ 
determined by }(\ _{s}^{\shortmid }\mathbf{g}\simeq \ _{\shortmid }^{H}%
\mathbf{\mathbf{g},}\ _{s}^{\shortmid }\mathbf{N}\simeq \ _{\shortmid }^{H}%
\mathbf{N),}  \notag
\end{eqnarray}%
The nonholonomic distorted data $(\ _{\shortmid }^{H}\mathbf{N};\
_{\shortmid }^{H}\mathbf{e}_{\alpha _{s}}, \ _{\shortmid }^{H}\mathbf{e}%
^{\alpha _{s}};\ _{\shortmid }^{H}\mathbf{g}_{\alpha _{s}\beta _{s}}, \
_{s\shortmid }^{H}\widehat{\mathbf{D}}),$ when a $H$-label is kept if $%
\widetilde{H}\rightarrow H$ for some general frame/coordinate frame
transforms on $\ ^{\shortmid }\mathcal{M}$, can be used for applying the
AFCDM. 

In general, Hamiltonian mechanics is not equivalent to Lagrangian mechanics.
For instance, in the first case, we consider symplectomorphisms as specific
symmetries related to almost symplectic and almost complex structures
studied in \cite%
{oproiu85,mhss2000,vmon3,vacaru07,vacaru10a,vacaru13,vacaru18,bnsvv24}. In
this work, we do not study Hamilton and almost symplectic models of (dual)
Finsler-Lagrange geometry and respective nonholonomic d-connection
structures. The system of nonlinear PDEs (\ref{cfeq3ad}) can be written in
equivalent form by using the LC-connection $\ ^{\shortmid }\nabla _{\gamma }$
with respective dual equations to (\ref{gfeq2a}), or (\ref{emtdist}). Such
equations can be reduced in nonholonomic and parametric form to the Einstein
equations in GR if $\ ^{\shortmid }\nabla _{\gamma }\rightarrow \ (\nabla
_{i_{1}},\nabla _{a_{2}})$ and the d-metric structure is transformed into (%
\ref{lqed}). 

\subsubsection{Canonical integrable nonmetric FLH deformed gravitational
phase space equations}

To decouple and solve in certain general off-diagonal forms systems of
nonlinear PDEs (\ref{cfeq3a}) and (\ref{cfeq3ad}) is not possible because of
various nonholonomic constraints and coupling conditions like (\ref{cfeq3b})
or, respectively, (\ref{cfeq3bd}). We have to re-define such FLH-modified
gravitational field equations in corresponding nonholonomic s-variables
which allows us to apply the AFCDM. Let us explain how such nonholonomic
transforms can be performed in explicit form:

On $\mathcal{M}$, we can use distortions (\ref{driccidist}) and write (\ref%
{gfeq2a}) in the form 
\begin{align}
\widehat{\mathbf{R}}_{\alpha \beta }& =\ _{Q}\widehat{\yen }_{\alpha \beta },%
\mbox{ for }  \label{cfeq4a} \\
\ _{Q}\widehat{\yen }_{\alpha \beta }& =\ ^{e}\widehat{\mathbf{Y}}_{\alpha
\beta }+\ ^{m}\widehat{\mathbf{Y}}_{\alpha \beta }.  \label{ceemt}
\end{align}%
The source $\ _{Q}\widehat{\yen }_{\alpha \beta }$ (\ref{ceemt}) for (\ref%
{cfeq4a}) is defined by two d-tensors: the first one with effective source
(e), $\ ^{e}\widehat{\mathbf{Y}}_{\alpha \beta }=\breve{Z}ic_{\alpha \beta }-%
\widehat{\mathbf{Z}}ic_{\alpha \beta }$, is of geometric distorting nature,
which can be computed in explicit form using (\ref{driccidist}). The second
one is the energy-momentum d-tensor $\ ^{m}\widehat{\mathbf{Y}}_{\alpha
\beta }$ of the matter fields $\breve{\Upsilon}_{\alpha \beta }$ encoding
also contributions of the nonmetricity scalar $\widehat{\mathbf{Q}}$ and
d-tensor $\widehat{\mathbf{P}}_{\ \ \alpha \beta }^{\gamma }$ defined
respectively by formulas (\ref{nmsc}), (\ref{nmcjdt}) and (\ref{nmsc}). For
the effective and matter field sources \ in above system of nonlinear PDEs,
we can consider any type nonholonomic transforms of d-metrics ($\mathbf{g}%
_{\alpha \beta }=e_{\ \alpha }^{\underline{\alpha }}\ e_{\ \beta }^{%
\underline{\beta }}\ g_{\underline{\alpha }\underline{\beta }}$ or s-adapted
ones,$\ \mathbf{g}_{\alpha _{s}\beta _{s}}=e_{\ \alpha _{s}}^{\underline{%
\alpha }}\ e_{\ \beta _{s}}^{\underline{\beta }}\ g_{\underline{\alpha }%
\underline{\beta }})$ and respective re-distributions of distortions, when
the s-adapted form of sources (\ref{ceemt}) is parameterized 
\begin{equation}
\ _{Q}\widehat{\yen }_{\ \ \beta _{s}}^{\alpha _{s}}=[\ _{1}\Upsilon
(x^{k_{1}})\delta _{\ \ j_{1}}^{i_{1}},\ _{2}\Upsilon
(x^{k_{1}},y^{c_{2}})\delta _{\ \ b_{2}}^{a_{2}},\ _{3}\Upsilon
(x^{k_{1}},y^{c_{2}},v^{c_{3}})\delta _{\ \ b_{3}}^{a_{3}},\ _{4}\Upsilon
(x^{k_{1}},y^{c_{2}},v^{c_{3}},v^{c_{4}})\delta _{\ \ b_{4}}^{a_{4}}],
\label{ssourc}
\end{equation}%
where the effective sources can be parameterized in conventional oriented
forms $\ _{s}\Upsilon (x^{k_{s-1}},v^{a_{s}}).$ Such s-adapted
parameterizations of $\widehat{\mathbf{R}}_{\alpha _{s}\beta _{s}}$ and $\
_{Q}\widehat{\yen }_{\alpha _{s}\beta _{s}}$ are important for explicit
integrations of nonmetric FLH gravitational equations. For $\widehat{\mathbf{%
Q}}\rightarrow 0,$ we shall be able to generate solutions with nontrivial $%
\widehat{\mathcal{T}}=\{\widehat{\mathbf{T}}_{\ \alpha \beta }^{\gamma }\},$
and then to extract LC-configurations $\nabla $ (as we prove in the next
section and Appendix).

We can define canonical d-equations on dual phase space $\ ^{\shortmid }%
\mathcal{M}$ encoding Hamilton generating functions and respective nonmetric
distortions: 
\begin{align}
\ ^{\shortmid }\widehat{\mathbf{R}}_{\alpha \beta }& =\ _{Q}^{\shortmid }%
\widehat{\yen }_{\alpha \beta },\mbox{ for }  \label{cfeq4ad} \\
\ _{Q}^{\shortmid }\widehat{\yen }_{\alpha \beta }& =\ _{\shortmid }^{e}%
\widehat{\mathbf{Y}}_{\alpha \beta }+\ _{\shortmid }^{m}\widehat{\mathbf{Y}}%
_{\alpha \beta },  \label{ceemtd}
\end{align}%
when the nonholonomic s-adapted parametrization of the sources is stated in
the form 
\begin{equation}
\ _{Q}^{\shortmid }\widehat{\yen }_{\ \ \beta _{s}}^{\alpha _{s}}=[\
_{1}^{\shortmid }\Upsilon (x^{k_{1}})\delta _{\ \ j_{1}}^{i_{1}},\
_{2}^{\shortmid }\Upsilon (x^{k_{1}},y^{c_{2}})\delta _{\ \ b_{2}}^{a_{2}},\
_{3}^{\shortmid }\Upsilon (x^{k_{1}},y^{c_{2}},p_{c_{3}})\delta _{a_{3}}^{\
b_{3}},\ _{4}^{\shortmid }\Upsilon
(x^{k_{1}},y^{c_{2}},p_{c_{3}},p_{c_{4}})\delta _{a_{4}}^{\ b_{4}}].
\label{ssourcd}
\end{equation}

\subsubsection{Modelling, distorting, or extracting gravitational field
equations in GR and FLH MGTs}

The systems of FLH equations (\ref{cfeq4a}) and (\ref{cfeq4ad}) with
respective sources $\ _{Q}\widehat{\yen }_{\ \ \beta _{s}}^{\alpha _{s}}$ (%
\ref{ssourc}) and $\ _{Q}\widehat{\yen }_{\alpha \beta }$ (\ref{ssourcd})
can be integrated in general forms as we shall prove in section \ref{sec4}.
In abstract geometric forms, the corresponding FLH MGTs can be stated by
such data:%
\begin{eqnarray}
(\ \widetilde{\mathcal{M}} &:&F\mbox{ or }L;\widetilde{\mathbf{g}}\simeq \
_{s}\widetilde{\mathbf{g}},\ \widetilde{\mathbf{N}}\simeq \ _{s}\ \widetilde{%
\mathbf{N}},\ _{s}\widehat{\mathbf{D}}=\widetilde{\mathbf{D}}+\ _{s}\mathbf{Z%
},\ _{\ }^{g}\widehat{\mathcal{L}}+\ _{\ }^{m}\widehat{\mathcal{L}},\ _{Q}%
\widehat{\yen }_{\alpha \beta },\widehat{\mathbf{Q}}\simeq \widetilde{%
\mathbf{Q}}=0,  \notag \\
&&\mbox{ for Lagrange - Finsler - Cartan configurations});  \label{classflh}
\\
(\ \ ^{C}\mathcal{M} &:&F\mbox{ or }L;\mathbf{g}\simeq \ _{s}\mathbf{g},\ \
^{C}\mathbf{N}\simeq \ _{s}^{C}\mathbf{N},\ _{s}^{C}\widehat{\mathbf{D}}=\
^{C}\mathbf{D}+\ _{s}^{C}\mathbf{Z},\ _{\ ^{C}}^{g}\widehat{\mathcal{L}}+\
_{\ ^{C}}^{m}\widehat{\mathcal{L}},\ _{Q}^{C}\widehat{\yen }_{\alpha \beta
},\ ^{C}\widehat{\mathbf{Q}}\simeq \ ^{C}\mathbf{Q}=0,  \notag \\
&&\mbox{ for Lagrange - Finsler - Chern configurations});  \notag \\
(\ ^{B}\mathcal{M} &:&F\mbox{ or }L;\mathbf{g}\simeq \ _{s}\mathbf{g},\ \
^{B}\mathbf{N}\simeq \ _{s}^{B}\ \mathbf{N},\ _{s}^{B}\widehat{\mathbf{D}}=\
^{B}\mathbf{D}+\ _{s}^{B}\mathbf{Z},\ _{\ ^{B}}^{g}\widehat{\mathcal{L}}+\
_{\ ^{B}}^{m}\widehat{\mathcal{L}},\ _{Q}^{B}\widehat{\yen }_{\alpha \beta
},\ ^{B}\widehat{\mathbf{Q}}\simeq \ ^{B}\mathbf{Q}=0,  \notag \\
&&\mbox{ for Lagrange - Finsler - Berwald configurations});  \notag \\
(\ \widehat{\mathcal{M}} &:&\mbox{ we can prescribe  }L;\mathbf{g}\simeq \
_{s}\mathbf{g},\ \mathbf{N}\simeq \ _{s}\ \mathbf{N},\ _{s}\widehat{\mathbf{D%
}}=\ \mathbf{\nabla }+\ _{s}\mathbf{Z},\ _{\ }^{g}\widehat{\mathcal{L}}+\
_{\ }^{m}\widehat{\mathcal{L}},\ _{Q}\widehat{\yen }_{\alpha \beta },\ 
\widehat{\mathbf{Q}}=0,\ \widehat{\mathbf{T}}\neq 0,  \notag \\
&&\mbox{ for phase space canonical configurations});  \notag \\
(\ _{\mathbf{\nabla }}\ \widehat{\mathcal{M}} &:&\mathbf{g}\simeq \ _{s}%
\mathbf{g},\ \mathbf{N}\simeq \ _{s}\ \mathbf{N},\ _{s}\widehat{\mathbf{D}}%
=\ \mathbf{\nabla }+\ _{s}\mathbf{Z},\ _{\ }^{g}\widehat{\mathcal{L}}+\ _{\
}^{m}\widehat{\mathcal{L}},\ _{Q}\widehat{\yen }_{\alpha \beta },\ \widehat{%
\mathbf{Q}}=0,\ \widehat{\mathbf{T}}=0,\widehat{\mathbf{D}}_{|\mathcal{T}%
=0}=\nabla  \notag \\
&&\mbox{ extracting phase space LC-configurations});  \notag \\
(\mathcal{M} &:&\mathbf{g}\simeq \ _{s}\mathbf{g},\ \mathbf{N}\simeq \ _{s}\ 
\mathbf{N},\ _{s}\widehat{\mathbf{D}}=\ \mathbf{D}+\ _{s}\mathbf{Z},\ _{\
}^{g}\widehat{\mathcal{L}}+\ _{\ }^{m}\widehat{\mathcal{L}},\ _{Q}\widehat{%
\yen }_{\alpha \beta },\ \widehat{\mathbf{Q}}\simeq \mathbf{Q}\neq 0,\ 
\widehat{\mathbf{T}}\simeq \mathbf{T}\neq 0),  \notag \\
&&\mbox{ nonholonomic metric-affine  phase spaces};  \notag \\
(\ _{\mathbf{Q}}\mathcal{M} &:&\mathbf{g}\simeq \ _{s}\mathbf{g},\ \mathbf{N}%
\simeq \ _{s}\ \mathbf{N},\ _{s}\widehat{\mathbf{D}}=\ \mathbf{D}+\ _{s}%
\mathbf{Z},\ _{\ }^{g}\widehat{\mathcal{L}}+\ _{\ }^{m}\widehat{\mathcal{L}}%
,\ _{Q}\widehat{\yen }_{\alpha \beta },\ \widehat{\mathbf{Q}}\simeq \mathbf{Q%
}\neq 0,\ \widehat{\mathbf{T}}\neq 0,\mathcal{T}\mathbf{[\nabla ]=0}), 
\notag \\
&&\mbox{ nonmetric  phase space}\ \widehat{f}(\widehat{\mathbf{Q}})%
\mbox{
gravity }.  \notag
\end{eqnarray}%
In the above formulas, we use "$\simeq $" to state equivalence up to a
certain frame and distortion of connections transforms.

The MGT theories defined above for $\mathcal{M}$ can be formulated in
similar forms on $\ ^{\shortmid }\mathcal{M}$ using momentum-like variables
and geometric objects and (effective) Lagrangians and sources labelled by "$%
\ ^{\shortmid }$".

Let us discuss how various types of FLH theories analyzed in paragraphs 1-7]
of section \ref{sec1} can be included, modelled, or generalized to MGTs on
(co) tangent Lorentz bundles classified in (\ref{classflh}):

Perhaps, the simplest analysis concerns the class of models and solutions
for Finsler-Randers-Sasaki theories studied in \cite%
{triantafy20,kapsabelis21,kapsabelis23,savvopoulos23,miliaresis25} using
Sasaki d-metrics of type $\widetilde{\mathbf{g}}_{\alpha \beta }(x,y)$ (\ref%
{cdms8}) and (typically) certain metric compatible Finsler d-connections $\
^{F}\mathbf{D}$. \ Performing respective s-adapted frame transforms with $%
\mathbf{g}_{\alpha _{s}\beta _{s}}=e_{\ \alpha _{s}}^{\alpha }\ e_{\ \beta
_{s}}^{\beta }\ \widetilde{\mathbf{g}}_{\alpha \beta }$ and distortion $\
_{s}\widehat{\mathbf{D}}=\ ^{F}\mathbf{D}+\ _{s}^{F}\mathbf{X}$ and
parameterizations of sources $\ _{Q}\widehat{\yen }_{\ \ \beta _{s}}^{\alpha
_{s}}$ (\ref{ssourc}) (if necessary, we consider $Q=0$), we can impose the
conditions that $\mathbf{g}_{\alpha _{s}\beta _{s}}$ is determined by a
solution of (\ref{cfeq4a}). For such constructions, we can embed respective
Finsler-Randers configurations into more general classes of FLH\ theories
discussed above. 

Considering models on $\mathcal{M},$ with respective Sasaki d-metrics,
effective sources and d-connections, we can model Barthel-Randers/ -
Koropina models \cite{hama21,boulai23,hama23} as some sub-classes of
solutions of (\ref{cfeq4a}). Of course, we have to adapt respectively to the
N-connection structures and restrict the classes of generating and
integration functions.

Then, we can model in our approach certain alternative variants and
extensions of the works \cite%
{pfeifer11a,pfeifer11b,hohmann13,barcaroli15,fuster18,voicu23,heefer24}
which are not necessarily formulated on (co) tangent Lorentz bundles and may
involve (nonmetric) Berwald d-connection, or metric Cartan d-connections,
respective sources, etc. The main assumption for such solutions is that we
do not enter into "exotic" locally anisotropic causality scenarios, but
perform all constructions on $\mathcal{M}$. The N-adapted variational
calculus and effective sources can be chosen as, for instance, in \cite%
{pfeifer11a,pfeifer11b}. Furthermore, noholonomic s-adapted frame transforms
and distortions of connections enable us to speculate on metric and
nonmetric FLH theories with EYMHD configurations, which can be defined and
studied if we follow only the original Pfeifer- Wohlarth particular model.
The variants with modified Einstein equations (\ref{cfeq4a}) and (\ref%
{cfeq4ad}) offer such possibilities even for some particular cases of
effective sources $\ _{Q}\widehat{\yen }_{\ \ \beta _{s}}^{\alpha _{s}}$ (%
\ref{ssourc}) and $\ _{Q}\widehat{\yen }_{\alpha \beta }$ (\ref{ssourcd}). 

Finally, in this subsection, we briefly discuss how we can reproduce in a
relativistic and self-consistent form the results and (numerical and
graphic) methods from \cite{nekouee23,nekouee24,praveen25}. In those works,
the lifts and deformations from a base spacetimes with extensions to certain 
$y$-variables are performed using Akbar-Zadekh constructions \cite%
{akbar88,akbar95} or the version of Finsler modified Einstein equations from 
\cite{xi14}. In principle, such models are incomplete and undetermined if we
do not introduce Sasaki-like lifts on total spaces, and do not perform a
(co) tangent Lorentz bundle formulation. We can construct arbitrary $y$%
-deformations and speculate on any types of local anisotropy and $y$%
-deformations as mentioned in 7.4] of section \ref{sec1}. Nevertheless, we
can consider similar models with Sasaki lifts to d-metrics, when solutions
of systems (\ref{cfeq4a}) and (\ref{cfeq4ad}) (for certain subclasses of
generating functions and sources will reproduce BH, WH and cosmological
scenarios proposed in certain phenomenological and numerical forms in the
mentioned works). They consist of particular cases, or nonholonomic
deformations and distortion of the d-metrics studied and reviewed in \cite%
{vacaru18,bubuianu18,partner06,bsssvv25,v13,v14,v01t,v01q,vacaruplb16}. 

\section{Metric and nonmetric geometric FLH-flows and Ricci solitons}

\label{sec3}The systems of nonlinear PDE equations (\ref{cfeq4a}) and (\ref%
{cfeq4ad}) consist examples of associative and commutative nonholonomic
Ricci solitons defined as particular cases of self-similar nonassociative
geometric flows on metric compatible 8-d phase spaces in our recent works 
\cite{bsv22,bsv23,partner06,bsssvv25,vacaru18,bubuianu18}. Such
constructions of generic off-diagonal solutions encoding nonmetricity were
also performed in \cite{bnsvv24,bvvz24,vacaru25b}. We shall consider
applications of the AFCDM and explicit 8-d examples encoding nonmetric FLH
configurations in sections \ref{sec4} and \ref{sec5}. 

The generic off-diagonal solutions constructed and studied in various MGTs
(including FLH models) are not characterized, in general, by certain
hypersurface or holographic conditions. As we explained in paragraph 6.4] of
section \ref{sec1}, the Bekenstein--Hawking BH thermodynamic paradigm \cite%
{bek2,haw2} does not apply even to off-diagonal classes of quasi-stationary
and regular BH solutions in GR \cite{vv25}. In the case of FLH deformations
of the Einstein equations, similar issues exist and this motivates the
elaboration of corresponding models of nonholonomic and Finsler-like
geometric flows. The goal of this section is to formulate and study such
metric and nonmetric FLH geometric flow theories and related statistical
thermodynamic models as generalizations of the results and methods from a
series of works on Riemannian geometric flows and various relativistic and
non-Riemannian modifications \cite%
{perelman1,hamilton82,monogrrf1,monogrrf2,monogrrf3,vacaru06dd,vv06,svnonh08,vacaru07ee,vacaru11,vacaru13,gheorghiuap16}%
. 

\subsection{Metric compatible FL and FH deformed geometric \ flows}

We note that in our works on nonholonomic relativistic and/or FLH (non)
metric geometric flows generalizations of the famous Poincar\'{e}--Thurston
conjecture are not formulated/ proved. This is a very difficult mathematical
problem which depends on the types of metric and N- and d-connection
structures, i.e. can't be formulated and proven in some general nonholonomic
metric-affine forms. Nevertheless, we use the concept of W-entropy \cite%
{perelman1}, which is useful for formulating statistical and geometric
thermodynamic models characterizing physical properties of FLH theories and
respective classes of nonholonomic geometric evolutions or dynamical
equations. Corresponding thermodynamic variables involve families of
distorted Ricci tensors and respective Ricci scalars defined typically by $%
\tau $-families of metrics $g_{\alpha \beta }(\tau ):=g_{\alpha \beta }(\tau
,u^{\gamma })$ and linear connections $\Gamma _{\ \alpha \beta
}^{\gamma}(\tau ):=\Gamma _{\ \alpha \beta }^{\gamma }(\tau ,u^{\gamma })$,
where $\tau $ is a positive flow parameter (treated as a conventional
temperature). Typically, we shall write for the geometric objects only the $%
\tau $-parametric dependencies if that will not result in ambiguities, for
instance, $\ _{s}\widehat{\mathbf{D}}(\tau ),\ ^{F}\mathbf{D}(\tau ),%
\widetilde{\mathbf{g}}(\tau )=\{\widetilde{\mathbf{g}}_{\alpha \beta
}(\tau)\},$ etc. The priority of our nonholonomic approach is that we apply
the AFCDM to decouple and solve in general off-diagonal forms respective
systems of nonlinear PDEs, which define FLH geometric flow evolution models
or, for fixed $\tau =\tau _{0},$ nonholonomic Ricci soliton configurations
which are equivalent to certain classes of FLH gravity theories. 

\subsubsection{F- and W-functionals for FL and FH flows}

Let us consider a FL phase space $\widetilde{\mathcal{M}}$ and FH phase
space $\ ^{\shortmid }\widetilde{\mathcal{M}}$ which admit respective
re-definitions in canonical 4+4 or nonholonomic dyadic variables as in (\ref%
{classflh}), with respective classes of nonholonomic equivalence, $%
\widetilde{\mathcal{M}}\simeq \widehat{\mathcal{M}}\simeq \ _{s}\widehat{%
\mathcal{M}}$ and $\ ^{\shortmid }\widetilde{\mathcal{M}}\simeq \
^{\shortmid }\widehat{\mathcal{M}}\simeq \ _{s}^{\shortmid }\widehat{%
\mathcal{M}}$. For simplicity, in this subsection (with metric-compatible
geometric constructions), we consider only "tilde" geometric objects. The
models with "hat" connections can be formulated in similar forms, when the
s-adapted variants result in respective FL and FH geometric flow equations,
which can be integrated in general off-diagonal form (we prove it in section %
\ref{sec4} in certain general forms with nonmetricity). 

We consider FL and FH flows on temperature like parameter $\tau $ (when $%
0\leq \tau \leq \tau _{0}$) of d-objects on, respective, $\widetilde{%
\mathcal{M}}$ and $\ ^{\shortmid }\widetilde{\mathcal{M}},$ by using $\tau $%
-flows of volume elements 
\begin{equation}
d\ \ \widetilde{\mathcal{V}}ol(\tau )=\sqrt{|\ \widetilde{\mathbf{g}}%
_{\alpha \beta }\ (\tau )|}\ \widetilde{\delta }^{8}\ u^{\gamma _{s}}(\tau )%
\mbox{ and }d\ \ ^{\shortmid }\widetilde{\mathcal{V}}ol(\tau )=\sqrt{|\ \
^{\shortmid }\widetilde{\mathbf{g}}_{\alpha \beta }\ (\tau )|}\ \
^{\shortmid }\widetilde{\delta }^{8}\ ^{\shortmid }u^{\gamma _{s}}(\tau ).
\label{volformfh}
\end{equation}%
Such a value is computed using N-elongated s-differentials, for instance, $\
^{\shortmid }\widetilde{\delta }^{8}\ ^{\shortmid }u^{\gamma _{s}}(\tau ),$
which are linear on $\ ^{\shortmid }\widetilde{N}_{\ i_{s}a_{s}}\ (\tau )$
as in $\ ^{\shortmid }\widetilde{\mathbf{e}}_{\alpha _{s}}(\tau ),$ see
formulas (\ref{ccnadapv}) and (\ref{ccnadapc}). Nonholonomic geometric flow
theories can be formulated for the geometric data $[\widetilde{\mathbf{g}}%
(\tau ),\widetilde{\mathbf{D}}(\tau )]$ and $[\ ^{\shortmid }\widetilde{%
\mathbf{g}}(\tau ),\ ^{\shortmid }\widetilde{\mathbf{D}}(\tau )],$ when the
Perelman type functionals are respectively postulated: 
\begin{eqnarray}
\ \widetilde{\mathcal{F}}(\tau ) &=&\int_{\ \widetilde{\Xi }}(\widetilde{%
\mathbf{R}}sc+|\widetilde{\mathbf{D}}\widetilde{f}|^{2})e^{-\ \widetilde{f}%
}\ d\widetilde{\mathcal{V}}ol(\tau ),  \label{naffunctfh} \\
\ \widetilde{\mathcal{W}}(\tau ) &=&\int_{\ \widetilde{\Xi }}\left( 4\pi
\tau \right) ^{-4}\ [\tau (\widetilde{\mathbf{R}}sc+|\widetilde{\mathbf{D}}%
\widetilde{f}|)^{2}+\widetilde{f}-8]e^{-\ \widetilde{f}}\ d\ \widetilde{%
\mathcal{V}}ol(\tau );  \label{nawfunctfh} \\
&&\mbox{ and }  \notag
\end{eqnarray}%
\begin{eqnarray}
\ ^{\shortmid }\widetilde{\mathcal{F}}(\tau ) &=&\int_{\ ^{\shortmid }%
\widetilde{\Xi }}(\ ^{\shortmid }\widetilde{\mathbf{R}}sc+|\ ^{\shortmid }%
\widetilde{\mathbf{D}}\ ^{\shortmid }\widetilde{f}|^{2})e^{-\ ^{\shortmid }%
\widetilde{f}}\ d\ ^{\shortmid }\widetilde{\mathcal{V}}ol(\tau ),  \notag \\
\ ^{\shortmid }\widetilde{\mathcal{W}}(\tau ) &=&\int_{\ ^{\shortmid }%
\widetilde{\Xi }}\left( 4\pi \tau \right) ^{-4}\ [\tau (\ ^{\shortmid }%
\widetilde{\mathbf{R}}sc+|\ ^{\shortmid }\widetilde{\mathbf{D}}\ ^{\shortmid
}\widetilde{f}|)^{2}+\ ^{\shortmid }\widetilde{f}-8]e^{-\ \ ^{\shortmid }%
\widetilde{f}}\ d\ ^{\shortmid }\widetilde{\mathcal{V}}ol(\tau ).  \notag
\end{eqnarray}%
The 8-d hypersurfrace integrals $\ \widetilde{\Xi }$ and for $\ ^{\shortmid }%
\widetilde{\Xi }$, such F- and W-functionals are determined by volume
elements (\ref{volformfh}). For instance, a h-c-normalizing function $\
^{\shortmid }\widetilde{f}(\tau ,\ ^{\shortmid }u)$ can be stated to satisfy
the condition 
\begin{equation}
\int_{\ ^{\shortmid }\widetilde{\Xi }}\ ^{\shortmid }\widetilde{\nu }\ \ d\
^{\shortmid }\widetilde{\mathcal{V}}ol(\tau ):=\int_{t_{1}}^{t_{2}}\int_{%
\widetilde{\Xi }_{t}}\ \int_{\ ^{\shortmid }\widetilde{\Xi }_{E}}\
^{\shortmid }\widetilde{\nu }\ \ d\ ^{\shortmid }\widetilde{\mathcal{V}}%
ol(\tau )=1.  \label{normcond}
\end{equation}%
In these formulas, where the integration measures $\ \ ^{\shortmid }%
\widetilde{\nu }=\left( 4\pi \tau \right) ^{-4}e^{-\ ^{\shortmid }\widetilde{%
f}}$ are parameterized for the h- and c-components, with shell further
parameterizations if necessary. For general topological considerations, such
conditions may not be considered. We can also choose certain variants of $%
\widetilde{f}$ or $\ ^{\shortmid }\widetilde{f}$ to simplify certain
formulas and computations. 

The FL and FH geometric flow evolution equations are postulated in the form%
\begin{eqnarray}
\partial _{\tau }\widetilde{\mathbf{g}}_{\alpha \beta }(\tau ) &=&-2%
\widetilde{\mathbf{R}}_{\ \alpha \beta }(\tau ),  \label{nonassocfhfl} \\
\partial _{\tau }\widetilde{f}(\tau ) &=&\widetilde{\mathbf{R}}sc(\tau )-\ 
\widetilde{\bigtriangleup }(\tau )\widetilde{f}(\tau )+(\widetilde{\mathbf{D}%
}(\tau )\widetilde{f}(\tau ))^{2}  \notag
\end{eqnarray}%
and 
\begin{eqnarray}
\partial _{\tau }\ ^{\shortmid }\widetilde{\mathbf{g}}_{\alpha \beta }(\tau
) &=&-2\ ^{\shortmid }\widetilde{\mathbf{R}}_{\ \alpha \beta }(\tau ), 
\notag \\
\partial _{\tau }\ ^{\shortmid }\widetilde{f}(\tau ) &=&\ ^{\shortmid }%
\widetilde{\mathbf{R}}sc(\tau )-\ ^{\shortmid }\widetilde{\bigtriangleup }%
(\tau )\ ^{\shortmid }\widetilde{f}(\tau )+(\ ^{\shortmid }\widetilde{%
\mathbf{D}}(\tau )\ ^{\shortmid }\widetilde{f}(\tau ))^{2}.  \notag
\end{eqnarray}%
For instance, $\ ^{\shortmid }\widetilde{\bigtriangleup }(\tau )=[\
^{\shortmid }\widetilde{\mathbf{D}}(\tau )]^{2}$ in (\ref{nonassocfhfl}) are
families of the Laplace d-operators computed for $\ ^{\shortmid }\widetilde{%
\mathbf{g}}_{\alpha \beta }(\tau ).$ Such nonlinear PDEs can be derived in
variational forms from the F- and W-potentials, respectively, (\ref%
{naffunctfh}) and (\ref{nawfunctfh}) generalizing the proofs provided in 
\cite{perelman1}, see details in monographs \cite%
{monogrrf1,monogrrf2,monogrrf3}, and, for various nonassociative or
nonholonomic non-Riemannian generalizations, \cite%
{bsv22,bsv23,partner06,bsssvv25,vacaru18,bubuianu18,bnsvv24,bvvz24,vacaru25b}%
.

Nonholonomic Ricci solitons for the FH Cartan d-connection $\ ^{\shortmid }%
\widetilde{\mathbf{D}}$ are defined as self-similar configurations of (dual)
gradient geometric flows (\ref{nonassocfhfl}) for a fixed parameter $%
\tau_{0}.$ So, on $\ ^{\shortmid }\widetilde{\mathcal{M}},$ the FH-Ricci
soliton d-equations are of type 
\begin{equation}
\ ^{\shortmid }\widetilde{\mathbf{R}}_{\ \alpha \beta }+\ ^{\shortmid }%
\widetilde{\mathbf{D}}_{\alpha }\ ^{\shortmid }\widetilde{\mathbf{D}}_{\beta
}\ ^{\shortmid }\widetilde{\varpi }(\ ^{\shortmid }u)=\ ^{\shortmid }\lambda
\ \ \ ^{\shortmid }\widetilde{\mathbf{g}}_{\alpha \beta },
\label{naricsolfh}
\end{equation}%
where $\ ^{\shortmid }\widetilde{\varpi }$ is a smooth potential function
and $\ ^{\shortmid }\lambda =const.$ The FH-modified Einstein equations
involving the Cartan d-connection consists of an example of nonholonomic
Ricci soliton ones (\ref{naricsolfh}). Even in abstract geometric form of \
F- and W-functional and nonholononmic geometric flow equations are very
similar on $\ \widetilde{\mathcal{M}}$ and $\ ^{\shortmid }\widetilde{%
\mathcal{M}}$, in general, they may involve different almost symplectic
formulations because the Lagrange and Hamilton approaches to mechanical and
classical and quantum field theories may be not equivalent.

The nonlinear systems of PDEs (\ref{nonassocfhfl}) or (\ref{naricsolfh})
written for the respective Cartan d-connections $\ \widetilde{\mathbf{D}}%
_{\alpha }$ and $\ ^{\shortmid }\widetilde{\mathbf{D}}_{\alpha }$ can't be
decoupled and integrated in general off-diagonal forms. To apply the AFCDM,
we have to distort such systems, for instance, $\ \widetilde{\mathbf{D}}%
_{\alpha }\rightarrow \ \widehat{\mathbf{D}}_{\alpha _{s}}.$ Then, imposing
additional nonholonomic constraints on a found class of solutions, we can
extract FL or FH Cartan configurations. 

\subsubsection{Thermodynamic models for FL and FH flows}

Let us consider such $\tau $-families of respective geometric data: \ $\left[
\widetilde{\mathbf{g}}_{\alpha \beta }\ (\tau), \widetilde{\mathbf{D}} (\tau
)\right] $ and $\left[ \ ^{\shortmid }\widetilde{\mathbf{g}}_{\alpha \beta
}\ (\tau ),\ ^{\shortmid }\widetilde{\mathbf{D}}(\tau )\right] ,$ used for
nonholonomic deformations with respective closed hypersurface $\widetilde{%
\Xi }\subset \widetilde{\mathcal{M}}$ and $\ ^{\shortmid }\widetilde{\Xi }$ $%
\subset \ ^{\shortmid }\widetilde{\mathcal{M}};$ and the corresponding
volume forms (\ref{volformfh}). We can introduce respective partition
functions for FL and FH phase spaces of dimension $n=8$, 
\begin{eqnarray}
\ \widetilde{\mathcal{Z}}(\tau ) &=&\exp [\int_{\widetilde{\Xi }}[-%
\widetilde{f}+4]\ \left( 4\pi \tau \right) ^{-4}e^{-\ \widetilde{f}}\ d\ 
\widetilde{\mathcal{V}}ol(\tau )\mbox{ and }  \label{spffh} \\
\ ^{\shortmid }\widetilde{\mathcal{Z}}(\tau ) &=&\exp [\int_{^{\shortmid }%
\widetilde{\Xi }}[-\ ^{\shortmid }\widetilde{f}+4]\ \left( 4\pi \tau \right)
^{-4}e^{-\ ^{\shortmid }\widetilde{f}}\ d\ ^{\shortmid }\widetilde{\mathcal{V%
}}ol(\tau ).  \notag
\end{eqnarray}%
Using standard statistical and geometric mechanics computations \cite%
{perelman1,monogrrf1,monogrrf2,monogrrf3,vacaru06dd,vv06,svnonh08,vacaru07ee,vacaru11,vacaru13,gheorghiuap16}
(or abstract geometric methods), we can define and compute such
thermodynamic variables: 
\begin{eqnarray}
\mbox{ average energy },\ \widetilde{\mathcal{E}}(\tau )\ &=&-\tau ^{2}\int_{%
\widetilde{\Xi }}\ \left( 4\pi \tau \right) ^{-4}\left( \ \widetilde{\mathbf{%
R}}sc+|\ \widetilde{\mathbf{D}}\ \widetilde{f}|^{2}-\frac{4}{\tau }\right)
e^{-\ \widetilde{f}}\ d\widetilde{\mathcal{V}}ol(\tau );
\label{nagthermodfh} \\
\mbox{ entropy },\ \widetilde{\mathcal{S}}(\tau )\ &=&-\int_{\widetilde{\Xi }%
}\left( 4\pi \tau \right) ^{-4}\left( \tau (\ \widetilde{\mathbf{R}}sc+|%
\widetilde{\mathbf{D}}\widetilde{f}|^{2})+\tilde{f}-8\right) e^{-\ 
\widetilde{f}}\ d\widetilde{\mathcal{V}}ol(\tau );  \notag \\
\mbox{ fluctuation },\widetilde{\sigma }(\tau ) &=&2\ \tau ^{4}\int_{%
\widetilde{\Xi }}\left( 4\pi \tau \right) ^{-4}|\ \widetilde{\mathbf{R}}%
_{\alpha \beta }+\widetilde{\mathbf{D}}_{\alpha }\ \widetilde{\mathbf{D}}%
_{\beta }\ \tilde{f}-\frac{1}{2\tau }\widetilde{\mathbf{g}}_{\alpha \beta
}|^{2}e^{-\ \widetilde{f}}\ d\ \widetilde{\mathcal{V}}ol(\tau );  \notag
\end{eqnarray}%
and 
\begin{eqnarray}
\ ^{\shortmid }\widetilde{\mathcal{E}}(\tau )\ &=&-\tau
^{2}\int_{^{\shortmid }\widetilde{\Xi }}\ \left( 4\pi \tau \right)
^{-4}\left( \ ^{\shortmid }\widetilde{\mathbf{R}}sc+|\ ^{\shortmid }%
\widetilde{\mathbf{D}}\ ^{\shortmid }\widetilde{f}|^{2}-\frac{4}{\tau }%
\right) e^{-\ ^{\shortmid }\widetilde{f}}\ d\ ^{\shortmid }\widetilde{%
\mathcal{V}}ol(\tau );  \notag \\
\ ^{\shortmid }\widetilde{\mathcal{S}}(\tau )\ &=&-\int_{^{\shortmid }%
\widetilde{\Xi }}\left( 4\pi \tau \right) ^{-4}\left( \tau (\ ^{\shortmid }%
\widetilde{\mathbf{R}}sc+|\ ^{\shortmid }\widetilde{\mathbf{D}}\ ^{\shortmid
}\widetilde{f}|^{2})+\ ^{\shortmid }\tilde{f}-8\right) e^{-\ ^{\shortmid }%
\widetilde{f}}\ d\ ^{\shortmid }\widetilde{\mathcal{V}}ol(\tau );  \notag \\
\ ^{\shortmid }\widetilde{\sigma }(\tau ) &=&2\ \tau ^{4}\int_{^{\shortmid }%
\widetilde{\Xi }}\left( 4\pi \tau \right) ^{-4}|\ \ ^{\shortmid }\widetilde{%
\mathbf{R}}_{\alpha \beta }+\ ^{\shortmid }\widetilde{\mathbf{D}}_{\alpha }\
^{\shortmid }\widetilde{\mathbf{D}}_{\beta }\ ^{\shortmid }\tilde{f}-\frac{1%
}{2\tau }\ ^{\shortmid }\widetilde{\mathbf{g}}_{\alpha \beta }|^{2}e^{-\
^{\shortmid }\widetilde{f}}\ d\ ^{\shortmid }\widetilde{\mathcal{V}}ol(\tau
).  \notag
\end{eqnarray}%
We note that, for instance, $\ ^{\shortmid }\widetilde{\mathcal{W}}(\tau)=-\
^{\shortmid }\widetilde{\mathcal{S}}(\tau )$ (\ref{nawfunctfh}). The
formulas allow us to compare and select thermodynamically different FL and
FH theories (\ref{nagthermodfh}). 

\subsection{Nonmetric geometric flow equations in canonical dyadic variables}

We elaborate on the theory of nonmetric FLH geometric flows in canonical
dyadic variables and families of geometric and physical data 
\begin{equation}
(\ \widehat{\mathcal{M}}\mathbf{,N}(\tau ),\mathbf{g}(\tau ),\mathbf{D}(\tau
)=\nabla (\tau )+\mathbf{L}(\tau )=\widehat{\mathbf{D}}(\tau )+\widehat{%
\mathbf{L}}(\tau ),\ ^{g}\widehat{\mathcal{L}}(\tau )+\ ^{m}\widehat{%
\mathcal{L}}(\tau ))  \label{canonicalmafdata}
\end{equation}%
parameterized by a real parameter $\tau ,0\leq \tau \leq \tau _{1}.$ The
distortion relations for any fixed $\tau _{0}$ are defined by formulas (\ref%
{disf}) and (\ref{driccidist}), when the conventional actions $\ ^{g}%
\widehat{\mathcal{L}}(\tau )$ and $\ ^{m}\widehat{\mathcal{L}}(\tau )$ for
phase space interactions are stated as in (\ref{actnmc}). We also consider
the classification (\ref{classflh}) and the possibility of defining dual N-
and s-adapted geometric models on $\ ^{\shortmid }\widehat{\mathcal{M}}$.

\subsubsection{$Q$-distorted R. Hamilton and D. Friedan equations}

Natural $\tau $-flows with 4+4 respective h- and v-splitting exist on $%
\widehat{\mathcal{M}},$ when $\mathbf{g}(\tau )=[g_{ij}(\tau),g_{ab}(\tau)]$
for respective $\tau $-families of $\mathbf{N}(\tau )=\{N_{i}^{a}(\tau)\}. $
We can consider similar dyadic s-decompositions with 4 shells, $s=1,2,3,4$
and respective splitting of indices $\alpha _{s}=(i_{s-1},a_{s}).$ In such
cases, we write $\ _{s}\widehat{\mathcal{M}}$ with $\mathbf{g}(\tau)\simeq \
_{s}\mathbf{g}(\tau )=[g_{i_{s-1}j_{s-1}}(\tau),g_{a_{s}b_{s}}(\tau )].$ The
nonholonomic and distortion of d-connection structures can be prescribed in
such forms that for any $\tau =\tau _{0}$ the canonical modified Einstein
equations (\ref{cfeq4a}) hold. The nonmetric geometric flow evolution
equations can be postulated: 
\begin{align}
\partial _{\tau }g_{ij}(\tau )& =-2[\widehat{\mathbf{R}}_{ij}(\tau )-\ _{Q}%
\widehat{\yen }_{ij}(\tau )];\   \label{ricciflowr2} \\
\partial _{\tau }g_{ab}(\tau )& =-2[\widehat{\mathbf{R}}_{ab}(\tau )-\ _{Q}%
\widehat{\yen }_{ab}(\tau )];  \notag \\
\widehat{\mathbf{R}}_{ia}(\tau )& =\widehat{\mathbf{R}}_{ai}(\tau )=0;%
\widehat{\mathbf{R}}_{ij}(\tau )=\widehat{\mathbf{R}}_{ji}(\tau );\widehat{%
\mathbf{R}}_{ab}(\tau )=\widehat{\mathbf{R}}_{ba}(\tau ),\   \notag
\end{align}%
where the h- and v-components of sources $\ _{Q}\widehat{\yen }_{\alpha
\beta }=[\ _{Q}\widehat{\yen }_{ij}(\tau ),\ _{Q}\widehat{\yen }_{ab}(\tau
)] $ (\ref{ssourc}) can be written in s-adapted forms. In (\ref{ricciflowr2}%
), $\widehat{\mathbf{R}}_{\alpha \beta }(\tau )$ is equivalent to $\widehat{%
\square }(\tau )=\widehat{\mathbf{D}}^{\alpha }(\tau )\widehat{\mathbf{D}}%
_{\alpha }(\tau )$ (the canonical d'Alambert operator, or phase space wave
operator) for small perturbations of the standard Minkowski tangent bundle.
The conditions $\widehat{\mathbf{R}}_{ia}(\tau )=\widehat{\mathbf{R}}%
_{ai}(\tau )=0$ have to be imposed for any fixed $\tau _{0},$ $\widehat{R}ic[%
\widehat{\mathbf{D}}]=\{\widehat{\mathbf{R}}_{\alpha \beta }=[\widehat{R}%
_{ij},\widehat{R}_{ia},\widehat{R}_{ai},\widehat{R}_{ab}]\}$ if we elaborate
on a phase space theory with symmetric d-metrics evolving under nonmetric
nonholonomic Ricci flows. 

On dual phase spaces $\ ^{\shortmid }\widehat{\mathcal{M}}$, the nonmetric
geometric flow evolution equations can be written in the form 
\begin{align}
\partial _{\tau }g_{ij}(\tau )& =-2[\widehat{\mathbf{R}}_{ij}(\tau )-\ _{Q}%
\widehat{\yen }_{ij}(\tau )];\   \label{ricciflowr2d} \\
\partial _{\tau }\ ^{\shortmid }g^{ab}(\tau )& =-2[\ ^{\shortmid }\widehat{%
\mathbf{R}}^{ab}(\tau )-\ _{Q}^{\shortmid }\widehat{\yen }^{ab}(\tau )]; 
\notag \\
\ ^{\shortmid }\widehat{\mathbf{R}}_{i}^{\ a}(\tau )& =\ ^{\shortmid }%
\widehat{\mathbf{R}}_{\ i}^{a}(\tau )=0;\widehat{\mathbf{R}}_{ij}(\tau )=%
\widehat{\mathbf{R}}_{ji}(\tau );\ ^{\shortmid }\widehat{\mathbf{R}}%
^{ab}(\tau )=\ ^{\shortmid }\widehat{\mathbf{R}}^{ab}(\tau ),\   \notag
\end{align}%
for respective h- and c-sources $\ _{Q}\widehat{\yen }_{\alpha \beta }=[\
_{Q}\widehat{\yen }_{ij}(\tau ),\ _{Q}\widehat{\yen }_{ab}(\tau )].$

The geometric flow equations (\ref{ricciflowr2}), involving flows on
velocity type variables and (\ref{ricciflowr2d}), involving momentum type
variables, consist of certain generalizations of the R. Hamilton equations 
\cite{hamilton82} postulated for $\nabla .$ We note that equivalent
equations were considered a few years before the mentioned mathematical
works D. Friedan was inspired to introduce geometric flow evolution
equations for research on string theory and condensed matter physics \cite%
{friedan2,friedan3}. Considering respective distortion relations, the above
nonholonomic geometric flow equations can be transformed into respective
tilde ones (for the Cartan d-connection) (\ref{nonassocfhfl}). We can derive
nonmetric variants of geometric flow equations considering an approach,
which is similar to the abstract geometric method from \cite{misner73}, when
certain $\tau $-running fundamental geometric objects Ricci tensors and
generalized sources are distorted to canonical nonholonomic data (\ref%
{canonicalmafdata}). 

\subsubsection{$\protect\tau $-running nonmetric Einstein equations for $%
f(Q) $ nonmetric geometric flows}

We can consider the terms $\partial _{\tau }\mathbf{g}_{\mu ^{\prime }\nu
^{\prime }}(\tau )$ in (\ref{ricciflowr2}) as additional effective sources
determined by $\tau $-running of geometric flows of the canonical Ricci
d-tensors. Using $\tau $-families of vierbein transforms $\mathbf{e}_{\
\mu^{\prime }}^{\mu }(\tau )=\mathbf{e}_{\ \mu ^{\prime }}^{\mu
}(\tau,u^{\gamma })$ and their dual transform $\mathbf{e}_{\nu }^{\ \nu
^{\prime}}(\tau ,u^{\gamma })$ with $\mathbf{e}_{\ }^{\mu }(\tau )=\mathbf{e}%
_{\ \mu^{\prime }}^{\mu }(\tau )du^{\mu ^{\prime }},$ we can introduce
N-adapted effective sources 
\begin{equation}
\ \quad \ _{Q}\mathbf{J}_{\ \nu }^{\mu }(\tau )=\mathbf{e}_{\ \mu }^{\mu
^{\prime }}(\tau )\mathbf{e}_{\nu }^{\ \nu ^{\prime }}(\tau )[~\ _{Q}\yen %
_{\mu ^{\prime }\nu ^{\prime }}(\tau )-\frac{1}{2}~\partial _{\tau }\mathbf{g%
}_{\mu ^{\prime }\nu ^{\prime }}(\tau )]=[\ _{Q}^{h}J(\tau ,{x}^{k})\delta
_{j}^{i},~\ \ _{Q}^{v}J(\tau ,x^{k},y^{a})\delta _{b}^{a}].
\label{dsourcparam}
\end{equation}%
The data $\ _{Q}\mathbf{J}_{\ \nu }^{\mu }(\tau )=[\ _{Q}^{h}J(\tau ), \
_{Q}^{v}J(\tau )]$ can be fixed as some generating functions for effective
matter sources encoding also contributions from $Q$-deformations.
Prescribing explicit values of $\ _{Q}^{h}J(\tau )$ and $\ _{Q}^{v}J(\tau),$
we impose certain nonholonomic constraints on the noncommutative geometric
flow scenarios. If $Q=0,$ such sources encode nonholonomic distortions of $\
\nabla (\tau )$ to $\widehat{\mathbf{D}}(\tau ).$ 

Considering nonholonomic dyadic s-frames and respective generating sources $%
\ _{Q}\mathbf{J}_{\ \nu }^{\mu }(\tau )$ (\ref{dsourcparam}) $\rightarrow \
_{Q}\mathbf{J}_{\ \nu _{s}}^{\mu _{s}}(\tau )$, we can write the nonmetric
geometric flow equations (\ref{ricciflowr2}) as $\tau $-running and $Q$%
-deformed Einstein equations for $\widehat{\mathbf{D}}_{\alpha _{s}}(\tau ),$
\begin{equation}
\widehat{\mathbf{R}}_{\ \ \beta _{s}}^{\alpha _{s}}(\tau )=\ \ _{Q}\mathbf{J}%
_{\ \ \beta _{s}}^{\alpha _{s}}(\tau ),\mbox{ i.e. }\ \widehat{\mathbf{R}}%
_{\ \ \gamma _{s}}^{\beta _{s}}(\tau )={\delta }_{\ \ \gamma _{s}}^{\beta
_{s}}\ \ _{Q}^{s}\mathbf{J}(\tau ),  \label{cfeq4af}
\end{equation}%
with s-adapting of sources as in (\ref{ssourcd}). Such systems of nonlinear
PDEs can be decoupled and integrated in very general forms by applying the
AFCDM as we show in the next section. Constraining nonholonomically the $%
\tau $-parametric equations (\ref{cfeq4af}) system for zero canonical
d-connections, we model nonmetric $\tau $-evolution scenarios in terms of
LC-connections: 
\begin{align}
\widehat{\mathbf{T}}_{\ \alpha _{s}\beta _{s}}^{\gamma _{s}}(\tau )& =0,%
\mbox{ for }\ ^{s}\widehat{\mathbf{D}}_{|\widehat{\mathcal{T}}=0}(\tau )=\
^{s}\nabla (\tau ),\mbox{
when }  \label{lccondf} \\
\breve{R}_{\beta _{s}\gamma _{s}}(\tau )& =\ \ \ \breve{J}_{\beta _{s}\gamma
_{s}}(\tau ),\mbox{ when }_{Q}\mathbf{J}_{\beta _{s}\gamma _{s}}(\tau
)\rightarrow \ \breve{J}_{\beta _{s}\gamma _{s}}(\tau ),\ \ ^{s}\widehat{%
\mathbf{Q}}=0\mbox{ for }\ ^{s}\widehat{\mathcal{T}}=0.  \label{gfeq2af}
\end{align}%
We emphasize that to model metric or nonmetric FL geometric flows, the
nonholonomic LC-constraints (\ref{lccondf}) or (\ref{gfeq2af}) are not
appropriate. For instance, we have to consider a nontrivial Cartan
d-connection $\widetilde{\mathcal{T}}\neq 0$ for $\ ^{s}\widehat{\mathbf{D}}%
\rightarrow \widetilde{\mathbf{D}}.$

If we work on dual phase spaces $\ \ _{s}^{\shortmid }\widehat{\mathcal{M}}$%
, a similar abstract s-adapted geometric formalism allows to re-write the
nonmetric geometric flow equations (\ref{ricciflowr2d}) into 
\begin{equation}
\ ^{\shortmid }\widehat{\mathbf{R}}_{\ \ \beta _{s}}^{\alpha _{s}}(\tau )=\
_{Q}^{\shortmid }\mathbf{J}_{\ \ \beta _{s}}^{\alpha _{s}}(\tau )%
\mbox{
i.e. }\ ^{\shortmid }\widehat{\mathbf{R}}_{\ \ \gamma _{s}}^{\beta
_{s}}(\tau )={\delta }_{\ \ \gamma _{s}}^{\beta _{s}}\ _{Q}^{s\shortmid }%
\mathbf{J}(\tau ).  \label{feq4afd}
\end{equation}%
Such systems of nonlinear PDEs involve momentum-like variables in a form
dual to (\ref{cfeq4af}). The respective N-adapted dyadic configurations can
be stated in certain forms when respective classes of generic off-diagonal
solutions are also dual. If necessary, we can impose on $\ _{s}^{\shortmid }%
\widehat{\mathcal{M}}$ nonholonomic constraints of type (\ref{lccondf}) or (%
\ref{gfeq2af}).

In \cite{bnsvv24,bvvz24}, the nonholonomic and nonmetric Ricci soliton
configurations were defined as self-similar ones for the corresponding
nonmetric geometric flow equations. Such constructions can be performed on $%
\widehat{\mathcal{M}}.$ Fixing $\tau =\tau _{0}$ in (\ref{ricciflowr2}), we
obtain the equations for the $\widehat{f}(\widehat{Q})$ Ricci solitons: 
\begin{align}
\widehat{\mathbf{R}}_{ij}& =\ _{Q}\widehat{\yen }_{ij},\ \widehat{\mathbf{R}}%
_{ab}=\ _{Q}\widehat{\yen }_{ab},  \label{canriccisol} \\
\widehat{\mathbf{R}}_{ia}& =\widehat{\mathbf{R}}_{ai}=0;\widehat{\mathbf{R}}%
_{ij}=\widehat{\mathbf{R}}_{ji};\widehat{\mathbf{R}}_{ab}=\widehat{\mathbf{R}%
}_{ba}.  \notag
\end{align}%
The nonholonomic variables can be chosen in such forms that (\ref%
{canriccisol}) are equivalent to $\tau $-families of nonmetric modified
Einstein equations (\ref{cfeq4af}) for $\ ^{s}\widehat{\mathbf{D}}^{\alpha
}(\tau _{0})$. For additional nonholonomic constraints (\ref{lccondf}) or (%
\ref{gfeq2af}), such equations define solutions of the nonmetric phase space
Einstein equations for $\nabla (\tau _{0}).$ On $\ ^{\shortmid }\widehat{%
\mathcal{M}}$, we can define nonmetric canonical Ricci solitons as solutions
of a nonlinear system of PDEs which are dual to (\ref{canriccisol}). 

Finally, in this subsection, we discuss the problem of formulating
conservation laws for $Q$-deformed systems. Using the canonical
d-connection, we can check that for systems of type (\ref{canriccisol}), and
related nonholonomic modified Einstein equations, the conditions 
\begin{equation*}
\widehat{\mathbf{D}}^{\beta }\widehat{\mathbf{E}}_{\ \ \beta }^{\alpha }=%
\widehat{\mathbf{D}}(\widehat{\mathbf{R}}_{\ \ \beta }^{\alpha }-\frac{1}{2}%
\ ^{s}\widehat{R}\delta _{\ \ \beta }^{\alpha })\neq 0\mbox{ and }\widehat{%
\mathbf{D}}^{\beta }\ _{Q}\yen _{\ \ \beta }^{\alpha }\neq 0,
\end{equation*}%
are satisfied. This is typical for nonholonomic systems with non-integrable
constraints, which requests additional geometric constructions and
restrictions on classes of generating functions and effective sources. The
issue of formulating conservation laws on $\ \widehat{\mathcal{M}}$ becomes
more sophisticate because of nonmetricities. Similar problems exist also in
nonholonomic mechanics and various versions of FLH geometries, when the
conservation laws are formulated by solving nonholonomic constraints or
introducing some Lagrange multipliers associated with certain classes of
nonholonomic constraints. In certain cases, we can solve the constraint
equations (they may depend on local coordinates, velocities or momentum
variables); in such cases, we can redefine the variables and formulate
conservation laws in some explicit or non-explicit forms. Such nonholonomic
variables allow us to introduce new effective (mechanical) Lagrangians and
Hamiltonians. This allows us to define conservation laws in certain standard
forms only if $\mathbf{Q}_{\alpha \beta \gamma }=0$ and nonholonomically
induced, or vanishing, torsions. 

\subsection{Statistical geometric thermodynamics for $f(Q)$ geometric flows}

\label{ssperelth}G. Perelman \cite{perelman1} introduced the so-called F-and
W-functionals as certain Lyapunov-type functionals, $\breve{F}(\tau,g,\nabla
,\breve{R}sc)$ and $\breve{W}(\tau ,g,\nabla ,\breve{R}sc)$ depending on a
temperature like parameter $\tau $ and fundamental geometric objects when $%
\breve{W}$ have properties of "minus entropy". Using $\breve{F} $ or $\breve{%
W},$ he elaborated a variational proof for geometric flow equations of
Riemannian metrics, which was applied to developing a strategy for proving
the Poincar\'{e}--Thorston conjecture. Respective details are provided in
mathematical monographs \cite{monogrrf1,monogrrf2,monogrrf3}. It is not
possible to formulate and prove in some general form such conjectures for
non-Riemannian geometries, including FLH theories. More than that, it is not
clear how, in a general form, we could define general metric-affine (with
nontrivial nonmetricity) generalizations of GR and standard particle
theories (with nonmetric variants of Dirac equations) \cite%
{vacaru10,vacaru18,vmon3}. Nevertheless, nonmetric geometric flow (including
certain Finsler-like variables) where studied in a series of works \cite%
{vacaru07a,svnonh08,vacaru11,vacaru13,vacaru25b,bsssvv25}. In such works,
the main goals are to consider modifications of G. Perelman's Ricci flows
thermodynamics and applications in modern gravity and cosmology.

\subsubsection{Nonholonomic F-/ W-functionals for $Q$-deformed geometric
flows and gravity}

Considering $\tau $-families of distortions (\ref{ccnadapv}), resulting in
respective changing of geometric data $[\widetilde{\mathbf{g}}(\tau ),%
\widetilde{\mathbf{D}}(\tau )]\rightarrow \lbrack \mathbf{g}(\tau ),\widehat{%
\mathbf{D}}(\tau )],$ we can distort (\ref{naffunctfh}) and (\ref{nawfunctfh}%
) to 
\begin{eqnarray}
\ \widehat{\mathcal{F}}(\tau ) &=&\int_{\ \widehat{\Xi }}(\widehat{\mathbf{R}%
}sc+|\widehat{\mathbf{D}}\widehat{f}|^{2})e^{-\ \widehat{f}}\ d\widehat{%
\mathcal{V}}ol(\tau ),  \label{fcanon} \\
\ \widehat{\mathcal{W}}(\tau ) &=&\int_{\ \widehat{\Xi }}\left( 4\pi \tau
\right) ^{-4}\ [\tau (\widehat{\mathbf{R}}sc+|\widehat{\mathbf{D}}\widehat{f}%
|)^{2}+\widehat{f}-8]e^{-\ \widehat{f}}\ d\ \widehat{\mathcal{V}}ol(\tau ).
\label{wcanon}
\end{eqnarray}%
The 8-d hypersurfrace integrals on $\widehat{\Xi }$ \ for F- and
W-functionals are determined corresponding by volume elements (\ref%
{volformfh}) redefined for respective transforms of $\tau $-families of
d-metrics, $\widetilde{\mathbf{g}}_{\alpha \beta }\ (\tau )$ (\ref{cdms8})
to $\mathbf{g}_{\alpha \beta }\ (\tau )$ (\ref{dmgener}), when 
\begin{equation}
d\ \ \widehat{\mathcal{V}}ol(\tau )=\sqrt{|\ \widehat{\mathbf{g}}_{\alpha
\beta }\ (\tau )|}\ \widehat{\delta }^{8}\ u^{\gamma }(\tau ).
\label{volumforma}
\end{equation}%
Such transforms can be adapted to a canonical normalizing function $\widehat{%
f}(\tau ,u)$ which, for instance, can be stated to satisfy the condition 
\begin{equation}
\int_{\ \widehat{\Xi }}\widehat{\nu }\ \ d\ \widehat{\mathcal{V}}ol(\tau
):=\int_{t_{1}}^{t_{2}}\int_{\widehat{\Xi }_{t}}\ \int_{\ \widehat{\Xi }_{E}}%
\widehat{\nu }\ \ d\ \widehat{\mathcal{V}}ol(\tau )=1.  \label{volformfc}
\end{equation}%
In these formulas, where the integration measures $\ \ \widehat{\nu }%
=\left(4\pi \tau \right) ^{-4}e^{-\ \widehat{f}}$ are parameterized for the
h- and v-components (or with shell further parameterizations if necessary).
For general topological considerations, such conditions may not be
considered. We can also choose certain variants of $\widehat{f}$ to simplify
certain formulas and computations. The formulas (\ref{fcanon}) - (\ref%
{volformfc}) are defined for canonical nonholonomical variables on $\ 
\widehat{\mathcal{M}}.$ Using corresponding abstract labels and nonholonomic
geometric objects, we can redefine them for canonical geometric flows on $\
^{\shortmid }\widehat{\mathcal{M}}\,.$ 

To introduce in explicit form certain terms with $\tau $-flows of matter and
effective (encoding distortions and nonmetricity) Lagrangians, we can
re-define correspondingly the normalizing function $\ \widehat{f}%
(\tau)\rightarrow \zeta (\tau ),$ when 
\begin{equation}
\partial _{\tau }\zeta (\tau )=-\widehat{\square }(\tau )[\zeta (\tau
)]+\left\vert \widehat{\mathbf{D}}\ \zeta (\tau )\right\vert ^{2}-\widehat{f}%
(\widehat{\mathbf{R}}sc(\tau ))-\ ^{e}\mathcal{L}(\tau )-\ ^{m}\widehat{%
\mathcal{L}}(\tau ).  \label{normconda}
\end{equation}%
This equation for $\zeta (\tau )$ can be N- and s-adapted in such forms when
the nonlinear partial differential operators of first and second order on $\ 
\widehat{\mathcal{M}}$ relate $\tau $-families of canonical Ricci scalars $\ 
\widehat{\mathbf{R}}sc(\tau )$ and nonmetric sources (\ref{ceemt}) for (\ref%
{cfeq4a}), for certain fixed $\tau _{0},$ or with general sources (\ref%
{dsourcparam}) for (\ref{cfeq4af}). The nonholonomic variables can be
considered in any form as stated in (\ref{classflh}). Such a nonlinear PDE
can't be solved in a general form which do not allow us to study in general
forms models of flow evolution of FLH and topological configurations
determined by arbitrary nonmetric structures and distributions of effective
and real matter fields. We can prescribe a topological structure for an
off-diagonal metric constructed as an exact/ parametric solution of
nonholonomic system of nonlinear PDEs (\ref{ricciflowr2}). Or we can use
transforms (\ref{normcond}) for (\ref{fcanon}), or (\ref{wcanon}), to
perform formal variational N-adapted or s-adapted proofs of the canonical
nonholonomic geometric flow equations. We can choose a convenient $\zeta
(\tau )$ (it can be prescribed to be a constant, or zero) which states
certain constraints on nonmetric geometric evolution but allows to simplify
certain formulas, or to generate some classes of solutions. Alternatively,
we can solve the equation (\ref{normcond}) in certain parametric forms and
then to redefine the constructions for arbitrary systems of reference. 

\subsubsection{Canonical thermodynamic variables for $f(Q)$ theories}

For nonmetric FL geometric flow models defined by $\tau $-families of
nonholonomic data $\ _{\mathbf{Q}}\mathcal{M}:\mathbf{g}\simeq \ _{s}\mathbf{%
g},\ \mathbf{N}\simeq \ _{s}\ \mathbf{N},\ _{s}\widehat{\mathbf{D}}=\ 
\mathbf{D}+\ _{s}\mathbf{Z}, \ _{\ }^{g}\widehat{\mathcal{L}}+\ _{\ }^{m}%
\widehat{\mathcal{L}},\ _{Q}\widehat{\yen }_{\alpha \beta },\ \widehat{%
\mathbf{Q}}\simeq \mathbf{Q}\neq 0,\ \widehat{\mathbf{T}}\neq 0,\mathcal{T}%
\mathbf{[\nabla ]=0}),$ when contributions for $\widehat{f}(\widehat{\mathbf{%
Q}})$ are encoded into $\zeta (\tau )\rightarrow \ _{\mathbf{Q}}\zeta (\tau) 
$ as in (\ref{normcond}), when $\ ^{e}\mathcal{L}(\tau )\rightarrow \
_{Q}^{e}\mathcal{L}(\tau )$ is determined by distortions of nonmetricity
fields. Conventionally we consider $\tau $-families of geometric data: \ $[ 
\mathbf{g}_{\alpha \beta }\ (\tau ),\widehat{\mathbf{D}}(\tau )=\ \mathbf{D}%
(\tau )+\mathbf{Z}(\tau )]$ and $[\ ^{\shortmid }\mathbf{g}_{\alpha \beta }\
(\tau ),\ ^{\shortmid }\widehat{\mathbf{D}}(\tau)=\ \ ^{\shortmid }\mathbf{D}%
(\tau )+\ ^{\shortmid }\mathbf{Z}(\tau )].$ Nonholonomic dyadic s-adapted
canonical variables can also be introduced for characterizing respective
classes of off-diagonal solutions, when $\widehat{\Xi }\subset \ _{\mathbf{Q}%
}\mathcal{M}$ and $\ ^{\shortmid }\widehat{\Xi }\subset \ _{\mathbf{Q}%
}^{\shortmid }\mathcal{M};$ and the corresponding volume forms of type (\ref%
{volformfc}). We can introduce respective partition functions for nonmetric
deformed FL and FH phase spaces of dimension $n=8$, 
\begin{eqnarray}
\ \ _{\mathbf{Q}}\widehat{\mathcal{Z}}(\tau ) &=&\exp [\int_{\widehat{\Xi }%
}[-\ \ _{\mathbf{Q}}{{{{\widehat{\zeta }}}}}+4]\ \left( 4\pi \tau \right)
^{-4}e^{-\ \ \ _{\mathbf{Q}}{{{{\widehat{\ \zeta }}}}}}\ d\widehat{\mathcal{V%
}}ol(\tau )\mbox{ and }  \label{spffhq} \\
\ _{\mathbf{Q}}^{\shortmid }\widehat{\mathcal{Z}}(\tau ) &=&\exp [\int_{\
^{\shortmid }\widehat{\Xi }}[-\ _{\mathbf{Q}}^{\shortmid }{{{{\widehat{\zeta 
}}}}}+4]\ \left( 4\pi \tau \right) ^{-4}e^{-\ _{\mathbf{Q}}^{\shortmid }{{{{%
\widehat{\zeta }}}}}}\ d\ ^{\shortmid }\widetilde{\mathcal{V}}ol(\tau ). 
\notag
\end{eqnarray}%
Similarly to (\ref{spffhq}) and (\ref{nagthermodfh}), for respective
distortions. Using standard statisical and geometric mechanics computations 
\cite%
{perelman1,monogrrf1,monogrrf2,monogrrf3,vacaru06dd,vv06,svnonh08,vacaru07ee,vacaru11,vacaru13,gheorghiuap16}
(or applying the abstract geometric methods) we can define and compute such
thermodynamic variables: 
\begin{eqnarray}
\ _{\mathbf{Q}}\widehat{\mathcal{E}}(\tau )\ &=&-\tau ^{2}\int_{\widehat{\Xi 
}}\ \left( 4\pi \tau \right) ^{-4}\left( \ \widehat{\mathbf{R}}sc+|\ 
\widehat{\mathbf{D}}\ _{\mathbf{Q}}{{{{\widehat{\zeta }}}}}|^{2}-\frac{4}{%
\tau }\right) e^{-\ _{\mathbf{Q}}{{{{\widehat{\zeta }}}}}}\ d\widehat{%
\mathcal{V}}ol(\tau );  \label{nagthermodfhq} \\
\ _{\mathbf{Q}}\widehat{\mathcal{S}}(\tau )\ &=&-\int_{\widehat{\Xi }}\left(
4\pi \tau \right) ^{-4}\left( \tau (\ \widehat{\mathbf{R}}sc+|\widehat{%
\mathbf{D}}\ _{\mathbf{Q}}{{{{\widehat{\zeta }}}}}|^{2})+\ _{\mathbf{Q}}{{{{%
\widehat{\zeta }}}}}-8\right) e^{-\ _{\mathbf{Q}}{{{{\widehat{\zeta }}}}}}\ d%
\widehat{\mathcal{V}}ol(\tau );  \notag \\
\ _{\mathbf{Q}}\widehat{\sigma }(\tau ) &=&2\ \tau ^{4}\int_{\widehat{\Xi }%
}\left( 4\pi \tau \right) ^{-4}|\ \widehat{\mathbf{R}}_{\alpha \beta }+%
\widehat{\mathbf{D}}_{\alpha }\ \widehat{\mathbf{D}}_{\beta }\ _{\mathbf{Q}}{%
{{{\widehat{\zeta }}}}}-\frac{1}{2\tau }\mathbf{g}_{\alpha \beta }|^{2}e^{-\
_{\mathbf{Q}}{{{{\widehat{\zeta }}}}}}\ d\widehat{\mathcal{V}}ol(\tau ); 
\notag
\end{eqnarray}
and 
\begin{eqnarray}
\ _{\mathbf{Q}}^{\shortmid }\widehat{\mathcal{E}}(\tau )\ &=&-\tau
^{2}\int_{^{\shortmid }\widehat{\Xi }}\ \left( 4\pi \tau \right) ^{-4}\left(
\ ^{\shortmid }\widehat{\mathbf{R}}sc+|\ ^{\shortmid }\widehat{\mathbf{D}}\
_{Q}^{\shortmid }\widehat{\zeta }|^{2}-\frac{4}{\tau }\right) e^{-\
_{Q}^{\shortmid }\widehat{\zeta }}\ d\ ^{\shortmid }\widehat{\mathcal{V}}%
ol(\tau );  \notag \\
\ _{\mathbf{Q}}^{\shortmid }\widehat{\mathcal{S}}(\tau )\
&=&-\int_{^{\shortmid }\widehat{\Xi }}\left( 4\pi \tau \right) ^{-4}\left(
\tau (\ ^{\shortmid }\widehat{\mathbf{R}}sc+|\ ^{\shortmid }\widehat{\mathbf{%
D}}\ _{Q}^{\shortmid }\widehat{\zeta }|^{2})+\ _{Q}^{\shortmid }\widehat{%
\zeta }-8\right) e^{-\ _{Q}^{\shortmid }\widehat{\zeta }}\ d\ ^{\shortmid }%
\widehat{\mathcal{V}}ol(\tau );  \notag \\
\ _{\mathbf{Q}}^{\shortmid }\widehat{\sigma }(\tau ) &=&2\ \tau
^{4}\int_{^{\shortmid }\widehat{\Xi }}\left( 4\pi \tau \right) ^{-4}|\ \
^{\shortmid }\widehat{\mathbf{R}}_{\alpha \beta }+\ ^{\shortmid }\widehat{%
\mathbf{D}}_{\alpha }\ ^{\shortmid }\widehat{\mathbf{D}}_{\beta }\
_{Q}^{\shortmid }\widehat{\zeta }-\frac{1}{2\tau }\ ^{\shortmid }\widetilde{%
\mathbf{g}}_{\alpha \beta }|^{2}e^{-\ _{Q}^{\shortmid }\widehat{\zeta }}\ d\
^{\shortmid }\widehat{\mathcal{V}}ol(\tau ).  \notag
\end{eqnarray}%
In a similar form, we can compute, compare and select thermodynamically
different nonmetric FL and FH theories. 

\subsection{Different thermodynamic formulations of (non) metric FLH theories%
}

The formulas (\ref{spffhq}) and (\ref{nagthermodfhq}) written in hat
geometric data allows to compute respective thermodynamic variables directly
for any class of solutions of $\,Q$-deformed Einstein equations (\ref%
{cfeq4af}) or (\ref{feq4afd}). Such thermodynamic values can be defined in
similar forms for any d-connections, d-metrics and respective distortions of
type (\ref{flh}), (\ref{canonicalscon}), and (\ref{classflh}).

For instance, can consider the Berwald, $\ ^{B}\mathbf{D},$ d-connection
from (\ref{flh}) and compute respectively 
\begin{eqnarray}
\ _{\mathbf{Q}}^{B}\mathcal{E}(\tau )\ &=&-\tau ^{2}\int_{\Xi }\ \left( 4\pi
\tau \right) ^{-4}\left( \ \ ^{B}\mathbf{R}sc+|\ \ ^{B}\mathbf{D}\ _{\mathbf{%
Q}}^{\ ^{B}}{\zeta }|^{2}-\frac{4}{\tau }\right) e^{-\ _{\mathbf{Q}}^{\ ^{B}}%
{\zeta }}\ d\ ^{B}\mathcal{V}ol(\tau );  \label{bervnonh} \\
\ _{\mathbf{Q}}^{B}\mathcal{S}(\tau )\ &=&-\int_{\Xi }\left( 4\pi \tau
\right) ^{-4}\left( \tau (\ \ \ ^{B}\mathbf{R}sc+|\ \ ^{B}\mathbf{D}\ _{%
\mathbf{Q}}^{B}{\zeta }|^{2})+\ _{\mathbf{Q}}^{B}{\zeta }-8\right) e^{-\ _{%
\mathbf{Q}}^{B}{\zeta }}\ d\ ^{B}\mathcal{V}ol(\tau );  \notag \\
&&...,  \notag
\end{eqnarray}%
Such formulas are determined by corresponding geometric d-objects with
abstract left label $B$ introduced in (\ref{nagthermodfhq}) and can be
defined for any $\tau $-family of Berwald-Finsler geometries $\left( \mathbf{%
g}(\tau )\mathbf{\simeq }\ ^{F}\mathbf{g}(\tau ), \ ^{B}\mathbf{D} (\tau
)\right) $ on $\ _{\mathbf{Q}}^{B}\mathcal{M}$. In non-explicit form, such
nonholonomic geometric thermodynamic configurations are characterized by
generalized R. Hamilton and D. Friedan equations of type 
\begin{eqnarray}
\partial _{\tau }\mathbf{g}_{\alpha \beta }(\tau ) &=&-2\ ^{B}\mathbf{R}_{\
\alpha \beta }(\tau ),  \label{hamiltbervnonh} \\
\partial _{\tau }\ ^{B}f(\tau ) &=&\ ^{B}\mathbf{R}sc(\tau )-\
^{B}\bigtriangleup (\tau )\ ^{B}f(\tau )+(\ ^{B}\mathbf{D}(\tau )\
^{B}f(\tau ))^{2},  \notag
\end{eqnarray}%
where $\ ^{B}\bigtriangleup (\tau )=[\ \ ^{B}\mathbf{D}(\tau )]^{2}$ are
families of the Laplace d-operators computed for $\ \mathbf{g}_{\alpha
\beta}(\tau ).$ Such nonlinear PDEs can be derived in N-adapted variational
forms from the corresponding F- and W-potentials, when $\ _{\mathbf{Q}}^{B}%
\mathcal{W}(\tau )\ =-\ _{\mathbf{Q}}^{B}\mathcal{S}(\tau )$ as it was
explained for (\ref{nonassocfhfl}). In equivalent form, such proofs can be
performed in abstract geometric form using respective distortions of the
results and methods from \cite{perelman1}, see details in monographs \cite%
{monogrrf1,monogrrf2,monogrrf3}. We have to consider nonholonomic s-adapted
hat variables and respective additional distortions with re-definition of
normalizing functions if we want to compute (\ref{bervnonh}) for explicit
solutions of (\ref{hamiltbervnonh}). Such systems of nonlinear PDEs can be
written in the form (\ref{cfeq4af}) for $\mathbf{g}\mathbf{\simeq } \ ^{F}%
\mathbf{g\simeq }\ _{s}\mathbf{\mathbf{g\simeq }\ }\{g_{\alpha \beta }\}%
\mathbf{\mathbf{\simeq }\ }\{g_{\alpha _{s}\beta _{s}}\}.$ In the next
section, we shall prove that such equations can be decoupled and integrated
in explicit forms. For a class of solutions for Berwald-Finsler geometric
flows, $\left(\ _{s}\mathbf{g}(\tau )\mathbf{\simeq }\ _{s}^{F}\mathbf{g}%
(\tau ), \ _{s}^{B}\widehat{\mathbf{D}}(\tau )\right) $ written in canonical
s-variables, we can compute: 
\begin{eqnarray}
\ _{\mathbf{Q}}^{B}\widehat{\mathcal{E}}(\tau )\ &=&-\tau ^{2}\int_{\widehat{%
\Xi }}\ \left( 4\pi \tau \right) ^{-4}\left( \ \ _{s}^{B}\widehat{\mathbf{R}}%
sc+|\ \ _{s}^{B}\widehat{\mathbf{D}}\ _{\mathbf{Q}}^{s}{{{{\widehat{\zeta }}}%
}}|^{2}-\frac{4}{\tau }\right) e^{-\ _{\mathbf{Q}}^{s}{{{{\widehat{\zeta }}}}%
}}\ d\ _{s}\widehat{\mathcal{V}}ol(\tau );  \label{berwfinsltherm} \\
\ _{\mathbf{Q}}^{B}\widehat{\mathcal{S}}(\tau )\ &=&-\int_{\widehat{\Xi }%
}\left( 4\pi \tau \right) ^{-4}\left( \tau (\ \ \ _{s}^{B}\widehat{\mathbf{R}%
}sc+|\ \ _{s}^{B}\mathbf{D}\ _{\mathbf{Q}}^{s}{\zeta }|^{2})+\ _{\mathbf{Q}%
}^{s}{\zeta }-8\right) e^{-\ _{\mathbf{Q}}^{s}{\zeta }}\ d\ \ _{s}\widehat{%
\mathcal{V}}ol(\tau );  \notag \\
&&.....  \notag
\end{eqnarray}

Alternatively, we can consider the Chern, $\ ^{C}\mathbf{D},$ d-connection
stated in (\ref{flh}) and compute above thermodynamic variables, labeled and
defined in functional forms as 
\begin{eqnarray}
\ _{\mathbf{Q}}^{C}\mathcal{E}(\tau ) &=&\ _{\mathbf{Q}}^{C}\mathcal{E}\left[
\tau ,\ ^{C}\mathbf{D}(\tau ),\ _{\mathbf{Q}}^{\ ^{C}}{\zeta }(\tau ),\ ^{C}%
\mathcal{V}ol(\tau )\right] ,\ _{\mathbf{Q}}^{C}\mathcal{S}(\tau )=\ _{%
\mathbf{Q}}^{C}\mathcal{S}\left[ \tau ,\ ^{C}\mathbf{D}(\tau ),\ _{\mathbf{Q}%
}^{\ ^{C}}{\zeta }(\tau ),\ ^{C}\mathcal{V}ol(\tau )\right] ;
\label{chernfinsltherm} \\
\ _{\mathbf{Q}}^{C}\widehat{\mathcal{E}}(\tau ) &=&\ _{\mathbf{Q}}^{C}%
\widehat{\mathcal{E}}\left[ \tau ,\ _{s}^{C}\widehat{\mathbf{D}}(\tau ),\ _{%
\mathbf{Q}}^{\ ^{C}}{\zeta }(\tau ),\ _{s}^{C}\mathcal{V}ol(\tau )\right] ,\
_{\mathbf{Q}}^{C}\widehat{\mathcal{S}}(\tau )=\ _{\mathbf{Q}}^{C}\widehat{%
\mathcal{S}}\left[ \tau ,\ _{s}^{C}\widehat{\mathbf{D}}(\tau ),\ _{\mathbf{Q}%
}^{\ ^{C}}{\zeta }(\tau ),\ _{s}^{C}\mathcal{V}ol(\tau )\right] ;  \notag \\
&&.....  \notag
\end{eqnarray}

The above thermodynamic FLH geometric flow variables are very important for
characterizing physical properties of certain models of Finsler-like gravity
and respective classes of exact/parametric solutions. For instance, if $_{%
\mathbf{Q}}^{B}\widehat{\mathcal{S}}(\tau )\ < \ _{\mathbf{Q}}^{C}\widehat{%
\mathcal{S}}(\tau )$ for certain prescribed common nonholonomic data, we can
conclude that such a Berwald-Finsler geometry is for a more probable
relativistic phase space theory than another Chern-Finsler geometry.
Perhaps, such an analysis should involve computations of analogous values
for the Cartan-Finsler geometry $\widetilde{\mathcal{S}}(\tau )$ (\ref%
{nagthermodfh}). Formulas (\ref{bervnonh}) can be used for computing G.
Perelman thermodynamic values for respective classes of exact/parametric
solutions (in section \ref{sec5}, we shall provide explicit examples for
physically important solutions). 

The FH thermodynamic variables for Berwald-like, Chern-like or other types
configurations on dual phase spaces $\ _{\mathbf{Q}}^{B\shortmid }\mathcal{M}
$ can be defined and computed in similar form by using geometric data
labeled additionally by "$\ ^{\shortmid }".$ Finally, we note that the
abstract energy characteristics are computed and analyzed for certain
functionals $\ _{\mathbf{Q}}^{B\shortmid }\widehat{\mathcal{E}}(\tau )\ < \
_{\mathbf{Q}}^{B\shortmid }\widehat{\mathcal{E}}(\tau ).$ Certain
relativistic FLH spaces, or FLH configurations (determined by a class of
solutions) can be energetically more convenient for a realization of DE or
DM model. 

\section{Decoupling and integration of (non) metric FLH-flows and modified
Einstein equations}

\label{sec4} In this section, we show how the FLH-fow canonically distorted
Einstein equations can be decoupled and integrated in general forms using
the AFCDM. We provide necessary N- and s-adapted coefficient formulas,
analyse respective nonlinear and dual symmetries and discuss the most
important parameterizations of such solutions. Proofs and technical results
are provided in Appendix \ref{appendixa}. Tables 1-13 from Appendix \ref%
{appendixb} consist of a summary of the AFCDM for constructing exact and
parametric solutions on nonmetric FLH deformed geometric flows and modified
Einstein equations. 

\subsection{Off-diagonal and s-adapted ansatz for metrics and canonical
d-connections}

Using general frame and coordinate transforms, $\mathbf{g}_{\alpha _{s}\beta
_{s}}=e_{\ \alpha _{s}}^{\alpha }e_{\ \beta _{s}}^{\beta } \mathbf{\mathring{%
g}}_{\alpha \beta }$ and $\ ^{\shortmid }\mathbf{g}_{\alpha _{s}\beta
_{s}}=\ ^{\shortmid }e_{\ \alpha _{s}}^{\alpha } \ ^{\shortmid }e_{\ \beta
_{s}}^{\beta }\ ^{\shortmid }\mathbf{\mathring{g}}_{\alpha \beta },$ any
d-metric structure (\ref{dmgener}) can be transformed, respectively, into a
canonical type ansatz of s-metrics defined canonically in s-adapted frames (%
\ref{ccnadapc}) or (\ref{ccnadapv}). The \textbf{prime} d-metrics $\mathbf{%
\mathring{g}}_{\alpha \beta }$ or $\ ^{\shortmid }\mathbf{\mathring{g}}%
_{\alpha \beta }$ up to frame transforms can be any FL or FH d-metrics
(Sasaki type or other ones which are well-defined on respective total phase
spaces). Here, we note that if necessary we can consider $\tau $-families of
respective \textbf{target} s- or d-metrics written as $\mathbf{g}_{\alpha
_{s}\beta _{s}}(\tau )$ and $\ ^{\shortmid }\mathbf{g}_{\alpha _{s}\beta
_{s}}(\tau ).$ We can also use left labels (for Berwald, Chern, or Cartan
configurations) as in (\ref{flh}) to emphasize that we investigate some
classes of solutions of respective type. For instance, we can write $\ ^{B}%
\mathbf{\mathring{g}}_{\alpha \beta }, \ ^{C}\mathbf{\mathring{g}}_{\alpha
\beta }, \widetilde{\mathbf{\mathring{g}}}_{\alpha \beta }....$ and to put
such labels for the target configurations $\ ^{B}\mathbf{g}_{\alpha
_{s}\beta _{s}}(\tau ),$ $^{C}\mathbf{g}_{\alpha _{s}\beta _{s}}(\tau ),%
\widetilde{\mathbf{g}}_{\alpha _{s}\beta _{s}}(\tau )...$. This emphasizes
that the target s-metrics are subjected to the conditions to satisfy a $\tau 
$-family of FL modified relativistic flow equations (\ref{cfeq4a}) and
"keeps in memory" the data of a prime FL-configuration. For simplicity, we
shall omit labels of type $\ _{s}^{B} \mathbf{g}(\tau )=\{\ ^{B}\mathbf{g}%
_{\alpha _{s}\beta _{s}}(\tau )\} $ and, for deriving exact and parametric
solutions, write only $\ _{s}\mathbf{g}(\tau )=\{\ \mathbf{g}_{\alpha
_{s}\beta _{s}}(\tau )\},$ etc. if that will not result in ambiguities. The
main conventions are those that we can always fix a $\tau =\tau _{0}$ and
relate the constructions to respective classes of nonholonomic Ricci
solitons and generalized Einstein equations in an MGT. For constructions of
dual phase spaces, we can put an abstract label "$\ ^{\shortmid }$".

In the simplest form, we can prove the decoupling properties of physically
important systems of nonlinear PDEs for off-diagonal anstatz with one
Killing space or time symmetry for $s=2;$ then on one phase
velocity/momentum coordinate symmetry for $s=3;$ and another forth velocity/
energy coordinate symmetry for $s=4.$ In correspondingly s-adapted forms of
reference, such canonical d-metrics are of type 
\begin{eqnarray}
d\widehat{s}^{2}(\tau ) &=&g_{i_{1}}(\tau )(dx^{i_{1}})^{2}+g_{a_{2}}(\tau )(%
\mathbf{e}^{a_{2}}(\tau ))^{2}+\ g_{a_{3}}(\tau )(\mathbf{e}^{a_{3}}(\tau
))^{2}+g_{a_{4}}(\tau )(\mathbf{e}^{a_{4}}(\tau ))^{2},\mbox{where }
\label{ssolutions} \\
\mathbf{e}^{a_{2}}(\tau ) &=&dy^{a_{2}}+N_{k_{1}}^{a_{2}}(\tau )dx^{k_{1}},\ 
\mathbf{e}^{a_{3}}(\tau )=dv^{a_{3}}+\ N_{k_{2}}^{a_{3}}(\tau )dx^{k_{2}},%
\mathbf{e}^{a_{4}}(\tau )=dv^{a_{4}}+N_{k_{3}}^{a_{4}}(\tau )d\ x^{k_{3}},%
\mbox{ or }\   \notag
\end{eqnarray}
\begin{eqnarray}
d\ ^{\shortmid }\widehat{s}^{2}(\tau ) &=&g_{i_{1}}(\tau
)(dx^{i_{1}})^{2}+g_{a_{2}}(\tau )(\mathbf{e}^{a_{2}}(\tau ))^{2}+\
^{\shortmid }g^{a_{3}}(\tau )(\ ^{\shortmid }\mathbf{e}_{a_{3}}(\tau
))^{2}+\ ^{\shortmid }g^{a_{4}}(\tau )(\ ^{\shortmid }\mathbf{e}%
_{a_{4}}(\tau ))^{2},\mbox{where }  \label{ssolutionsd} \\
\mathbf{e}^{a_{2}}(\tau ) &=&dy^{a_{2}}+N_{k_{1}}^{a_{2}}(\tau )dx^{k_{1}},\
^{\shortmid }\mathbf{e}_{a_{3}}(\tau )=dp_{a_{3}}+\ ^{\shortmid
}N_{a_{3}k_{2}}(\tau )dx^{k_{2}},\ ^{\shortmid }\mathbf{e}_{a_{4}}(\tau
)=dp_{a_{4}}+\ ^{\shortmid }N_{a_{4}k_{3}}(\tau )d\ ^{\shortmid }x^{k_{3}}.\ 
\notag
\end{eqnarray}

For instance, the s-metric and N-connection coefficients of (\ref%
{ssolutionsd}) are parameterized in the form 
\begin{equation*}
\frame{{\scriptsize $%
\begin{array}{ccccc}
\begin{array}{c}
{\ g}_{i_{1}}{\ (\tau ,x}^{k_{1}}) \\ 
{\ =e}^{\psi (\hbar ,\kappa ;\tau ,x^{k_{1}})}%
\end{array}
& 
\begin{array}{cc}
\begin{array}{c}
\begin{array}{c}
{g}_{a_{2}}{(\tau ,x}^{i_{1}},y^{3}) \\ 
{N}_{k_{1}}^{a_{2}}{(\tau ,x}^{i_{1}},y^{3})%
\end{array}
\\ 
\end{array}
& 
\begin{array}{c}
\mbox{quasi-} \\ 
\mbox{stationary}%
\end{array}
\\ 
\begin{array}{c}
\begin{array}{c}
{g}_{a_{2}}(\tau ,x^{i_{1}},t) \\ 
{N}_{k_{1}}^{a_{2}}(\tau ,x^{i_{1}},t)%
\end{array}
\\ 
\end{array}
& 
\begin{array}{c}
\mbox{locally anisotropic} \\ 
\mbox{cosmology}%
\end{array}%
\end{array}
& 
\begin{array}{c}
\begin{array}{c}
\ ^{\shortmid }{\ g}^{a_{3}}(\tau ,x^{i_{2}},p_{5}) \\ 
\ ^{\shortmid }N_{a_{3}k_{2}}(\tau ,x^{i_{2}},p_{5})%
\end{array}
\\ 
\\ 
\begin{array}{c}
\ ^{\shortmid }{g}^{a_{3}}(\tau ,x^{i_{2}},p_{6}) \\ 
\ ^{\shortmid }{N}_{a_{3}k_{2}}(\tau ,x^{i_{2}},p_{6})%
\end{array}
\\ 
\end{array}
& 
\begin{array}{c}
\ ^{\shortmid }{g}^{a_{4}}(\tau ,\ ^{\shortmid }x^{i_{3}},p_{7}) \\ 
\ ^{\shortmid }{N}_{a_{4}k_{3}}(\tau ,\ ^{\shortmid }{x}^{i_{3}},p_{7})%
\end{array}
& 
\begin{array}{c}
\mbox{fixed} \\ 
p_{8}=E_{0}%
\end{array}
\\ 
\begin{array}{c}
\tau \mbox{-flows of 2-d} \\ 
\mbox{Poisson eqs} \\ 
{\partial }_{11}^{2}{\psi +\partial }_{22}^{2}{\psi =} \\ 
{2\ \ _{1}^{\shortmid }\Upsilon (\tau ,x}^{k_{1}})%
\end{array}
& 
\begin{array}{cc}
\begin{array}{c}
\begin{array}{c}
{g}_{a_{2}}(\tau ,x^{i_{1}},y^{3}) \\ 
{N}_{k_{1}}^{a_{2}}(\tau ,x^{i_{1}},y^{3})%
\end{array}
\\ 
\end{array}
& 
\begin{array}{c}
\mbox{quasi-} \\ 
\mbox{stationary}%
\end{array}
\\ 
\begin{array}{c}
\begin{array}{c}
{g}_{a_{2}}(\tau ,x^{i_{1}},t) \\ 
{N}_{k_{1}}^{a_{2}}(\tau ,x^{i_{1}},t)%
\end{array}
\\ 
\end{array}
& 
\begin{array}{c}
\mbox{locally anisotropic} \\ 
\mbox{cosmology}%
\end{array}%
\end{array}
& 
\begin{array}{c}
\begin{array}{c}
\\ 
\ ^{\shortmid }{g}^{a_{3}}(\tau ,x^{i_{2}},p_{5}) \\ 
\ ^{\shortmid }N_{a_{3}k_{2}}(\tau ,x^{i_{2}},p_{5})%
\end{array}
\\ 
\\ 
\begin{array}{c}
\ ^{\shortmid }{g}^{a_{3}}(\tau ,x^{i_{2}},p_{6}) \\ 
\ ^{\shortmid }{N}_{a_{3}k_{2}}(\tau ,x^{i_{2}},p_{6})%
\end{array}
\\ 
\end{array}
& 
\begin{array}{c}
\ ^{\shortmid }{g}^{a_{4}}(\tau ,\ ^{\shortmid }x^{i_{3}},E) \\ 
\ ^{\shortmid }{N}_{a_{4}k_{3}}(\tau ,\ ^{\shortmid }x^{i_{3}},E)%
\end{array}
& 
\begin{array}{c}
\mbox{rainbow} \\ 
\mbox{s-metrics} \\ 
\mbox{variable } \\ 
p_{8}=E%
\end{array}%
\end{array}%
$}}.
\end{equation*}%
Similar s-parameterizations can be written for $\tau $-families of s-metrics
(\ref{ssolutions}) by changing momenta into velocities for respective labels
and s-indices.

The AFCDM method for ansatz of type (\ref{ssolutions}) and (\ref{ssolutionsd}%
) is summarized in Tabels 1-13 from Appendix \ref{appendixb}.

\subsection{Decoupling of nonmetric FLH geometric flow and modified Einstein
equations}

For a quasi-stationary ansatz (\ref{qstsm}) on $\ _{\mathbf{Q}}^{s}\mathcal{M%
}$ with a respectively computed canonical Ricci s-tensor $\widehat{R}_{\
\beta _{s}}^{\alpha _{s}}$ with coefficients (\ref{hcdric}), (\ref{vhcdric3}%
), (\ref{vhcdric4}) and (\ref{vcdric}), the system (\ref{cfeq4af}) of $\tau $%
-running and $Q$-deformed Einstein equations for $\widehat{\mathbf{D}}%
_{\alpha _{s}}(\tau )$ decouples in such a general s-adapted form: 
\begin{eqnarray}
\psi ^{\bullet \bullet }+\psi ^{\prime \prime } &=&2\ _{Q}^{1}\mathbf{J}%
(\tau ),  \label{eq1} \\
(\varpi )^{\ast }h_{4}^{\ast } &=&2h_{3}h_{4}\ \ _{Q}^{2}\mathbf{J}(\tau ),
\label{e2a} \\
\beta w_{j}-\alpha _{j} &=&0,  \label{e2b} \\
\ n_{k}^{\ast \ast }+\gamma n_{k}^{\ast } &=&0;  \label{e2c}
\end{eqnarray}%
and extending on shells $s=3$,4, we obtain similar systems of nonlinear PDEs
involving additional coordinates $v^{a_{3}},v^{a_{3}}:$%
\begin{eqnarray}
(\ ^{3}\varpi )^{\ast _{3}}h_{6}^{\ast _{3}} &=&2h_{5}h_{6}\ \ _{Q}^{3}%
\mathbf{J}(\tau ),  \label{eq3a} \\
\ ^{3}\beta w_{j_{2}}-\alpha _{j_{2}} &=&0,  \label{eq3b} \\
\ n_{k_{2}}^{\ast _{3}\ast _{3}}+\ ^{3}\gamma n_{k_{2}}^{\ast _{3}} &=&0;
\label{eq3c}
\end{eqnarray}%
\begin{eqnarray}
(\ ^{4}\varpi )^{\ast _{4}}h_{8}^{\ast _{4}} &=&2h_{7}h_{8}\ \ _{Q}^{4}%
\mathbf{J}(\tau ),  \label{eq4a} \\
\ ^{4}\beta w_{j_{3}}-\alpha _{j_{3}} &=&0,  \label{eq4b} \\
\ n_{k_{3}}^{\ast _{4}\ast _{4}}+\ ^{4}\gamma n_{k_{3}}^{\ast _{4}} &=&0.
\label{eq4c}
\end{eqnarray}%
To emphasize how the abstract geometric calculus can be used for extending
such equations from the shell $s=2$ to $s=3$ and $s=4,$ we introduced
certain additional notations for some partial derivatives, for instance, $%
h_{6}^{\ast _{3}}=\partial _{5}h_{6}$ and $h_{8}^{\ast _{4}}=\partial
_{7}h_{8}.$ We can write conventionally $h_{4}^{\ast }=h_{4}^{\ast
_{2}}=\partial _{3}h_{4}$ to show certain compatibility with the notations
in our former works on 4-d nonholonomic manifolds. Such notations allow to
understand in the simplified form how to define certain important systems of
equations with coefficients on $s=1,2$ (using $\ast ,$ omitting some labels
if that does not result in ambiguities) and then to extend the constructions
in abstract symbolic form for $s=3,4$ (using $\ast _{3}$ and $\ast _{4}$).
In the above system of nonlinear PDEs, we introduced, respectively, such
coefficients and generating functions: 
\begin{eqnarray}
g_{i}(\tau ) &=&g_{i_{1}}(\tau )=e^{\psi (\tau ,x^{k})},\mbox{ in }(\ref{eq1}%
)\mbox{ such a generating function is determined by Poisson equations on }%
s=1;  \notag \\
\alpha _{i} &=&h_{4}^{\ast _{2}}\partial _{i}(\varpi ),\beta =h_{4}^{\ast
_{2}}(\varpi )^{\ast _{2}},\mbox{ equivalently }\alpha _{i_{1}}=h_{4}^{\ast
_{2}}\partial _{i_{1}}(\varpi ),\mbox{ in }(\ref{eq3a})-(\ref{e2c}),  \notag
\\
\beta &=&\ ^{2}\beta =h_{4}^{\ast _{2}}(\varpi )^{\ast _{2}}\mbox{  and }%
\gamma =(\ln \frac{|h_{4}|^{3/2}}{|h_{3}|})^{\ast _{2}},\mbox{ for }\varpi
=\ ^{2}\varpi =\ln |\frac{h_{4}^{\ast _{2}}}{\sqrt{|h_{3}h_{4}}}|,
\label{coeffs} \\
&& \mbox{ where }\Psi =\ ^{2}\Psi =\exp (\varpi ) 
\mbox{ is a generating
function on } s=2;  \notag
\end{eqnarray}%
\begin{eqnarray*}
\alpha _{i_{2}} &=&h_{6}^{\ast _{3}}\partial _{i_{2}}(\ ^{3}\varpi ),%
\mbox{
in }(\ref{e2a})-(\ref{eq3c}), \\
\ ^{3}\beta &=&h_{6}^{\ast _{3}}(\ ^{3}\varpi )^{\ast _{3}}\mbox{  and }\
^{3}\gamma =(\ln \frac{|h_{6}|^{3/2}}{|h_{5}|})^{\ast _{3}},\mbox{ for }\
^{3}\varpi =\ln |\frac{h_{6}^{\ast _{3}}}{\sqrt{|h_{5}h_{6}}}|, \\
&& \mbox{ where }\ ^{3}\Psi = \exp (\ ^{3}\varpi ) 
\mbox{ is a generating
function on } s=3;
\end{eqnarray*}%
\begin{eqnarray*}
\alpha _{i_{3}} &=&h_{8}^{\ast _{4}}\partial _{i_{3}}(\ ^{4}\varpi ),%
\mbox{
in }(\ref{eq4a})-(\ref{eq4c}), \\
\ ^{4}\beta &=&h_{8}^{\ast _{4}}(\ ^{4}\varpi )^{\ast _{4}}\mbox{  and }\
^{4}\gamma =(\ln \frac{|h_{8}|^{3/2}}{|h_{7}|})^{\ast _{4}},\mbox{ for }\
^{4}\varpi =\ln |\frac{h_{8}^{\ast _{4}}}{\sqrt{|h_{7}h_{8}}}|, \\
&& \mbox{ where } \ ^{4}\Psi =\exp (\ ^{4}\varpi ) 
\mbox{ is a generating
function on }s=4.
\end{eqnarray*}%
Such coefficients (\ref{coeffs}) are related respectively to $\tau $%
-families of generating sources (\ref{dsourcparam}) with additional shell
spitting. We note that all such coefficients are for nonmetric geometric
flows (for instance, we can write $\alpha _{i_{s}}(\tau ),\ ^{s}\beta (\tau)$%
, etc.). For a fixed $\tau =\tau _{0},$ the system (\ref{eq1}) - (\ref{eq4c}%
) for decoupling the nonmetric Ricci soliton configurations and related FL
modified Einstein equations. 

Let us explain the general decoupling property of the above systems of
equations for quasi-stationary configurations. For simplicity, explain this
for the shells $s=1,2$ because the same properties hold true for higher
shells: The equation (\ref{eq1}) is a standard 2-d Poisson equation with
source $\tau $-parametric source $2\ _{Q}^{1}\mathbf{J}(\tau ).$ It can be a 
$\tau $-family of 2-d wave equation if we consider h-metrics with signature,
for instance, $(+,-).$ Prescribing any data $(h_{3}(\tau ),\ _{Q}^{2}\mathbf{%
J}(\tau )),$ we can search $h_{4}(\tau )$ as a $\tau $-family of solutions
of second order on $\partial _{3}$ nonlinear PDEs (\ref{e2a}). Contrary, we
can consider an inverse problem with prescribed data $(h_{4}(\tau ),\
_{Q}^{2}\mathbf{J}(\tau ))$ when $h_{3}(\tau )$ are solutions of a
corresponding $\tau $-family of first-order nonlinear PDE. Having defined in
some general forms $h_{3}(\tau ,x^{k},y^{3})$ and $h_{4}(\tau ,x^{k},y^{3}),$
we can compute respective coefficients $\alpha _{i_{1}}(\tau )$ and $\
^{2}\beta (\tau )$ for (\ref{e2b}). Such $\tau $-families of linear
equations for $w_{j}(\tau ,x^{k},y^{3})$ can be solved in general form. So,
we can conclude that such equations and respective unknown functions are
decoupled from the rest of the system of equations. Then, we can solve (\ref%
{e2c}) and find $n_{k}(\tau ,x^{k},y^{3}).$ Then, we can perform two general
integrations on $y^{3}$ for any $\gamma (\tau ,x^{k},y^{3})$ determined by $%
h_{3}(\tau ,x^{k},y^{3})$ and $h_{4}(\tau,x^{k},y^{3}).$ So, we can generate
off-diagonal solutions of (modified) Einstein equations written in canonical
d-connection variables by solving step-by-step four equations (\ref{eq1}) - (%
\ref{e2c}). 

Above formulas with decoupling were stated for the quasi-stationary
off-diagonal metric ansatz (\ref{qstsm}) and respective generating sources.
In a similar form, we can prove general decoupling properties for locally
anisotropic cosmological d-metrics (\ref{lcsm}). In generic form, respective
coefficients depend on $y^{4}=t$ \ and respective symbols are underlined,
for instance, in the form $(\underline{h}_{3}(\tau ,x^{k},t),\ \ _{Q}^{2} 
\underline{\mathbf{J}}(\tau )(\tau ,x^{k},t))$ for $^{2}\underline{\Psi }%
(\tau ,x^{k},t);\underline{\alpha }_{i_{s-1}}$ and $\ ^{s}\underline{\beta }%
, $ etc., where we underline certain symbols to emphasize that they are
considered for locally anisotropic cosmological configurations with generic
dependence on $t$-coordinate. Using duality transforms and abstract
geometric computations, the above system of equations can be defined for $%
\tau $-families of quasi-stationary or locally anisotropic configurations on 
$\ _{s}^{\shortmid }\mathcal{M}.$

\subsection{Integration of nonmetric FLH geometric flow equations}

The geometric constructions provided in appendix \ref{appendixab} consist of
examples of extension of the AFCDM for FLH geometric flow and MGTs. They
allow us to generate exact and parametric solutions of the nonlinear systems
of PDEs (\ref{cfeq4af}). In this subsection, we provide and discuss the main
properties of such generic off-diagonal solutions defining quasi-stationary
configurations with s-metrics (\ref{ssolutions}).

\subsubsection{Different forms of quasi-stationary solutions and their
nonlinear symmetries}

Taking the values of s-adapted coefficients (\ref{auxscoef}), we construct
d-metrics (\ref{qstsm}) as general quasi-stationary solutions of the $\tau $%
-parametric FL-modified Einstein equations (\ref{cfeq4af}). The
corresponding quadratic element can be written in the form%
\begin{eqnarray}
d\widehat{s}^{2}(\tau ) &=&e^{\psi (\tau ,x^{k},\ \ \ _{Q}^{1}\mathbf{J}%
)}[(dx^{1})^{2}+(dx^{2})^{2}]+  \label{qeltors} \\
&&\sum\nolimits_{s=2}^{s=4}[\frac{[(\ ^{s}\Psi )^{\ast _{s}}]^{2}}{4(\
_{Q}^{s}\mathbf{J})^{2}\{g_{2s}^{[0]}-\int du^{2s-1}[(\ ^{s}\Psi
)^{2}]^{\ast _{s}}/4(\ _{Q}^{s}\mathbf{J})\}}[du^{2s-1}+\frac{\partial
_{i_{s}}(\ ^{s}\Psi )}{(\ ^{s}\Psi )^{\ast _{s}}}dx^{i_{s}}]^{2}+  \notag
\end{eqnarray}%
\begin{equation*}
\{g_{2s}^{[0]}-\int du^{2s-1}\frac{[(\ ^{s}\Psi )^{2}]^{\ast _{s}}}{4(\
_{Q}^{s}\mathbf{J})}\}\{du^{2s}+[\ _{1}n_{k_{s}}+\ _{2}n_{k_{s}}\int
du^{2s-1}\frac{[(\ ^{s}\Psi )^{2}]^{\ast _{s}}}{4(\ _{Q}^{s}\mathbf{J}%
)^{2}|g_{2s}^{[0]}-\int du^{2s-1}[(\ ^{s}\Psi )^{2}]^{\ast _{s}}/4(\ _{Q}^{s}%
\mathbf{J})|^{5/2}}]dx^{k_{s}}\}].
\end{equation*}%
In these formulas, we use: 
\begin{eqnarray}
\mbox{generating functions: } &&\psi (\tau )\simeq \psi (\tau ,x^{k_{1}});\
_{s}\Psi (\tau )\simeq \ _{s}\Psi (\tau ,x^{k_{s}},y^{s+1});
\label{integrfunctrffh} \\
\mbox{generating sources: } &&\ \ _{Q}^{1}\mathbf{J}\mathcal{(\tau )}\simeq
\ _{Q}^{1}\mathbf{J}(\tau ,x^{k_{1}});\ \ _{Q}^{s}\mathbf{J}\mathcal{(\tau )}%
\simeq \ _{Q}^{s}\mathbf{J}(\tau ,x^{k_{s}},y^{s+1});  \notag \\
\mbox{integrating functions:  } &&\ g_{2s}^{[0]}(\tau ) \simeq
g_{2s}^{[0]}(\tau ,x^{k_{s}}),\ _{1}n_{k_{s}}\mathcal{(\tau )}\simeq \
_{1}n_{k_{s}}(\tau ,x^{j_{s}}),\ _{2}n_{k_{s}}\mathcal{(\tau )}\simeq \
_{2}n_{k_{s}}(\tau ,x^{j_{s}}).  \notag
\end{eqnarray}

\begin{enumerate}
\item Above classes of such solutions are with nontrivial geometric flows of
nonholonomic torsion, which is not zero for hat variables. We can define
certain classes of nonholonomic frame transforms and distortions of the
canonical s-variables when the FL geometric evolution is described by
families of LC-connections $\ _{s}\nabla (\tau ).$

\item We can compute necessary thermodynamic variables (\ref{nagthermodfhq})
associated with canonical quasi-stationary solutions, or their time dual
ones defined as locally anisotropic cosmological solutions with additional
cosmological flow. In the next section, we shall provide such examples for
nonassociative BH and WH configurations.

\item The solutions for FL Ricci soliton equations (\ref{canriccisol})
consist self-similar configurations of (\ref{cfeq4af}) with $\tau =\tau
_{0}. $ We can construct such quasi-stationary solutions directly or after a
class of generic off-diagonal solutions was constructed for FL geometric
evolution flows. Such nonholonomic Ricci soliton configurations can be
generated equivalently by solutions constructed using the AFCDM as it is
outlined in Appendix \ref{appendixb}.

\item Finally, we note that $\tau $-families of nonholonomic FLH
quasi-stationary solutions can be generated using Tables 5, 6, 10 and 11
(see respective ansatz (\ref{qst8d7}), (\ref{qst8d8}), (\ref{qstd8d7}) and (%
\ref{qstrain8d8})) when the s- and N-coefficients are considered with
additional $\tau $-dependence and the generating sources (\ref%
{integrfunctrffh}) are correspondingly redefined for FH distortions and $\
_{Q}^{s\shortmid }\mathbf{J}$.
\end{enumerate}

\paragraph{Nonlinear symmetries of quasi-stationary configurations \newline
}

The off-diagonal solutions (\ref{qeltors}) are described by some nonlinear
symmetries which allow us to transform different classes of generating
functions and effective sources into other types of generating functions
with effective cosmological constants. By tedious computations (see similar
details in \cite{vacaru18,vbubuianu17,bsssvv25,sv11}), we can prove that
such solutions admit a change of the generating data, $(\ ^{s}\Psi ,\ \ \
_{Q}^{s}\mathbf{J})\leftrightarrow (\ ^{s}\Phi ,\ ^{s}\Lambda =const\neq 0)$
on $\ _{s}^{Q}\mathcal{M}.$ The quasi-stationary configurations can be
modelled for $\tau $-families of generating data when $(\ ^{s}\Psi (\tau ),
\ _{Q}^{s}\mathbf{J}(\tau ))\leftrightarrow (\ ^{s}\Phi (\tau ),\
^{s}\Lambda (\tau )).$ For such transforms, we can consider different shell
cosmological constants $\ ^{s}\Lambda $ which may be different from a
h-cosmological constant $\ ^{h}\Lambda .$ For projections or nonholonomic
constraints, or small parametric limits to GR, we can consider $\
^{h}\Lambda =\ ^{v}\Lambda = \Lambda ,$ and $\ ^{s}\Lambda =\Lambda .$ For
such nonlinear transforms, the quasi-stationary solutions $\ ^{s}\mathbf{%
\hat{g}}[\ ^{s}\Psi ]$ (\ref{qeltors}) of $\widehat{\mathbf{R}}_{\ \ \beta
_{s}}^{\alpha _{s}}[\ ^{s}\Psi ]=\ _{Q}\mathbf{J}_{\ \ \beta _{s}}^{\alpha
_{s}}$ (\ref{cfeq4af}) can be expressed in an equivalent class of solutions
of 
\begin{equation}
\widehat{\mathbf{R}}_{\ \ \beta _{s}}^{\alpha _{s}}[\ ^{s}\Phi (\tau )]=\
^{s}\Lambda (\tau )\mathbf{\delta }_{\ \ \beta _{s}}^{\alpha _{s}}.
\label{cfeq4afc}
\end{equation}%
Such equivalent systems of nonlinear PDEs involve effective cosmological
constants $\ ^{s}\Lambda $ and possible (temperature like) $\tau $-running.
The generating data $(\ ^{s}\Phi (\tau ),\ ^{s}\Lambda (\tau )),$ or $(\
^{s}\Psi (\tau ),\ \ \ _{Q}^{s}\mathbf{J}(\tau ))$ can be chosen, for
instance, to describe DE and DM configurations in accelerating cosmology and
study of possible astrophysical effects (we provide examples in the next
section). 

The quasi-stationary configurations (\ref{qeltors}) transform into certain
quasi-stationary solutions of (\ref{cfeq4afc}) if there are satisfied such
differential or integral equations (for simplicity, we do not write in the
formulas below, in this subsection, the dependence on $\tau $-parameter): 
\begin{eqnarray}
\frac{\lbrack \ ^{s}\Psi ^{2}]^{\ast _{s}}}{\ _{Q}^{s}\mathbf{J}} &=&\frac{%
[\ ^{s}\Phi ^{2}]^{\ast _{s}}}{\ ^{s}\Lambda },%
\mbox{ which can be
integrated as  }  \label{ntransf1} \\
\ ^{s}\Phi ^{2} &=&\ \ ^{s}\Lambda \int du^{s+1}(\ _{Q}^{s}\mathbf{J}%
)^{-1}[\ ^{s}\Psi ^{2}]^{\ast _{s}}\mbox{ and/or }\ ^{s}\Psi ^{2}=\
^{s}\Lambda ^{-1}\int du^{s+1}(\ _{Q}^{s}\mathbf{J})[\ ^{s}\Phi ^{2}]^{\ast
_{s}}.  \label{ntransf2}
\end{eqnarray}%
Using (\ref{ntransf1}), we can simplify the formula (\ref{g4}) and extend it
for $s=3,4:$ 
\begin{equation*}
h_{4}=h_{4}^{[0]}-\frac{\ ^{2}\Phi ^{2}}{4\ ^{2}\Lambda },h_{6}=h_{6}^{[0]}-%
\frac{\ ^{3}\Phi ^{2}}{4\ ^{3}\Lambda },h_{8}=h_{8}^{[0]}-\frac{\ ^{4}\Phi
^{2}}{4\ ^{4}\Lambda }\mbox{ i. e. }h_{2s}=h_{2s}^{[0]}-\frac{\ ^{s}\Phi ^{2}%
}{4\ ^{s}\Lambda }.
\end{equation*}%
This allows us to express the formulas (\ref{g3}) and (\ref{gn}) in terms of
new generating data and extend on (co) fiber shells. For such transforms, we
can write $(\ ^{s}\Psi )^{\ast _{s}}/\ \ _{Q}^{s}\mathbf{J}$ in terms of
such $(\ ^{s}\Phi ,\ ^{s}\Lambda )$ and write (\ref{ntransf1}) and (\ref%
{ntransf2}) in the form:%
\begin{equation*}
\frac{\ ^{s}\Psi (\ \ ^{s}\Psi )^{\ast _{s}}}{\ _{Q}^{s}\mathbf{J}}=\frac{(\
^{s}\Phi ^{2})^{\ast _{s}}}{2\ ^{s}\Lambda }\mbox{ and }\ \ ^{s}\Psi =|\
^{s}\Lambda |^{-1/2}\sqrt{|\int du^{s+1}\ \ _{Q}^{s}\mathbf{J}\ (\ ^{s}\Phi
^{2})^{\ast _{s}}|}.
\end{equation*}%
Introducing $\ ^{s}\Psi $ from the above second equation in the first
equation, we redefine $\ ^{s}\Psi ^{\ast _{s}}$ in terms of generating data $%
(\ _{Q}^{s}\mathbf{J},\ ^{s}\Phi ,\ ^{s}\Lambda )$ on respective shells,
when 
\begin{equation*}
\frac{\ ^{s}\Psi ^{\ast _{s}}}{\ _{Q}^{s}\mathbf{J}}=\frac{[\ ^{s}\Phi
^{2}]^{\ast _{s}}}{2\sqrt{|\ \ ^{s}\Lambda \int du^{s+1}(\ _{Q}^{s}\mathbf{J}%
)[\ ^{s}\Phi ^{2}]^{\ast _{s}}|}}.
\end{equation*}

We conclude that any quasi-stationary solution (\ref{qeltors}) possess
important nonlinear symmetries of type (\ref{ntransf1}) and (\ref{ntransf2})
which are trivial or do not exist for diagonal ansatz.

Similar nonlinear symmetries can be defined for quasi-stationary solutions
on $\ _{s}^{\shortmid Q}\mathcal{M}.$ Using an abstract geometric calculus,
we write the formulas for such nonlinear transforms $(\ _{\shortmid}^{s}\Psi
,\ _{Q}^{s\shortmid }\mathbf{J}) \leftrightarrow (\ _{\shortmid}^{s}\Phi ,\
_{\shortmid }^{s}\Lambda =const\neq 0)$ and respective nonlinear symmetries,
for instance,%
\begin{equation*}
\ _{\shortmid }^{s}\Phi ^{2}=\ \ _{\shortmid }^{s}\Lambda \int d\
^{\shortmid }u^{s+1}(\ _{Q}^{s\shortmid }\mathbf{J})^{-1}[\ _{\shortmid
}^{s}\Psi ^{2}]^{\ast _{s}}.
\end{equation*}%
Such formulas involve momentum like variables $p_{5},p_{6},p_{7}$ for fixed $%
p_{8}=E_{(0)}$. In abstract geometric form, similar quasi-stationary
solutions (\ref{qeltors}) and nonlinear symmetries can be generated on $\
_{s}^{\shortmid Q}\mathcal{M}$ for the conventional momentum variables ($%
p_{5},p_{6},p_{7}=p_{7(0)},E),$ when the h-part coefficients do not depend
on $y^{4}=t.$

\paragraph{Quasi-stationary solutions with effective cosmological constants 
\newline
}

Using above-stated nonlinear symmetries, the quadratic element for
quasi-stationary solutions (\ref{qeltors}) can be written in an equivalent
form for generating data $(\ _{Q}^{s}\mathbf{J}(\tau ),\ ^{s}\Phi (\tau ),\
^{s}\Lambda (\tau )),$ 
\begin{equation}
d\widehat{s}^{2}(\tau )=\widehat{g}_{\alpha _{s}\beta _{s}}(\tau ,\ _{Q}^{s}%
\mathbf{J},\ ^{s}\Phi ,\ ^{s}\Lambda )du^{\alpha _{s}}du^{\beta
_{s}}=e^{\psi (\tau ,x^{k},\ \ \ _{Q}^{1}\mathbf{J}%
)}[(dx^{1})^{2}+(dx^{2})^{2}]+  \label{offdiagcosmcsh}
\end{equation}%
\begin{eqnarray*}
&&\sum\nolimits_{s=2}^{s=4}[-\frac{\ ^{s}\Phi ^{2}[\ ^{s}\Phi ^{\ast
_{s}}]^{2}}{|\ ^{s}\Lambda (\tau )\int du^{2s-1}\ \ _{Q}^{s}\mathbf{J}[\
^{s}\Phi ^{2}]^{\ast _{s}}|[h_{2s}^{[0]}-\ ^{s}\Phi ^{2}/4\ ^{s}\Lambda
(\tau )]}\{du^{2s-1}+\frac{\partial _{i_{s}}\ \int du^{2s-1}\ \ _{Q}^{s}%
\mathbf{J}\ [\ ^{s}\Phi ^{2}]^{\ast _{s}}}{\ _{Q}^{s}\mathbf{J}\ [(\
^{s}\Phi )^{2}]^{\ast _{s}}}dx^{i_{s}}\}^{2}+ \\
&&\{h_{2s}^{[0]}-\frac{\ ^{s}\Phi ^{2}}{4\ ^{s}\Lambda (\tau )}%
\}\{du^{2s-1}+[\ _{1}n_{k_{s}}+\ _{2}n_{k_{s}}\int \frac{du^{2s-1}\ ^{s}\Phi
^{2}[\ ^{s}\Phi ^{\ast _{s}}]^{2}}{|\ ^{s}\Lambda (\tau )\int du^{2s-1}\ \
_{Q}^{s}\mathbf{J}[\ ^{s}\Phi ^{2}]^{\ast _{s}}|[h_{2s}^{[0]}-\ ^{s}\Phi
^{2}/4\ ^{s}\Lambda (\tau )]^{5/2}}]dx^{k_{s}}\}].
\end{eqnarray*}%
We emphasize that the quasi-stationary solutions represented in the form (%
\ref{offdiagcosmcsh}) "disperse" into respective off-diagonal forms the
prescribed generating data $(\ ^{s}\Psi (\tau ),\ _{Q}^{s}\mathbf{J}(\tau ))$
transforming them into another type ones $(\ ^{s}\Phi (\tau ),\ ^{s}\Lambda
(\tau )),$ with effective cosmological constants. The contributions of
generating sources (for effective and matter fields on phase space) $\
_{Q}^{s}\mathbf{J}$ are not completely transformed into $\tau $-running
cosmological constants $\ ^{s}\Lambda (\tau ).$ The coefficients of
d-metrics (\ref{offdiagcosmcsh}) keep certain memory about the sources $\
_{Q}^{s}\mathbf{J}(\tau )\,\ $ stated in (\ref{qeltors}). Considering
effective $\ ^{s}\Lambda (\tau ),$ we can simplify the method of computing
G. Perelman thermodynamic variables as we show in the next section. 

\paragraph{Solutions when some d-metric coefficients are used as generating
functions \newline
}

Taking the partial derivative on $y^{3}$ of formula (\ref{g4}) and acting
similarly on fiber shells allows us to write 
\begin{equation*}
h_{2s}^{\ast _{s}}(\tau )=-[\ ^{s}\Psi ^{2}(\tau )]^{\ast _{s}}/4\ _{Q}^{s}%
\mathbf{J}(\tau ).
\end{equation*}%
Prescribing data for $h_{2s}(\tau ,x^{i_{s}},u^{i_{s}+1})$ and $\ _{Q}^{s}%
\mathbf{J}(\tau ,x^{i_{s}},u^{i_{s}+1})$ , we can compute (up to an
integration function) a generating function $\ ^{s}\Psi (\tau )$ which
satisfies $[\ ^{s}\Psi ^{2}]^{\ast _{s}}=\int du^{s+1}\ _{Q}^{s}\mathbf{J\ }
h_{2s}^{\ast _{s}}$ and define off-diagonal solutions of type (\ref{qeltors}%
). So, considering the generating data $(h_{2s}(\tau ),\ _{Q}^{s}\mathbf{J}%
(\tau )),$ we can re-write the quadratic elements for a quasi-stationary
d-metric (\ref{qeltors}) as 
\begin{equation}
d\widehat{s}^{2}(\tau )=\widehat{g}_{\alpha _{s}\beta _{s}}(\tau
,x^{i_{s}},u^{i_{s}+1},\ _{Q}^{s}\mathbf{J},\ h_{2s})du^{\alpha
_{s}}du^{\beta _{s}}=e^{\psi (\tau ,x^{k},\ _{Q}^{1}\mathbf{J}%
)}[(dx^{1})^{2}+(dx^{2})^{2}]+  \label{offdsolgenfgcosmc}
\end{equation}%
\begin{eqnarray*}
&&\sum\nolimits_{s=2}^{s=4}[-\frac{(h_{2s}^{\ast _{s}})^{2}}{|\int
du^{2s-1}[\ \ _{Q}^{s}\mathbf{J}h_{2s}]^{\ast _{s}}|\ h_{2s}}\{du^{2s-1}+%
\frac{\partial _{i_{s}}[\int du^{2s-1}(\ _{Q}^{s}\mathbf{J})\ h_{2s}^{\ast
_{s}}]}{\ _{Q}^{s}\mathbf{J}\ h_{2s}^{\ast _{s}}}dx^{i_{s}}\}^{2}+ \\
&&h_{2s}\{du^{2s}+[\ _{1}n_{k_{s}}+\ _{2}n_{k_{s}}\int du^{2s-1}\frac{%
(h_{2s}^{\ast _{s}})^{2}}{|\int du^{2s-1}[\ \ _{Q}^{s}\mathbf{J}%
h_{2s}]^{\ast _{s}}|\ (h_{2s})^{5/2}}]dx^{k_{s}}\}].
\end{eqnarray*}

The nonlinear symmetries (\ref{ntransf1}) and (\ref{ntransf2}) allow us to
perform similar computations related to (\ref{offdiagcosmcsh}). Expressing $%
\ ^{s}\Phi ^{2}=-4\ \ ^{s}\Lambda h_{2s},$ we can eliminate $\ ^{s}\Phi $
from the nonlinear element and generate solutions of type (\ref%
{offdsolgenfgcosmc}) which are determined by the generating data $(h_{2s};\
^{s}\Lambda ,\ _{Q}^{s}\mathbf{J}).$ 

\paragraph{Quasi-stationary gravitational polarizations of prime s-metrics 
\newline
}

The above-generated off-diagonal solutions and their nonlinear symmetries
can be parameterized in certain forms that describe nonholonomic
deformations of certain FL prime metrics (which may be, or not, solutions of
other or the same nonholonomic geometric flow or modified gravitational
field equations). The main condition is that the target s-metrics define
quasi-stationary configurations as solutions of (\ref{cfeq4af}).

We denote a $\tau $-family of \textbf{prime} s-metric as 
\begin{equation}
\ ^{s}\mathbf{\mathring{g}}(\tau )\mathbf{=}[\mathring{g}_{\alpha _{s}}(\tau
),\mathring{N}_{i_{s-1}}^{a_{s}}(\tau )]  \label{offdiagpm}
\end{equation}%
and transform it into a $\tau $-family of \textbf{target} s-metrics $\ ^{s}%
\mathbf{g,}$ 
\begin{equation}
\ ^{s}\mathbf{\mathring{g}}(\tau )\rightarrow \ ^{s}\mathbf{g}(\tau
)=[g_{\alpha _{s}}(\tau )=\eta _{\alpha _{s}}(\tau )\mathring{g}_{\alpha
_{s}}(\tau ),N_{i_{s-1}}^{a_{s}}(\tau )=\eta _{i_{s-1}}^{a_{s}}(\tau )\ 
\mathring{N}_{i_{s-1}}^{a_{s}}(\tau )].  \label{offdiagdefr}
\end{equation}%
$\ ^{s}\mathbf{g}(\tau )$ are quasi-stationary s-metrics of type (\ref%
{qeltors}) (which can be also formulated in equivalent forms as (\ref%
{offdiagcosmcsh}) or (\ref{offdsolgenfgcosmc})). The functions $\eta
_{\alpha _{s}}(\tau ,x^{k_{s-1}},v^{k_{s-1}+1})$ and $\eta
_{i_{s-1}}^{a_{s}}(\tau ,x^{k_{s-1}},v^{k_{s-1}+1})$ from (\ref{offdiagdefr}%
) are called phase space gravitational polarization (in brief, $\eta $%
-polarization) functions. To generate exact or parametric solutions we can
consider that the nonlinear symmetries (\ref{ntransf1}) are parameterized in
the form (in general, we can consider $\tau $-running $\ ^{s}\Lambda (\tau),$
$\ ^{s}\Psi (\tau ),$ etc.) 
\begin{eqnarray}
(\ ^{s}\Psi ,\ _{Q}^{s}\mathbf{J}) &\leftrightarrow &(\ ^{s}\mathbf{g},\
_{Q}^{s}\mathbf{J})\leftrightarrow (\eta _{\alpha _{s}}\ \mathring{g}%
_{\alpha _{s}}\sim (\zeta _{\alpha _{s}}(1+\epsilon \chi _{\alpha _{s}})%
\mathring{g}_{\alpha _{s}},\ _{Q}^{s}\mathbf{J})\leftrightarrow
\label{nonlintrsmalp} \\
(\ ^{s}\Phi ,\ \ ^{s}\Lambda ) &\leftrightarrow &(\ ^{s}\mathbf{g},\ \
^{s}\Lambda )\leftrightarrow (\eta _{\alpha _{s}}\ \mathring{g}_{\alpha
_{s}}\sim (\zeta _{\alpha _{s}}(1+\epsilon \chi _{\alpha _{s}})\mathring{g}%
_{\alpha _{s}},\ \ ^{s}\Lambda ),  \notag
\end{eqnarray}%
where $\ ^{s}\Lambda $ are effective shell cosmological constants and $%
\epsilon $ is a small parameter satisfying the condition: $0\leq \epsilon
<1; $ $\zeta _{\alpha _{s}}(x^{k_{s-1}},v^{k_{s-1}+1})$ and $\chi _{\alpha
_{s}}(x^{k_{s-1}},v^{k_{s-1}+1})$ are respective polarization functions.
Such nonholonomic transforms have to result in a target metric $\ ^{s}%
\mathbf{g}$ defined as a solution of type (\ref{qeltors}) or, equivalently, (%
\ref{offdiagcosmcsh}), if the $\eta $- and/or $\chi $-polarizations are
subjected to the conditions (\ref{ntransf2}) written in the form: 
\begin{eqnarray}
\partial _{2s-1}[\ ^{s}\Psi ^{2}] &=&-\int du^{2s-1}\ \ _{Q}^{s}\mathbf{J\ }%
\partial _{2s-1}h_{2s}\simeq -\int du^{2s-1}\ \ _{Q}^{s}\mathbf{J\ }\partial
_{2s-1}(\eta _{2s}\ \mathring{g}_{2s})\simeq  \notag \\
&&-\int \int du^{2s-1}\ \ _{Q}^{s}\mathbf{J\ }\partial _{2s-1}[\zeta
_{2s}(1+\epsilon \chi _{2s})\ \mathring{g}_{2s}],  \notag \\
\ ^{s}\Phi ^{2} &=&-4\ \ ^{s}\Lambda h_{2s}\simeq -4\ ^{s}\Lambda \eta
_{2s}\ \mathring{g}_{2s}\simeq -4\ ^{s}\Lambda \ \zeta _{2s}(1+\epsilon \chi
_{2s})\ \mathring{g}_{2s}.  \label{nonlinsymrex}
\end{eqnarray}%
%
%
%
%

Off-diagonal $\eta $-transforms resulting in d-metrics (\ref{offdiagdefr})
can be parameterized to be generated for $\psi $- and $\eta $-polarizations, 
\begin{equation}
\psi (\tau )\simeq \psi (\tau ,x^{k_{1}}),\eta _{4}(\tau )\ \simeq \eta
_{4}(\tau ,x^{k_{1}},y^{3}),\eta _{6}(\tau )\simeq \eta _{6}(\tau
,x^{k_{2}},v^{5}),\eta _{8}(\tau )\simeq \eta _{8}(\tau ,x^{k_{3}},v^{7}),
\label{etapolgen}
\end{equation}%
in a form equivalent to (\ref{offdsolgenfgcosmc}) if the quasi-stationary
quadratic element can be written in the form 
\begin{equation}
d\widehat{s}^{2}(\tau )=\widehat{g}_{\alpha _{s}\beta _{s}}(\tau ,\mathring{g%
}_{\alpha _{s}};\psi ,\eta _{2s};\ \ ^{s}\Lambda ,\ _{Q}^{s}\mathbf{J}%
)du^{\alpha _{s}}du^{\beta _{s}}=e^{\psi }[(dx^{1})^{2}+(dx^{2})^{2}] +
\label{offdiagpolfr}
\end{equation}%
\begin{eqnarray*}
&&\sum\nolimits_{s=2}^{s=4}[-\frac{[\partial _{2s-1}(\eta _{2s}\ \mathring{g}%
_{2s})]^{2}}{|\int du^{2s-1}\ \ _{Q}^{s}\mathbf{J\ }\partial _{2s-1}(\eta
_{2s}\ \mathring{g}_{2s})|\ (\eta _{2s}\ \mathring{g}_{2s})}\{du^{2s-1}+%
\frac{\partial _{i_{s}}[\int du^{2s-1}\ \ _{Q}^{s}\mathbf{J\ }\partial
_{2s-1}(\eta _{2s}\ \mathring{g}_{2s})]}{\ _{Q}^{s}\mathbf{J\ }\partial
_{2s-1}(\eta _{2s}\ \mathring{g}_{2s})}dx^{i_{s}}\}^{2} \\
&&+\eta _{2s}\ \mathring{g}_{2s}\{du^{2s}+[\ _{1}n_{k_{s}}+\
_{2}n_{k_{s}}\int du^{2s-1}\frac{[\partial _{2s-1}(\eta _{2s}\ \mathring{g}%
_{2s})]^{2}}{|\int du^{2s-1}\ \ _{Q}^{s}\mathbf{J\ }\partial _{2s-1}(\eta
_{2s}\ \mathring{g}_{2s})|\ (\eta _{2s}\ \mathring{g}_{2s})^{5/2}}%
]dx^{k_{s}}\}^{2}].
\end{eqnarray*}%
We can relate a solution of type (\ref{offdiagcosmcsh}) to an another one in
the form (\ref{offdiagpolfr}) if $\ ^{s}\Phi ^{2}=-4\ \ ^{s}\Lambda h_{2s}$
and the $\eta $-polarizations are determined by the generating data $%
(h_{2s}=\eta _{2s}\ \mathring{g}_{2s};\ ^{s}\Lambda , \ _{Q}^{s}\mathbf{J}).$

Many important applications can be considered for solutions of type (\ref%
{offdiagpolfr}) with small $\chi \,$-polarizations (\ref{nonlinsymrex}) used
instead of generating functions (\ref{etapolgen}). They allow to study, for
instance, small deformations of BHs in GR into BE configurations in FLH
theories and other types of physically important solutions. The $\epsilon $%
-deformations consist of a more special case when physically important
solutions in GR and MGTs can be transformed into FLH configurations with
almost similar, but locally anisotropic, properties. In Appendix \ref%
{appendixac}, the off-diagonal quasi-stationary solutions are provided in
terms of $\chi$-polarization functions. 

\subsubsection{Dualities of space and time, and space momenta and energy,
for off-diagonal solutions}

We mentioned above the existence of a specific space and time duality
between ansatz ansatz of type (\ref{ssolutions}) and (\ref{ssolutionsd}). A
corresponding duality principle can be formulated for generic off-diagonal
solutions. It allows us to not repeat all computations presented for
quasi-stationary metrics with nontrivial partial derivatives $\partial _{3}$
for locally anisotropic cosmological solutions with nontrivial partial
derivatives $\partial _{4}=\partial _{t}$. For (co) tangent Lorentz bundles,
similar properties exist for the duality of partial derivatives $\partial
_{7}$ and $\partial _{8},$ or $\ ^{\shortmid }\partial ^{7}$ and $\
^{\shortmid }\partial ^{8}=\partial _{E}.$ All formulas for quasi-stationary
solutions from the previous subsection and Appendices \ref{appendixab} and %
\ref{appendixac} can be re-defined by constructing locally anisotropic
cosmological solutions.

In abstract symbolic form, the \textbf{principle of space and time duality }
of generic off-diagonal configurations with one Killing symmetry on a
space-like $\partial _{3}$ or time-like $\partial _{t}$ on a Lorentz base
spacetime manifold is formulated: 
\begin{eqnarray*}
y^{3} &\longleftrightarrow &y^{4}=t,h_{3}(\tau
,x^{k_{1}},y^{3})\longleftrightarrow \underline{h}_{4}(\tau
,x^{k_{1}},t),h_{4}(\tau ,x^{k_{1}},y^{3})\longleftrightarrow \underline{h}%
_{3}(\tau ,x^{k_{1}},t), \\
N_{i_{1}}^{3} &=&w_{i_{1}}(\tau ,x^{k_{1}},y^{3})\longleftrightarrow
N_{i_{1}}^{4}=\underline{n}_{i_{1}}(\tau
,x^{k_{1}},t),N_{i_{1}}^{4}=n_{i_{1}}(\tau
,x^{k_{1}},y^{3})\longleftrightarrow N_{i_{1}}^{3}=\underline{w}%
_{i_{1}}(\tau ,x^{k_{1}},t).
\end{eqnarray*}%
Such duality principles can be extended for phase space extensions of
Lorentz manifolds. For constructing explicit classes of solutions, the above
duality conditions have to be stated also for prime d-metrics, and
extensions to s-metrics, and respective generating functions, generating
sources and gravitational polarization functions (and in certain cases, for
the integration functions). Such details on the duality of the generic
off-diagonal solutions are given by 4-d configurations: $\ _{Q}\mathbf{J}_{\
\nu }^{\mu}(\tau )$ (\ref{dsourcparam})$\rightarrow \ _{Q}\mathbf{J}_{\ \nu
_{s}}^{\mu _{s}}(\tau )$ 
\begin{equation*}
\ _{Q}\mathbf{J}_{~3}^{3}=\ _{Q}\mathbf{J}_{~4}^{4}=\ _{Q}^{2}\mathbf{J}%
(x^{k_{1}},y^{3})\longleftrightarrow \ _{Q}\underline{\mathbf{J}}_{~4}^{4}=\
_{Q}\underline{\mathbf{J}}_{~3}^{3}=\ _{Q}^{2}\underline{\mathbf{J}}%
(x^{k},t),\mbox{ see }\ _{Q}\mathbf{J}_{\ \nu }^{\mu }(\tau )\ (\ref%
{dsourcparam})\rightarrow \ _{Q}\mathbf{J}_{\ \nu _{s}}^{\mu _{s}}(\tau );
\end{equation*}%
\begin{equation}
\begin{array}{ccc}
\begin{array}{c}
(\ \ ^{s}\Psi ,\ _{Q}^{s}\mathbf{J})\leftrightarrow (\ ^{s}\mathbf{g},\
_{Q}^{s}\mathbf{J})\leftrightarrow \\ 
(\eta _{\alpha _{s}}\ \mathring{g}_{\alpha _{s}}\sim (\zeta _{\alpha
_{s}}(1+\epsilon \chi _{\alpha _{s}})\mathring{g}_{\alpha _{s}},\ _{Q}^{s}%
\mathbf{J})\leftrightarrow%
\end{array}
& \Longleftrightarrow & 
\begin{array}{c}
(\ ^{s}\underline{\Psi },\ \ _{Q}^{s}\underline{\mathbf{J}})\leftrightarrow
(\ ^{s}\underline{\mathbf{g}},\ _{Q}^{s}\underline{\mathbf{J}}%
)\leftrightarrow \\ 
(\underline{\eta }_{\alpha _{s}}\ \underline{\mathring{g}}_{\alpha _{s}}\sim
(\underline{\zeta }_{\alpha _{s}}(1+\epsilon \underline{\chi }_{\alpha _{s}})%
\underline{\mathring{g}}_{\alpha _{s}},\ _{Q}^{s}\underline{\mathbf{J}}%
)\leftrightarrow%
\end{array}
\\ 
\begin{array}{c}
(\ ^{s}\Phi ,\ ^{s}\Lambda )\leftrightarrow (\ ^{s}\mathbf{g},\ ^{s}\Lambda
)\leftrightarrow \\ 
(\eta _{\alpha _{s}}\ \mathring{g}_{\alpha _{s}}\sim (\zeta _{\alpha
_{s}}(1+\epsilon \chi _{\alpha _{s}})\mathring{g}_{\alpha _{s}},\
^{s}\Lambda ),%
\end{array}
& \Longleftrightarrow & 
\begin{array}{c}
(\ ^{s}\underline{\Phi },\ ^{s}\underline{\Lambda })\leftrightarrow (\ ^{s}%
\underline{\mathbf{g}},\ ^{s}\underline{\Lambda })\leftrightarrow \\ 
(\underline{\eta }_{\alpha _{s}}\ \underline{\mathring{g}}_{\alpha _{s}}\sim
(\underline{\zeta }_{\alpha _{s}}(1+\epsilon \underline{\chi }_{\alpha _{s}})%
\underline{\mathring{g}}_{\alpha _{s}},\ ^{s}\underline{\Lambda }).%
\end{array}%
\end{array}
\label{dualnonltr}
\end{equation}%
The duality conditions are extended also to the corresponding systems of
nonlinear PDE with decoupling (see (\ref{eq1}) - (\ref{e2c}) and respective
coefficients): 
\begin{equation}
\begin{array}{ccc}
\ ^{s}\Psi ^{\ast _{s}}h_{2s}^{\ast _{s}}=2h_{2s-1}h_{2s}\ \ _{Q}^{s}\mathbf{%
J}\ ^{s}\Psi , & \longleftrightarrow & \sqrt{|\underline{h}_{2s-1}\underline{%
h}_{2s}|}\ ^{s}\underline{\Psi }=\underline{h}_{2s-1}^{\diamond _{s}}, \\ 
\sqrt{|h_{2s-1}h_{2s}|}\ ^{s}\Psi =h_{2s}^{\ast _{s}}, & \longleftrightarrow
& \ ^{s}\underline{\Psi }^{\diamond _{s}}\underline{h}_{2s-1}^{\diamond
_{s}}=2\underline{h}_{2s-1}\underline{h}_{2s}\ \ \ _{Q}^{s}\underline{%
\mathbf{J}}\ ^{s}\underline{\Psi }, \\ 
\ ^{s}\Psi ^{\ast _{s}}w_{i_{s}}-\partial _{i_{s}}\ ^{s}\Psi =\ 0, & 
\longleftrightarrow & \underline{n}_{i_{s}}^{\diamond _{s}\diamond
_{s}}+\left( \ln \frac{|\underline{h}_{2s-1}|^{3/2}}{|\underline{h}_{2s}|}%
\right) ^{\diamond _{s}}\underline{n}_{i_{s}}^{\diamond _{2}}=0, \\ 
\ n_{i_{s}}^{\ast _{s}\ast _{s}}+\left( \ln \frac{|h_{2s}|^{3/2}}{|h_{2s-1}|}%
\right) ^{\ast _{s}}n_{i_{s}}^{\ast _{s}}=0 & \longleftrightarrow & \ ^{s}%
\underline{\Psi }^{\diamond _{s}}\underline{w}_{i_{s}}-\partial _{i_{s}}\
^{s}\underline{\Psi }=\ 0,%
\end{array}%
\mbox{ see }(\ref{auxa1})-(\ref{aux1ac}).  \label{dualcosm}
\end{equation}%
In these formulas on $\ _{Q}^{s}\mathcal{M}$, $\ ^{s}\underline{\Psi }
^{\diamond _{s}}=\partial _{2s}\ ^{s}\underline{\Psi },$ $\ ^{s}\Psi ^{\ast
_{s}}=\partial _{2s-1}\ ^{s}\Psi ,$ etc., for $s=2,3,4.$ These formulas may
involve a $\tau $-parameter. Using abstract geometric notations, we can
write the formulas (\ref{dualcosm}) on $\ _{Q}^{\shortmid s}\mathcal{M}$ in
the form:%
\begin{equation}
\begin{array}{ccc}
\ _{\shortmid }^{s}\Psi ^{\ast _{s}}\ ^{\shortmid }h_{2s}^{\ast _{s}}=2\
^{\shortmid }h_{2s-1}\ ^{\shortmid }h_{2s}\ \ _{Q}^{\ \shortmid s}\mathbf{J}%
\ _{\shortmid }^{s}\Psi , & \longleftrightarrow & \sqrt{|\ ^{\shortmid }%
\underline{h}_{2s-1}\ ^{\shortmid }\underline{h}_{2s}|}\ _{\shortmid }^{s}%
\underline{\Psi }=\ ^{\shortmid }\underline{h}_{2s-1}^{\diamond _{s}}, \\ 
\sqrt{|\ ^{\shortmid }h_{2s-1}\ ^{\shortmid }h_{2s}|}\ _{\shortmid }^{s}\Psi
=\ ^{\shortmid }h_{2s}^{\ast _{s}}, & \longleftrightarrow & \ _{\shortmid
}^{s}\underline{\Psi }^{\diamond _{s}}\ ^{\shortmid }\underline{h}%
_{2s-1}^{\diamond _{s}}=2\ ^{\shortmid }\underline{h}_{2s-1}\ ^{\shortmid }%
\underline{h}_{2s}\ \ \ _{Q}^{s}\underline{\mathbf{J}}\ _{\shortmid }^{s}%
\underline{\Psi }, \\ 
\ _{\shortmid }^{s}\Psi ^{\ast _{s}}\ ^{\shortmid }w_{i_{s}}-\ ^{\shortmid
}\partial _{i_{s}}\ _{\shortmid }^{s}\Psi =\ 0, & \longleftrightarrow & \
^{\shortmid }\underline{n}_{i_{s}}^{\diamond _{s}\diamond _{s}}+\left( \ln 
\frac{|\ ^{\shortmid }\underline{h}_{2s-1}|^{3/2}}{|\ ^{\shortmid }%
\underline{h}_{2s}|}\right) ^{\diamond _{s}}\ ^{\shortmid }\underline{n}%
_{i_{s}}^{\diamond }=0, \\ 
\ \ ^{\shortmid }n_{i_{s}}^{\ast _{s}\ast _{s}}+\left( \ln \frac{|\
^{\shortmid }h_{2s}|^{3/2}}{|\ ^{\shortmid }h_{2s-1}|}\right) ^{\ast _{s}}\
^{\shortmid }n_{i_{s}}^{\ast _{s}}=0 & \longleftrightarrow & \ _{\shortmid
}^{s}\underline{\Psi }^{\diamond _{s}}\ ^{\shortmid }\underline{w}_{i_{s}}-\
^{\shortmid }\partial _{i_{s}}\ _{\ \shortmid }^{s}\underline{\Psi }=\ 0.%
\end{array}%
.  \label{dualcosmd}
\end{equation}%
We note that (\ref{dualcosm}) and (\ref{dualcosmd}) are not completely dual
if we introduces symplectomorphisms on $\ _{Q}^{\shortmid s}\mathcal{M}$.
This is because the FL and FH geometries are not completely equivalent/ dual
in such cases (similarly to the fact that the Lagrange mechanics is not
completely equivalent to the Hamilton mechanics and respective almost
symplectic generalizations). 

The nonlinear symmetries (\ref{ntransf1}) and (\ref{ntransf2}) are written
in respective dual forms for locally anisotropic cosmological solutions $\
_{Q}^{\shortmid s}\mathcal{M}$. For instance: 
\begin{eqnarray*}
\frac{\lbrack \ ^{s}\underline{\Psi }^{2}]^{\diamond _{s}}}{\ \ _{Q}^{s}%
\underline{\mathbf{J}}} &=&\frac{[\ ^{s}\underline{\Phi }^{2}]^{\diamond
_{s}}}{\ ^{s}\underline{\Lambda }},\mbox{ which can be
integrated as  } \\
\ ^{s}\underline{\Phi }^{2} &=&\ \ ^{s}\underline{\Lambda }\int du^{2s}(\
_{Q}^{s}\underline{\mathbf{J}})^{-1}[\ ^{s}\underline{\Psi }^{2}]^{\diamond
_{s}}\mbox{ and/or
}\ ^{s}\underline{\Psi }^{2}=(\ ^{s}\underline{\Lambda })^{-1}\int du^{2s}(\
\ _{Q}^{s}\underline{\mathbf{J}})[\ ^{s}\underline{\Phi }^{2}]^{\diamond
_{s}}.
\end{eqnarray*}%
These nonlinear symmetries allow us to redefine for different types of
cosmological models the quasi-stationary d-metrics (\ref{qeltors}), (\ref%
{offdiagcosmcsh}), (\ref{offdsolgenfgcosmc}), (\ref{offdiagpolfr}) and (\ref%
{offdncelepsilon}). The corresponding locally anisotropic cosmological
analogues also define exact or parametric solutions of the FLH geometric
flow deformed Einstein equations (\ref{cfeq4af}) or, respectively, (\ref%
{feq4afd}). As an example of applications of such an abstract symbolic
calculus for deriving off-diagonal solutions, we provide the formula for the
dualized s-metric (\ref{qeltors}): 
\begin{equation}
d\underline{s}^{2}(\tau )=e^{\psi (\tau
,x^{k_{1}})}[(dx^{1})^{2}+(dx^{2})^{2}]+  \label{qeltorsc}
\end{equation}%
\begin{eqnarray*}
&&\sum\nolimits_{s=2}^{s=4}[\{g_{2s+1}^{[0]}-\int du^{2s}\frac{[\ ^{s}%
\underline{\Psi }^{2}]^{\diamond _{s}}}{4~(\ _{Q}^{s}\underline{\mathbf{J}})}%
\}\{du^{2s+1}+[\ _{1}n_{k_{s}}+\ _{2}n_{k_{s}}\int \frac{du^{2s}[(\ ^{s}%
\underline{\Psi })^{2}]^{\diamond _{s}}}{4(\ _{Q}^{s}\underline{\mathbf{J}}%
)^{2}|g_{2s+1}^{[0]}-\int dt[\underline{\Psi }^{2}]^{\diamond _{s}}/4\ (\
_{Q}^{s}\underline{\mathbf{J}})|^{5/2}}]dx^{k_{s}}\} \\
&&+\frac{[\ ^{s}\underline{\Psi }^{\diamond _{s}}]^{2}}{4(\ _{Q}^{s}%
\underline{\mathbf{J}})^{2}\{g_{2s+1}^{[0]}-\int du^{2s}[\ ^{s}\underline{%
\Psi }^{2}]^{\diamond _{s}}/4\ (\ _{Q}^{s}\underline{\mathbf{J}})\}}(du^{2s}+%
\frac{\partial _{i_{s}}\ ^{s}\underline{\Psi }}{\ ^{s}\underline{\Psi }%
^{\diamond _{s}}}dx^{i_{s}})^{2}].
\end{eqnarray*}%
Using (\ref{dualcosmd}), the solutions (\ref{qeltorsc}) can be re-defined on 
$\ _{Q}^{\shortmid s}\mathcal{M}$.

Let us conclude the constructions related to the above-stated space and time
duality principles: locally anisotropic cosmological s-metrics can be
derived in abstract dual form using some quasi-stationary solutions by
changing corresponding indices 3 into 4, 4 into 3. We correspondingly
underline the cosmological generating functions, effective sources and
gravitational polarizations for dependencies on $(x^{i},t);$ the v-partial
derivatives are changed in the form: $\ast \rightarrow \diamond $, i.e. $%
\partial _{3}\rightarrow \partial _{4}$. \ Such abstract index transforms
can be performed on all shells for respective velocity/ momentum-like
variables.%

\subsubsection{Constraints on generating functions and sources for
extracting LC configurations}

The generic off--diagonal solutions from the previous subsections are
constructed for auxiliary canonical d--connections $\widehat{\mathbf{D}}, \
^{s}\widehat{\mathbf{D}},\ _{\shortmid }^{s}\widehat{\mathbf{D}},$ etc. In
general, such solutions are characterized by nonholonomically induced
d--torsion coefficients $\widehat{\mathbf{T}}_{\ \alpha \beta }^{\gamma }$ (%
\ref{nontrtors}) completely defined by the N--connection and s--metric
structures. To generate exact and parametric solutions on base spacetime as
in GR we have to solve additional anholonomic constraints of type (\ref%
{lccondf}). Here we emphasize that FLH theories can be described in terms of
LC-connections on phase spaces, but the formulas for the geometric objects
and geometric/ physically important systems of nonlinear PDEs are not
adapted to N- and s-connections structures. 

We can extract zero torsion LC configurations in explicit form if we impose
additionally zero conditions (\ref{lccondf}) after we constructed a class of
quasi-stationary (\ref{qeltors}) or locally anisotropic cosmological
solutions (\ref{offdiagcosmcsh}). Corresponding computations for
quasi-stationary configurations state that all d-torsion coefficients vanish
if the coefficients of the N--adapted frames and the components of
s--metrics are subjected to the conditions: 
\begin{eqnarray}
\ w_{i_{1}}^{\ast _{2}}(x^{k_{1}},y^{3}) &=&\mathbf{e}_{i_{1}}\ln \sqrt{|\
h_{3}(x^{k_{1}},y^{3})|},\mathbf{e}_{i_{1}}\ln \sqrt{|\
h_{4}(x^{k_{1}},y^{3})|}=0,\partial _{i_{1}}w_{j_{1}}=\partial
_{j_{1}}w_{i_{{}}}\mbox{ and }n_{i_{1}}^{\ast _{2}}=0;  \notag \\
n_{k_{1}}(x^{i_{1}}) &=&0\mbox{ and }\partial
_{i_{1}}n_{j_{1}}(x^{k_{1}})=\partial _{j_{1}}n_{i_{1}}(x_{1}^{k});
\label{zerot1} \\
\ w_{i_{2}}^{\ast _{3}}(x^{k_{2}},v^{5}) &=&\mathbf{e}_{i_{2}}\ln \sqrt{|\
h_{5}(x^{k_{2}},y^{5})|},\mathbf{e}_{i_{2}}\ln \sqrt{|\
h_{6}(x^{k_{2}},y^{5})|}=0,\partial _{i_{2}}w_{j_{2}}=\partial
_{j_{2}}w_{i_{2}}\mbox{ and }n_{i_{2}}^{\ast _{3}}=0;  \notag \\
n_{k_{2}}(x^{i_{2}}) &=&0\mbox{ and }\partial
_{i_{2}}n_{j_{2}}(x^{k_{2}})=\partial _{j_{2}}n_{i_{2}}(x^{k_{2}});  \notag
\\
w_{i_{3}}^{\ast _{4}}(x^{k_{3}},v^{7}) &=&\mathbf{e}_{i_{3}}\ln \sqrt{|\
h_{7}(x^{k_{3}},y^{7})|},\mathbf{e}_{i_{s}}\ln \sqrt{|\
h_{8}(x^{k_{3}},y^{7})|}=0,\partial _{i_{3}}w_{j_{3}}=\partial
_{j_{3}}w_{i_{3}}\mbox{ and }n_{i_{3}}^{\ast _{4}}=0.  \notag
\end{eqnarray}%
The solutions for such $w$- and $n$-functions depend on the class of vacuum
or non--vacuum metrics which we aim to generate. To solve this problem, we
can follow such steps:

Let us consider how we can solve the equations (\ref{zerot1}) for $s=2.$ If
we prescribe a generating function $\ ^{2}\Psi =\ ^{2}\check{\Psi}%
(x^{i_{1}},y^{3})$ for which $[\partial _{i_{1}}(\ ^{2}\check{\Psi})]^{\ast
_{2}}=\partial _{i_{1}}(\ \ ^{2}\check{\Psi})^{\ast _{2}},$ we can solve the
equations for $w_{j_{1}}$ from (\ref{zerot1}). This is possible in explicit
form if $\ _{Q}^{2}\mathbf{J}=const,$ or if the effective source is
expressed as a functional $\ _{Q}^{2}\mathbf{J}(x^{i_{1}},y^{3})=\ _{Q}^{2}%
\mathbf{J}[\ ^{2}\check{\Psi}].$ Then, we can solve the third conditions $%
\partial _{i_{1}}w_{j_{1}}=\partial _{j_{1}}w_{i_{1}}$ if we chose a
generating function $\ \ ^{2}\check{A}=\ ^{2}\check{A}(x^{k_{1}},y^{3})$ and
define 
\begin{equation*}
w_{i_{1}}(x^{k_{1}},y^{3})=\check{w}_{i_{1}}(x^{k_{1}},y^{3})=\partial
_{i_{1}}\ ^{2}\check{\Psi}/(\ ^{2}\check{\Psi})^{\ast _{2}}=\partial
_{i_{1}}\ ^{2}\check{A}(x^{k},y^{3}).
\end{equation*}%
The equations for $n$-functions in (\ref{zerot1}) are solved by any $%
n_{i_{1}}(x^{k_{1}})=\partial _{i_{1}}[\ ^{2}n(x^{k_{1}})].$ 

The above formulas allow us to write the quadratic element for
quasi-stationary solutions with zero canonical d-torsion in a form similar
to (\ref{qeltors}), 
\begin{eqnarray}
d\check{s}^{2}(\tau ) &=&\check{g}_{\alpha _{s}\beta
_{s}}(x^{k_{s}},v^{k_{s}+1})du^{\alpha _{s}}du^{\beta _{s}}=e^{\psi
}[(dx^{1})^{2}+(dx^{2})^{2}]+  \label{qellc} \\
&&\sum\nolimits_{s=2}^{s=4}[\frac{[\ ^{s}\check{\Psi}^{\ast _{s}}]^{2}}{4(\
_{Q}^{s}\mathbf{J}[\check{\Psi}])^{2}\{h_{4}^{[0]}-\int du^{2s-1}[\ ^{s}%
\check{\Psi}]^{\ast _{s}}/4\ \ _{Q}^{s}\mathbf{J}[\ ^{s}\check{\Psi}]\}}%
\{du^{2s-1}+[\partial _{i_{s}}(\ ^{s}\check{A})]dx^{i_{s}}\}^{2}  \notag \\
&&+\{h_{4}^{[0]}-\int du^{2s-1}\frac{[\ ^{s}\check{\Psi}^{2}]^{\ast _{s}}}{%
4(\ \ \ _{Q}^{s}\mathbf{J}[\ ^{s}\check{\Psi}])}\}\{du^{2s-1}+\partial
_{i_{s}}[\ ^{s}n]dx^{i_{s}}\}^{2}].  \notag
\end{eqnarray}%
Similar constraints on generation functions as in (\ref{zerot1}), with
re-defined nonlinear symmetries, allow us to extract LC configurations for
all classes of quasi-stationary or locally anisotropic cosmological models
with FLH geometric flows and off-diagonal deformations. This is always
possible if for some generic off-diagonal metrics with nontrivial canonical
d-torsion we chose respective (more special) conditions for generating data,
for instance, of type $(\ ^{s}\check{\Psi}(x^{i_{s}},v^{i_{s}+1}), \ _{Q}^{s}%
\mathbf{J}[\ ^{s}\check{\Psi}],\ ^{s}\check{A}(x^{i_{s}},v^{i_{s}+1})).$
Dualizing the coefficient formulas as in (\ref{dualcosm}), we transform (\ref%
{qellc}) into locally anisotropic cosmological solutions of the FLH
geometric flow and off-diagonal modifications of the Einstein equations in
GR. 

\subsection{Mutual transforms of FLH geometric flow and MGTs and their
classes of solutions}

Off-diagonal geometric flows and (or) nonholonomic interactions and
distortion of connections result in such new nonlinear geometric effects:

\begin{enumerate}
\item Certain FLH models may transform equivalently, or partially, in other
types of FLH theories (which can be metric compatible or not).

\item A class of off-diagonal \ exact/parametric solutions can be
transformed in another class of off-diagonal solutions of the same FLH MGT
model. In particular, we can chose a prime s-metric which is not a solution
of some FLH-modified Einstein equation and transform it in a target s-metric
which is a solution of FLH geometric flow, or nonholonomic Ricci soliton,
equations.
\end{enumerate}

Any transform 1 or 2 can be characterized by respective nonlinear symmetries
(\ref{nonlintrsmalp}) and (\ref{nonlinsymrex}) and modified G. Perelman
thermodynamic variables (\ref{nagthermodfhq}).

\subsubsection{Thermodynamic variables for FLH geometric flows of nonmetric
quasi-stationary configurations}

Let us consider a $\tau $-family of quasi-stationary s-metrics $\widehat{%
\mathbf{g}}_{\alpha _{s}}[\ _{Q}^{s}\Phi (\tau )]$ (\ref{offdiagcosmcsh}) as
solutions of (\ref{cfeq4afc}) and respective nonlinear symmetries (\ref%
{ntransf1}) and (\ref{ntransf2}). For such nonholonomic canonical geometric
data on $\ _{Q}^{s}\mathcal{M}$, we can compute the canonical thermodynamic
variables (\ref{spffhq}) and (\ref{nagthermodfhq}) by expressing 
\begin{equation*}
\ _{Q}^{s}\widehat{\mathbf{R}}sc(\tau )=8(\ _{Q}^{h}\Lambda (\tau )+\
_{Q}^{v}\Lambda (\tau )),
\end{equation*}%
where different $\tau $-running of cosmological constant are considered on
the typical base and fiber spaces for $\ _{Q}^{h}\Lambda (\tau )=\
^{1}\Lambda (\tau )=\ ^{2}\Lambda (\tau )$ and $\ _{Q}^{v}\Lambda (\tau )=\
^{3}\Lambda (\tau )=\ ^{4}\Lambda (\tau )$. We do not present in this work
more cumbersome technical results for computing $\widehat{\sigma }(\tau )$
involving the canonical Ricci s-tensors $\ _{Q}^{s}\widehat{\mathbf{R}}%
c(\tau )$. Here we note that introducing effective cosmological constants we
simplify the procedure of computing thermodynamic variables, when the
contributions of $Q\,$-fields are encoded into nonlinear symmetries and
respective volume forms.

Such assumptions allow us to write the phase space statistical partition
function and thermodynamic variables in the form: 
\begin{eqnarray}
\ \ _{Q}^{s}\widehat{Z}(\tau ) &=&\exp [\int_{\widehat{\Xi }}\frac{1}{\left(
4\pi \tau \right) ^{4}}\ \delta \ \ _{Q}^{s}\mathcal{V}(\tau )],\ \   \notag
\\
_{Q}^{s}\widehat{\mathcal{E}}\ (\tau ) &=&-\tau ^{2}\int_{\widehat{\Xi }}\ 
\frac{8}{\left( 4\pi \tau \right) ^{4}}[\ _{Q}^{h}\Lambda (\tau )+\
_{Q}^{v}\Lambda (\tau )-\frac{1}{2\tau }]\ \delta \ \ _{Q}^{s}\mathcal{V}%
(\tau ),  \notag \\
\ \ _{Q}^{s}\widehat{S}(\tau ) &=&-\ \ \ _{Q}^{s}\widehat{W}(\tau )=-\int_{%
\widehat{\Xi }}\frac{8}{\left( 4\pi \tau \right) ^{2}}[\tau (\
_{Q}^{h}\Lambda (\tau )+\ _{Q}^{v}\Lambda (\tau ))-\frac{1}{2}]\delta \ \
_{Q}^{s}\mathcal{V}(\tau ).  \label{thermvar1}
\end{eqnarray}%
To simplify the computations, we have stated that the normalizing function
is subjected to the conditions: $-\ _{\mathbf{Q}}{\widehat{\zeta }} +4=1,\
^{s}\widehat{\mathbf{D}}\ _{\mathbf{Q}}{\widehat{\zeta }=0}$ and omit the
constant multiple $e^{-\ _{\mathbf{Q}}{\widehat{\ \zeta }}}.$ Such
conditions prescribe a "scale" for the nonholonomic FL evolution, which
allows us to study certain thermodynamic properties for a class of
topological configurations. We can consider arbitrary frame and coordinate
transforms and recompute the geometric/ physical variables for arbitrary
normalizing functions after the models have been elaborated.

Here we note that in (\ref{thermvar1}) the the information about a
quasi-stationary solution (the label $q$ is used for quasi-stationary $\tau $%
-running configurations) is embed into the volume element (\ref{volumforma}) 
\begin{equation*}
\delta \ _{Q}^{q}\mathcal{V}(\tau )=\sqrt{|\ \ _{Q}^{q}\mathbf{g}(\tau )|}\
dx^{1}dx^{2}\delta y^{3}\delta y^{4}\delta v^{5}\delta v^{6}\delta
v^{7}\delta v^{8}.
\end{equation*}%
To simplify further computations awe can consider trivial integration
functions $\ _{1}n_{k}=0$ and $\ _{2}n_{k}=0$ (such conditions change for
arbitrary frame and coordinate transforms). Using the formulas (\ref%
{nonlinsymrex}), we compute and write 
\begin{eqnarray*}
\ \delta \ \ _{Q}^{q}\mathcal{V} &=&\delta \mathcal{V}[\tau ,\ \widehat{%
\mathbf{J}}(\tau ),\ \ _{Q}^{h}\Lambda (\tau ),\ _{Q}^{v}\Lambda (\tau
);\psi (\tau ),\ g_{2s}(\tau )=\eta _{2s}(\tau )\ \mathring{g}_{2s}] \\
&=&\frac{1}{|\ \ _{Q}^{h}\Lambda (\tau )\times \ _{Q}^{v}\Lambda (\tau )|}\
\delta \ _{\eta }^{Q}\mathcal{V},\mbox{ where }\ \delta \ _{\eta }\mathcal{V}%
=\ \delta \ _{\eta }^{1}\mathcal{V}\times \delta \ _{\eta }^{2}\mathcal{V}%
\times \delta \ _{\eta }^{3}\mathcal{V}\times \delta \ _{\eta }^{4}\mathcal{V%
},\mbox{ for }
\end{eqnarray*}%
\begin{eqnarray}
\delta \ _{\eta }^{1}\mathcal{V} &=&\delta \ _{\eta }^{1}\mathcal{V}[\
_{Q}^{1}\widehat{\mathbf{J}}(\tau ),\eta _{1}(\tau )\ \mathring{g}_{1}]
\label{volumfuncts} \\
&=&e^{\widetilde{\psi }(\tau )}dx^{1}dx^{2}=\sqrt{|\ _{Q}^{1}\widehat{%
\mathbf{J}}(\tau )|}e^{\psi (\tau )}dx^{1}dx^{2},\mbox{ for }\psi (\tau )%
\mbox{ being a
solution of  }(\ref{eq1}),\mbox{ with sources }\ _{Q}^{1}\widehat{\mathbf{J}}%
(\tau );  \notag \\
\delta \ _{\eta }^{s}\mathcal{V} &=&\delta \ _{\eta }^{s}\mathcal{V}[\
_{Q}^{s}\widehat{\mathbf{J}}(\tau ),\eta _{2s}(\tau ),\ \mathring{g}_{2s}] 
\notag \\
&=&\frac{\partial _{2s-1}|\ \eta _{2s}(\tau )\ \mathring{g}_{2s}|^{3/2}}{\ 
\sqrt{|\int du^{2s-1}\ \ _{Q}^{2}\widehat{\mathbf{J}}(\tau )\{\partial
_{2s-1}|\ \eta _{2s}(\tau )\ \mathring{g}_{2s}|\}^{2}|}}  \notag \\
&&\lbrack du^{2s-1}+\frac{\partial _{i_{s-1}}\left( \int du^{2s-1}\ _{Q}^{s}%
\widehat{\mathbf{J}}(\tau )\partial _{2s-1}|\ \eta _{2s}(\tau )\ \mathring{g}%
_{2s}|\right) dx^{i_{s}-1}}{\ \ \ _{Q}^{s}\widehat{\mathbf{J}}(\tau
)\partial _{3}|\ \eta _{4}(\tau )\mathring{g}_{4}|}du^{2s}].  \notag
\end{eqnarray}%
Integrating such products of forms from (\ref{volumfuncts}) on a closed
hypersurface $\widehat{\Xi }\subset \ _{s}^{Q}\mathcal{M},$ we obtain a
running phase space volume functional 
\begin{equation}
\ _{\eta }^{\mathbf{J}}\mathcal{V[}\ _{Q}^{q}\mathbf{g}(\tau )]=\int_{\ 
\widehat{\Xi }}\delta \ _{\eta }^{Q}\mathcal{V}(\ _{s}^{s}\widehat{\mathbf{J}%
}(\tau ),\ \eta _{\alpha _{s}}(\tau ),\mathring{g}_{\alpha _{s}}).
\label{volumf1}
\end{equation}

Using the volume functional (\ref{volumf1}), we obtain such formulas for
nonholonomic thermodynamic variables (\ref{thermvar1}): 
\begin{eqnarray}
\ _{Q}^{q}\widehat{Z}(\tau ) &=&\exp \left[ \frac{1}{(4\pi \tau )^{4}}\ \
_{\eta }^{\mathbf{J}}\mathcal{V}[\ _{Q}^{q}\mathbf{g}(\tau )]\right] ,
\label{thermvar2} \\
\ _{Q}^{q}\widehat{\mathcal{E}}\ (\tau ) &=&\frac{1}{64\pi ^{4}\tau ^{3}}\
\left( 1-2\tau (\ _{Q}^{h}\Lambda (\tau )+\ _{Q}^{v}\Lambda (\tau ))\right)
\ \ \ _{\eta }^{\mathbf{J}}\mathcal{V[}\ _{Q}^{q}\mathbf{g}(\tau )],  \notag
\\
\ \ \ \ _{Q}^{q}\widehat{S}(\tau ) &=&-\ _{Q}^{q}\widehat{W}(\tau )=\frac{2}{%
(4\pi \tau )^{4}}(1-4(\ _{Q}^{h}\Lambda (\tau )+\ _{Q}^{v}\Lambda (\tau )))\
\ _{\eta }^{\mathbf{J}}\mathcal{V[}\ _{Q}^{q}\mathbf{g}(\tau )].  \notag
\end{eqnarray}

The formulas (\ref{thermvar2}) can be used for defining thermodynamic
characteristics of FL Ricci soliton quasi-stationary configurations for $%
\tau =\tau _{0}.$ This is for off-diagonal quasi-stationary solutions of
nonmetric FL deformed Einstein equations. 

\subsubsection{Nonholonomic distortions and equivalence of FLH theories}

Let us consider on a phase space $\ _{s}\mathcal{M}$ two classes of
nonmetric FL geometric flow theories, for instance, given by prime geometric
data $\left( \ _{s}^{B}\mathbf{\mathring{g}}(\tau )\mathbf{\simeq }\ _{s}^{%
\mathring{F}}\mathbf{\mathring{g}}(\tau ), \ _{s}^{B}\widehat{\mathbf{%
\mathring{D}}}(\tau )\right) ,$ of Berwald-Finsler type, and target
geometric data $\ \left( \ _{s}^{C}\mathbf{g}(\tau )\mathbf{\simeq }\
_{s}^{F}\mathbf{g}(\tau ), \ _{s}^{C}\widehat{\mathbf{D}}(\tau)\right) ,$ of
Chern-Finsler type. We can chose $_{s}\mathbf{\mathring{g}}(\tau )\mathbf{%
\simeq }\ ^{s}\mathbf{\hat{g}}[\ ^{s}\mathring{\Psi}]$ to be of type (\ref%
{qeltors}) as a solution of $\widehat{\mathbf{\mathring{R}}}_{\ \ \beta
_{s}}^{\alpha _{s}}[\ ^{s}\mathring{\Psi}]=\ _{Q}\mathbf{\mathring{J}}_{\ \
\beta _{s}}^{\alpha _{s}}$ (as for (\ref{cfeq4af})), or of $\widehat{\mathbf{%
\mathring{R}}}_{\ \ \beta _{s}}^{\alpha _{s}}[\ ^{s}\mathring{\Phi}(\tau )]=
\ ^{s}\mathring{\Lambda}(\tau )\mathbf{\delta }_{\ \ \beta _{s}}^{\alpha
_{s}}$(\ref{cfeq4afc}). Such prime nonmetric FL configurations are
characterized by respective nonlinear symmetries (\ref{ntransf1}), (\ref%
{ntransf2}) and (\ref{nonlintrsmalp}), (\ref{nonlinsymrex}) with a
corresponding labelling of geometric objects by a circle symbol. Similar
formulas, with different effective sources $\ _{Q}\mathbf{J}_{\ \ \beta
_{s}}^{\alpha _{s}}(\tau )$ and $\tau $-running effective cosmological
constants $\ ^{s}\Lambda (\tau ),$ hold true for the Finsler-Chern data. 

Using gravitational polarization functions as in (\ref{offdiagdefr}), we can
define nonholonomic transforms and distortions 
\begin{eqnarray}
\ _{s}^{B}\mathbf{\mathring{g}}(\tau ) &\rightarrow &\ _{s}^{C}\mathbf{g}%
(\tau )=[g_{\alpha _{s}}(\tau )=\eta _{\alpha _{s}}(\tau )\mathring{g}%
_{\alpha _{s}}(\tau ),N_{i_{s-1}}^{a_{s}}(\tau )=\eta
_{i_{s-1}}^{a_{s}}(\tau )\ \mathring{N}_{i_{s-1}}^{a_{s}}(\tau )],%
\mbox{ and
}\   \notag \\
\ _{s}^{C}\widehat{\mathbf{D}}(\tau ) &=&\ _{s}^{B}\widehat{\mathbf{%
\mathring{D}}}(\tau )+\ _{s}^{BC}\widehat{\mathbf{Z}}(\tau ),
\label{auxtranfflh}
\end{eqnarray}%
where $\ _{s}^{BC}\widehat{\mathbf{Z}}(\tau )$ is used for a family of
distortions s-tensors from $\ _{s}^{B}\widehat{\mathbf{\mathring{D}}}(\tau )$
to $\ _{s}^{C}\widehat{\mathbf{D}}(\tau ).$ The $\eta $-polarizations are
subjected to the conditions that they relate two types of nonmetric FL
theories. We can speculate that a Chern-Finsler model is more preferred
thermodynamically than a Berwald-Finsler one (or inversely) if, for
instance, the generalized G. Perelman entropy from $_{\mathbf{Q}}^{C}%
\widehat{\mathcal{S}}(\tau )$ (\ref{chernfinsltherm}) is smaller (bigger)
than similar values from $_{\mathbf{Q}}^{B}\widehat{\mathcal{S}}(\tau)\ $ (%
\ref{berwfinsltherm}). In certain cases, some Chern-Finsler configurations
can flow geometrically very close to other Berwal-Finsler ones if we
consider small $\epsilon$-polarizations. 

We can consider other types of nonholonomic transforms which are different
from (\ref{auxtranfflh}). For certain nonholonomic configurations, a
subclass of equivalent nonmetric FL theories can be modelled as metric ones,
for instance, of Cartan-Finsler type. Such geometric evolution scenarios
depend on the prescribed generating functions and generating sources, and
respective integration data. 

\subsubsection{Equivalent modelling and different classes of FLH solutions}

Gravitational polarization functions can be defined for a case when both the
prime and target s-data are given for the same class of nonmetric FL
geometric flow theory, or MGT. Respective nonholonomic transforms and
distortions are parameterized in the form (for instance, for Chern-Finsler
configurations) 
\begin{equation}
\ _{s}^{C}\mathbf{\mathring{g}}(\tau )\rightarrow \ _{s}^{C}\mathbf{g}(\tau
)=[g_{\alpha _{s}}(\tau )=\eta _{\alpha _{s}}(\tau )\mathring{g}_{\alpha
_{s}}(\tau ),N_{i_{s-1}}^{a_{s}}(\tau )=\eta _{i_{s-1}}^{a_{s}}(\tau )\ 
\mathring{N}_{i_{s-1}}^{a_{s}}(\tau )].\   \notag
\end{equation}%
We can chose that $\ _{s}^{C}\mathbf{\mathring{g}}(\tau )$ is a solution of
a system of nonlinear PDEs (\ref{cfeq4af}), but positively impose that the
target configuration $\ _{s}^{C}\mathbf{g}(\tau )$ is defined by a class of
solutions of such a system with prescribed effective sources $\ _{Q}^{C}%
\mathbf{J}_{\ \ \beta _{s}}^{\alpha _{s}}(\tau )$ and $\tau $-running
effective cosmological constants $\ ^{s}\Lambda (\tau ).$ This way, we can
state certain nonholonomic conditions when certain arbitrary prime FL data
(not defining a geometric flow or a nonholonomic Ricci soliton
configuration) transform because of the geometric evolution and/or
off-diagonal interactions into FL deformed geometric flows or MGTs. For
instance, we can prescribe that the target $\left(\ _{s}^{C}\mathbf{g}(\tau )%
\mathbf{\simeq }\ _{s}^{F}\mathbf{g}(\tau ), \ _{s}^{C}\widehat{\mathbf{D}}%
(\tau )\right) $ are solutions of (\ref{cfeq4afc}). 

If both $\ _{s}^{C}\mathbf{\mathring{g}}(\tau )$ \ and $\ _{s}^{C}\mathbf{g}%
(\tau )$ are solutions of the same system of nonlinear PDEs, we can analyze
which class of such solutions is more convenient thermodynamically. For
instance, we compute both $_{\mathbf{Q}}^{C}\widehat{\mathcal{\mathring{S}}}%
(\tau )\ \ $and $_{\mathbf{Q}}^{C}\widehat{\mathcal{S}}(\tau )$ for a chosen
closed phase space region and say the prime configuration is more probable
if $_{\mathbf{Q}}^{C}\widehat{\mathcal{\mathring{S}}}(\tau )\ < \ _{\mathbf{Q%
}}^{C}\widehat{\mathcal{S}}(\tau ).$

Finally, in this subsection, we note that we can consider $\eta $%
-polarizations for $\ ^{B}\mathbf{\mathring{g}}(\tau )\rightarrow \ _{s}^{B}%
\mathbf{g}(\tau ),$ or for metric compatible FL models, etc., and analyze
which model is \ more convenient, for instance, energetically or with less
quadratic dispersion. 

\section{Examples of FLH-modified off-diagonal solutions}

\label{sec5}In this section, we show how the AFCDM (see reviews \cite%
{vacaru18,partner06,vbubuianu17,partner02,bsssvv25} and generalizations for
FLH theories in previous section and Appendices \ref{appendixa} and \ref%
{appendixb}) \ can be applied for constructing four classes of physically
important solutions of (\ref{cfeq4af}), (\ref{feq4afd}), or (\ref{cfeq4afc}%
). The first three are defined by quasi-stationary off-diagonal metrics and
may describe FLH geometric flows of solutions in GR and MGTs: 1]
nonholonomic BH-like solutions with distortions to BE configurations; 2]
locally anisotropic wormhole, WH, FLH solutions; 3] some systems of black
torus (BT) FLH solutions. We also provide an example of locally anisotropic
cosmological solutions describing FLH geometric evolutions of cosmological
solitonic and spheroid deformations involving 2-d vertices. 

In this section, we show how to compute in explicit form G. Perelman's
thermodynamic variables for four classes of physically important
off-diagonal solutions in nonmetric FLH geometric flow and MGTs. We
emphasize that only in a special case of rotoid deformations of KdS BHs (for
instance, with an ellipsoid generating function, we can introduce
hypersurface (ellipsoid type) configurations. This allows us to apply the
Bekenstein-Hawking thermodynamic paradigm \cite{bek2,haw2}. Many examples
are studied and reviewed in \cite{vacaru18,bsssvv25}. For different types of
(nonassociative, noncommutative, supersymmetric, algebroid etc.)
off-diagonal deformations of KdS BHs, WHs, BTs and locally anisotropic
cosmological solutions, respective thermodynamic characterizations are
possible if we consider a relativistic generalization of the concept of
W-entropy \cite{perelman1,svnonh08,gheorghiuap16,vacaru18,bsssvv25}.

\subsection{FLH geometric flows of new Kerr de Sitter BHs to (double)
spheroidal configurations}

In a series of works on MGTs \cite%
{vacaru18,vbubuianu17,partner02,partner06,bsssvv25}, effective contributions
from (non) associative/ commutative sources in string theory and geometric
information flows), nonholonomic off-diagonal deformations of the Kerr and
Schwarzschild - (a) de Sitter, K(a)dS, BH metrics were studied. For
spherical rotating configurations of KdS in GR, such metrics can be
described by various families of rotating diagonal metrics involving, or
not, certain warping effects of curvature \cite{ovalle21}. In this
subsection, we show how new classes of solutions of FLH geometric flow
equations can be constructed as off-diagonal deformations of some primary
KdS metrics in GR. Such $\tau $-families of rotating BHs can be deformed to
parametric quasi-stationary s-metrics of type (\ref{offdncelepsilon}). We
show how to compute in explicit form spheroidal rotoid deformations. 

\subsubsection{Prime new KdS metrics and gravitational polarizations}

We consider a prime quadratic s-metric (\ref{offdiagpm}) involving 4-d
spherical coordinates parameterized in the form $x^{1}=r,x^{2}=\varphi
,y^{3}=\theta ,y^{4}=t.$ Such spherical coordinates can be considered for
(co) fibers on phase spaces $\ \mathcal{M}$ or $\ ^{\shortmid }\mathcal{M}$
for respective velocity or momentum type variables. On the base spacetime
Lorentz manifold with shells $s=1$ and $s=2$, the quadratic line element can
be written in the form 
\begin{equation}
d\breve{s}^{2}=\breve{g}_{\alpha _{2}}(r,\varphi ,\theta )(\mathbf{\breve{e}}%
^{\alpha _{2}})^{2}.  \label{offdiagpm1}
\end{equation}%
The corresponding nontrivial coefficients of the prime s-metric and
N-connection are 
\begin{eqnarray}
\breve{g}_{1} &=&\frac{\breve{\rho}^{2}}{\triangle _{\Lambda }},\breve{g}%
_{2}=\frac{\sin ^{2}\theta }{\breve{\rho}^{2}}[\Sigma _{\Lambda }-\frac{%
(r^{2}+a^{2}-\triangle _{\Lambda })^{2}}{a^{2}\sin ^{2}\theta -\triangle
_{\Lambda }}],\breve{g}_{3}=\breve{\rho}^{2},\breve{g}_{4}=\frac{a^{2}\sin
^{2}\theta -\triangle _{\Lambda }}{\breve{\rho}^{2}},\mbox{ and }  \notag \\
\breve{N}_{2}^{4} &=&\breve{n}_{2}=-a\sin \theta \frac{r^{2}+a^{2}-\triangle
_{\Lambda }}{a^{2}\sin ^{2}\theta -\triangle _{\Lambda }}.  \label{prdat1}
\end{eqnarray}%
Any (\ref{offdiagpm1}) can be extended \ to a 8-d phase space metric of type
(\ref{lqe}) or (\ref{lqed}), see below; and, in more general cases (\ref%
{ssolutions}) or (\ref{ssolutionsd}). A new KdS solution in GR (see \cite%
{ovalle21} and references in that work, but we emphasize that we follow a
different system of notations) is generated if the functions and parameters
are chosen in the form 
\begin{eqnarray}
\Sigma _{\Lambda } &=&(r^{2}+a^{2})^{2}-\triangle _{\Lambda }a^{2}\sin
^{2}\theta ,\triangle _{\Lambda }=r^{2}-2Mr+a^{2}-\frac{\Lambda _{0}}{3}%
r^{4},  \notag \\
\breve{\rho}^{2} &=&r^{2}+a^{2}\cos ^{2}\theta ,\mbox{ for constants  }%
a=J/M=const.  \label{prdat2}
\end{eqnarray}%
In these formulas, $J$ is the angular momentum, $M$ is the total mass of the
system, and the cosmological constant $\Lambda _{0}>0.$ We emphasize that
the solution (\ref{offdiagpm1}) is different from the standard KdS metrics,
defining $\Lambda $-vacuum solutions, because the scalar curvature $%
R(r,\theta )=4\widetilde{\Lambda }(r,\theta )=4\Lambda _{0}\frac{r^{2}}{%
\breve{\rho}^{2}}\neq 4\Lambda _{0}.$ So, the above formulas define a new
KdS solution which possesses a warped effect when the curvature is warped
everywhere except the equatorial plane. This is a rotating configuration of
a BH with an effective polarization $(r,\theta )$ of a cosmological constant 
$\Lambda _{0}.$ It shows a rotational effect on the vacuum energy in GR with
a cosmological constant. Such an effect disappears for $r\gg a.$ In the next
subsections, we prove that different types of polarizations are possible for
FLH geometric flows and off-diagonal interactions, in general, involving
nonmetricity fields.

A d-metric (\ref{offdiagpm1}) defines a rotating version of the
Schwarzschild de Sitter metric and represents a new solution describing the
exterior of a BH with cosmological constant. To get a BH like solution
certain bond conditions for $M(a,\Lambda _{0})$ have to be imposed.
Corresponding, the upper, $M_{\max }:=M_{+}$ and lower, $M_{\min }:=M,$
bounds are computed 
\begin{equation}
18\Lambda _{0}M_{\pm }^{2}=1+12\Lambda _{0}a^{2}\pm (1-4\Lambda
_{0}a^{2})^{3/2}.  \label{bonds}
\end{equation}%
A solution (\ref{offdiagpm1}) defines a LC-configuration for the Einstein
equations in GR with fluid type energy momentum tensor 
\begin{equation}
\breve{T}_{\alpha _{2}\beta _{2}}(r,\theta )=diag[p_{r},p_{\varphi
}=p_{\theta },p_{\theta }=\rho -2\Lambda _{0}r^{2}/\mathring{\rho}^{2},\rho
=-p_{r}=\widetilde{\Lambda }^{2}/\Lambda _{0}].  \label{efmt1}
\end{equation}%
Such primary s-metrics have clear physical interpretations: 1) they are
defined as solutions of some vacuum locally anisotropic polarizations on $%
(r,\theta )$ of the cosmological constant, $\Lambda _{0}\rightarrow 
\widetilde{\Lambda }(r,\theta );$ or 2) consist a result of some locally
anisotropic energy-momentum tensors of type $\breve{T}_{\alpha
\beta}(r,\theta )$, or more general (effective) sources. For simplicity, we
study in this subsection only a prime s-metric when the target s-metrics are
generated as $\tau $-families for nonmetric FLH geometric flow or MGTs. 

\subsubsection{Nonmetric FLH geometric flow off-diagonal deformations of KdS
metrics}

In this subsection, we study more general off-diagonal deformations of the
standard Kerr solution when there are involved $\tau $-running gravitational
polarizations and effective cosmological constants. For $\tau =\tau _{0},$
such target quasi-stationary s-metric are defined by coefficients depending
on all space coordinates $(r,\varphi ,\theta )$, not only on $(r,\theta )$
as we considered for above prime s-metrics. The new classes of
quasi-stationary FLH deformed spacetimes possess nonlinear symmetries of
type (\ref{ntransf1}) and (\ref{ntransf2}), defined by respective classes of
distorted s-connections and nonholonomic constraints. So, we generate target
solutions of type (\ref{ssolutions}) when $\mathbf{\hat{g}}(\tau ,r,\varphi
,\theta )$ are defined equivalently by generating sources of type $\ _{Q}%
\mathbf{J}_{\ \nu }^{\mu }(\tau )$ (\ref{dsourcparam})$\rightarrow \ _{Q}%
\mathbf{\breve{J}}_{\ \nu _{s}}^{\mu _{s}}(\tau )$, when the s-adapted $%
s=1,2 $ components $\ _{Q}\mathbf{\breve{J}}_{\ \nu _{2}}^{\mu _{2}}(\tau )$
are related via frame transforms (\ref{ntransf1}), 
\begin{eqnarray}
\ _{Q}\mathbf{\breve{J}}_{\ \nu _{s}}^{\mu _{s}}(\tau ) &=&\ _{Q}\mathbf{%
\breve{J}}_{\ \nu _{s}}^{\mu _{s}}(\tau ,u^{\alpha _{s}})  \label{sourcbh} \\
&=&[\ ^{1}\mathbf{\breve{J}}(\tau ,r,\varphi )\delta _{\ \ j_{1}}^{i_{1}},\
^{2}\mathbf{\breve{J}}(\tau ,r,\varphi ,\theta )\delta _{\ \
b_{2}}^{a_{2}},\ ^{3}\mathbf{\breve{J}}(\tau ,r,\varphi ,\theta
,t,v^{c_{3}})\delta _{\ \ b_{3}}^{a_{3}},\ ^{4}\mathbf{\breve{J}}(\tau
,r,\varphi ,\theta ,t,v^{e_{3}},v^{c_{4}})\delta _{\ \ b_{4}}^{a_{4}}], 
\notag
\end{eqnarray}%
where $v^{c_{3}}=v^{5},$ or $v^{6};$ and $v^{c_{4}}=v^{7},$ or $v^{8}.$ To
generate "pure" quasi-stationary configurations, the sources $\ ^{3}\mathbf{%
\breve{J}}$ and $\ ^{4}\mathbf{\breve{J}}$ in (\ref{sourcbh}) must be
prescribed in some forms not containing dependencies on the time-like
variable $t.$

Using Tables 4-6 from Appendix \ref{appendixb}, but for corresponding local
coordinates and generating sources $\ _{Q}\mathbf{\breve{J}}_{\ \nu
_{s}}^{\mu _{s}}(\tau )$ (\ref{sourcbh}), we can construct off-diagonal
solutions with $\eta $-polarization functions as in (\ref{offdiagpolfr}), 
\begin{eqnarray}
d\widehat{s}^{2}(\tau ) &=&\widehat{g}_{\alpha \beta }(\tau ,r,\varphi
,y^{3}=\theta ;\breve{g}_{\alpha _{s}};\psi ,\eta _{2s};\ ^{s}\Lambda ,\ ^{s}%
\mathbf{\breve{J}})du^{\alpha }du^{\beta }=e^{\psi (r,\varphi
)}[(dx^{1}(r,\varphi ))^{2}+(dx^{2}(r,\varphi ))^{2}]  \label{nkernew} \\
&&-\frac{[\partial _{\theta }(\eta _{4}\ \breve{g}_{4})]^{2}}{|\int d\theta
\ ^{2}\mathbf{\breve{J}}\partial _{\theta }(\eta _{4}\ \breve{g}_{4})|\ \eta
_{4}\breve{g}_{4}}\{d\theta +\frac{\partial _{i_{1}}[\int d\theta \ ^{2}%
\mathbf{\breve{J}}\ \partial _{\theta }(\eta _{4}\breve{g}_{4})]}{\ \ ^{2}%
\mathbf{\breve{J}}\partial _{\theta }(\eta _{4}\breve{g}_{4})}%
dx^{i_{1}}\}^{2}  \notag \\
&&+\eta _{4}\breve{g}_{4}\{dt+[\ _{1}n_{k_{1}}(r,\varphi )+\
_{2}n_{k_{1}}(r,\varphi )\int d\theta \frac{\lbrack \partial _{\theta }(\eta
_{4}\breve{g}_{4})]^{2}}{|\int d\theta \ \ ^{2}\mathbf{\breve{J}}\partial
_{3}(\eta _{4}\breve{g}_{4})|\ (\eta _{4}\breve{g}_{4})^{5/2}}%
]dx^{k_{1}}\}^{2}  \notag \\
&&-\frac{[\partial _{2s+1}(\eta _{2s}\ \breve{g}_{2s})]^{2}}{|\int
dv^{2s+1}\ ^{s}\mathbf{\breve{J}}\partial _{2s+1}(\eta _{2s}\ \breve{g}%
_{2s})|\ \eta _{2s}\breve{g}_{2s}}\{dv^{2s+1}+\frac{\partial _{i_{2s}}[\int
dv^{2s+1}\ \ ^{s}\mathbf{\breve{J}}\ \partial _{2s+1}(\eta _{2s}\breve{g}%
_{2s})]}{\ \ ^{s}\mathbf{\breve{J}}\partial _{2s+1}(\eta _{2s}\breve{g}_{2s})%
}dx^{i_{2s}}\}^{2}  \notag \\
&&+\eta _{2s}\breve{g}_{2s}\{dv^{2s+2}+[\ _{1}n_{k_{s}}(v^{i_{s}})+\
_{2}n_{k_{s}}(v^{i_{s}})\int dv^{2s+1}\frac{[\partial _{2s+1}(\eta _{2s}%
\breve{g}_{2s})]^{2}}{|\int dv^{2s+1}\ ^{s}\mathbf{\breve{J}}\partial
_{2s+1}(\eta _{2s}\breve{g}_{2s})|\ (\eta _{2s}\breve{g}_{2s})^{5/2}}%
]dx^{k_{s}}\}^{2},  \notag \\
\mbox{ for }s &=&3,4\mbox{ (in this subsection)}.  \notag
\end{eqnarray}%
The $\tau $-family of off-diagonal solutions $\ _{Q}^{KdS}\mathbf{g}$ (\ref%
{nkernew}) is determined by a generating function $\eta _{4}(\tau)=\eta
_{4}(\tau ,r,\varphi ,\theta ),$ $\eta _{2s}(\tau )=\eta
_{2s}(\tau,r,\varphi ,\theta )$ and respective integration functions $\
_{1}n_{k_{1}}(\tau ,r,\varphi ),\ _{2}n_{k_{1}}(\tau ,r,\varphi )$ and $\
_{1}n_{k_{s-1}}(\tau ,r,\varphi ,v^{2s-1}),\newline
$ $\ _{2}n_{k_{s-1}}(\tau ,r,\varphi ,v^{2s-1}),$ for $s=3,4.$ The locally
anisotropic vacuum effects in such a quasi-stationary s-metric are very
complex, and it is difficult to state well-defined and general conditions
when the solutions define BH configurations. We need additional assumptions
to generate BH solutions in a non-trivial gravitational vacuum. A
corresponding additional stability analysis and additional nonholonomic
constraints are necessary for some explicit generating and integrating data
if we try to construct stable configurations. Here we note that non-stable
solutions may also have physical importance, for instance, describing some
evolution, or structure formation, phase transitions, for a period of time
or under certain temperature regimes.

Quasi-stationary s-metric (\ref{nkernew}) are characterized by nonlinear
symmetries of type (\ref{nonlinsymrex}), 
\begin{eqnarray}
\partial _{\theta }[\ ^{2}\Psi ^{2}] &=&-\int d\theta \ ^{2}\mathbf{\breve{J}%
}\partial _{\theta }h_{4}\simeq -\int d\theta \ \ ^{2}\mathbf{\breve{J}}%
\partial _{\theta }(\eta _{4}\ \breve{g}_{4})\simeq -\int d\theta \ \ ^{2}%
\mathbf{\breve{J}}\partial _{\theta }[\zeta _{4}(1+\epsilon \ \chi _{4})\ 
\breve{g}_{4}],  \label{nlims2} \\
\ ^{2}\Psi &=&|\ \widetilde{\Lambda }|^{-1/2}\sqrt{|\int d\theta \ ^{2}%
\mathbf{\breve{J}}\ (\ ^{2}\Phi ^{2})^{\ast _{2}}|},  \notag \\
\ ^{2}\Phi ^{2} &=&-4\ \widetilde{\Lambda }h_{4}\simeq -4\ \ \widetilde{%
\Lambda }\eta _{4}\breve{g}_{4}\simeq -4\ \widetilde{\Lambda }\ \zeta
_{4}(1+\epsilon \chi _{4})\ \breve{g}_{4},  \notag \\
\partial _{2s+1}[\ ^{s}\Psi ^{2}] &=&-\int dv^{2s+1}\ ^{s}\mathbf{\breve{J}}%
\partial _{2s+1}h_{2s}\simeq -\int dv^{2s+1}\ ^{s}\mathbf{\breve{J}}\partial
_{2s+1}(\eta _{2s}\ \breve{g}_{2s})  \notag \\
&\simeq &-\int dv^{2s+1}\ ^{2}\mathbf{\breve{J}}\partial _{2s+1}[\zeta
_{2s}(1+\epsilon \ \chi _{2s})\ \breve{g}_{2s}],  \notag \\
\ ^{s}\Psi &=&|\ ^{s}\widetilde{\Lambda }|^{-1/2}\sqrt{|\int dv^{2s+1}\ ^{s}%
\mathbf{\breve{J}}\ (\ ^{s}\Phi ^{2})^{\ast _{s}}|},  \notag \\
\ ^{s}\Phi ^{2} &=&-4\ \ ^{s}\widetilde{\Lambda }h_{2s}\simeq -4\ \ ^{s}%
\widetilde{\Lambda }\eta _{2s}\breve{g}_{2s}\simeq -4\ ^{s}\widetilde{%
\Lambda }\ \zeta _{2s}(1+\epsilon \chi _{2s})\ \breve{g}_{2s},  \notag
\end{eqnarray}%
where $\widetilde{\Lambda }$ is extended to a $\tau $-family $\ ^{s}%
\widetilde{\Lambda }(\tau ,r,\theta )$ on $\ _{s}\mathcal{M}$. In a next
subsection, we shall compute respective G. Perelman's thermodynamic
variables.

We note that in a series of our former works on MGTs \cite%
{vacaru18,partner02,partner06,bsssvv25} K(a)dS and other type BH solutions
were nonholonomically deformed for $y^{3}=\varphi .$ In those papers,
effective sources are generated by certain extra dimension (super) string
contributions, nonassociative and/ or noncommutative terms, metric and
nonmetric generalized Finsler or other types of modified dispersion
deformations. The AFCDM can be applied in similar forms for FLH theories on
(co) tangent Lorentz bundles. In the nonholonomic geometric flow and various
types of MGTs and GR, we can state explicit conditions when off-diagonal $%
\varphi $-, or $\theta $-, deformations (with general dependence on
velocity/momentum like coordinates) may result in black ellipsoid, BE,
configurations.

\subsubsection{Off-diagonal solutions with small parametric deformations of
KdS d-metrics}

Considering small parametric decompositions with $\epsilon $-linear terms as
in (\ref{epsilongenfdecomp}), we can provide a physical interpretation of
off-diagonal quasi-stationary solutions (\ref{nkernew}). We avoid singular
off-diagonal frame or coordinate deformations if we use a new system of
coordinates with nontrivial terms of a prime N-connection. Respectively, for 
$a=3,$ with some $\breve{N}_{i}^{3}=\breve{w}_{i}(r,\varphi ,\theta ),$
which can be zero in certain rotation frames; and, for $a=4,$ $\breve{N}%
_{i}^{4}=\breve{n}_{i}(r,\varphi ,\theta )$ which may be with a nontrivial $%
\breve{n}_{2}=-a\sin \theta (r^{2}+a^{2}-\triangle _{\Lambda })/(a^{2}\sin
^{2}\theta -\triangle _{\Lambda }).$ We construct a s-metric of type (\ref%
{offdncelepsilon}) determined by $\chi $-generating functions: 
\begin{equation*}
d\ \widehat{s}^{2}(\tau )=\widehat{g}_{\alpha _{s}\beta _{s}}(r,\varphi
,\theta ,u^{2s+1};\psi ,g_{2s};\ ^{s}\mathbf{\breve{J}})du^{\alpha
_{s}}du^{\beta _{s}}=e^{\psi _{0}}(1+\kappa \ ^{\psi }\chi
)[(dx^{1}(r,\varphi ))^{2}+(dx^{2}(r,\varphi ))^{2}]
\end{equation*}%
\begin{eqnarray*}
&&-\{\frac{4[\partial _{\theta }(|\zeta _{4}\breve{g}_{4}|^{1/2})]^{2}}{%
\breve{g}_{3}|\int d\theta \lbrack \ ^{2}\mathbf{\breve{J}}\partial
_{3}(\zeta _{4}\breve{g}_{4})]|}-\epsilon \lbrack \frac{\partial _{\theta
}(\chi _{4}|\zeta _{4}\breve{g}_{4}|^{1/2})}{4\partial _{\theta }(|\zeta _{4}%
\breve{g}_{4}|^{1/2})}-\frac{\int d\theta \{\ ^{2}\mathbf{\breve{J}}\partial
_{\theta }[(\zeta _{4}\breve{g}_{4})\chi _{4}]\}}{\int d\theta \lbrack \ ^{2}%
\mathbf{\breve{J}}\partial _{\theta }(\zeta _{4}\breve{g}_{4})]}]\}\breve{g}%
_{3} \\
&&\{d\theta +[\frac{\partial _{i_{1}}\ \int d\theta \ \ ^{2}\mathbf{\breve{J}%
}\ \partial _{\theta }\zeta _{4}}{(\breve{N}_{i}^{3})\ \ ^{2}\mathbf{\breve{J%
}}\partial _{\theta }\zeta _{4}}+\epsilon (\frac{\partial _{i_{1}}[\int
d\theta \ \ ^{2}\mathbf{\breve{J}}\ \partial _{\theta }(\zeta _{4}\chi _{4})]%
}{\partial _{i_{1}}\ [\int d\theta \ \ ^{2}\mathbf{\breve{J}}\partial
_{\theta }\zeta _{4}]}-\frac{\partial _{\theta }(\zeta _{4}\chi _{4})}{%
\partial _{\theta }\zeta _{4}})]\breve{N}_{k_{1}}^{3}dx^{k_{1}}\}^{2}
\end{eqnarray*}%
\begin{eqnarray}
&&+\zeta _{4}(1+\epsilon \ \chi _{4})\ \breve{g}_{4}\{dt+[(\breve{N}%
_{k_{1}}^{4})^{-1}[\ _{1}n_{k_{1}}+16\ _{2}n_{k_{1}}[\int d\theta \frac{%
\left( \partial _{\theta }[(\zeta _{4}\breve{g}_{4})^{-1/4}]\right) ^{2}}{%
|\int d\theta \partial _{\theta }[\ ^{2}\mathbf{\breve{J}}(\zeta _{4}\breve{g%
}_{4})]|}]  \label{offdnceleps1} \\
&&+\epsilon \frac{16\ _{2}n_{k_{1}}\int d\theta \frac{\left( \partial
_{\theta }[(\zeta _{4}\breve{g}_{4})^{-1/4}]\right) ^{2}}{|\int d\theta
\partial _{\theta }[\ ^{2}\mathbf{\breve{J}}(\zeta _{4}\breve{g}_{4})]|}(%
\frac{\partial _{\theta }[(\zeta _{4}\breve{g}_{4})^{-1/4}\chi _{4})]}{%
2\partial _{\theta }[(\zeta _{4}\breve{g}_{4})^{-1/4}]}+\frac{\int d\theta
\partial _{\theta }[\ ^{2}\mathbf{\breve{J}}(\zeta _{4}\chi _{4}\breve{g}%
_{4})]}{\int d\theta \partial _{\theta }[\ ^{2}\mathbf{\breve{J}}(\zeta _{4}%
\breve{g}_{4})]})}{\ _{1}n_{k_{1}}+16\ _{2}n_{k_{1}}[\int d\theta \frac{%
\left( \partial _{\theta }[(\zeta _{4}\breve{g}_{4})^{-1/4}]\right) ^{2}}{%
|\int d\theta \partial _{\theta }[\ ^{2}\mathbf{\breve{J}}(\zeta _{4}\breve{g%
}_{4})]|}]}]\breve{N}_{k_{1}}^{4}dx^{k_{1}}\}^{2}  \notag
\end{eqnarray}%
\begin{eqnarray*}
&&+\zeta _{2s}(1+\epsilon \ \chi _{2s})\ \breve{g}_{2s}\{dv^{2s}+[(\breve{N}%
_{k_{2}}^{2s})^{-1}[\ _{1}n_{k_{s=1}}+16\ _{2}n_{k_{s-1}}[\int dv^{2s+1}%
\frac{\left( \partial _{2s+1}[(\zeta _{2s}\breve{g}_{2s})^{-1/4}]\right) ^{2}%
}{|\int dv^{2s+1}\partial _{2s+1}[\ ^{s}\mathbf{\breve{J}}(\zeta _{2s}\breve{%
g}_{2s})]|}] \\
&&+\epsilon \frac{16\ _{2}n_{k_{s}}\int dv^{2s+1}\frac{\left( \partial
_{2s+1}[(\zeta _{2s}\breve{g}_{2s})^{-1/4}]\right) ^{2}}{|\int
dv^{2s+1}\partial _{2s+1}[\ ^{s}\mathbf{\breve{J}}(\zeta _{2s}\breve{g}%
_{2s})]|}(\frac{\partial _{2s+1}[(\zeta _{2s}\breve{g}_{2s})^{-1/4}\chi
_{2s})]}{2\partial _{2s+1}[(\zeta _{2s}\breve{g}_{2s})^{-1/4}]}+\frac{\int
dv^{2s+1}\partial _{2s+1}[\ ^{s}\mathbf{\breve{J}}(\zeta _{2s}\chi _{2s}%
\breve{g}_{2s})]}{\int dv^{2s+1}\partial _{2s+1}[\ ^{s}\mathbf{\breve{J}}%
(\zeta _{2s}\breve{g}_{2s})]})}{\ _{1}n_{k_{s-1}}+16\ _{2}n_{k_{s-1}}[\int
dv^{2s+1}\frac{\left( \partial _{2s+1}[(\zeta _{2s}\breve{g}%
_{2s})^{-1/4}]\right) ^{2}}{|\int dv^{2s+1}\partial _{2s+1}[\ ^{s}\mathbf{%
\breve{J}}(\zeta _{2s}\breve{g}_{2s})]|}]}]\breve{N}%
_{k_{s-1}}^{a_{s}}dx^{k_{s-1}}\}^{2},
\end{eqnarray*}%
where $s=3,4.$ The polarization functions $\zeta _{4}(r,\varphi ,\theta
),\zeta _{2s}(r,\varphi ,\theta ,v^{2s-1})$ and $\chi _{4}(r,\varphi ,\theta
),\chi _{2s}(r,\varphi ,\theta ,v^{2s-1})$ in (\ref{offdnceleps1}) can be
prescribed to be of a necessary smooth class. Such a s-metric $\ _{\epsilon
Q}^{KdS}\mathbf{g}$ describes small $\epsilon $-parametric deformations of a
new KdS d-metric when the coefficients are additionally anisotropic on the $%
\varphi $-coordinate, or on other velocity-type coordinates.

We generate additional ellipsoidal deformations on $\theta ,v^{2s-1}$ using (%
\ref{offdnceleps1}) if we chose 
\begin{equation}
\chi _{4}(r,\varphi ,\theta )=\underline{\chi }_{4}(r,\varphi )\sin (\omega
_{0}\theta +\theta _{0}),\mbox{ and/or }\chi _{2s}(r,\varphi ,\theta
,v^{2s-1})=\underline{\chi }_{2s}(r,\varphi ,v^{2s-2})\sin (\omega _{\lbrack
0,2s-2]}v^{2s-1}+\theta _{\lbrack 0,2s-2]}).  \label{rotoid}
\end{equation}%
In these formulas, $\underline{\chi }_{4}(r,\varphi ),\underline{\chi }%
_{2s}(r,\varphi ,v^{2s-2})$ are smooth functions and $\omega _{0},\omega
_{\lbrack 0,2s-2]}$ and $\theta _{0},\theta _{\lbrack 0,2s-2]}$ are some
constants. For such generating polarization functions and $\zeta
_{4}(r,\varphi ,\theta )\neq 0,$ we obtain that 
\begin{equation*}
(1+\epsilon \ \chi _{4})\ \breve{g}_{4}\simeq a^{2}\sin ^{2}\theta
-\triangle _{\Lambda }+\epsilon \ \chi _{4}=0.
\end{equation*}%
For instance, for small $a$ and $\frac{\Lambda _{0}}{3},$ we can approximate 
$r=2M/(1+\theta _{0}\ \chi _{4}),$ which is a parametric equation for a
rotoid configuration. The parameter $\epsilon $ can be used as an
eccentricity parameter and generating function (\ref{rotoid}). 

We can prescribe polarization functions generating KdS BH embedded into a
nontrivial nonholonomic quasi-stationary background for FL MGTs. For small
ellipsoidal deformations of type (\ref{rotoid}), we model black ellipsoid,
BE, objects as generic off-diagonal solutions of the Einstein equations (for
projections on the base Lorentz manifold), or FLH deformed gravitational
field equations. 

\subsubsection{Double BE solutions in nonmetric FLH geometric flow and MGTs}

We can prescribe the nonholonomic s-structure in (\ref{offdnceleps1}) to
generate double BE solutions (the first BH one being a rotoid configurations
on base Lorentz manifold and the second one defined by other constants and
prescribing data in the typical fiber/velocity space. The coordinates $%
v^{5}=\ ^{r}v,v^{6}=\ ^{\varphi }v$ and $v^{7}=\ ^{\theta }v$ are considered
for respective angular velocities, when the quasi-stationary configurations
do not depend on $y^{4}=t$ and on $v^{8}=E$.

The primary data for such double phase space BH solutions are prescribed in
the form $\breve{N}_{i_{2}}^{7}=\breve{w}_{i}(r,\varphi ,\theta , \
^{\theta}v),$ which can be zero in certain rotation frames; and $\breve{N}%
_{i_{3}}^{8}=\breve{n}_{i_{3}}(r,\varphi ,\ ^{\theta }v)$ which may be with
a nontrivial 
\begin{equation*}
\breve{n}_{8}=-\ ^{v}a\sin \ ^{v}\theta (\ ^{v}r^{2}+\ ^{v}a^{2}-\
^{v}\triangle _{\Lambda })/(\ ^{v}a^{2}\sin ^{2}\ ^{v}\theta -\
^{v}\triangle _{\Lambda }).
\end{equation*}%
In such formulas, the left label $v$ states that the variables and constants
are considered on a typical fiber space with velocity variables. For such
primary double BH configurations and rotoid configurations similar to (\ref%
{rotoid}), the target quasi-stationary s-metrics (\ref{offdnceleps1})
describe double BE configurations.

We extend on phase space the primary metric (\ref{offdiagpm1}) and KdS BH
data (\ref{prdat1}) and (\ref{prdat2}), respectively, into 
\begin{eqnarray}
d\breve{s}^{2} &=&\breve{g}_{\alpha }(r,\varphi ,\theta ,\ ^{r}v,\ ^{\varphi
}v,\ ^{\theta }v)(\mathbf{\breve{e}}^{\alpha })^{2}  \label{offdiagpm1a} \\
&=&\breve{g}_{\alpha _{2}}(r,\varphi ,\theta )(\mathbf{\breve{e}}^{\alpha
_{2}})^{2}+\breve{g}_{a_{3}}(r,\varphi ,\theta ,\ ^{r}v,\ ^{\varphi }v,\
^{\theta }v)(\mathbf{\breve{e}}^{a_{3}})^{2}+\breve{g}_{a_{4}}(r,\varphi
,\theta ,\ ^{r}v,\ ^{\varphi }v,\ ^{\theta }v)(\mathbf{\breve{e}}^{a4})^{2},
\notag
\end{eqnarray}
nontrivial $s=3,4$ coefficients of the prime s-metric and N-connection are 
\begin{eqnarray*}
\breve{g}_{5} &=&\frac{\ ^{v}\breve{\rho}^{2}}{\ ^{v}\triangle _{\Lambda }},%
\breve{g}_{6}=\frac{\sin ^{2}\ ^{v}\theta }{\ ^{v}\breve{\rho}^{2}}[\
^{v}\Sigma _{\Lambda }-\frac{(\ ^{v}r^{2}+\ ^{v}a^{2}-\ ^{v}\triangle
_{\Lambda })^{2}}{\ ^{v}a^{2}\sin ^{2}\ ^{v}\theta -\ ^{v}\triangle
_{\Lambda }}],\breve{g}_{7}=\ ^{v}\breve{\rho}^{2},\breve{g}_{8}=\frac{\
^{v}a^{2}\sin ^{2}\ ^{v}\theta -\triangle _{\Lambda }}{\breve{\rho}^{2}},%
\mbox{ and } \\
\breve{N}_{6}^{8} &=&\ ^{4}\breve{n}_{6}=-\ ^{v}a\sin \ ^{v}\theta \frac{\
^{v}r^{2}+\ ^{v}a^{2}-\ ^{v}\triangle _{\Lambda }}{\ ^{v}a^{2}\sin ^{2}\
^{v}\theta -\ ^{v}\triangle _{\Lambda }},\mbox{ for }
\end{eqnarray*}
\begin{eqnarray*}
\ ^{v}\Sigma _{\Lambda } &=&(\ ^{v}r^{2}+\ ^{v}a^{2})^{2}-\ ^{v}\triangle
_{\Lambda }\ ^{v}a^{2}\sin ^{2}\theta ,\ ^{v}\triangle _{\Lambda }=\
^{v}r^{2}-2\ ^{v}M\ ^{v}r+\ ^{v}a^{2}-\frac{\ ^{v}\Lambda _{0}}{3}\
^{v}r^{4}, \\
\ ^{v}\breve{\rho}^{2} &=&\ ^{v}r^{2}+\ ^{v}a^{2}\cos ^{2}\ ^{v}\theta ,%
\mbox{ for constants  }\ ^{v}a=\ ^{v}J/\ ^{v}M=const.
\end{eqnarray*}%
In these formulas, we use a left label "v" stating that the constants are
respectively stated for a typical fiber space where $\ ^{v}J$ is the angular
momentum, $\ ^{v}M$ is the total mass of the system, and the cosmological
constant $\ ^{v}\Lambda _{0}>0.$

In general, a primary s-metric (\ref{offdiagpm1a}) is not a solution of some
FL modified Einstein equations even certain phase space gravitational field
equations can be postulated in holonomic variables, for which the $s=1,2$
and $s=3,4.$ But introducing above coefficients into (\ref{offdiagpm1}) with
s-adapted rotoid distributions (\ref{rotoid}), we generate double BE
solutions $\ _{\epsilon Q}^{2BE}\mathbf{g}(\tau )$ of FL distorted geometric
flow equations (\ref{cfeq4af}). In a similar form, we can construct $\tau $%
-families of double BE configurations in the framework of FH theories on $\
^{\shortmid }\mathcal{M}$.

\subsubsection{Perelman thermodynamic variables for general off-diagonal
deformed KdS BHs}

In general, FLH geometric flow and off-diagonal deformed KdS BH
configurations do not possess closed horizons and do not involve any
duality/ holographic properties. To characterize the physical properties of
the corresponding off-diagonal, we have to change the Bekenstein-Hawking
thermodynamic paradigm \cite{bek2,haw2} and (for respective quasi-stationary
configurations) use FLH geometric flow thermodynamic variables (\ref%
{thermvar2}). In explicit form, we have to compute the volume forms $\
_{\eta }^{\mathbf{J}}\mathcal{V[}\ _{Q}^{q}\mathbf{g}(\tau )]$ (\ref{volumf1}%
) when $\ _{Q}^{q}\mathbf{g}(\tau )$ is defined by corresponding $\tau $%
-families of quasi-stationary solutions. We can consider three types of
relativistic geometric flow thermodynamic models depending on introducing or
not small parameters:

\begin{enumerate}
\item[{a]}] $\ _{Q}^{q}\mathbf{g}(\tau )\rightarrow \ _{Q}^{KdS}\mathbf{g}%
(\tau )$ (\ref{nkernew}), when the gravitational $\eta $-polarizations are
defined by $\tau $-running s-metrics which allow us to compute the volume
functional $\ _{\eta }^{J}\mathcal{V}[\ \ ^{KdS}\mathbf{g}(\tau )],$ using
formulas (\ref{volumf1}).

\item[{b]}] $\ _{Q}^{q}\mathbf{g}(\tau )\rightarrow \ _{\epsilon Q}^{KdS}%
\mathbf{g}(\tau )$ (\ref{offdnceleps1}), when the gravitational $\chi $%
-polarizations are defined by $\tau $-running s-metrics involving a small
parameter $\epsilon .$ The corresponding volume functional $\ _{\chi }^{J}%
\mathcal{V}[\ _{\epsilon Q}^{KdS}\mathbf{g}(\tau )]$ can be computed in $%
\epsilon $-parametric form.

\item[{c]}] $\ _{Q}^{q}\mathbf{g}(\tau )\rightarrow \ _{\epsilon Q}^{BE}%
\mathbf{g}(\tau )$ when the prime data (\ref{offdiagpm1a}) are chosen to
define double BE configurations. We have possibilities to define and compute
a volume functional (\ref{volumf1}) on the 8-d phase space, $\ _{\chi }^{J}%
\mathcal{V}[\ _{\epsilon Q}^{2BE}\mathbf{g}(\tau )],$ and a similar
projection for geometric flows on base Lorentz manifold, $\ _{\chi }^{J}%
\mathcal{V}[\ _{\epsilon Q}^{2BE}h\mathbf{g}(\tau )],$ where the
h-projection is defined by $s=1,2.$ A similar volume form can be computed
for the typical fiber $\ _{\chi }^{J}\mathcal{V}[\ _{\epsilon Q}^{2BE}v%
\mathbf{g}(\tau )]$ with $s=3,4,$ which may result in a "nonstandard"
thermodynamic model for velocity type variables.
\end{enumerate}

For general $\eta $-polarizations as in a], the thermodynamic variables (\ref%
{thermvar2}) are defined and computed: 
\begin{eqnarray}
\ _{Q}^{KdS}\widehat{Z}(\tau ) &=&\exp \left[ \frac{1}{(4\pi \tau )^{4}}\ \
_{\eta }^{\mathbf{J}}\mathcal{V}[\ _{Q}^{KdS}\mathbf{g}(\tau )]\right] ,
\label{thermvar3} \\
\ _{Q}^{KdS}\widehat{\mathcal{E}}\ (\tau ) &=&\frac{1}{64\pi ^{4}\tau ^{3}}\
\left( 1-2\tau (\ _{Q}^{h}\Lambda (\tau )+\ _{Q}^{v}\Lambda (\tau ))\right)
\ \ _{\eta }^{\mathbf{J}}\mathcal{V}[\ _{Q}^{KdS}\mathbf{g}(\tau )],  \notag
\\
\ \ _{Q}^{KdS}\widehat{S}(\tau ) &=&-\ _{Q}^{q}\widehat{W}(\tau )=\frac{2}{%
(4\pi \tau )^{4}}(1-4(\ _{Q}^{h}\Lambda (\tau )+ \ _{Q}^{v}\Lambda (\tau
)))\ \ _{\eta }^{\mathbf{J}}\mathcal{V}[\ _{Q}^{KdS}\mathbf{g}(\tau )]. 
\notag
\end{eqnarray}%
Similar values can be computed using $\ _{\chi }^{J}\mathcal{V}[\ _{\epsilon
Q}^{KdS}\mathbf{g}(\tau )]$ or $\ _{\chi }^{J}\mathcal{V}[\ _{\epsilon
Q}^{2BE}\mathbf{g}(\tau )].$

We note that for h-flows defined in paragraph c], we have different formulas
as in 4-d MGTs \cite{perelman1,gheorghiuap16,bubuianu18,vv25,bsssvv25}: 
\begin{eqnarray}
\ \ _{\ \ \chi }^{J}\widehat{Z}(\tau ) &=&\exp \left[ \frac{1}{8\pi ^{2}\tau
^{2}}\ \ _{\chi }^{J}\mathcal{V[}\ \ \ _{\epsilon Q}^{2BE}h\mathbf{g}(\tau )]%
\right] ,\ \ _{\chi }^{J}\widehat{\mathcal{E}}\ (\tau )=\ \frac{1-2\tau \ \
_{Q}^{h}\Lambda (\tau )}{8\pi ^{2}\tau }\ \ _{\chi }^{J}\mathcal{V[}\ \ \
_{\epsilon Q}^{2BE}h\mathbf{g}(\tau )],  \notag \\
\ \ \ \ \ _{\chi }^{J}\widehat{S}(\tau ) &=&-\ \ _{\chi }^{J}\widehat{W}%
(\tau )=\frac{1-\ _{Q}^{h}\Lambda (\tau )}{4\pi ^{2}\tau ^{2}}\ _{\chi }^{J}%
\mathcal{V[}\ \ \ _{\epsilon Q}^{2BE}h\mathbf{g}(\tau )]\ .
\label{thermvarkds}
\end{eqnarray}%
The dependencies on temperature-like parameter $\tau $ are different for
thermodynamic values (\ref{thermvar3}) and (\ref{thermvarkds}). This can be
used for distinguishing different BH solutions in different FLH theories.

Both types of quasi-stationary configurations (\ref{thermvarkds}) are
described by a similar behaviour under a $\tau $-running cosmological
constants $\ _{Q}^{h}\Lambda (\tau )$ and$\ _{Q}^{v}\Lambda (\tau ).$ So,
such configurations can exist in an off-diagonal phase space background
determined by the FLH distribution with evolution on effective temperature $%
\tau .$ For a fixed $\tau _{0},$ we obtain thermodynamic models of certain
FLH Ricci soliton configurations. The class a] of thermodynamic models (\ref%
{thermvarkds}) is appropriate for describing general $\eta $-deformations
(for instance, for certain nonlinear waves and solitionic hierarchies) of
KdS BHs. In such models, a BH can be stable if certain stability conditions
are satisfied, see as a review \cite{vacaru18}, and references therein.

For $\epsilon $-parametric deformations of KdS BHs, for instance, as double
BE configurations, we can speculate on $\chi $-polarizations of the
gravitational vacuum, of some effective parameters, or of certain $\chi $%
-polarizations of physical constants. In more special cases, we can generate
rotoid deformations of horizons (\ref{rotoid}) and describe such FLH MGTs
systems alternatively the Bekenstein-Hawking thermodynamics. The G. Perelman
thermodynamic variables can be defined and computed in all cases. This is
possible even a KdS BHs can be unstable and "slow dissipate" on spacetime or
momentum like coordinates into another types of quasi-stationary solutions
with less known physical properties. For certain nonholonomic distributions,
the quasi-stationary solutions can be transformed into locally anisotropic
cosmological ones, and inversely. In many cases, we can prescribe certain
FLH distributions when BHs transform into BEs, with some polarized horizons
and effective constants. In such cases, the physical interpretation of
off-diagonal target solutions is quite similar to the prime BH ones.
Corresponding relativistic FLH geometric flow, or nonholononmic Ricci flow
models depend on the type of volume form (\ref{volumf1}) which can be
computed exactly for respective generating and integrating data. 

\subsubsection{Discussion of alternative BH solutions in Finsler-like MGTs}

In paragraph 7.4] of section \ref{sec1}, we discussed the direction related
to constructing BH solutions in Finsler-like gravity theories. A series of
recent works \cite{nekouee23,nekouee24,praveen25} contain numeric and
graphic results on BH and WH solutions constructed as lifts with dependence
on anisotropic $y$-coordinates using the models of Finsler gravity \cite%
{xi14} and the definition of a (incomplete) variant of Finsler-Ricci tensors
proposed in \cite{akbar88,akbar95}. Those constructions and found solutions
have not generated in a rigorous mathematical form on (co) tangent Lorentz
bundles and do not involve Sasaki-type lifts \cite{akbar88,akbar95} to
d-metric structures adapted to an N-connection structure. The issues of
general covariance and distortions of connections in Finsler gravity
theories were not studied in those models, depending on the type of Finsler
generating function and Finsler connections. A series of more general former
results on nonholonomic off-diagonal and Finsler BH and WH were not cited
and not discussed using geometric approaches and the AFCDM \cite%
{v01t,v01q,vmon3,v13,v14,vacaru00,vacaru09b,vacaru12a,gheorghiuap16,biv16,vbubuianu17,vacaru18}%
. In priciple, using arbitrary lifts and extensions on $y$-coordinates we
can deform on such variables and solution in GR and model various types of
anisotropic cosmological (DM and DE) structures and exotic astrophysical
objects.

Nevertheless, the numerical and graphical results from \cite%
{nekouee23,nekouee24,praveen25} can be included as certain important
physical examples of FLH deformed geometric flow and Einstein equations
studied in rigorous mathematical form in this work. For instance, we can
prescribe such generating functions and generating sources, and respective
integrating functions (\ref{integrfunctrffh}) for s-metrics (\ref{qeltors})
which reproduce certain formulas and graphs provided by other authors. In
our approach, the solutions are for FLH geometric flow and MGTs extending in
a self-consistent form certain BH and WH in GR. New classes of generic
off-diagonal FLH BH, BE, WH and other type solutions are encoded in
nonassociative and noncommutative versions, and nonmetric extensions in our
recent works \cite{bsv22,bsv23,partner06,bnsvv24,vv25,vacaru25b}. 

\subsection{Off-diagonal FLH geometric flow deformed WHs}

Nonholonomic deformations of WH solutions \cite{morris88,bron20} to locally
anisotropic quasi-stationary configurations were studied in \cite{v13,v14},
with generalizations to MGTs \cite{vacaru18,partner06,bsssvv25}. In
paragraph 7.4] of section \ref{sec1}, we discussed some new WH solutions in
Finsler-like gravity theories \cite{nekouee23,nekouee24,praveen25}. In this
subsection, we generate new classes of off-diagonal quasi-stationary
solutions of FL modified geometric flow and Einstein equations (\ref{cfeq4af}%
). For constructing explicit solutions and computing thermodynamic
variables, it is convenient to use gravitational polarizations as in (\ref%
{offdiagpolfr}) (in this case, of some primary WH metrics).

\subsubsection{FL quasi-stationary gravitational polarizations of WHs in GR
and lifts to phase spaces}

We begin with the main formulas defining the generic Morris-Thorne WH 4-d
solution \cite{morris88}: 
\begin{equation*}
d\mathring{s}^{2}=(1-\frac{b(r)}{r})^{-1}dr^{2}+r^{2}d\theta ^{2}+r^{2}\sin
^{2}\theta d\varphi ^{2}-e^{2\Phi (r)}dt^{2},
\end{equation*}%
In this diagonal metric, $e^{2\Phi (r)}$ is a red-shift function and $b(r)$
is the shape function defined in spherically polar coordinates $u^{\alpha
_{2}}=(r,\theta ,\varphi ,t).$ A usual Ellis-Bronnikov, EB, WH is defined
for $\Phi (r)=0$ and $b(r)=\ _{0}b^{2}/r$ which state a zero tidal WH with $%
\ _{0}b$ the throat radius. We cite \cite{kar94,roy20,souza22,bsssvv25} for
details and recent reviews, including nonassociative and noncommutative FLH
WHs. A generalized EB configuration is characterized by considering even
integers $2k$ (with $k=1,2,...$), where $r(l)=(l^{2k}+\ _{0}b^{2k})^{1/2k}$
is a proper radial distance (tortoise coordinate) and the cylindrical
angular coordinate is $\phi \in \lbrack 0,2\pi ).$ In such coordinates, $%
-\infty <l<\infty $ a prime metric can be written as 
\begin{equation*}
d\mathring{s}^{2}=dl^{2}+r^{2}(l)d\theta ^{2}+r^{2}(l)\sin ^{2}\theta
d\varphi ^{2}-dt^{2},
\end{equation*}%
when $dl^{2}=(1-\frac{b(r)}{r})^{-1}dr^{2}$ and $b(r)=r-r^{3(1-k)}(r^{2k}-\
_{0}b^{2k})^{(2-1/k))}.$ We can perform some frame transforms to a
parametrization with trivial N-connection coefficients $\check{N}%
_{i_{1}}^{a_{2}}=$ $\check{N}_{i_{1}}^{a_{2}}(u^{\alpha _{2}}(l,\theta
,\varphi ,t))$ and $\check{g}_{\beta _{2}}(u^{j}(l,\theta ,\varphi
),u^{3}(l,\theta ,\varphi )),$ which allows us to avoid off-diagonal
deformations with singularities. On a 4-d base spacetime Lorentz manifold,
we can introduce new coordinates $u^{1}=x^{1}=l,u^{2}=\theta ,$ and $%
u^{3}=y^{3}=\varphi +\ ^{3}B(l,\theta ),u^{4}=y^{4}=t+\ ^{4}B(l,\theta ),$
when 
\begin{eqnarray*}
\mathbf{\check{e}}^{3} &=&d\varphi =du^{3}+\check{N}_{i}^{3}(l,\theta
)dx^{i}=du^{3}+\check{N}_{1}^{3}(l,\theta )dl+\check{N}_{2}^{3}(l,\theta
)d\theta , \\
\mathbf{\check{e}}^{4} &=&dt=du^{4}+\check{N}_{i}^{4}(l,\theta
)dx^{i}=du^{4}+\check{N}_{1}^{4}(l,\theta )dl+\check{N}_{2}^{4}(l,\theta
)d\theta ,
\end{eqnarray*}%
are defined for $\mathring{N}_{i}^{3}=-\partial \ ^{3}B/\partial x^{i}$ and $%
\mathring{N}_{i}^{4}=-\partial \ ^{4}B/\partial x^{i}.$ So, the quadratic
line elements for WH solutions can be parameterized as a prime d-metric, 
\begin{equation}
d\mathring{s}^{2}=\check{g}_{\alpha _{2}}(l,\theta ,\varphi )[\mathbf{\check{%
e}}^{\alpha _{2}}(l,\theta ,\varphi )]^{2},  \label{pmwh}
\end{equation}%
where $\check{g}_{1}=1,\check{g}_{2}=r^{2}(l),\check{g}_{3}=r^{2}(l)\sin
^{2}\theta $ and $\check{g}_{4}=-1.$

On a nonmetric 8-d phase space $\ _{Q}^{s}\mathcal{M}$, a (\ref{pmwh}) can
be extended as 
\begin{eqnarray}
d\mathring{s}^{2} &=&\check{g}_{\alpha _{2}}(l,\theta ,\varphi )[\mathbf{%
\check{e}}^{\alpha _{2}}(l,\theta ,\varphi )]^{2}  \label{pmwh8} \\
&&+\check{g}_{a_{3}}(\ ^{v}l,\ ^{v}\theta ,\ ^{v}\varphi )[\mathbf{\check{e}}%
^{a_{3}}(\ ^{v}l,\ ^{v}\theta ,\ ^{v}\varphi )]^{2}+\check{g}_{a_{4}}(\
^{v}l,\ ^{v}\theta ,\ ^{v}\varphi )[\mathbf{\check{e}}^{a_{4}}(\ ^{v}l,\
^{v}\theta ,\ ^{v}\varphi )]^{2},  \notag
\end{eqnarray}%
where the velocity type coordinates $\ v^{a_{3}}=(\ ^{v}l,\ ^{v}\theta )$
and $\ v^{a_{4}}=(\ ^{v}\varphi ,E).$ We can consider two physically
important sets of s-coefficients in (\ref{pmwh8}):%
\begin{eqnarray}
\mbox{horizontal prime WH with flat fiber} &:&\check{g}_{5}=1,\check{g}%
_{6}=1,\check{g}_{7}=1,\check{g}_{8}=-1;  \label{primdata1} \\
\mbox{prime h WH and v WH} &:&\check{g}_{5}=1,\check{g}_{6}=\ ^{v}r^{2}(\
^{v}l),\check{g}_{7}=\ ^{v}r^{2}(\ ^{v}l)\sin ^{2}\ ^{v}\theta ,\check{g}%
_{8}=-1.  \label{primdata2}
\end{eqnarray}%
Here we note that such prime s-metrics may be not solutions of (\ref{cfeq4af}%
) if we do not consider phase space LC-configurations.

We can perform geometric flow off-diagonal quasi-stationary deformations of
WHs (\ref{pmwh8}) by introducing nontrivial sources $\ _{Q}^{s}\mathbf{J}%
(\tau )$ (\ref{dsourcparam}) with%
\begin{eqnarray*}
\ _{Q}^{1}\mathbf{J}(\tau ,l,\theta ) &=&\ _{Q}^{1wh}\mathbf{J}(\tau ),\
_{Q}^{2}\mathbf{J}(\tau ,l,\theta ,\varphi )=\ \ \ _{Q}^{2wh}\mathbf{J}(\tau
), \\
\ _{Q}^{3}\mathbf{J}(\tau ,l,\theta ,\varphi ,\ ^{v}l,\ ^{v}\theta ) &=&\ \
\ _{Q}^{3wh}\mathbf{J}(\tau ),\ _{Q}^{4}\mathbf{J}(\tau ,l,\theta ,\varphi
,\ ^{v}l,\ ^{v}\theta ,\ ^{v}\varphi )=\ \ \ _{Q}^{4wh}\mathbf{J}(\tau ).
\end{eqnarray*}%
related via nonlinear symmetries (\ref{nonlinsymrex}) to (effective) $\tau $%
-running cosmological constants $\ ^{s}\Lambda (\tau ).$ Using gravitational 
$\eta $-polarization functions, we construct such $\tau $-families of target
quasi-stationary metrics $\ _{Q\eta }^{wh}\mathbf{g}(\tau )$ and respective
quadratic elements: 
\begin{eqnarray}
d\widehat{s}^{2}(\tau ) &=&\widehat{g}_{\alpha _{s}\beta _{s}}(\tau
,l,\theta ,\varphi ,\ ^{v}l,\ ^{v}\theta ,\ ^{v}\varphi ;\psi ,\eta _{2s};\
\ ^{s}\Lambda (\tau ),\ _{Q}^{swh}\mathbf{J}(\tau ),\ \check{g}_{\alpha
_{s}})du^{\alpha _{s}}du^{\beta _{s}}  \notag \\
&=&e^{\psi (\tau ,l,\theta ,\ _{Q}^{1wh}\mathbf{J}(\tau ))}[(dx^{1}(l,\theta
))^{2}+(dx^{2}(l,\theta ))^{2}]  \label{whpolf} \\
&&-\frac{[\partial _{\varphi }(\eta _{4}\ \check{g}_{4})]^{2}}{|\int
d\varphi \ \ _{Q}^{2wh}\mathbf{J}(\tau )\partial _{\varphi }(\eta _{4}\ 
\breve{g}_{4})|\ \eta _{4}\ \check{g}_{4}}\{d\varphi +\frac{\partial
_{i_{1}}[\int d\varphi \ \ _{Q}^{2wh}\mathbf{J}(\tau )\ \partial _{\varphi
}(\eta _{4}\ \check{g}_{4})]}{\ _{Q}^{2wh}\mathbf{J}(\tau )\partial
_{\varphi }(\eta _{4}\ \check{g}_{4})}dx^{i_{1}}\}^{2}  \notag \\
&&+\eta _{4}\breve{g}_{4}\{dt+[\ _{1}n_{k_{1}}(l,\theta )+\
_{2}n_{k_{1}}(l,\theta )\int d\varphi \frac{\lbrack \partial _{\varphi
}(\eta _{4}\ \breve{g}_{4})]^{2}}{|\int d\varphi \ \ _{Q}^{2wh}\mathbf{J}%
(\tau )\partial _{\varphi }(\eta _{4}\ \breve{g}_{4})|\ (\eta _{4}\ \breve{g}%
_{4})^{5/2}}]dx^{k_{1}}\}  \notag
\end{eqnarray}%
\begin{eqnarray*}
&&-\frac{[\partial _{5}(\eta _{6}\ \check{g}_{6})]^{2}}{|\int d\ ^{v}l\ \
_{Q}^{3wh}\mathbf{J}(\tau )\partial _{5}(\eta _{6}\ \breve{g}_{6})|\ \eta
_{6}\ \check{g}_{6}}\{d\ ^{v}l+\frac{\partial _{i_{2}}[\int d\ ^{v}l\ \
_{Q}^{3wh}\mathbf{J}(\tau )\ \partial _{5}(\eta _{6}\ \check{g}_{6})]}{\
_{Q}^{3wh}\mathbf{J}(\tau )\partial _{5}(\eta _{6}\ \check{g}_{6})}%
dx^{i_{2}}\}^{2} \\
&&+\eta _{6}\breve{g}_{6}\{d\ ^{v}\theta +[\ _{1}n_{k_{2}}(\tau ,l,\theta
,\varphi )+\ _{2}n_{k_{2}}(\tau ,l,\theta ,\varphi )\int d\ ^{v}l\frac{%
[\partial _{5}(\eta _{6}\ \breve{g}_{6})]^{2}}{|\int d\ ^{v}l\ \ _{Q}^{3wh}%
\mathbf{J}(\tau )\partial _{5}(\eta _{6}\ \breve{g}_{6})|\ (\eta _{6}\ 
\breve{g}_{6})^{5/2}}]dx^{k_{2}}\}
\end{eqnarray*}%
\begin{eqnarray*}
&&-\frac{[\partial _{7}(\eta _{8}\ \check{g}_{8})]^{2}}{|\int d\ ^{v}\varphi
\ \ _{Q}^{4wh}\mathbf{J}(\tau )\partial _{7}(\eta _{8}\ \breve{g}_{8})|\
\eta _{8}\ \check{g}_{8}}\{d\ ^{v}\varphi +\frac{\partial _{i_{3}}[\int d\
^{v}\varphi \ \ _{Q}^{4wh}\mathbf{J}(\tau )\ \partial _{7}(\eta _{8}\ \check{%
g}_{8})]}{\ _{Q}^{4wh}\mathbf{J}(\tau )\partial _{7}(\eta _{8}\ \check{g}%
_{8})}dx^{i_{3}}\}^{2}+\eta _{8}\breve{g}_{8}\{dE \\
&&+[\ _{1}n_{k_{3}}(\tau ,l,\theta ,\varphi ,\ ^{v}l,\ ^{v}\theta )+\
_{2}n_{k_{3}}(\tau ,l,\theta ,\varphi ,\ ^{v}l,\ ^{v}\theta )\int d\
^{v}\varphi \frac{\lbrack \partial _{7}(\eta _{8}\ \breve{g}_{8})]^{2}}{%
|\int d\ ^{v}\varphi \ \ _{Q}^{4wh}\mathbf{J}(\tau )\partial _{7}(\eta _{8}\ 
\breve{g}_{8})|\ (\eta _{8}\ \breve{g}_{8})^{5/2}}]dx^{k_{3}}\}.
\end{eqnarray*}%
Such parameterizations can be considered in various FLH-theories and
projections to 4-d MGTs and GR, when the physical interpretation is
different because of different effective and matter sources. This
quasi-stationary solutions (\ref{whpolf}) \ are determined by three
generating functions $\eta _{s}(\tau )=\eta _{s}(\tau ,l,\theta ,\varphi ,\
^{v}l,\ ^{v}\theta ,\ ^{v}\varphi )$ and integration functions $\
_{1}n_{k_{s}}(\tau )$ and $\ _{2}n_{k_{s}}(\tau ).$ The functions $\psi
(\tau ,l,\theta )$ are defined as solutions of 2-d Poisson equation $%
\partial _{11}^{2}\psi (\tau )+\partial _{22}^{2}\psi (\tau )=2\ _{Q}^{1wh}%
\mathbf{J}(\tau ,l,\theta ).$

Finally, we emphasize that the target s-metrics (\ref{whpolf}) do not
describe $\tau $-evolution of exact WH-like s-objects for general FLH
deforms and general classes of generating and integrating data. General $%
\tau $-flows and off-diagonal deformations may "annihilate or dissipate" a
WH spacetime object or certain double WH configurations for the base and
typical fiber subspaces. The above formulas can be written in corresponding
momentum variables on phase spaces $\ _{Q}^{s\shortmid }\mathcal{M}.$

\subsubsection{Small parametric FLH quasi-stationary deformations of WH
d-metrics}

We can define locally anisotropic FLH (double) WH configurations encoding
nonmetric data if we consider for small parametric geometric flow of
off-diagonal deformations of prime metrics of type (\ref{pmwh8}). In terms
of $\chi $-polarization functions in $_{Q\chi }^{wh}\mathbf{g}(\tau )$, the
quadratic linear elements are computed 
\begin{eqnarray*}
d\ \widehat{s}^{2}(\tau ) &=&\widehat{g}_{\alpha _{s}\beta _{s}}(\tau
,l,\theta ,\varphi ,\ ^{v}l,\ ^{v}\theta ,\ ^{v}\varphi ;\psi ,\chi _{2s};\
\ ^{s}\Lambda (\tau ),\ _{Q}^{swh}\mathbf{J}(\tau ),\ \check{g}_{\alpha
_{s}})du^{\alpha _{s}}du^{\beta _{s}} \\
&=&e^{\psi _{0}(\tau ,l,\theta )}[1+\epsilon \ ^{\psi (\tau ,l,\theta )}\chi
(\tau ,l,\theta )][(dx^{1}(l,\theta ))^{2}+(dx^{2}(l,\theta ))^{2}]
\end{eqnarray*}%
\begin{eqnarray*}
&&-\{\frac{4[\partial _{\varphi }(|\zeta _{4}\ \breve{g}_{4}|^{1/2})]^{2}}{%
\breve{g}_{3}|\int d\varphi \{\ \ _{Q}^{2wh}\mathbf{J}(\tau )\partial
_{\varphi }(\zeta _{4}\ \breve{g}_{4})\}|}-\epsilon \lbrack \frac{\partial
_{\varphi }(\chi _{4}|\zeta _{4}\breve{g}_{4}|^{1/2})}{4\partial _{\varphi
}(|\zeta _{4}\ \breve{g}_{4}|^{1/2})}-\frac{\int d\varphi \{\ \ _{Q}^{2wh}%
\mathbf{J}(\tau )\partial _{\varphi }[(\zeta _{4}\ \breve{g}_{4})\chi _{4}]\}%
}{\int d\varphi \{\ \ _{Q}^{2wh}\mathbf{J}(\tau )\partial _{\varphi }(\zeta
_{4}\ \breve{g}_{4})\}}]\}\ \breve{g}_{3} \\
&&\{d\varphi +[\frac{\partial _{i_{1}}\ \int d\varphi \ \ _{Q}^{2wh}\mathbf{J%
}(\tau )\ \partial _{\varphi }\zeta _{4}}{(\check{N}_{i_{1}}^{3})\ \ \
_{Q}^{2wh}\mathbf{J}(\tau )\partial _{\varphi }\zeta _{4}}+\epsilon (\frac{%
\partial _{i_{1}}[\int d\varphi \ \ \ _{Q}^{2wh}\mathbf{J}(\tau )\ \partial
_{\varphi }(\zeta _{4}\chi _{4})]}{\partial _{i_{1}}\ [\int d\varphi \ \ \
_{Q}^{2wh}\mathbf{J}(\tau )\partial _{\varphi }\zeta _{4}]}-\frac{\partial
_{\varphi }(\zeta _{4}\chi _{4})}{\partial _{\varphi }\zeta _{4}})]\check{N}%
_{i_{1}}^{3}dx^{i_{1}}\}^{2}
\end{eqnarray*}%
\begin{eqnarray}
&&+\zeta _{4}(1+\epsilon \ \chi _{4})\ \breve{g}_{4}\{dt+[(\check{N}%
_{k_{1}}^{4})^{-1}[\ _{1}n_{k_{1}}+16\ _{2}n_{k_{1}}[\int d\varphi \frac{%
\left( \partial _{\varphi }[(\zeta _{4}\ \breve{g}_{4})^{-1/4}]\right) ^{2}}{%
|\int d\varphi \partial _{\varphi }[\ \ \ _{Q}^{2wh}\mathbf{J}(\tau )(\zeta
_{4}\ \breve{g}_{4})]|}]  \notag \\
&&+\epsilon \frac{16\ _{2}n_{k_{1}}\int d\varphi \frac{\left( \partial
_{\varphi }[(\zeta _{4}\ \breve{g}_{4})^{-1/4}]\right) ^{2}}{|\int d\varphi
\partial _{\varphi }[\ \ \ _{Q}^{2wh}\mathbf{J}(\tau )(\zeta _{4}\ \breve{g}%
_{4})]|}(\frac{\partial _{\varphi }[(\zeta _{4}\ \breve{g}_{4})^{-1/4}\chi
_{4})]}{2\partial _{\varphi }[(\zeta _{4}\ ^{cy}g)^{-1/4}]}+\frac{\int
d\varphi \partial _{\varphi }[\ \ _{Q}^{2wh}\mathbf{J}(\tau )(\zeta _{4}\chi
_{4}\ \breve{g}_{4})]}{\int d\varphi \partial _{\varphi }[\ \ _{Q}^{2wh}%
\mathbf{J}(\tau )(\zeta _{4}\ \breve{g}_{4})]})}{\ _{1}n_{k_{1}}+16\
_{2}n_{k_{1}}[\int d\varphi \frac{\left( \partial _{\varphi }[(\zeta _{4}\ 
\breve{g}_{4})^{-1/4}]\right) ^{2}}{|\int d\varphi \partial _{\varphi }[\ \
\ _{Q}^{2wh}\mathbf{J}(\tau )(\zeta _{4}\ \breve{g}_{4})]|}]}]\check{N}%
_{k_{1}}^{4}dx^{k_{1}}\}^{2}  \label{whpolf1}
\end{eqnarray}%
\begin{eqnarray*}
&&-\{\frac{4[\partial _{5}(|\zeta _{6}\ \breve{g}_{6}|^{1/2})]^{2}}{\breve{g}%
_{5}|\int d\ ^{v}l\{\ \ _{Q}^{3wh}\mathbf{J}(\tau )\partial _{5}(\zeta _{6}\ 
\breve{g}_{6})\}|}-\epsilon \lbrack \frac{\partial _{5}(\chi _{6}|\zeta _{6}%
\breve{g}_{6}|^{1/2})}{4\partial _{5}(|\zeta _{6}\ \breve{g}_{6}|^{1/2})}-%
\frac{\int d\ ^{v}l\{\ \ _{Q}^{3wh}\mathbf{J}(\tau )\partial _{5}[(\zeta
_{6}\ \breve{g}_{6})\chi _{6}]\}}{\int d\ ^{v}l\{\ \ _{Q}^{3wh}\mathbf{J}%
(\tau )\partial _{5}(\zeta _{6}\ \breve{g}_{6})\}}]\}\ \breve{g}_{5} \\
&&\{d\ ^{v}l+[\frac{\partial _{i_{2}}\ \int d\ ^{v}l\ \ _{Q}^{3wh}\mathbf{J}%
(\tau )\ \partial _{5}\zeta _{6}}{(\check{N}_{i}^{5})\ \ \ _{Q}^{3wh}\mathbf{%
J}(\tau )\partial _{5}\zeta _{6}}+\epsilon (\frac{\partial _{i_{2}}[\int d\
^{v}l\ \ \ _{Q}^{3wh}\mathbf{J}(\tau )\ \partial _{5}(\zeta _{6}\chi _{6})]}{%
\partial _{i_{2}}\ [\int d\ ^{v}l\ \ \ _{Q}^{3wh}\mathbf{J}(\tau )\partial
_{5}\zeta _{6}]}-\frac{\partial _{5}(\zeta _{6}\chi _{6})}{\partial
_{5}\zeta _{6}})]\check{N}_{i_{2}}^{5}dx^{i_{2}}\}^{2}
\end{eqnarray*}%
\begin{eqnarray*}
&&+\zeta _{6}(1+\epsilon \ \chi _{6})\ \breve{g}_{6}\{d\ ^{v}\theta +[(%
\check{N}_{k_{2}}^{6})^{-1}[\ _{1}n_{k_{2}}+16\ _{2}n_{k_{2}}[\int d\ ^{v}l%
\frac{\left( \partial _{5}[(\zeta _{6}\ \breve{g}_{6})^{-1/4}]\right) ^{2}}{%
|\int d\ ^{v}l\partial _{5}[\ \ \ _{Q}^{3wh}\mathbf{J}(\tau )(\zeta _{6}\ 
\breve{g}_{6})]|}] \\
&&+\epsilon \frac{16\ _{2}n_{k_{2}}\int d\ ^{v}l\frac{\left( \partial
_{5}[(\zeta _{6}\ \breve{g}_{6})^{-1/4}]\right) ^{2}}{|\int d\ ^{v}l\partial
_{5}[\ \ \ _{Q}^{3wh}\mathbf{J}(\tau )(\zeta _{6}\ \breve{g}_{6})]|}(\frac{%
\partial _{5}[(\zeta _{6}\ \breve{g}_{6})^{-1/4}\chi _{6})]}{2\partial
_{5}[(\zeta _{6}\ \ \breve{g}_{6})^{-1/4}]}+\frac{\int d\ ^{v}l\partial
_{5}[\ \ _{Q}^{3wh}\mathbf{J}(\tau )(\zeta _{6}\chi _{6}\ \breve{g}_{6})]}{%
\int d\ ^{v}l\partial _{5}[\ \ _{Q}^{3wh}\mathbf{J}(\tau )(\zeta _{6}\ 
\breve{g}_{6})]})}{\ _{1}n_{k_{2}}+16\ _{2}n_{k_{2}}[\int d\ ^{v}l\frac{%
\left( \partial _{5}[(\zeta _{6}\ \breve{g}_{6})^{-1/4}]\right) ^{2}}{|\int
d\ ^{v}l\partial _{5}[\ \ \ _{Q}^{3wh}\mathbf{J}(\tau )(\zeta _{6}\ \breve{g}%
_{6})]|}]}]\check{N}_{k_{2}}^{6}dx^{k_{2}}\}^{2}
\end{eqnarray*}%
\begin{eqnarray*}
&&-\{\frac{4[\partial _{7}(|\zeta _{8}\ \breve{g}_{8}|^{1/2})]^{2}}{\breve{g}%
_{7}|\int d\ ^{v}\varphi \{\ \ _{Q}^{4wh}\mathbf{J}(\tau )\partial
_{7}(\zeta _{8}\ \breve{g}_{8})\}|}-\epsilon \lbrack \frac{\partial
_{7}(\chi _{8}|\zeta _{8}\breve{g}_{8}|^{1/2})}{4\partial _{7}(|\zeta _{8}\ 
\breve{g}_{8}|^{1/2})}-\frac{\int d\ ^{v}\varphi \{\ \ _{Q}^{4wh}\mathbf{J}%
(\tau )\partial _{7}[(\zeta _{8}\ \breve{g}_{8})\chi _{8}]\}}{\int d\
^{v}\varphi \{\ \ _{Q}^{4wh}\mathbf{J}(\tau )\partial _{7}(\zeta _{8}\ 
\breve{g}_{8})\}}]\}\ \breve{g}_{7} \\
&&\{d\ ^{v}\varphi +[\frac{\partial _{i_{3}}\ \int d\ ^{v}\varphi \ \
_{Q}^{4wh}\mathbf{J}(\tau )\ \partial _{7}\zeta _{8}}{(\check{N}_{i}^{7})\ \
\ _{Q}^{4wh}\mathbf{J}(\tau )\partial _{7}\zeta _{8}}+\epsilon (\frac{%
\partial _{i_{3}}[\int d\ ^{v}\varphi \ \ \ _{Q}^{4wh}\mathbf{J}(\tau )\
\partial _{7}(\zeta _{8}\chi _{8})]}{\partial _{i_{3}}\ [\int d\ ^{v}\varphi
\ \ \ _{Q}^{4wh}\mathbf{J}(\tau )\partial _{7}\zeta _{8}]}-\frac{\partial
_{7}(\zeta _{8}\chi _{8})}{\partial _{7}\zeta _{8}})]\check{N}%
_{i_{3}}^{7}dx^{i_{3}}\}^{2}
\end{eqnarray*}%
\begin{eqnarray*}
&&+\zeta _{8}(1+\epsilon \ \chi _{8})\ \breve{g}_{8}\{dE+[(\check{N}%
_{k_{3}}^{8})^{-1}[\ _{1}n_{k_{3}}+16\ _{2}n_{k_{3}}[\int d\ ^{v}\varphi 
\frac{\left( \partial _{7}[(\zeta _{8}\ \breve{g}_{8})^{-1/4}]\right) ^{2}}{%
|\int d\ ^{v}\varphi \partial _{7}[\ \ \ _{Q}^{4wh}\mathbf{J}(\tau )(\zeta
_{8}\ \breve{g}_{8})]|}] \\
&&+\epsilon \frac{16\ _{2}n_{k_{3}}\int d\ ^{v}\varphi \frac{\left( \partial
_{7}[(\zeta _{8}\ \breve{g}_{8})^{-1/4}]\right) ^{2}}{|\int d\ ^{v}\varphi
\partial _{7}[\ \ \ _{Q}^{4wh}\mathbf{J}(\tau )(\zeta _{8}\ \breve{g}_{8})]|}%
(\frac{\partial _{7}[(\zeta _{8}\ \breve{g}_{8})^{-1/4}\chi _{8})]}{%
2\partial _{7}[(\zeta _{8}\ \ \breve{g}_{8})^{-1/4}]}+\frac{\int d\
^{v}\varphi \partial _{7}[\ \ _{Q}^{4wh}\mathbf{J}(\tau )(\zeta _{8}\chi
_{8}\ \breve{g}_{8})]}{\int d\ ^{v}\varphi \partial _{7}[\ \ _{Q}^{4wh}%
\mathbf{J}(\tau )(\zeta _{8}\ \breve{g}_{8})]})}{\ _{1}n_{k_{3}}+16\
_{2}n_{k_{3}}[\int d\ ^{v}\varphi \frac{\left( \partial _{7}[(\zeta _{8}\ 
\breve{g}_{8})^{-1/4}]\right) ^{2}}{|\int d\ ^{v}\varphi \partial _{7}[\ \ \
_{Q}^{4wh}\mathbf{J}(\tau )(\zeta _{8}\ \breve{g}_{8})]|}]}]\check{N}%
_{k_{3}}^{8}dx^{k_{3}}\}^{2}.
\end{eqnarray*}

In detail, analogues of the formula (\ref{whpolf1}) was derived in part I of 
\cite{bsssvv25}, for 4-d nonholonomic configurations and generalized for
nonassociative 8-d FLH WHs in the part II of that work. In this subsection,
we analyze 8-d configurations which extend 4-d off-diagonal modifications of
GR to phase spaces. We can model elliptic deformations of WHs in GR and 4-d
MGTs as particular cases of target d-metrics determined by generating
functions of type $\chi _{4}(l,\theta,\varphi )=\underline{\chi }(l,\theta
)\sin (\omega _{0}\varphi +\varphi _{0}).$ These are cylindric-elliptic
configurations with $\varphi $-anisotropy. In a similar way, we can generate
elliptic configurations on (co) fibers using respective cylindrical
variables associated to velocity/momentum coordinates.

In geometric symbolic form, we can generate doulbe 4d+4d FLH WH metics $\
_{Q\chi }^{2wh}\mathbf{g}(\tau ),$ defined by prime s-metric data (\ref%
{primdata2}) used in (\ref{whpolf1}). Other types of FLH WH configurations
can be generated by (\ref{primdata1}) when a 4-d spacetime WH is extended to
a total 8-d phase space in certain forms when h-projections encode certain
nonmetricity FLH data.

\subsubsection{G. Perelman thermodynamic variables for FLH off-diagonal
deformed WHs}

For general $\eta $-polarizations, the thermodynamic variables (\ref%
{thermvar2}) are defined and computed: 
\begin{eqnarray}
\ _{Q}^{wh}\widehat{Z}(\tau ) &=&\exp \left[ \frac{1}{(4\pi \tau )^{4}}\ \
_{\eta }^{\mathbf{J}}\mathcal{V}[\ \ _{Q}^{wh}\mathbf{g}(\tau )]\right] ,
\label{thermvar3wh} \\
\ _{Q}^{wh}\widehat{\mathcal{E}}\ (\tau ) &=&\frac{1}{64\pi ^{4}\tau ^{3}}\
\left( 1-2\tau (\ _{Q}^{h}\Lambda (\tau )+\ _{Q}^{v}\Lambda (\tau ))\right)
\ \ \ _{\eta }^{\mathbf{J}}\mathcal{V}[\ \ _{Q}^{wh}\mathbf{g}(\tau )], 
\notag \\
\ \ \ \ _{Q}^{wh}\widehat{S}(\tau ) &=&-\ _{Q}^{wh}\widehat{W}(\tau )=\frac{2%
}{(4\pi \tau )^{4}}(1-4(\ _{Q}^{h}\Lambda (\tau )+\ _{Q}^{v}\Lambda (\tau
)))\ \ _{\eta }^{\mathbf{J}}\mathcal{V[}\ \ _{Q}^{wh}\mathbf{g}(\tau )]. 
\notag
\end{eqnarray}%
Similar values can be computed using $\ _{\chi }^{J}\mathcal{V}[\ _{\epsilon
Q}^{wh}\mathbf{g}(\tau )]$ or $\ _{\chi }^{J}\mathcal{V}[\ \ _{\epsilon
Q}^{2wh}\mathbf{g}(\tau )].$

We note that for h-flows, we have different formulas as in 4-d MGTs \cite%
{perelman1,gheorghiuap16,bubuianu18,vv25,bsssvv25}: 
\begin{eqnarray}
\ \ _{\ \ \chi }^{Jwh}\widehat{Z}(\tau ) &=&\exp \left[ \frac{1}{8\pi
^{2}\tau ^{2}}\ \ _{\chi }^{J}\mathcal{V[}\ _{\epsilon Q}^{2wh}h\mathbf{g}%
(\tau )]\right] ,\ \ _{\chi }^{Jwh}\widehat{\mathcal{E}}\ (\tau )=\ \frac{%
1-2\tau \ \ _{Q}^{h}\Lambda (\tau )}{8\pi ^{2}\tau }\ \ _{\chi }^{J}\mathcal{%
V[}\ _{\epsilon Q}^{2wh}h\mathbf{g}(\tau )],  \notag \\
\ \ \ \ \ _{\chi }^{Jwh}\widehat{S}(\tau ) &=&-\ \ _{\chi }^{Jwh}\widehat{W}%
(\tau )=\frac{1-\ _{Q}^{h}\Lambda (\tau )}{4\pi ^{2}\tau ^{2}}\ _{\chi }^{J}%
\mathcal{V[}\ _{\epsilon Q}^{2wh}h\mathbf{g}(\tau )]\ .
\label{thermvarkdswk}
\end{eqnarray}%
These dependencies on temperature-like parameter $\tau $ are different for
respective thermodynamic values computed for FLH BEs and BHs (\ref%
{thermvar3wh}) and (\ref{thermvarkdswk}). Such formulas can be used for
distinguishing different WH and BH solutions in different FLH theories.

The $\epsilon $-parametric models with geometric entropy $\ _{\chi }^{Jwh}%
\widehat{S}(\tau )$ (\ref{thermvarkdswk}) allow us to construct off-diagonal
quasi-stationary metrics which really describe WH configurations for FLH
theories. For instance, such nonholonomic WHs can be with "small"
polarization of physical constants (in particular, with rotoid-type throats
etc.) in phase spaces. Certain FLH geometric or off-diagonal deformations
may open or close certain WH throats. We can impose LC conditions and
generate such locally anisotropic WHs in the framework of GR, encoding
certain nonmetric data. They can be characterized thermodynamically using G.
Perelman's W-entropy generalized for FLH theories, but not in the frameworks
of the Bekenstein-Hawking paradigm. 

\subsection{Nonholonomic toroid configurations and black torus, BT}

Different classes of black torus, BT, and black ring solutions were
constructed in GR and MGTs, see \cite{lemos01,peca98,emparan02,emparan08}
for reviews of results. Nonholonomic off-diagonal deformations of toroidal
BHs and double systems of BE and BT configurations were studied in \cite%
{v01t,v01q} using the AFCDM and integration of nonlinear PDEs. Seven years
later, another class of so-called black Saturn configurations was
constructed \cite{evslin08,yazadjiev08} by transforming (modified) Einstein
equations into systems of nonlinear ODEs. We analyze an example when the
AFCDM is applied for generating $\tau $-families of quasi-stationary locally
anisotropic solutions using prime BT metrics considered in \cite{astorino17}%
. For simplicity, we provide only the formulas for small parametric
deformations when the physical interpretation of new classes of solutions is
very similar to some primary holonomic/ diagonalizable metric ansatz. For
general nonholonomic deformations, the physical interpretation of such
generic off-diagonal solutions remains unclear.

On a 4-d Lorentz manifold, we consider a prime quadratic line element 
\begin{eqnarray}
d\tilde{s}^{2} &=&f^{-1}(\tilde{r})d\tilde{r}^{2}+\tilde{r}^{2}(\tilde{k}%
_{1}^{2}dx^{2}+\tilde{k}_{2}^{2}dy^{2})-f(\tilde{r})d\tilde{t}^{2}
\label{prmtor1} \\
&=&\tilde{g}_{\alpha _{2}}(\tilde{x}^{1})(d\tilde{u}^{\alpha _{2}})^{2},%
\mbox{ for }f(\tilde{r})=-\epsilon ^{2}b^{2}-\tilde{\mu}/\tilde{r}-\Lambda 
\tilde{r}^{2}/3  \notag
\end{eqnarray}%
constructed and studied in section 3.1 of \cite{astorino17}. The $s=1,2$
coordinates in this diagonal metric are related via the re-scaling parameter 
$^{h}\epsilon $ to standard toroid "normalized" coordinates. In the above
formulas, $r$ is a radial coordinate, with $\theta =2\pi k_{1}x$ and $%
\varphi =2\pi k_{2}y$ (when $x,y\in \lbrack 0,1]$) and re-scaling $k_{1}=\
^{h}\epsilon \tilde{k}_{1},k_{2}=\ ^{h}\epsilon \tilde{k}_{2},\mu
\rightarrow \frac{\mu }{(2\pi )^{3}}=\tilde{\mu}/(\ ^{h}\epsilon ) ^{3};$ $%
r\rightarrow \frac{r}{2\pi }=\tilde{r}/\ ^{h}\epsilon ,$ $t\rightarrow 2\pi
t=\ ^{h}\epsilon \tilde{r}.$ In (\ref{prmtor1}), the parameter $b$ is a
coupling constant as in the energy-momentum tensor for the nonlinear SU(2)
sigma model, 
\begin{equation}
T_{\mu _{2}\nu _{2}}=\frac{b^{2}\ ^{h}\epsilon ^{2}}{8\pi G\tilde{r}^{2}}[f(%
\tilde{r})\delta _{\mu _{2}}^{4}\delta _{\nu _{2}}^{4}-f^{-1}(\tilde{r}%
)\delta _{\mu _{2}}^{1}\delta _{\nu _{2}}^{1}].  \label{emtsm}
\end{equation}%
In this formula, $\mu $ is an integration constant which can be fixed as a
mass parameter. The value $\ ^{h}\epsilon =0$ allows us to recover in a
formal way certain toroid vacuum solutions found in \cite{lemos01,peca98}.
The toroid metric (\ref{prmtor1}) defines an exact static solution of the
Einstein equations for the LC connection and energy-momentum tensor (\ref%
{emtsm}). We can extend the constructions on phase space $_{Q}^{s}\mathcal{M}
$ if we redefine indices in above formulas as $\mu _{2}\rightarrow i_{2}$
and introduce (co) fiber indices $a_{3}$ and $a_{4},$ or $a=(a_{3},a_{4}).$

Above formulas generate an AdS BH with a toroidal horizon in 4-d Einstein
gravity and a corresponding nonlinear $\sigma $-model. For 8-d
constructions, we consider a prime s-metric 
\begin{eqnarray*}
d\tilde{s}^{2} &=&f^{-1}(\tilde{r})d\tilde{r}^{2}+\tilde{r}^{2}(\tilde{k}%
_{1}^{2}dx^{2}+\tilde{k}_{2}^{2}dy^{2})-f(\tilde{r})d\tilde{t}^{2}+\  \\
&&^{v}f^{-1}(\ ^{v}\tilde{r})d\ ^{v}\tilde{r}^{2}+\ ^{v}\tilde{r}^{2}(\ ^{v}%
\tilde{k}_{1}^{2}d\ ^{v}x^{2}+\ ^{v}\tilde{k}_{2}^{2}d\ ^{v}y^{2})-\ ^{v}f(\
^{v}\tilde{r})d\widetilde{E}^{2} \\
&=&\tilde{g}_{i_{2}}(\tilde{x}^{1})(d\tilde{u}^{i_{2}})^{2}+\tilde{g}%
_{a_{3}}(\ ^{v}\tilde{x}^{5})(d\tilde{u}^{a_{3}})^{2}+\tilde{g}_{a_{4}}(\
^{v}\tilde{x}^{5})(d\tilde{u}^{a_{3}})^{2}, \\
&& \mbox{ for }f(\tilde{r}) =-\ ^{h}\epsilon ^{2}b^{2}-\tilde{\mu}/\tilde{r}%
-\ ^{h}\Lambda \tilde{r}^{2}/3\mbox{ and }\ ^{v}f(\ ^{v}\tilde{r})=-\
^{v}\epsilon ^{2}\ ^{v}b^{2}-\ ^{v}\tilde{\mu}/\ ^{v}\tilde{r}-\ ^{v}\Lambda
\ ^{v}\tilde{r}^{2}/3; \\
&&\mbox{ for coordinates }u^{\alpha _{s}} =(\tilde{r},x,y,t;\ ^{v}\tilde{r}%
,\ ^{v}x,\ ^{v}y,E),
\end{eqnarray*}%
where the left label "v" is used for dubbing respectively the constants for $%
s=3,4.$ The matter source (\ref{emtsm}) is extended on the typical fiber
space:%
\begin{eqnarray}
T_{a_{3}b_{3}} &=&-\frac{\ ^{v}b^{2}\ ^{v}\epsilon ^{2}}{8\pi \ ^{v}G\ ^{v}%
\tilde{r}^{2}}\ ^{v}f^{-1}(\ ^{v}\tilde{r})\delta _{a_{3}}^{5}\delta
_{b_{3}}^{5}\mbox{ and }T_{a_{4}b_{4}}=\frac{\ ^{v}b^{2}\ ^{v}\epsilon ^{2}}{%
8\pi \ ^{v}G\ ^{v}\tilde{r}^{2}}\ ^{v}f(\ ^{v}\tilde{r})\delta
_{a_{4}}^{8}\delta _{a_{4}}^{8},  \notag \\
&& \mbox{ for }\mathbf{T}_{\alpha _{s}\beta _{s}} = \{T_{\mu _{2}\nu
_{2}},T_{a_{3}b_{3}},T_{a_{4}b_{4}}\}.  \label{emtsmt}
\end{eqnarray}

To apply the AFCDM without frame and coordinate singularity transforms, we
can consider frame transforms to an off-diagonal parametrization of primary
d- and s-metrics. For instance, we can transforms (\ref{prmtor1}) to a form
with trivial N-connection coefficients $\tilde{N}_{i_{1}}^{a_{2}}=$ $\tilde{N%
}_{i_{1}}^{a_{2}}(u^{\alpha _{2}}(\tilde{r},x,y,t))$ and $\tilde{g}_{\alpha
_{1}\beta _{1}}(u^{j_{1}}(\tilde{r},x,y),u^{3}(\tilde{r},x,y)).$ Such
transforms are defined in any form which do not involve singular frame
transforms and off-diagonal deformations. This is possible if we introduce
new coordinates $u^{1}=x^{1}=\tilde{r},u^{2}=x,$ and $u^{3}=y^{3}=y+\ ^{3}B(%
\tilde{r},x),u^{4}=y^{4}=t+\ ^{4}B(\tilde{r},x),$ when for $\tilde{N}%
_{i_{1}}^{3}=-\partial \ ^{3}B/\partial x^{i_{1}}$ and $\tilde{N}%
_{i_{1}}^{4}=-\partial \ ^{4}B/\partial x^{i_{1}}:$ 
\begin{eqnarray*}
\mathbf{\tilde{e}}^{3} &=&dy=du^{3}+\tilde{N}_{i_{1}}^{3}(\tilde{r}%
,x)dx^{i}=du^{3}+\tilde{N}_{1}^{3}(\tilde{r},x)dr+\tilde{N}_{2}^{3}(\tilde{r}%
,x)dz, \\
\mathbf{\tilde{e}}^{4} &=&dt=du^{4}+\tilde{N}_{i_{1}}^{4}(\tilde{r}%
,x)dx^{i}=du^{4}+\tilde{N}_{1}^{4}(\tilde{r},x)dr+\tilde{N}_{2}^{4}(\tilde{r}%
,x)dz.
\end{eqnarray*}%
In new nonlinear coordinates, the diagonal toroid metric (\ref{prmtor1})
transforms into an off-diagonal toroid s-metric 
\begin{equation}
d\tilde{s}^{2}=\tilde{g}_{\alpha _{2}}(\tilde{r},x,y)[\mathbf{\tilde{e}}%
^{\alpha _{2}}(\tilde{r},x,y)]^{2},  \label{prmtor2}
\end{equation}%
where $\tilde{g}_{1}=f^{-1}(x^{1}),\tilde{g}_{2}=(x^{1})^{2}\tilde{k}%
_{1}^{2},\tilde{g}_{3}=(x^{2})^{2}\tilde{k}_{2}^{2}$ and $\tilde{g}%
_{4}=f(x^{1}).$ Extending (\ref{prmtor2}) on total phase space, we define a
prime s-metric $\mathbf{\tilde{g}=\{\tilde{g}}_{\alpha _{s}\beta _{s}}\}$
defines a s-adapted quadratic element: 
\begin{eqnarray}
d\tilde{s}^{2} &=&\tilde{g}_{\alpha _{2}}(\tilde{r},x,y)[\mathbf{\tilde{e}}%
^{\alpha _{2}}(\tilde{r},x,y)]^{2}  \label{prmtor2tot} \\
&&+\tilde{g}_{a_{3}}(\ ^{v}\tilde{r},\ ^{v}x,\ ^{v}y)[\mathbf{\tilde{e}}%
^{\alpha _{3}}(\ ^{v}\tilde{r},\ ^{v}x,\ ^{v}y)]^{2}+\tilde{g}_{\alpha
_{4}}(\ ^{v}\tilde{r},\ ^{v}x,\ ^{v}y)[\mathbf{\tilde{e}}^{\alpha _{4}}(\
^{v}\tilde{r},\ ^{v}x,\ ^{v}y)]^{2},  \notag
\end{eqnarray}%
where $\tilde{g}_{5}=\ ^{v}f^{-1}(\ ^{v}\tilde{r}),\tilde{g}_{6}=(\
^{v}x)^{2}\ ^{v}\tilde{k}_{1}^{2},\tilde{g}_{7}=(\ ^{v}y)^{2}\ ^{v}\tilde{k}%
_{2}^{2}$ and $\tilde{g}_{4}=\ ^{v}f(\ ^{v}\tilde{r})$.

At the next step, we can generate new classes of FL toroid solutions if we
construct small parametric quasi-stationary deformations of prime metrics (%
\ref{prmtor2tot}) defined by an effective source $\ ^{tor}\mathbf{Y}[\mathbf{%
g,}\widehat{\mathbf{D}}]\simeq \{-\Lambda \mathbf{g}_{\alpha _{s}\beta _{s}}+%
\mathbf{T}_{\alpha _{s}\beta _{s}}\},$ for (\ref{emtsmt}). The left label
"tor" is used for toroid configurations when additional $Q$ labels have to
be introduced for respective $\tau $-families of nonmetric fields, when 
\begin{equation*}
\ ^{tor}\mathbf{Y}_{\alpha _{s}\beta _{s}}=-\Lambda \mathbf{g}_{\alpha
_{s}\beta _{s}}+\mathbf{T}_{\alpha _{s}\beta _{s}}+\ ^{z}\mathbf{T}_{\alpha
_{s}\beta _{s}},
\end{equation*}%
where $\ ^{z}\mathbf{T}_{\alpha _{s}\beta _{s}}$ is determined by the
distortions of the Ricci s-tensor under distortions of a chosen
metric-affine s-connection, see formulas (\ref{driccidist}) and (\ref%
{emtdist}). On corresponding shells, we can consider generating sources
parameterized in s-adapted form (after a corresponding redefinition of the
nonholonomic structure and double toroid coordinates on phase space): $\
_{Q}^{tor}\mathbf{Y}_{\beta _{s}}^{\alpha _{s}}=\left[ \ _{1Q}^{tor}\Upsilon
(\tilde{r},x),\ _{2Q}^{tor}\Upsilon (\tilde{r},x,y),\ _{3Q}^{tor}\Upsilon (\
^{v}\tilde{r},\ ^{v}x),\ _{4Q}^{tor}\Upsilon (\ ^{v}\tilde{r},\ ^{v}x,\
^{v}y)\right] $. For target $\tau $-families of distorted toroid s-metrics,
we also have to consider the terms $\partial _{\tau }\mathbf{g}_{\mu
^{\prime }\nu ^{\prime }}(\tau )$ in (\ref{ricciflowr2}) as additional
effective sources determined by $\tau $-running of the canonical Ricci
s-tensors, cosmological constant running and FL flows of matter fields (\ref%
{emtsmt}). This way, we can introduce s-adapted effective sources as in $\
_{Q}\mathbf{J}_{\ \nu }^{\mu }(\tau )$ (\ref{dsourcparam}) $\rightarrow \
_{Q}\mathbf{J}_{\ \nu _{s}}^{\mu _{s}}(\tau ),$ when 
\begin{eqnarray}
\ \quad \ _{Q}^{tor}\mathbf{J}_{\ \nu }^{\mu }(\tau ) &=&\mathbf{e}_{\
}^{\mu \mu ^{\prime }}(\tau )\mathbf{e}_{\nu }^{\ \nu ^{\prime }}(\tau )[~\
_{Q}^{tor}\mathbf{Y}_{\mu ^{\prime }\nu ^{\prime }}(\tau )-\frac{1}{2}%
~\partial _{\tau }\mathbf{g}_{\mu ^{\prime }\nu ^{\prime }}(\tau )]
\label{torsourcrf} \\
&=&[\ _{1Q}^{tor}J(\tau )\delta _{j_{1}}^{i_{1}},\ \ _{2Q}^{tor}J(\tau
)\delta _{b_{2}}^{a_{2}},\ \ _{3Q}^{tor}J(\tau )\delta _{b_{3}}^{a_{3}},\ \
_{4Q}^{tor}J(\tau )\delta _{b_{4}}^{a_{4}}],  \notag
\end{eqnarray}%
when the toroidal coordinates a correspondingly s-adapted on phase space. We
shall define a class of parametric nonmetric geometric flow deformations $%
\mathbf{\tilde{g}}$ (\ref{prmtor2tot}) to $\tau $-families of
quasi-stationary double FLH deformed toroid configurations $\mathbf{g}_{\mu
_{s}\nu _{s}}(\tau ).$

The corresponding nonlinear symmetries (\ref{nonlinsymrex}), for the
generating sources (\ref{torsourcrf}) related to $\tau $-running
cosmological constants $[\Lambda (\tau )+\ _{Q}^{htor}\Lambda (\tau),\Lambda
(\tau )+\ _{Q}^{vtor}\Lambda (\tau )],$ are written: 
\begin{eqnarray}
\partial _{y}[\ ^{2}\Psi ^{2}(\tau ,\tilde{r},x,y)] &=&-\int dy\ \ \
_{2Q}^{tor}J(\tau )\partial _{y}g_{4}\simeq -\int dy\ \ \ _{2Q}^{tor}J(\tau ,%
\tilde{r},x,y)\partial _{y}[\eta _{4}(\tau ,\tilde{r},x,y)\ \tilde{g}_{4}(%
\tilde{r})]  \label{nsymtor} \\
&\simeq &-\int dy\ \ _{2Q}^{tor}J(\tau ,\tilde{r},x,y)\partial _{y}[\zeta
_{4}(\tau ,\tilde{r},x,y)(1+\epsilon \ \chi _{4}(\tau ,\tilde{r},x,y))\ 
\tilde{g}_{4}(\tilde{r})],  \notag \\
\ ^{2}\Psi (\tau ,\tilde{r},x,y) &=&|\ \ \Lambda (\tau )+\ ^{htor}\Lambda
(\tau )|^{-1/2}\sqrt{|\int dy\ \ \ _{2Q}^{tor}J(\tau ,\tilde{r},x,y)\
\partial _{y}(\ ^{2}\Phi ^{2})|},  \notag \\
(\ ^{2}\Phi (\tau ,\tilde{r},x,y))^{2} &=&-4\ (\ \Lambda (\tau )+\
^{htor}\Lambda (\tau ))g_{4}(\tau ,\tilde{r},x,y)\simeq -4\ (\ \Lambda (\tau
)+\ ^{htor}\Lambda (\tau ))\eta _{4}(\tau ,\tilde{r},x,y)\ \ \tilde{g}_{4}(%
\tilde{r})  \notag \\
&\simeq &-4(\ \Lambda (\tau )+\ ^{htor}\Lambda (\tau ))\ \zeta _{4}(\tau ,%
\tilde{r},x,y)(1+\epsilon \chi _{4}(\tau ,\tilde{r},x,y))\ \tilde{g}_{4}(%
\tilde{r}).  \notag
\end{eqnarray}%
\begin{eqnarray*}
\partial _{\ ^{v}\tilde{r}}[\ ^{3}\Psi ^{2}(\tilde{r},x,y,\ ^{v}\tilde{r})]
&=&-\int d\ ^{v}\tilde{r}\ \ \ _{3Q}^{tor}J(\tau )\partial _{\ ^{v}\tilde{r}%
}g_{6} \\
&\simeq &-\int d\ ^{v}\tilde{r}\ \ \ _{3Q}^{tor}J(\tau ,\tilde{r},x,y,\ ^{v}%
\tilde{r})\partial _{\ ^{v}\tilde{r}}[\eta _{6}(\tau ,\tilde{r},x,y,\ ^{v}%
\tilde{r})\ \tilde{g}_{6}(\ ^{v}\tilde{r})] \\
&\simeq &-\int d\ ^{v}\tilde{r}\ \ _{3Q}^{tor}J(\tau ,\tilde{r},x,y,\ ^{v}%
\tilde{r})\partial _{\ ^{v}\tilde{r}}[\zeta _{6}(\tau ,\tilde{r},x,y,\ ^{v}%
\tilde{r})(1+\epsilon \ \chi _{6}(\tau ,\tilde{r},x,y,\ ^{v}\tilde{r}))\ 
\tilde{g}_{6}(\ ^{v}\tilde{r})], \\
\ ^{3}\Psi (\tau ,\tilde{r},x,y,\ ^{v}\tilde{r}) &=&|\ \ \Lambda (\tau )+\
^{vtor}\Lambda (\tau )|^{-1/2}\sqrt{|\int d\ ^{v}\tilde{r}\ \ \ \
_{3Q}^{tor}J(\tau ,\tilde{r},x,y,\ ^{v}\tilde{r})\ \partial _{\ ^{v}\tilde{r}%
}(\ ^{3}\Phi ^{2})|}, \\
(\ ^{3}\Phi (\tau ,\tilde{r},x,y,\ ^{v}\tilde{r}))^{2} &=&-4\ (\ \Lambda
(\tau )+\ ^{vtor}\Lambda (\tau ))g_{6}(\tau ,\tilde{r},x,y,\ ^{v}\tilde{r})
\\
&\simeq &-4\ (\ \Lambda (\tau )+\ ^{vtor}\Lambda (\tau ))\eta _{6}(\tau ,%
\tilde{r},x,y,\ ^{v}\tilde{r})\ \tilde{g}_{6}(\ ^{v}\tilde{r}) \\
&\simeq &-\int d\ ^{v}\tilde{r}\ \ _{3Q}^{tor}J(\tau ,\tilde{r},x,y,\ ^{v}%
\tilde{r})\partial _{\ ^{v}\tilde{r}}[\zeta _{6}(\tau ,\tilde{r},x,y,\ ^{v}%
\tilde{r})(1+\epsilon \ \chi _{6}(\tau ,\tilde{r},x,y,\ ^{v}\tilde{r}))\ 
\tilde{g}_{6}(\ ^{v}\tilde{r})],
\end{eqnarray*}%
\begin{eqnarray*}
\partial _{\ ^{v}y}[\ ^{4}\Psi ^{2}(\tilde{r},x,y,\ ^{v}\tilde{r},\ ^{v}x,\
^{v}y)] &=&-\int d\ ^{v}y\ \ \ _{4Q}^{tor}J(\tau )\partial _{\ ^{v}y}g_{8} \\
&\simeq &-\int d\ ^{v}y\ \ \ _{4Q}^{tor}J(\tau ,\tilde{r},x,y,\ ^{v}\tilde{r}%
,\ ^{v}x,\ ^{v}y)\partial _{\ ^{v}y}[\eta _{8}(\tau ,\tilde{r},x,y,\ ^{v}%
\tilde{r},\ ^{v}x,\ ^{v}y)\ \tilde{g}_{8}(\ ^{v}\tilde{r})] \\
&\simeq &-\int d\ ^{v}y\ \ _{4Q}^{tor}J(\tau ,\tilde{r},x,y,\ ^{v}\tilde{r}%
,\ ^{v}x,\ ^{v}y)\partial _{\ ^{v}y}[\zeta _{8}(\tau ,\tilde{r},x,y,\ ^{v}%
\tilde{r},\ ^{v}x,\ ^{v}y) \\
&&(1+\epsilon \ \chi _{8}(\tau ,\tilde{r},x,y,\ ^{v}\tilde{r},\ ^{v}x,\
^{v}y))\ \tilde{g}_{8}(\ ^{v}\tilde{r})], \\
\ ^{4}\Psi (\tau ,\tilde{r},x,y,\ ^{v}\tilde{r},\ ^{v}x,\ ^{v}y) &=&|\ \
\Lambda (\tau )+\ ^{vtor}\Lambda (\tau )|^{-1/2}\sqrt{|\int d\ ^{v}y\ \ \ \
_{4Q}^{tor}J(\tau ,\tilde{r},x,y,\ ^{v}\tilde{r},\ ^{v}x,\ ^{v}y)\ \partial
_{\ ^{v}y}(\ ^{4}\Phi ^{2})|}, \\
(\ ^{4}\Phi (\tau ,\tilde{r},x,y,\ ^{v}\tilde{r},\ ^{v}x,\ ^{v}y))^{2}
&=&-4\ (\ \Lambda (\tau )+\ ^{vtor}\Lambda (\tau ))g_{8}(\tau ,\tilde{r}%
,x,y,\ ^{v}\tilde{r},\ ^{v}x,\ ^{v}y) \\
&\simeq &-4\ (\ \Lambda (\tau )+\ ^{vtor}\Lambda (\tau ))\eta _{8}(\tau ,%
\tilde{r},x,y,\ ^{v}\tilde{r},\ ^{v}x,\ ^{v}y)\ \tilde{g}_{8}(\ ^{v}\tilde{r}%
) \\
&\simeq &-\int d\ ^{v}y\ \ _{4Q}^{tor}J(\tau ,\tilde{r},x,y,\ ^{v}\tilde{r}%
,\ ^{v}x,\ ^{v}y)\partial _{\ ^{v}y}[\zeta _{6}(\tau ,\tilde{r},x,y,\ ^{v}%
\tilde{r},\ ^{v}x,\ ^{v}y) \\
&&(1+\epsilon \ \chi _{8}(\tau ,\tilde{r},x,y,\ ^{v}\tilde{r},\ ^{v}x,\
^{v}y))\ \tilde{g}_{8}(\ ^{v}\tilde{r})],
\end{eqnarray*}%
In these formulas, we use $\ ^{tor}\Lambda (\tau )=[$ $^{htor}\Lambda
(\tau),\ ^{vtor}\Lambda (\tau )]$ as $\tau $-running effective cosmological
constants related via nonlinear symmetries to an energy-momentum tensor (\ref%
{emtsmt}). Such $\ ^{tor}\Lambda (\tau )$ can be different from a prescribed
cosmological constant associated to other types of gravitational and matter
interactions. In this subsection, we write $\widetilde{\Lambda }(\tau
)=\Lambda (\tau )+\ ^{tor}\Lambda (\tau ).$

For small parametric deformations with $\chi $-polarization functions, the
quadratic linear elements for FL toroid solutions $\ _{\epsilon Q}^{tor}%
\mathbf{g}(\tau )$ are computed: 
\begin{eqnarray*}
d\ \widehat{s}^{2}(\tau ) &=&\widehat{g}_{\alpha _{s}\beta _{s}}(\tau ,%
\tilde{r},x,y,^{v}\tilde{r},\ ^{v}x,\ ^{v}y;\psi ,\eta _{4};\widetilde{%
\Lambda }=\Lambda +\ ^{tor}\Lambda ,\ \ \ _{sQ}^{tor}J,\ \tilde{g}_{\alpha
_{s}})du^{\alpha _{s}}du^{\beta _{s}} \\
&=&e^{\psi _{0}(\tilde{r},x)}[1+\epsilon \ ^{\psi (\tilde{r},x)}\chi (\tilde{%
r},x)][(dx^{1}(\tilde{r},x))^{2}+(dx^{2}(\tilde{r},x))^{2}]-
\end{eqnarray*}%
\begin{eqnarray*}
&&\{\frac{4[\partial _{y}(|\zeta _{4}\ \tilde{g}_{4}|^{1/2})]^{2}}{\tilde{g}%
_{3}|\int dy\{\ _{2Q}^{tor}J\partial _{y}(\zeta _{4}\ \tilde{g}_{4})\}|}%
-\epsilon \lbrack \frac{\partial _{y}(\chi _{4}|\zeta _{4}\tilde{g}%
_{4}|^{1/2})}{4\partial _{y}(|\zeta _{4}\tilde{g}_{4}|^{1/2})}-\frac{\int
dy\{\ _{2Q}^{tor}J\partial _{y}[(\zeta _{4}\ \tilde{g}_{4})\chi _{4}]\}}{%
\int dy\{\ _{2Q}^{tor}J\partial _{y}(\zeta _{4}\ \tilde{g}_{4})\}}]\}\ 
\tilde{g}_{3} \\
&&\{dy+[\frac{\partial _{i_{1}}\ \int dy\ \ _{2Q}^{tor}J\ \partial _{y}\zeta
_{4}}{(\tilde{N}_{i_{1}}^{3})\ \ _{2Q}^{tor}J\partial _{y}\zeta _{4}}%
+\epsilon (\frac{\partial _{i_{1}}[\int dy\ \ _{2Q}^{tor}J\ \partial
_{y}(\zeta _{4}\chi _{4})]}{\partial _{i_{1}}\ [\int dy\ \
_{2Q}^{tor}J\partial _{y}\zeta _{4}]}-\frac{\partial _{y}(\zeta _{4}\chi
_{4})}{\partial _{y}\zeta _{4}})]\tilde{N}_{i_{1}}^{3}dx^{i_{1}}\}^{2}+
\end{eqnarray*}%
\begin{eqnarray}
&&\zeta _{4}(1+\epsilon \ \chi _{4})\ \tilde{g}_{4}\{dt+[(\tilde{N}%
_{k_{1}}^{4})^{-1}[\ _{1}n_{k_{1}}+16\ _{2}n_{k_{1}}[\int dy\frac{\left(
\partial _{y}[(\zeta _{4}\ \tilde{g}_{4})^{-1/4}]\right) ^{2}}{|\int
dy\partial _{y}[\ \ _{2Q}^{tor}J(\zeta _{4}\ \tilde{g}_{4})]|}]+  \notag \\
&&\epsilon \frac{16\ _{2}n_{k_{1}}\int dy\frac{\left( \partial _{y}[(\zeta
_{4}\ \tilde{g}_{4})^{-1/4}]\right) ^{2}}{|\int dy\partial _{y}[\
_{2Q}^{tor}J(\zeta _{4}\ \tilde{g}_{4})]|}(\frac{\partial _{y}[(\zeta _{4}\ 
\tilde{g}_{4})^{-1/4}\chi _{4})]}{2\partial _{y}[(\zeta _{4}\ \tilde{g}%
_{4})^{-1/4}]}+\frac{\int dy\partial _{y}[\ _{2Q}^{tor}J(\zeta _{4}\chi
_{4}\ \tilde{g}_{4})]}{\int dy\partial _{y}[\ \ _{2Q}^{tor}J(\zeta _{4}\ 
\tilde{g}_{4})]})}{\ _{1}n_{k_{1}}+16\ _{2}n_{k_{1}}[\int dy\frac{\left(
\partial _{y}[(\zeta _{4}\ \tilde{g}_{4})^{-1/4}]\right) ^{2}}{|\int
dy\partial _{y}[\ \ _{2Q}^{tor}J(\zeta _{4}\ \tilde{g}_{4})]|}]}]\tilde{N}%
_{k_{1}}^{4}dx^{k_{1}}\}^{2}+  \label{tortpol2}
\end{eqnarray}%
\begin{eqnarray*}
&&\{\frac{4[\partial _{\ ^{v}\tilde{r}}(|\zeta _{6}\ \tilde{g}%
_{6}|^{1/2})]^{2}}{\tilde{g}_{5}|\int d\ ^{v}\tilde{r}\{\
_{3Q}^{tor}J\partial _{\ ^{v}\tilde{r}}(\zeta _{6}\ \tilde{g}_{6})\}|}%
-\epsilon \lbrack \frac{\partial _{\ ^{v}\tilde{r}}(\chi _{6}|\zeta _{6}%
\tilde{g}_{6}|^{1/2})}{4\partial _{\ ^{v}\tilde{r}}(|\zeta _{6}\tilde{g}%
_{6}|^{1/2})}-\frac{\int d\ ^{v}\tilde{r}\{\ _{3Q}^{tor}J\partial _{\ ^{v}%
\tilde{r}}[(\zeta _{6}\ \tilde{g}_{6})\chi _{6}]\}}{\int d\ ^{v}\tilde{r}\{\
_{3Q}^{tor}J\partial _{\ ^{v}\tilde{r}}(\zeta _{6}\ \tilde{g}_{6})\}}]\}\ 
\tilde{g}_{5} \\
&&\{d\ ^{v}\tilde{r}+[\frac{\partial _{i_{2}}\ \int d\ ^{v}\tilde{r}\ \
_{3Q}^{tor}J\ \partial _{\ ^{v}\tilde{r}}\zeta _{6}}{(\tilde{N}%
_{i_{2}}^{5})\ \ _{3Q}^{tor}J\partial _{\ ^{v}\tilde{r}}\zeta _{6}}+\epsilon
(\frac{\partial _{i_{2}}[\int d\ ^{v}\tilde{r}\ \ _{3Q}^{tor}J\ \partial _{\
^{v}\tilde{r}}(\zeta _{6}\chi _{6})]}{\partial _{i_{2}}\ [\int d\ ^{v}\tilde{%
r}\ \ _{3Q}^{tor}J\partial _{\ ^{v}\tilde{r}}\zeta _{6}]}-\frac{\partial _{\
^{v}\tilde{r}}(\zeta _{6}\chi _{6})}{\partial _{\ ^{v}\tilde{r}}\zeta _{6}})]%
\tilde{N}_{i_{2}}^{5}dx^{i_{2}}\}^{2}+
\end{eqnarray*}%
\begin{eqnarray*}
&&\zeta _{6}(1+\epsilon \chi _{6})\ \tilde{g}_{6}\{d\ ^{v}x+[(\tilde{N}%
_{k_{2}}^{6})^{-1}[\ _{1}n_{k_{2}}+16\ _{2}n_{k_{2}}[\int d\ ^{v}\tilde{r}%
\frac{\left( \partial _{\ ^{v}\tilde{r}}[(\zeta _{6}\ \tilde{g}%
_{6})^{-1/4}]\right) ^{2}}{|\int d\ ^{v}\tilde{r}\partial _{\ ^{v}\tilde{r}%
}[\ \ _{3Q}^{tor}J(\zeta _{6}\ \tilde{g}_{6})]|}]+ \\
&&\epsilon \frac{16\ _{2}n_{k_{2}}\int d\ ^{v}\tilde{r}\frac{\left( \partial
_{\ ^{v}\tilde{r}}[(\zeta _{6}\ \tilde{g}_{6})^{-1/4}]\right) ^{2}}{|\int d\
^{v}\tilde{r}\partial _{\ ^{v}\tilde{r}}[\ _{3Q}^{tor}J(\zeta _{6}\ \tilde{g}%
_{6})]|}(\frac{\partial _{\ ^{v}\tilde{r}}[(\zeta _{6}\ \tilde{g}%
_{6})^{-1/4}\chi _{6})]}{2\partial _{\ ^{v}\tilde{r}}[(\zeta _{6}\ \tilde{g}%
_{6})^{-1/4}]}+\frac{\int d\ ^{v}\tilde{r}\partial _{\ ^{v}\tilde{r}}[\
_{3Q}^{tor}J(\zeta _{6}\chi _{6}\ \tilde{g}_{6})]}{\int d\ ^{v}\tilde{r}%
\partial _{\ ^{v}\tilde{r}}[\ \ _{3Q}^{tor}J(\zeta _{6}\ \tilde{g}_{6})]})}{%
\ _{1}n_{k_{2}}+16\ _{2}n_{k_{2}}[\int d\ ^{v}\tilde{r}\frac{\left( \partial
_{\ ^{v}\tilde{r}}[(\zeta _{6}\ \tilde{g}_{6})^{-1/4}]\right) ^{2}}{|\int d\
^{v}\tilde{r}\partial _{\ ^{v}\tilde{r}}[\ \ _{3Q}^{tor}J(\zeta _{6}\ \tilde{%
g}_{6})]|}]}]\tilde{N}_{k_{2}}^{6}dx^{k_{2}}\}^{2}+
\end{eqnarray*}%
\begin{eqnarray*}
&&\{\frac{4[\partial _{\ ^{v}y}(|\zeta _{8}\ \tilde{g}_{8}|^{1/2})]^{2}}{%
\tilde{g}_{7}|\int d\ ^{v}y\{\ _{4Q}^{tor}J\partial _{\ ^{v}y}(\zeta _{8}\ 
\tilde{g}_{8})\}|}-\epsilon \lbrack \frac{\partial _{\ ^{v}y}(\chi
_{8}|\zeta _{8}\tilde{g}_{8}|^{1/2})}{4\partial _{\ ^{v}y}(|\zeta _{8}\tilde{%
g}_{8}|^{1/2})}-\frac{\int d\ ^{v}y\{\ _{4Q}^{tor}J\partial _{\
^{v}y}[(\zeta _{8}\ \tilde{g}_{8})\chi _{8}]\}}{\int d\ ^{v}y\{\
_{4Q}^{tor}J\partial _{\ ^{v}y}(\zeta _{8}\ \tilde{g}_{8})\}}]\}\ \tilde{g}%
_{7} \\
&&\{d\ ^{v}y+[\frac{\partial _{i_{3}}\ \int d\ ^{v}y\ \ _{4Q}^{tor}J\
\partial _{\ ^{v}y}\zeta _{8}}{(\tilde{N}_{i_{3}}^{7})\ \
_{4Q}^{tor}J\partial _{\ ^{v}y}\zeta _{8}}+\epsilon (\frac{\partial
_{i_{3}}[\int d\ ^{v}y\ \ _{4Q}^{tor}J\ \partial _{\ ^{v}y}(\zeta _{8}\chi
_{8})]}{\partial _{i_{3}}\ [\int d\ ^{v}y\ \ _{4Q}^{tor}J\partial _{\
^{v}y}\zeta _{8}]}-\frac{\partial _{\ ^{v}y}(\zeta _{8}\chi _{8})}{\partial
_{\ ^{v}y}\zeta _{8}})]\tilde{N}_{i_{3}}^{7}dx^{i_{3}}\}^{2}+
\end{eqnarray*}%
\begin{eqnarray*}
&&\zeta _{8}(1+\epsilon \chi _{8})\ \tilde{g}_{8}\{dE+[(\tilde{N}%
_{k_{3}}^{8})^{-1}[\ _{1}n_{k_{3}}+16\ _{2}n_{k_{3}}[\int d\ ^{v}y\frac{%
\left( \partial _{\ ^{v}y}[(\zeta _{8}\ \tilde{g}_{8})^{-1/4}]\right) ^{2}}{%
|\int d\ ^{v}y\partial _{\ ^{v}y}[\ \ _{4Q}^{tor}J(\zeta _{8}\ \tilde{g}%
_{8})]|}]+ \\
&&\epsilon \frac{16\ _{2}n_{k_{3}}\int d\ ^{v}y\frac{\left( \partial _{\
^{v}y}[(\zeta _{8}\ \tilde{g}_{8})^{-1/4}]\right) ^{2}}{|\int d\
^{v}y\partial _{\ ^{v}y}[\ _{4Q}^{tor}J(\zeta _{8}\ \tilde{g}_{8})]|}(\frac{%
\partial _{\ ^{v}8}[(\zeta _{8}\ \tilde{g}_{8})^{-1/4}\chi _{8})]}{2\partial
_{\ ^{v}y}[(\zeta _{8}\ \tilde{g}_{8})^{-1/4}]}+\frac{\int d\ ^{v}y\partial
_{\ ^{v}y}[\ _{4Q}^{tor}J(\zeta _{8}\chi _{8}\ \tilde{g}_{8})]}{\int d\
^{v}y\partial _{\ ^{v}y}[\ \ _{4Q}^{tor}J(\zeta _{8}\ \tilde{g}_{8})]})}{\
_{1}n_{k_{3}}+16\ _{2}n_{k_{3}}[\int d\ ^{v}y\frac{\left( \partial _{\
^{v}y}[(\zeta _{8}\ \tilde{g}_{8})^{-1/4}]\right) ^{2}}{|\int d\
^{v}y\partial _{\ ^{v}y}[\ \ _{4Q}^{tor}J(\zeta _{8}\ \tilde{g}_{8})]|}]}]%
\tilde{N}_{k_{3}}^{8}dx^{k_{3}}\}^{2}.
\end{eqnarray*}%
The parametric solution (\ref{tortpol2}) describe elliptic deformations if
we chose generating functions 
\begin{eqnarray*}
\chi _{4}(\tau ,\tilde{r},x,y) &=&\underline{\chi }_{4}(\tau ,\tilde{r}%
,x)\sin (\omega _{0}y+y_{0}),\chi _{6}(\tau ,\tilde{r},x,y,\ ^{v}\tilde{r})=%
\underline{\chi }_{6}(\tau ,\tilde{r},x,y)\sin (\ ^{r}\omega _{0}\ ^{v}%
\tilde{r}+\ ^{v}\tilde{r}_{0}), \\
&& \chi _{8}(\tau ,\tilde{r},x,y,\ ^{v}\tilde{r},^{v}x,\ ^{v}y) =\underline{%
\chi }_{8}(\tau ,\tilde{r},x,y,^{v}x)\sin (\ ^{y}\omega _{0}\ \ ^{v}y+ \
^{v}y_{0}).
\end{eqnarray*}
In such cases, we generate a family of toroid configurations with
ellipsoidal deformations on $y$ coordinate. Similarly, we can construct
solutions with ellipsoidal deformations on other types space and fiber
coordinates. Using abstract geometric calculus, the above formulas and
solutions can be defined on phase spaces with momentum-like variables.

Finally, we note that we can consider more sophisticated classes of FLH
geometric flow deformations which transform a toroid prime s-metric into
"spagetti" quasi-stationary configurations. Under nonholonomic geometric
evolution and for off-diagonal interactions result in modification of
certain sections of geometric configurations, curved and waved, possible
interruptions, singularities etc. Such configurations can be embedded into
locally anisotropic gravitational vacuum media, which are polarized on
velocity/momentum variable. Parametric solutions (\ref{tortpol2}) can be
used for modelling DM quasi-stationary FLH configurations. For such
solutions and nonlinear symmetries \ (\ref{nsymtor}), the effective
cosmological constant $\widetilde{\Lambda }=\Lambda +\ ^{tor}\Lambda $ can
be considered as a phenomenological sum of parameters, which can be related
to DE and encoding toroid configurations. Such parameters for rotoid and
toroid configurations may explain certain observational data. 

\subsubsection{Relativistic Ricci flow thermodynamic variables for FLH BTs}

The FLH geometric thermodynamic models studied in the previous sections for
nonmetric quasi-stationary off-diagonal generalizations of KdS BH and WH
solutions can be re-defined respectively in an abstract geometric language
which allows us to describe physically important properties of $\tau $%
-families of nonholonomic BT solutions (\ref{tortpol2}). For off-diagonal
solutions with $\chi $-polarizations $\ _{\epsilon Q}^{tor}\mathbf{g}(\tau )$
(\ref{tortpol2}), the thermodynamic variables (\ref{thermvar2}) are defined
and computed: 
\begin{eqnarray*}
\ _{\epsilon Q}^{tor}\widehat{Z}(\tau ) &=&\exp \left[ \frac{1}{(4\pi \tau
)^{4}}\ \ _{\chi }^{\mathbf{J}}\mathcal{V}[\ \ _{\epsilon Q}^{tor}\mathbf{g}%
(\tau )]\right] , \\
\ \ _{\epsilon Q}^{tor}\mathcal{E}\ (\tau ) &=&\frac{1}{64\pi ^{4}\tau ^{3}}%
\ \left( 1-2\tau (\ _{Q}^{h}\Lambda (\tau )+\ _{Q}^{v}\Lambda (\tau
))\right) \ \ \ _{\chi }^{\mathbf{J}}\mathcal{V[}\ \ _{\epsilon Q}^{tor}%
\mathbf{g}(\tau )], \\
\ \ \ \ \ _{\epsilon Q}^{tor}\widehat{S}(\tau ) &=&-\ \ _{\epsilon Q}^{tor}%
\widehat{W}(\tau )=\frac{2}{(4\pi \tau )^{4}}(1-4(\ _{Q}^{h}\Lambda (\tau
)+\ _{Q}^{v}\Lambda (\tau )))\ \ _{\chi }^{\mathbf{J}}\mathcal{V[}\ \
_{\epsilon Q}^{tor}\mathbf{g}(\tau )].
\end{eqnarray*}%
Similar values can be computed using $\ \ _{\eta }^{J}\mathcal{V}[\
_{Q}^{tor}\mathbf{g}(\tau )]$ or $\ _{\eta }^{J}\mathcal{V}[ \ _{Q}^{2tor}%
\mathbf{g}(\tau )].$

We note that for FL h-flows, we have BT formulas as in 4-d MGTs \cite%
{perelman1,gheorghiuap16,bubuianu18,vv25,bsssvv25}: 
\begin{eqnarray*}
\ \ _{\ \ \chi }^{Jtor}\widehat{Z}(\tau ) &=&\exp \left[ \frac{1}{8\pi
^{2}\tau ^{2}}\ \ _{\chi }^{J}\mathcal{V[}\ _{\epsilon Q}^{2tor}h\mathbf{g}%
(\tau )]\right] ,\ \ _{\chi }^{Jtor}\widehat{\mathcal{E}}\ (\tau )=\ \frac{%
1-2\tau \ \ _{Q}^{h}\Lambda (\tau )}{8\pi ^{2}\tau }\ \ _{\chi }^{J}\mathcal{%
V[}\ _{\epsilon Q}^{2tor}h\mathbf{g}(\tau )], \\
\ \ \ \ \ _{\chi }^{Jtor}\widehat{S}(\tau ) &=&-\ \ _{\chi }^{Jtor}\widehat{W%
}(\tau )=\frac{1-\ _{Q}^{h}\Lambda (\tau )}{4\pi ^{2}\tau ^{2}}\ _{\chi }^{J}%
\mathcal{V[}\ _{\epsilon Q}^{2tor}h\mathbf{g}(\tau )]\ .
\end{eqnarray*}%
These dependencies on temperature like parameter $\tau $ are different for
respective thermodynamic values computed for FLH BEs and BHs (\ref%
{thermvar3wh}) and (\ref{thermvarkdswk}). This can be used for
distinguishing different WH and BH solutions in different FLH theories.

\subsection{Off-diagonal 4-d cosmological solitonic and spheroid
cosmological solutions involving 2-d vertices}

The goal of this subsection is to study some physically important examples
of locally anisotropic cosmological solutions (\ref{qeltorsc}) and their
equivalents when the gravitational $\eta $- and $\chi $-polarizations depend
on a time-like coordinate. Such solutions can be generic off-diagonal and
characterized by nonlinear symmetries. 

\subsubsection{Off-diagonal transforms of cosmological models with
spheroidal symmetry and voids}

Let us consider a 4-d Minkowski spacetime endowed with \textbf{prolate}
spheroidal coordinates $u^{\alpha }=(r,\theta ,\phi ,t)$. Respective
Cartesian coordinates can be defined in the form $u^{\alpha }=(x=r\sin
\theta \cos \phi ,y=r\sin \theta \sin \phi ,z=\sqrt{r^{2}+r_{\lozenge }^{2}}%
\cos \theta ,t)$, where the constant parameter $r_{\lozenge }$ has the
meaning of the distance of the foci from the origin of the coordinate
system. For any fixed $r=\ _{0}r$, we can define a prolate spheroid (i.e. a
rotoid, or ellipsoid) with the foci along the $z$-axis, when 
\begin{equation*}
\frac{x^{2}+y^{2}}{(\ _{0}r)^{2}}+\frac{z^{2}}{(\ _{0}r)^{2}+r_{\lozenge
}^{2}}=1.
\end{equation*}%
Such a $\ _{0}r$ corresponds to the length of its minor radius and the size
of its major radius is $\sqrt{(\ _{0}r)^{2}+r_{\lozenge }^{2}}.$ In prolate
coordinates, the flat Minkowski spacetime metric can be written 
\begin{equation*}
ds^{2}=(r^{2}+r_{\lozenge }^{2}\sin ^{2}\theta )(\frac{dr^{2}}{%
r^{2}+r_{\lozenge }^{2}}+d\theta ^{2})+r^{2}\sin ^{2}\theta d\phi -dt^{2}.
\end{equation*}

In a similar form, we can introduce \textbf{oblate} coordinates, when $x=%
\sqrt{r^{2}+r_{\lozenge }^{2}}\sin \theta \cos \phi ,$\newline
$y=\sqrt{r^{2}+r_{\lozenge }^{2}}\sin \theta \sin \phi ,z=r\cos \theta .$
So, for a fixed $r=\ _{0}r,$ an oblate spheroid with a $z$ symmetric axis, 
\begin{equation*}
\frac{x^{2}+y^{2}}{(\ _{0}r)^{2}+r_{\lozenge }^{2}}+\frac{z^{2}}{(\
_{0}r)^{2}}=1,
\end{equation*}%
can be defined. For such a hypersurface, the value $\sqrt{%
r^{2}+r_{\lozenge}^{2}}$ corresponds to the major radius and $_{0}r$ is the
minor one. Correspondingly, the flat Minkowki spacetime metric can be
written in the form 
\begin{equation*}
ds^{2}=(r^{2}+r_{\lozenge }^{2}\cos ^{2}\theta )(\frac{dr^{2}}{%
r^{2}+r_{\lozenge }^{2}}+d\theta ^{2})+r^{2}\sin ^{2}\theta d\phi -dt^{2}.
\end{equation*}

We consider a quadratic element introduced in \cite{boero16}: 
\begin{eqnarray}
d\underline{s}^{2} &=&\frac{a^{2}(t)}{[1+\frac{\varsigma }{4}%
(r^{2}+r_{\lozenge }^{2}\cos ^{2}\theta )]^{2}}[(r^{2}+r_{\lozenge }^{2}\sin
^{2}\theta )(\frac{dr^{2}}{r^{2}-\frac{M(r)}{r}(r^{2}+r_{\lozenge }^{2}\sin
^{2}\theta )+r_{\lozenge }^{2}}+d\theta ^{2})  \notag \\
&&+r^{2}\sin ^{2}\theta d\phi ]-B(r)dt^{2},%
\mbox{ with prolate spheroidal
symmetry};  \label{cosmvoidm1}
\end{eqnarray}%
\begin{eqnarray*}
d\underline{s}^{2} &=&\frac{a^{2}(t)}{[1+\frac{\varsigma }{4}%
(r^{2}+r_{\lozenge }^{2}\sin ^{2}\theta )]^{2}}[(r^{2}+r_{\lozenge }^{2}\sin
^{2}\theta )(\frac{dr^{2}}{r^{2}-\frac{M(r)}{r}(r^{2}+r_{\lozenge }^{2}\cos
^{2}\theta )+r_{\lozenge }^{2}}+d\theta ^{2}) \\
&&+(r^{2}+r_{\lozenge }^{2})\sin ^{2}\theta d\phi ]-B(r)dt^{2},%
\mbox{with
oblate spheroidal symmetry}.
\end{eqnarray*}%
The conditions For $B(r)=1$ and $M(r)=0$ are used if the above formulas
define respective FLRW cosmological quadratic line elements, when $\varsigma
=1,0,-1 $ refer respectively to a positive curved, flat, hyperbolic space
geometry. 

The mass profile function $M(r)$ from (\ref{cosmvoidm1}) can be specified as
in \cite{amendola98} (for simplicity, we can state $B(r)=1$),%
\begin{equation*}
M(r)=\left\{ 
\begin{array}{cc}
\frac{4\pi }{3}\rho _{int}r^{3}, & \mbox{ for }r<\ _{v}r; \\ 
M(\ _{v}r)+\frac{4\pi }{3}\rho _{bor}(r^{3}-\ _{v}r^{3}), & \mbox{ for }\
_{v}r\leq r<\ _{v}r+\ _{w}r; \\ 
0 & \mbox{ for }\ _{v}r+\ _{w}r\leq r.%
\end{array}%
\right.
\end{equation*}%
Two important constants have such meaning: $\ _{v}r$ is associated with the
radius of the void and $\ _{w}r$ is related to the size of the wall. We can
model such a profile in a form that the border compensates for the amount
missing in the void (i.e. it models a compensated void) by choosing the
spherical symmetry. The internal density of the matter, $\rho _{int},$ and
border density of matter, $\rho _{bor},$ are related to the mean density
outside the void, $\rho _{0}$. Using the formulas%
\begin{equation}
\rho _{int}=-\rho _{0}\xi \mbox{ and }\rho _{bor}=\rho _{0}\xi /[(1+\
_{w}r/\ _{v}r)^{3}-1],  \label{emtvoid}
\end{equation}%
for a constant parameter $\xi <1$, a cosmological metric (\ref{cosmvoidm1})
is a solution of the Einstein equations in GR if $a(t)$ is defined by the
Friedman equations, $\frac{3}{a^{2}(t)}[\frac{da}{dt}+\varsigma ]=8\pi \rho
_{0}.$ Parameterizing $\ B(r)=B_{0}[B_{1}+\ln (\frac{r}{r_{\lozenge }}%
)]^{2}, $ for some constants $B_{0}$ and $B_{1}$, we can use such solutions
to explain certain phenomenology for astrophysical systems with DM as in 
\cite{galo12}. The value of $B_{1}$ can be fixed in a form that the
component $T_{r}^{r}=T_{1}^{1}$ of the energy-momentum tensor remains of the
same order as $\rho _{0}$ (they fix $B_{1}=10^{7}$). Choosing
phenomenological parameters $\ _{w}r=0.3\ _{v}r, \xi =0.1, r_{\lozenge
}=0.1\ _{v}r$, when a radius $\ _{v}r$ corresponds to a physical size of
22Mpc. 

We can re-define (\ref{cosmvoidm1}) using some local coordinates with
non-trivial N-connection coefficients $\underline{\mathring{N}}_{i}^{a}= 
\underline{\mathring{N}}_{i}^{a}(u^{\alpha }(r,\theta ,\phi ,t)) $ and 
\underline{$\mathring{g}$}$_{\alpha \beta }(u^{j}(r,\theta ,\phi
,t),u^{4}(r,\theta ,\phi ,t))$. Such coordinate transforms can be defined in
any form not involving singular frame transforms and off-diagonal
deformations. For such conditions, we can apply the AFCDM to generate new
classes of locally anisotropic cosmological solutions. Such new coordinates
are defined $u^{1}=x^{1}=r,u^{2}=\theta ,$ and $u^{3}=y^{3}=y^{3}(r,\theta
,\phi )$ and $u^{4}=y^{4}=t+\ ^{4}B(r,\theta),$ when (for $\underline{%
\mathring{N}}_{i}^{3}=-\partial \ y^{3}/\partial x^{i}$ and $\underline{%
\mathring{N}}_{i}^{4}=-\partial \ ^{4}B/\partial x^{i}):$ 
\begin{eqnarray*}
\underline{\mathbf{\mathring{e}}}^{3} &=&du^{3}+\underline{\mathring{N}}%
_{i}^{3}(r,\theta )dx^{i}=du^{3}+\underline{\mathring{N}}_{1}^{3}(r,\theta
)dr+\underline{\mathring{N}}_{2}^{3}(r,\theta )d\theta , \\
\underline{\mathbf{\mathring{e}}}^{4} &=&du^{4}+\underline{\mathring{N}}%
_{i}^{4}(r,\theta )dx^{i}=du^{4}+\underline{\mathring{N}}_{1}^{4}(r,\theta
)dr+\underline{\mathring{N}}_{2}^{4}(r,\theta )dz.
\end{eqnarray*}
This way, we obtain an off-diagonal spheroid-type cosmological metric
parameterized as a d-metric, 
\begin{eqnarray}
d\underline{\mathring{s}}^{2} &=&\underline{\mathring{g}}_{\alpha }(r,\theta
,t)[\underline{\mathbf{\mathring{e}}}^{\alpha }(r,\theta ,t)]^{2},%
\mbox{where for }\left\{ 
\begin{array}{c}
\mbox{ prolate }: \\ 
\mbox{ oblate }:%
\end{array}%
\right.  \label{cosmvoidm2} \\
\underline{\mathring{g}}_{1}(r,\theta ,t) &=&\left\{ 
\begin{array}{c}
\frac{a^{2}(t)(r^{2}+r_{\lozenge }^{2}\sin ^{2}\theta )}{[1+\frac{\varsigma 
}{4}(r^{2}+r_{\lozenge }^{2}\cos ^{2}\theta )]^{2}[r^{2}-\frac{M(r)}{r}%
(r^{2}+r_{\lozenge }^{2}\sin ^{2}\theta )+r_{\lozenge }^{2}]} \\ 
\frac{a^{2}(t)(r^{2}+r_{\lozenge }^{2}\sin ^{2}\theta )}{[1+\frac{\varsigma 
}{4}(r^{2}+r_{\lozenge }^{2}\sin ^{2}\theta )]^{2}[r^{2}-\frac{M(r)}{r}%
(r^{2}+r_{\lozenge }^{2}\cos ^{2}\theta )+r_{\lozenge }^{2}]}%
\end{array}%
\right. ,  \notag \\
\underline{\mathring{g}}_{2}(r,\theta ,t) &=&\left\{ 
\begin{array}{c}
\frac{a^{2}(t)}{[1+\frac{\varsigma }{4}(r^{2}+r_{\lozenge }^{2}\cos
^{2}\theta )]^{2}} \\ 
\frac{a^{2}(t)}{[1+\frac{\varsigma }{4}(r^{2}+r_{\lozenge }^{2}\sin
^{2}\theta )]^{2}}%
\end{array}%
\right. ,\ \underline{\mathring{g}}_{3}(r,\theta ,t)=\left\{ 
\begin{array}{c}
\frac{a^{2}(t)r^{2}\sin ^{2}\theta }{[1+\frac{\varsigma }{4}%
(r^{2}+r_{\lozenge }^{2}\cos ^{2}\theta )]^{2}} \\ 
\frac{a^{2}(t)(r^{2}+r_{\lozenge }^{2})\sin ^{2}\theta }{[1+\frac{\varsigma 
}{4}(r^{2}+r_{\lozenge }^{2}\sin ^{2}\theta )]^{2}}%
\end{array}%
\right. ,\underline{\ \mathring{g}}_{4}(r)=-B(r).  \notag
\end{eqnarray}

\subsubsection{Off-diagonal cosmological solitonic FLH evolution encoding
2-d vertices}

To generate such configurations, we consider nonholonomic deformations of
data $(\underline{\mathring{g}}_{\alpha},\underline{\mathring{N}}%
_{i}^{a})\rightarrow (\underline{g}_{\alpha }=\underline{\eta }_{\alpha }%
\underline{\mathring{g}}_{\alpha },$ $\underline{N}_{i}^{a}=\underline{\eta }%
_{i}^{a}\underline{\mathring{N}}_{i}^{a})$ using underlined versions of
formulas (\ref{offdiagpm}), (\ref{offdiagdefr}) with nonlinear symmetries (%
\ref{dualnonltr}). The gravitational polarizations $\underline{\eta }%
_{i}(r,\theta,t)=a^{-2}(t)\eta _{i}(r,\theta ),\underline{\eta }%
_{3}(r,\theta,t)=a^{-2}(t)\underline{\eta } (r,\theta ,t)$ and $\underline{%
\eta }_{4}(r,\theta ,t)$ will be prescribed or computed in such forms that 
\begin{eqnarray}
\underline{\mathbf{g}} &=&(g_{i},g_{b},\underline{N}_{i}^{3}=\underline{n}%
_{i},\underline{N}_{i}^{4}=\underline{w}_{i})=g_{i}(r,\theta )dx^{i}\otimes
dx^{i}+\underline{h}_{3}(r,\theta ,t)\underline{\mathbf{e}}^{3}\otimes 
\underline{\mathbf{e}}^{3}+\underline{h}_{4}(r,\theta ,t)\underline{\mathbf{e%
}}^{4}\otimes \underline{\mathbf{e}}^{4},  \label{lacosm1} \\
&&\underline{\mathbf{e}}^{3}=d\phi +\underline{n}_{i}(r,\theta ,t)dx^{i},\ 
\underline{\mathbf{e}}^{4}=dt+\underline{w}_{i}(r,\theta ,t)dx^{i},  \notag
\end{eqnarray}%
with Killing symmetry on the angular coordinate $\varphi ,$ when $\partial
_{\varphi }$ transforms into zero the N-adapted coefficients of such a
d-metric. 

In terms of $\eta $-polarization functions, we can consider an off-diagonal
cosmological ansatz stated in a $t$-dual form to (\ref{cosm1d}), see also
Table 3 in the Appendix, when 
\begin{eqnarray}
d\widehat{\underline{s}}^{2} &=&\ ^{c}\widehat{\underline{g}}_{\alpha \beta
}(r,\theta ,t;\underline{\mathring{g}}_{\alpha };\psi ,\eta _{3};\ 
\underline{\Lambda },\ _{Q}^{1}\underline{J},\ _{Q}^{2}\underline{J}%
)du^{\alpha }du^{\beta }=e^{\psi }[(dx^{1})^{2}+(dx^{2})^{2}]
\label{lacosm2} \\
&&+(\underline{\eta }\underline{\mathring{g}}_{3})\{d\phi +[\ _{1}n_{k}+\
_{2}n_{k}\int dt\frac{[\partial _{t}(\underline{\eta }\underline{\mathring{g}%
}_{3})]^{2}}{|\int dt\ \ _{Q}^{2}\underline{J}\partial _{t}(\underline{\eta }%
\underline{\mathring{g}}_{3})|\ (\underline{\eta }\underline{\mathring{g}}%
_{3})^{5/2}}]dx^{k}\}^{2}  \notag \\
&&-\frac{[\partial _{t}(\underline{\eta }\ \underline{\mathring{g}}_{3})]^{2}%
}{|\int dt\ \ _{Q}^{2}\underline{J}\partial _{t}(\underline{\eta }\underline{%
\mathring{g}}_{3})|\ \eta \mathring{g}_{3}}\{dt+\frac{\partial _{i}[\int dt\
\ _{Q}^{2}\underline{J}\ \partial _{t}(\underline{\eta }\underline{\mathring{%
g}}_{3})]}{\ _{Q}^{2}\underline{J}\partial _{t}(\underline{\eta }\underline{%
\mathring{g}}_{3})}dx^{i}\}^{2}.  \notag
\end{eqnarray}%
Using $\underline{\Phi }^{2}=-4\ \underline{\Lambda }\underline{g}_{3},$ we
can transform (\ref{lacosm2}) in a variant of (\ref{qeltorsc}) with
underlined $\eta $-polarizations determined by the generating data $(%
\underline{g}_{3};\ \underline{\Lambda },\ _{Q}^{2}\underline{J}).$ The
effective cosmological constant $\underline{\Lambda }$ is chosen as the
effective one related via nonlinear symmetries (\ref{dualnonltr}) to an
energy-momentum tensor (\ref{emtvoid}) in a fluid type form. For such
cosmological configurations, the respective generating sources $(\ \ _{Q}^{1}%
\underline{J},\ _{Q}^{2}\underline{J})$ are related to a $T_{\alpha \beta }$
via respective frame or coordinate transforms. Locally anisotropic
cosmological scenarios with nonholonomic evolution from a primary void
configuration (\ref{cosmvoidm2}) are determined by corresponding classes of
generating polarization functions $\psi \simeq \psi (x^{k})$ and $\underline{%
\eta }\ \simeq \underline{\eta }(x^{k},t).$ 

Let us consider such an explicit example: We prescribe that the h-part of a
s-metric (\ref{lacosm2}) must satisfy the generalized Taubes equation for
vortices on a curved background 2-d surface, 
\begin{equation}
_{h}\nabla ^{2}\psi =\Omega _{0}(C_{0}-C_{1}e^{2\psi }).  \label{taubeq}
\end{equation}%
In (\ref{taubeq}), the position-dependent conformal factor $\Omega _{0}$ and
the effective source $(C_{0}-C_{1}e^{2\psi })$ are prescribed as respective
generating h-function $\psi (x^{k})$ and generating h-source $\ ^{h}%
\underline{\Upsilon }(x^{k}).$ We can re-scale both constants $C_{0}$ and $%
C_{1}$ to take standard values $-1,0,$ or 1, but there are only five
combinations of these values that allow vortex solutions $\psi \lbrack
vortex]$ without singularities \cite{manton17}. In a different form, the
v-part of (\ref{lacosm2}) can be modelled as a solitonic wave when 
\begin{equation}
\underline{\eta }\ \simeq \left\{ 
\begin{array}{ccc}
_{r}^{sol}\underline{\eta }(r,t) & 
\mbox{ as a solution of the modified KdV
equation }\frac{\partial \underline{\eta }}{\partial t}-6\underline{\eta }%
^{2}\frac{\partial \underline{\eta }}{\partial r}+\frac{\partial ^{3}%
\underline{\eta }}{\partial r^{3}}=0, & \mbox{  radial solitons}; \\ 
_{\theta }^{sol}\underline{\eta }(\theta ,t) & 
\mbox{ as a solution of the
modified KdV equation }\frac{\partial \underline{\eta }}{\partial \theta }-6%
\underline{\eta }^{2}\frac{\partial \underline{\eta }}{\partial \theta }+%
\frac{\partial ^{3}\underline{\eta }}{\partial \,^{3}}=0, & 
\mbox{ angular
solitons}.%
\end{array}%
\right.  \label{solitonw}
\end{equation}%
The references \cite{vacaru18,bsssvv25,doikou20} contains many examples of
such solitonic wave equations in MGTs. 

So, we conclude that the generic off-diagonal metrics (\ref{lacosm2}) can
describe nonholonomic cosmological evolution scenarios with conventional h-
and v-splitting for a (2+2)-configuration. In the above example, a primary
2-d metric with prolate/oblate rotoid void transforms into a vertex
h-configuration (\ref{taubeq}). Differently, the v-part is defined by a
solitonic wave evolution of type (\ref{solitonw}). On a base 4-d Lorentz
spacetime cosmological manifold, this describes a geometric evolution with
gravitational polarizations and for respective generating sources. Such a
nonholonomic cosmological evolution results also in solitonic configurations
for the N-connection coefficients. Corresponding 4-d spacetime cosmological
solitonic waves on $t$-variable can be with a radial space variable, $r,$ or
with an angular variable, $\theta .$ In a series of our and co-authors
works, there were constructed more general classes of generic off-diagonal
cosmological and quasi-stationary solutions with 3-d solitonic waves and
solitonic hierarchies in GR and MGTs, see reviews \cite{vacaru18,partner06}.
In \cite{bsssvv25} such solitionic cosmological solutions are considered for
nonassociative Finsler-like MGTs for modelling DM quasi-periodic and
pattern-forming structures. In this subsection, those results were modified
in such forms that analogous cosmological solutions are constructed in the
framework of the Einstein gravity theory. 

\subsection{Double off-diagonal (4+4)-d cosmological solitonic and spheroid
cosmological involving 2-d vertices}

The d-metrics (\ref{cosmvoidm2}) and (\ref{lacosm2}) can be extended on 8-d
phase spaces to describe FLH cosmological configurations involving
respectively base spacetime and typical (co) fiber solitons and vertices. In
this subsection, we consider respective s-adapted prolate and oblate
coordinates for or $\ ^{\shortmid }u^{\alpha _{s}}=(r,\theta ,\phi ,t,\
^{\shortmid }r,\ ^{\shortmid }\theta ,\ ^{\shortmid }\phi ,E)$ on $\
^{\shortmid s}\mathcal{M}$ are 
\begin{equation}
d\ ^{\shortmid }\underline{\mathring{s}}^{2}=\ ^{\shortmid }\underline{%
\mathring{g}}_{\alpha _{s}}(r,\theta ,t,\ ^{\shortmid }r,\ \ ^{\shortmid
}\theta ,\ \ ^{\shortmid }\phi ,\ E)[\underline{\mathbf{\mathring{e}}}%
^{\alpha _{s}}(r,\theta ,t,\ ^{\shortmid }r,\ \ ^{\shortmid }\theta ,\ \
^{\shortmid }\phi ,\ E)]^{2},\mbox{where for }\left\{ 
\begin{array}{c}
\mbox{ prolate }: \\ 
\mbox{ oblate }:%
\end{array}%
\right. ,  \label{cosmvoidm2f}
\end{equation}%
where 
\begin{eqnarray*}
&&\underline{\mathring{g}}_{1}(r,\theta ,t),\underline{\mathring{g}}%
_{2}(r,\theta ,t),\underline{\mathring{g}}_{3}(r,\theta ,t),\underline{\ 
\mathring{g}}_{4}(r)\mbox{ as in }(\ref{cosmvoidm2})\mbox{ and } \\
\ ^{\shortmid }\underline{\mathring{g}}^{5}(\ ^{\shortmid }r,\ \ ^{\shortmid
}\theta ,\ \ E) &=&\left\{ 
\begin{array}{c}
\frac{\ ^{\shortmid }a^{2}(E)(\ ^{\shortmid }r^{2}+\ ^{\shortmid
}r_{\lozenge }^{2}\sin ^{2}\ ^{\shortmid }\theta )}{[1+\frac{\ ^{\shortmid
}\varsigma }{4}(\ ^{\shortmid }r^{2}+\ ^{\shortmid }r_{\lozenge }^{2}\cos
^{2}\ ^{\shortmid }\theta )]^{2}[\ ^{\shortmid }r^{2}-\frac{\ ^{\shortmid
}M(\ ^{\shortmid }r)}{\ ^{\shortmid }r}(\ ^{\shortmid }r^{2}+\ ^{\shortmid
}r_{\lozenge }^{2}\sin ^{2}\ ^{\shortmid }\theta )+\ ^{\shortmid
}r_{\lozenge }^{2}]} \\ 
\frac{\ ^{\shortmid }a^{2}(E)(\ ^{\shortmid }r^{2}+\ ^{\shortmid
}r_{\lozenge }^{2}\sin ^{2}\ ^{\shortmid }\theta )}{[1+\frac{\ ^{\shortmid
}\varsigma }{4}(\ ^{\shortmid }r^{2}+\ ^{\shortmid }r_{\lozenge }^{2}\sin
^{2}\ ^{\shortmid }\theta )]^{2}[\ ^{\shortmid }r^{2}-\frac{\ ^{\shortmid
}M(\ ^{\shortmid }r)}{\ ^{\shortmid }r}(\ ^{\shortmid }r^{2}+\ ^{\shortmid
}r_{\lozenge }^{2}\cos ^{2}\ ^{\shortmid }\theta )+\ ^{\shortmid
}r_{\lozenge }^{2}]}%
\end{array}%
\right. ,
\end{eqnarray*}%
\begin{equation*}
\ ^{\shortmid }\underline{\mathring{g}}^{6}(\ ^{\shortmid }r,\ ^{\shortmid
}\theta ,E)=\left\{ 
\begin{array}{c}
\frac{\ ^{\shortmid }a^{2}(t)}{[1+\frac{\ ^{\shortmid }\varsigma }{4}(\
^{\shortmid }r^{2}+\ ^{\shortmid }r_{\lozenge }^{2}\cos ^{2}\ ^{\shortmid
}\theta )]^{2}} \\ 
\frac{\ ^{\shortmid }a^{2}(t)}{[1+\frac{\ ^{\shortmid }\varsigma }{4}(\
^{\shortmid }r^{2}+\ ^{\shortmid }r_{\lozenge }^{2}\sin ^{2}\ ^{\shortmid
}\theta )]^{2}}%
\end{array}%
\right. ,\ ^{\shortmid }\underline{\mathring{g}}^{7}(\ ^{\shortmid }r,\
^{\shortmid }\theta ,E)=\left\{ 
\begin{array}{c}
\frac{\ ^{\shortmid }a^{2}(E)r^{2}\sin ^{2}\ ^{\shortmid }\theta }{[1+\frac{%
\ ^{\shortmid }\varsigma }{4}(\ ^{\shortmid }r^{2}+\ ^{\shortmid
}r_{\lozenge }^{2}\cos ^{2}\ ^{\shortmid }\theta )]^{2}} \\ 
\frac{\ ^{\shortmid }a^{2}(E)(\ ^{\shortmid }r^{2}+\ ^{\shortmid
}r_{\lozenge }^{2})\sin ^{2}\ ^{\shortmid }\theta }{[1+\frac{\ ^{\shortmid
}\varsigma }{4}(\ ^{\shortmid }r^{2}+\ ^{\shortmid }r_{\lozenge }^{2}\sin
^{2}\ ^{\shortmid }\theta )]^{2}}%
\end{array}%
\right. ,\ ^{\shortmid }\underline{\mathring{g}}^{8}(\ ^{\shortmid }r)=-\
^{\shortmid }B(\ ^{\shortmid }r).
\end{equation*}%
Such a prime s-metric is can be written in a similar form on $\ ^{s}\mathcal{%
M}$ using analogous velocity variables.

We generate 8-d extensions of (\ref{lacosm2}) using nonholonomic s-adapted
deformations 
\begin{equation*}
(\ ^{\shortmid }\underline{\mathring{g}}_{\alpha _{s}},\ ^{\shortmid }%
\underline{\mathring{N}}_{i_{s-1}}^{a_{s}})\rightarrow (\ ^{\shortmid }%
\underline{g}_{\alpha _{s}}=\ ^{\shortmid }\underline{\eta }_{\alpha _{s}}\
^{\shortmid }\underline{\mathring{g}}_{\alpha _{s}},\ ^{\shortmid }%
\underline{N}_{i_{s-1}}^{a_{s}}=\ ^{\shortmid }\underline{\eta }%
_{i_{s-1}}^{a_{s}}\ ^{\shortmid }\underline{\mathring{N}}_{i_{s-1}}^{a_{s}})
\end{equation*}%
of prime s-metric (\ref{cosmvoidm2f}). The gravitational polarizations in
the total phase space are parameterized $\underline{\eta }_{i}(r,\theta
,t)=a^{-2}(t)\eta _{i}(r,\theta ),\underline{\eta }_{3}(r,\theta
,t)=a^{-2}(t)\underline{\eta }(r,\theta ,t)$ and $\underline{\eta }%
_{4}(r,\theta ,t);$ and, on co-fiber space, \newline
$\ ^{\shortmid }\underline{\eta }^{a_{3}}(\ ^{\shortmid }r,\ ^{\shortmid
}\theta ,\ \ E)=\ ^{\shortmid }a^{-2}(E)\eta ^{a_{3}}(r,\theta ),\underline{%
\eta }^{7}(\ ^{\shortmid }r,\ \ ^{\shortmid }\theta ,\ \ E)=\ ^{\shortmid
}a^{-2}(E)\ ^{\shortmid }\underline{\eta }(\ ^{\shortmid }r,\ \ ^{\shortmid
}\theta ,\ \ E)$ and $\ ^{\shortmid }\underline{\eta }^{4}(\ ^{\shortmid
}r,\ \ ^{\shortmid }\theta ,\ \ E).$ The generating functions $\underline{%
\eta }_{3}(r,\theta ,t),$ $\ ^{\shortmid }\underline{\eta }^{6}(\
^{\shortmid }r,\ ^{\shortmid }\theta ,\ \ E)$ and $\ ^{\shortmid }\underline{%
\eta }^{7}(\ ^{\shortmid }r,\ ^{\shortmid }\theta ,\ \ E)$ define
cosmological solutions of type (\ref{lc8cstp8}), see Table 13 in Appendix.
The corresponding off-diagonal nonmetric cosmological rainbow metrics can be
written in the form: 
\begin{eqnarray}
&&d\widehat{\underline{s}}_{[8d]}^{2}(\tau )=\ ^{c}\widehat{\underline{g}}%
_{\alpha _{s}\beta _{s}}(x^{k},t,p_{5},E;\underline{h}_{3},\mathbf{\
^{\shortmid }}\underline{g}^{6},\mathbf{\ ^{\shortmid }}\underline{g}^{7};\
_{1}^{Q}J,\ _{2}^{Q}\underline{J},\ _{3}^{\shortmid Q}J,\ _{4}^{\shortmid Q}%
\underline{J};\ _{1}\Lambda ,\ _{2}\underline{\Lambda },\ _{3}^{\shortmid
}\Lambda ,\ _{4}^{\shortmid }\underline{\Lambda })d\mathbf{\ ^{\shortmid }}%
u^{\alpha _{s}}d\mathbf{\ ^{\shortmid }}u^{\beta _{s}}=  \label{lacosm2ext8d}
\\
&&e^{\psi (x^{k},\ \ _{1}^{Q}J)}[(dx^{1})^{2}+(dx^{2})^{2}]+\underline{h}%
_{3}[dy^{3}+(\ _{1}n_{k_{1}}+4\ _{2}n_{k_{1}}\int dt\frac{((a^{-2}\underline{%
\eta }\underline{\mathring{g}}_{3}){}^{\diamond _{2}})^{2}}{|\int dt\ \
_{2}^{Q}\underline{J}(a^{-2}\underline{\eta }\underline{\mathring{g}}%
_{3}){}^{\diamond _{2}}|(a^{-2}\underline{\eta }\underline{\mathring{g}}%
_{3})^{5/2}})dx^{k_{1}}]+  \notag \\
&&\frac{((a^{-2}\underline{\eta }\underline{\mathring{g}}_{3}){}^{\diamond
_{2}})^{2}}{|\int dt\ \ _{2}^{Q}\underline{J}(a^{-2}\underline{\eta }%
\underline{\mathring{g}}_{3}){}^{\diamond _{2}}|\ a^{-2}\underline{\eta }%
\underline{\mathring{g}}_{3}}[dt+\frac{\partial _{i_{1}}(\int dt\ \ _{2}^{Q}%
\underline{J}\ (a^{-2}\underline{\eta }\underline{\mathring{g}}%
_{3}){}^{\diamond _{2}}])}{\ \ \ _{2}^{Q}\underline{J}\ (a^{-2}\underline{%
\eta }\underline{\mathring{g}}_{3}){}^{\diamond _{2}}}dx^{i_{1}}]+  \notag \\
&&\frac{(\mathbf{\ ^{\shortmid }}\partial ^{5}(\ ^{\shortmid }a^{-2}(E)\
^{\shortmid }\underline{\eta }\ ^{\shortmid }\underline{\mathring{g}}%
^{6}))^{2}}{|\int dp_{5}\mathbf{\ ^{\shortmid }}\partial ^{5}[\ \
_{3}^{\shortmid Q}J\mathbf{\ }(\ ^{\shortmid }a^{-2}(E)\ ^{\shortmid }%
\underline{\eta }\ ^{\shortmid }\underline{\mathring{g}}^{6})]|\ \
^{\shortmid }a^{-2}(E)\ ^{\shortmid }\underline{\eta }\ ^{\shortmid }%
\underline{\mathring{g}}^{6}}\{dp_{5}+\frac{\partial _{i_{2}}[\int dp_{5}(\
_{3}^{\shortmid Q}J)\mathbf{\ ^{\shortmid }}\partial ^{5}(\ ^{\shortmid
}a^{-2}(E)\ ^{\shortmid }\underline{\eta }\ ^{\shortmid }\underline{%
\mathring{g}}^{6})]}{\ \ _{3}^{\shortmid Q}J\mathbf{\ ^{\shortmid }}\partial
^{5}(\ ^{\shortmid }a^{-2}(E)\ ^{\shortmid }\underline{\eta }\ ^{\shortmid }%
\underline{\mathring{g}}^{6})}dx^{i_{2}}\}^{2}+  \notag \\
&&\ \ ^{\shortmid }a^{-2}(E)\ ^{\shortmid }\underline{\eta }\ ^{\shortmid }%
\underline{\mathring{g}}^{6}\{dp_{5}+[\ _{1}n_{k_{2}}+\ _{2}n_{k_{2}}\int
dp_{5}\frac{(\mathbf{\ ^{\shortmid }}\partial ^{5}\mathbf{\ }(\ ^{\shortmid
}a^{-2}(E)\ ^{\shortmid }\underline{\eta }\ ^{\shortmid }\underline{%
\mathring{g}}^{6}))^{2}}{|\int dp_{5}\mathbf{\ ^{\shortmid }}\partial ^{5}[\
_{3}^{\shortmid Q}J\mathbf{\ }(\ ^{\shortmid }a^{-2}(E)\ ^{\shortmid }%
\underline{\eta }\ ^{\shortmid }\underline{\mathring{g}}^{6})]|\ ((\
^{\shortmid }a^{-2}(E)\ ^{\shortmid }\underline{\eta }\ ^{\shortmid }%
\underline{\mathring{g}}^{6}))^{5/2}}]dx^{k_{2}}\}+  \notag \\
&&\mathbf{\ }\ ^{\shortmid }a^{-2}(E)\ ^{\shortmid }\underline{\eta }\
^{\shortmid }\underline{\mathring{g}}^{7}\{dp_{7}+[\ _{1}^{\shortmid }%
\underline{n}_{k_{3}}+\ _{2}^{\shortmid }\underline{n}_{k_{3}}\int dp_{7}%
\frac{(\mathbf{\ ^{\shortmid }}\underline{\partial }^{8}\mathbf{\ (}\
^{\shortmid }a^{-2}(E)\ ^{\shortmid }\underline{\eta }\ ^{\shortmid }%
\underline{\mathring{g}}^{7}))^{2}}{|\int dE\mathbf{\ ^{\shortmid }}%
\underline{\partial }^{8}[\ \ _{4}^{\shortmid Q}\underline{J}\mathbf{\ (}\
^{\shortmid }a^{-2}(E)\ ^{\shortmid }\underline{\eta }\ ^{\shortmid }%
\underline{\mathring{g}}^{7})]|\ (\ ^{\shortmid }a^{-2}(E)\ ^{\shortmid }%
\underline{\eta }\ ^{\shortmid }\underline{\mathring{g}}^{7})^{5/2}}]d%
\mathbf{\ ^{\shortmid }}x^{k_{3}}\}+  \notag \\
&&\frac{(\mathbf{\ ^{\shortmid }}\underline{\partial }^{8}\mathbf{\ (}\
^{\shortmid }a^{-2}(E)\ ^{\shortmid }\underline{\eta }\ ^{\shortmid }%
\underline{\mathring{g}}^{7}))^{2}}{|\int dE\mathbf{\ ^{\shortmid }}%
\underline{\partial }^{8}[\ _{4}^{\shortmid Q}\underline{J}\mathbf{\ (}\
^{\shortmid }a^{-2}(E)\ ^{\shortmid }\underline{\eta }\ ^{\shortmid }%
\underline{\mathring{g}}^{7})]|\ \mathbf{(}\ ^{\shortmid }a^{-2}(E)\
^{\shortmid }\underline{\eta }\ ^{\shortmid }\underline{\mathring{g}}^{7})}%
\{dE+\frac{\partial _{i_{3}}[\int dE(\ \ _{4}^{\shortmid Q}\underline{J})\ 
\mathbf{\ ^{\shortmid }}\underline{\partial }^{8}\mathbf{\ ^{\shortmid }(}\
^{\shortmid }a^{-2}(E)\ ^{\shortmid }\underline{\eta }\ ^{\shortmid }%
\underline{\mathring{g}}^{7})]}{\ \ _{4}^{\shortmid Q}\underline{J}\ \mathbf{%
\ ^{\shortmid }}\underline{\partial }^{8}\mathbf{\ (}\ ^{\shortmid
}a^{-2}(E)\ ^{\shortmid }\underline{\eta }\ ^{\shortmid }\underline{%
\mathring{g}}^{7})}d\mathbf{\ ^{\shortmid }}x^{i_{3}}\}^{2}.  \notag
\end{eqnarray}%
Similar 8-d cosmological configurations can be generated for another sets of
phase space local coordinates or in velocity-type variables as in Tables 12,
8, or 7. In (\ref{lacosm2ext8d}), the solitonic waves from the v-part of (%
\ref{lacosm2}) are completed with solitonic distributions on momentum-like
variables, 
\begin{equation*}
\ ^{\shortmid }\underline{\eta }\ \simeq \left\{ 
\begin{array}{ccc}
_{\ ^{\shortmid }r\ }^{sol}\underline{\eta }(\ ^{\shortmid }r,E) & 
\mbox{ as a solution of the modified KdV
equation }\frac{\partial \ ^{\shortmid }\underline{\eta }}{\partial E}-6\
^{\shortmid }\underline{\eta }^{2}\frac{\partial \ ^{\shortmid }\underline{%
\eta }}{\partial \ ^{\shortmid }r}+\frac{\partial ^{3}\ ^{\shortmid }%
\underline{\eta }}{\partial \ ^{\shortmid }r^{3}}=0; &  \\ 
_{\ ^{\shortmid }\theta }^{sol}\underline{\eta }(\ ^{\shortmid }\theta ,E) & 
\mbox{ as a solution of the
modified KdV equation }\frac{\partial \ ^{\shortmid }\underline{\eta }}{%
\partial E}-6\ ^{\shortmid }\underline{\eta }^{2}\frac{\partial \
^{\shortmid }\underline{\eta }}{\partial \ ^{\shortmid }\theta }+\frac{%
\partial ^{3}\ ^{\shortmid }\underline{\eta }}{\partial \,^{3}\ ^{\shortmid
}\theta }=0. & 
\end{array}%
\right.
\end{equation*}%
Such cosmological momentum type momentum solutions are different from those
studied in \cite{vacaru18,bsssvv25}. 

\subsubsection{Geometric flow thermodynamics for off-diagonal FLH
cosmological solutions}

Locally anisotropic cosmological solutions on base spacetime Lorentz
manifold (which can also be considered in GR) are defined by off-diagonal
metrics with conventional underlined coefficients can be generated as we
summarized in Table 3 from Appendix \ref{appendixb}. A typical ansatz of
d-metrics (\ref{lacosm2}) in a dual on a time-variant of (\ref{qeltorsc}) as
stated by formulas (\ref{dualnonltr}). For instance, the generating
functions are related by formulas $\underline{\Phi }^{2}=-4\ \underline{%
\Lambda }\underline{g}_{3},$ with underlined $\eta $-polarizations
determined by the generating data $(\ ^{c}\underline{g}_{3}=\underline{\eta }%
_{3}\underline{\mathring{g}}_{3};\ \underline{\Lambda },\ \ _{Q}\yen _{\ \
\beta }^{\alpha }),$ see (\ref{canriccisol}). For relativistic geometric
flows, such formulas are generalized for $\tau $-families, for instance,
written as $\underline{\Phi }^{2}(\tau )=-4\ \underline{\Lambda }(\tau )\
^{c}\underline{g}_{3}(\tau )$ and $(\ ^{c}\underline{g}_{3}(\tau )=%
\underline{\eta }_{3}(\tau )\underline{\mathring{g}}_{3};\ \underline{%
\Lambda }(\tau ),\ _{Q}^{v}\underline{J}(\tau )),$ see respective systems of
PDEs and sources (\ref{cfeq4af}) and (\ref{feq4afd}). 

For time dual transforms,$\ ^{q}\mathbf{g}(\tau )\rightarrow \ ^{c}%
\underline{\mathbf{g}}(\tau ),$ of (\ref{thermvar2}) and (\ref{volumf1}), we
obtain such formulas for geometric thermodynamic variables of $\tau $%
-families of locally anisotropic cosmological solutions $\ ^{c}\underline{%
\mathbf{g}}(\tau )$ (\ref{lacosm2}): 
\begin{eqnarray}
\ ^{c}\widehat{\underline{Z}}(\tau ) &=&\exp \left[ \frac{1}{8\pi ^{2}\tau
^{2}}\ \ _{\eta }^{\mathbf{J}}\underline{\mathcal{V}}[\ ^{c}\underline{%
\mathbf{g}}(\tau )]\right] ,\ ^{c}\widehat{\underline{\mathcal{E}}}\ (\tau
)=\ \frac{1-2\tau \ \underline{\Lambda }(\tau )}{8\pi ^{2}\tau }\ \ \ _{\eta
}^{\mathbf{J}}\underline{\mathcal{V}}\mathcal{[}\ ^{c}\underline{\mathbf{g}}%
(\tau )],  \label{thermovar2c} \\
\ \ \ \ ^{c}\widehat{\underline{S}}(\tau ) &=&-\ ^{c}\widehat{\underline{W}}%
(\tau )=\frac{1-\underline{\Lambda }(\tau )}{4\pi ^{2}\tau ^{2}}\ \ _{\eta
}^{\mathbf{J}}\underline{\mathcal{V}}\mathcal{[}\ ^{c}\underline{\mathbf{g}}%
(\tau )],\mbox{ where }  \notag \\
\ _{\eta }^{\mathbf{J}}\underline{\mathcal{V}}\mathcal{[}\ ^{c}\underline{%
\mathbf{g}}(\tau )] &=&\int_{\ \widehat{\Xi }}\delta \ _{\eta }\underline{%
\mathcal{V}}(\ ^{v}\widehat{\underline{\mathbf{J}}}(\tau ),\ \underline{\eta 
}_{\alpha }(\tau ),\underline{\mathring{g}}_{\alpha }).  \notag
\end{eqnarray}%
These formulas can be used for defining and computing the thermodynamic
characteristics of nonholonomic Ricci soliton cosmological configurations
constructed for $\tau =\tau _{0}$. They encode nonmetric deformations in the
effective sources and nonlinear symmetries relating via off-diagonal
configurations the generating data $\ \underline{\Lambda }$ and$\ \ _{Q}\yen %
_{\ \ \beta }^{\alpha }.$

For general $\eta $-polarizations on a phase space $\ _{Q}^{\shortmid s}%
\mathcal{M}$ determined by cosmological s-metrics $\ _{sQ}^{c}\underline{%
\mathbf{g}}(\tau )$ (\ref{lacosm2ext8d}), the thermodynamic variables (\ref%
{thermovar2c}) are generalized in 8-d form and computed: 
\begin{eqnarray*}
\ _{Q}^{c}\widehat{\underline{Z}}(\tau ) &=&\exp \left[ \frac{1}{(4\pi \tau
)^{4}}\ \ _{\eta }^{\mathbf{J}}\underline{\mathcal{V}}[\ \ _{sQ}^{c}%
\underline{\mathbf{g}}(\tau )]\right] , \\
\ _{Q}^{c}\widehat{\underline{\mathcal{E}}}\ (\tau ) &=&\frac{1}{64\pi
^{4}\tau ^{3}}\ \left( 1-2\tau (\ _{Q}^{h}\underline{\Lambda }(\tau )+\
_{Q}^{v}\underline{\Lambda }(\tau ))\right) \ \ \ _{\eta }^{\mathbf{J}}%
\underline{\mathcal{V}}[\ _{sQ}^{c}\underline{\mathbf{g}}(\tau )], \\
\ \ \ \ _{Q}^{c}\widehat{\underline{S}}(\tau ) &=&-\ _{Q}^{c}\widehat{%
\underline{W}}(\tau )=\frac{2}{(4\pi \tau )^{4}}(1-4(\ _{Q}^{h}\underline{%
\Lambda }(\tau )+\ _{Q}^{v}\underline{\Lambda }(\tau )))\ \ _{\eta }^{%
\mathbf{J}}\underline{\mathcal{V}}\mathcal{[}\ \ _{sQ}^{c}\underline{\mathbf{%
g}}(\tau )].
\end{eqnarray*}%
Similar values can be computed using decompositions on a small $\kappa $%
-parameter, $\ _{\chi }^{J}\widehat{\underline{Z}}[\ _{\kappa Q}^{c}%
\underline{\mathbf{g}}(\tau )]$ or $\ _{\chi }^{J}\underline{\mathcal{V}}[\
\ _{\kappa Q}^{c}\underline{\mathbf{g}}(\tau )].$ 

\section{Conclusions and perspectives}

\label{sec6} This is a review article on metric and nonmetric geometric
Finsler-Lagrange-Hamilton (FLH) flows and off-diagonal deformations of
Einstein gravity, with applications and original results in modified gravity
theories (MGTs). Such nonholonomic geometric models are constructed on
relativistic phase spaces modelled as (co) tangent Lorentz bundles. We
elaborate on natural and physically motivated (non) metric FLH extensions of
the general relativity (GR) theory and standard particle physics models
using conventional base spacetime and (co) fiber (momentum) velocity
variables and coordinates. This work focuses on mathematical gravity and
accelerating cosmology, and modern astrophysics of MGTs, developing a new
advanced geometric techniques for generating new classes of exact and
parametric solutions. It is also a progress report on the anholonomic frame
and connection deformation method (AFCDM) for constructing generic
off-diagonal exact and parametric solutions in FLH MGTs. A series of ideas,
methods and results was proposed many years ago beginning in 1986 and was
developed by the author and his co-authors (master and postgraduate
students) in Eastern Europe. Later, such directions in mathematical physics
and geometry were supported by many NATO, CERN, Fulbright and Scholars at
Risk research grants beginning in 1999. The main goals of this article ares
to summarize transfer of knowledge, with a review and critical discussion of
new results on FLH gravity, and to provide detailed proofs and new
applications of the AFCDM to FLH geometric flow and MGTs. Substantially new
are the results and methods related to nonmetric geometric and gravitational
models, generalized G. Perelman thermodynamics for nonmetric theories, and a
study of new classes of off-diagonal Finsler-like deformed black hole (BH),
wormhole (WH), black torus (BT) and locally anisotropic cosmological
solutions. This work complements in nonmetric commutative and associative
FLH forms the review \cite{vacaru18} (based on metric compatible generalized
Finsler theories) and is different from its nonassociative partner article 
\cite{bsssvv25}, which also allowed to study also nonsymmetric metrics but
considering only metric compatible structures. All geometric constructions
and off-diagonal solutions are performed for 8-d nonholonomic phase spaces
in a form when the projections to base Lorentz spacetime manifolds are
related to a series of similar results in GR and possible nonmetric 4-d
MGTs, and applications in modern cosmology and classical and quantum
information and geometric flow theories \cite%
{vbubuianu17,bubuianu18,bsv22,bsv23,bvvz24,vacaru25,vacaru25b,vv25,vv25a}.

\vskip4pt

Let us conclude and discuss the main results solving the aims (in explicit,
form the objectives, \textbf{Obj 1- Obj 9}) of this work:

We critically discussed in subsection 1.1 of the Introduction seven most
important steps 1-7] in formulating and further developments of nonholonomic
FLH geometry and physics, which are important for Finsler-like
generalizations of GR, various nonmetric MGTs, and the purposes of this
review article. Such a critical analysis is very important because many
Finsler gravity theories and applications in modern cosmology and dark
energy (DE) and dark matter (DM) physics were elaborated but the methods and
results were formulated in a non-rigorous geometric formalism, and using
incomplete and not rigorous solutions, or undetermined causal and
relativistic physical models. There were proposed many Finsler
generalizations of the Einstein equations which are difficult to decouple
and solve in general forms, and which numerically/ graphically contain many
parameters, etc. Our AFCDM allows us to construct generic off-diagonal
solutions in all such metric and nonmetric MGTs, and the approach allows an
axiomatic formulation as in \cite{vacaru18}. Using respective classes of
quasi-stationary or cosmological FLH solutions, and respective generalized
G. Perelman thermodynamic variables, we can speculate on the viability of
certain MGTs and Finsler-like modifications of physically important
solutions.

\vskip4pt The first aim, i.e. \textbf{Obj1}, of this work was achieved as a
formulation of metric and nonmetric FLH theories as certain nonholonomic
metric-affine gravity (MAG) theories constructed on (co) tangent Lorentz
bundles. Considering distortions of d- and s-connections (which may result
in metric compatible or noncompatible structures, in general with nontrivial
s-torsions), we solved \textbf{Obj2. }It allows us to construct FLH
modifications of GR in axiomatic form as we stated in section 2 and \cite%
{vacaru18}. All geometric constructions were performed in an abstract
geometric language, with necessary s-adapted indices, and adapted to
respective N-connection, d-connection and s-connection structures with
conventional 2+2 and 2+2+2+2 nonholonomic shell (s) splitting of geometric
objects.

\vskip4pt The theory of nonmetric FLH geometric flows was first formulated
in relativistic form on (co) tangent Lorentz bundles in section 3, when 
\textbf{Obj3 } was achieved by defining respective nonmetric s-adapted
versions of F-and W-functionals, and respective nonmetric FLH geometric flow
equations. We solved \textbf{Obj4} by deriving in abstract geometric and
s-adapted variational form, using distortions of s-connections, of nonmetric
versions of G. Perelman's thermodynamic variables. We concluded that
nonholonomic geometric and statistical thermodynamic objects allow us to
state a new thermodynamic paradigm for general classes of solutions of
nonmetric geometric flows and MGTs even in the cases when the
Bekenstein-Hawking paradigm is not applicable.

\vskip4pt The generalization of the AFCDM for constructing generic
off-diagonal solutions in nonmetric FLH geometric flow and MGTs was
performed in section 4, and using proofs from Appendix A, for \textbf{Obj5}.
This geometric method is summarized in Tables 1-13 from Appendix B. Such
tables and respective ansatz are different from Tables 1-16 in \cite%
{bsssvv25}, where the effective sources encode nonassociative and
noncommutative geometric data. The generating data and formulas from this
work allow us to generate various classes of quasi-stationary and
cosmological solutions using generic off-diagonal data and generating
functions/ effective sources. For FLH MGTs, such constructing can be used
for deriving self-consistent and complete phase space modifications of GR
and 4-d nonmetric theories using Finsler-like variables. We proved that such
generic off-diagonal solutions are characterized by specific nonlinear
symmetries relating effective sources to certain geometric flow running
effective cosmological constants. This simplifies substantially the
computation of FLH modified G. Perelman thermodynamic variables, which
provide a solution for \textbf{Obj6}. Then, a new \textbf{Obj7} was solved
by the formulas, allowing to study how nonholonomic geometric flows and
off-diagonal interactions, and distortions of s-connections, may result, and
relate, different classes of FLH theories and respective classes of exact/
parametric solutions.

\vskip4pt The last two \textbf{Obj8 }and \textbf{Obj9 }are solved in section
5 by elaborating on respective classes of phase space FLH black hole,
wormhole, toroid and cosmological solutions. We note that similar classes of
solutions were constructed and analyzed in detail in \cite{bsssvv25} but for
different effective sources (in this work, we consider only commutative and
associative FLH data).

\vskip4pt The solutions provided for the above-stated main goals
(objectives) motivate the hypotheses H1-H3 from the Introduction section of
this article. So, this way, we formulated a general mathematical physics
approach to (non) metric FLH geometric flows and MGTs, and such theories are
generally integrable in off-diagonal forms and provide many applications in
modern cosmology and astrophysics and elaborating geometric and quantum
information models. Further partner works will be devoted to the objectives
related to H4 and H5. Such hypotheses concern the possibility of
self-consistent Finsler-like metric and nonmetric generalizations of the
theory of EYMHD interactions. And such theories can be quantized using DQ
methods and the BFV formalism. We note that \cite%
{vacaru96,vacaru96c,vacaru01,vd00,vacaru01,vacaru07,vacaru08,vacaru12,gvv15,bsvv24,vacaru25,vacaru25b}
contain necessary geometric methods and a number of preliminary results, see
also recent our publications \cite%
{vbubuianu17,bubuianu18,bsv22,bsv23,bvvz24,vacaru25,vacaru25b,vv25,vv25a}.

\vskip4pt Our approach to the theory of relativistic FLH geometric flows
(see also a series of previous results in \cite%
{svnonh08,gheorghiuap16,vacaru18,bsvv24}) has both natural and pragmatic
motivations in GR and MGTs because it allows us to provide a geometric
thermodynamic formalism for all classes of diagonal or off-diagonal
solutions. It provides a new paradigm due to G. Perelman's concept of
W-entropy \cite{perelman1} which we extended in nonmetric form on (co)
tangent Lorentz bundles. Even though we do not propose in this work to
formulate/ prove any FLH versions of the Thorston-Poincar\'{e} conjecture,
we show that the thermodynamic part of the G. Perelman work \cite{perelman1}
can be extended in certain Finsler-relativistic-ways. Applying the AFCDM,
the corresponding FLH geometric flow equations (in particular, the
generalized Finsler-Ricci soliton equations including the nonmetric FLH
modified Einstein equations), can be decoupled and solved in certain general
off-diagonal forms. For such models, relativistic FLH extensions of G.
Perelman's thermodynamic variables can be computed in very general forms in
terms of effective cosmological constants and respective volume forms.

\vskip4pt Finally, we conclude that the results and methods of this work can
be used for constructing quasi-stationary and cosmological off-diagonal
solutions describing FLH geometric flow and metric or nonmetric EYMHD
systems with potential applications for elaborating on DE and DM models, in
classical and quantum flow information theories and quantum gravity (we cite
relevant partner works \cite{partner06,bsvv24,vacaru25,bsssvv25,bsvv24}).

\vskip5pt \textbf{Acknowledgement:}\ This is a progress report containing also
a series of new and original results and physically important solutions on
FLH geometric flow and MGTs. It is also a review author's research beginning
1988, oriented to transfer to Western Countries of knowledge and methods
elaborated in Eastern Europe. This motivates an increased number of
self-citations of the author's (and co-authors') works published in the
former USSR and Romania. The work is supported by a visiting fellowship for
the Kocaeli University in T\"{u}rkiye (not covering publication fees) and
extends former volunteer research programs at California State University at
Fresno, the USA, CAS LMU Munich, Germany, and Taras Shevchenko National
University of Kyiv, Ukraine.

\vskip5pt \textbf{Data Availability Statement:} Data sharing does not apply to this article (such data are absent) as no data sets were generated or analysed during the current study. 


\newpage \appendix\setcounter{equation}{0} 
\renewcommand{\theequation}
{A.\arabic{equation}} \setcounter{subsection}{0} 
\renewcommand{\thesubsection}
{A.\arabic{subsection}}

\section{Proofs of decoupling and integrating nonmetric FLH-geometric flow
equations}

\label{appendixa}

In this Appendix, we outline and adapt for FLH geometries the computations
and proofs provided in sections 3.1 and 3.2\ of \cite{bsssvv25} and used for
nonassociative MGTs. Certain results and methods but without technical
details for the AFCDM were reviewed and discussed in \cite%
{vacaru18,sv11,vbubuianu17,partner02}, see also references therein (on MGTs,
noncommutative and supersymmetric geometries, etc.).

\subsection{Canonical Ricci s-tensors for quasi-stationary configurations}

The main formulas and solutions will be proved for quasi-stationary
s-metrics. Time-like and energy-like dual symmetries will be used for
deriving off-diagonal cosmological solutions.

\subsubsection{Conventions on quasi-stationary ansatz and their dualization
to locally anisotropic cosmological configurations}

We compute in explicit form the N-adapted coefficients of the
quasi-stationary canonical d-connections; N-connection curvature; canonical
d-torsion and LC-conditions; and canonical Ricci d-tensor for canonical
d-connections. To simplify computations we use brief notations of partial
derivatives on phase space $\ _{s}\mathcal{M}$ : for instance, $\partial
_{1}q(u^{\alpha })=q^{\bullet },$ $\partial _{2}q(u^{\alpha })=q^{\prime }$, 
$\partial _{3}q(u^{\alpha })=q^{\ast }$ and $\partial
_{4}q(u^{\alpha})=q^{\diamond };$ on $s=3,$ we use only partial derivatives $%
\partial _{5}q(u^{\alpha })$ and $\partial _{6}q(u^{\alpha });$ on $s=4,$ we
use only partial derivatives $\partial _{7}q(u^{\alpha })$ and $\partial
_{6}q(u^{\alpha }),$ for $u^{\alpha }=u^{\alpha _{4}}=(x^{i_{1}}, y^{a_{2}},
v^{a_{3}}, v^{a_{4}})= (u^{\alpha _{3}}=u^{i_{3}},v^{a_{4}})=(u^{\alpha
_{2}}=u^{i_{2}},v^{a_{3}},v^{a_{4}}).$ Such notations are slightly modified
for $s=3$ and 4 on $\ _{s}^{\shortmid }\mathcal{M}$ : on $s=3,$ we use only
partial derivatives $\ ^{\shortmid}\partial ^{5}\ ^{\shortmid }q(\
^{\shortmid }u^{\alpha })$ and $\ ^{\shortmid }\partial ^{6}\ ^{\shortmid
}q(\ ^{\shortmid }u^{\alpha });$ on $s=4,$ we use only partial derivatives $%
\ ^{\shortmid }\partial ^{7}\ ^{\shortmid }q(\ ^{\shortmid }u^{\alpha })$
and $\ ^{\shortmid }\partial ^{8}\ ^{\shortmid }q(\ ^{\shortmid }u^{\alpha
}),$ for $\ ^{\shortmid}u^{\alpha }=\ ^{\shortmid }u^{\alpha
_{4}}=(x^{i_{1}},y^{a_{2}},p_{a_{3}},p_{a_{4}})= (\ ^{\shortmid }u^{\alpha
_{3}}=\ ^{\shortmid }u^{i_{3}},p_{a_{4}}=\{p_{7},p_{8}=E\})=(u^{\alpha
_{2}}=u^{i_{2}},p_{3},p_{7},E).$

The \textbf{quasi-stationary configurations} determined by s-metrics of type
(\ref{ssolutions}) on $\ _{s}\mathcal{M}$ are described by generic
off-diagonal ansatz 
\begin{eqnarray}
d\widehat{s}_{q}^{2}(\tau ) &=&g_{i_{1}}(\tau
)(dx^{i_{1}})^{2}+h_{a_{2}}(\tau )(\mathbf{e}^{a_{2}}(\tau ))^{2}+\
h_{a_{3}}(\tau )(\mathbf{e}^{a_{3}}(\tau ))^{2}+h_{a_{4}}(\tau )(\mathbf{e}%
^{a_{4}}(\tau ))^{2},\mbox{where }  \label{qstsm} \\
\mathbf{e}^{a_{2}}(\tau ) &=&dy^{a_{2}}+\left[ w_{k_{1}}^{a_{2}}(\tau
)+n_{k_{1}}^{a_{2}}(\tau )\right] dx^{k_{1}},\   \notag \\
\mathbf{e}^{a_{3}}(\tau ) &=&dv^{a_{3}}+\left[ w_{k_{2}}^{a_{3}}(\tau
)+n_{k_{2}}^{a_{3}}(\tau )\right] \ dx^{k_{2}},\mathbf{e}^{a_{4}}(\tau
)=dv^{a_{4}}+\left[ w_{k_{3}}^{a_{4}}(\tau )+n_{k_{3}}^{a_{4}}(\tau )\right]
d\ x^{k_{3}},\ \mbox{ for }  \notag
\end{eqnarray}
\begin{eqnarray*}
g_{i_{1}}(\tau ) &=&\exp \left[ \psi (\tau ,x^{i_{1}})\right]
,g_{a_{2}}(\tau )=h_{a_{2}}(\tau ,x^{i_{1}},y^{3}), \\
g_{a_{3}}(\tau ) &=&h_{a_{3}}(\tau
,x^{i_{1}},y^{a_{2}},v^{5}),g_{a_{4}}(\tau )=h_{a_{4}}(\tau
,x^{i_{1}},y^{a_{2}},v^{a_{3}},v^{a_{7}}); \\
N_{k_{1}}^{3}(\tau ) &=&w_{k_{1}}(\tau ,x^{i_{1}},y^{3}),N_{k_{1}}^{4}(\tau
)=n_{k_{1}}(\tau ,x^{i_{1}},y^{3}),N_{k_{2}}^{5}(\tau )=w_{k_{2}}(\tau
,x^{i_{1}},y^{a_{2}},v^{5}), \\
N_{k_{2}}^{6}(\tau ) &=&n_{k_{2}}(\tau
,x^{i_{1}},y^{a_{2}},v^{5}),N_{k_{3}}^{7}(\tau )=w_{k_{3}}(\tau
,x^{i_{2}},v^{a_{3}},v^{7}),N_{k_{2}}^{8}(\tau )=n_{k_{2}}(\tau
,x^{i_{2}},v^{a_{3}},v^{7}).
\end{eqnarray*}
The s-coefficients of such an ansatz do not depend on the time like variable 
$y^{4}=4$ on $s=2,$ but dependence on $t$ may exist on upper shells $s=3,4;$
the coefficients on $s=3$ do not depend on $v^{6},$ but on the upper shell $%
s=4$ such a dependence on $v^{6}$ may exist. Then, the coefficients of (\ref%
{qstsm}) do not depend on $v^{8}$ on $s=4.$ Here we emphasize that we can
consider other types of s-adapted quasi-stationary dependence if we consider
explicit dependencies on $v^{6}$ (but not on $v^{5}$), and on $v^{8} $ but
not on $v^{7}.$ For simplicity, can analyze only s-metrics with Killing
symmetry on $\partial _{8}$ (on all shells), when the Killing symmetry on $%
\partial _{6}$ is for the first three shells $(s=1,2,3);$ and the Killing
symmetry on the time-like coordinate $\partial _{4}=\partial _{t} $ (the
quasi-stationarity conditions) exists on the first two shells $(s=1,2).$ 

The previous ansatz can be transformed into locally anisotropic cosmological
(generic off-diagonal) ansatz on $\ _{s}\mathcal{M}$ if we consider generic
dependencies of the s-coefficients on the time like coordinate $y^{4}=t$ but
not on $y^{3}$ and change respectively the $w$-coefficients into $n$%
-coefficients and inverse. So, in this work, we consider such \textbf{%
locally anisotropic cosmological s-metrics}: 
\begin{eqnarray}
d\widehat{s}_{lc}^{2}(\tau ) &=&g_{i_{1}}(\tau )(dx^{i_{1}})^{2}+\underline{h%
}_{a_{2}}(\tau )(\underline{\mathbf{e}}^{a_{2}}(\tau ))^{2}+\ \underline{h}%
_{a_{3}}(\tau )(\underline{\mathbf{e}}^{a_{3}}(\tau ))^{2}+\underline{h}%
_{a_{4}}(\tau )(\underline{\mathbf{e}}^{a_{4}}(\tau ))^{2},\mbox{where }
\label{lcsm} \\
\underline{\mathbf{e}}^{a_{2}}(\tau ) &=&dy^{a_{2}}+\left[ \underline{n}%
_{k_{1}}^{a_{2}}(\tau )+\underline{w}_{k_{1}}^{a_{2}}(\tau )\right]
dx^{k_{1}},\   \notag \\
\underline{\mathbf{e}}^{a_{3}}(\tau ) &=&dv^{a_{3}}+\left[ \underline{n}%
_{k_{2}}^{a_{3}}(\tau )+\underline{w}_{k_{2}}^{a_{3}}(\tau )\right] \
dx^{k_{2}},\underline{\mathbf{e}}^{a_{4}}(\tau )=dv^{a_{4}}+\left[ 
\underline{n}_{k_{3}}^{a_{4}}(\tau )+\underline{w}_{k_{3}}^{a_{4}}(\tau )%
\right] d\ x^{k_{3}},\ \mbox{ for }  \notag
\end{eqnarray}%
\begin{eqnarray*}
g_{i_{1}}(\tau ) &=&\exp \left[ \psi (\tau ,x^{i_{1}})\right] ,\underline{g}%
_{a_{2}}(\tau )=\underline{h}_{a_{2}}(\tau ,x^{i_{1}},t), \\
\underline{g}_{a_{3}}(\tau ) &=&\underline{h}_{a_{3}}(\tau
,x^{i_{1}},y^{a_{2}},v^{5}),\underline{g}_{a_{4}}(\tau )=\underline{h}%
_{a_{4}}(\tau ,x^{i_{1}},y^{a_{2}},v^{a_{3}},v^{a_{7}}); \\
\underline{N}_{k_{1}}^{3}(\tau ) &=&\underline{n}_{k_{1}}(\tau ,x^{i_{1}},t),%
\underline{N}_{k_{1}}^{4}(\tau )=\underline{w}_{k_{1}}(\tau ,x^{i_{1}},t),%
\underline{N}_{k_{2}}^{5}(\tau )=\underline{n}_{k_{2}}(\tau
,x^{i_{1}},y^{a_{2}},v^{5}), \\
\underline{N}_{k_{2}}^{6}(\tau ) &=&\underline{w}_{k_{2}}(\tau
,x^{i_{1}},y^{a_{2}},v^{5}),\underline{N}_{k_{3}}^{7}(\tau )=\underline{n}%
_{k_{3}}(\tau ,x^{i_{2}},v^{a_{3}},v^{7}),\underline{N}_{k_{2}}^{8}(\tau )=%
\underline{w}_{k_{2}}(\tau ,x^{i_{2}},v^{a_{3}},v^{7}).
\end{eqnarray*}%
In the above formulas (\ref{lcsm}), we underline respective symbols to
emphasize that they are considered for locally anisotropic cosmological
configurations with generic dependence on $t$-coordinate.

$\tau $-families of quasi-stationary configurations on $\ _{s}^{\shortmid }%
\mathcal{M}$ are defined by such ansatz: 
\begin{eqnarray}
d\ ^{\shortmid }\widehat{s}_{q}^{2}(\tau ) &=&g_{i_{1}}(\tau
)(dx^{i_{1}})^{2}+h_{a_{2}}(\tau )(\mathbf{e}^{a_{2}}(\tau ))^{2}+\
^{\shortmid }h^{a_{3}}(\tau )(\ ^{\shortmid }\mathbf{e}_{a_{3}}(\tau
))^{2}+\ ^{\shortmid }h^{a_{4}}(\tau )(\ ^{\shortmid }\mathbf{e}%
_{a_{4}}(\tau ))^{2},\mbox{where }  \label{qstsmd} \\
\mathbf{e}^{a_{2}}(\tau ) &=&dy^{a_{2}}+\left[ w_{k_{1}}^{a_{2}}(\tau
)+n_{k_{1}}^{a_{2}}(\tau )\right] dx^{k_{1}},\   \notag \\
\ ^{\shortmid }\mathbf{e}^{a_{3}}(\tau ) &=&dp_{a_{3}}+\left[ \ ^{\shortmid
}w_{k_{2}a_{3}}(\tau )+\ ^{\shortmid }n_{k_{2}a_{3}}(\tau )\right] \ d\
^{\shortmid }x^{k_{2}},\ ^{\shortmid }\mathbf{e}_{a_{4}}(\tau )=dp_{a_{4}}+%
\left[ \ ^{\shortmid }w_{k_{3}a_{4}}(\tau )+\ ^{\shortmid
}n_{k_{3a_{4}}}(\tau )\right] d\ \ ^{\shortmid }x^{k_{3}},\ \mbox{ for } 
\notag
\end{eqnarray}%
\begin{eqnarray*}
g_{i_{1}}(\tau ) &=&\exp \left[ \psi (\tau ,x^{i_{1}})\right]
,g_{a_{2}}(\tau )=h_{a_{2}}(\tau ,x^{i_{1}},y^{3}), \\
\ ^{\shortmid }g_{a_{3}}(\tau ) &=&\ ^{\shortmid }h_{a_{3}}(\tau
,x^{i_{1}},y^{a_{2}},p_{5}),\ ^{\shortmid }g_{a_{4}}(\tau )=\ ^{\shortmid
}h_{a_{4}}(\tau ,x^{i_{1}},y^{a_{2}},p_{a_{3}},p_{a_{4}}); \\
\ ^{\shortmid }N_{k_{1}}^{3}(\tau ) &=&\ ^{\shortmid }w_{k_{1}}(\tau
,x^{i_{1}},y^{3}),\ ^{\shortmid }N_{k_{1}}^{4}(\tau )=\ ^{\shortmid
}n_{k_{1}}(\tau ,x^{i_{1}},y^{3}),\ ^{\shortmid }N_{k_{2}5}(\tau )=\
^{\shortmid }w_{k_{2}}(\tau ,x^{i_{1}},y^{a_{2}},p_{5}), \\
\ ^{\shortmid }N_{k_{2}6}(\tau ) &=&\ ^{\shortmid }n_{k_{2}}(\tau
,x^{i_{1}},y^{a_{2}},p_{5}),\ ^{\shortmid }N_{k_{3}7}(\tau )=\ ^{\shortmid
}w_{k_{3}}(\tau ,x^{i_{2}},v^{a_{3}},p_{7}),\ ^{\shortmid }N_{k_{2}8}(\tau
)=n_{k_{2}}(\tau ,x^{i_{2}},v^{a_{3}},p_{7}).
\end{eqnarray*}%
These formulas are similar to (\ref{qstsm}) but written on the dual phase
space when the velocity-type variables/indices are changed respectively into
the momentum-type variables/indices. We use corresponding labels $\
"^{\shortmid }"$ to state that the coefficients depend on momentum-like
coordinates.

In this subsection, we provide the ansatz for $\tau $-families of locally
anisotropic configurations on $\ _{s}^{\shortmid }\mathcal{M},$ which is a
respective $^{\shortmid }$-transform of (\ref{lcsm}), also preserving
generic dependence on the time-like coordinate $y^{4}=t$ but not on $y^{3}$.
We can use such formulas: 
\begin{eqnarray}
d\ ^{\shortmid }\widehat{s}_{lc}^{2}(\tau ) &=&g_{i_{1}}(\tau
)(dx^{i_{1}})^{2}+\underline{h}_{a_{2}}(\tau )(\underline{\mathbf{e}}%
^{a_{2}}(\tau ))^{2}+\ \ ^{\shortmid }\underline{h}^{a_{3}}(\tau )(\
^{\shortmid }\underline{\mathbf{e}}_{a_{3}}(\tau ))^{2}+\ ^{\shortmid }%
\underline{h}^{a_{3}}(\tau )(\ ^{\shortmid }\underline{\mathbf{e}}%
_{a_{4}}(\tau ))^{2},\mbox{where }  \label{lcsmd} \\
\underline{\mathbf{e}}^{a_{2}}(\tau ) &=&dy^{a_{2}}+\left[ \underline{n}%
_{k_{1}}^{a_{2}}(\tau )+\underline{w}_{k_{1}}^{a_{2}}(\tau )\right]
dx^{k_{1}},\   \notag \\
\ ^{\shortmid }\underline{\mathbf{e}}_{a_{3}}(\tau ) &=&dp_{a_{3}}+\left[ \
^{\shortmid }\underline{n}_{k_{2}a_{3}}(\tau )+\ ^{\shortmid }\underline{w}%
_{k_{2}a_{3}}(\tau )\right] \ d\ ^{\shortmid }x^{k_{2}},\ ^{\shortmid }%
\underline{\mathbf{e}}_{a_{4}}(\tau )=dp_{a_{4}}+\left[ \ ^{\shortmid }%
\underline{n}_{k_{3}a_{4}}(\tau )+\ ^{\shortmid }\underline{w}%
_{k_{3}a_{4}}(\tau )\right] d\ \ ^{\shortmid }x^{k_{3}},\ \mbox{ for } 
\notag
\end{eqnarray}%
\begin{eqnarray*}
g_{i_{1}}(\tau ) &=&\exp \left[ \psi (\tau ,x^{i_{1}})\right] ,\underline{g}%
_{a_{2}}(\tau )=\underline{h}_{a_{2}}(\tau ,x^{i_{1}},t), \\
\ ^{\shortmid }\underline{g}^{a_{3}}(\tau ) &=&\ ^{\shortmid }\underline{h}%
^{a_{3}}(\tau ,x^{i_{1}},y^{a_{2}},p_{5}),\ ^{\shortmid }\underline{g}%
^{a_{4}}(\tau )=\ ^{\shortmid }\underline{h}_{a_{4}}(\tau
,x^{i_{1}},y^{a_{2}},p_{a_{3}},p_{7}); \\
\underline{N}_{k_{1}}^{3}(\tau ) &=&\ ^{\shortmid }\underline{n}%
_{k_{1}}(\tau ,x^{i_{1}},t),\underline{N}_{k_{1}}^{4}(\tau )=\ ^{\shortmid }%
\underline{w}_{k_{1}}(\tau ,x^{i_{1}},t),\ ^{\shortmid }\underline{N}%
_{k_{2}5}(\tau )=\ ^{\shortmid }\underline{n}_{k_{2}}(\tau
,x^{i_{1}},y^{a_{2}},p_{5}), \\
\ ^{\shortmid }\underline{N}_{k_{2}6}(\tau ) &=&\ ^{\shortmid }\underline{w}%
_{k_{2}}(\tau ,x^{i_{1}},y^{a_{2}},p_{5}),\ ^{\shortmid }\underline{N}%
_{k_{3}7}(\tau )=\ ^{\shortmid }\underline{n}_{k_{3}}(\tau
,x^{i_{2}},p_{a_{3}},p_{7}),\ ^{\shortmid }\underline{N}_{k_{2}8}(\tau )=\
^{\shortmid }\underline{w}_{k_{2}}(\tau ,x^{i_{2}},p_{a_{3}},p_{7}).
\end{eqnarray*}

Hereafter, we shall provide explicit commutations and proofs for
quasi-stationary ansatz (\ref{qstsm}). Similar constructions for the generic
off-diagonal ansatz (\ref{lcsm}), (\ref{qstsmd}), or (\ref{lcsmd}) can be
performed in similar forms. In principle, they can be defined as certain
phase space and/or time-like dual transforms of the results obtained for (%
\ref{qstsm}).

\subsubsection{Computing the coefficients of \ the canonical s-connection}

The nontrivial coefficients of $\ \widehat{\mathbf{D}}=\{\widehat{\Gamma }%
_{\ \alpha \beta }^{\gamma }\simeq \widehat{\Gamma }_{\ \alpha _{s}\beta
_{s}}^{\gamma _{s}}\}$ (\ref{cdc}) computed for quasi-stationary d-metrics (%
\ref{qstsm}) are computed in such forms: 
\begin{eqnarray}
\widehat{L}_{11}^{1} &=&\frac{g_{1}^{\bullet }}{2g_{1}}=\frac{\partial
_{1}g_{1}}{2g_{1}},\ \widehat{L}_{12}^{1}=\frac{g_{1}^{\prime }}{2g_{1}}=%
\frac{\partial _{2}g_{1}}{2g_{1}},\widehat{L}_{22}^{1}=-\frac{g_{2}^{\bullet
}}{2g_{1}},\ \widehat{L}_{11}^{2}=\frac{-g_{1}^{\prime }}{2g_{2}},\ \widehat{%
L}_{12}^{2}=\frac{g_{2}^{\bullet }}{2g_{2}},\ \widehat{L}_{22}^{2}=\frac{%
g_{2}^{\prime }}{2g_{2}},  \label{nontrdc} \\
\widehat{L}_{4k}^{4} &=&\frac{\mathbf{\partial }_{k}(h_{4})}{2h_{4}}-\frac{%
w_{k}h_{4}^{\ast }}{2h_{4}},\widehat{L}_{3k}^{3}=\frac{\mathbf{\partial }%
_{k}h_{3}}{2h_{3}}-\frac{w_{k}h_{3}^{\ast }}{2h_{3}},\widehat{L}_{4k}^{3}=-%
\frac{h_{4}}{2h_{3}}n_{k}^{\ast },  \notag \\
\widehat{L}_{3k}^{4} &=&\frac{1}{2}n_{k}^{\ast }=\frac{1}{2}\partial
_{3}n_{k},\widehat{C}_{33}^{3}=\frac{h_{3}^{\ast }}{2h_{3}},\widehat{C}%
_{44}^{3}=-\frac{h_{4}^{\ast }}{h_{3}},\ \widehat{C}_{33}^{4}=0,~\widehat{C}%
_{34}^{4}=\frac{h_{4}^{\ast }}{2h_{4}},\widehat{C}_{44}^{4}=0,  \notag
\end{eqnarray}%
\begin{eqnarray*}
\widehat{L}_{6k_{2}}^{6} &=&\frac{\mathbf{\partial }_{k_{2}}(h_{6})}{2h_{6}}-%
\frac{w_{k_{2}}\partial _{5}h_{6}}{2h_{6}},\widehat{L}_{5k_{2}}^{5}=\frac{%
\mathbf{\partial }_{k_{2}}h_{5}}{2h_{5}}-\frac{w_{k_{2}}\partial _{5}h_{5}}{%
2h_{5}},\widehat{L}_{6k_{2}}^{5}=-\frac{h_{6}}{2h_{5}}\partial _{5}n_{k_{2}},%
\mbox{ for }k_{2}=1,2,...,4; \\
\widehat{L}_{5k_{2}}^{6} &=&\frac{1}{2}\partial _{5}n_{k_{2}},\widehat{C}%
_{55}^{5}=\frac{\partial _{5}h_{5}}{2h_{5}},\widehat{C}_{66}^{5}=-\frac{%
\partial _{5}h_{6}}{h_{5}},\ \widehat{C}_{55}^{6}=0,~\widehat{C}_{56}^{6}=%
\frac{\partial _{5}h_{6}}{2h_{6}},\widehat{C}_{66}^{6}=0;
\end{eqnarray*}%
\begin{eqnarray*}
\widehat{L}_{8k_{3}}^{8} &=&\frac{\mathbf{\partial }_{k_{2}}(h_{8})}{2h_{8}}-%
\frac{w_{k_{3}}\partial _{7}h_{8}}{2h_{8}},\widehat{L}_{7k_{3}}^{7}=\frac{%
\mathbf{\partial }_{k_{2}}h_{7}}{2h_{7}}-\frac{w_{k_{3}}\partial _{7}h_{7}}{%
2h_{7}},\widehat{L}_{8k_{3}}^{7}=-\frac{h_{8}}{2h_{7}}\partial _{7}n_{k_{3}},%
\mbox{ for }k_{3}=1,2,...,6; \\
\widehat{L}_{7k_{3}}^{8} &=&\frac{1}{2}\partial _{7}n_{k_{2}},\widehat{C}%
_{77}^{7}=\frac{\partial _{7}h_{7}}{2h_{7}},\widehat{C}_{88}^{7}=-\frac{%
\partial _{7}h_{8}}{h_{7}},\ \widehat{C}_{77}^{8}=0,~\widehat{C}_{78}^{8}=%
\frac{\partial _{7}h_{8}}{2h_{8}},\widehat{C}_{88}^{8}=0.
\end{eqnarray*}%
We also have compute the values 
\begin{eqnarray}
\ \widehat{C}_{3} &=&\widehat{C}_{33}^{3}+\widehat{C}_{34}^{4}=\frac{%
h_{3}^{\ast }}{2h_{3}}+\frac{h_{4}^{\ast }}{2h_{4}},\widehat{C}_{4}=\widehat{%
C}_{43}^{3}+\widehat{C}_{44}^{4}=0,  \label{aux3} \\
\widehat{C}_{5} &=&\widehat{C}_{55}^{5}+\widehat{C}_{56}^{6}=\frac{\partial
_{5}h_{5}}{2h_{5}}+\frac{\partial _{5}h_{6}}{2h_{6}},\widehat{C}_{6}=%
\widehat{C}_{65}^{5}+\widehat{C}_{66}^{6}=0,  \notag \\
\widehat{C}_{7} &=&\widehat{C}_{77}^{7}+\widehat{C}_{78}^{8}=\frac{\partial
_{7}h_{7}}{2h_{7}}+\frac{\partial _{7}h_{8}}{2h_{8}},\widehat{C}_{8}=%
\widehat{C}_{87}^{7}+\widehat{C}_{88}^{8}=0,  \notag
\end{eqnarray}%
which are necessary together with the set of coefficients (\ref{nontrdc})
for computing the s-adapted coefficients of the canonical torsion and
canonical Ricci and Einstein s-tensors.

Introducing the N-connection coefficients in (\ref{qstsm}), we compute the
coefficients of the N-connection curvature $\widehat{\Omega }%
_{i_{s-1}j_{s-1}}^{a_{s}}=\widehat{\mathbf{e}}_{j_{s-1}}\left( \widehat{N}%
_{i_{s-1}}^{a_{s}}\right) -\widehat{\mathbf{e}}_{i_{s-1}}(\widehat{N}%
_{j_{s-1}}^{a_{s}}),$ see similar formulas (\ref{anholcond8}). We obtain 
\begin{eqnarray*}
\widehat{\Omega }_{i_{1}j_{1}}^{a_{2}} &=&\mathbf{\partial }_{j_{1}}\left( 
\widehat{N}_{i_{1}}^{a_{2}}\right) -\partial _{i_{1}}(\widehat{N}%
_{j_{1}}^{a_{2}})-w_{i_{1}}\partial _{3}(\widehat{N}%
_{j_{1}}^{a_{2}})+w_{j_{1}}\partial _{3}(\widehat{N}_{i_{1}}^{a_{2}}), \\
\widehat{\Omega }_{i_{2}j_{2}}^{a_{3}} &=&\mathbf{\partial }_{j_{2}}\left( 
\widehat{N}_{i_{2}}^{a_{3}}\right) -\partial _{i_{2}}(\widehat{N}%
_{j_{2}}^{a_{3}})-w_{i_{2}}\partial _{5}(\widehat{N}%
_{j_{2}}^{a_{3}})+w_{j_{2}}\partial _{5}(\widehat{N}_{i_{2}}^{a_{3}}), \\
\widehat{\Omega }_{i_{3}j3}^{a_{4}} &=&\mathbf{\partial }_{j_{3}}\left( 
\widehat{N}_{i_{3}}^{a_{4}}\right) -\partial _{i_{3}}(\widehat{N}%
_{j_{3}}^{a_{4}})-w_{i_{3}}\partial _{7}(\widehat{N}%
_{j_{3}}^{a_{4}})+w_{j_{3}}\partial _{7}(\widehat{N}_{i_{3}}^{a_{4}}).
\end{eqnarray*}%
These formulas result in such nontrivial values: 
\begin{eqnarray}
\widehat{\Omega }_{12}^{3} &=&-\widehat{\Omega }_{21}^{3}=\mathbf{\partial }%
_{2}w_{1}-\partial _{1}w_{2}-w_{1}w_{2}^{\ast }+w_{2}w_{1}^{\ast
}=w_{1}^{\prime }-w_{2}^{\bullet }-w_{1}w_{2}{}^{\ast }+w_{2}w_{1}^{\ast }{};
\notag \\
\widehat{\Omega }_{12}^{4} &=&-\widehat{\Omega }_{21}^{4}=\mathbf{\partial }%
_{2}n_{1}-\partial _{1}n_{2}-w_{1}n_{2}^{\ast }+w_{2}n_{1}^{\ast
}=n_{1}^{\prime }-n_{2}^{\bullet }-w_{1}n_{2}^{\ast }{}+w_{2}n_{1}^{\ast }{}.
\label{omeg}
\end{eqnarray}
\begin{eqnarray*}
\widehat{\Omega }_{i_{2}j_{2}}^{5} &=&-\widehat{\Omega }_{j_{2}i_{2}}^{5}=%
\mathbf{\partial }_{j_{2}}w_{i_{2}}-\partial
_{i_{2}}w_{j_{2}}-w_{i_{2}}\partial _{5}w_{j_{2}}+w_{j_{2}}\partial
_{5}w_{i_{2}}=\mathbf{\partial }_{j_{2}}w_{i_{2}}-\mathbf{\partial }%
_{i_{2}}w_{j_{2}}-w_{i_{2}}\partial _{5}w_{j_{2}}{}+w_{j_{2}}\partial
_{5}w_{i_{2}}{}; \\
\widehat{\Omega }_{i_{2}j_{2}}^{6} &=&-\widehat{\Omega }_{i_{2}j_{2}}^{6}=%
\mathbf{\partial }_{j_{2}}n_{i_{2}}-\partial
_{i_{2}}n_{j_{2}}-w_{i_{2}}\partial _{5}n_{j_{2}}+w_{j_{2}}\partial
_{5}n_{i_{2}}=\mathbf{\partial }_{j_{2}}n_{i_{2}}-\mathbf{\partial }%
_{i_{2}}n_{j_{2}}-w_{i_{2}}\partial _{5}n_{j_{2}}{}+w_{j_{2}}\partial
_{5}n_{i_{2}};
\end{eqnarray*}%
\begin{eqnarray*}
\widehat{\Omega }_{i_{3}j_{3}}^{7} &=&-\widehat{\Omega }_{j_{3}i_{3}}^{7}=%
\mathbf{\partial }_{j_{3}}w_{i_{3}}-\partial
_{i_{3}}w_{j_{3}}-w_{i_{3}}\partial _{7}w_{j_{3}}+w_{j_{3}}\partial
_{7}w_{i_{3}}=\mathbf{\partial }_{j_{3}}w_{i_{3}}-\mathbf{\partial }%
_{i_{3}}w_{j_{3}}-w_{i_{3}}\partial _{7}w_{j_{3}}{}+w_{j_{3}}\partial
_{7}w_{i_{3}}{}; \\
\widehat{\Omega }_{i_{3}j_{3}}^{8} &=&-\widehat{\Omega }_{i_{3}j_{32}}^{8}=%
\mathbf{\partial }_{j_{3}}n_{i_{3}}-\partial
_{i_{3}}n_{j_{3}}-w_{i_{3}}\partial _{7}n_{j_{3}}+w_{j_{3}}\partial
_{7}n_{i_{3}}=\mathbf{\partial }_{j_{3}}n_{i_{3}}-\mathbf{\partial }%
_{i_{3}}n_{j_{3}}-w_{i_{3}}\partial _{7}n_{j_{3}}{}+w_{j_{3}}\partial
_{7}n_{i_{3}}.
\end{eqnarray*}

Using (\ref{omeg}), we can compute the nontrivial coefficients of the
canonical version of s--torsion (\ref{stors}). Details on such component
formulas (for various dimensions and quite different nonholonomic
distributions) are provided in \cite{bsssvv25,sv11}. We have such nontrivial
coefficients $\widehat{T}_{\ j_{s=1}i_{s=1}}^{a_{s}}=-\Omega _{\
j_{s=1}i_{s=1}}^{a_{s}}$ and $\widehat{T}_{a_{s}j_{s-1}}^{c_{s}}=\widehat{L}%
_{a_{s}j_{s=1}}^{c_{s}}-e_{a_{s}}(\widehat{N}_{j_{s=1}}^{c_{s}}).$ Other
subsets of the nontrivial coefficients are computed:%
\begin{eqnarray*}
\widehat{T}_{\ j_{s-1}k_{s-1}}^{i_{s-1}} &=&\widehat{L}_{\
j_{s-1}k_{s-1}}^{i_{s-1}}-\widehat{L}_{\ k_{s-1}j_{s-1}}^{i_{s-1}}=0,~%
\widehat{T}_{\ j_{s-1}a_{s}}^{i_{s-1}}=\widehat{C}_{\
j_{s-1}a_{s}}^{i_{s-1}}=0,~\widehat{T}_{\ b_{s}c_{s}}^{a_{s}}=\ \widehat{C}%
_{\ b_{s}c_{s}}^{a_{s}}-\ \widehat{C}_{\ c_{s}b_{s}}^{a_{s}}=0, \\
\widehat{T}_{3k_{1}}^{3} &=&\widehat{L}_{3k_{1}}^{3}-e_{3}(\widehat{N}%
_{k_{1}}^{3})=\frac{\mathbf{\partial }_{k_{1}}h_{3}}{2h_{3}}-w_{k_{1}}\frac{%
\partial _{3}h_{3}}{2h_{3}}-\partial _{3}w_{k_{1}}{},\widehat{T}%
_{4k_{1}}^{3}=\widehat{L}_{4k_{1}}^{3}-e_{4}(\widehat{N}_{k_{1}}^{3})=-\frac{%
h_{4}}{2h_{3}}\partial _{3}n_{k_{1}}, \\
\widehat{T}_{5k_{2}}^{5} &=&\widehat{L}_{5k_{2}}^{5}-e_{5}(\widehat{N}%
_{k_{2}}^{5})=\frac{\mathbf{\partial }_{k_{2}}h_{5}}{2h_{5}}-w_{k_{2}}\frac{%
\partial _{5}h_{5}}{2h_{5}}-\partial _{5}w_{k_{2}}{},\widehat{T}%
_{6k_{2}}^{5}=\widehat{L}_{6k_{2}}^{5}-e_{6}(\widehat{N}_{k_{2}}^{5})=-\frac{%
h_{6}}{2h_{5}}\partial _{5}n_{k_{2}},\  \\
\widehat{T}_{7k_{3}}^{7} &=&\widehat{L}_{7k_{3}}^{7}-e_{7}(\widehat{N}%
_{k_{3}}^{7})=\frac{\mathbf{\partial }_{k_{3}}h_{7}}{2h_{7}}-w_{k_{3}}\frac{%
\partial _{7}h_{7}}{2h_{7}}-\partial _{7}w_{k_{3}}{},\widehat{T}%
_{8k_{3}}^{7}=\widehat{L}_{8k_{3}}^{7}-e_{8}(\widehat{N}_{k_{3}}^{7})=-\frac{%
h_{8}}{2h_{7}}\partial _{7}n_{k_{3}},
\end{eqnarray*}%
\begin{eqnarray}
\widehat{T}_{3k_{1}}^{4} &=&~\widehat{L}_{3k_{1}}^{4}-e_{3}(\widehat{N}%
_{k_{1}}^{4})=\frac{1}{2}n_{k_{1}}^{\ast }-n_{k_{1}}^{\ast }=-\frac{1}{2}%
n_{k_{1}}^{\ast },\widehat{T}_{4k_{1}}^{4}=\widehat{L}%
_{4k_{1}}^{4}-e_{4}(N_{k_{1}}^{4})=\frac{\mathbf{\partial }_{k_{1}}h_{4}}{%
2h_{4}}-w_{k_{1}}\frac{h_{4}^{\ast }}{2h_{4}},  \notag \\
-\widehat{T}_{12}^{3} &=&w_{1}^{\prime }-w_{2}^{\bullet }-w_{1}w_{2}^{\ast
}{}+w_{2}w_{1}^{\ast },\ -\widehat{T}_{12}^{4}=n_{1}^{\prime
}-n_{2}^{\bullet }-w_{1}n_{2}^{\ast }{}+w_{2}n_{1}^{\ast }{},
\label{nontrtors} \\
\widehat{T}_{5k_{2}}^{6} &=&~\widehat{L}_{5k_{2}}^{6}-e_{5}(\widehat{N}%
_{k_{2}}^{6})=\frac{1}{2}\partial _{5}n_{k_{2}}-\partial _{5}n_{k_{2}}=-%
\frac{1}{2}\partial _{5}n_{k_{2}},\widehat{T}_{6k_{2}}^{6}=\widehat{L}%
_{6k_{2}}^{6}-e_{6}(N_{k_{2}}^{6})=\frac{\mathbf{\partial }_{k_{2}}h_{6}}{%
2h_{6}}-w_{k_{2}}\frac{\partial _{5}h_{6}}{2h_{6}};  \notag \\
-\widehat{T}_{i_{2}j_{2}}^{5} &=&\partial _{j_{2}}w_{i_{2}}-\partial
_{i_{2}}w_{j_{2}}-w_{i_{2}}\partial _{5}w_{j_{2}}{}+w_{j_{2}}\partial
_{5}w_{i_{2}},\ -\widehat{T}_{i_{2}j_{2}}^{6}=\partial
_{j_{2}}n_{i_{2}}-\partial _{i_{2}}n_{j_{2}}-w_{i_{2}}\partial
_{5}n_{j_{2}}{}+w_{j_{2}}\partial _{5}n_{i_{2}}{},\mbox{ for }i_{2}\neq
j_{2};  \notag
\end{eqnarray}%
\begin{eqnarray*}
\widehat{T}_{7k_{3}}^{8} &=&~\widehat{L}_{7k_{3}}^{8}-e_{7}(\widehat{N}%
_{k_{3}}^{8})=\frac{1}{2}\partial _{7}n_{k_{3}}-\partial _{7}n_{k_{3}}=-%
\frac{1}{2}\partial _{7}n_{k_{3}},\widehat{T}_{8k_{3}}^{8}=\widehat{L}%
_{8k_{3}}^{8}-e_{8}(N_{k_{3}}^{8})=\frac{\mathbf{\partial }_{k_{3}}h_{8}}{%
2h_{8}}-w_{k_{3}}\frac{\partial _{7}h_{8}}{2h_{8}}; \\
-\widehat{T}_{i_{3}j_{3}}^{7} &=&\partial _{j_{3}}w_{i_{3}}-\partial
_{i_{3}}w_{j_{3}}-w_{i_{3}}\partial _{7}w_{j_{3}}{}+w_{j_{3}}\partial
_{7}w_{i_{3}},\ -\widehat{T}_{i_{3}j_{3}}^{8}=\partial
_{j_{3}}n_{i_{3}}-\partial _{i_{3}}n_{j_{3}}-w_{i_{3}}\partial
_{7}n_{j_{3}}{}+w_{j_{3}}\partial _{7}n_{i_{3}}{},\mbox{ for }i_{3}\neq
j_{3}.
\end{eqnarray*}

The conditions that the canonical d-torsions (\ref{nontrtors}) are zero
which allows us to extracting LC-configurations are satisfied if 
\begin{equation}
\widehat{L}_{a_{s}j_{s=1}}^{c_{s}}=e_{a_{s}}(\widehat{N}_{j_{s-1}}^{c_{s}}),%
\ \widehat{C}_{j_{s-1}b_{s}}^{i_{s-1}}=0,\ \widehat{\Omega }_{\
j_{s-1}i_{s=1}}^{a_{s}}=0,  \label{lcconstr}
\end{equation}%
when in N-adapted frames we can state $\widehat{\mathbf{\Gamma }}_{\ \alpha
_{s}\beta _{s}}^{\gamma _{s}}=\Gamma _{\ \alpha _{s}\beta _{s}}^{\gamma
_{s}} $ even, in general, $_{s}\widehat{\mathbf{D}}\neq \nabla $. This is
possible because two different linear connections have different
transformation laws under general frame/ coordinate transforms
(d-connections are not (d) tensor objects). For LC-configurations, all
coefficients (\ref{nontrtors}) must be zero. Nontrivial off-diagonal
solutions can be chosen for $h_{4}^{\ast }\neq 0$ and $w_{k_{1}}^{\ast }\neq
0,$ then $\partial _{5}h_{6}\neq 0$ and $\partial _{5}w_{k_{2}}\neq 0,$ then 
$\partial _{7}h_{8}\neq 0$ and $\partial _{7}w_{k_{3}}\neq 0.$ We also state
for other subsets of N-connection coefficients: $n_{k_{1}}^{\ast }=0,$ for $%
w_{k_{1}}=\mathbf{\partial }_{k_{1}}h_{4}/h_{4}^{\ast };\partial
_{5}n_{k_{2}}=0,$ for $w_{k_{2}}=\mathbf{\partial }_{k_{2}}h_{6}/\partial
_{5}h_{6};$ and $\partial _{7}n_{k_{3}}=0,$ for $w_{k_{3}}=\mathbf{\partial }
_{k_{3}}h_{8}/\partial _{7}h_{8}.$ In principle, we can search for other
types of LC-configurations when $n_{k}^{\ast }\neq 0$ and/or $%
h_{3}^{\ast}\neq 0.$ We note that conditions of type (\ref{lcconstr}) can be
imposed after a general class of quasi-stationary off-diagonal metrics (\ref%
{qstsm}) is constructed in a general off-diagonal form involving a
nonholonomic d-torsion structure. In FLH theories, LC-configurations are not
considered on total phase spaces, but they could be important for analysing
projections on the Lorentz base spacetime manifold.

The s-coefficients on $s=1$ of a canonical Ricci d-tensor (see formulas (\ref%
{dcurv}) and (\ref{driccic}) for (\ref{cdc}) and similar details in \cite%
{vacaru18,vbubuianu17,bsssvv25,sv11}) are computed for respective
contractions of indices, when $\widehat{R}_{i_{1}j_{1}}= \widehat{R}_{\
i_{1}j_{1}k_{1}}^{k_{1}}$, for 
\begin{eqnarray}
\widehat{R}_{\ h_{1}j_{1}k_{1}}^{i_{1}} &=&\mathbf{e}_{k_{1}}\widehat{L}%
_{.h_{1}j_{1}}^{i_{1}}-\mathbf{e}_{j_{1}}\widehat{L}_{h_{1}k_{1}}^{i_{1}}+%
\widehat{L}_{h_{1}j_{1}}^{m_{1}}\widehat{L}_{m_{1}k_{1}}^{i_{1}}-\widehat{L}%
_{h_{1}k_{1}}^{m_{1}}\widehat{L}_{m_{1}j_{1}}^{i_{1}}-\widehat{C}%
_{h_{1}a_{2}}^{i_{1}}\widehat{\Omega }_{j_{1}k_{1}}^{a_{2}}  \notag \\
&=&\mathbf{\partial }_{k_{1}}\widehat{L}_{.h_{1}j_{1}}^{i_{1}}-\partial
_{j_{1}}\widehat{L}_{h_{1}k_{1}}^{i_{1}}+\widehat{L}_{h_{1}j_{1}}^{m_{1}}%
\widehat{L}_{m_{1}k_{1}}^{i_{1}}-\widehat{L}_{h_{1}k_{1}}^{m_{1}}\widehat{L}%
_{m_{1}j_{1}}^{i_{1}}.  \label{auxcurvh}
\end{eqnarray}%
We note that these formulas are considered for a quasi-stationary ansatz (%
\ref{qstsm}) and values (\ref{nontrdc}). The conditions $\widehat{C}_{\
h_{s-1}a_{s}}^{i_{s-1}}=0$ and $s=1$ formulas 
\begin{equation*}
e_{k_{1}}\widehat{L}_{h_{1}j_{1}}^{i_{1}}=\partial _{k_{1}}\widehat{L}%
_{h_{1}j_{1}}^{i_{1}}+N_{k_{1}}^{a_{2}}\partial _{a_{2}}\widehat{L}%
_{h_{1}j_{1}}^{i_{1}}=\partial _{k_{1}}\widehat{L}%
_{h_{1}j_{1}}^{i_{1}}+w_{k_{1}}(\widehat{L}_{hj}^{i})^{\ast }+n_{k}(\widehat{%
L}_{hj}^{i})^{\diamond }=\partial _{k}\widehat{L}_{hj}^{i}
\end{equation*}%
can be used because $\widehat{L}_{h_{1}j_{1}}^{i_{1}}$ depend only on
coordinates $x^{i_{1}}$. Taking respective derivatives of (\ref{nontrdc}),
we obtain 
\begin{eqnarray}
\partial _{1}\widehat{L}_{\ 11}^{1} &=&(\frac{g_{1}^{\bullet }}{2g_{1}}%
)^{\bullet }=\frac{g_{1}^{\bullet \bullet }}{2g_{1}}-\frac{\left(
g_{1}^{\bullet }\right) ^{2}}{2\left( g_{1}\right) ^{2}},\ \partial _{1}%
\widehat{L}_{\ 12}^{1}=(\frac{g_{1}^{\prime }}{2g_{1}})^{\bullet }=\frac{%
g_{1}^{\prime \bullet }}{2g_{1}}-\frac{g_{1}^{\bullet }g_{1}^{\prime }}{%
2\left( g_{1}\right) ^{2}},\   \label{auxfa1} \\
\partial _{1}\widehat{L}_{\ 22}^{1} &=&(-\frac{g_{2}^{\bullet }}{2g_{1}}%
)^{\bullet }=-\frac{g_{2}^{\bullet \bullet }}{2g_{1}}+\frac{g_{1}^{\bullet
}g_{2}^{\bullet }}{2\left( g_{1}\right) ^{2}},\ \partial _{1}\widehat{L}_{\
11}^{2}=(-\frac{g_{1}^{\prime }}{2g_{2}})^{\bullet }=-\frac{g_{1}^{\prime
\bullet }}{2g_{2}}+\frac{g_{1}^{\bullet }g_{2}^{\prime }}{2\left(
g_{2}\right) ^{2}},  \notag \\
\partial _{1}\widehat{L}_{\ 12}^{2} &=&(\frac{g_{2}^{\bullet }}{2g_{2}}%
)^{\bullet }=\frac{g_{2}^{\bullet \bullet }}{2g_{2}}-\frac{\left(
g_{2}^{\bullet }\right) ^{2}}{2\left( g_{2}\right) ^{2}},\ \partial _{1}%
\widehat{L}_{\ 22}^{2}=(\frac{g_{2}^{\prime }}{2g_{2}})^{\bullet }=\frac{%
g_{2}^{\prime \bullet }}{2g_{2}}-\frac{g_{2}^{\bullet }g_{2}^{\prime }}{%
2\left( g_{2}\right) ^{2}},  \notag
\end{eqnarray}%
\begin{eqnarray*}
\partial _{2}\widehat{L}_{\ 11}^{1} &=&(\frac{g_{1}^{\bullet }}{2g_{1}}%
)^{\prime }=\frac{g_{1}^{\bullet \prime }}{2g_{1}}-\frac{g_{1}^{\bullet
}g_{1}^{\prime }}{2\left( g_{1}\right) ^{2}},~\partial _{2}\widehat{L}_{\
12}^{1}=(\frac{g_{1}^{\prime }}{2g_{1}})^{\prime }=\frac{g_{1}^{\prime
\prime }}{2g_{1}}-\frac{\left( g_{1}^{\prime }\right) ^{2}}{2\left(
g_{1}\right) ^{2}}, \\
\partial _{2}\widehat{L}_{\ 22}^{1} &=&(-\frac{g_{2}^{\bullet }}{2g_{1}}%
)^{\prime }=-\frac{g_{2}^{\bullet ^{\prime }}}{2g_{1}}+\frac{g_{2}^{\bullet
}g_{1}^{^{\prime }}}{2\left( g_{1}\right) ^{2}},\ \partial _{2}\widehat{L}%
_{\ 11}^{2}=(-\frac{g_{1}^{\prime }}{2g_{2}})^{\prime }=-\frac{g_{1}^{\prime
\prime }}{2g_{2}}+\frac{g_{1}^{\bullet }g_{1}^{\prime }}{2\left(
g_{2}\right) ^{2}}, \\
\partial _{2}\widehat{L}_{\ 12}^{2} &=&(\frac{g_{2}^{\bullet }}{2g_{2}}%
)^{\prime }=\frac{g_{2}^{\bullet \prime }}{2g_{2}}-\frac{g_{2}^{\bullet
}g_{2}^{\prime }}{2\left( g_{2}\right) ^{2}},\partial _{2}\widehat{L}_{\
22}^{2}=(\frac{g_{2}^{\prime }}{2g_{2}})^{\prime }=\frac{g_{2}^{\prime
\prime }}{2g_{2}}-\frac{\left( g_{2}^{\prime }\right) ^{2}}{2\left(
g_{2}\right) ^{2}}.
\end{eqnarray*}%
We can always chose s-adapted distributions and parameterizations of
s-adapted coefficients (\ref{qstsm}) when on shells $s=3,4$ all such
derivatives $\partial _{j_{s-1}}~\widehat{L}_{\ i_{s-1}n_{s-1}}^{k_{s=1}}=0.$

\subsubsection{Computing the coefficients of the canonical Ricci and
Einstein s-tensors}

Introducing values (\ref{auxfa1}) in (\ref{auxcurvh}), we obtain two of
nontrivial components 
\begin{eqnarray*}
\widehat{R}_{\ 212}^{1} &=&\frac{g_{2}^{\bullet \bullet }}{2g_{1}}-\frac{%
g_{1}^{\bullet }g_{2}^{\bullet }}{4\left( g_{1}\right) ^{2}}-\frac{\left(
g_{2}^{\bullet }\right) ^{2}}{4g_{1}g_{2}}+\frac{g_{1}^{\prime \prime }}{%
2g_{1}}-\frac{g_{1}^{\prime }g_{2}^{\prime }}{4g_{1}g_{2}}-\frac{\left(
g_{1}^{\prime }\right) ^{2}}{4\left( g_{1}\right) ^{2}}, \\
\widehat{R}_{\ 112}^{2} &=&-\frac{g_{2}^{\bullet \bullet }}{2g_{2}}+\frac{%
g_{1}^{\bullet }g_{2}^{\bullet }}{4g_{1}g_{2}}+\frac{\left( g_{2}^{\bullet
}\right) ^{2}}{4(g_{2})^{2}}-\frac{g_{1}^{\prime \prime }}{2g_{2}}+\frac{%
g_{1}^{\prime }g_{2}^{\prime }}{4(g_{2})^{2}}+\frac{\left( g_{1}^{\prime
}\right) ^{2}}{4g_{1}g_{2}}.
\end{eqnarray*}%
Because of the anti-symmetry of the last two indices, there are four
nontrivial such terms. By definition, $\widehat{R}_{11}=-\widehat{R}_{\
112}^{2}$ and $\widehat{R}_{22}=\widehat{R}_{\ 212}^{1},$ for $g^{i}=1/g_{i}$
and $\widehat{R}_{j}^{j}=g^{j}\widehat{R}_{jj}$ (in these formulas, we do
not summarize on repeating indices). As a result, we compute 
\begin{equation}
\widehat{R}_{1}^{1}=\widehat{R}_{2}^{2}=-\frac{1}{2g_{1}g_{2}}%
[g_{2}^{\bullet \bullet }-\frac{g_{1}^{\bullet }g_{2}^{\bullet }}{2g_{1}}-%
\frac{\left( g_{2}^{\bullet }\right) ^{2}}{2g_{2}}+g_{1}^{\prime \prime }-%
\frac{g_{1}^{\prime }g_{2}^{\prime }}{2g_{2}}-\frac{(g_{1}^{\prime })^{2}}{%
2g_{1}}].  \label{hcdric}
\end{equation}

Let us compute the nontrivial canonical Ricci d-tensor components on shells $%
s=1$ and $s=2$ involving vertical indices. For simplicity, we use in the
next formulas labels of type $(i_{1},a_{2})\rightarrow (i,a),$ when $%
i_{1},i=1,2$ and $a_{2},a=3,4.$ For the N-adapted coefficients with mixed h-
and v-indices of the canonical Ricci d-tensor. Considering other groups of
coefficients, we write 
\begin{equation*}
\widehat{R}_{\ bka}^{c}=\frac{\partial \widehat{L}_{bk}^{c}}{\partial y^{a}}-%
\widehat{C}_{~ba|k}^{c}+\widehat{C}_{~bd}^{c}\widehat{T}_{~ka}^{d}=\frac{%
\partial \widehat{L}_{bk}^{c}}{\partial y^{a}}-(\frac{\partial \widehat{C}%
_{ba}^{c}}{\partial x^{k}}+\widehat{L}_{dk}^{c\,}\widehat{C}_{ba}^{d}-%
\widehat{L}_{bk}^{d}\widehat{C}_{da}^{c}-\widehat{L}_{ak}^{d}\widehat{C}%
_{bd}^{c})+\widehat{C}_{bd}^{c}\widehat{T}_{ka}^{d}.
\end{equation*}%
Contracting the indices, we obtain 
\begin{equation*}
\widehat{R}_{bk}=\widehat{R}_{\ bka}^{a}=\frac{\partial L_{bk}^{a}}{\partial
y^{a}}-\widehat{C}_{ba|k}^{a}+\widehat{C}_{bd}^{a}\widehat{T}_{ka}^{d},
\end{equation*}%
where $\widehat{C}_{b}:=\widehat{C}_{ba}^{c}$ are given by formulas (\ref%
{aux3}). We have respectively:%
\begin{equation*}
\widehat{C}_{b|k}=\mathbf{e}_{k}\widehat{C}_{b}-\widehat{L}_{\ bk}^{d\,}%
\widehat{C}_{d}=\partial _{k}\widehat{C}_{b}-N_{k}^{e}\partial _{e}\widehat{C%
}_{b}-\widehat{L}_{\ bk}^{d\,}\widehat{C}_{d}=\partial _{k}\widehat{C}%
_{b}-w_{k}\widehat{C}_{b}^{\ast }-n_{k}\widehat{C}_{b}^{\diamond }-\widehat{L%
}_{\ bk}^{d\,}\widehat{C}_{d}.
\end{equation*}%
Let us introduce a conventional splitting $\widehat{R}_{bk}=\ _{[1]}R_{bk}+\
_{[2]}R_{bk}+\ _{[3]}R_{bk},$ where%
\begin{eqnarray*}
\ _{[1]}R_{bk} &=&(\widehat{L}_{bk}^{3})^{\ast }+(\widehat{L}%
_{bk}^{4})^{\diamond },\ _{[2]}R_{bk}=-\partial _{k}\widehat{C}_{b}+w_{k}%
\widehat{C}_{b}^{\ast }+n_{k}\widehat{C}_{b}^{\diamond }+\widehat{L}_{\
bk}^{d\,}\widehat{C}_{d}, \\
\ _{[3]}R_{bk} &=&\widehat{C}_{bd}^{a}\widehat{T}_{ka}^{d}=\widehat{C}%
_{b3}^{3}\widehat{T}_{k3}^{3}+\widehat{C}_{b4}^{3}\widehat{T}_{k3}^{4}+%
\widehat{C}_{b3}^{4}\widehat{T}_{k4}^{3}+\widehat{C}_{b4}^{4}\widehat{T}%
_{k4}^{4}.
\end{eqnarray*}%
We use formulas (\ref{nontrdc}), (\ref{nontrtors}) and (\ref{aux3}) to
compute the values 
\begin{eqnarray}
\ _{[1]}R_{3k} &=&\left( \widehat{L}_{3k}^{3}\right) ^{\ast }+\left( 
\widehat{L}_{3k}^{4}\right) ^{\diamond }=\left( \frac{\mathbf{\partial }%
_{k}h_{3}}{2h_{3}}-w_{k}\frac{h_{3}^{\ast }}{2h_{3}}\right) ^{\ast
}=-w_{k}^{\ast }\frac{h_{3}^{\ast }}{2h_{3}}-w_{k}\left( \frac{h_{3}^{\ast }%
}{2h_{3}}\right) ^{\ast }+\frac{1}{2}\left( \frac{\mathbf{\partial }_{k}h_{3}%
}{h_{3}}\right) ^{\ast },  \notag \\
\ _{[2]}R_{3k} &=&-\partial _{k}\widehat{C}_{3}+w_{k}\widehat{C}_{3}^{\ast
}+n_{k}\widehat{C}_{3}^{\diamond }+\widehat{L}_{\ 3k}^{3\,}\widehat{C}_{3}+%
\widehat{L}_{\ 3k}^{4\,}\widehat{C}_{4}=  \label{aux4a} \\
&=&w_{k}[\frac{h_{3}^{\ast \ast }}{2h_{3}}-\frac{3}{4}\frac{(h_{3}^{\ast
})^{2}}{(h_{3})^{2}}+\frac{h_{4}^{\ast \ast }}{2h_{4}}-\frac{1}{2}\frac{%
(h_{4}^{\ast })^{2}}{(h_{4})^{2}}-\frac{1}{4}\frac{h_{3}^{\ast }}{h_{3}}%
\frac{h_{4}^{\ast }}{h_{4}}]+\frac{\mathbf{\partial }_{k}h_{3}}{2h_{3}}(%
\frac{h_{3}^{\ast }}{2h_{3}}+\frac{h_{4}^{\ast }}{2h_{4}})-\frac{1}{2}%
\partial _{k}(\frac{h_{3}^{\ast }}{h_{3}}+\frac{h_{4}^{\ast }}{h_{4}}), 
\notag \\
\ _{[3]}R_{3k} &=&\widehat{C}_{33}^{3}\widehat{T}_{k3}^{3}+\widehat{C}%
_{34}^{3}\widehat{T}_{k3}^{4}+\widehat{C}_{33}^{4}\widehat{T}_{k4}^{3}+%
\widehat{C}_{34}^{4}\widehat{T}_{k4}^{4}  \notag \\
&=&w_{k}\left( \frac{(h_{3}^{\ast })^{2}}{4(h_{3})^{2}}+\frac{(h_{4}^{\ast
})^{2}}{4(h_{4})^{2}}\right) +w_{k}^{\ast }\frac{h_{3}^{\ast }}{2h_{3}}-%
\frac{h_{3}^{\ast }}{2h_{3}}\frac{\mathbf{\partial }_{k}h_{3}}{2h_{3}}-\frac{%
h_{4}^{\ast }}{2h_{4}}\frac{\mathbf{\partial }_{k}h_{4}}{2h_{4}}.  \notag
\end{eqnarray}%
Putting together formulas (\ref{aux4a}) and returning the $s=1$ shall
indices $k_{1}$, and then considering indices of type $k_{2}$ and $k_{3}$
with respective $a_{3},a_{4},$ we express 
\begin{eqnarray}
\ \widehat{R}_{3k_{1}} &=&w_{k_{1}}[\frac{h_{4}^{\ast \ast }}{2h_{4}}-\frac{1%
}{4}\frac{(h_{4}^{\ast })^{2}}{(h_{4})^{2}}-\frac{1}{4}\frac{h_{3}^{\ast }}{%
h_{3}}\frac{h_{4}^{\ast }}{h_{4}}]+\frac{h_{4}^{\ast }}{2h_{4}}\frac{\mathbf{%
\partial }_{k_{1}}h_{3}}{2h_{3}}-\frac{1}{2}\frac{\partial _{k}h_{4}^{\ast }%
}{h_{4}}+\frac{1}{4}\frac{h_{4}^{\ast }\partial _{k_{1}}h_{4}}{(h_{4})^{2}} 
\notag \\
&=&\frac{w_{k_{1}}}{2h_{4}}[h_{4}^{\ast \ast }-\frac{(h_{4}^{\ast })^{2}}{%
2h_{4}}-\frac{h_{3}^{\ast }h_{4}^{\ast }}{2h_{3}}]+\frac{h_{4}^{\ast }}{%
4h_{4}}(\frac{\mathbf{\partial }_{k_{1}}h_{3}}{h_{3}}+\frac{\partial
_{k_{1}}h_{4}^{\ast }}{h_{4}})-\frac{1}{2}\frac{\partial _{k_{1}}h_{4}^{\ast
}}{h_{4}},  \label{vhcdric3}
\end{eqnarray}%
\begin{eqnarray*}
\ \widehat{R}_{5k_{2}} &=&\frac{w_{k_{2}}}{2h_{6}}[\partial _{5}(\partial
_{5}h_{6})-\frac{(\partial _{5}h_{6})^{2}}{2h_{6}}-\frac{(\partial
_{5}h_{5})(\partial _{5}h_{6})}{2h_{5}}]+\frac{\partial _{5}h_{6}}{4h_{6}}(%
\frac{\mathbf{\partial }_{k_{2}}h_{5}}{h_{5}}+\frac{\partial
_{k_{2}}(\partial _{5}h_{6})}{h_{6}})-\frac{1}{2}\frac{\partial
_{k_{2}}(\partial _{5}h_{6})}{h_{6}}, \\
\ \widehat{R}_{7k_{3}} &=&\frac{w_{k_{3}}}{2h_{8}}[\partial _{7}(\partial
_{7}h_{8})-\frac{(\partial _{7}h_{8})^{2}}{2h_{8}}-\frac{(\partial
_{7}h_{7})(\partial _{7}h_{8})}{2h_{8}}]+\frac{\partial _{7}h_{8}}{4h_{8}}(%
\frac{\mathbf{\partial }_{k_{3}}h_{7}}{h_{7}}+\frac{\partial
_{k_{3}}(\partial _{7}h_{8})}{h_{8}})-\frac{1}{2}\frac{\partial
_{k_{3}}(\partial _{7}h_{8})}{h_{8}}.
\end{eqnarray*}

We can compute $\ \widehat{R}_{4k}=\ _{[1]}R_{4k}+\ _{[2]}R_{4k}+\
_{[3]}R_{4k}$ for%
\begin{eqnarray*}
\ _{[1]}R_{4k} &=&(\widehat{L}_{4k}^{3})^{\ast }+(\widehat{L}%
_{4k}^{4})^{\diamond },\ _{[2]}R_{4k}=-\partial _{k}\widehat{C}_{4}+w_{k}%
\widehat{C}_{4}^{\ast }+n_{k}\widehat{C}_{4}^{\diamond }+\widehat{L}_{\
4k}^{3\,}\widehat{C}_{3}+\widehat{L}_{\ 4k}^{4\,}\widehat{C}_{4}, \\
_{\lbrack 3]}R_{4k} &=&\widehat{C}_{4d}^{a}\widehat{T}_{ka}^{d}=\widehat{C}%
_{43}^{3}\widehat{T}_{k3}^{3}+\widehat{C}_{44}^{3}\widehat{T}_{k3}^{4}+%
\widehat{C}_{43}^{4}\widehat{T}_{k4}^{3}+\widehat{C}_{44}^{4}\widehat{T}%
_{k4}^{4}.
\end{eqnarray*}%
Using $\widehat{L}_{4k}^{3}$ and $\widehat{L}_{4k}^{4}$ from (\ref{nontrdc}%
), we obtain%
\begin{equation*}
\ _{[1]}R_{4k}=(\widehat{L}_{4k}^{3})^{\ast }+(\widehat{L}%
_{4k}^{4})^{\diamond }=(-\frac{h_{4}}{2h_{3}}n_{k}^{\ast })^{\ast
}=-n_{k}^{\ast \ast }\frac{h_{4}}{2h_{3}}-n_{k}^{\ast }\frac{h_{4}^{\ast
}h_{3}-h_{4}h_{3}^{\ast }}{2(h_{3})^{2}},
\end{equation*}%
where the second term follows from $\widehat{C}_{3}$ and $\widehat{C}_{4},$
see (\ref{aux3}). Using again $\ \widehat{L}_{4k}^{3}$ and $\widehat{L}%
_{4k}^{4}$ (\ref{nontrdc}), we compute the next term: 
\begin{equation*}
\ _{[2]}R_{4k}=-\partial _{k}\widehat{C}_{4}+w_{k}\widehat{C}_{4}^{\ast
}+n_{k}\widehat{C}_{4}^{\diamond }+\widehat{L}_{\ 4k}^{3\,}\widehat{C}_{3}+%
\widehat{L}_{\ 4k}^{4\,}\widehat{C}_{4}=-n_{k}^{\ast }{}\frac{h_{4}}{2h_{3}}(%
\frac{h_{3}^{\ast }}{2h_{3}}+\frac{h_{4}^{\ast }}{2h_{4}}).
\end{equation*}%
Then, considering $\widehat{C}_{43}^{3},\widehat{C}_{44}^{3},\widehat{C}%
_{43}^{4},\widehat{C}_{44}^{4},$ see (\ref{nontrdc}), (then, similarly, for $%
\widehat{C}_{65}^{5},\widehat{C}_{66}^{5},\widehat{C}_{65}^{6},\widehat{C}%
_{66}^{6};$ then for $\widehat{C}_{87}^{7},\widehat{C}_{88}^{7},\widehat{C}%
_{87}^{8},\widehat{C}_{88}^{8}$), and $\widehat{T}_{k3}^{3},\widehat{T}%
_{k3}^{4},\widehat{T}_{k4}^{3},\widehat{T}_{k4}^{4},$ see (\ref{nontrtors}),
(similarly, for $\widehat{T}_{k_{1}3}^{3},\widehat{T}_{k_{1}3}^{4},\widehat{T%
}_{k_{1}4}^{3},\widehat{T}_{k_{1}4}^{4};$ $\widehat{T}_{k_{2}5}^{5},\widehat{%
T}_{k_{2}5}^{6},\widehat{T}_{k_{2}6}^{5},\widehat{T}_{k_{2}6}^{6}$; $%
\widehat{T}_{k_{3}7}^{7},\widehat{T}_{k_{3}7}^{8},\widehat{T}_{k_{3}8}^{7},%
\widehat{T}_{k_{3}8}^{8}$) we compute corresponding third terms:%
\begin{equation*}
_{\lbrack 3]}R_{4k}=\widehat{C}_{43}^{3}\widehat{T}_{k3}^{3}+\widehat{C}%
_{44}^{3}\widehat{T}_{k3}^{4}+\widehat{C}_{43}^{4}\widehat{T}_{k4}^{3}+%
\widehat{C}_{44}^{4}\widehat{T}_{k4}^{4}=0\mbox{ and similarly for }%
k_{1},k_{2},k_{3}.
\end{equation*}%
Summarizing above three terms (introducing the $s=1$ index $k_{1},$ and
respective $k_{2},k_{3}$ and $a_{3},a_{4}$), we express%
\begin{eqnarray}
\widehat{R}_{4k_{1}} &=&-n_{k_{1}}^{\ast \ast }{}\frac{h_{4}}{2h_{3}}%
+n_{k_{1}}^{\ast }{}\left( -\frac{h_{4}^{\ast }}{2h_{3}}+\frac{h_{4}^{\ast
}h_{3}^{\ast }}{2(h_{3})^{\ast }}-\frac{h_{4}^{\ast }h_{3}^{\ast }}{%
4(h_{3})^{\ast }}-\frac{h_{4}^{\ast }}{4h_{3}}\right) ,  \label{vhcdric4} \\
\widehat{R}_{6k_{2}} &=&-\partial _{5}(\partial _{5}n_{k_{2}}){}\frac{h_{6}}{%
2h_{5}}+\partial _{5}n_{k_{2}}{}\left( -\frac{\partial _{5}h_{6}}{2h_{5}}+%
\frac{(\partial _{5}h_{6})(\partial _{5}h_{5})}{2\partial _{5}h_{5}}-\frac{%
\partial _{5}h_{6}\partial _{5}h_{5}}{4\partial _{5}h_{5}}-\frac{\partial
_{5}h_{6}}{4h_{5}}\right) ,  \notag \\
\widehat{R}_{8k_{3}} &=&-\partial _{7}(\partial _{7}n_{k_{3}}){}\frac{h_{8}}{%
2h_{7}}+\partial _{7}n_{k_{3}}{}\left( -\frac{\partial _{7}h_{8}}{2h_{7}}+%
\frac{(\partial _{7}h_{8})(\partial _{7}h_{7})}{2\partial _{7}h_{7}}-\frac{%
\partial _{7}h_{8}\partial _{7}h_{7}}{4\partial _{7}h_{7}}-\frac{\partial
_{7}h_{8}}{4h_{7}}\right) .  \notag
\end{eqnarray}

In a similar form, we can compute another group of N-adapted coefficients: 
\begin{equation*}
\widehat{R}_{\ jka}^{i}=\frac{\partial \widehat{L}_{jk}^{i}}{\partial y^{k}}%
-(\frac{\partial \widehat{C}_{ja}^{i}}{\partial x^{k}}+\widehat{L}_{lk}^{i}%
\widehat{C}_{ja}^{l}-\widehat{L}_{jk}^{l}\widehat{C}_{la}^{i}-\widehat{L}%
_{ak}^{c}\widehat{C}_{jc}^{i})+\widehat{C}_{jb}^{i}\widehat{T}_{ka}^{b}.
\end{equation*}%
Such coefficients are zero because $\widehat{C}_{jb}^{i}=0$ and $\widehat{L}%
_{jk}^{i}$ do not depend on $y^{k}.$ Correspondingly, we obtain $\widehat{R}%
_{ja}=\widehat{R}_{\ jia}^{i}=0,$ or $\widehat{R}_{j_{s-1}a_{s}}=\widehat{R}%
_{\ j_{s-1}i_{s-1}a_{s}}^{i_{s-1}}=0$

At the next step, we contract the indices in $\widehat{R}_{\ bcd}^{a},$ when
the Ricci $s=2$ v-coefficients are computed 
\begin{equation*}
\widehat{R}_{bc}=\frac{\partial \widehat{C}_{bc}^{d}}{\partial y^{d}}-\frac{%
\partial \widehat{C}_{bd}^{d}}{\partial y^{c}}+\widehat{C}_{bc}^{e}\widehat{C%
}_{e}-\widehat{C}_{bd}^{e}\widehat{C}_{ec}^{d}.
\end{equation*}%
Summarizing indices (and for $s=2,3,4$), we obtain 
\begin{eqnarray*}
\widehat{R}_{b_{2}c_{2}} &=&(\widehat{C}_{b_{2}c_{2}}^{3})^{\ast }+(\widehat{%
C}_{b_{2}c_{2}}^{4})^{\diamond }-\partial _{c_{2}}\widehat{C}_{b_{2}}+%
\widehat{C}_{b_{2}c_{2}}^{3}\widehat{C}_{3}+\widehat{C}_{b_{2}c_{2}}^{4}%
\widehat{C}_{4}-\widehat{C}_{b_{2}3}^{3}\widehat{C}_{3c_{2}}^{3}-\widehat{C}%
_{b_{2}4}^{3}\widehat{C}_{3c_{2}}^{4}-\widehat{C}_{b_{2}3}^{4}\widehat{C}%
_{4c_{2}}^{3}-\widehat{C}_{b_{2}4}^{4}\widehat{C}_{4c_{2}}^{4}, \\
\widehat{R}_{b_{3}c_{3}} &=&\partial _{5}(\widehat{C}_{b_{3}c_{3}}^{5})+%
\partial _{6}(\widehat{C}_{b_{3}c_{3}}^{6})-\partial _{c_{3}}\widehat{C}%
_{b_{3}}+\widehat{C}_{b_{3}c_{3}}^{5}\widehat{C}_{5}+\widehat{C}%
_{b_{3}c_{3}}^{6}\widehat{C}_{6}-\widehat{C}_{b_{3}5}^{5}\widehat{C}%
_{5c_{3}}^{5}-\widehat{C}_{b_{3}6}^{5}\widehat{C}_{5c_{3}}^{6}-\widehat{C}%
_{b_{3}5}^{6}\widehat{C}_{6c_{3}}^{5}-\widehat{C}_{b_{3}6}^{6}\widehat{C}%
_{6c_{3}}^{6}, \\
&&\mbox{ and similar for }b_{4},c_{4}.
\end{eqnarray*}%
From these formulas, we compute such nontrivial $s=2$ adapted coefficients: 
\begin{eqnarray*}
\widehat{R}_{33} &=&\left( \widehat{C}_{33}^{3}\right) ^{\ast }+\left( 
\widehat{C}_{33}^{4}\right) ^{\diamond }-\widehat{C}_{3}^{\ast }+\widehat{C}%
_{33}^{3}\widehat{C}_{3}+\widehat{C}_{33}^{4}\widehat{C}_{4}-\widehat{C}%
_{33}^{3}\widehat{C}_{33}^{3}-2\widehat{C}_{34}^{3}\widehat{C}_{33}^{4}-%
\widehat{C}_{34}^{4}\widehat{C}_{43}^{4} \\
&=&-\frac{1}{2}\frac{h_{4}^{\ast \ast }}{h_{4}}+\frac{1}{4}\frac{%
(h_{4}^{\ast })^{2}}{(h_{4})^{2}}+\frac{1}{4}\frac{h_{3}^{\ast }}{h_{3}}%
\frac{h_{4}^{\ast }}{h_{4}}, \\
\widehat{R}_{44} &=&\left( \widehat{C}_{44}^{3}\right) ^{\ast }+\left( 
\widehat{C}_{44}^{4}\right) ^{\diamond }-\partial _{4}\widehat{C}_{4}+%
\widehat{C}_{44}^{3}\widehat{C}_{3}+\widehat{C}_{44}^{4}\widehat{C}_{4}-%
\widehat{C}_{43}^{3}\widehat{C}_{34}^{3}-2\widehat{C}_{44}^{3}\widehat{C}%
_{34}^{4}-\widehat{C}_{44}^{4}\widehat{C}_{44}^{4} \\
&=&-\frac{1}{2}\frac{h_{4}^{\ast \ast }}{h_{3}}+\frac{1}{4}\frac{h_{3}^{\ast
}h_{4}^{\ast }}{(h_{3})^{2}}+\frac{1}{4}\frac{h_{4}^{\ast }}{h_{3}}\frac{%
h_{4}^{\ast }}{h_{4}}.
\end{eqnarray*}%
For further applications, such formulas for $s=2,3,4$ can be written
equivalently in the form%
\begin{eqnarray}
\widehat{R}_{~3}^{3} &=&\frac{1}{h_{3}}\widehat{R}_{33}=\frac{1}{2h_{3}h_{4}}%
(-h_{4}^{\ast \ast }+\frac{(h_{4}^{\ast })^{2}}{2h_{4}}+\frac{h_{3}^{\ast
}h_{4}^{\ast }}{2h_{3}}),  \label{vcdric} \\
\widehat{R}_{~4}^{4} &=&\frac{1}{h_{4}}\widehat{R}_{44}=\frac{1}{2h_{3}h_{4}}%
(-h_{4}^{\ast \ast }+\frac{(h_{4}^{\ast })^{2}}{2h_{4}}+\frac{h_{3}^{\ast
}h_{4}^{\ast }}{2h_{3}});  \notag \\
\widehat{R}_{~5}^{5} &=&\frac{1}{h_{5}}\widehat{R}_{55}=\frac{1}{2h_{5}h_{6}}%
(-\partial _{5}(\partial _{5}h_{6})+\frac{(\partial _{5}h_{6})^{2}}{2h_{6}}+%
\frac{(\partial _{5}h_{5})(\partial _{5}h_{6})}{2h_{5}}),  \notag \\
\widehat{R}_{~6}^{6} &=&\frac{1}{h_{6}}\widehat{R}_{66}=\frac{1}{2h_{5}h_{6}}%
(-\partial _{5}(\partial _{5}h_{6})+\frac{(\partial _{5}h_{6})^{2}}{2h_{6}}+%
\frac{(\partial _{5}h_{5})(\partial _{5}h_{6})}{2h_{5}});  \notag \\
\widehat{R}_{~7}^{7} &=&\frac{1}{h_{7}}\widehat{R}_{77}=\frac{1}{2h_{7}h_{8}}%
(-\partial _{7}(\partial _{7}h_{8})+\frac{(\partial _{7}h_{8})^{2}}{2h_{8}}+%
\frac{(\partial _{7}h_{7})(\partial _{7}h_{8})}{2h_{7}}),  \notag \\
\widehat{R}_{~8}^{8} &=&\frac{1}{h_{8}}\widehat{R}_{88}=\frac{1}{2h_{7}h_{8}}%
(-\partial _{7}(\partial _{7}h_{8})+\frac{(\partial _{7}h_{8})^{2}}{2h_{8}}+%
\frac{(\partial _{7}h_{7})(\partial _{7}h_{8})}{2h_{7}}),  \notag
\end{eqnarray}%
Here we note that originally such computations were provided in \cite{sv11})
for 4-d nonholonomic gravitational models; more details with abstract
geometric extensions on higher dimensions are provided in \cite{bsssvv25}.

So, a quasi-stationary d-metric ansatz (\ref{qstsm}) is characterized by
such nontrivial s-adapted canonical Ricci coefficients $\widehat{R}_{1}^{1}=%
\widehat{R}_{2}^{2},$ see (\ref{hcdric}); $\ \widehat{R}_{3k_{1}},$ $%
\widehat{R}_{5k_{2}},$ $\widehat{R}_{7k_{3}},$ see (\ref{vhcdric3}); $%
\widehat{R}_{4k_{1}},\widehat{R}_{6k_{2}},\widehat{R}_{8k_{3}},$ see (\ref%
{vhcdric4}); and $\widehat{R}_{~3}^{3}=\widehat{R}_{~4}^{4},$ $\widehat{R}%
_{~5}^{5}=\widehat{R}_{~6}^{6},$ $\widehat{R}_{~7}^{7}=\widehat{R}_{~8}^{8},$
see (\ref{vcdric}). For such an ansatz, other classes of coefficients are
trivial with respect to N-adapted frames: $\ \widehat{R}_{k_{s-1}a_{s}}%
\equiv 0.$ Such values may be not zero in other systems of reference or
coordinates.

We compute the canonical Ricci d-scalar using above N-adapted nontrivial
coefficients of the canonical Ricci s-tensor, 
\begin{equation*}
\ _{s}\widehat{R}sc:=\widehat{\mathbf{g}}^{\alpha _{s}\beta _{s}}\widehat{%
\mathbf{R}}_{\ \alpha _{s}\beta _{s}}=\widehat{g}^{i_{1}j_{1}}\widehat{R}%
_{i_{1}j_{1}}+\widehat{g}^{a_{2}b_{2}}\widehat{R}_{a_{2}b_{2}}+\widehat{g}%
^{a_{3}b_{3}}\widehat{R}_{a_{3}b_{3}}=\widehat{R}_{~i_{1}}^{i_{1}}+\widehat{R%
}_{~a_{2}}^{a_{2}}+\widehat{R}_{~a_{3}}^{a_{3}}+\widehat{R}%
_{~a_{4}}^{a_{4}}=2(\widehat{R}_{2}^{2}+\widehat{R}_{~4}^{4}+\widehat{R}%
_{~6}^{6}+\widehat{R}_{~8}^{8}).
\end{equation*}%
In this formula, we consider nontrivial (\ref{hcdric}) and (\ref{vcdric}).
We can compute also the nontrivial components of the canonical Einstein
d-tensor, 
\begin{eqnarray*}
\widehat{\mathbf{E}}n &:=&\{\widehat{\mathbf{R}}_{\ \gamma }^{\beta }-\frac{1%
}{2}\delta _{\gamma }^{\beta }\ \widehat{R}sc\}=\{\widehat{E}_{~2}^{2}=-(%
\widehat{R}_{~4}^{4}+\widehat{R}_{~6}^{6}+\widehat{R}_{~8}^{8}),\widehat{E}%
_{~4}^{4}=-(\widehat{R}_{~2}^{2}+\widehat{R}_{~6}^{6}+\widehat{R}_{~8}^{8}),
\\
&& \widehat{E}_{~6}^{6} = -(\widehat{R}_{~2}^{2}+\widehat{R}_{~4}^{4}+%
\widehat{R}_{~8}^{8}), \widehat{E}_{~8}^{8}=-(\widehat{R}_{~2}^{2}+\widehat{R%
}_{~4}^{4}+\widehat{R}_{~6}^{6}), \widehat{R}_{a_{s}k_{s-1}}, \widehat{R}%
_{k_{s-1}a_{s}}\equiv 0\}.
\end{eqnarray*}%
Such symmetries are important for a general decoupling and integration of
the Einstein equations written in canonical dyadic variables. 

\subsection{ Off-diagonal integration of decoupled FLH geometric flow
modified Einstein equations}

\label{appendixab}We prove that the system of nonlinear PDEs (\ref{eq1})-(%
\ref{eq4c}) can be integrated (i.e. solved) in general forms in terms of
generating and integration functions and generating sources. The details of
such a proof are provided for the shells $s=1,2$ (i.e. for (\ref{eq1})-(\ref%
{e2c})) when similar constructions for $s=3,4$ can be performed by abstract
geometric extensions.

The first two coefficient $g_{i^{\prime }}(\tau )=e^{\psi (\tau ,x^{k_{1}})}$
of a s-metric (\ref{qstsm}) \ are defined in general form as $\tau $%
-families of solutions 2-d Poisson equation (\ref{eq1}) with a generating
source $\ _{Q}^{1}\mathbf{J}(\tau ,x^{k_{1}}).$ Further computations are
possible if we prescribe such sources in explicit forms and fix certain
systems of reference/coordinates.

Introducing $h_{3}(\tau ,y^{3})$ and $h_{4}(\tau ,y^{3})$ in explicit form
in the coefficients (\ref{coeffs}) from (\ref{e2a})-(\ref{e2c}) and for a
generating source , we obtain such a nonlinear system: 
\begin{eqnarray}
\ ^{2}\Psi ^{\ast }h_{4}^{\ast } &=&2h_{3}h_{4}\ \ _{Q}^{2}\mathbf{J}\ (\
^{2}\Psi ),  \label{auxa1} \\
\sqrt{|h_{3}h_{4}|}\ ^{2}\Psi &=&h_{4}^{\ast },  \label{auxa2} \\
\ (\ ^{2}\Psi )^{\ast }\ w_{i_{1}}-\partial _{i_{1}}\ ^{2}\Psi &=&\ 0,
\label{aux1ab} \\
\ n_{i_{1}}^{\ast \ast }+\left( \ln \frac{|h_{4}|^{3/2}}{|h_{3}|}\right)
^{\ast }n_{i_{1}}^{\ast } &=&0.\   \label{aux1ac}
\end{eqnarray}%
Prescribing a generating function, $\ ^{2}\Psi (x^{i_{1}},y^{3}),$ and a
generating source, $\ _{Q}^{2}\mathbf{J}(\tau ,x^{k_{1}},y^{3}),$ we can
integrate recurrently these equations if $h_{4}^{\ast }\neq 0$ and $\
_{Q}^{2}\mathbf{J}\neq 0.$ If such conditions are not satisfied in some
points of a phase space with $\tau $-running, more special analytic methods
have to be applied. We do not consider such cases because we can always
choose certain s-adapted frames of reference when "good" conditions (without
coordinate singularities) allow us to find necessary smooth class solutions. 

Defining 
\begin{equation}
\rho ^{2}:=-h_{3}h_{4},  \label{rho}
\end{equation}%
we re-write (\ref{auxa1}) and (\ref{auxa2}), respectively, as a system of
two nonlinear PDEs 
\begin{equation}
\ ^{2}\Psi ^{\ast }h_{4}^{\ast }=-2\rho ^{2}\ \ _{Q}^{2}\mathbf{J}\ (\
^{2}\Psi )\mbox{
and }h_{4}^{\ast }=\rho \ \ ^{2}\Psi .  \label{auxa3a}
\end{equation}%
So, we can substitute the value of $h_{4}^{\ast }(\tau )$ from the second
equation into the first equation and express 
\begin{equation}
\rho =-\ ^{2}\Psi ^{\ast }/2\ \ \ _{Q}^{2}\mathbf{J}.  \label{rho1}
\end{equation}%
This $\rho $ can be considered for the second equation in (\ref{auxa3a}) and
integrate on $y^{3},$ 
\begin{equation}
\ h_{4}(\tau ,x^{k_{1}},y^{3})=h_{4}^{[0]}(\tau ,x^{k_{1}})-\int dy^{3}[\
^{2}\Psi ^{2}]^{\ast }/4(\ \ _{Q}^{2}\mathbf{J}).  \label{g4}
\end{equation}%
Then, introducing this coefficient in (\ref{rho}) and (\ref{rho1}), we
compute%
\begin{equation}
h_{3}(\tau ,x^{k_{1}},y^{3})=-\frac{1}{4h_{4}}\left( \frac{\ ^{2}\Psi ^{\ast
}}{\ _{Q}^{2}\mathbf{J}}\right) ^{2}=-\left( \frac{\ ^{2}\Psi ^{\ast }}{2\ \
_{Q}^{2}\mathbf{J}}\right) ^{2}\left( h_{4}^{[0]}(\tau ,x^{k_{1}})-\int
dy^{3}\frac{[(\ ^{2}\Psi )^{2}]^{\ast }}{4\ \ \ \ _{Q}^{2}\mathbf{J}}\right)
^{-1}.  \label{g3}
\end{equation}

Using above formulas for $h_{3}(\tau )$ (\ref{g3}) and $h_{4}(\tau )$(\ref%
{g4}), we can integrate two times on $y^{3}$ and generate solutions of (\ref%
{aux1ac}): 
\begin{eqnarray}
n_{k_{1}}(\tau ,x^{i_{1}},y^{3}) &=&\ _{1}n_{k}(\tau )+\ _{2}n_{k}(\tau
)\int dy^{3}\ \frac{h_{3}}{|\ h_{4}|^{3/2}}=\ _{1}n_{k}+\ _{2}n_{k}\int
dy^{3}\left( \frac{\ ^{2}\Psi ^{\ast }}{2\ \ _{Q}^{2}\mathbf{J}}\right)
^{2}|\ h_{4}|^{-5/2}  \notag \\
&=&\ _{1}n_{k_{1}}(\tau )+\ _{2}n_{k_{1}}(\tau )\int dy^{3}\left( \frac{\
^{2}\Psi ^{\ast }}{2\ \ _{Q}^{2}\mathbf{J}}\right) ^{2}\left\vert
h_{4}^{[0]}(x^{k_{1}})-\int dy^{3}[(\ ^{2}\Psi )^{2}]^{\ast }/4\ \ _{Q}^{2}%
\mathbf{J}\right\vert ^{-5/2}.  \label{gn}
\end{eqnarray}%
In (\ref{gn}), we consider two integration functions $\
_{1}n_{k_{1}}(\tau)=\ _{1}n_{k_{1}}(\tau ,x^{i_{1}})$ and (re-defining
introducing certain coefficients) $\ _{2}n_{k_{1}}(\tau )=\
_{2}n_{k_{1}}(\tau ,x^{i_{1}}).$

The linear on $w_{i_{1}}$ algebraic system (\ref{aux1ab}) allows us to
compute%
\begin{equation}
w_{i_{1}}(\tau ,x^{k_{1}},y^{3})=\partial _{i_{1}}\ \ ^{2}\Psi /(\ ^{2}\Psi
)^{\ast }.  \label{gw}
\end{equation}

Putting together the above values for the coefficients of the d-metric and
N-connection (as defined by formulas (\ref{g3}),(\ref{g4}) and (\ref{gw}), (%
\ref{gn}) and together with $\tau $-families of solution of 2-d Poisson
equations for $\psi (\tau ,x^{k_{1}})$), we can generate quasi-stationary
generic off-diagonal solutions of $\tau $-families of FLH geometric flow
modified Einstein equations written in canonical nonholonomic variables.

Above procedure can be performed on shells $s=3$ and $s=4,$ with
corresponding extensions on velocity type variables and using, for
quasi-stationary configurations, the partial derivatives $\ast _{3}$ and $%
\ast _{4}$ and respective classes of generating and integration functions
and generating sources (for respective left labels $3$ and 4). Using such a
geometric formalism for the systems of nonlinear PDEs (\ref{eq3a})-(\ref%
{eq3c}) and (\ref{eq4a})-(\ref{eq4c}), we define and compute such s-adapted
coefficients for $\tau $-families of quasi-stationary s-metrics (\ref{qstsm}%
). For convenience (and to outline how the abstract geometric calculus can
be performed by analogy on all shells), we summarize such formulas 
\begin{eqnarray*}
g_{i^{\prime }}(\tau ) &=&e^{\psi (\tau ,x^{k_{1}})}, \\
h_{3}(\tau ,x^{k_{1}},y^{3}) &=&-\left( \frac{\ ^{2}\Psi ^{\ast }}{2\ \
_{Q}^{2}\mathbf{J}}\right) ^{2}\left( h_{4}^{[0]}(\tau ,x^{k_{1}})-\int
dy^{3}\frac{[(\ ^{2}\Psi )^{2}]^{\ast _{2}}}{4\ \ \ \ _{Q}^{2}\mathbf{J}}%
\right) ^{-1}, \\
h_{4}(\tau ,x^{k_{1}},y^{3}) &=&h_{4}^{[0]}(\tau ,x^{k_{1}})-\int dy^{3}[\
^{2}\Psi ^{2}]^{\ast _{2}}/4(\ \ _{Q}^{2}\mathbf{J}), \\
w_{i_{1}}(\tau ,x^{k_{1}},y^{3}) &=&\partial _{i_{1}}\ \ ^{2}\Psi /(\
^{2}\Psi )^{\ast _{2}}, \\
n_{k_{1}}(\tau ,x^{i_{1}},y^{3}) &=&\ _{1}n_{k_{1}}(\tau )+\
_{2}n_{k_{1}}(\tau )\int dy^{3}\left( \frac{\ ^{2}\Psi ^{\ast }}{2\ \
_{Q}^{2}\mathbf{J}}\right) ^{2}\left\vert h_{4}^{[0]}(x^{k_{1}})-\int
dy^{3}[(\ ^{2}\Psi )^{2}]^{\ast _{2}}/4\ \ _{Q}^{2}\mathbf{J}\right\vert
^{-5/2};
\end{eqnarray*}%
\begin{eqnarray}
h_{5}(\tau ,x^{k_{2}},y^{5}) &=&-\left( \frac{\ ^{3}\Psi ^{\ast _{3}}}{2\ \
_{Q}^{3}\mathbf{J}}\right) ^{2}\left( h_{6}^{[0]}(\tau ,x^{k_{2}})-\int
dy^{5}\frac{[(\ ^{3}\Psi )^{2}]^{\ast _{3}}}{4\ \ \ \ _{Q}^{3}\mathbf{J}}%
\right) ^{-1},  \label{auxscoef} \\
\ h_{6}(\tau ,x^{k_{2}},y^{5}) &=&h_{6}^{[0]}(\tau ,x^{k_{2}})-\int dy^{5}[\
^{3}\Psi ^{2}]^{\ast _{3}}/4(\ \ _{Q}^{3}\mathbf{J}),  \notag \\
w_{i_{2}}(\tau ,x^{k_{2}},y^{5}) &=&\partial _{i_{2}}\ ^{3}\Psi /(\ ^{3}\Psi
)^{\ast _{3}},  \notag \\
n_{k_{2}}(\tau ,x^{i_{2}},y^{5}) &=&\ _{1}n_{k_{2}}(\tau )+\
_{2}n_{k_{2}}(\tau )\int dy^{5}\left( \frac{\ ^{3}\Psi ^{\ast _{3}}}{2\ \
_{Q}^{3}\mathbf{J}}\right) ^{2}\left\vert h_{6}^{[0]}(\tau ,x^{k_{2}})-\int
dy^{5}[(\ ^{3}\Psi )^{2}]^{\ast 3}/4\ \ _{Q}^{3}\mathbf{J}\right\vert
^{-5/2};  \notag
\end{eqnarray}%
\begin{eqnarray*}
h_{7}(\tau ,x^{k_{3}},y^{7}) &=&-\left( \frac{\ ^{4}\Psi ^{\ast _{4}}}{2\ \
_{Q}^{4}\mathbf{J}}\right) ^{2}\left( h_{8}^{[0]}(\tau ,x^{k_{3}})-\int
dy^{7}\frac{[(\ ^{4}\Psi )^{2}]^{\ast _{4}}}{4\ \ \ \ _{Q}^{4}\mathbf{J}}%
\right) ^{-1}, \\
\ h_{8}(\tau ,x^{k_{3}},y^{7}) &=&h_{8}^{[0]}(\tau ,x^{k_{3}})-\int dy^{7}[\
^{4}\Psi ^{2}]^{\ast _{4}}/4(\ _{Q}^{4}\mathbf{J}), \\
w_{i_{3}}(\tau ,x^{k_{3}},y^{7}) &=&\partial _{i_{3}}\ ^{4}\Psi /(\ ^{4}\Psi
)^{\ast _{4}}, \\
n_{k_{3}}(\tau ,x^{i_{3}},y^{7}) &=&\ _{1}n_{k_{3}}(\tau )+\
_{2}n_{k_{3}}(\tau )\int dy^{7}\left( \frac{\ ^{4}\Psi ^{\ast _{4}}}{2\ \
_{Q}^{4}\mathbf{J}}\right) ^{2}\left\vert h_{8}^{[0]}(\tau ,x^{k_{3}})-\int
dy^{7}[(\ ^{4}\Psi )^{2}]^{\ast _{4}}/4\ \ _{Q}^{4}\mathbf{J}\right\vert
^{-5/2}.
\end{eqnarray*}

In a similar form, for underlined v-coefficients depending generically on $%
y^{4}=t,$ the above formulated integration procedure can be performed for
generating locally anisotropic cosmological solutions. We omit such
incremental computations and cumbersome formulas. They can be obtained by a
respective abstract geometric calculus and dualizations on $\ _{s}^{Q}%
\mathcal{M}$ and $\ _{s}^{Q\shortmid }\mathcal{M}$. In Appendix \ref%
{appendixb}, such applications of the AFCDM are summarized in Tables 4-13.

\subsection{Off-diagonal quasi-stationary solutions with small parameters}

\label{appendixac}We can consider $\epsilon $-linear nonlinear transforms (%
\ref{nonlintrsmalp}) with generating functions involving $\chi $%
-polarizations as in (\ref{offdiagpolfr}). This defines small nonholonomic
deformations of a prime s-metric $\ ^{s}\mathbf{\mathring{g}}$ into
so-called $\epsilon $-parametric solutions with $\zeta $- and $\chi $%
-coefficients derived for such approximations: 
\begin{eqnarray}
\psi (\tau ) &\simeq &\psi (\tau ,x^{k_{s}})\simeq \psi _{0}(\tau
,x^{k_{s}})(1+\epsilon \ _{\psi }\chi (\tau ,x^{k_{1}})),\mbox{ for }\ 
\label{epsilongenfdecomp} \\
\ \eta _{2}(\tau ) &\simeq &\eta _{2}(\tau ,x^{k_{1}})\simeq \zeta _{2}(\tau
,x^{k})(1+\epsilon \chi _{2}(\tau ,x^{k})),\mbox{ we can consider }\ \eta
_{2}(\tau )=\ \eta _{1}(\tau );  \notag \\
\eta _{2s}(\tau ) &\simeq &\eta _{2s}(\tau ,x^{k_{s}},u^{k_{s}+1})\simeq
\zeta _{2s}(\tau ,x^{k_{s}},u^{k_{s}+1})(1+\epsilon \chi _{2s}(\tau
,x^{k_{s}},u^{k_{s}+1})).  \notag
\end{eqnarray}%
In these formulas, $\psi $ and $\eta _{2}=\ \eta _{1}$ are chosen to be
related to the solutions of the 2-d Poisson equation $\partial _{11}^{2}\psi
+\partial _{22}^{2}\psi =2\ \ _{Q}^{1}\mathbf{J}(\tau ,x^{k}),$ see (\ref%
{eq1}).

We compute $\epsilon $-parametric deformations to quasi-stationary d-metrics
with $\chi $-generating functions by introducing formulas (\ref%
{epsilongenfdecomp}) for respective coefficients of d-metrics: 
\begin{equation*}
d\ \widehat{s}^{2}=\widehat{g}_{\alpha _{s}\beta _{s}}(h_{2s}=(1+\epsilon
\chi _{2s})\ \mathring{g}_{2s};\ ^{s}\Lambda ,\ _{Q}^{s}\mathbf{J}%
))du^{\alpha _{s}}du^{\beta _{s}}=e^{\psi _{0}}(1+\kappa \ ^{\psi }\chi
)[(dx^{1})^{2}+(dx^{2})^{2}]+
\end{equation*}%
\begin{eqnarray*}
&&\sum\nolimits_{s=2}^{s=4}[-\{\frac{4[\partial _{2s-1}(|\zeta _{2s}%
\mathring{g}_{2s}|^{1/2})]^{2}}{\mathring{g}_{2s}|\int du^{2s-1}\{\ _{Q}^{s}%
\mathbf{J\ }\partial _{2s-1}(\zeta _{2s}\mathring{g}_{2s})\}|} \\
&&-\epsilon \lbrack \frac{\partial _{2s-1}(\chi _{2s}|\zeta _{2s}\mathring{g}%
_{2s}|^{1/2})}{4\partial _{2s-1}(|\zeta _{2s}\mathring{g}_{2s}|^{1/2})}-%
\frac{\int du^{2s-1}\{\ _{Q}^{s}\mathbf{J\ }\partial _{2s-1}[(\zeta _{2s}%
\mathring{g}_{2s})\chi _{2s}]\}}{\int du^{2s-1}\{\ _{Q}^{s}\mathbf{J\ }%
\partial _{2s-1}(\zeta _{4}\mathring{g}_{4})\}}]\}\mathring{g}_{2s-1} \\
&&\{du^{2s+1}+[\frac{\partial _{i_{s}}\ \int du^{2s-1}\ \ _{Q}^{s}\mathbf{J}%
\ \partial _{2s-1}\zeta _{2s}}{(\mathring{N}_{i_{s-1}}^{i_{s-1}+1})\ \
_{Q}^{s}\mathbf{J}\partial _{2s-1}\zeta _{2s}}+\epsilon (\frac{\partial
_{i_{s}}[\int du^{2s-1}\ \ _{Q}^{s}\mathbf{J}\ \partial _{2s-1}(\zeta
_{2s}\chi _{2s})]}{\partial _{i_{s}}\ [\int du^{2s-1}\ \ _{Q}^{s}\mathbf{J\ }%
\partial _{2s-1}\zeta _{2s}]}-\frac{\partial _{2s-1}(\zeta _{2s}\chi _{2s})}{%
\partial _{2s-1}\zeta _{2s}})]\mathring{N}_{i_{s}}^{i_{s}+1}dx^{i_{s}}\}^{2}
\end{eqnarray*}%
\begin{eqnarray}
&&+\zeta _{2s}(1+\epsilon \chi _{2s})\ \mathring{g}_{2s}\{du^{2s-1}+[(%
\mathring{N}_{k_{s}}^{k_{s}+2})^{-1}[\ _{1}n_{k_{s}}+16\ _{2}n_{k_{s}}[\int
du^{2s-1}\frac{\left( \partial _{2s-1}[(\zeta _{2s}\mathring{g}%
_{2s})^{-1/4}]\right) ^{2}}{|\int du^{2s-1}\partial _{2s}[\ _{Q}^{s}\mathbf{J%
}(\zeta _{2s}\mathring{g}_{2s})]|}]  \label{offdncelepsilon} \\
&&+\epsilon \frac{16\ _{2}n_{k_{s}}\int du^{2s-1}\frac{\left( \partial
_{2s-1}[(\zeta _{2s}\mathring{g}_{2s})^{-1/4}]\right) ^{2}}{|\int
dy^{3}\partial _{2s-1}[\ _{Q}^{s}\mathbf{J}(\zeta _{2s}\mathring{g}_{2s})]|}(%
\frac{\partial _{2s-1}[(\zeta _{2s}\mathring{g}_{2s})^{-1/4}\chi _{2s})]}{%
2\partial _{2s-1}[(\zeta _{2s}\mathring{g}_{2s})^{-1/4}]}+\frac{\int
dy^{3}\partial _{2s-1}[\ _{Q}^{s}\mathbf{J}(\zeta _{2s}\chi _{2s}\mathring{g}%
_{2s})]}{\int dy^{3}\partial _{2s-1}[\ _{Q}^{s}\mathbf{J}(\zeta _{2s}%
\mathring{g}_{2s})]})}{\ _{1}n_{k_{s}}+16\ _{2}n_{k_{s}}[\int du^{2s-1}\frac{%
\left( \partial _{2s-1}[(\zeta _{2s}\mathring{g}_{2s})^{-1/4}]\right) ^{2}}{%
|\int du^{2s-1}\partial _{2s-1}[\ _{Q}^{s}\mathbf{J}(\zeta _{2s}\mathring{g}%
_{2s})]|}]}]\mathring{N}_{k_{s}}^{k_{s+2}}dx^{k_{s}}\}^{2}].  \notag
\end{eqnarray}%
We can relate a solution of type (\ref{offdncelepsilon}) to an another one
in the form (\ref{offdiagpolfr}) if $\ ^{s}\Phi ^{2}=-4\ \ ^{s}\Lambda
(1+\epsilon \chi _{2s})\ \mathring{g}_{2s}$ and the $\eta $-polarizations
are determined by the generating data $(h_{2s}=(1+\epsilon \chi _{2s})\ 
\mathring{g}_{2s};\ ^{s}\Lambda ,\ _{Q}^{s}\mathbf{J}).$ 

\newpage \setcounter{equation}{0} 
\renewcommand{\theequation}
{B.\arabic{equation}} \setcounter{subsection}{0} 
\renewcommand{\thesubsection}
{B.\arabic{subsection}}

\section{Tables 1-13 for generating off-diagonal solutions in FLH theories}

\label{appendixb}

In this Appendix, we summarize the AFCDM for constructing exact and
parametric solutions for 4-d and 8-d relativistic geometric flow and
nonholonomic Ricci soliton equations encoding FLH distortions of the
Einstein equations in GR. Various relativistic phase space theories and FLH
models can be elaborated on tangent bundle, $T\mathbf{V,}$ and cotangent
bundle, $T^{\ast }\mathbf{V,}$ where $\mathbf{V}$ is a Lorentz manifold as
we explained in Section \ref{sec4} and \cite%
{vmon3,vacaru18,vbubuianu17,bsssvv25,sv11}. In this Appendix, we study
models with $\dim \mathbf{V}=4,$ when $\dim T\mathbf{V}=8$ and $\dim T^{\ast}%
\mathbf{V}=8$. 

\subsection{Nonmetric 4-d off-diagonal quasi-stationary and cosmological
solutions, Tables 1-3}

The first three tables are reproduced from our partner work \cite{bsssvv25},
with extension of sources $\ _{Q}\widehat{\mathbf{J}}(\tau )=(\
_{Q}^{1}J(\tau ),\ \ _{Q}^{2}J(\tau ))$ to encode nonmetric distortions. For
relativistic geometric flows, such generating sources can be related via
nonlinear symmetries to some $\tau $-running effective cosmological
constants $\Lambda (\tau )=(\ _{1}\Lambda (\tau ),\ _{2}\Lambda (\tau )).$
They summarize the main steps on how to use 2+2 nonholonomic variables and
corresponding ansatz for metrics to generate quasi-stationary and, for
respective $t$-dual symmetries, locally anisotropic cosmological solutions
in GR and 4-d toy models of FLH theories. Details on definition of geometric
d-objects and notations can be found in Sections \ref{sec2} - \ref{sec4} and
Appendix \ref{appendixa}, for shells $s=1$ and 2.

\subsubsection{Ansatz for 4-d metrics and d-metrics and systems of nonlinear
ODEs and PDEs}

Table 1 outlines main formulas on parameterizations of frames/coordinates
for Lorentz manifolds with N-connection h- and v-splitting of geometric
objects and generating of (effective) sources. Two types of generic
off-diagonal metric ansatz are considered. The first one is for generating
quasi-stationary metrics, with dependence only on space coordinates and the
second one, for so-called locally anisotropic cosmological solutions, is
with dependence on the time-like coordinate and possible dependencies on two
other space-like coordinates.

General decoupling properties can be proven in explicit form for a generic
off-diagonal ansatz with Killing symmetry on $\partial _{4},$ for
quasi-stationary configurations, or on $\partial _{3},$ for locally
anisotropic cosmological models, see respectively the shells $s=1$ and $s=2$%
\ (\ref{ssolutions}) and (\ref{ssolutionsd}).



{\scriptsize 
\begin{eqnarray*}
&&%
\begin{tabular}{l}
\hline\hline
\begin{tabular}{lll}
& {\ \textsf{Table 1:\ Diagonal and off-diagonal ansatz resulting in systems
of nonlinear ODEs and PDEs} } &  \\ 
& the Anholonomic Frame and Connection Deformation Method, \textbf{AFCDM}, & 
\\ 
& \textit{for constructing }$\tau $-families of\textit{\ generic
off-diagonal exact and parametric solutions on Lorentz manifold \ }$\mathbf{V%
}$ & 
\end{tabular}%
\end{tabular}
\\
&&{%
\begin{tabular}{lll}
\hline
diagonal ansatz: PDEs $\rightarrow $ \textbf{ODE}s &  & AFCDM: \textbf{PDE}s 
\textbf{with decoupling; \ generating functions} \\ 
radial coordinates $u^{\alpha }=(r,\theta ,\varphi ,t)$ & $u=(x,y):$ & 
\mbox{ nonholonomic 2+2
splitting, } $u^{\alpha }=(x^{1},x^{2},y^{3},y^{4}=t)$ \\ 
LC-connection $\mathring{\nabla}$ & [connections] & $%
\begin{array}{c}
\mathbf{N}:T\mathbf{V}=hT\mathbf{V}\oplus vT\mathbf{V,}\mbox{ locally }%
\mathbf{N}=\{N_{i}^{a}(x,y)\} \\ 
\mbox{ canonical connection distortion }\widehat{\mathbf{D}}=\nabla +%
\widehat{\mathbf{Z}};\widehat{\mathbf{D}}\mathbf{g=0,} \\ 
\widehat{\mathcal{T}}[\mathbf{g,N,}\widehat{\mathbf{D}}]%
\mbox{ canonical
d-torsion}%
\end{array}%
$ \\ 
$%
\begin{array}{c}
\mbox{ diagonal ansatz  }g_{\alpha \beta }(u) \\ 
=\left( 
\begin{array}{cccc}
\mathring{g}_{1} &  &  &  \\ 
& \mathring{g}_{2} &  &  \\ 
&  & \mathring{g}_{3} &  \\ 
&  &  & \mathring{g}_{4}%
\end{array}%
\right)%
\end{array}%
$ & $\mathbf{g}(\tau )\Leftrightarrow $ & $%
\begin{array}{c}
g_{\alpha \beta }(\tau )=%
\begin{array}{c}
g_{\alpha \beta }(\tau ,x^{i},y^{a})\mbox{ general frames / coordinates} \\ 
\left[ 
\begin{array}{cc}
g_{ij}+N_{i}^{a}N_{j}^{b}h_{ab} & N_{i}^{b}h_{cb} \\ 
N_{j}^{a}h_{ab} & h_{ac}%
\end{array}%
\right] ,\mbox{ 2 x 2 blocks }%
\end{array}
\\ 
\mathbf{g}_{\alpha \beta }(\tau )=[g_{ij},h_{ab}],\mathbf{g}(\tau )=\mathbf{g%
}_{i}(x^{k})dx^{i}\otimes dx^{i}+\mathbf{g}_{a}(x^{k},y^{b})\mathbf{e}%
^{a}\otimes \mathbf{e}^{b}%
\end{array}%
$ \\ 
$\mathring{g}_{\alpha \beta }=\left\{ 
\begin{array}{cc}
\mathring{g}_{\alpha }(r) & \mbox{ for BHs} \\ 
\mathring{g}_{\alpha }(t) & \mbox{ for FLRW }%
\end{array}%
\right. $ & [coord.frames] & $g_{\alpha \beta }=\left\{ 
\begin{array}{cc}
g_{\alpha \beta }(x^{i},y^{3}) & \mbox{ quasi-stationary configurations} \\ 
\underline{g}_{\alpha \beta }(x^{i},y^{4}=t) & 
\mbox{locally anisotropic
cosmology}%
\end{array}%
\right. $ \\ 
&  &  \\ 
$%
\begin{array}{c}
\mbox{coord. transf. }e_{\alpha }=e_{\ \alpha }^{\alpha ^{\prime }}\partial
_{\alpha ^{\prime }}, \\ 
e^{\beta }=e_{\beta ^{\prime }}^{\ \beta }du^{\beta ^{\prime }},\mathring{g}%
_{\alpha \beta }=\mathring{g}_{\alpha ^{\prime }\beta ^{\prime }}e_{\ \alpha
}^{\alpha ^{\prime }}e_{\ \beta }^{\beta ^{\prime }} \\ 
\begin{array}{c}
\mathbf{\mathring{g}}_{\alpha }(x^{k},y^{a})\rightarrow \mathring{g}_{\alpha
}(r),\mbox{ or }\mathring{g}_{\alpha }(t), \\ 
\mathring{N}_{i}^{a}(x^{k},y^{a})\rightarrow 0.%
\end{array}%
\end{array}%
$ & [N-adapt. fr.] & $\left\{ 
\begin{array}{cc}
\begin{array}{c}
\mathbf{g}_{i}(\tau ,x^{k}),\mathbf{g}_{a}(\tau ,x^{k},y^{3}), \\ 
\mbox{ or }\mathbf{g}_{i}(\tau ,x^{k}),\underline{\mathbf{g}}_{a}(\tau
,x^{k},t),%
\end{array}
& \mbox{ d-metrics } \\ 
\begin{array}{c}
N_{i}^{3}(\tau )=w_{i}(\tau ,x^{k},y^{3}),N_{i}^{4}=n_{i}(\tau ,x^{k},y^{3}),
\\ 
\mbox{ or }\underline{N}_{i}^{3}=\underline{n}_{i}(\tau ,x^{k},t),\underline{%
N}_{i}^{4}=\underline{w}_{i}(\tau ,x^{k},t),%
\end{array}
& 
\end{array}%
\right. $ \\ 
$\mathring{\nabla},$ $Ric=\{\mathring{R}_{\ \beta \gamma }\}$ & Ricci tensors
& $\widehat{\mathbf{D}}(\tau ),\ \widehat{\mathcal{R}}ic(\tau )=\{\widehat{%
\mathbf{R}}_{\ \beta \gamma }(\tau )\}$ \\ 
$~^{m}\mathcal{L[\mathbf{\phi }]\rightarrow }\ ^{m}\mathbf{T}_{\alpha \beta }%
\mathcal{[\mathbf{\phi }]}$ & 
\begin{tabular}{l}
generating \\ 
sources%
\end{tabular}
& $%
\begin{array}{cc}
\ _{Q}\widehat{\mathbf{J}}_{\ \nu }^{\mu }(\tau )=\mathbf{e}_{\ \mu ^{\prime
}}^{\mu }\mathbf{e}_{\nu }^{\ \nu ^{\prime }}\ _{Q}\mathbf{J}_{\ \nu
^{\prime }}^{\mu ^{\prime }}[\ ^{m}\mathcal{L}\mathbf{,}T_{\mu \nu },^{e}%
\mathcal{L}], & \ ^{1}\Lambda (\tau ),\ ^{2}\Lambda (\tau ) \\ 
=diag[\ _{Q}^{1}J(x^{i})\delta _{j}^{i},\ \ _{Q}^{2}J(x^{i},y^{3})\delta
_{b}^{a}], & \mbox{ quasi-stationary configurations} \\ 
=diag[\ \ _{Q}^{1}J(x^{i})\delta _{j}^{i},\ _{Q}^{2}\underline{J}%
(x^{i},t)\delta _{b}^{a}], & \mbox{ locally anisotropic cosmology}%
\end{array}%
$ \\ 
trivial equations for $\mathring{\nabla}$-torsion & LC-conditions & $%
\widehat{\mathbf{D}}_{\mid \widehat{\mathcal{T}}\rightarrow 0}(\tau )=%
\mathbf{\nabla }(\tau )\mbox{ extracting new classes of solutions in GR}$ \\ 
\hline\hline
\end{tabular}%
}
\end{eqnarray*}%
}


\subsubsection{Decoupling and integration of (modified) Einstein eqs \&
quasi-stationary configurations}

We provide a summary of formulas for a general decoupling and integrating of
generalized Einstein equations with generic off-diagonal quasi-stationary
and locally anisotropic cosmological metrics in 4-d gravity theories
outlined below in Tables 1-3.

{\scriptsize 
\begin{eqnarray*}
&&%
\begin{tabular}{l}
\hline\hline
\begin{tabular}{lll}
& {\large \textsf{Table 2:\ Off-diagonal nonmetric quasi-stationary
configurations}} &  \\ 
& Exact solutions of $\widehat{\mathbf{R}}_{\mu _{2}\nu _{2}}=\ _{Q}\widehat{%
\mathbf{J}}_{\mu _{2}\nu _{2}}(\tau )$ (\ref{cfeq4af}) (for $s-1,2$)
transformed into a system of nonlinear PDEs (\ref{eq1})-(\ref{e2c}) & 
\end{tabular}
\\ 
\end{tabular}
\\
&&%
\begin{tabular}{lll}
\hline\hline
&  &  \\ 
$%
\begin{array}{c}
\\ 
\end{array}%
\begin{array}{c}
\mbox{d-metric ansatz with} \\ 
\mbox{Killing symmetry }\partial _{4}=\partial _{t} \\ 
\mbox{general or spherical coordinates}%
\end{array}%
$ &  & $%
\begin{array}{c}
ds^{2}(\tau )=g_{i_{1}}(\tau ,x^{k_{1}})(dx^{i_{1}})^{2}+g_{a_{2}}(\tau
,x^{k_{1}},y^{3})(dy^{a_{2}}+N_{i_{1}}^{a_{2}}(\tau
,x^{k_{1}},y^{3})dx^{i_{1}})^{2},\mbox{ for } \\ 
g_{i_{1}}=e^{\psi {(x}^{k_{1}}{)}%
},g_{a_{2}}=h_{a_{2}}(x^{k_{1}},y^{3}),N_{i_{1}}^{3}=w_{i_{1}}(x^{k_{1}},y^{3}),N_{i_{1}}^{4}=n_{i_{1}}(x^{k_{1}},y^{3});
\\ 
g_{i_{1}}=e^{\psi {(r,\theta )}},\,\,\,\,g_{a_{2}}=h_{a_{2}}({r,\theta }%
,\varphi ),\ N_{i_{1}}^{3}=w_{i_{1}}({r,\theta },\varphi
),\,\,\,\,N_{i_{1}}^{4}=n_{i_{1}}({r,\theta },\varphi ),%
\end{array}%
$ \\ 
&  &  \\ 
Effective matter sources &  & $\ _{Q}\widehat{\mathbf{J}}_{\ \nu }^{\mu }=[\
_{Q}^{1}J({r,\theta })\delta _{j}^{i},\ _{Q}^{2}J({r,\theta },\varphi
)\delta _{b}^{a}],\mbox{ if }x^{1}=r,x^{2}=\theta ,y^{3}=\varphi ,y^{4}=t$
\\ \hline
Nonlinear PDEs (\ref{eq1})-(\ref{e2c}) &  & $%
\begin{array}{c}
\psi ^{\bullet \bullet }+\psi ^{\prime \prime }=2\ \ _{Q}^{1}J; \\ 
\ ^{2}\varpi ^{\ast _{2}}\ h_{4}^{\ast _{2}}=2h_{3}h_{4}\ _{Q}^{2}J; \\ 
\ ^{2}\beta w_{i_{1}}-\alpha _{i_{1}}=0; \\ 
n_{k_{1}}^{\ast _{2}\ast _{2}}+\ ^{2}\gamma n_{k_{1}}^{\ast _{2}}=0;%
\end{array}%
$ for $%
\begin{array}{c}
\ ^{2}\varpi {=\ln |\partial _{3}h_{4}/\sqrt{|h_{3}h_{4}|}|,} \\ 
\alpha _{i}=(\partial _{3}h_{4})\ (\partial _{i}\ ^{2}\varpi ),\ ^{2}\beta
=(\partial _{3}h_{4})\ (\partial _{3}\ ^{2}\varpi ),\  \\ 
\ \ ^{2}\gamma =\partial _{3}\left( \ln |h_{4}|^{3/2}/|h_{3}|\right) , \\ 
\partial _{1}q=q^{\bullet },\partial _{2}q=q^{\prime },\partial
_{3}q=\partial q/\partial \varphi =q^{\ast _{2}}%
\end{array}%
$ \\ \hline
$%
\begin{array}{c}
\mbox{ Generating functions:}\ h_{3}(\tau ,x^{k_{1}},y^{3}), \\ 
\ ^{2}\Psi (\tau ,x^{k_{1}},y^{3})=e^{\ ^{2}\varpi },\ ^{2}\Phi (\tau
,x^{k_{1}},y^{3}); \\ 
\mbox{integration functions:}\ h_{4}^{[0]}(\tau ,x^{k_{1}}),\  \\ 
_{1}n_{k_{1}}(\tau ,x^{i_{1}}),\ _{2}n_{k_{1}}(\tau ,x^{i_{1}}); \\ 
\mbox{\& nonlinear symmetries}%
\end{array}%
$ &  & $%
\begin{array}{c}
\ (\ ^{2}\Psi ^{2})^{\ast _{2}}=-\int dy^{3}\ \ \ _{Q}^{2}Jh_{4}^{\ \ast
_{2}}, \\ 
\ ^{2}\Phi ^{2}=-4\ _{2}\Lambda h_{4},\mbox{ see }(\ref{nonlinsymrex}); \\ 
h_{4}=h_{4}^{[0]}-\ ^{2}\Phi ^{2}/4\ _{2}\Lambda ,h_{4}^{\ast _{2}}\neq 0,\
_{2}\Lambda \neq 0=const%
\end{array}%
$ \\ \hline
Off-diag. solutions, $%
\begin{array}{c}
\mbox{d--metric} \\ 
\mbox{N-connec.}%
\end{array}%
$ &  & $%
\begin{array}{c}
\ g_{i}=e^{\ \psi }\mbox{ as a solution of 2-d Poisson eqs. }\psi ^{\bullet
\bullet }+\psi ^{\prime \prime }=2~\ \ _{Q}^{1}J; \\ 
h_{3}=-(\ ^{2}\Psi ^{\ast _{2}})^{2}/4\ \ _{Q}^{2}Jh_{4},\mbox{ see }(\ref%
{g3}),(\ref{g4}); \\ 
h_{4}=h_{4}^{[0]}-\int dy^{3}(\ ^{2}\Psi ^{2})^{\ast _{2}}/4\ \
_{Q}^{2}J=h_{4}^{[0]}-\ ^{2}\Phi ^{2}/4\ _{2}\Lambda ; \\ 
\\ 
w_{i_{1}}=\partial _{i_{1}}\ ^{2}\Psi /\ \partial _{3}\ ^{2}\Psi =\partial
_{i_{1}}\ ^{2}\Psi ^{2}/\ \partial _{3}\ ^{2}\Psi ^{2}|; \\ 
n_{k_{1}}=\ _{1}n_{k_{1}}+\ _{2}n_{k_{1}}\int dy^{3}(\ ^{2}\Psi ^{\ast
_{2}})^{2}/\ \ _{Q}^{2}J^{2}|h_{4}^{[0]}-\int dy^{3}(\ ^{2}\Psi ^{2})^{\ast
_{2}}/4\ \ _{Q}^{2}J^{2}|^{5/2}. \\ 
\\ 
\end{array}%
$ \\ \hline
LC-configurations (\ref{zerot1}) &  & $%
\begin{array}{c}
\partial _{\varphi }w_{i_{1}}=(\partial _{i_{1}}-w_{i_{1}}\partial _{3})\ln 
\sqrt{|h_{3}|},(\partial _{i_{1}}-w_{i_{1}}\partial _{3})\ln \sqrt{|h_{4}|}%
=0, \\ 
\partial _{k_{1}}w_{i_{1}}=\partial _{i_{1}}w_{k_{1}},\partial
_{3}n_{i_{1}}=0,\partial _{i_{1}}n_{k_{1}}=\partial _{k_{1}}n_{i_{1}}; \\ 
\mbox{ see d-metric }(\ref{qellc})\mbox{ for } \\ 
\ ^{2}\Psi =\ ^{2}\check{\Psi}(x^{i_{1}},y^{3}),(\partial _{i_{1}}\ ^{2}%
\check{\Psi})^{\ast _{2}}=\partial _{i_{1}}(\check{\Psi}^{\ast _{2}})%
\mbox{
and } \\ 
\ _{Q}^{2}J(x^{i_{1}},\varphi )=\ _{Q}^{2}J[\check{\Psi}]=\ _{Q}^{2}\check{J}%
,\mbox{ or }\ _{Q}^{2}J=const. \\ 
\end{array}%
$ \\ \hline
N-connections, zero torsion &  & $%
\begin{array}{c}
w_{i_{1}}=\partial _{i_{1}}\ ^{2}\check{A}=\left\{ 
\begin{array}{c}
\partial _{i_{1}}(\int dy^{3}\ \ _{Q}^{2}\check{J}\ \check{h}_{4}{}^{\ast
_{2}}])/\ _{Q}^{2}\check{J}\ \check{h}_{4}{}^{\ast _{2}}; \\ 
\partial _{i_{1}}\ ^{2}\check{\Psi}/\ ^{2}\check{\Psi}^{\ast _{2}}; \\ 
\partial _{i_{1}}(\int dy^{3}\ \ _{Q}^{2}\check{J}(\ ^{2}\check{\Phi}%
^{2})^{\ast _{2}})/(\ ^{2}\check{\Phi})^{\ast _{2}}\ _{Q}^{2}\check{J};%
\end{array}%
\right. \\ 
\mbox{ and }n_{k_{1}}=\check{n}_{k_{1}}=\partial _{k_{1}}n(x^{i_{1}}).%
\end{array}%
$ \\ \hline
$%
\begin{array}{c}
\mbox{polarization functions} \\ 
\mathbf{\mathring{g}}\rightarrow \widehat{\mathbf{g}}\mathbf{=}[g_{\alpha
_{2}}=\eta _{\alpha _{2}}\mathring{g}_{\alpha _{2}},\ \eta _{i_{1}}^{a_{2}}%
\mathring{N}_{i_{1}}^{a_{2}}]%
\end{array}%
$ &  & $%
\begin{array}{c}
\\ 
ds^{2}=\eta _{1}(r,\theta )\mathring{g}_{1}(r,\theta )[dx^{1}(r,\theta
)]^{2}+\eta _{2}(r,\theta )\mathring{g}_{2}(r,\theta )[dx^{2}(r,\theta
)]^{2}+ \\ 
\eta _{3}(r,\theta ,\varphi )\mathring{g}_{3}(r,\theta )[d\varphi +\eta
_{i}^{3}(r,\theta ,\varphi )\mathring{N}_{i}^{3}(r,\theta )dx^{i}(r,\theta
)]^{2}+ \\ 
\eta _{4}(r,\theta ,\varphi )\mathring{g}_{4}(r,\theta )[dt+\eta
_{i}^{4}(r,\theta ,\varphi )\mathring{N}_{i}^{4}(r,\theta )dx^{i}(r,\theta
)]^{2}, \\ 
\end{array}%
$ \\ \hline
Prime metric defines a BH &  & $%
\begin{array}{c}
\\ 
\lbrack \mathring{g}_{i}(r,\theta ),\mathring{g}_{a}=\mathring{h}%
_{a}(r,\theta );\mathring{N}_{k}^{3}=\mathring{w}_{k}(r,\theta ),\mathring{N}%
_{k}^{4}=\mathring{n}_{k}(r,\theta )] \\ 
\mbox{diagonalizable by frame/ coordinate transforms.} \\ 
\end{array}%
$ \\ 
Example of a prime metric &  & $%
\begin{array}{c}
\\ 
\mathring{g}_{1}=(1-r_{g}/r)^{-1},\mathring{g}_{2}=r^{2},\mathring{h}%
_{3}=r^{2}\sin ^{2}\theta ,\mathring{h}_{4}=(1-r_{g}/r),r_{g}=const \\ 
\mbox{the Schwarzschild solution, or any BH solution.} \\ 
\mbox{ for new KdS solutions }(\ref{offdiagpm1})\mbox{ with }\mathbf{%
\mathring{g}\simeq \breve{g}}(x^{i_{1}},y^{3})\mathbf{=}(\breve{g}_{\alpha
_{2}};\breve{N}_{i_{1}}^{a_{2}}); \\ 
\end{array}%
$ \\ \hline
Solutions for polarization funct. &  & $%
\begin{array}{c}
\eta _{i_{1}}=e^{\ \psi }/\mathring{g}_{i_{1}};\eta _{3}\mathring{h}_{3}=-%
\frac{4[(|\eta _{4}\mathring{h}_{4}|^{1/2})^{\ast _{2}}]^{2}}{|\int dy^{3}\
\ _{Q}^{2}J[(\eta _{4}\mathring{h}_{4})]^{\ast _{2}}|\ }; \\ 
\eta _{4}=\eta _{4}(x^{k},y^{3})\mbox{ as a generating
function}; \\ 
\ \eta _{i_{1}}^{3}\ \mathring{N}_{i_{1}}^{3}=\frac{\partial _{i}\ \int
dy^{3}\ _{Q}^{2}J(\eta _{4}\ \mathring{h}_{4})^{\ast _{2}}}{\ _{Q}^{2}J\
(\eta _{4}\ \mathring{h}_{4})^{\ast _{2}}}; \\ 
\eta _{k_{1}}^{4}\ \mathring{N}_{k_{1}}^{4}=\ _{1}n_{k_{1}}+16\ \
_{2}n_{k_{1}}\int dy^{3}\frac{\left( [(\eta _{4}\mathring{h}%
_{4})^{-1/4}]^{\ast _{2}}\right) ^{2}}{|\int dy^{3}\ _{Q}^{2}J[(\eta _{4}\ 
\mathring{h}_{4})]^{\ast _{2}}|\ }%
\end{array}%
;$ \\ \hline
Polariz. funct. with zero torsion &  & $%
\begin{array}{c}
\eta _{i_{1}}=e^{\ \psi }/\mathring{g}_{i_{1}};\eta _{4}=\check{\eta}%
_{4}(x^{k_{1}},y^{3})\mbox{ as a generating function}; \\ 
\eta _{3}=-\frac{4[(|\eta _{4}\mathring{h}_{4}|^{1/2})^{\ast _{2}}]^{2}}{%
\mathring{g}_{3}|\int dy^{3}\ _{Q}^{2}\check{J}[(\check{\eta}_{4}\mathring{h}%
_{4})]^{\ast _{2}}|\ };\eta _{i_{1}}^{3}=\partial _{i_{1}}\ ^{2}\check{A}/%
\mathring{w}_{k_{1}},\eta _{k_{1}}^{4}=\frac{\ \partial _{k_{1}}\ ^{2}n}{%
\mathring{n}_{k_{1}}}. \\ 
\end{array}%
$ \\ \hline\hline
\end{tabular}%
\end{eqnarray*}%
%
} 

Using the AFCDM, we are able to investigate off-diagonal nonlinear
gravitational and (effective) matter field interactions and construct
respective classes of solutions in explicit form. This is more general than
in the case when the (modified) Einstein equations are transformed in
systems of nonlinear ODEs.

\subsubsection{Decoupling and integration of 4-d gravitational PDEs
generating cosmological s-metrics}

In Table 3, we summarize the main steps for constructing 4-d off-diagonal
locally anisotropic cosmological solutions of FLH deformed geometric flow
and modified Einstein equations.

{\scriptsize 
\begin{eqnarray*}
&&%
\begin{tabular}{l}
\hline\hline
\begin{tabular}{lll}
& {\large \textsf{Table 3:\ Off-diagonal locally anisotropic nonmetric
cosmological models}} &  \\ 
& Exact solutions of $\widehat{\mathbf{R}}_{\mu _{2}\nu _{2}}(\tau )=\ _{Q}%
\widehat{\underline{\mathbf{J}}}_{\mu _{2}\nu _{2}}(\tau )$ (\ref{cfeq4af})
transformed into a system of nonlinear PDEs (\ref{dualcosm}) & 
\end{tabular}
\\ 
\end{tabular}
\\
&&%
\begin{tabular}{lll}
\hline\hline
$%
\begin{array}{c}
\mbox{d-metric ansatz with} \\ 
\mbox{Killing symmetry }\partial _{3}=\partial _{\varphi }%
\end{array}%
$ &  & $%
\begin{array}{c}
d\underline{s}^{2}(\tau )=g_{i_{1}}(x^{k_{1}})(dx^{i_{1}})^{2}+\underline{g}%
_{a_{2}}(x^{k_{1}},y^{4})(dy^{a_{2}}+\underline{N}%
_{i_{1}}^{a_{2}}(x^{k_{1}},y^{4})dx^{i_{2}})^{2},\mbox{ for }g_{i_{1}}(\tau )
\\ 
=e^{\psi {(x}^{k_{1}}{)}},\,\,\,\,\underline{g}_{a_{2}}(\tau )=\underline{h}%
_{a_{2}}({x}^{k_{1}},t),\ \underline{N}_{i_{1}}^{3}(\tau )=\underline{n}%
_{i_{1}}({x}^{k_{1}},t),\,\,\,\underline{\,N}_{i_{1}}^{4}(\tau )=\underline{w%
}_{i_{1}}({x}^{k_{1}},t)%
\end{array}%
$ \\ 
&  &  \\ 
Effective matter sources &  & $\ \ _{Q}\widehat{\underline{\mathbf{J}}}_{\
\nu _{2}}^{\mu _{2}}(\tau )=[\ _{Q}^{1}J(\tau ,{x}^{k_{1}})\delta
_{j_{1}}^{i_{1}},\ _{Q}^{2}\underline{J}(\tau ,{x}^{k_{1}},t)\delta
_{b_{2}}^{a_{2}}]$ \\ \hline
Nonlinear PDEs &  & $%
\begin{array}{c}
\psi ^{\bullet \bullet }+\psi ^{\prime \prime }=2\ \ _{Q}^{1}J; \\ 
\ ^{2}\underline{\varpi }^{\diamond _{2}}\ \underline{h}_{3}^{\diamond
_{2}}=2\underline{h}_{3}\underline{h}_{4}\ \ _{Q}^{2}\underline{J}; \\ 
\underline{n}_{k_{2}}^{\diamond _{2}\diamond _{2}}+\ ^{2}\underline{\gamma }%
\underline{n}_{k_{2}}^{\diamond _{2}}=0; \\ 
\ ^{2}\underline{\beta }\underline{w}_{i_{1}}-\underline{\alpha }_{i_{1}}=0;%
\end{array}%
$ for $%
\begin{array}{c}
\ ^{2}\underline{\varpi }(\tau ){=\ln |\partial _{t}\underline{{h}}_{3}/%
\sqrt{|\underline{h}_{3}\underline{h}_{4}|}|,} \\ 
\underline{\alpha }_{i_{1}}=(\partial _{t}\underline{h}_{3})\ (\partial
_{i_{1}}\ ^{2}\underline{\varpi }),\ \ ^{2}\underline{\beta }=(\partial _{t}%
\underline{h}_{3})\ (\partial _{t}\ ^{2}\underline{\varpi }), \\ 
\ \ ^{2}\underline{\gamma }=\partial _{t}\left( \ln |\underline{h}%
_{3}|^{3/2}/|\underline{h}_{4}|\right) , \\ 
\partial _{1}q=q^{\bullet },\partial _{2}q=q^{\prime },\partial
_{4}q=\partial q/\partial t=q^{\diamond }%
\end{array}%
$ \\ \hline
$%
\begin{array}{c}
\mbox{ Generating functions:}\ \underline{h}_{4}(\tau ,{x}^{k},t), \\ 
^{2}\underline{\Psi }(\tau ,x^{k},t)=e^{^{2}\underline{\varpi }},^{2}%
\underline{\Phi }(\tau ,{x}^{k},t); \\ 
\mbox{integr. func.:}\ h_{4}^{[0]}(\tau ,x^{k}),\ _{1}n_{k_{1}}(\tau
,x^{i_{1}}),\  \\ 
_{2}n_{k_{1}}(\tau ,x^{i_{1}});\mbox{\& nonlinear sym.}%
\end{array}%
$ &  & $%
\begin{array}{c}
\ (\ ^{2}\underline{\Psi }^{2})^{\diamond _{2}}=-\int dt\ \ _{Q}^{2}%
\underline{J}\underline{h}_{3}^{\diamond }, \\ 
\ ^{2}\underline{\Phi }^{2}=-4\ _{2}\underline{\Lambda }(\tau )\underline{h}%
_{3}; \\ 
\underline{h}_{3}=\underline{h}_{3}^{[0]}-\ ^{2}\underline{\Phi }^{2}/4\ _{2}%
\underline{\Lambda },\underline{h}_{3}^{\diamond _{2}}\neq 0,\ _{2}%
\underline{\Lambda }(\tau _{0})\neq 0=const%
\end{array}%
$ \\ \hline
Off-diag. solutions, $%
\begin{array}{c}
\mbox{d--metric} \\ 
\mbox{N-connec.}%
\end{array}%
$ &  & $%
\begin{array}{c}
\ g_{i_{1}}=e^{\ \psi }\mbox{ as a solution of 2-d Poisson eqs. }\psi
^{\bullet \bullet }+\psi ^{\prime \prime }=2\ \ _{Q}^{1}J; \\ 
\underline{h}_{4}=-(\ ^{2}\underline{\Psi }^{2})^{\diamond _{2}}/4\ \
_{Q}^{2}\underline{J}^{2}\underline{h}_{3}; \\ 
\underline{h}_{3}=\underline{h}_{3}^{[0]}-\int dt(\ ^{2}\underline{\Psi }%
^{2})^{\diamond _{2}}/4\ \ _{Q}^{2}\underline{J}=\underline{h}_{3}^{[0]}-\
^{2}\underline{\Phi }^{2}/4\ _{2}\underline{\Lambda }; \\ 
\underline{n}_{k_{1}}=\ _{1}n_{k_{1}}+\ _{2}n_{k_{1}}\int dt(\ ^{2}%
\underline{\Psi }^{\diamond _{2}})^{2}/\ _{Q}^{2}\underline{J}^{2}\ |%
\underline{h}_{3}^{[0]}-\int dt(\ ^{2}\underline{\Psi }^{2})^{\diamond
_{2}}/4\ \ _{Q}^{2}\underline{J}|^{5/2}; \\ 
\underline{w}_{i_{1}}=\partial _{i_{1}}\ \ ^{2}\underline{\Psi }/\ \partial
_{t}\ ^{2}\underline{\Psi }=\partial _{i_{1}}\ ^{2}\underline{\Psi }^{2}/\
\partial _{t}\ ^{2}\underline{\Psi }^{2}. \\ 
\end{array}%
$ \\ \hline
LC-configurations &  & $%
\begin{array}{c}
\partial _{t}\underline{w}_{i_{1}}=(\partial _{i_{1}}-\underline{w}%
_{i_{1}}\partial _{t})\ln \sqrt{|\underline{h}_{4}|},(\partial _{i_{1}}-%
\underline{w}_{i_{1}}\partial _{4})\ln \sqrt{|\underline{h}_{3}|}=0, \\ 
\partial _{k_{1}}\underline{w}_{i_{1}}=\partial _{i_{1}}\underline{w}%
_{k_{1}},\partial _{t}\underline{n}_{i_{1}}=0,\partial _{i_{1}}\underline{n}%
_{k_{1}}=\partial _{k_{1}}\underline{n}_{i_{1}}; \\ 
\ \ ^{2}\underline{\Psi }=\ \ ^{2}\underline{\check{\Psi}}(\tau
,x^{i_{1}},t),(\partial _{i_{1}}\ ^{2}\underline{\check{\Psi}})^{\diamond
_{2}}=\partial _{i_{1}}(\ ^{2}\underline{\check{\Psi}}^{\diamond _{2}})%
\mbox{ and } \\ 
\ \ \ _{Q}^{2}\underline{J}(\tau ,x^{i_{1}},t)=\underline{J}[\ ^{2}%
\underline{\check{\Psi}}]=\underline{\check{J}},\mbox{ or }\underline{\check{%
J}}=const. \\ 
\end{array}%
$ \\ \hline
N-connections, zero torsion &  & $%
\begin{array}{c}
\underline{n}_{k_{1}}=\underline{\check{n}}_{k_{1}}=\partial _{k_{1}}\ ^{2}%
\underline{n}(x^{i_{1}}) \\ 
\mbox{ and }\underline{w}_{i_{1}}=\partial _{i_{1}}\ ^{2}\underline{\check{A}%
}=\left\{ 
\begin{array}{c}
\partial _{i_{1}}(\int dt\ \underline{\check{J}}\ \underline{\check{h}}%
_{3}^{\diamond _{2}}])/\underline{\check{J}}\ \underline{\check{h}}%
_{3}^{\diamond _{2}}{}; \\ 
\partial _{i_{k}}\ ^{2}\underline{\check{\Psi}}/\ ^{2}\underline{\check{\Psi}%
}^{\diamond _{2}}; \\ 
\partial _{i}(\int dt\ \ \underline{\check{J}}(\ ^{2}\underline{\check{\Phi}}%
^{2})^{\diamond _{2}})/\underline{\check{\Phi}}^{\diamond _{2}}\ \underline{%
\check{J}};%
\end{array}%
\right. .%
\end{array}%
$ \\ \hline
$%
\begin{array}{c}
\mbox{polarization functions} \\ 
\mathbf{\mathring{g}}\rightarrow \underline{\widehat{\mathbf{g}}}(\tau )%
\mathbf{=} \\ 
\lbrack \underline{g}_{\alpha _{2}}(\tau )=\underline{\eta }_{\alpha
_{2}}(\tau )\mathring{g}_{\alpha _{2}},\underline{\eta }_{i_{1}}^{a_{2}}(%
\tau )\underline{\mathring{N}}_{i_{1}}^{a_{2}}]%
\end{array}%
$ &  & $%
\begin{array}{c}
ds^{2}(\tau )=\underline{\eta }_{i_{1}}(x^{k_{1}},t)\underline{\mathring{g}}%
_{i_{1}}(x^{k_{1}},t)[dx^{i_{1}}]^{2} \\ 
+\underline{\eta }_{3}(x^{k_{1}},t)\underline{\mathring{h}}%
_{3}(x^{k_{1}},t)[dy^{3}+\underline{\eta }_{i_{1}}^{3}(x^{k_{1}},t)%
\underline{\mathring{N}}_{i_{1}}^{3}(x^{k_{1}},t)dx^{i_{1}}]^{2} \\ 
+\underline{\eta }_{4}(x^{k_{1}},t)\underline{\mathring{h}}%
_{4}(x^{k_{1}},t)[dt+\underline{\eta }_{i_{1}}^{4}(x^{k_{1}},t)\underline{%
\mathring{N}}_{i_{1}}^{4}(x^{k_{1}},t)dx^{i_{1}}]^{2},%
\end{array}%
$ \\ \hline
$%
\begin{array}{c}
\mbox{ Prime metric defines } \\ 
\mbox{ a cosmological solution}%
\end{array}%
$ &  & $%
\begin{array}{c}
\lbrack \underline{\mathring{g}}_{i_{1}}(x^{k_{1}},t),\underline{\mathring{g}%
}_{a_{2}}=\underline{\mathring{h}}_{a_{2}}(x^{k_{1}},t);\underline{\mathring{%
N}}_{k_{1}}^{3}=\underline{\mathring{w}}_{k_{1}}(x^{i_{1}},t),\underline{%
\mathring{N}}_{k_{1}}^{4}=\underline{\mathring{n}}_{k_{1}}(x^{k_{1}},t)] \\ 
\mbox{diagonalizable by frame/ coordinate transforms.} \\ 
\end{array}%
$ \\ 
$%
\begin{array}{c}
\mbox{Example of a prime } \\ 
\mbox{ cosmological metric }%
\end{array}%
$ &  & $%
\begin{array}{c}
\mathring{g}_{1}=a^{2}(t)/(1-kr^{2}),\mathring{g}_{2}=a^{2}(t)r^{2}, \\ 
\underline{\mathring{h}}_{3}=a^{2}(t)r^{2}\sin ^{2}\theta ,\underline{%
\mathring{h}}_{4}=c^{2}=const,k=\pm 1,0; \\ 
\mbox{ any frame transform of a FLRW or a Bianchi metrics} \\ 
\end{array}%
$ \\ \hline
Solutions for polariz. funct. &  & $%
\begin{array}{c}
\eta _{i_{1}}=e^{\ \psi }/\mathring{g}_{i_{1}};\underline{\eta }_{4}%
\underline{\mathring{h}}_{4}=-\frac{4[(|\underline{\eta }_{3}\underline{%
\mathring{h}}_{3}|^{1/2})^{\diamond _{2}}]^{2}}{|\int dt\ \ _{Q}^{2}%
\underline{J}[(\underline{\eta }_{3}\underline{\mathring{h}}_{3})]^{\diamond
_{2}}|\ };\mbox{ gener. f. }\underline{\eta }_{3}=\underline{\eta }_{3}(\tau
,x^{i_{1}},t);\underline{\eta }_{k_{1}}^{3}\ \underline{\mathring{N}}%
_{k_{1}}^{3} \\ 
=\ _{1}n_{k_{1}}+16\ \ _{2}n_{k_{1}}\int dt\frac{\left( [(\underline{\eta }%
_{3}\underline{\mathring{h}}_{3})^{-1/4}]^{\diamond _{2}}\right) ^{2}}{|\int
dt\ \ _{Q}^{2}\underline{J}[(\underline{\eta }_{3}\underline{\mathring{h}}%
_{3})]^{\diamond _{2}}|\ };\ \underline{\eta }_{i_{1}}^{4}\ \underline{%
\mathring{N}}_{i_{1}}^{4}=\frac{\partial _{i_{1}}\ \int dt\ \ _{Q}^{2}%
\underline{J}(\underline{\eta }_{3}\underline{\mathring{h}}_{3})^{\diamond
_{2}}}{\ \ _{Q}^{2}\underline{J}(\underline{\eta }_{3}\underline{\mathring{h}%
}_{3})^{\diamond _{2}}}%
\end{array}%
$ \\ \hline
Polariz. funct. with zero torsion &  & $%
\begin{array}{c}
\eta _{i_{1}}=e^{\ \psi }/\mathring{g}_{i_{1}};\underline{\eta }_{4}=-\frac{%
4[(|\underline{\eta }_{3}\underline{\mathring{h}}_{3}|^{1/2})^{\diamond
_{2}}]^{2}}{\underline{\mathring{g}}_{4}|\int dt\ \ _{Q}^{2}\underline{J}[(%
\underline{\eta }_{3}\underline{\mathring{h}}_{3})]^{\diamond _{2}}|\ };%
\mbox{ gener.
funct. }\underline{\eta }_{3}=\underline{\check{\eta}}_{3}({x}^{i_{1}},t);
\\ 
\underline{\eta }_{k_{1}}^{4}=\partial _{k_{1}}\ ^{2}\underline{\check{A}}/%
\mathring{w}_{k_{1}};\underline{\eta }_{k_{1}}^{3}=(\partial _{k}\ ^{2}%
\underline{n})/\ ^{2}\mathring{n}_{k}. \\ 
\end{array}%
$ \\ \hline\hline
\end{tabular}%
\end{eqnarray*}%
%
%
} 

Applying a nonholonomic deformation procedure as described in this Table 3
(when, for simplicity, the d-metrics are determined by a generating function 
$\underline{h}_{4}({x}^{k_{1}},t)),$ we construct a class of generic
off--diagonal cosmological solutions of nonholonomic Ricci soltions with
Killing symmetry on $\partial _{3}$ determined by effective sources, $\
_{Q}^{1}\underline{J}$ and$\ _{Q}^{2}\underline{J},$ and a nontrivial
cosmological constant, $_{2}\underline{\Lambda },$ 
\begin{eqnarray}
ds^{2} &=&e^{\ \psi (x^{k_{1}},\ _{Q}^{1}\underline{J}%
)}[(dx^{1})^{2}+(dx^{2})^{2}]+\underline{h}_{3}[dy^{3}+(\ _{1}n_{k_{1}}+4\
_{2}n_{k_{1}}\int dt\frac{(\underline{h}_{3}{}^{\diamond _{2}})^{2}}{|\int
dt\ \ _{Q}^{2}\underline{J}\underline{h}_{3}{}^{\diamond _{2}}|(\underline{h}%
_{3})^{5/2}})dx^{k_{1}}]  \notag \\
&&-\frac{(\underline{h}_{3}{}^{\diamond _{2}})^{2}}{|\int dt\ \ _{Q}^{2}%
\underline{J}h_{3}{}^{\diamond _{2}}|\ \underline{h}_{3}}[dt+\frac{\partial
_{i_{1}}(\int dt\ \ _{Q}^{2}\underline{J}\ \underline{h}_{3}{}^{\diamond
_{2}}])}{\ \ _{Q}^{2}\underline{J}\ \underline{h}_{3}{}^{\diamond _{2}}}%
dx^{i_{1}}].  \label{cosm1d}
\end{eqnarray}%
Such a d-metric is equivalent to (\ref{qeltorsc}) like (\ref%
{offdsolgenfgcosmc}) is equivalent to (\ref{qeltors}). In similar forms, the
d-metric (\ref{cosm1d}) can be written in terms of gravitational $\eta $%
-polarization and/or $\chi $-polarization functions. 

\subsection{Off-diagonal velocity depending quasi-stationary or cosmological
FL solutions}

FL geometric flow and MGTs are modelled on tangent bundle $T\mathbf{V}$ to a
nonholonomic Lorentz manifold $\mathbf{V}.$ The typical signature of total
metrics is $(+++-;+++-)$ for a Lorentz base with signature $(+++-).$ To
apply the AFCDM we need four shells of dyads (when $s=1,2,3,4$) with a
corresponding (2+2)+(2+2) nonholonomic splitting of the total dimension. The
formulas are quite similar to those provided in the previous subsection when 
$y^{a_{s}}=v^{a_{s}},$ for $s=3$ and 4. If the phase space solutions are
with Killing symmetry on $\partial _{8},$ we can fix $v^{8}=v_{[0]}^{8},$
and elaborate on phase space models with space like velocity hypersurfaces.
Another class of solutions can be with variable $v^{8}$ but a fixed, for
instance, velocity $v^{7}=v_{[0]}^{7},$ which consists examples of
"velocity-rainbow" metrics in phase gravity. Both types of s-metrics with
the mentioned behaviour in the velocity typical fiber may have a Killing
symmetry on $\partial _{4}$ (for locally anisotropic cosmological
solutions), or, for instance, on $\partial _{3}$, for quasi-stationary
solutions. As result, we obtain 4 different types of velocity-phase
s-metrics with typical quadratic elements and applications of the AFCDM
stated in subsections below and respective Tables 4-8. 

\subsubsection{Diagonal and off-diagonal ansatz for velocity phase spaces
and FL geometric flows}

The parametrization of local coordinates, N-connection and canonical
d-connection structures and s-metrics for velocity-phase spaces are sated in
Table 4.

Such parameterizations, with respective polarization functions and
generating sources, can be considered for generalized relativistic Finsler
spaces encoding data for metric and nonmetric nonassociative /
noncommutative / supersymmetric theories etc. The generating and integration
functions can be restricted to define LC-configurations. The solutions are
determined by respective generating sources $\ _{Q}^{s}\mathbf{J}(\tau )$ (%
\ref{dsourcparam}) in the FL-deformed geometric flow equations (\ref{cfeq4af}%
). For simplicity, we do not write in the formulas of Table 4 the dependence
of geometric objects on $\tau $-parameter. It can be introduce additionally
to the phase space coordinates if the generating functions, generating
sources and effective cosmological constant involve such a temperature-like
dependence.

\newpage

{\scriptsize 
\begin{eqnarray*}
&&%
\begin{tabular}{lll}
& {\ \textsf{Table 4:\ Diagonal and off-diagonal ansatz for FL theories on
8-d tangent Lorentz bundles} } &  \\ 
& and the Anholonomic Frame and Connection Deformation Method, \textbf{AFCDM}%
, &  \\ 
& \textit{for constructing generic off-diagonal exact and parametric
solutions} & 
\end{tabular}
\\
&&{%
\begin{tabular}{lll}
\hline
diagonal ansatz: PDEs $\rightarrow $ \textbf{ODE}s &  & AFCDM: \textbf{PDE}s 
\textbf{with decoupling; } \\ 
\begin{tabular}{l}
coordinates \\ 
$u^{\alpha _{s}}=(x^{1},x^{2},y^{3},y^{4}=t,$ \\ 
$v^{5},v^{6},v^{7},v^{8})$%
\end{tabular}
& $%
\begin{array}{c}
\ ^{s}u=(\ ^{s-1}x,\ ^{s}y) \\ 
s=1,2,3,4;%
\end{array}%
$ & $%
\begin{tabular}{l}
nonholonomic 2+2+2+2 splitting; shels $s=1,2,3,4$ \\ 
$u^{\alpha _{s}}=(x^{1},x^{2},y^{3},y^{4}=t,y^{5},y^{6},y^{7},y^{8});$ \\ 
$u^{\alpha _{s}}=(x^{i_{1}},y^{a_{2}},y^{a_{3}},y^{a_{4}});u^{\alpha
_{s}}=(x^{i_{s-1}},y^{a_{s}});$ \\ 
$\ $ $i_{1}=1,2;a_{2}=3,4;a_{3}=5,6;a_{4}=7,8;$%
\end{tabular}%
$ \\ 
LC-connection $\mathring{\nabla}$ & 
\begin{tabular}{l}
N-connection; \\ 
canonical \\ 
d-connection%
\end{tabular}
& $%
\begin{array}{c}
\ ^{s}\mathbf{N}:T\ ^{s}\mathbf{V}=hT\mathbf{V}\oplus \ ^{2}hT\mathbf{V}%
\oplus \ ^{3}v\mathbf{T\mathbf{V}\oplus }\ ^{4}v\mathbf{T\mathbf{V},} \\ 
\mbox{ locally }\ ^{s}\mathbf{N}%
=\{N_{i_{s-1}}^{a_{s}}(x,v)=N_{i_{s-1}}^{a_{s}}(\ ^{s-1}x,\
^{s}y)=N_{i_{s-1}}^{a_{s}}(\ ^{s}u)\} \\ 
\ ^{s}\widehat{\mathbf{D}}=(\ ^{1}h\widehat{\mathbf{D}},\ ^{2}v\widehat{%
\mathbf{D}},\ ^{3}v\widehat{\mathbf{D}},\ ^{4}v\widehat{\mathbf{D}}%
)=\{\Gamma _{\ \beta _{s}\gamma _{s}}^{\alpha _{s}}\}; \\ 
\mbox{ canonical connection distortion }\ ^{s}\widehat{\mathbf{D}}=\nabla +\
^{s}\widehat{\mathbf{Z}};\ ^{s}\widehat{\mathbf{D}}\ ^{s}\mathbf{g=0,} \\ 
\ ^{s}\widehat{\mathcal{T}}[\ ^{s}\mathbf{g,}\ ^{s}\mathbf{N,}\ ^{s}\widehat{%
\mathbf{D}}]\mbox{ canonical
d-torsion}%
\end{array}%
$ \\ 
$%
\begin{array}{c}
\mbox{ diagonal ansatz  } \\ 
\ ^{2}\mathring{g}=\mathring{g}_{\alpha _{2}\beta _{2}}(\ ^{s}u)= \\ 
\left( 
\begin{array}{cccc}
\mathring{g}_{1} &  &  &  \\ 
& \mathring{g}_{2} &  &  \\ 
&  & \mathring{g}_{3} &  \\ 
&  &  & \mathring{g}_{4}%
\end{array}%
\right) ; \\ 
\ ^{s}g=\mathring{g}_{\alpha _{s}\beta _{s}}(\ ^{s}u)= \\ 
\left( 
\begin{array}{cccc}
\ ^{2}\mathring{g} &  &  &  \\ 
& \mathring{g}_{5} &  &  \\ 
&  & \ddots &  \\ 
&  &  & \mathring{g}_{8}%
\end{array}%
\right)%
\end{array}%
$ & $\mathbf{g}\Leftrightarrow $ & $%
\begin{tabular}{l}
$g_{\alpha _{2}\beta _{2}}=%
\begin{array}{c}
g_{\alpha _{2}\beta _{2}}(x^{i_{1}},y^{a_{2}})%
\mbox{ general frames /
coordinates} \\ 
\left[ 
\begin{array}{cc}
g_{i_{1}j_{1}}+N_{i_{1}}^{a_{2}}N_{j_{1}}^{b_{2}}h_{a_{2}b_{2}} & 
N_{i_{1}}^{b_{2}}h_{c_{2}b_{2}} \\ 
N_{j_{1}}^{a_{2}}h_{a_{2}b_{2}} & h_{a_{2}c_{2}}%
\end{array}%
\right] ,%
\end{array}%
$ \\ 
$\ ^{2}\mathbf{g=\{g}_{\alpha _{2}\beta
_{2}}=[g_{i_{1}j_{1}},h_{a_{2}b_{2}}]\},$ \\ 
$\ ^{2}\mathbf{g}=\mathbf{g}_{i_{1}}(x^{k_{1}})dx^{i_{1}}\otimes dx^{i_{1}}+%
\mathbf{g}_{a_{2}}(x^{k_{2}},y^{b_{2}})\mathbf{e}^{a_{2}}\otimes \mathbf{e}%
^{b_{2}}$ \\ 
$\vdots $ \\ 
$g_{\alpha _{s}\beta _{s}}=%
\begin{array}{c}
g_{\alpha _{s}\beta _{s}}(x^{i_{s-1}},y^{a_{s}})%
\mbox{ general frames /
coordinates} \\ 
\left[ 
\begin{array}{cc}
g_{i_{s}j_{s}}+N_{i_{s-1}}^{a_{s}}N_{j_{s-1}}^{b_{s}}h_{a_{s}b_{s}} & 
N_{i_{s-1}}^{b_{s}}h_{c_{s}b_{s}} \\ 
N_{j_{s-1}}^{a_{s}}h_{a_{s}b_{s}} & h_{a_{s}c_{s}}%
\end{array}%
\right] ,%
\end{array}%
$ \\ 
$\ ^{s}\mathbf{g=\{g}_{\alpha _{s}\beta
_{s}}=[g_{i_{s-1}j_{s-1}},h_{a_{s}b_{s}}]$ \\ 
$=[g_{i_{1}j_{1}},h_{a_{2}b_{2}},h_{a_{3}b_{3}},h_{a_{4}b_{4}}]\}$ \\ 
$\ ^{s}\mathbf{g}=\mathbf{g}_{i_{s-1}}(x^{k_{s-1}})dx^{i_{s-1}}\otimes
dx^{i_{s-1}}+$ \\ 
$\mathbf{g}_{a_{s}}(x^{k_{s-1}},y^{b_{s}})\mathbf{e}^{a_{s}}\otimes \mathbf{e%
}^{b_{s}}$ \\ 
$=\mathbf{g}_{i_{1}}(x^{k_{1}})dx^{i_{1}}\otimes dx^{i_{1}}+\mathbf{g}%
_{a_{2}}(x^{k_{1}},y^{b_{2}})\mathbf{e}^{a_{2}}\otimes \mathbf{e}^{b_{2}}+$
\\ 
$\mathbf{g}_{a_{3}}(x^{k_{1}},y^{b_{2}},v^{b_{3}})\mathbf{e}^{a_{3}}\otimes 
\mathbf{e}^{b_{3}}+\mathbf{g}%
_{a_{4}}(x^{k_{1}},y^{b_{2}},v^{b_{3}},v^{b_{4}})\mathbf{e}^{a_{4}}\otimes 
\mathbf{e}^{b_{4}};$%
\end{tabular}%
$ \\ 
$\mathring{g}_{\alpha _{2}\beta _{2}}=\left\{ 
\begin{array}{cc}
\mathring{g}_{\alpha _{2}}(\ ^{2}r) & \mbox{ for BHs} \\ 
\mathring{g}_{\alpha _{2}}(t) & \mbox{ for FLRW }%
\end{array}%
\right. $ & [coord.frames] & $g_{\alpha _{2}\beta _{2}}=\left\{ 
\begin{array}{cc}
g_{\alpha _{2}\beta _{2}}(x^{i},y^{3}) & \mbox{ quasi-stationary config. }
\\ 
\underline{g}_{\alpha _{2}\beta _{2}}(x^{i},y^{4}=t) & 
\mbox{locally anisotropic
cosmology}%
\end{array}%
\right. $ \\ 
$\mathring{g}_{\alpha _{s}\beta _{s}}=\left\{ 
\begin{array}{cc}
\mathring{g}_{\alpha _{s}}(\ ^{s}r) & \mbox{ for BHs} \\ 
\mathring{g}_{\alpha _{s}}(t) & \mbox{ for FLRW }%
\end{array}%
\right. $ &  & $g_{\alpha _{5}\beta _{5}}=\left\{ 
\begin{array}{cc}
g_{\alpha _{5}\beta _{5}}(x^{i_{3}},v^{7}) &  \\ 
\underline{g}_{\alpha _{s}\beta _{s}}(x^{i_{3}},y^{8}) & 
\end{array}%
\right. $ \\ 
$%
\begin{array}{c}
\mbox{coord. transf. }e_{\alpha _{s}}=e_{\ \alpha _{s}}^{\alpha _{s}^{\prime
}}\partial _{\alpha _{s}^{\prime }}, \\ 
e^{\beta _{s}}=e_{\beta _{s}^{\prime }}^{\ \beta _{s}}du^{\beta _{s}^{\prime
}}, \\ 
\mathring{g}_{\alpha _{s}\beta _{s}}=\mathring{g}_{\alpha _{s}^{\prime
}\beta _{s}^{\prime }}e_{\ \alpha _{s}}^{\alpha _{s}^{\prime }}e_{\ \beta
_{s}}^{\beta _{s}^{\prime }} \\ 
\begin{array}{c}
\mathbf{\mathring{g}}_{\alpha _{s}}(x^{k_{s-1}},y^{a_{s}})\rightarrow 
\mathring{g}_{\alpha _{s}}(\ ^{s}r),\mbox{ or } \\ 
\mathring{g}_{\alpha _{s}}(t),\mathring{N}%
_{i_{s-1}}^{a_{s}}(x^{k_{s-1}},y^{a_{s}})\rightarrow 0.%
\end{array}%
\end{array}%
$ & [N-adapt. fr.] & 
\begin{tabular}{l}
$\left\{ 
\begin{array}{cc}
\begin{array}{c}
\mathbf{g}_{i_{1}}(x^{k_{1}}),\mathbf{g}_{a_{2}}(x^{k_{1}},y^{3}), \\ 
\mbox{ or }\mathbf{g}_{i_{1}}(x^{k_{1}}),\underline{\mathbf{g}}%
_{a_{2}}(x^{k_{1}},t),%
\end{array}
& \mbox{ d-metrics } \\ 
\begin{array}{c}
N_{i_{1}}^{3}=w_{i_{1}}(x^{k},y^{3}),N_{i_{1}}^{4}=n_{i_{1}}(x^{k},y^{3}),
\\ 
\mbox{ or }\underline{N}_{i_{1}}^{3}=\underline{n}_{i_{1}}(x^{k_{1}},t),%
\underline{N}_{i_{1}}^{4}=\underline{w}_{i_{1}}(x^{k_{1}},t),%
\end{array}
& 
\end{array}%
\right. $ \\ 
$\vdots $ \\ 
$\left\{ 
\begin{array}{cc}
\begin{array}{c}
\mathbf{g}_{i_{3}}(x^{k_{3}}),\mathbf{g}_{a_{4}}(x^{k_{3}},v^{7}), \\ 
\mbox{ or }\mathbf{g}_{i_{3}}(x^{k_{1}}),\underline{\mathbf{g}}%
_{a_{4}}(x^{k_{3}},v^{8}),%
\end{array}
&  \\ 
\begin{array}{c}
N_{i_{3}}^{7}=w_{i_{3}}(x^{k_{3}},v^{7}),N_{i_{3}}^{8}=n_{i_{3}}(x^{k_{3}},v^{7}),
\\ 
\mbox{ or }\underline{N}_{i_{3}}^{8}=\underline{n}_{i_{3}}(x^{k_{3}},v^{8}),%
\underline{N}_{i_{3}}^{8}=\underline{w}_{i_{3}}(x^{k_{3}},v^{8}),%
\end{array}
& 
\end{array}%
\right. $%
\end{tabular}
\\ 
$\ ^{s}\mathring{\nabla},$ $\ ^{s}Ric=\{\mathring{R}_{\ \beta _{s}\gamma
_{s}}\}$ & Ricci tensors & $\ ^{s}\widehat{\mathbf{D}},\ \ ^{s}\widehat{%
\mathcal{R}}ic=\{\widehat{\mathbf{R}}_{\ \beta _{s}\gamma _{s}}\}$ \\ 
$~^{m}\mathcal{L[\mathbf{\phi }]\rightarrow }\ ^{m}\mathbf{T}_{\alpha
_{s}\beta _{s}}\mathcal{[\mathbf{\phi }]},~^{e}\mathcal{L}[Q,...]$ & 
\begin{tabular}{l}
generating \\ 
sources%
\end{tabular}
& $%
\begin{array}{cc}
\ _{Q}^{s}\mathbf{J}_{\ \nu _{s}}^{\mu _{s}}=\mathbf{e}_{\ \mu _{s}^{\prime
}}^{\mu _{s}}\mathbf{e}_{\nu _{s}}^{\ \nu _{s}^{\prime }}\ _{Q}\mathbf{J}_{\
\nu _{s}^{\prime }}^{\mu _{s}^{\prime }}[\ ^{m}\mathcal{L}\mathbf{,}\ ^{e}%
\mathcal{L},T_{\mu _{s}\nu _{s}},\ ^{s}\Lambda ] &  \\ 
\begin{array}{c}
=diag[\ _{1}^{Q}J(x^{i_{1}})\delta _{j_{1}}^{i_{1}},\
_{2}^{Q}J(x^{i_{1}},y^{3})\delta _{b_{2}}^{a_{2}}, \\ 
\ _{3}^{Q}J(x^{i_{2}},v^{5})\delta _{b_{3}}^{a_{3}},\
_{4}^{Q}J(x^{i_{3}},v^{7})\delta _{b_{4}}^{a_{4}}], \\ 
\mbox{ quasi-stationary configurations};%
\end{array}
&  \\ 
\begin{array}{c}
=diag[\ _{1}^{Q}J(x^{i_{1}})\delta _{j_{1}}^{i_{1}},\ _{2}^{Q}\underline{J}%
(x^{i_{1}},t)\delta _{b_{2}}^{a_{2}}, \\ 
\ _{3}^{Q}\underline{J}(x^{i_{2}},v^{6})\delta _{b_{3}}^{a_{3}},\ _{4}^{Q}%
\underline{J}(x^{i_{3}},v^{8})\delta _{b_{4}}^{a_{4}}], \\ 
\mbox{ locally anisotropic cosmology};%
\end{array}
& 
\end{array}%
$ \\ 
trivial eqs for $\ ^{s}\mathring{\nabla}$-torsion & LC-conditions & $\ ^{s}%
\widehat{\mathbf{D}}_{\mid \ ^{s}\widehat{\mathcal{T}}\rightarrow 0}=\ ^{s}%
\mathbf{\nabla .}$ \\ \hline\hline
\end{tabular}%
}
\end{eqnarray*}%
}

\subsubsection{Quasi-stationary solutions for nonmetric FL geometric flows
with fixed light velocity}

Such quasi-stationary solutions are constructed as nonholonomic
generalizations and extensions on tangent Lorentz bundles with $v^{8}=const,$
when the velocity phase space involves space-like hypersurfaces. 
{\scriptsize 
\begin{eqnarray*}
&&%
\begin{tabular}{l}
\hline\hline
\begin{tabular}{lll}
& {\large \textsf{Table 5:\ Off-diagonal nonmetric quasi-stationary
spacetime and space velocity configurations}} &  \\ 
& Exact solutions of $\ \widehat{\mathbf{R}}_{\mu _{s}\nu _{s}}(\tau )=\ _{Q}%
\widehat{\mathbf{J}}_{\mu _{s}\nu _{s}}(\tau )$ (\ref{cfeq4af}) on $TV$
transformed into a shall system of nonlinear PDEs (\ref{eq1})-(\ref{eq4c}) & 
\end{tabular}
\\ 
\end{tabular}
\\
&&%
\begin{tabular}{lll}
\hline\hline
&  &  \\ 
$%
\begin{array}{c}
\mbox{d-metric ansatz with} \\ 
\mbox{Killing symmetry }\partial _{4}=\partial _{t},\partial _{8}%
\end{array}%
$ &  & $%
\begin{array}{c}
ds^{2}(\tau
)=g_{i_{1}}(x^{k_{1}})(dx^{i_{1}})^{2}+g_{a_{2}}(x^{k_{1}},y^{3})(dy^{a_{2}}+N_{i_{1}}^{a_{2}}(x^{k_{1}},y^{3})dx^{i_{1}})^{2}
\\ 
+g_{a_{3}}(x^{k_{2}},v^{5})(dy^{a_{3}}+N_{i_{2}}^{a_{3}}(x^{k_{2}},v^{5})dx^{i_{2}})^{2}
\\ 
+g_{a_{4}}(x^{k_{3}},v^{7})(dy^{a_{4}}+N_{i_{3}}^{a_{4}}(x^{k_{3}},v^{7})dx^{i_{3}})^{2},%
\mbox{ for }g_{i_{1}}=e^{\psi {(x}^{k_{1}}{)}}, \\ 
g_{a_{2}}=h_{a_{2}}(x^{k_{1}},y^{3}),N_{i_{1}}^{3}=\
^{2}w_{i_{1}}=w_{i_{1}}(x^{k_{1}},y^{3}),N_{i_{1}}^{4}=\
^{2}n_{i_{1}}=n_{i_{1}}(x^{k_{1}},y^{3}), \\ 
g_{a_{3}}=h_{a_{3}}(x^{k_{2}},v^{5}),N_{i_{2}}^{5}=\
^{3}w_{i_{2}}=w_{i_{2}}(x^{k_{2}},v^{5}),N_{i_{2}}^{6}=\
^{3}n_{i_{2}}=n_{i_{2}}(x^{k_{2}},v^{5}), \\ 
g_{a_{4}}=h_{a_{4}}(x^{k_{3}},v^{7}),N_{i_{3}}^{7}=\
^{4}w_{i_{3}}=w_{i_{3}}(x^{k_{3}},v^{7}),N_{i_{3}}^{8}=\
^{4}n_{i_{3}}=n_{i_{3}}(x^{k_{3}},v^{7}),%
\end{array}%
$ \\ 
Effective matter sources &  & $\ _{Q}\widehat{\mathbf{J}}_{\ \nu _{s}}^{\mu
_{s}}=[\ _{1}^{Q}J({x}^{k_{1}})\delta _{j_{1}}^{i_{1}},\ \ _{2}^{Q}J({x}%
^{k_{1}},y^{3})\delta _{b_{2}}^{a_{2}},\ \ _{3}^{Q}J({x}^{k_{2}},v^{5})%
\delta _{b_{3}}^{a_{3}},\ _{4}^{Q}J({x}^{k_{3}},v^{7})\delta
_{b_{4}}^{a_{4}},],$ \\ \hline
Nonlinear PDEs (\ref{eq1})-(\ref{e2c}) &  & $%
\begin{tabular}{lll}
$%
\begin{array}{c}
\psi ^{\bullet \bullet }+\psi ^{\prime \prime }=2\ \ \ _{1}^{Q}J; \\ 
\ ^{2}\varpi ^{\ast }\ h_{4}^{\ast _{2}}=2h_{3}h_{4}\ \ _{2}^{Q}J; \\ 
\ ^{2}\beta \ ^{2}w_{i_{1}}-\ ^{2}\alpha _{i_{1}}=0; \\ 
\ ^{2}n_{k_{1}}^{\ast _{2}\ast _{2}}+\ ^{2}\gamma \ ^{2}n_{k_{1}}^{\ast
_{2}}=0;%
\end{array}%
$ &  & $%
\begin{array}{c}
\ ^{2}\varpi {=\ln |\partial _{3}h_{4}/\sqrt{|h_{3}h_{4}|}|,} \\ 
\ ^{2}\alpha _{i_{1}}=(\partial _{3}h_{4})\ (\partial _{i_{1}}\ ^{2}\varpi ),
\\ 
\ ^{2}\beta =(\partial _{3}h_{4})\ (\partial _{3}\ ^{2}\varpi ),\  \\ 
\ \ ^{2}\gamma =\partial _{3}\left( \ln |h_{4}|^{3/2}/|h_{3}|\right) , \\ 
\partial _{1}q=q^{\bullet },\partial _{2}q=q^{\prime },\partial
_{3}q=q^{\ast }%
\end{array}%
$ \\ 
$%
\begin{array}{c}
\partial _{5}(\ ^{3}\varpi )\ \partial _{5}h_{6}=2h_{5}h_{6}\ \ _{3}^{Q}J;
\\ 
\ ^{3}\beta \ ^{3}w_{i_{2}}-\ ^{3}\alpha _{i_{2}}=0; \\ 
\partial _{5}(\partial _{5}\ ^{3}n_{k_{2}})+\ ^{3}\gamma \partial _{5}(\
^{3}n_{k_{2}})=0;%
\end{array}%
$ &  & $%
\begin{array}{c}
\\ 
\ ^{3}\varpi {=\ln |\partial _{5}h_{6}/\sqrt{|h_{5}h_{6}|}|,} \\ 
\ ^{3}\alpha _{i_{2}}=(\partial _{5}h_{6})\ (\partial _{i_{2}}\ ^{3}\varpi ),
\\ 
\ ^{3}\beta =(\partial _{5}h_{6})\ (\partial _{5}\ ^{3}\varpi ),\  \\ 
\ \ ^{3}\gamma =\partial _{5}\left( \ln |h_{6}|^{3/2}/|h_{5}|\right) ,%
\end{array}%
$ \\ 
$%
\begin{array}{c}
\partial _{7}(\ ^{4}\varpi )\ \partial _{7}h_{8}=2h_{7}h_{8}\ \ _{4}^{Q}J;
\\ 
\ ^{4}\beta \ ^{4}w_{i_{3}}-\ ^{4}\alpha _{i_{3}}=0; \\ 
\partial _{7}(\partial _{7}\ ^{4}n_{k_{3}})+\ ^{4}\gamma \partial _{7}(\
^{4}n_{k_{3}})=0;%
\end{array}%
$ &  & $%
\begin{array}{c}
\\ 
\ ^{4}\varpi {=\ln |\partial _{7}h_{8}/\sqrt{|h_{7}h_{8}|}|,} \\ 
\ ^{4}\alpha _{i}=(\partial _{7}h_{8})\ (\partial _{i}\ ^{4}\varpi ), \\ 
\ ^{4}\beta =(\partial _{7}h_{8})\ (\partial _{7}\ ^{4}\varpi ),\  \\ 
\ \ ^{4}\gamma =\partial _{7}\left( \ln |h_{8}|^{3/2}/|h_{7}|\right) ,%
\end{array}%
$%
\end{tabular}%
$ \\ \hline
$%
\begin{array}{c}
\mbox{ Gener.  functs:}\ h_{3}(x^{k_{1}},y^{3}), \\ 
\ ^{2}\Psi (x^{k_{1}},y^{3})=e^{\ ^{2}\varpi },\ ^{2}\Phi (x^{k_{1}},y^{3}),
\\ 
\mbox{integr. functs:}\ h_{4}^{[0]}(x^{k_{1}}),\  \\ 
_{1}n_{k_{1}}(x^{i_{1}}),\ _{2}n_{k_{1}}(x^{i_{1}}); \\ 
\mbox{ Gener.  functs:}h_{5}(x^{k_{2}},v^{5}), \\ 
\ ^{3}\Psi (x^{k_{2}},v^{5})=e^{\ ^{3}\varpi },\ ^{3}\Phi (x^{k_{2}},v^{5}),
\\ 
\mbox{integr. functs:}\ h_{6}^{[0]}(x^{k_{2}}),\  \\ 
_{1}^{3}n_{k_{2}}(x^{i_{2}}),\ _{2}^{3}n_{k_{2}}(x^{i_{2}}); \\ 
\mbox{ Gener.  functs:}h_{7}(x^{k_{3}},v^{7}), \\ 
\ ^{5}\Psi (x^{k_{2}},v^{7})=e^{\ ^{4}\varpi },\ ^{4}\Phi (x^{k_{3}},v^{7}),
\\ 
\mbox{integr. functs:}\ h_{8}^{[0]}(x^{k_{3}}),\  \\ 
_{1}^{4}n_{k_{3}}(x^{i_{3}}),\ _{2}^{4}n_{k_{3}}(x^{i_{4}}); \\ 
\mbox{\& nonlinear symmetries}%
\end{array}%
$ &  & $%
\begin{array}{c}
\ ((\ ^{2}\Psi )^{2})^{\ast _{2}}=-\int dy^{3}\ \ _{2}^{Q}Jh_{4}^{\ \ast
_{2}}, \\ 
(\ ^{2}\Phi )^{2}=-4\ _{2}\Lambda h_{4},\mbox{ see }(\ref{nonlinsymrex}), \\ 
h_{4}=h_{4}^{[0]}-(\ ^{2}\Phi )^{2}/4\ _{2}\Lambda ,h_{4}^{\ast _{2}}\neq
0,\ _{2}\Lambda \neq 0=const; \\ 
\\ 
\partial _{5}((\ ^{3}\Psi )^{2})=-\int dv^{5}\ \ _{3}^{Q}J\partial
_{5}h_{6}^{\ }, \\ 
(\ ^{3}\Phi )^{2}=-4\ _{3}\Lambda h_{6}, \\ 
h_{6}=h_{6}^{[0]}-(\ ^{3}\Phi )^{2}/4\ _{3}\Lambda ,\partial _{5}h_{6}\neq
0,\ _{3}\Lambda \neq 0=const; \\ 
\\ 
\partial _{7}((\ ^{4}\Psi )^{2})=-\int dv^{7}\ \ _{4}^{Q}J\partial
_{7}h_{8}^{\ }, \\ 
(\ ^{4}\Phi )^{2}=-4\ _{4}\Lambda h_{8}, \\ 
h_{8}=h_{8}^{[0]}-(\ ^{4}\Phi )^{2}/4\ _{4}\Lambda ,\partial _{7}h_{8}\neq
0,\ _{4}\Lambda \neq 0=const;%
\end{array}%
$ \\ \hline
Off-diag. solutions, $%
\begin{array}{c}
\mbox{d--metric} \\ 
\mbox{N-connec.}%
\end{array}%
$ &  & $%
\begin{tabular}{l}
$%
\begin{array}{c}
\ g_{i}=e^{\ \psi (x^{k})}\mbox{ as a solution of 2-d Poisson eqs. }\psi
^{\bullet \bullet }+\psi ^{\prime \prime }=2~\ \ _{1}^{Q}J; \\ 
h_{3}=-(\ ^{2}\Psi ^{\ast _{2}})^{2}/4\ \ _{2}^{Q}Jh_{4},\mbox{ see }(\ref%
{g3}),(\ref{g4}); \\ 
h_{4}=h_{4}^{[0]}-\int dy^{3}(\ ^{2}\Psi ^{2})^{\ast _{2}}/4\ \
_{2}^{Q}J=h_{4}^{[0]}-\ ^{2}\Phi ^{2}/4\ _{2}\Lambda ; \\ 
w_{i_{2}}=\partial _{i_{2}}\ \ ^{2}\Psi /\ \partial _{3}\ ^{2}\Psi =\partial
_{i_{2}}\ \ ^{2}\Psi ^{2}/\ \partial _{3}\ ^{2}\Psi ^{2}|; \\ 
n_{k}=\ _{1}n_{k}+\ _{2}n_{k}\int dy^{3}(\ ^{2}\Psi ^{\ast _{2}})^{2}/\ \
_{2}^{Q}J^{2}|h_{4}^{[0]}-\int dy^{3}(\ ^{2}\Psi ^{2})^{\ast _{2}}/4\ \
_{2}^{Q}J^{2}|^{5/2};%
\end{array}%
$ \\ 
$%
\begin{array}{c}
h_{5}=-(\partial _{5}\ ^{3}\Psi )^{2}/4\ \ _{3}^{Q}J^{2}h_{6}; \\ 
h_{6}=h_{6}^{[0]}-\int dv^{5}\partial _{5}((\ ^{3}\Psi )^{2})/4\ \
_{2}^{Q}J=h_{6}^{[0]}-(\ ^{3}\Phi )^{2}/4\ _{3}\Lambda ; \\ 
w_{i_{2}}=\partial _{i_{2}}(\ ^{3}\Psi )/\ \partial _{5}(\ ^{3}\Psi
)=\partial _{i_{2}}(\ ^{3}\Psi )^{2}/\ \partial _{5}(\ ^{3}\Psi )^{2}|; \\ 
n_{k_{2}}=\ _{1}n_{k_{2}}+\ _{2}n_{k_{2}}\int dv^{5}(\partial _{5}\ ^{3}\Psi
)^{2}/\ \ _{3}^{Q}J^{2}|h_{6}^{[0]}-\int dv^{5}\partial _{5}((\ ^{3}\Psi
)^{2})/4\ \ _{3}^{Q}J^{2}|^{5/2};%
\end{array}%
$ \\ 
$%
\begin{array}{c}
h_{7}=-(\partial _{7}\ ^{3}\Psi )^{2}/4\ \ _{4}^{Q}J^{2}h_{8}; \\ 
h_{8}=h_{8}^{[0]}-\int dv^{7}\partial _{9}((\ ^{4}\Psi )^{2})/4\ \
_{4}^{Q}J=h_{8}^{[0]}-(\ ^{4}\Phi )^{2}/4\ _{4}\Lambda ; \\ 
w_{i_{3}}=\partial _{i_{3}}(\ ^{4}\Psi )/\ \partial _{7}(\ ^{4}\Psi
)=\partial _{i_{3}}(\ ^{4}\Psi )^{2}/\ \partial _{7}(\ ^{4}\Psi )^{2}|; \\ 
n_{k_{3}}=\ _{1}n_{k_{3}}+\ _{2}n_{k_{3}}\int dv^{7}(\partial _{7}\ ^{4}\Psi
)^{2}/\ \ _{4}^{Q}J^{2}|h_{8}^{[0]}-\int dv^{7}\partial _{7}((\ ^{4}\Psi
)^{2})/4\ \ _{4}^{Q}J^{2}|^{5/2}.%
\end{array}%
$%
\end{tabular}%
$ \\ \hline\hline
\end{tabular}%
\end{eqnarray*}%
}As an example of 8-d quasi-stationary quadratic element with $v^{8}=const$
on $T\mathbf{V},$ we provide 
\begin{eqnarray}
d\widehat{s}_{[8d]}^{2} &=&\widehat{g}_{\alpha _{s}\beta
_{s}}(x^{k},y^{3},v^{5},v^{7};h_{4},h_{6},h_{8,};\ \ _{s}^{Q}J;\ _{s}\Lambda
)du^{\alpha _{s}}du^{\beta _{s}}  \label{qst8d7} \\
&=&e^{\psi (x^{k},\ \ _{s}^{Q}J)}[(dx^{1})^{2}+(dx^{2})^{2}]-\frac{%
(h_{4}^{\ast _{2}})^{2}}{|\int dy^{3}[\ \ _{2}^{Q}Jh_{4}]^{\ast _{2}}|\ h_{4}%
}\{dy^{3}+\frac{\partial _{i_{1}}[\int dy^{3}(\ \ \ _{2}^{Q}J)\ h_{4}^{\ast
_{2}}]}{\ _{2}^{Q}J\ h_{4}^{\ast _{2}}}dx^{i_{1}}\}^{2}+  \notag \\
&&h_{4}\{dt+[\ _{1}n_{k_{1}}+\ _{2}n_{k_{1}}\int dy^{3}\frac{(h_{4}^{\ast
_{2}})^{2}}{|\int dy^{3}[\ \ _{2}^{Q}Jh_{4}]^{\ast _{2}}|\ (h_{4})^{5/2}}%
]dx^{k_{2}}\}+  \notag \\
&&\frac{(\partial _{5}h_{6})^{2}}{|\int dy^{5}\partial _{5}[\ \
_{3}^{Q}Jh_{6}]|\ h_{6}}\{dv^{5}+\frac{\partial _{i_{2}}[\int dy^{5}(\
_{3}^{Q}J)\ \partial _{5}h_{6}]}{\ \ _{3}^{Q}J\ \partial _{5}h_{6}}%
dx^{i_{2}}\}^{2}+  \notag \\
&&h_{6}\{dv^{5}+[\ _{1}n_{k_{2}}+\ _{2}n_{k_{2}}\int dv^{5}\frac{(\partial
_{5}h_{6})^{2}}{|\int dy^{5}\partial _{5}[\ _{3}^{Q}Jh_{6}]|\ (h_{6})^{5/2}}%
]dx^{k_{2}}\}+  \notag \\
&&\frac{(\partial _{7}h_{8})^{2}}{|\int dv^{7}\partial _{7}[\
_{4}^{Q}Jh_{8}]|\ h_{8}}\{dv^{7}+\frac{\partial _{i_{3}}[\int dv^{7}(\
_{4}^{Q}J)\ \partial _{7}h_{8}]}{\ _{4}^{Q}J\ \partial _{7}h_{8}}%
dx^{i_{3}}\}^{2}+  \notag \\
&&h_{8}\{dv^{8}+[\ _{1}n_{k_{3}}+\ _{2}n_{k_{3}}\int dv^{7}\frac{(\partial
_{7}h_{8})^{2}}{|\int dv^{7}\partial _{7}[\ _{4}^{Q}Jh_{8}]|\ (h_{8})^{5/2}}%
]dx^{k_{3}}\}.  \notag
\end{eqnarray}%
This class of s-metrics possesses nonlinear symmetries which allow to
redefine the generating functions and generating sources and related them to
conventional cosmological constants and their $\tau $-parametric flows.
Solutions with $\tau $-families of gravitational $\eta $- and $\chi $%
-polarizations can be defined for respective off-diagonal deformations of
prime s-metrics into target ones.

\subsubsection{ Quasi-stationary nonmetric solutions with variable light
velocity}

Another class of quasi-stationary extensions of a Lorentz manifold, $\mathbf{%
V},$ metrics is for $\tau $-families of quadratic line elements with $%
v^{7}=const$ which provide examples of velocity rainbow s-metrics on $T%
\mathbf{V}.$ Considering a $v^{8}\leftrightarrow v^{7}$ changing of velocity
phase coordinates in (\ref{qst8d7}), we construct an example of 8-d
quasi-stationary quadratic element with $v^{7}=const$ on $T\mathbf{V}$
defining an example of a velocity rainbow s-metric, 
\begin{eqnarray}
d\widehat{s}_{[8d]}^{2} &=&\widehat{g}_{\alpha _{s}\beta
_{s}}(x^{k},y^{3},v^{5},v^{8};h_{4},h_{6},h_{8,};\ _{s}^{Q}J;\ _{s}\Lambda
(\tau ))du^{\alpha _{s}}du^{\beta _{s}}  \label{qst8d8} \\
&=&e^{\psi (x^{k},\ _{1}^{Q}J)}[(dx^{1})^{2}+(dx^{2})^{2}]-\frac{%
(h_{4}^{\ast _{2}})^{2}}{|\int dy^{3}[\ \ _{2}^{Q}Jh_{4}]^{\ast _{2}}|\ h_{4}%
}\{dy^{3}+\frac{\partial _{i_{1}}[\int dy^{3}(\ _{2}^{Q}J)\ h_{4}^{\ast
_{2}}]}{\ \ _{2}^{Q}J\ h_{4}^{\ast _{2}}}dx^{i_{1}}\}^{2}+  \notag \\
&&h_{4}\{dt+[\ _{1}n_{k_{1}}+\ _{2}n_{k_{1}}\int dy^{3}\frac{(h_{4}^{\ast
_{2}})^{2}}{|\int dy^{3}[\ \ _{2}^{Q}Jh_{4}]^{\ast _{2}}|\ (h_{4})^{5/2}}%
]dx^{k_{1}}\}+  \notag \\
&&\frac{(\partial _{5}h_{6})^{2}}{|\int dy^{5}\partial _{5}[\ \
_{3}^{Q}Jh_{6}]|\ h_{6}}\{dv^{5}+\frac{\partial _{i_{2}}[\int dy^{5}(\ \
_{3}^{Q}J)\ \partial _{5}h_{6}]}{\ \ _{3}^{Q}J\ \partial _{5}h_{6}}%
dx^{i_{2}}\}^{2}+  \notag \\
&&h_{6}\{dy^{6}+[\ _{1}n_{k_{2}}+\ _{2}n_{k_{2}}\int dv^{5}\frac{(\partial
_{5}h_{6})^{2}}{|\int dv^{5}\partial _{5}[\ \ \ _{3}^{Q}Jh_{6}]|\
(h_{6})^{5/2}}]dx^{k_{2}}\}+  \notag \\
&&\underline{h}_{7}\{dv^{7}+[\ _{1}n_{k_{3}}+\ _{2}n_{k_{3}}\int dv^{8}\frac{%
(\partial _{8}\underline{h}_{7})^{2}}{|\int dv^{8}\partial _{8}[\ \ \
_{4}^{Q}\underline{J}\Upsilon \underline{h}_{7}]|\ (\underline{h}_{7})^{5/2}}%
]dx^{k_{3}}\}+  \notag \\
&&\frac{(\partial _{8}\underline{h}_{7})^{2}}{|\int dv^{8}\partial _{8}[\ \
_{4}^{Q}\underline{J}\underline{h}_{7}]|\ \underline{h}_{7}}\{dv^{8}+\frac{%
\partial _{i_{3}}[\int dv^{8}(\ \ _{4}^{Q}\underline{J})\ \partial _{8}%
\underline{h}_{7}]}{\ _{4}^{Q}\underline{J}\ \partial _{8}\underline{h}_{7}}%
dx^{i_{3}}\}^{2}.  \notag
\end{eqnarray}%
The principles of generating such quasi-stationary and rainbow solutions are
summarized in Table 6.

{\scriptsize 
\begin{eqnarray*}
&&%
\begin{tabular}{l}
\hline\hline
\begin{tabular}{lll}
& {\large \textsf{Table 6:\ Off-diagonal nonmetric quasi-stationary
spacetimes with velocity rainbows}} &  \\ 
& Exact solutions of $\ \widehat{\mathbf{R}}_{\mu _{s}\nu _{s}}(\tau )=\ _{Q}%
\widehat{\mathbf{J}}_{\mu _{s}\nu _{s}}(\tau )$ (\ref{cfeq4af}) on $TV$
transformed into a shall system of nonlinear PDEs (\ref{eq1})-(\ref{e2c}) & 
\end{tabular}
\\ 
\end{tabular}
\\
&&%
\begin{tabular}{lll}
\hline\hline
&  &  \\ 
$%
\begin{array}{c}
\mbox{d-metric ansatz with} \\ 
\mbox{Killing symmetry }\partial _{4}=\partial _{t},\partial _{7}%
\end{array}%
$ &  & $%
\begin{array}{c}
ds^{2}=g_{i_{1}}(x^{k_{1}})(dx^{i_{1}})^{2}+g_{a_{2}}(x^{k_{1}},y^{3})(dy^{a_{2}}+N_{i_{1}}^{a_{2}}(x^{k_{1}},y^{3})dx^{i_{1}})^{2}
\\ 
+g_{a_{3}}(x^{k_{2}},v^{5})(dy^{a_{3}}+N_{i_{2}}^{a_{3}}(x^{k_{2}},v^{5})dx^{i_{2}})^{2}
\\ 
+\underline{g}_{a_{4}}(x^{k_{3}},v^{8})(dy^{a_{4}}+\underline{N}%
_{i_{3}}^{a_{4}}(x^{k_{3}},v^{8})dx^{i_{3}})^{2},\mbox{ for }%
g_{i_{1}}=e^{\psi {(x}^{k_{1}}{)}}, \\ 
g_{a_{2}}=h_{a_{2}}(x^{k_{1}},y^{3}),N_{i_{1}}^{3}=\
^{2}w_{i_{1}}=w_{i_{1}}(x^{k_{1}},y^{3}),N_{i_{1}}^{4}=\
^{2}n_{i_{1}}=n_{i_{1}}(x^{k_{1}},y^{3}), \\ 
g_{a_{3}}=h_{a_{3}}(x^{k_{2}},v^{5}),N_{i_{2}}^{5}=\
^{3}w_{i_{2}}=w_{i_{2}}(x^{k_{2}},v^{5}),N_{i_{2}}^{6}=\
^{3}n_{i_{2}}=n_{i_{2}}(x^{k_{2}},v^{5}), \\ 
\underline{g}_{a_{4}}=\underline{h}_{a_{4}}(x^{k_{3}},v^{8}),\underline{N}%
_{i_{3}}^{7}=\ ^{4}\underline{n}_{i_{3}}=\underline{n}%
_{i_{3}}(x^{k_{3}},v^{8}),\underline{N}_{i_{3}}^{8}=\ ^{4}\underline{w}%
_{i_{3}}=\underline{w}_{i_{3}}(x^{k_{3}},v^{8}),%
\end{array}%
$ \\ 
Effective matter sources &  & $\ _{Q}\widehat{\mathbf{J}}_{\ \nu _{s}}^{\mu
_{s}}=[\ \ _{1}^{Q}J({x}^{k_{1}})\delta _{j_{1}}^{i_{1}},\ \ _{2}^{Q}J({x}%
^{k_{1}},y^{3})\delta _{b_{2}}^{a_{2}},\ \ _{3}^{Q}J({x}^{k_{2}},v^{5})%
\delta _{b_{3}}^{a_{3}},\ _{4}^{Q}\underline{J}({x}^{k_{3}},v^{8})\delta
_{b_{4}}^{a_{4}},],$ \\ \hline
Nonlinear PDEs (\ref{eq1})-(\ref{e2c}) &  & $%
\begin{tabular}{lll}
$%
\begin{array}{c}
\psi ^{\bullet \bullet }+\psi ^{\prime \prime }=2\ \ _{1}^{Q}J; \\ 
\ ^{2}\varpi ^{\ast }\ h_{4}^{\ast _{2}}=2h_{3}h_{4}\ \ _{2}^{Q}J; \\ 
\ ^{2}\beta \ ^{2}w_{i_{1}}-\ ^{2}\alpha _{i_{1}}=0; \\ 
\ ^{2}n_{k_{1}}^{\ast _{2}\ast _{2}}+\ ^{2}\gamma \ ^{2}n_{k_{1}}^{\ast
_{2}}=0;%
\end{array}%
$ &  & $%
\begin{array}{c}
\ ^{2}\varpi {=\ln |\partial _{3}h_{4}/\sqrt{|h_{3}h_{4}|}|,} \\ 
\ ^{2}\alpha _{i_{1}}=(\partial _{3}h_{4})\ (\partial _{i_{1}}\ ^{2}\varpi ),
\\ 
\ ^{2}\beta =(\partial _{3}h_{4})\ (\partial _{3}\ ^{2}\varpi ),\  \\ 
\ \ ^{2}\gamma =\partial _{3}\left( \ln |h_{4}|^{3/2}/|h_{3}|\right) , \\ 
\partial _{1}q=q^{\bullet },\partial _{2}q=q^{\prime },\partial
_{3}q=q^{\ast _{2}}%
\end{array}%
$ \\ 
$%
\begin{array}{c}
\partial _{5}(\ ^{3}\varpi )\ \partial _{5}h_{6}=2h_{5}h_{6}\ \ _{3}^{Q}J;
\\ 
\ ^{3}\beta \ ^{3}w_{i_{2}}-\ ^{3}\alpha _{i_{2}}=0; \\ 
\partial _{5}(\partial _{5}\ ^{3}n_{k_{2}})+\ ^{3}\gamma \partial _{5}(\
^{3}n_{k_{2}})=0;%
\end{array}%
$ &  & $%
\begin{array}{c}
\\ 
\ ^{3}\varpi {=\ln |\partial _{5}h_{6}/\sqrt{|h_{5}h_{6}|}|,} \\ 
\ ^{3}\alpha _{i_{2}}=(\partial _{5}h_{6})\ (\partial _{i_{2}}\ ^{3}\varpi ),
\\ 
\ ^{3}\beta =(\partial _{5}h_{6})\ (\partial _{5}\ ^{3}\varpi ),\  \\ 
\ \ ^{3}\gamma =\partial _{5}\left( \ln |h_{6}|^{3/2}/|h_{5}|\right) ,%
\end{array}%
$ \\ 
$%
\begin{array}{c}
\partial _{8}(\ ^{4}\underline{\varpi })\ \partial _{8}\underline{h}_{7}=2%
\underline{h}_{7}\underline{h}_{8}\ \ _{4}^{Q}\underline{J}; \\ 
\partial _{8}(\partial _{8}\ ^{4}\underline{n}_{k_{3}})+\ ^{4}\underline{%
\gamma }\partial _{8}(\ ^{4}\underline{n}_{k_{3}})=0; \\ 
\ ^{4}\underline{\beta }\ ^{4}\underline{w}_{i_{3}}-\ ^{4}\underline{\alpha }%
_{i_{3}}=0;%
\end{array}%
$ &  & $%
\begin{array}{c}
\\ 
\ ^{4}\underline{\varpi }{=\ln |\partial _{8}\underline{h}_{7}/\sqrt{|%
\underline{h}_{7}\underline{h}_{8}|}|,} \\ 
\ ^{4}\underline{\alpha }_{i}=(\partial _{8}\underline{h}_{7})\ (\partial
_{i}\ ^{4}\underline{\varpi }), \\ 
\ ^{4}\underline{\beta }=(\partial _{8}\underline{h}_{7})\ (\partial _{8}\
^{4}\underline{\varpi }),\  \\ 
\ \ ^{4}\underline{\gamma }=\partial _{8}\left( \ln |\underline{h}%
_{7}|^{3/2}/|\underline{h}_{8}|\right) ,%
\end{array}%
$%
\end{tabular}%
$ \\ \hline
$%
\begin{array}{c}
\mbox{ Gener.  functs:}\ h_{3}(x^{k_{1}},y^{3}), \\ 
\ ^{2}\Psi (x^{k_{1}},y^{3})=e^{\ ^{2}\varpi },\ ^{2}\Phi (x^{k_{1}},y^{3}),
\\ 
\mbox{integr. functs:}\ h_{4}^{[0]}(x^{k_{1}}),\  \\ 
_{1}n_{k_{1}}(x^{i_{1}}),\ _{2}n_{k_{1}}(x^{i_{1}}); \\ 
\mbox{ Gener.  functs:}h_{5}(x^{k_{2}},v^{5}), \\ 
\ ^{3}\Psi (x^{k_{2}},v^{5})=e^{\ ^{3}\varpi },\ ^{3}\Phi (x^{k_{2}},v^{5}),
\\ 
\mbox{integr. functs:}\ h_{6}^{[0]}(x^{k_{2}}),\  \\ 
_{1}^{3}n_{k_{2}}(x^{i_{2}}),\ _{2}^{3}n_{k_{2}}(x^{i_{2}}); \\ 
\mbox{ Gener.  functs:}\underline{h}_{8}(x^{k_{3}},v^{8}), \\ 
\ ^{4}\underline{\Psi }(x^{k_{2}},v^{8})=e^{\ ^{4}\underline{\varpi }},\ ^{4}%
\underline{\Phi }(x^{k_{3}},v^{8}), \\ 
\mbox{integr. functs:}\ h_{8}^{[0]}(x^{k_{3}}),\  \\ 
_{1}^{4}n_{k_{3}}(x^{i_{3}}),\ _{2}^{4}n_{k_{3}}(x^{i_{4}}); \\ 
\mbox{\& nonlinear symmetries}%
\end{array}%
$ &  & $%
\begin{array}{c}
\ ((\ ^{2}\Psi )^{2})^{\ast _{2}}=-\int dy^{3}\ \ _{2}^{Q}Jh_{4}^{\ \ast
_{2}}, \\ 
(\ ^{2}\Phi )^{2}=-4\ _{2}\Lambda h_{4},\mbox{ see }(\ref{nonlinsymrex}), \\ 
h_{4}=h_{4}^{[0]}-(\ ^{2}\Phi )^{2}/4\ _{2}\Lambda ,h_{4}^{\ast _{2}}\neq
0,\ _{2}\Lambda \neq 0=const; \\ 
\\ 
\partial _{5}((\ ^{3}\Psi )^{2})=-\int dv^{5}\ \ _{3}^{Q}J\partial
_{5}h_{6}^{\ }, \\ 
(\ ^{3}\Phi )^{2}=-4\ _{3}\Lambda h_{6}, \\ 
h_{6}=h_{6}^{[0]}-(\ ^{3}\Phi )^{2}/4\ _{3}\Lambda ,\partial _{5}h_{6}\neq
0,\ _{3}\Lambda \neq 0=const; \\ 
\\ 
\partial _{8}((\ ^{4}\underline{\Psi })^{2})=-\int dv^{8}\ \ _{4}^{Q}%
\underline{J}\partial _{8}\underline{h}_{7}^{\ }, \\ 
(\ ^{4}\underline{\Phi })^{2}=-4\ _{4}\underline{\Lambda }\underline{h}_{7},
\\ 
\underline{h}_{7}=h_{7}^{[0]}-(\ ^{4}\underline{\Phi })^{2}/4\ _{4}%
\underline{\Lambda },\partial _{8}\underline{h}_{7}\neq 0,\ _{4}\underline{%
\Lambda }\neq 0=const;%
\end{array}%
$ \\ \hline
Off-diag. solutions, $%
\begin{array}{c}
\mbox{d--metric} \\ 
\mbox{N-connec.}%
\end{array}%
$ &  & $%
\begin{tabular}{l}
$%
\begin{array}{c}
\ g_{i}=e^{\ \psi (x^{k})}\mbox{ as a solution of 2-d Poisson eqs. }\psi
^{\bullet \bullet }+\psi ^{\prime \prime }=2~\ \ _{1}^{Q}J; \\ 
h_{3}=-(\ ^{2}\Psi ^{\ast _{2}})^{2}/4\ \ \ _{2}^{Q}J^{2}h_{4},\mbox{ see }(%
\ref{g3}),(\ref{g4}); \\ 
h_{4}=h_{4}^{[0]}-\int dy^{3}(\ ^{2}\Psi ^{2})^{\ast _{2}}/4\ \ \
_{2}^{Q}J=h_{4}^{[0]}-\ ^{2}\Phi ^{2}/4\ _{2}\Lambda ; \\ 
w_{i_{2}}=\partial _{i_{2}}\ \ ^{2}\Psi /\ \partial _{3}\ ^{2}\Psi =\partial
_{i_{2}}\ \ ^{2}\Psi ^{2}/\ \partial _{3}\ ^{2}\Psi ^{2}|; \\ 
n_{k_{1}}=\ _{1}n_{k_{1}}+\ _{2}n_{k}\int dy^{3}(\ ^{2}\Psi ^{\ast
_{2}})^{2}/\ \ \ _{2}^{Q}J^{2}|h_{4}^{[0]}-\int dy^{3}(\ ^{2}\Psi
^{2})^{\ast _{2}}/4\ \ \ _{2}^{Q}J^{2}|^{5/2};%
\end{array}%
$ \\ 
$%
\begin{array}{c}
h_{5}=-(\partial _{5}\ ^{3}\Psi )^{2}/4\ \ \ _{3}^{Q}J^{2}h_{6}; \\ 
h_{6}=h_{6}^{[0]}-\int dv^{5}\partial _{5}((\ ^{3}\Psi )^{2})/4\ \ \
_{3}^{Q}J=h_{6}^{[0]}-(\ ^{3}\Phi )^{2}/4\ _{3}\Lambda ; \\ 
w_{i_{2}}=\partial _{i_{2}}(\ ^{3}\Psi )/\ \partial _{5}(\ ^{3}\Psi
)=\partial _{i_{2}}(\ ^{3}\Psi )^{2}/\ \partial _{5}(\ ^{3}\Psi )^{2}|; \\ 
n_{k_{2}}=\ _{1}n_{k_{2}}+\ _{2}n_{k_{2}}\int dv^{5}(\partial _{5}\ ^{3}\Psi
)^{2}/\ \ \ _{3}^{Q}J^{2}|h_{6}^{[0]}-\int dv^{5}\partial _{5}((\ ^{3}\Psi
)^{2})/4\ \ \ _{3}^{Q}J^{2}|^{5/2};%
\end{array}%
$ \\ 
$%
\begin{array}{c}
\underline{h}_{7}=\underline{h}_{7}^{[0]}-\int dv^{8}\partial _{8}((\ ^{4}%
\underline{\Psi })^{2})/4\ _{4}^{Q}\underline{J}=\underline{h}_{7}^{[0]}-(\
^{4}\underline{\Phi })^{2}/4\ _{4}\underline{\Lambda }; \\ 
\underline{h}_{8}=-(\partial _{8}\ ^{4}\underline{\Psi })^{2}/4\ \ _{4}^{Q}%
\underline{J}^{2}\underline{h}_{7}; \\ 
n_{k_{3}}=\ _{1}n_{k_{3}}+\ _{2}n_{k_{3}}\int dv^{8}(\partial _{8}\ ^{4}%
\underline{\Psi })^{2}/\ \ _{4}^{Q}\underline{J}^{2}|\underline{h}%
_{7}^{[0]}-\int dv^{8}\partial _{8}((\ ^{4}\underline{\Psi })^{2})/4\ \
_{4}^{Q}\underline{J}^{2}|^{5/2}; \\ 
w_{i_{3}}=\partial _{i_{3}}(\ ^{4}\underline{\Psi })/\ \partial _{8}(\ ^{4}%
\underline{\Psi })=\partial _{i_{3}}(\ ^{4}\underline{\Psi })^{2}/\ \partial
_{8}(\ ^{4}\underline{\Psi })^{2}|.%
\end{array}%
$%
\end{tabular}%
$ \\ \hline\hline
\end{tabular}%
\end{eqnarray*}%
}

We can construct other types of quasi-stationary and velocity rainbow
solutions by using nonlinear transforms of generating functions,
gravitational polarizations and constraints to metric (in particular, with
induced torsion) or LC-configurations. All nonmetric or metric nonholonomic
geometric constructions involve respective abstract geometric proofs and
modifications/ generalizations of formulas.

\subsubsection{Nonmetric locally anisotropic cosmological solutions with
phase space velocities}

Such 8-d cosmological models are defined by cosmological s-metrics with a
fixed $v^{8}=const,$ i.e. when the solutions do not depend on this
conventional coordinate. Respective classes of generic off-diagonal
s-metrics are constructed following the steps outlined below in Table 7.

{\scriptsize 
\begin{eqnarray*}
&&%
\begin{tabular}{l}
\hline\hline
\begin{tabular}{lll}
& {\large \textsf{Table 7:\ Off-diagonal nonmetric cosmological spacetimes
with space velocity configurations}} &  \\ 
& Exact solutions of \ $\ \widehat{\mathbf{R}}_{\mu _{s}\nu _{s}}(\tau )=\
_{Q}\widehat{\underline{\mathbf{J}}}_{\mu _{s}\nu _{s}}(\tau )$ (\ref%
{cfeq4af}) on $TV$ transformed into a shall system of nonlinear PDEs (\ref%
{eq1})-(\ref{e2c}) & 
\end{tabular}
\\ 
\end{tabular}
\\
&&%
\begin{tabular}{lll}
\hline\hline
&  &  \\ 
$%
\begin{array}{c}
\mbox{d-metric ansatz with} \\ 
\mbox{Killing symmetry }\partial _{4}=\partial _{t},\partial _{8}%
\end{array}%
$ &  & $%
\begin{array}{c}
ds^{2}=g_{i_{1}}(x^{k_{1}})(dx^{i_{1}})^{2}+\underline{g}%
_{a_{2}}(x^{k_{1}},t)(dy^{a_{2}}+\underline{N}%
_{i_{1}}^{a_{2}}(x^{k_{1}},t)dx^{i_{1}})^{2} \\ 
+g_{a_{3}}(x^{k_{2}},v^{5})(dy^{a_{3}}+N_{i_{2}}^{a_{3}}(x^{k_{2}},v^{5})dx^{i_{2}})^{2}
\\ 
+g_{a_{4}}(x^{k_{3}},v^{7})(dy^{a_{4}}+N_{i_{3}}^{a_{4}}(x^{k_{3}},v^{7})dx^{i_{3}})^{2},%
\mbox{ for }g_{i_{1}}=e^{\psi {(x}^{k_{1}}{)}}, \\ 
\underline{g}_{a_{2}}=\underline{h}_{a_{2}}(x^{k_{1}},t),\underline{N}%
_{i_{1}}^{3}=\ ^{2}\underline{n}_{i_{1}}=\underline{n}_{i_{1}}(x^{k_{1}},t),%
\underline{N}_{i_{1}}^{4}=\ ^{2}\underline{w}_{i_{1}}=\underline{w}%
_{i_{1}}(x^{k_{1}},t), \\ 
g_{a_{3}}=h_{a_{3}}(x^{k_{2}},v^{5}),N_{i_{2}}^{5}=\
^{3}w_{i_{2}}=w_{i_{2}}(x^{k_{2}},v^{5}),N_{i_{2}}^{6}=\
^{3}n_{i_{2}}=n_{i_{2}}(x^{k_{2}},v^{5}), \\ 
g_{a_{4}}=h_{a_{4}}(x^{k_{3}},v^{7}),N_{i_{3}}^{7}=\
^{4}w_{i_{3}}=w_{i_{3}}(x^{k_{3}},v^{7}),N_{i_{3}}^{8}=\
^{4}n_{i_{3}}=n_{i_{3}}(x^{k_{3}},v^{7}),%
\end{array}%
$ \\ 
Effective matter sources &  & $\ _{Q}\widehat{\mathbf{J}}_{\ \nu _{s}}^{\mu
_{s}}=[\ _{1}^{Q}J({x}^{k_{1}})\delta _{j_{1}}^{i_{1}},\ \ _{2}^{Q}%
\underline{J}({x}^{k_{1}},t)\delta _{b_{2}}^{a_{2}},\ \ _{3}^{Q}J({x}%
^{k_{2}},v^{5})\delta _{b_{3}}^{a_{3}},\ _{4}^{Q}J({x}^{k_{3}},v^{7})\delta
_{b_{4}}^{a_{4}},],$ \\ \hline
Nonlinear PDEs (\ref{eq1})-(\ref{e2c}) &  & $%
\begin{tabular}{lll}
$%
\begin{array}{c}
\psi ^{\bullet \bullet }+\psi ^{\prime \prime }=2\ \ _{1}^{Q}J; \\ 
\ ^{2}\underline{\varpi }^{\diamond _{2}}\ \underline{h}_{3}^{\diamond
_{2}}=2\underline{h}_{3}\underline{h}_{4}\ \ _{2}^{Q}\underline{J}; \\ 
\ ^{2}\underline{n}_{k_{1}}^{\diamond _{2}\diamond _{2}}+\ ^{2}\underline{%
\gamma }\ ^{2}\underline{n}_{k_{1}}^{\diamond _{2}}=0; \\ 
\ ^{2}\underline{\beta }\ ^{2}\underline{w}_{i_{1}}-\ ^{2}\underline{\alpha }%
_{i_{1}}=0;%
\end{array}%
$ &  & $%
\begin{array}{c}
\ ^{2}\underline{\varpi }{=\ln |\partial _{4}\underline{{h}}_{4}/\sqrt{|%
\underline{h}_{3}\underline{h}_{4}|}|,} \\ 
\ ^{2}\underline{\alpha }_{i_{1}}=(\partial _{4}\underline{h}_{3})\
(\partial _{i_{1}}\ ^{2}\underline{\varpi }), \\ 
\ ^{2}\underline{\beta }=(\partial _{4}\underline{h}_{4})\ (\partial _{3}\
^{2}\underline{\varpi }),\  \\ 
\ \ ^{2}\underline{\gamma }=\partial _{4}\left( \ln |\underline{h}%
_{3}|^{3/2}/|\underline{h}_{4}|\right) , \\ 
\partial _{1}q=q^{\bullet },\partial _{2}q=q^{\prime },\partial
_{4}q=\partial _{t}q=q^{\diamond _{2}}%
\end{array}%
$ \\ 
$%
\begin{array}{c}
\partial _{5}(\ ^{3}\varpi )\ \partial _{5}h_{6}=2h_{5}h_{6}\ \ _{3}^{Q}J;
\\ 
\ ^{3}\beta \ ^{3}w_{i_{2}}-\ ^{3}\alpha _{i_{2}}=0; \\ 
\partial _{5}(\partial _{5}\ ^{3}n_{k_{2}})+\ ^{3}\gamma \partial _{5}(\
^{3}n_{k_{2}})=0;%
\end{array}%
$ &  & $%
\begin{array}{c}
\\ 
\ ^{3}\varpi {=\ln |\partial _{5}h_{6}/\sqrt{|h_{5}h_{6}|}|,} \\ 
\ ^{3}\alpha _{i_{2}}=(\partial _{5}h_{6})\ (\partial _{i_{2}}\ ^{3}\varpi ),
\\ 
\ ^{3}\beta =(\partial _{5}h_{6})\ (\partial _{5}\ ^{3}\varpi ),\  \\ 
\ \ ^{3}\gamma =\partial _{5}\left( \ln |h_{6}|^{3/2}/|h_{5}|\right) ,%
\end{array}%
$ \\ 
$%
\begin{array}{c}
\partial _{7}(\ ^{4}\varpi )\ \partial _{7}h_{8}=2h_{7}h_{8}\ \ _{4}^{Q}J;
\\ 
\ ^{4}\beta \ ^{4}w_{i_{3}}-\ ^{4}\alpha _{i_{3}}=0; \\ 
\partial _{7}(\partial _{7}\ ^{4}n_{k_{3}})+\ ^{4}\gamma \partial _{7}(\
^{4}n_{k_{3}})=0;%
\end{array}%
$ &  & $%
\begin{array}{c}
\\ 
\ ^{4}\varpi {=\ln |\partial _{7}h_{8}/\sqrt{|h_{7}h_{8}|}|,} \\ 
\ ^{4}\alpha _{i}=(\partial _{7}h_{8})\ (\partial _{i}\ ^{4}\varpi ), \\ 
\ ^{4}\beta =(\partial _{7}h_{8})\ (\partial _{7}\ ^{4}\varpi ),\  \\ 
\ \ ^{4}\gamma =\partial _{7}\left( \ln |h_{8}|^{3/2}/|h_{7}|\right) ,%
\end{array}%
$%
\end{tabular}%
$ \\ \hline
$%
\begin{array}{c}
\mbox{ Gener.  functs:}\ \underline{h}_{4}(x^{k_{1}},t), \\ 
\ ^{2}\underline{\Psi }(x^{k_{1}},t)=e^{\ ^{2}\underline{\varpi }},\ ^{2}%
\underline{\Phi }(x^{k_{1}},t), \\ 
\mbox{integr. functs:}\ \underline{h}_{3}^{[0]}(x^{k_{1}}),\  \\ 
_{1}\underline{n}_{k_{1}}(x^{i_{1}}),\ _{2}\underline{n}_{k_{1}}(x^{i_{1}});
\\ 
\mbox{ Gener.  functs:}h_{5}(x^{k_{2}},v^{5}), \\ 
\ ^{3}\Psi (x^{k_{2}},v^{5})=e^{\ ^{3}\varpi },\ ^{3}\Phi (x^{k_{2}},v^{5}),
\\ 
\mbox{integr. functs:}\ h_{6}^{[0]}(x^{k_{2}}),\  \\ 
_{1}^{3}n_{k_{2}}(x^{i_{2}}),\ _{2}^{3}n_{k_{2}}(x^{i_{2}}); \\ 
\mbox{ Gener.  functs:}h_{7}(x^{k_{3}},v^{7}), \\ 
\ ^{4}\Psi (x^{k_{2}},v^{7})=e^{\ ^{4}\varpi },\ ^{4}\Phi (x^{k_{3}},v^{7}),
\\ 
\mbox{integr. functs:}\ h_{8}^{[0]}(x^{k_{3}}),\  \\ 
_{1}^{4}n_{k_{3}}(x^{i_{3}}),\ _{2}^{4}n_{k_{3}}(x^{i_{4}}); \\ 
\mbox{\& nonlinear symmetries}%
\end{array}%
$ &  & $%
\begin{array}{c}
\ ((\ ^{2}\underline{\Psi })^{2})^{\diamond _{2}}=-\int dt\ \ _{2}^{Q}%
\underline{J}\underline{h}_{3}^{\ \diamond _{2}}, \\ 
(\ ^{2}\underline{\Phi })^{2}=-4\ _{2}\underline{\Lambda }\underline{h}_{3},
\\ 
h_{3}=h_{3}^{[0]}-(\ ^{2}\underline{\Phi })^{2}/4\ _{2}\underline{\Lambda },%
\underline{h}_{3}^{\diamond _{2}}\neq 0,\ _{2}\underline{\Lambda }\neq
0=const; \\ 
\\ 
\partial _{5}((\ ^{3}\Psi )^{2})=-\int dv^{5}\ \ _{3}^{Q}J\partial
_{5}h_{6}^{\ }, \\ 
(\ ^{3}\Phi )^{2}=-4\ _{3}\Lambda h_{6}, \\ 
h_{6}=h_{6}^{[0]}-(\ ^{3}\Phi )^{2}/4\ _{3}\Lambda ,\partial _{5}h_{6}\neq
0,\ _{3}\Lambda \neq 0=const; \\ 
\\ 
\partial _{7}((\ ^{4}\Psi )^{2})=-\int dv^{7}\ \ _{4}^{Q}J\partial
_{7}h_{8}^{\ }, \\ 
(\ ^{4}\Phi )^{2}=-4\ _{4}\Lambda h_{8}, \\ 
h_{8}=h_{8}^{[0]}-(\ ^{4}\Phi )^{2}/4\ _{4}\Lambda ,\partial _{7}h_{8}\neq
0,\ _{4}\Lambda \neq 0=const;%
\end{array}%
$ \\ \hline
Off-diag. solutions, $%
\begin{array}{c}
\mbox{d--metric} \\ 
\mbox{N-connec.}%
\end{array}%
$ &  & $%
\begin{tabular}{l}
$%
\begin{array}{c}
\ g_{i}=e^{\ \psi (x^{k})}\mbox{ as a solution of 2-d Poisson eqs. }\psi
^{\bullet \bullet }+\psi ^{\prime \prime }=2~\ \ _{1}^{Q}J; \\ 
\underline{h}_{4}=-(\underline{\Psi }^{\diamond _{2}})^{2}/4\ \ _{2}^{Q}%
\underline{J}^{2}\underline{h}_{3}; \\ 
\underline{h}_{3}=\underline{h}_{3}^{[0]}-\int dt(\underline{\Psi }%
^{2})^{\diamond _{2}}/4\ \ _{2}^{Q}\underline{J}=\underline{h}_{3}^{[0]}-%
\underline{\Phi }^{2}/4\ _{2}\underline{\Lambda }; \\ 
\underline{w}_{i_{1}}=\partial _{i_{1}}\ \underline{\Psi }/\ \partial 
\underline{\Psi }^{\diamond _{2}}=\partial _{i_{1}}\ \underline{\Psi }^{2}/\
\partial _{t}\underline{\Psi }^{2}|; \\ 
\underline{n}_{k_{1}}=\ _{1}n_{k_{1}}+\ _{2}n_{k_{1}}\int dt(\underline{\Psi 
}^{\diamond _{2}})^{2}/\ \ _{2}^{Q}\underline{J}^{2}|\underline{h}%
_{3}^{[0]}-\int dt(\underline{\Psi }^{2})^{\diamond _{2}}/4\ \ _{2}^{Q}%
\underline{J}^{2}|^{5/2};%
\end{array}%
$ \\ 
$%
\begin{array}{c}
h_{5}=-(\partial _{5}\ ^{3}\Psi )^{2}/4\ \ _{3}^{Q}J^{2}h_{6}; \\ 
h_{6}=h_{6}^{[0]}-\int dv^{5}\partial _{5}((\ ^{3}\Psi )^{2})/4\ \
_{3}^{Q}J=h_{6}^{[0]}-(\ ^{3}\Phi )^{2}/4\ _{3}\Lambda ; \\ 
w_{i_{2}}=\partial _{i_{2}}(\ ^{3}\Psi )/\ \partial _{5}(\ ^{3}\Psi
)=\partial _{i_{2}}(\ ^{3}\Psi )^{2}/\ \partial _{5}(\ ^{3}\Psi )^{2}|; \\ 
n_{k_{2}}=\ _{1}n_{k_{2}}+\ _{2}n_{k_{2}}\int dv^{5}(\partial _{5}\ ^{3}\Psi
)^{2}/\ \ _{3}^{Q}J^{2}|h_{6}^{[0]}-\int dv^{5}\partial _{5}((\ ^{3}\Psi
)^{2})/4\ \ _{3}^{Q}J^{2}|^{5/2};%
\end{array}%
$ \\ 
$%
\begin{array}{c}
h_{7}=-(\partial _{7}\ ^{3}\Psi )^{2}/4\ \ _{4}^{Q}J^{2}h_{8}; \\ 
h_{8}=h_{8}^{[0]}-\int dv^{7}\partial _{9}((\ ^{4}\Psi )^{2})/4\ \ \
_{4}^{Q}J=h_{8}^{[0]}-(\ ^{4}\Phi )^{2}/4\ _{4}\Lambda ; \\ 
w_{i_{3}}=\partial _{i_{3}}(\ ^{4}\Psi )/\ \partial _{7}(\ ^{4}\Psi
)=\partial _{i_{3}}(\ ^{4}\Psi )^{2}/\ \partial _{7}(\ ^{4}\Psi )^{2}|; \\ 
n_{k_{3}}=\ _{1}n_{k_{3}}+\ _{2}n_{k_{3}}\int dv^{7}(\partial _{7}\ ^{4}\Psi
)^{2}/\ \ \ _{4}^{Q}J^{2}|h_{8}^{[0]}-\int dv^{7}\partial _{7}((\ ^{4}\Psi
)^{2})/4\ \ \ _{4}^{Q}J^{2}|^{5/2}.%
\end{array}%
$%
\end{tabular}%
$ \\ \hline\hline
\end{tabular}%
\end{eqnarray*}%
}

Similar classes of locally cosmological phase velocity space solutions (in
general, encoding nonmetric geometric flow data) can be derived for the same
Killing symmetries on $\partial _{3}$ and $\partial _{8}$ using respective
nonlinear symmetries and generating and integration functions.

\subsubsection{Nonmetric cosmological solutions with phase space rainbow
symmetries}

The locally anisotropic nonmetric cosmological models from previous Table 7
can be re-defined by phase space rainbow symmetries with the shells $s=3,4$
part as in Table 6. The procedure of constructing such classes of solutions
with Killing symmetries on $\partial _{3}$ and $\partial _{7}$ is summarized
below in Table 8. As an example of 8-d locally anisotropic cosmological
quadratic element with $v^{7}=const$ on $T\mathbf{V},$ and re-defining
rainbow configurations as for $s=3,4$ in (\ref{qst8d8}) but with
dependencies on another fiber variables, we provide 
\begin{eqnarray}
d\widehat{s}_{[8d]}^{2} &=&\widehat{g}_{\alpha _{s}\beta
_{s}}(x^{k},t,v^{5},v^{8};\underline{h}_{3},h_{6},\underline{h}_{7,};\
_{1}^{Q}J\ ,\ \ _{2}^{Q}\underline{J},\ \ _{3}^{Q}J,\ \ _{4}^{Q}\underline{J}%
;\ _{1}\Lambda ,\ _{2}\underline{\Lambda },\ _{3}\Lambda ,\ _{4}\underline{%
\Lambda })du^{\alpha _{s}}du^{\beta _{s}}  \label{lc8d8} \\
&=&e^{\psi (x^{k},\ \ _{1}^{Q}J)}[(dx^{1})^{2}+(dx^{2})^{2}]+\underline{h}%
_{3}[dy^{3}+(\ _{1}n_{k_{1}}+4\ _{2}n_{k_{1}}\int dt\frac{(\underline{h}%
_{3}{}^{\diamond _{2}})^{2}}{|\int dt\ _{2}\underline{J}\underline{h}%
_{3}{}^{\diamond _{2}}|(\underline{h}_{3})^{5/2}})dx^{k_{1}}]  \notag \\
&&-\frac{(\underline{h}_{3}{}^{\diamond _{2}})^{2}}{|\int dt\ _{2}\underline{%
J}\underline{h}_{3}{}^{\diamond _{2}}|\ \overline{h}_{3}}[dt+\frac{\partial
_{i_{1}}(\int dt\ _{2}\underline{J}\ \underline{h}_{3}{}^{\diamond _{2}}])}{%
\ \ _{2}\underline{J}\ \underline{h}_{3}{}^{\diamond _{2}}}dx^{i_{1}}]+ 
\notag \\
&&\frac{(\partial _{5}h_{6})^{2}}{|\int dv^{5}\partial _{5}[\ _{3}Jh_{6}]|\
h_{6}}\{dv^{5}+\frac{\partial _{i_{2}}[\int dv^{5}(\ _{3}J)\ \partial
_{5}h_{6}]}{\ _{3}J\ \partial _{5}h_{6}}dx^{i_{2}}\}^{2}+  \notag \\
&&h_{6}\{dv^{5}+[\ _{1}n_{k_{2}}+\ _{2}n_{k_{2}}\int dv^{5}\frac{(\partial
_{5}h_{6})^{2}}{|\int dv^{5}\partial _{5}[\ _{3}Jh_{6}]|\ (h_{6})^{5/2}}%
]dx^{k_{2}}\}+  \notag \\
&&\underline{h}_{7}\{dv^{7}+[\ _{1}n_{k_{3}}+\ _{2}n_{k_{3}}\int dv^{8}\frac{%
(\partial _{8}\underline{h}_{7})^{2}}{|\int dv^{8}\partial _{8}[\ _{4}%
\underline{J}h_{7}]|\ (\underline{h}_{7})^{5/2}}]dx^{k_{3}}\}+  \notag \\
&&\frac{(\partial _{8}\underline{h}_{7})^{2}}{|\int dv^{8}\partial _{8}[\
_{4}\underline{J}\underline{h}_{7}]|\ \underline{h}_{7}}\{dv^{8}+\frac{%
\partial _{i_{3}}[\int dv^{8}(\ _{4}\underline{J})\ \partial _{8}\underline{h%
}_{7}]}{\ _{4}\underline{J}\ \partial _{8}\underline{h}_{7}}dx^{i_{3}}\}^{2}.
\notag
\end{eqnarray}%
The AFCDM for generating nonmetric solutions for abouve mentioned type
geometric data are described as follow:

\newpage 
{\scriptsize 
\begin{eqnarray*}
&&%
\begin{tabular}{l}
\hline\hline
\begin{tabular}{lll}
& {\large \textsf{Table 8:\ Off-diagonal nonmetric cosmological spacetimes
with velocity rainbow symmetries}} &  \\ 
& Exact solutions of $\ \widehat{\mathbf{R}}_{\mu _{s}\nu _{s}}(\tau )=\ _{Q}%
\widehat{\underline{\mathbf{J}}}_{\mu _{s}\nu _{s}}(\tau )$ (\ref{cfeq4af})
on $TV$ transformed into a shall system of nonlinear PDEs (\ref{eq1})-(\ref%
{e2c}) & 
\end{tabular}
\\ 
\end{tabular}
\\
&&%
\begin{tabular}{lll}
\hline\hline
&  &  \\ 
$%
\begin{array}{c}
\mbox{d-metric ansatz with} \\ 
\mbox{Killing symmetry }\partial _{3}=\partial _{t},\partial _{8}%
\end{array}%
$ &  & $%
\begin{array}{c}
ds^{2}(\tau )=g_{i_{1}}(x^{k_{1}})(dx^{i_{1}})^{2}+\underline{g}%
_{a_{2}}(x^{k_{1}},t)(dy^{a_{2}}+\underline{N}%
_{i_{1}}^{a_{2}}(x^{k_{1}},t)dx^{i_{1}})^{2} \\ 
+g_{a_{3}}(x^{k_{2}},v^{5})(dy^{a_{3}}+N_{i_{2}}^{a_{3}}(x^{k_{2}},v^{5})dx^{i_{2}})^{2}
\\ 
+\underline{g}_{a_{4}}(x^{k_{3}},v^{8})(dy^{a_{4}}+\underline{N}%
_{i_{3}}^{a_{4}}(x^{k_{3}},v^{8})dx^{i_{3}})^{2},\mbox{ for }%
g_{i_{1}}=e^{\psi {(x}^{k_{1}}{)}}, \\ 
\underline{g}_{a_{2}}=\underline{h}_{a_{2}}(x^{k_{1}},t),\underline{N}%
_{i_{1}}^{3}=\ ^{2}\underline{n}_{i_{1}}=\underline{n}_{i_{1}}(x^{k_{1}},t),%
\underline{N}_{i_{1}}^{4}=\ ^{2}\underline{w}_{i_{1}}=\underline{w}%
_{i_{1}}(x^{k_{1}},t), \\ 
g_{a_{3}}=h_{a_{3}}(x^{k_{2}},v^{5}),N_{i_{2}}^{5}=\
^{3}w_{i_{2}}=w_{i_{2}}(x^{k_{2}},v^{5}),N_{i_{2}}^{6}=\
^{3}n_{i_{2}}=n_{i_{2}}(x^{k_{2}},v^{5}), \\ 
\underline{g}_{a_{4}}=\underline{h}_{a_{4}}(x^{k_{3}},v^{8}),\underline{N}%
_{i_{3}}^{7}=\ ^{4}\underline{n}_{i_{3}}=\underline{n}%
_{i_{3}}(x^{k_{3}},v^{8}),\underline{N}_{i_{3}}^{8}=\ ^{4}\underline{w}%
_{i_{3}}=\underline{w}_{i_{3}}(x^{k_{3}},v^{8}),%
\end{array}%
$ \\ 
Effective matter sources &  & $\widehat{\underline{\mathbf{J}}}_{\ \nu
_{s}}^{\mu _{s}}=[\ \ _{1}^{Q}J({x}^{k_{1}})\delta _{j_{1}}^{i_{1}},\
_{2}^{Q}\underline{J}({x}^{k_{1}},t)\delta _{b_{2}}^{a_{2}},\ _{3}J({x}%
^{k_{2}},v^{5})\delta _{b_{3}}^{a_{3}},\ \ _{4}^{Q}\underline{J}({x}%
^{k_{3}},v^{8})\delta _{b_{4}}^{a_{4}},],$ \\ \hline
Nonlinear PDEs (\ref{eq1})-(\ref{e2c}) &  & $%
\begin{tabular}{lll}
$%
\begin{array}{c}
\psi ^{\bullet \bullet }+\psi ^{\prime \prime }=2\ \ _{1}^{Q}J; \\ 
\ ^{2}\underline{\varpi }^{\diamond }\ \underline{h}_{3}^{\diamond }=2%
\underline{h}_{3}\underline{h}_{4}\ _{2}^{Q}\underline{J}; \\ 
\ ^{2}\underline{n}_{k_{1}}^{\diamond \diamond }+\ ^{2}\underline{\gamma }\
^{2}\underline{n}_{k_{1}}^{\diamond }=0; \\ 
\ ^{2}\underline{\beta }\ ^{2}\underline{w}_{i_{1}}-\ ^{2}\underline{\alpha }%
_{i_{1}}=0;%
\end{array}%
$ &  & $%
\begin{array}{c}
\ ^{2}\underline{\varpi }{=\ln |\partial _{4}\underline{{h}}_{4}/\sqrt{|%
\underline{h}_{3}\underline{h}_{4}|}|,} \\ 
\ ^{2}\underline{\alpha }_{i_{1}}=(\partial _{4}\underline{h}_{3})\
(\partial _{i_{1}}\ ^{2}\underline{\varpi }), \\ 
\ ^{2}\underline{\beta }=(\partial _{4}\underline{h}_{4})\ (\partial _{3}\
^{2}\underline{\varpi }),\  \\ 
\ \ ^{2}\underline{\gamma }=\partial _{4}\left( \ln |\underline{h}%
_{3}|^{3/2}/|\underline{h}_{4}|\right) , \\ 
\partial _{1}q=q^{\bullet },\partial _{2}q=q^{\prime },\partial
_{4}q=\partial _{t}q=q^{\diamond }%
\end{array}%
$ \\ 
$%
\begin{array}{c}
\partial _{5}(\ ^{3}\varpi )\ \partial _{5}h_{6}=2h_{5}h_{6}\ \ _{3}^{Q}J;
\\ 
\ ^{3}\beta \ ^{3}w_{i_{2}}-\ ^{3}\alpha _{i_{2}}=0; \\ 
\partial _{5}(\partial _{5}\ ^{3}n_{k_{2}})+\ ^{3}\gamma \partial _{5}(\
^{3}n_{k_{2}})=0;%
\end{array}%
$ &  & $%
\begin{array}{c}
\\ 
\ ^{3}\varpi {=\ln |\partial _{5}h_{6}/\sqrt{|h_{5}h_{6}|}|,} \\ 
\ ^{3}\alpha _{i_{2}}=(\partial _{5}h_{6})\ (\partial _{i_{2}}\ ^{3}\varpi ),
\\ 
\ ^{3}\beta =(\partial _{5}h_{6})\ (\partial _{5}\ ^{3}\varpi ),\  \\ 
\ \ ^{3}\gamma =\partial _{5}\left( \ln |h_{6}|^{3/2}/|h_{5}|\right) ,%
\end{array}%
$ \\ 
$%
\begin{array}{c}
\partial _{8}(\ ^{4}\underline{\varpi })\ \partial _{8}\underline{h}_{7}=2%
\underline{h}_{7}\underline{h}_{8}\ _{4}^{Q}\underline{J}; \\ 
\partial _{8}(\partial _{8}\ ^{4}\underline{n}_{k_{3}})+\ ^{4}\underline{%
\gamma }\partial _{8}(\ ^{4}\underline{n}_{k_{3}})=0; \\ 
\ ^{4}\underline{\beta }\ ^{4}\underline{w}_{i_{3}}-\ ^{4}\underline{\alpha }%
_{i_{3}}=0;%
\end{array}%
$ &  & $%
\begin{array}{c}
\\ 
\ ^{4}\underline{\varpi }{=\ln |\partial _{8}\underline{h}_{7}/\sqrt{|%
\underline{h}_{7}\underline{h}_{8}|}|,} \\ 
\ ^{4}\underline{\alpha }_{i}=(\partial _{8}\underline{h}_{7})\ (\partial
_{i}\ ^{4}\underline{\varpi }), \\ 
\ ^{4}\underline{\beta }=(\partial _{8}\underline{h}_{7})\ (\partial _{8}\
^{4}\underline{\varpi }),\  \\ 
\ \ ^{4}\underline{\gamma }=\partial _{8}\left( \ln |\underline{h}%
_{7}|^{3/2}/|\underline{h}_{8}|\right) ,%
\end{array}%
$%
\end{tabular}%
$ \\ \hline
$%
\begin{array}{c}
\mbox{ Gener.  functs:}\ \underline{h}_{4}(x^{k_{1}},t), \\ 
\ ^{2}\underline{\Psi }(x^{k_{1}},t)=e^{\ ^{2}\underline{\varpi }},\ ^{2}%
\underline{\Phi }(x^{k_{1}},t), \\ 
\mbox{integr. functs:}\ \underline{h}_{3}^{[0]}(x^{k_{1}}),\  \\ 
_{1}\underline{n}_{k_{1}}(x^{i_{1}}),\ _{2}\underline{n}_{k_{1}}(x^{i_{1}});
\\ 
\mbox{ Gener.  functs:}h_{5}(x^{k_{2}},v^{5}), \\ 
\ ^{3}\Psi (x^{k_{2}},v^{5})=e^{\ ^{3}\varpi },\ ^{3}\Phi (x^{k_{2}},v^{5}),
\\ 
\mbox{integr. functs:}\ h_{6}^{[0]}(x^{k_{2}}),\  \\ 
_{1}^{3}n_{k_{2}}(x^{i_{2}}),\ _{2}^{3}n_{k_{2}}(x^{i_{2}}); \\ 
\mbox{ Gener.  functs:}h_{7}(x^{k_{3}},v^{7}), \\ 
\ ^{4}\underline{\Psi }(x^{k_{2}},v^{8})=e^{\ ^{4}\underline{\varpi }},\ ^{4}%
\underline{\Phi }(x^{k_{3}},v^{8}), \\ 
\mbox{integr. functs:}\ h_{8}^{[0]}(x^{k_{3}}),\  \\ 
_{1}^{4}n_{k_{3}}(x^{i_{3}}),\ _{2}^{4}n_{k_{3}}(x^{i_{4}}); \\ 
\mbox{\& nonlinear symmetries}%
\end{array}%
$ &  & $%
\begin{array}{c}
\ ((\ ^{2}\underline{\Psi })^{2})^{\diamond _{2}}=-\int dt\ \ _{2}^{Q}%
\underline{J}\underline{h}_{3}^{\ \diamond _{2}}, \\ 
(\ ^{2}\underline{\Phi })^{2}=-4\ _{2}\underline{\Lambda }\underline{h}_{3},
\\ 
h_{3}=h_{3}^{[0]}-(\ ^{2}\underline{\Phi })^{2}/4\ _{2}\underline{\Lambda },%
\underline{h}_{3}^{\diamond _{2}}\neq 0,\ _{2}\underline{\Lambda }\neq
0=const; \\ 
\\ 
\partial _{5}((\ ^{3}\Psi )^{2})=-\int dv^{5}\ _{3}^{Q}J\partial
_{5}h_{6}^{\ }, \\ 
(\ ^{3}\Phi )^{2}=-4\ _{3}\Lambda h_{6}, \\ 
h_{6}=h_{6}^{[0]}-(\ ^{3}\Phi )^{2}/4\ _{3}\Lambda ,\partial _{5}h_{6}\neq
0,\ _{3}\Lambda \neq 0=const; \\ 
\\ 
\partial _{8}((\ ^{4}\underline{\Psi })^{2})=-\int dv^{8}\ _{4}^{Q}%
\underline{J}\partial _{8}\underline{h}_{7}^{\ }, \\ 
(\ ^{4}\underline{\Phi })^{2}=-4\ _{4}\underline{\Lambda }\underline{h}_{7},
\\ 
\underline{h}_{7}=h_{7}^{[0]}-(\ ^{4}\underline{\Phi })^{2}/4\ _{4}%
\underline{\Lambda },\partial _{8}\underline{h}_{7}\neq 0,\ _{4}\underline{%
\Lambda }\neq 0=const;%
\end{array}%
$ \\ \hline
Off-diag. solutions, $%
\begin{array}{c}
\mbox{d--metric} \\ 
\mbox{N-connec.}%
\end{array}%
$ &  & $%
\begin{tabular}{l}
$%
\begin{array}{c}
\ g_{i}=e^{\ \psi (x^{k})}\mbox{ as a solution of 2-d Poisson eqs. }\psi
^{\bullet \bullet }+\psi ^{\prime \prime }=2~\ \ _{1}^{Q}J; \\ 
\underline{h}_{4}=-(\underline{\Psi }^{\diamond })^{2}/4\ \ _{2}^{Q}%
\underline{J}^{2}\underline{h}_{3}; \\ 
\underline{h}_{3}=\underline{h}_{3}^{[0]}-\int dt(\underline{\Psi }%
^{2})^{\diamond _{2}}/4\ \ _{2}^{Q}\underline{J}=\underline{h}_{3}^{[0]}-%
\underline{\Phi }^{2}/4\ _{2}\underline{\Lambda }; \\ 
\underline{w}_{i_{1}}=\partial _{i_{1}}\ \underline{\Psi }/\ \partial 
\underline{\Psi }^{\diamond _{2}}=\partial _{i_{1}}\ \underline{\Psi }^{2}/\
\partial _{t}\underline{\Psi }^{2}|; \\ 
\underline{n}_{k_{1}}=\ _{1}n_{k_{1}}+\ _{2}n_{k_{1}}\int dt(\underline{\Psi 
}^{\diamond _{2}})^{2}/\ \ _{2}^{Q}\underline{J}^{2}|\underline{h}%
_{3}^{[0]}-\int dt(\underline{\Psi }^{2})^{\diamond _{2}}/4\ \ _{2}^{Q}%
\underline{J}^{2}|^{5/2};%
\end{array}%
$ \\ 
$%
\begin{array}{c}
h_{5}=-(\partial _{5}\ ^{3}\Psi )^{2}/4\ \ _{3}^{Q}J^{2}h_{6}; \\ 
h_{6}=h_{6}^{[0]}-\int dv^{5}\partial _{5}((\ ^{3}\Psi )^{2})/4\ \
_{3}^{Q}J=h_{6}^{[0]}-(\ ^{3}\Phi )^{2}/4\ _{3}\Lambda ; \\ 
w_{i_{2}}=\partial _{i_{2}}(\ ^{3}\Psi )/\ \partial _{5}(\ ^{3}\Psi
)=\partial _{i_{2}}(\ ^{3}\Psi )^{2}/\ \partial _{5}(\ ^{3}\Psi )^{2}|; \\ 
n_{k_{2}}=\ _{1}n_{k_{2}}+\ _{2}n_{k_{2}}\int dv^{5}(\partial _{5}\ ^{3}\Psi
)^{2}/\ \ _{3}^{Q}J^{2}|h_{6}^{[0]}-\int dv^{5}\partial _{5}((\ ^{3}\Psi
)^{2})/4\ \ _{3}^{Q}J^{2}|^{5/2};%
\end{array}%
$ \\ 
$%
\begin{array}{c}
\underline{h}_{7}=\underline{h}_{7}^{[0]}-\int dv^{8}\partial _{8}((\ ^{4}%
\underline{\Psi })^{2})/4\ \ _{4}^{Q}\underline{J}=\underline{h}%
_{7}^{[0]}-(\ ^{4}\underline{\Phi })^{2}/4\ _{4}\underline{\Lambda }; \\ 
\underline{h}_{8}=-(\partial _{8}\ ^{4}\underline{\Psi })^{2}/4\ \ \ _{4}^{Q}%
\underline{J}^{2}\underline{h}_{7}; \\ 
n_{k_{3}}=\ _{1}n_{k_{3}}+\ _{2}n_{k_{3}}\int dv^{8}(\partial _{8}\ ^{4}%
\underline{\Psi })^{2}/\ \ \ _{4}^{Q}\underline{J}^{2}|\underline{h}%
_{7}^{[0]}-\int dv^{8}\partial _{8}((\ ^{4}\underline{\Psi })^{2})/4\ \ \
_{4}^{Q}\underline{J}^{2}|^{5/2}; \\ 
w_{i_{3}}=\partial _{i_{3}}(\ ^{4}\underline{\Psi })/\ \partial _{8}(\ ^{4}%
\underline{\Psi })=\partial _{i_{3}}(\ ^{4}\underline{\Psi })^{2}/\ \partial
_{8}(\ ^{4}\underline{\Psi })^{2}|.%
\end{array}%
$%
\end{tabular}%
$ \\ \hline\hline
\end{tabular}%
\end{eqnarray*}%
}

Velocity rainbow s-metrics (\ref{lc8d8}) can be also considered just for
Finsler spaces if we use the respective generating and distortions
functions. We can impose homogeneity and other type conditions in order to
define more special classes of relativistic generalized Finsler geometries.
Such models can be redefined for momentum variables, for Cartan-Finsler
models on cotangent Lorentz bundles as in the next subsection. 

\subsection{Momentum depending quasi-stationary and cosmological solutions}

\label{tab1216}A series of recent works on nonassociative geometric and
quantum information flows, nonassociative and noncommutative gravity and FH
geometry and gravity have been elaborated on nonholonomic phases spaces
modeled on a cotangent Lorentz bundle, $\ ^{\shortmid }\mathcal{M}=T^{\ast }%
\mathbf{V},$ see reviews and original results in \cite%
{bsssvv25,partner06,vmon3,vacaru18,bubuianu18,vacaru25b,bnsvv24,bvvz24}. In
this subsection, we modify those constructions for nonmetric FH flows on
relativistic 8-d phase spaces with conventional dyadic splitting
(2+2)+(2+2). For such theories, the local coordinates on shells $s=3$ and $%
s=4$ are momentum type $p_{a}$ and the local coordinates on the total space
are labeled $\ ^{\shortmid }u=(x,p)=\{\ ^{\shortmid }u^{\alpha
}=(x^{i},p_{a})\}=\{\ ^{\shortmid }u^{\alpha
_{s}}=(x^{i_{1}},y^{a_{2}},p_{a_{3}},p_{a_{4}})\}$ for $\ ^{\shortmid
}p=p=(p_{a_{3}},p_{7},p_{8}=E),$ where $E$ is an energy type variable. For
mechanical like models on cotangent bundles, the momentum like variables $%
(p_{a_{3}},p_{a_{4}})$ can be related to velocity type variables $%
(v^{b_{3}},v^{b_{4}})$, considered in previous subsection, using Legendre
transforms.

\subsubsection{Off-diagonal ansatz for nonmetric geometric flows on momentum
phase spaces}

The parametrization of local coordinates, N-connection and canonical
d-connection structures and s-metrics are sated in Table 9, when the AFCDM
is modified for generating solutions for nonmetric geometric flow equations (%
\ref{feq4afd}) on$\ _{s}^{\shortmid }\mathcal{M}$.

\newpage

{\scriptsize 
\begin{eqnarray*}
&&%
\begin{tabular}{lll}
& {\ \textsf{Table 9:\ Diagonal and off-diagonal ansatz for FH geometric
flows on 8-d cotangent Lorentz bundles} } &  \\ 
& and the Anholonomic Frame and Connection Deformation Method, \textbf{AFCDM}%
, &  \\ 
& \textit{for constructing generic off-diagonal exact and parametric
solutions} & 
\end{tabular}
\\
&&{%
\begin{tabular}{lll}
\hline
diagonal ansatz: PDEs $\rightarrow $ \textbf{ODE}s &  & AFCDM: \textbf{PDE}s 
\textbf{with decoupling; } \\ 
\begin{tabular}{l}
coordinates \\ 
$\ ^{\shortmid }u^{\alpha _{s}}=(x^{1},x^{2},y^{3},y^{4}=t,$ \\ 
$p_{5},p_{6},p_{7},p_{8}=E)$%
\end{tabular}
& $%
\begin{array}{c}
\ _{s}^{\shortmid }u=(\ _{s-1}x,\ _{s}^{\shortmid }y) \\ 
s=1,2,3,4;%
\end{array}%
$ & $%
\begin{tabular}{l}
nonholonomic 2+2++2+2 splitting; shels $s=1,2,3,4$ \\ 
$\ ^{\shortmid }u^{\alpha
_{s}}=(x^{1},x^{2},y^{3},y^{4}=t,p_{5},p_{6},p_{7},p_{8}=E);$ \\ 
$\ ^{\shortmid }u^{\alpha _{s}}=(x^{i_{1}},y^{a_{2}},p_{a_{3}},p_{a_{4}});\
^{\shortmid }u^{\alpha _{s}}=(x^{i_{s-1}},\ ^{\shortmid }y^{a_{s}});$ \\ 
$\ $ $i_{1}=1,2;a_{2}=3,4;a_{3}=5,6;a_{4}=7,8;\tau -\mbox{parameter}$%
\end{tabular}%
$ \\ 
LC-connection $\ ^{\shortmid }\mathring{\nabla}$ & 
\begin{tabular}{l}
N-connection; \\ 
canonical \\ 
d-connection%
\end{tabular}
& $%
\begin{array}{c}
\ \ _{s}^{\shortmid }\mathbf{N}:T\ _{s}^{\ast }\mathbf{V}=hT^{\ast }\mathbf{V%
}\oplus \ ^{2}vT^{\ast }\mathbf{V}\oplus \ ^{3}c\mathbf{T}^{\ast }\mathbf{%
\mathbf{V}\oplus }\ ^{4}c\mathbf{T}^{\ast }\mathbf{\mathbf{V},} \\ 
\mbox{ locally }\ \ _{s}^{\shortmid }\mathbf{N}=\{\ ^{\shortmid
}N_{i_{s-1}}^{a_{s}}(x,p)= \\ 
\ ^{\shortmid }N_{i_{s-1}}^{a_{s}}(\ _{s-1}x,\ _{s}^{\shortmid }y)=\
^{\shortmid }N_{i_{s-1}}^{a_{s}}(\ _{s}^{\shortmid }u)\} \\ 
\ \ _{s}^{\shortmid }\widehat{\mathbf{D}}=(\ ^{1}h\ ^{\shortmid }\widehat{%
\mathbf{D}},\ ^{2}v\ ^{\shortmid }\widehat{\mathbf{D}},\ ^{3}c\ ^{\shortmid }%
\widehat{\mathbf{D}},\ ^{4}c\ ^{\shortmid }\widehat{\mathbf{D}})=\{\
^{\shortmid }\Gamma _{\ \beta _{s}\gamma _{s}}^{\alpha _{s}}\}; \\ 
\mbox{ canonical s-connection distortion }\  \\ 
\ \ _{s}^{\shortmid }\widehat{\mathbf{D}}=\ ^{\shortmid }\nabla +\
_{s}^{\shortmid }\ \widehat{\mathbf{Z}};\ \ \ _{s}^{\shortmid }\widehat{%
\mathbf{D}}\ \ _{s}^{\shortmid }\mathbf{g=0,} \\ 
\ \ \ _{s}^{\shortmid }\widehat{\mathcal{T}}[\ _{s}^{\shortmid }\mathbf{g,}\
\ _{s}^{\shortmid }\mathbf{N,}\ \ _{s}^{\shortmid }\widehat{\mathbf{D}}]%
\mbox{ canonical
d-torsion}%
\end{array}%
$ \\ 
$%
\begin{array}{c}
\mbox{ diagonal ansatz  } \\ 
\ ^{2}\mathring{g}=\mathring{g}_{\alpha _{2}\beta _{2}}(\ ^{s}u)= \\ 
\left( 
\begin{array}{cccc}
\mathring{g}_{1} &  &  &  \\ 
& \mathring{g}_{2} &  &  \\ 
&  & \mathring{g}_{3} &  \\ 
&  &  & \mathring{g}_{4}%
\end{array}%
\right) ; \\ 
\ ^{s}g=\mathring{g}_{\alpha _{s}\beta _{s}}(\ ^{s}u)= \\ 
\left( 
\begin{array}{cccc}
\ ^{2}\mathring{g} &  &  &  \\ 
& \ ^{\shortmid }\mathring{g}^{5} &  &  \\ 
&  & \ddots &  \\ 
&  &  & \ ^{\shortmid }\mathring{g}^{8}%
\end{array}%
\right)%
\end{array}%
$ & $\ ^{\shortmid }\mathbf{g(\tau )}\Leftrightarrow $ & $%
\begin{tabular}{l}
$g_{\alpha _{2}\beta _{2}}=%
\begin{array}{c}
g_{\alpha _{2}\beta _{2}}(x^{i_{1}},y^{a_{2}})%
\mbox{ general frames /
coordinates} \\ 
\left[ 
\begin{array}{cc}
g_{i_{1}j_{1}}+N_{i_{1}}^{a_{2}}N_{j_{1}}^{b_{2}}h_{a_{2}b_{2}} & 
N_{i_{1}}^{b_{2}}h_{c_{2}b_{2}} \\ 
N_{j_{1}}^{a_{2}}h_{a_{2}b_{2}} & h_{a_{2}c_{2}}%
\end{array}%
\right]%
\end{array}%
$ \\ 
$\ ^{2}\mathbf{g=\{g}_{\alpha _{2}\beta
_{2}}=[g_{i_{1}j_{1}},h_{a_{2}b_{2}}]\},$ \\ 
$\ ^{2}\mathbf{g}=\mathbf{g}_{i_{1}}(x^{k_{1}})dx^{i_{1}}\otimes dx^{i_{1}}+%
\mathbf{g}_{a_{2}}(x^{k_{2}},y^{b_{2}})\mathbf{e}^{a_{2}}\otimes \mathbf{e}%
^{b_{2}}$ \\ 
$\vdots $ \\ 
$\ ^{\shortmid }g_{\alpha _{s}\beta _{s}}=$ \\ 
$%
\begin{array}{c}
\ ^{\shortmid }g_{\alpha _{s}\beta _{s}}(x^{i_{s-1}},\ ^{\shortmid
}y^{a_{s}})\mbox{ general frames / coordinates} \\ 
\left[ 
\begin{array}{cc}
\ ^{\shortmid }g_{i_{s}j_{s}}+\ ^{\shortmid }N_{i_{s-1}}^{a_{s}}\
^{\shortmid }N_{j_{s-1}}^{b_{s}}\ ^{\shortmid }h_{a_{s}b_{s}} & \
^{\shortmid }N_{i_{s-1}}^{b_{s}}\ ^{\shortmid }h_{c_{s}b_{s}} \\ 
\ ^{\shortmid }N_{j_{s-1}}^{a_{s}}\ ^{\shortmid }h_{a_{s}b_{s}} & \
^{\shortmid }h_{a_{s}c_{s}}%
\end{array}%
\right]%
\end{array}%
$ \\ 
$\ _{s}^{\shortmid }\mathbf{g=\{\ ^{\shortmid }g}_{\alpha _{s}\beta _{s}}=[\
^{\shortmid }g_{i_{s-1}j_{s-1}},\ ^{\shortmid }h_{a_{s}b_{s}}]$ \\ 
$=[g_{i_{1}j_{1}},h_{a_{2}b_{2}},\ ^{\shortmid }h^{a_{3}b_{3}},\ ^{\shortmid
}h^{a_{4}b_{4}}]\}$ \\ 
$\ \ _{s}^{\shortmid }\mathbf{g}=\ _{s}^{\shortmid }\mathbf{g}%
_{i_{s-1}}(x^{k_{s-1}})dx^{i_{s-1}}\otimes dx^{i_{s-1}}+$ \\ 
$\ _{s}^{\shortmid }\mathbf{g}_{a_{s}}(x^{k_{s-1}},y^{b_{s}})\mathbf{e}%
^{a_{s}}\otimes \mathbf{e}^{b_{s}}$ \\ 
$=\ \mathbf{g}_{i_{1}}(x^{k_{1}})dx^{i_{1}}\otimes dx^{i_{1}}+\mathbf{g}%
_{a_{2}}(x^{k_{1}},y^{b_{2}})\mathbf{e}^{a_{2}}\otimes \mathbf{e}^{a_{2}}+$
\\ 
$\mathbf{\ ^{\shortmid }g}^{a_{3}}(x^{k_{1}},y^{b_{2}},p_{b_{3}})\mathbf{\
^{\shortmid }e}_{a_{3}}\otimes \mathbf{\ ^{\shortmid }e}_{a_{3}}$ \\ 
$+\mathbf{\ ^{\shortmid }g}_{a_{4}}(x^{k_{1}},y^{b_{2}},p_{b_{3}},p_{b_{4}})%
\mathbf{\ ^{\shortmid }e}_{a_{4}}\otimes \mathbf{\ ^{\shortmid }e}_{a_{4}};$%
\end{tabular}%
$ \\ 
$\mathring{g}_{\alpha _{2}\beta _{2}}=\left\{ 
\begin{array}{cc}
\mathring{g}_{\alpha _{2}}(\ ^{2}r) & \mbox{ for BHs} \\ 
\mathring{g}_{\alpha _{2}}(t) & \mbox{ for FLRW }%
\end{array}%
\right. $ & [coord.frames] & $g_{\alpha _{2}\beta _{2}}=\left\{ 
\begin{array}{cc}
g_{\alpha _{2}\beta _{2}}(x^{i},y^{3}) &  \\ 
\underline{g}_{\alpha _{2}\beta _{2}}(x^{i},y^{4}=t) & 
\end{array}%
\right. $ \\ 
$\mathbf{\ ^{\shortmid }}\mathring{g}_{\alpha _{s}\beta _{s}}=\left\{ 
\begin{array}{cc}
\mathbf{\ ^{\shortmid }}\mathring{g}_{\alpha _{s}}(\ _{s}^{\shortmid }r) & %
\mbox{ for BHs} \\ 
\mathbf{\ ^{\shortmid }}\mathring{g}_{\alpha _{s}}(t) & \mbox{ for FLRW }%
\end{array}%
\right. $ &  & $\mathbf{\ ^{\shortmid }}g_{\alpha _{s}\beta _{s}}=\left\{ 
\begin{array}{cc}
\mathbf{\ ^{\shortmid }}g_{\alpha _{s}\beta _{s}}(x^{i_{3}},p_{7}) &  \\ 
\mathbf{\ ^{\shortmid }}\underline{g}_{\alpha _{s}\beta _{s}}(x^{i_{3}},E) & 
\end{array}%
\right. $ \\ 
$%
\begin{array}{c}
\mbox{coord. transf. }\mathbf{\ ^{\shortmid }}e_{\alpha _{s}}=\mathbf{\
^{\shortmid }}e_{\ \alpha _{s}}^{\alpha _{s}^{\prime }}\mathbf{\ ^{\shortmid
}}\partial _{\alpha _{s}^{\prime }}, \\ 
\mathbf{\ ^{\shortmid }}e^{\beta _{s}}=\mathbf{\ ^{\shortmid }}e_{\beta
_{s}^{\prime }}^{\ \beta _{s}}d\mathbf{\ ^{\shortmid }}u^{\beta _{s}^{\prime
}}, \\ 
\mathbf{\ ^{\shortmid }}\mathring{g}_{\alpha _{s}\beta _{s}}=\mathbf{\
^{\shortmid }}\mathring{g}_{\alpha _{s}^{\prime }\beta _{s}^{\prime }}%
\mathbf{\ ^{\shortmid }}e_{\ \alpha _{s}}^{\alpha _{s}^{\prime }}\mathbf{\
^{\shortmid }}e_{\ \beta _{s}}^{\beta _{s}^{\prime }} \\ 
\begin{array}{c}
\mathbf{\ ^{\shortmid }\mathring{g}}_{\alpha _{s}}(\mathbf{\ ^{\shortmid }}%
x^{k_{s-1}},\mathbf{\ ^{\shortmid }}y^{a_{s}})\rightarrow \mathbf{\
^{\shortmid }}\mathring{g}_{\alpha _{s}}(\ _{s}^{\shortmid }r), \\ 
\mathbf{\ ^{\shortmid }}\mathring{g}_{\alpha _{s}}(t),\mathbf{\ ^{\shortmid }%
}\mathring{N}_{i_{s-1}}^{a_{s}}(x^{k_{s-1}},\mathbf{\ ^{\shortmid }}%
y^{a_{s}})\rightarrow 0.%
\end{array}%
\end{array}%
$ & [N-adapt. fr.] & 
\begin{tabular}{l}
$\left\{ 
\begin{array}{cc}
\begin{array}{c}
\mathbf{g}_{i_{1}}(x^{k_{1}}),\mathbf{g}_{a_{2}}(x^{k_{1}},y^{3}), \\ 
\mbox{ or }\mathbf{g}_{i_{1}}(x^{k_{1}}),\underline{\mathbf{g}}%
_{a_{2}}(x^{k_{1}},t),%
\end{array}
&  \\ 
\begin{array}{c}
N_{i_{1}}^{3}=w_{i_{1}}(x^{k},y^{3}),N_{i_{1}}^{4}=n_{i_{1}}(x^{k},y^{3}),
\\ 
\mbox{ or }\underline{N}_{i_{1}}^{3}=\underline{n}_{i_{1}}(x^{k_{1}},t),%
\underline{N}_{i_{1}}^{4}=\underline{w}_{i_{1}}(x^{k_{1}},t),%
\end{array}
& 
\end{array}%
\right. $ \\ 
$\vdots $ \\ 
$\left\{ 
\begin{array}{cc}
\begin{array}{c}
\mathbf{\ ^{\shortmid }g}_{i_{3}}(\mathbf{\ ^{\shortmid }}x^{k_{3}}),\mathbf{%
\ ^{\shortmid }g}_{a_{4}}(\mathbf{\ ^{\shortmid }}x^{k_{3}},p_{7}), \\ 
\mbox{ or }\mathbf{\ ^{\shortmid }g}_{i_{3}}(\mathbf{\ ^{\shortmid }}%
x^{k_{1}}),\mathbf{\ ^{\shortmid }}\underline{\mathbf{g}}_{a_{4}}(\mathbf{\
^{\shortmid }}x^{k_{3}},E),%
\end{array}
&  \\ 
\begin{array}{c}
\ ^{\shortmid }N_{i_{3}7}=\ ^{\shortmid }w_{i_{3}}(\ ^{\shortmid
}x^{k_{3}},p_{7}),\ ^{\shortmid }N_{i_{3}8}=\ ^{\shortmid }n_{i_{3}}\
^{\shortmid }x^{k_{3}},p_{7}) \\ 
\ ^{\shortmid }\underline{N}_{i_{3}8}=\ ^{\shortmid }\underline{n}_{i_{3}}\
^{\shortmid }x^{k_{3}},E),\ ^{\shortmid }\underline{N}_{i_{3}8}=\
^{\shortmid }\underline{w}_{i_{3}}\ ^{\shortmid }x^{k_{3}},E)%
\end{array}
& 
\end{array}%
\right. $%
\end{tabular}
\\ 
$\ _{s}^{\shortmid }\mathring{\nabla},$ $\ \ _{s}^{\shortmid }Ric=\{\mathbf{%
\ ^{\shortmid }}\mathring{R}_{\ \beta _{s}\gamma _{s}}\}$ & Ricci tensors & $%
\ \ _{s}^{\shortmid }\widehat{\mathbf{D}},\ \ \ _{s}^{\shortmid }\widehat{%
\mathcal{R}}ic=\{\mathbf{\ ^{\shortmid }}\widehat{\mathbf{R}}_{\ \beta
_{s}\gamma _{s}}\}$ \\ 
$%
\begin{array}{c}
\ _{s}^{\shortmid m}\mathcal{L[\mathbf{\phi }]\rightarrow }\ \
_{s}^{\shortmid m}\mathbf{T}_{\alpha _{s}\beta _{s}}\mathcal{[\mathbf{\phi }]%
} \\ 
\ _{s}^{\shortmid e}\mathcal{L[\mathbf{Z}]\rightarrow }\ \ _{s}^{\shortmid e}%
\mathbf{T}_{\alpha _{s}\beta _{s}}\mathcal{[\mathbf{Z}]}%
\end{array}%
$ & 
\begin{tabular}{l}
generating \\ 
sources%
\end{tabular}
& $%
\begin{array}{cc}
\mathbf{\ ^{\shortmid }J}_{\ \nu _{s}}^{\mu _{s}}=\mathbf{\ ^{\shortmid }e}%
_{\ \mu _{s}^{\prime }}^{\mu _{s}}\mathbf{\ ^{\shortmid }e}_{\nu _{s}}^{\
\nu _{s}^{\prime }}\mathbf{\ ^{\shortmid }J}_{\ \nu _{s}^{\prime }}^{\mu
_{s}^{\prime }}[\ ^{m}\mathcal{L}\mathbf{,}\ ^{e}\mathcal{L},\mathbf{\
^{\shortmid }}T_{\mu _{s}\nu _{s}},\ \ _{s}^{\shortmid }\Lambda ] &  \\ 
\begin{array}{c}
=diag[\ \ _{1}^{Q}J(x^{i_{1}})\delta _{j_{1}}^{i_{1}},\
_{2}^{Q}J(x^{i_{1}},y^{3})\delta _{b_{2}}^{a_{2}}, \\ 
\ _{3}^{\shortmid Q}J(x^{i_{2}},p_{5})\delta _{b_{3}}^{a_{3}},\
_{4}^{\shortmid Q}J(x^{i_{3}},p_{7})\delta _{b_{4}}^{a_{4}}], \\ 
\mbox{ quasi-stationary configurations};%
\end{array}
&  \\ 
\begin{array}{c}
=diag[\ \ _{1}^{Q}J(x^{i_{1}})\delta _{j_{1}}^{i_{1}},\ _{2}^{Q}\underline{J}%
(x^{i_{1}},t)\delta _{b_{2}}^{a_{2}}, \\ 
\ _{3}^{\shortmid Q}\underline{J}(x^{i_{2}},p_{6})\delta _{b_{3}}^{a_{3}},\
_{4}^{\shortmid Q}\underline{J}(x^{i_{3}},E)\delta _{b_{4}}^{a_{4}}], \\ 
\mbox{ locally anisotropic cosmology};%
\end{array}
& 
\end{array}%
$ \\ 
trivial eqs for $\ \ _{s}^{\shortmid }\mathring{\nabla}$-torsion & 
LC-conditions & $\ \ _{s}^{\shortmid }\widehat{\mathbf{D}}_{\mid \ \
_{s}^{\shortmid }\widehat{\mathcal{T}}\rightarrow 0}=\ \ _{s}^{\shortmid }%
\mathbf{\nabla .}$ \\ \hline\hline
\end{tabular}%
}
\end{eqnarray*}%
}

Parameterizations of geometric s-objects on different shells $s=2,3,4$
depend on the type of shell Killing symmetries we prescribe for such
nonholonomic phase spaces with momentum-like variables.

\subsubsection{Quasi-stationary FH nonmetric solutions with fixed energy
parameter}

Such quasi-stationary solutions are nonholonomic momentum-type phase
configurations modeled on cotangent Lorentz bundles with $p_{8}=E=const,$
when the momentum phase space involves space-like hypersurfaces. 
{\scriptsize 
\begin{eqnarray*}
&&%
\begin{tabular}{l}
\hline\hline
\begin{tabular}{lll}
& {\large \textsf{Table 10:\ Off-diagonal FH quasi-stationary and pase space
configurations with fixed energy}} &  \\ 
& Exact solutions of $\ ^{\shortmid }\widehat{\mathbf{R}}_{\ \ \gamma
_{s}}^{\beta _{s}}(\tau )={\delta }_{\ \ \gamma _{s}}^{\beta _{s}}\ \
_{Q}^{s\shortmid }\mathbf{J}(\tau )\mathbf{\ }$ (\ref{feq4afd}) on $%
T_{s}^{\ast }V$ transformed into a momentum version of nonlinear PDEs (\ref%
{eq1})-(\ref{e2c}) & 
\end{tabular}
\\ 
\end{tabular}
\\
&&%
\begin{tabular}{lll}
\hline\hline
&  &  \\ 
$%
\begin{array}{c}
\mbox{d-metric ansatz with} \\ 
\mbox{Killing symmetry }\partial _{4}=\partial _{t},\mathbf{\ ^{\shortmid }}%
\partial ^{8}%
\end{array}%
$ &  & $%
\begin{array}{c}
ds^{2}(\tau
)=g_{i_{1}}(x^{k_{1}})(dx^{i_{1}})^{2}+g_{a_{2}}(x^{k_{1}},y^{3})(dy^{a_{2}}+N_{i_{1}}^{a_{2}}(x^{k_{1}},y^{3})dx^{i_{1}})^{2}
\\ 
+\mathbf{\ ^{\shortmid }}g^{a_{3}}(x^{k_{2}},p_{5})(dp_{a_{3}}+\mathbf{\
^{\shortmid }}N_{i_{2}a_{3}}(x^{k_{2}},p_{5})dx^{i_{2}})^{2} \\ 
+\mathbf{\ ^{\shortmid }}g^{a_{4}}(\mathbf{\ ^{\shortmid }}%
x^{k_{3}},p_{7})(dp_{a_{4}}+\mathbf{\ ^{\shortmid }}N_{i_{3}a_{4}}(\mathbf{\
^{\shortmid }}x^{k_{3}},p_{7})d\mathbf{\ ^{\shortmid }}x^{i_{3}})^{2},%
\mbox{
for }g_{i_{1}}=e^{\psi {(x}^{k_{1}}{)}}, \\ 
g_{a_{2}}=h_{a_{2}}(x^{k_{1}},y^{3}),N_{i_{1}}^{3}=\
^{2}w_{i_{1}}=w_{i_{1}}(x^{k_{1}},y^{3}),N_{i_{1}}^{4}=\
^{2}n_{i_{1}}=n_{i_{1}}(x^{k_{1}},y^{3}), \\ 
\mathbf{\ ^{\shortmid }}g^{a_{3}}=\mathbf{\ ^{\shortmid }}%
h^{a_{3}}(x^{k_{2}},p_{5}),\mathbf{\ ^{\shortmid }}N_{i_{2}5}=\ _{\shortmid
}^{3}w_{i_{2}}=\mathbf{\ ^{\shortmid }}w_{i_{2}}(x^{k_{2}},p_{5}), \\ 
\mathbf{\ ^{\shortmid }}N_{i_{2}6}=\ _{\shortmid }^{3}n_{i_{2}}=\mathbf{\
^{\shortmid }}n_{i_{2}}(x^{k_{2}},p_{5}), \\ 
\mathbf{\ ^{\shortmid }}g^{a_{4}}=\mathbf{\ ^{\shortmid }}h^{a_{4}}(\mathbf{%
\ ^{\shortmid }}x^{k_{3}},p_{7}),\mathbf{\ ^{\shortmid }}N_{i_{3}7}=\ \
_{\shortmid }^{4}w_{i_{3}}=\mathbf{\ ^{\shortmid }}%
w_{i_{3}}(x^{k_{3}},p_{7}), \\ 
\mathbf{\ ^{\shortmid }}N_{i_{3}8}=\ _{\shortmid }^{4}n_{i_{3}}=\mathbf{\
^{\shortmid }}n_{i_{3}}(x^{k_{3}},p_{7}),%
\end{array}%
$ \\ 
Effective matter sources &  & $\mathbf{\ _{Q}^{\shortmid }J}_{\ \nu
_{s}}^{\mu _{s}}(\tau )\mathbf{\ }=[\ _{1}^{Q}J({x}^{k_{1}})\delta
_{j_{1}}^{i_{1}},\ \ _{2}^{Q}J({x}^{k_{1}},y^{3})\delta _{b_{2}}^{a_{2}},\
_{3}^{\shortmid Q}J({x}^{k_{2}},p_{5})\delta _{b_{3}}^{a_{3}},\
_{4}^{\shortmid Q}J({x}^{k_{3}},p_{7})\delta _{b_{4}}^{a_{4}}],$ \\ \hline
Nonlinear PDEs (\ref{eq1})-(\ref{e2c}) &  & $%
\begin{tabular}{lll}
$%
\begin{array}{c}
\psi ^{\bullet \bullet }+\psi ^{\prime \prime }=2\ \ _{1}^{Q}J; \\ 
\ ^{2}\varpi ^{\ast _{2}}\ h_{4}^{\ast _{2}}=2h_{3}h_{4}\ _{2}^{Q}J; \\ 
\ ^{2}\beta \ ^{2}w_{i_{1}}-\ ^{2}\alpha _{i_{1}}=0; \\ 
\ ^{2}n_{k_{1}}^{\ast _{2}\ast _{2}}+\ ^{2}\gamma \ ^{2}n_{k_{1}}^{\ast
_{2}}=0;%
\end{array}%
$ &  & $%
\begin{array}{c}
\ ^{2}\varpi {=\ln |\partial _{3}h_{4}/\sqrt{|h_{3}h_{4}|}|,} \\ 
\ ^{2}\alpha _{i_{1}}=(\partial _{3}h_{4})\ (\partial _{i_{1}}\ ^{2}\varpi ),
\\ 
\ ^{2}\beta =(\partial _{3}h_{4})\ (\partial _{3}\ ^{2}\varpi ),\  \\ 
\ \ ^{2}\gamma =\partial _{3}\left( \ln |h_{4}|^{3/2}/|h_{3}|\right) , \\ 
\partial _{1}q=q^{\bullet },\partial _{2}q=q^{\prime },\partial
_{3}q=q^{\ast _{2}}%
\end{array}%
$ \\ 
$%
\begin{array}{c}
\mathbf{\ ^{\shortmid }}\partial ^{5}(\ _{\shortmid }^{3}\varpi )\ \mathbf{\
^{\shortmid }}\partial ^{5}\mathbf{\ ^{\shortmid }}h^{6}=2\mathbf{\
^{\shortmid }}h^{5}\mathbf{\ ^{\shortmid }}h^{6}\ _{3}^{\shortmid Q}J; \\ 
\ _{\shortmid }^{3}\beta \ _{\shortmid }^{3}w_{i_{2}}-\ _{\shortmid
}^{3}\alpha _{i_{2}}=0; \\ 
\mathbf{\ ^{\shortmid }}\partial ^{5}(\mathbf{\ ^{\shortmid }}\partial ^{5}\
_{\shortmid }^{3}n_{k_{2}})+\ _{\shortmid }^{3}\gamma \mathbf{\ ^{\shortmid }%
}\partial ^{5}(\ _{\shortmid }^{3}n_{k_{2}})=0;%
\end{array}%
$ &  & $%
\begin{array}{c}
\\ 
\ _{\shortmid }^{3}\varpi {=\ln |\mathbf{\ ^{\shortmid }}\partial ^{5}%
\mathbf{\ ^{\shortmid }}h^{6}/\sqrt{|\mathbf{\ ^{\shortmid }}h^{5}\mathbf{\
^{\shortmid }}h^{6}|}|,} \\ 
\ _{\shortmid }^{3}\alpha _{i_{2}}=(\mathbf{\ ^{\shortmid }}\partial ^{5}%
\mathbf{\ ^{\shortmid }}h^{6})\ (\partial _{i_{2}}\ _{\shortmid }^{3}\varpi
), \\ 
\ _{\shortmid }^{3}\beta =(\mathbf{\ ^{\shortmid }}\partial ^{5}\mathbf{\
^{\shortmid }}h^{6})\ (\mathbf{\ ^{\shortmid }}\partial ^{5}\ _{\shortmid
}^{3}\varpi ),\  \\ 
\ \ _{\shortmid }^{3}\gamma =\mathbf{\ ^{\shortmid }}\partial ^{5}\left( \ln
|\mathbf{\ ^{\shortmid }}h^{6}|^{3/2}/|\mathbf{\ ^{\shortmid }}h^{5}|\right)
,%
\end{array}%
$ \\ 
$%
\begin{array}{c}
\mathbf{\ ^{\shortmid }}\partial ^{7}(\ _{\shortmid }^{4}\varpi )\ \mathbf{\
^{\shortmid }}\partial ^{7}\ \mathbf{\ ^{\shortmid }}h^{8}=2\mathbf{\
^{\shortmid }}h^{7}\mathbf{\ ^{\shortmid }}h^{8}\ \ _{4}^{\shortmid Q}J; \\ 
\ _{\shortmid }^{4}\beta \ _{\shortmid }^{4}w_{i_{3}}-\ _{\shortmid
}^{4}\alpha _{i_{3}}=0; \\ 
\mathbf{\ ^{\shortmid }}\partial ^{7}(\mathbf{\ ^{\shortmid }}\partial ^{7}\
_{\shortmid }^{4}n_{k_{3}})+\ _{\shortmid }^{4}\gamma \mathbf{\ ^{\shortmid }%
}\partial ^{7}(\ _{\shortmid }^{4}n_{k_{3}})=0;%
\end{array}%
$ &  & $%
\begin{array}{c}
\\ 
\ _{\shortmid }^{4}\varpi {=\ln |\mathbf{\ ^{\shortmid }}\partial ^{7}%
\mathbf{\ ^{\shortmid }}h^{8}/\sqrt{|\mathbf{\ ^{\shortmid }}h^{7}\mathbf{\
^{\shortmid }}h^{8}|}|,} \\ 
\ _{\shortmid }^{4}\alpha _{i_{3}}=(\mathbf{\ ^{\shortmid }}\partial ^{7}%
\mathbf{\ ^{\shortmid }}h^{8})\ (\mathbf{\ ^{\shortmid }}\partial _{i_{3}}\
_{\shortmid }^{4}\varpi ), \\ 
\ _{\shortmid }^{4}\beta =(\mathbf{\ ^{\shortmid }}\partial ^{7}\mathbf{\
^{\shortmid }}h^{8})\ (\mathbf{\ ^{\shortmid }}\partial ^{7}\ _{\shortmid
}^{4}\varpi ),\  \\ 
\ \ _{\shortmid }^{4}\gamma =\mathbf{\ ^{\shortmid }}\partial ^{7}\left( \ln
|\mathbf{\ ^{\shortmid }}h^{8}|^{3/2}/|\mathbf{\ ^{\shortmid }}h^{7}|\right)
,%
\end{array}%
$%
\end{tabular}%
$ \\ \hline
$%
\begin{array}{c}
\mbox{ Gener.  functs:}\ h_{3}(x^{k_{1}},y^{3}), \\ 
\ ^{2}\Psi (x^{k_{1}},y^{3})=e^{\ ^{2}\varpi },\ ^{2}\Phi (x^{k_{1}},y^{3}),
\\ 
\mbox{integr. functs:}\ h_{4}^{[0]}(x^{k_{1}}),\  \\ 
_{1}n_{k_{1}}(x^{i_{1}}),\ _{2}n_{k_{1}}(x^{i_{1}}); \\ 
\mbox{ Gener.  functs:}\mathbf{\ ^{\shortmid }}h^{5}(x^{k_{2}},p_{5}), \\ 
\ \ _{\shortmid }^{3}\Psi (x^{k_{2}},p_{5})=e^{\ \ _{\shortmid }^{3}\varpi
},\ \ \ _{\shortmid }^{3}\Phi (x^{k_{2}},p_{5}), \\ 
\mbox{integr. functs:}\ h_{6}^{[0]}(x^{k_{2}}),\  \\ 
_{1}^{3}n_{k_{2}}(x^{i_{2}}),\ _{2}^{3}n_{k_{2}}(x^{i_{2}}); \\ 
\mbox{ Gener.  functs:}\mathbf{\ ^{\shortmid }}h^{7}(\mathbf{\ ^{\shortmid }}%
x^{k_{3}},p_{7}), \\ 
\ \ _{\shortmid }^{4}\Psi (x^{k_{2}},p_{7})=e^{\ \ \ _{\shortmid }^{4}\varpi
},\ \ \ _{\shortmid }^{4}\Phi (\mathbf{\ ^{\shortmid }}x^{k_{3}},p_{7}), \\ 
\mbox{integr. functs:}\ h_{8}^{[0]}(\mathbf{\ ^{\shortmid }}x^{k_{3}}),\  \\ 
_{1}^{4}n_{k_{3}}(\mathbf{\ ^{\shortmid }}x^{i_{3}}),\ _{2}^{4}n_{k_{3}}(%
\mathbf{\ ^{\shortmid }}x^{i_{3}}); \\ 
\mbox{\& nonlinear symmetries}%
\end{array}%
$ &  & $%
\begin{array}{c}
\ ((\ ^{2}\Psi )^{2})^{\ast _{2}}=-\int dy^{3}\ _{2}^{Q}Jh_{4}^{\ \ast _{2}},
\\ 
(\ ^{2}\Phi )^{2}=-4\ _{2}\Lambda h_{4},\mbox{ see }(\ref{nonlinsymrex}), \\ 
h_{4}=h_{4}^{[0]}-(\ ^{2}\Phi )^{2}/4\ _{2}\Lambda ,h_{4}^{\ast }\neq 0,\
_{2}\Lambda \neq 0=const; \\ 
\\ 
\mathbf{\ ^{\shortmid }}\partial ^{5}((\ \ _{\shortmid }^{3}\Psi
)^{2})=-\int dp_{5}\ \ \ _{3}^{\shortmid Q}J\mathbf{\ ^{\shortmid }}\partial
^{5}\mathbf{\ ^{\shortmid }}h^{6}, \\ 
(\ \ _{\shortmid }^{3}\Phi )^{2}=-4\ _{3}^{\shortmid }\Lambda \mathbf{\
^{\shortmid }}h^{6}, \\ 
\mathbf{\ ^{\shortmid }}h^{6}=\mathbf{\ ^{\shortmid }}h_{[0]}^{6}-(\ \
_{\shortmid }^{3}\Phi )^{2}/4\ _{3}\Lambda ,\mathbf{\ ^{\shortmid }}\partial
^{5}\mathbf{\ ^{\shortmid }}h^{6}\neq 0,\ _{3}^{\shortmid }\Lambda \neq
0=const; \\ 
\\ 
\mathbf{\ ^{\shortmid }}\partial ^{7}((\ _{\shortmid }^{4}\Psi )^{2})=-\int
dp_{7}\ \ \ _{4}^{\shortmid Q}J\mathbf{\ ^{\shortmid }}\partial ^{7}\mathbf{%
\ ^{\shortmid }}h^{8}, \\ 
(\ _{\shortmid }^{4}\Phi )^{2}=-4\ _{4}^{\shortmid }\Lambda \mathbf{\
^{\shortmid }}h^{8}, \\ 
\mathbf{\ ^{\shortmid }}h^{8}=\mathbf{\ ^{\shortmid }}h_{[0]}^{8}-(\
_{\shortmid }^{4}\Phi )^{2}/4\ _{4}^{\shortmid }\Lambda ,\mathbf{\
^{\shortmid }}\partial ^{7}\mathbf{\ ^{\shortmid }}h^{8}\neq 0,\
_{4}^{\shortmid }\Lambda \neq 0=const;%
\end{array}%
$ \\ \hline
Off-diag. solutions, $%
\begin{array}{c}
\mbox{d--metric} \\ 
\mbox{N-connec.}%
\end{array}%
$ &  & $%
\begin{tabular}{l}
$%
\begin{array}{c}
\ g_{i}=e^{\ \psi (x^{k})}\mbox{ as a solution of 2-d Poisson eqs. }\psi
^{\bullet \bullet }+\psi ^{\prime \prime }=2~\ _{1}^{Q}J; \\ 
h_{3}=-(\ ^{2}\Psi ^{\ast _{2}})^{2}/4\ \ _{2}^{Q}J^{2}h_{4},\mbox{ see }(%
\ref{g3}),(\ref{g4}); \\ 
h_{4}=h_{4}^{[0]}-\int dy^{3}(\ ^{2}\Psi ^{2})^{\ast _{2}}/4\ \ \
_{2}^{Q}J=h_{4}^{[0]}-\ ^{2}\Phi ^{2}/4\ _{2}\Lambda ; \\ 
w_{i_{2}}=\partial _{i_{2}}\ \ ^{2}\Psi /\ \partial _{3}\ ^{2}\Psi =\partial
_{i_{2}}\ \ ^{2}\Psi ^{2}/\ \partial _{3}\ ^{2}\Psi ^{2}|; \\ 
n_{k}=\ _{1}n_{k}+\ _{2}n_{k}\int dy^{3}(\ ^{2}\Psi ^{\ast _{2}})^{2}/\ \ \
_{2}^{Q}J^{2}|h_{4}^{[0]}-\int dy^{3}(\ ^{2}\Psi ^{2})^{\ast _{2}}/4\ \ \
_{2}^{Q}J^{2}|^{5/2};%
\end{array}%
$ \\ 
$%
\begin{array}{c}
\mathbf{\ ^{\shortmid }}h^{5}=-(\mathbf{\ ^{\shortmid }}\partial ^{5}\
_{\shortmid }^{3}\Psi )^{2}/4\ _{3}^{\shortmid Q}J^{2}\mathbf{\ ^{\shortmid }%
}h^{6}; \\ 
\mathbf{\ ^{\shortmid }}h^{6}=\mathbf{\ ^{\shortmid }}h_{[0]}^{6}-\int dp_{5}%
\mathbf{\ ^{\shortmid }}\partial ^{5}((\ \ _{\shortmid }^{3}\Psi )^{2})/4\ \
\ _{3}^{\shortmid Q}J=\mathbf{\ ^{\shortmid }}h_{[0]}^{6}-(\ \ _{\shortmid
}^{3}\Phi )^{2}/4\ _{3}^{\shortmid }\Lambda ; \\ 
w_{i_{2}}=\partial _{i_{2}}(\ _{\shortmid }^{3}\Psi )/\mathbf{\ ^{\shortmid }%
}\partial ^{5}(\ _{\shortmid }^{3}\Psi )=\partial _{i_{2}}(\ _{\shortmid
}^{3}\Psi )^{2}/\ \mathbf{\ ^{\shortmid }}\partial ^{5}(\ _{\shortmid
}^{3}\Psi )^{2}|; \\ 
n_{k_{2}}=\ _{1}n_{k_{2}}+\ _{2}n_{k_{2}}\int dp_{5}(\mathbf{\ ^{\shortmid }}%
\partial ^{5}\ _{\shortmid }^{3}\Psi )^{2}/\ \ _{3}^{\shortmid Q}J^{2}|%
\mathbf{\ ^{\shortmid }}h_{[0]}^{6}- \\ 
\int dp_{5}\mathbf{\ ^{\shortmid }}\partial ^{5}((\ _{\shortmid }^{3}\Psi
)^{2})/4\ \ _{3}^{\shortmid Q}J^{2}|^{5/2}%
\end{array}%
$ \\ 
$%
\begin{array}{c}
\mathbf{\ ^{\shortmid }}h^{7}=-(\mathbf{\ ^{\shortmid }}\partial ^{7}\
_{\shortmid }^{4}\Psi )^{2}/4\ _{4}^{\shortmid Q}J^{2}\mathbf{\ ^{\shortmid }%
}h^{8}; \\ 
\mathbf{\ ^{\shortmid }}h^{8}=\mathbf{\ ^{\shortmid }}h_{[0]}^{8}-\int dp_{7}%
\mathbf{\ ^{\shortmid }}\partial ^{7}((\ _{\shortmid }^{4}\Psi )^{2})/4\
_{4}^{\shortmid Q}J=h_{8}^{[0]}-(\ \ _{\shortmid }^{4}\Phi )^{2}/4\ \
_{4}^{\shortmid }\Lambda ; \\ 
\mathbf{\ ^{\shortmid }}w_{i_{3}}=\mathbf{\ ^{\shortmid }}\partial
_{i_{3}}(\ \ _{\shortmid }^{4}\Psi )/\ \mathbf{\ ^{\shortmid }}\partial
^{7}(\ _{\shortmid }^{4}\Psi )=\mathbf{\ ^{\shortmid }}\partial _{i_{3}}(\ \
_{\shortmid }^{4}\Psi )^{2}/\ \mathbf{\ ^{\shortmid }}\partial ^{7}(\
_{\shortmid }^{4}\Psi )^{2}|; \\ 
\mathbf{\ ^{\shortmid }}n_{k_{3}}=\ _{1}^{\shortmid }n_{k_{3}}+\
_{2}^{\shortmid }n_{k_{3}}\int dp_{7}(\ _{\shortmid }^{4}\Psi )^{2}/\ \
_{4}^{\shortmid Q}J^{2}|h_{8}^{[0]}-\int dp_{7}\mathbf{\ ^{\shortmid }}%
\partial ^{7}((\ \ _{\shortmid }^{4}\Psi )^{2})/4\ \ _{4}^{\shortmid
Q}J^{2}|^{5/2}%
\end{array}%
$%
\end{tabular}%
$ \\ \hline\hline
\end{tabular}%
\end{eqnarray*}%
}

As a FH $T^{\ast }\mathbf{V}$ analogue of the nonlinear quadratic element (%
\ref{qst8d7}) for FLH configurations, with $v^{8}=const,$ and data from
Table 8 we provide an example of 8-d quasi-stationary quadratic element with 
$p_{8}=E=const,$ 
\begin{eqnarray}
d\widehat{s}_{[8d]}^{2}(\tau ) &=&\widehat{g}_{\alpha _{s}\beta
_{s}}(x^{k},y^{3},p_{5},p_{7};h_{4},\mathbf{\ ^{\shortmid }}h^{6},\mathbf{\
^{\shortmid }}h^{8};\ _{s}^{\shortmid Q}J\ ;\ _{s}^{\shortmid }\Lambda (\tau
))d\mathbf{\ ^{\shortmid }}u^{\alpha _{s}}d\mathbf{\ ^{\shortmid }}u^{\beta
_{s}}  \label{qstd8d7} \\
&=&e^{\psi (x^{k},\ \ _{1}^{Q}J\ )}[(dx^{1})^{2}+(dx^{2})^{2}]-\frac{%
(h_{4}^{\ast _{2}})^{2}}{|\int dy^{3}[\ \ _{2}^{Q}Jh_{4}]^{\ast _{2}}|\ h_{4}%
}\{dy^{3}+\frac{\partial _{i_{1}}[\int dy^{3}\ _{2}^{Q}J\ h_{4}^{\ast _{2}}]%
}{\ \ _{2}^{Q}J\ h_{4}^{\ast _{2}}}dx^{i_{1}}\}^{2}+  \notag \\
&&h_{4}\{dt+[\ _{1}n_{k_{1}}+\ _{2}n_{k_{1}}\int dy^{3}\frac{(h_{4}^{\ast
_{2}})^{2}}{|\int dy^{3}[\ \ _{2}^{Q}Jh_{4}]^{\ast _{2}}|\ (h_{4})^{5/2}}%
]dx^{k_{1}}\}+  \notag \\
&&\frac{(\mathbf{\ ^{\shortmid }}\partial ^{5}\mathbf{\ ^{\shortmid }}%
h^{6})^{2}}{|\int dp_{5}\mathbf{\ ^{\shortmid }}\partial ^{5}[\ \
_{3}^{\shortmid Q}J\mathbf{\ ^{\shortmid }}h^{6}]|\ \mathbf{\ ^{\shortmid }}%
h^{6}}\{dp_{5}+\frac{\partial _{i_{2}}[\int dp_{5}\ _{3}^{\shortmid Q}J%
\mathbf{\ ^{\shortmid }}\partial ^{5}\mathbf{\ ^{\shortmid }}h^{6}]}{\ \
_{3}^{\shortmid Q}J\mathbf{\ ^{\shortmid }}\partial ^{5}\ \mathbf{%
^{\shortmid }}h^{6}}dx^{i_{2}}\}^{2}+  \notag \\
&&\mathbf{\ ^{\shortmid }}h^{6}\{dp_{5}+[\ _{1}n_{k_{2}}+\ _{2}n_{k_{2}}\int
dp_{5}\frac{(\mathbf{\ ^{\shortmid }}\partial ^{5}\mathbf{\ ^{\shortmid }}%
h^{6})^{2}}{|\int dp_{5}\mathbf{\ ^{\shortmid }}\partial ^{5}[\ \
_{3}^{\shortmid Q}J\mathbf{\ ^{\shortmid }}h^{6}]|\ (\mathbf{\ ^{\shortmid }}%
h^{6})^{5/2}}]dx^{k_{2}}\}+  \notag \\
&&\frac{(\mathbf{\ ^{\shortmid }}\partial ^{7}\mathbf{\ ^{\shortmid }}%
h^{8})^{2}}{|\int dp_{7}\mathbf{\ ^{\shortmid }}\partial ^{7}[\ \ \
_{4}^{\shortmid Q}J\mathbf{\ ^{\shortmid }}h^{8}]|\ \mathbf{\ ^{\shortmid }}%
h^{8}}\{dp_{7}+\frac{\partial _{i_{3}}[\int dp_{7}\ \ _{4}^{\shortmid Q}J\ 
\mathbf{\ ^{\shortmid }}\partial ^{7}\mathbf{\ ^{\shortmid }}h^{8}]}{\ \ \
_{4}^{\shortmid Q}J\ \mathbf{\ ^{\shortmid }}\partial ^{7}\mathbf{\
^{\shortmid }}h^{8}}d\mathbf{\ ^{\shortmid }}x^{i_{3}}\}^{2}+  \notag \\
&&\mathbf{\ ^{\shortmid }}h^{8}\{dE+[\ _{1}^{\shortmid }n_{k_{3}}+\
_{2}^{\shortmid }n_{k_{3}}\int dp_{7}\frac{(\mathbf{\ ^{\shortmid }}\partial
^{7}\mathbf{\ ^{\shortmid }}h^{8})^{2}}{|\int dp_{7}\mathbf{\ ^{\shortmid }}%
\partial ^{7}[\ \ _{4}^{\shortmid Q}J\mathbf{\ ^{\shortmid }}h^{8}]|\ (%
\mathbf{\ ^{\shortmid }}h^{8})^{5/2}}]d\mathbf{\ ^{\shortmid }}x^{k_{3}}\}. 
\notag
\end{eqnarray}%
Such $\tau $-families of s-metrics possess nonlinear symmetries in phase
spaces, which allow us to redefine the generating functions and generating
sources and related them to conventional $\tau $-running cosmological
constants $\ _{s}^{\shortmid }\Lambda (\tau ).$ 

\subsubsection{Quasi-stationary and rainbow phase space solutions}

Chronologically, rainbow s-metrics in generalized Finsler-Lagrange and dual
Cartan-Hamilton forms were constructed following different nonholonomic
parameterizations in \cite{vmon3,vacaru18}. The cosmological scenarios can
be re-defined on $T_{s}^{\ast }V$ and exploited as some alternative models
of dark matter and dark energy theories when the structure formation and
phase space dynamics depend on certain $E$ type variables/ coordinates.

\newpage

{\scriptsize 
\begin{eqnarray*}
&&%
\begin{tabular}{l}
\hline\hline
\begin{tabular}{lll}
& {\large \textsf{Table 11:\ Off-diagonal nonmetric quasi-stationary pase
space configurations with variable energy}} &  \\ 
& Exact solutions of $\ ^{\shortmid }\widehat{\mathbf{R}}_{\ \ \gamma
_{s}}^{\beta _{s}}(\tau )={\delta }_{\ \ \gamma _{s}}^{\beta _{s}}\ \
_{Q}^{s\shortmid }\mathbf{J}(\tau )\mathbf{\ }$ (\ref{feq4afd}) on $%
T_{s}^{\ast }V$ transformed into a momentum version of nonlinear PDEs (\ref%
{eq1})-(\ref{e2c}) & 
\end{tabular}
\\ 
\end{tabular}
\\
&&%
\begin{tabular}{lll}
\hline\hline
&  &  \\ 
$%
\begin{array}{c}
\mbox{d-metric ansatz with} \\ 
\mbox{Killing symmetry }\partial _{4}=\partial _{t},\mathbf{\ ^{\shortmid }}%
\partial ^{7}%
\end{array}%
$ &  & $%
\begin{array}{c}
ds^{2}=g_{i_{1}}(x^{k_{1}})(dx^{i_{1}})^{2}+g_{a_{2}}(x^{k_{1}},y^{3})(dy^{a_{2}}+N_{i_{1}}^{a_{2}}(x^{k_{1}},y^{3})dx^{i_{1}})^{2}
\\ 
+\mathbf{\ ^{\shortmid }}g^{a_{3}}(x^{k_{2}},p_{5})(dp_{a_{3}}+\mathbf{\
^{\shortmid }}N_{i_{2}a_{3}}(x^{k_{2}},p_{5})dx^{i_{2}})^{2} \\ 
+\mathbf{\ ^{\shortmid }}g^{a_{4}}(\mathbf{\ ^{\shortmid }}%
x^{k_{3}},p_{7})(dp_{a_{4}}+\mathbf{\ ^{\shortmid }}N_{i_{3}a_{4}}(\mathbf{\
^{\shortmid }}x^{k_{3}},p_{7})d\mathbf{\ ^{\shortmid }}x^{i_{3}})^{2},%
\mbox{
for }g_{i_{1}}=e^{\psi {(x}^{k_{1}}{)}}, \\ 
g_{a_{2}}=h_{a_{2}}(x^{k_{1}},y^{3}),N_{i_{1}}^{3}=\
^{2}w_{i_{1}}=w_{i_{1}}(x^{k_{1}},y^{3}), \\ 
N_{i_{1}}^{4}=\ ^{2}n_{i_{1}}=n_{i_{1}}(x^{k_{1}},y^{3}), \\ 
\mathbf{\ ^{\shortmid }}g^{a_{3}}=\mathbf{\ ^{\shortmid }}%
h^{a_{3}}(x^{k_{2}},p_{5}),\mathbf{\ ^{\shortmid }}N_{i_{2}5}=\ _{\shortmid
}^{3}w_{i_{2}}=\mathbf{\ ^{\shortmid }}w_{i_{2}}(x^{k_{2}},p_{5}), \\ 
\mathbf{\ ^{\shortmid }}N_{i_{2}6}=\ _{\shortmid }^{3}n_{i_{2}}=\mathbf{\
^{\shortmid }}n_{i_{2}}(x^{k_{2}},p_{5}), \\ 
\mathbf{\ ^{\shortmid }}\underline{g}^{a_{4}}=\mathbf{\ ^{\shortmid }}%
\underline{h}^{a_{4}}(\mathbf{\ ^{\shortmid }}x^{k_{3}},E),\mathbf{\
^{\shortmid }}\underline{N}_{i_{3}7}=\ \ _{\shortmid }^{4}\underline{n}%
_{i_{3}}=\mathbf{\ ^{\shortmid }}\underline{n}_{i_{3}}(x^{k_{3}},E), \\ 
\mathbf{\ ^{\shortmid }}\underline{N}_{i_{3}8}=\ _{\shortmid }^{4}\underline{%
w}_{i_{3}}=\mathbf{\ ^{\shortmid }}\underline{w}_{i_{3}}(x^{k_{3}},E),%
\end{array}%
$ \\ 
Effective matter sources &  & $\mathbf{\ _{Q}^{\shortmid }J}_{\ \nu
_{s}}^{\mu _{s}}(\tau )\mathbf{\ }=[\ _{1}^{Q}J({x}^{k_{1}})\delta
_{j_{1}}^{i_{1}},\ \ _{2}^{Q}J({x}^{k_{1}},y^{3})\delta _{b_{2}}^{a_{2}},\
_{3}^{\shortmid Q}J({x}^{k_{2}},p_{5})\delta _{b_{3}}^{a_{3}},\
_{4}^{\shortmid Q}\underline{J}({x}^{k_{3}},E)\delta _{b_{4}}^{a_{4}}]$ \\ 
\hline
Nonlinear PDEs (\ref{eq1})-(\ref{e2c}) &  & $%
\begin{tabular}{lll}
$%
\begin{array}{c}
\psi ^{\bullet \bullet }+\psi ^{\prime \prime }=2\ _{1}^{Q}J; \\ 
\ ^{2}\varpi ^{\ast }\ h_{4}^{\ast _{2}}=2h_{3}h_{4}\ _{2}^{Q}J; \\ 
\ ^{2}\beta \ ^{2}w_{i_{1}}-\ ^{2}\alpha _{i_{1}}=0; \\ 
\ ^{2}n_{k_{1}}^{\ast _{2}\ast _{2}}+\ ^{2}\gamma \ ^{2}n_{k_{1}}^{\ast
_{2}}=0;%
\end{array}%
$ &  & $%
\begin{array}{c}
\ ^{2}\varpi {=\ln |\partial _{3}h_{4}/\sqrt{|h_{3}h_{4}|}|,} \\ 
\ ^{2}\alpha _{i_{1}}=(\partial _{3}h_{4})\ (\partial _{i_{1}}\ ^{2}\varpi ),
\\ 
\ ^{2}\beta =(\partial _{3}h_{4})\ (\partial _{3}\ ^{2}\varpi ),\  \\ 
\ \ ^{2}\gamma =\partial _{3}\left( \ln |h_{4}|^{3/2}/|h_{3}|\right) , \\ 
\partial _{1}q=q^{\bullet },\partial _{2}q=q^{\prime },\partial
_{3}q=q^{\ast _{2}}%
\end{array}%
$ \\ 
$%
\begin{array}{c}
\mathbf{\ ^{\shortmid }}\partial ^{5}(\ _{\shortmid }^{3}\varpi )\ \mathbf{\
^{\shortmid }}\partial ^{5}\mathbf{\ ^{\shortmid }}h^{6}=2\mathbf{\
^{\shortmid }}h^{5}\mathbf{\ ^{\shortmid }}h^{6}\ \ _{3}^{\shortmid Q}J; \\ 
\ _{\shortmid }^{3}\beta \ _{\shortmid }^{3}w_{i_{2}}-\ _{\shortmid
}^{3}\alpha _{i_{2}}=0; \\ 
\mathbf{\ ^{\shortmid }}\partial ^{5}(\mathbf{\ ^{\shortmid }}\partial ^{5}\
_{\shortmid }^{3}n_{k_{2}})+\ _{\shortmid }^{3}\gamma \mathbf{\ ^{\shortmid }%
}\partial ^{5}(\ _{\shortmid }^{3}n_{k_{2}})=0;%
\end{array}%
$ &  & $%
\begin{array}{c}
\\ 
\ _{\shortmid }^{3}\varpi {=\ln |\mathbf{\ ^{\shortmid }}\partial ^{5}%
\mathbf{\ ^{\shortmid }}h^{6}/\sqrt{|\mathbf{\ ^{\shortmid }}h^{5}\mathbf{\
^{\shortmid }}h^{6}|}|,} \\ 
\ _{\shortmid }^{3}\alpha _{i_{2}}=(\mathbf{\ ^{\shortmid }}\partial ^{5}%
\mathbf{\ ^{\shortmid }}h^{6})\ (\partial _{i_{2}}\ _{\shortmid }^{3}\varpi
), \\ 
\ _{\shortmid }^{3}\beta =(\mathbf{\ ^{\shortmid }}\partial ^{5}\mathbf{\
^{\shortmid }}h^{6})\ (\mathbf{\ ^{\shortmid }}\partial ^{5}\ _{\shortmid
}^{3}\varpi ),\  \\ 
\ \ _{\shortmid }^{3}\gamma =\mathbf{\ ^{\shortmid }}\partial ^{5}\left( \ln
|\mathbf{\ ^{\shortmid }}h^{6}|^{3/2}/|\mathbf{\ ^{\shortmid }}h^{5}|\right)
,%
\end{array}%
$ \\ 
$%
\begin{array}{c}
\mathbf{\ ^{\shortmid }}\underline{\partial }^{8}(\ _{\shortmid }^{4}%
\underline{\varpi })\ \mathbf{\ ^{\shortmid }}\underline{\partial }^{8}%
\mathbf{\ ^{\shortmid }}\underline{h}^{7}=2\ \mathbf{^{\shortmid }}%
\underline{h}^{7}\mathbf{\ ^{\shortmid }}\underline{h}^{8}\ _{4}^{\shortmid
Q}\underline{J}; \\ 
\mathbf{\ ^{\shortmid }}\underline{\partial }^{8}(\mathbf{\ ^{\shortmid }}%
\underline{\partial }^{8}\ _{\shortmid }^{4}\underline{n}_{k_{3}})+\
_{\shortmid }^{4}\underline{\gamma }\mathbf{\ ^{\shortmid }}\underline{%
\partial }^{8}(\ _{\shortmid }^{4}\underline{n}_{k_{3}})=0; \\ 
\ _{\shortmid }^{4}\underline{\beta }\ _{\shortmid }^{4}\underline{w}%
_{i_{3}}-\ _{\shortmid }^{4}\underline{\alpha }_{i_{3}}=0;%
\end{array}%
$ &  & $%
\begin{array}{c}
\\ 
\ _{\shortmid }^{4}\underline{\varpi }{=\ln |\mathbf{\ ^{\shortmid }}%
\underline{{\partial }}^{8}\mathbf{\ ^{\shortmid }}\underline{{h}}^{7}/\sqrt{%
|\mathbf{\ ^{\shortmid }}\underline{h}^{7}\mathbf{\ ^{\shortmid }}\underline{%
h}^{8}|}|,} \\ 
\ _{\shortmid }^{4}\underline{\alpha }_{i_{3}}=(\mathbf{\ ^{\shortmid }}%
\underline{\partial }^{8}\mathbf{\ ^{\shortmid }}\underline{h}^{7})\ (%
\mathbf{\ ^{\shortmid }}\partial _{i_{3}}\ _{\shortmid }^{4}\underline{%
\varpi }), \\ 
\ _{\shortmid }^{4}\underline{\beta }=(\mathbf{\ ^{\shortmid }}\underline{%
\partial }^{8}\mathbf{\ ^{\shortmid }}\underline{h}^{7})\ (\mathbf{\
^{\shortmid }}\underline{\partial }^{8}\ _{\shortmid }^{4}\underline{\varpi }%
),\  \\ 
\ \ _{\shortmid }^{4}\underline{\gamma }=\mathbf{\ ^{\shortmid }}\underline{%
\partial }^{8}\left( \ln |\mathbf{\ ^{\shortmid }}\underline{h}^{7}|^{3/2}/|%
\mathbf{\ ^{\shortmid }}\underline{h}^{8}|\right) ,%
\end{array}%
$%
\end{tabular}%
$ \\ \hline
$%
\begin{array}{c}
\mbox{ Gener.  functs:}\ h_{3}(x^{k_{1}},y^{3}), \\ 
\ ^{2}\Psi (x^{k_{1}},y^{3})=e^{\ ^{2}\varpi },\ ^{2}\Phi (x^{k_{1}},y^{3}),
\\ 
\mbox{integr. functs:}\ h_{4}^{[0]}(x^{k_{1}}),\  \\ 
_{1}n_{k_{1}}(x^{i_{1}}),\ _{2}n_{k_{1}}(x^{i_{1}}); \\ 
\mbox{ Gener.  functs:}\mathbf{\ ^{\shortmid }}h^{5}(x^{k_{2}},p_{5}), \\ 
\ \ _{\shortmid }^{3}\Psi (x^{k_{2}},p_{5})=e^{\ \ _{\shortmid }^{3}\varpi
},\ \ \ _{\shortmid }^{3}\Phi (x^{k_{2}},p_{5}), \\ 
\mbox{integr. functs:}\ h_{6}^{[0]}(x^{k_{2}}),\  \\ 
_{1}^{3}n_{k_{2}}(x^{i_{2}}),\ _{2}^{3}n_{k_{2}}(x^{i_{2}}); \\ 
\mbox{ Gener.  functs:}\mathbf{\ ^{\shortmid }}h^{7}(\mathbf{\ ^{\shortmid }}%
x^{k_{3}},p_{7}), \\ 
\ \ _{\shortmid }^{4}\underline{\Psi }(x^{k_{2}},E)=e^{\ \ \ _{\shortmid
}^{4}\underline{\varpi }},\ \ \ _{\shortmid }^{4}\underline{\Phi }(\mathbf{\
^{\shortmid }}x^{k_{3}},E), \\ 
\mbox{integr. functs:}\ \underline{h}_{7}^{[0]}(\mathbf{\ ^{\shortmid }}%
x^{k_{3}}),\  \\ 
_{1}^{4}\underline{n}_{k_{3}}(\mathbf{\ ^{\shortmid }}x^{i_{3}}),\ _{2}^{4}%
\underline{n}_{k_{3}}(\mathbf{\ ^{\shortmid }}x^{i_{3}}); \\ 
\mbox{\& nonlinear symmetries}%
\end{array}%
$ &  & $%
\begin{array}{c}
\ ((\ ^{2}\Psi )^{2})^{\ast _{2}}=-\int dy^{3}\ _{2}^{Q}Jh_{4}^{\ \ast _{2}},
\\ 
(\ ^{2}\Phi )^{2}=-4\ _{2}\Lambda h_{4},\mbox{ see }(\ref{nonlinsymrex}), \\ 
h_{4}=h_{4}^{[0]}-(\ ^{2}\Phi )^{2}/4\ _{2}\Lambda ,h_{4}^{\ast _{2}}\neq
0,\ _{2}\Lambda \neq 0=const; \\ 
\\ 
\mathbf{\ ^{\shortmid }}\partial ^{5}((\ \ _{\shortmid }^{3}\Psi
)^{2})=-\int dp_{5}\ \ _{3}^{\shortmid Q}J\mathbf{\ ^{\shortmid }}\partial
^{5}\mathbf{\ ^{\shortmid }}h^{6}, \\ 
(\ \ _{\shortmid }^{3}\Phi )^{2}=-4\ _{3}^{\shortmid }\Lambda \mathbf{\
^{\shortmid }}h^{6}, \\ 
\mathbf{\ ^{\shortmid }}h^{6}=\mathbf{\ ^{\shortmid }}h_{[0]}^{6}-(\ \
_{\shortmid }^{3}\Phi )^{2}/4\ _{3}\Lambda ,\mathbf{\ ^{\shortmid }}\partial
^{5}\mathbf{\ ^{\shortmid }}h^{6}\neq 0,\ _{3}^{\shortmid }\Lambda \neq
0=const; \\ 
\\ 
\mathbf{\ ^{\shortmid }}\underline{\partial }^{8}((\ _{\shortmid }^{4}%
\underline{\Psi })^{2})=-\int dE\ _{4}^{\shortmid Q}\underline{J}\mathbf{\
^{\shortmid }}\underline{\partial }^{8}\mathbf{\ ^{\shortmid }}\underline{h}%
^{7}, \\ 
(\ _{\shortmid }^{4}\underline{\Phi })^{2}=-4\ _{4}^{\shortmid }\underline{%
\Lambda }\mathbf{\ ^{\shortmid }}\underline{h}^{7}, \\ 
\mathbf{\ ^{\shortmid }}\underline{h}^{7}=\mathbf{\ ^{\shortmid }}\underline{%
h}_{[0]}^{7}-(\ _{\shortmid }^{4}\underline{\Phi })^{2}/4\ _{4}^{\shortmid }%
\underline{\Lambda },\mathbf{\ ^{\shortmid }}\underline{\partial }^{8}%
\mathbf{\ ^{\shortmid }}\underline{h}^{7}\neq 0,\ _{4}^{\shortmid }%
\underline{\Lambda }\neq 0=const;%
\end{array}%
$ \\ \hline
Off-diag. solutions, $%
\begin{array}{c}
\mbox{d--metric} \\ 
\mbox{N-connec.}%
\end{array}%
$ &  & $%
\begin{tabular}{l}
$%
\begin{array}{c}
\ g_{i}=e^{\ \psi (x^{k})}\mbox{ as a solution of 2-d Poisson eqs. }\psi
^{\bullet \bullet }+\psi ^{\prime \prime }=2~\ _{1}^{Q}J; \\ 
h_{3}=-(\ ^{2}\Psi ^{\ast _{2}})^{2}/4\ \ _{2}^{Q}J^{2}h_{4},\mbox{ see }(%
\ref{g3}),(\ref{g4}); \\ 
h_{4}=h_{4}^{[0]}-\int dy^{3}(\ ^{2}\Psi ^{2})^{\ast _{2}}/4\ \
_{2}^{Q}J=h_{4}^{[0]}-\ ^{2}\Phi ^{2}/4\ _{2}\Lambda ; \\ 
w_{i_{2}}=\partial _{i_{2}}\ ^{2}\Psi /\ \partial _{3}\ ^{2}\Psi =\partial
_{i_{2}}\ \ ^{2}\Psi ^{2}/\ \partial _{3}\ ^{2}\Psi ^{2}|; \\ 
n_{k_{2}}=\ _{1}n_{k_{2}}+\ _{2}n_{k_{2}}\int dy^{3}(\ ^{2}\Psi ^{\ast
_{2}})^{2}/\ \ _{2}^{Q}J^{2}|h_{4}^{[0]}-\int dy^{3}(\ ^{2}\Psi ^{2})^{\ast
_{2}}/4\ _{2}^{Q}J^{2}|^{5/2};%
\end{array}%
$ \\ 
$%
\begin{array}{c}
\mathbf{\ ^{\shortmid }}h^{5}=-(\mathbf{\ ^{\shortmid }}\partial ^{5}\
_{\shortmid }^{3}\Psi )^{2}/4\ \ _{3}^{\shortmid Q}J\mathbf{\ }^{2}\mathbf{\
^{\shortmid }}h^{6}; \\ 
\mathbf{\ ^{\shortmid }}h^{6}=\mathbf{\ ^{\shortmid }}h_{[0]}^{6}-\int dp_{5}%
\mathbf{\ ^{\shortmid }}\partial ^{5}((\ \ _{\shortmid }^{3}\Psi )^{2})/4\
_{3}^{\shortmid Q}J\mathbf{\ }=\mathbf{\ ^{\shortmid }}h_{[0]}^{6}-(\ \
_{\shortmid }^{3}\Phi )^{2}/4\ _{3}^{\shortmid }\Lambda ; \\ 
w_{i_{2}}=\partial _{i_{2}}(\ _{\shortmid }^{3}\Psi )/\mathbf{\ ^{\shortmid }%
}\partial ^{5}(\ _{\shortmid }^{3}\Psi )=\partial _{i_{2}}(\ _{\shortmid
}^{3}\Psi )^{2}/\ \mathbf{\ ^{\shortmid }}\partial ^{5}(\ _{\shortmid
}^{3}\Psi )^{2}|; \\ 
n_{k_{2}}=\ _{1}n_{k_{2}}+\ _{2}n_{k_{2}}\int dp_{5}(\mathbf{\ ^{\shortmid }}%
\partial ^{5}\ _{\shortmid }^{3}\Psi )^{2}/\ \ _{3}^{\shortmid Q}J\mathbf{\ }%
^{2}|\mathbf{\ ^{\shortmid }}h_{[0]}^{6}- \\ 
\int dp_{5}\mathbf{\ ^{\shortmid }}\partial ^{5}((\ _{\shortmid }^{3}\Psi
)^{2})/4\ \ _{3}^{\shortmid Q}J\mathbf{\ }^{2}|^{5/2}%
\end{array}%
$ \\ 
$%
\begin{array}{c}
\mathbf{\ ^{\shortmid }}\underline{h}^{8}=-(\mathbf{\ ^{\shortmid }}%
\underline{\partial }^{8}\ _{\shortmid }^{4}\underline{\Psi })^{2}/4\ \
_{4}^{\shortmid Q}\underline{J}^{2}\mathbf{\ ^{\shortmid }}\underline{h}^{7};
\\ 
\mathbf{\ ^{\shortmid }}\underline{h}^{7}=\mathbf{\ ^{\shortmid }}\underline{%
h}_{[0]}^{7}-\int dE\mathbf{\ ^{\shortmid }}\underline{\partial }^{8}((\
_{\shortmid }^{4}\underline{\Psi })^{2})/4\ \ _{4}^{\shortmid Q}\underline{J}%
=\underline{h}_{[0]}^{7}-(\ _{\shortmid }^{4}\underline{\Phi })^{2}/4\ \
_{4}^{\shortmid }\underline{\Lambda }; \\ 
\mathbf{\ ^{\shortmid }}\underline{n}_{k_{3}}=\ _{1}^{\shortmid }\underline{n%
}_{k_{3}}+\ _{2}^{\shortmid }\underline{n}_{k_{3}}\int dE(\ _{\shortmid }^{4}%
\underline{\Psi })^{2}/\ \ _{4}^{\shortmid Q}\underline{J}^{2}|\underline{h}%
_{[0]}^{7}- \\ 
\int dE\mathbf{\ ^{\shortmid }}\underline{\partial }^{8}((\ _{\shortmid }^{4}%
\underline{\Psi })^{2})/4\ \ _{4}^{\shortmid Q}\underline{J}^{2}|^{5/2}; \\ 
\mathbf{\ ^{\shortmid }}\underline{w}_{i_{3}}=\mathbf{\ ^{\shortmid }}%
\partial _{i_{3}}(\ \ _{\shortmid }^{4}\underline{\Psi })/\ \mathbf{\
^{\shortmid }}\partial ^{8}(\ _{\shortmid }^{4}\underline{\Psi })=\mathbf{\
^{\shortmid }}\partial _{i_{3}}(\ _{\shortmid }^{4}\underline{\Psi })^{2}/%
\mathbf{\ ^{\shortmid }}\partial ^{8}(\ _{\shortmid }^{4}\underline{\Psi }%
)^{2}|.%
\end{array}%
$%
\end{tabular}%
$ \\ \hline\hline
\end{tabular}%
\end{eqnarray*}%
}Corresponding s-metrics can be parameterized in the form: 
\begin{eqnarray}
d\widehat{s}_{[8d]}^{2}(\tau ) &=&\widehat{g}_{\alpha _{s}\beta
_{s}}(x^{k},y^{3},p_{5},E;h_{4},\ \mathbf{^{\shortmid }}h^{6},\mathbf{\
^{\shortmid }}\underline{h}^{7};\ _{1}^{Q}J,\ _{2}^{Q}J,\ _{3}^{\shortmid
Q}J,\ _{4}^{\shortmid Q}\underline{J};\ _{1}\Lambda ,\ _{2}\Lambda ,\
_{3}^{\shortmid }\Lambda ,\ _{4}^{\shortmid }\underline{\Lambda })d\mathbf{\
^{\shortmid }}u^{\alpha _{s}}d\mathbf{\ ^{\shortmid }}u^{\beta _{s}}
\label{qstrain8d8} \\
&=&e^{\psi (x^{k},\ \ _{1}^{Q}J)}[(dx^{1})^{2}+(dx^{2})^{2}]-\frac{%
(h_{4}^{\ast _{2}})^{2}}{|\int dy^{3}[\ _{2}^{Q}Jh_{4}]^{\ast _{2}}|\ h_{4}}%
\{dy^{3}+\frac{\partial _{i_{1}}[\int dy^{3}(\ _{2}^{Q}J)\ h_{4}^{\ast _{2}}]%
}{\ \ _{2}^{Q}J\ h_{4}^{\ast _{2}}}dx^{i_{1}}\}^{2}+  \notag \\
&&h_{4}\{dt+[\ _{1}n_{k_{1}}+\ _{2}n_{k_{1}}\int dy^{3}\frac{(h_{4}^{\ast
_{2}})^{2}}{|\int dy^{3}[\ _{2}^{Q}Jh_{4}]^{\ast _{2}}|\ (h_{4})^{5/2}}%
]dx^{k_{1}}\}+  \notag \\
&&\frac{(\mathbf{\ ^{\shortmid }}\partial ^{5}\mathbf{\ ^{\shortmid }}%
h^{6})^{2}}{|\int dp_{5}\mathbf{\ ^{\shortmid }}\partial ^{5}[\
_{3}^{\shortmid Q}J\mathbf{\ ^{\shortmid }}h^{6}]|\ \mathbf{\ ^{\shortmid }}%
h^{6}}\{dp_{5}+\frac{\partial _{i_{2}}[\int dp_{5}(\ \ _{3}^{\shortmid Q}J)%
\mathbf{\ ^{\shortmid }}\partial ^{5}\mathbf{\ ^{\shortmid }}h^{6}]}{\ \
_{3}^{\shortmid Q}J\mathbf{\ ^{\shortmid }}\partial ^{5}\ \mathbf{%
^{\shortmid }}h^{6}}dx^{i_{2}}\}^{2}+  \notag \\
&&\mathbf{\ ^{\shortmid }}h^{6}\{dp_{5}+[\ _{1}n_{k_{2}}+\ _{2}n_{k_{2}}\int
dp_{5}\frac{(\mathbf{\ ^{\shortmid }}\partial ^{5}\mathbf{\ ^{\shortmid }}%
h^{6})^{2}}{|\int dp_{5}\mathbf{\ ^{\shortmid }}\partial ^{5}[\ \
_{3}^{\shortmid Q}J\mathbf{\ ^{\shortmid }}h^{6}]|\ (\mathbf{\ ^{\shortmid }}%
h^{6})^{5/2}}]dx^{k_{2}}\}+  \notag \\
&&\mathbf{\ ^{\shortmid }}\underline{h}^{7}\{dp_{7}+[\ _{1}^{\shortmid }%
\underline{n}_{k_{3}}+\ _{2}^{\shortmid }\underline{n}_{k_{3}}\int dp_{7}%
\frac{(\mathbf{\ ^{\shortmid }}\underline{\partial }^{8}\mathbf{\
^{\shortmid }}\underline{h}^{7})^{2}}{|\int dE\mathbf{\ ^{\shortmid }}%
\underline{\partial }^{8}[\ \ _{4}^{\shortmid Q}\underline{J}\mathbf{\
^{\shortmid }}\underline{h}^{7}]|\ (\mathbf{\ ^{\shortmid }}\underline{h}%
^{7})^{5/2}}]d\mathbf{\ ^{\shortmid }}x^{k_{3}}\}+  \notag \\
&&\frac{(\mathbf{\ ^{\shortmid }}\underline{\partial }^{8}\mathbf{\
^{\shortmid }}\underline{h}^{7})^{2}}{|\int dE\mathbf{\ ^{\shortmid }}%
\underline{\partial }^{8}[\ _{4}^{\shortmid Q}\underline{J}\mathbf{\
^{\shortmid }}\underline{h}^{7}]|\ \mathbf{\ ^{\shortmid }}\underline{h}^{7}}%
\{dE+\frac{\partial _{i_{3}}[\int dE(\ \ _{4}^{\shortmid Q}\underline{J})\ 
\mathbf{\ ^{\shortmid }}\underline{\partial }^{8}\mathbf{\ ^{\shortmid }}%
\underline{h}^{7}]}{\ \ _{4}^{\shortmid Q}\underline{J}\mathbf{\ ^{\shortmid
}}\underline{\partial }^{8}\mathbf{\ ^{\shortmid }}\underline{h}^{7}}d%
\mathbf{\ ^{\shortmid }}x^{i_{3}}\}^{2}.  \notag
\end{eqnarray}%
Such $\tau $-families of rainbow s-metrics can be re-parameterized for other
types of generating functions and/or with gravitational polarization
functions using respective nonlinear symmetries and effective sources
encoding nonmetricity.

\subsubsection{Locally anisotropic nonmetric cosmological solutions with
fixed energy parameter}

For dual fiber to cofiber transforms, the procedure for constructing locally
anisotropic nonmetric cosmological phase space solutions described in Table
7 transforms into a method of generating such solutions with off-diagonal
dependence on momentum like variables. Such generalizations and applications
of the AFCDM are summarized in Table 12. As a $T^{\ast }\mathbf{V}$ analog
of the nonlinear quadratic element (\ref{qst8d7}), with $v^{8}=const$ and
data from Table 8, we provide an example of $\tau $-families of 8-d
quasi-stationary quadratic element with $p_{8}=E=const,$ 
\begin{eqnarray}
d\widehat{s}_{[8d]}^{2}(\tau ) &=&\widehat{g}_{\alpha _{s}\beta
_{s}}(x^{k},t,p_{5},p_{7};\underline{h}_{3},\mathbf{\ ^{\shortmid }}h^{6},%
\mathbf{\ ^{\shortmid }}h^{8};\ _{1}^{Q}J,\ _{2}^{Q}\underline{J},\
_{3}^{Q}J,\ _{4}^{Q}J;\ _{1}\Lambda ,\ _{2}\underline{\Lambda },\
_{3}^{\shortmid }\Lambda ,\ _{4}^{\shortmid }\Lambda )d\mathbf{\ ^{\shortmid
}}u^{\alpha _{s}}d\mathbf{\ ^{\shortmid }}u^{\beta _{s}}  \label{lc8cstp7} \\
&=&e^{\psi (x^{k},\ _{1}^{Q}J)}[(dx^{1})^{2}+(dx^{2})^{2}]+\underline{h}%
_{3}[dy^{3}+(\ _{1}n_{k_{1}}+4\ _{2}n_{k_{1}}\int dt\frac{(\underline{h}%
_{3}{}^{\diamond _{2}})^{2}}{|\int dt\ _{2}^{Q}\underline{J}\underline{h}%
_{3}{}^{\diamond _{2}}|(\underline{h}_{3})^{5/2}})dx^{k_{1}}]  \notag \\
&&-\frac{(\underline{h}_{3}{}^{\diamond _{2}})^{2}}{|\int dt\ _{2}^{Q}%
\underline{J}\underline{h}_{3}{}^{\diamond _{2}}|\ \underline{h}_{3}}[dt+%
\frac{\partial _{i_{1}}(\int dt\ \ _{2}^{Q}\underline{J}\ \underline{h}%
_{3}{}^{\diamond _{2}}])}{\ \ _{2}^{Q}\underline{J}\ \underline{h}%
_{3}{}^{\diamond _{2}}}dx^{i_{1}}]+  \notag \\
&&\frac{(\mathbf{\ ^{\shortmid }}\partial ^{5}\mathbf{\ ^{\shortmid }}%
h^{6})^{2}}{|\int dp_{5}\mathbf{\ ^{\shortmid }}\partial ^{5}[\ _{3}^{Q}J%
\mathbf{\ ^{\shortmid }}h^{6}]|\ \mathbf{\ ^{\shortmid }}h^{6}}\{dp_{5}+%
\frac{\partial _{i_{2}}[\int dp_{5}(\ _{3}^{Q}J)\mathbf{\ ^{\shortmid }}%
\partial ^{5}\mathbf{\ ^{\shortmid }}h^{6}]}{\ \ _{3}^{Q}J\mathbf{\
^{\shortmid }}\partial ^{5}\ \mathbf{^{\shortmid }}h^{6}}dx^{i_{2}}\}^{2}+ 
\notag \\
&&\mathbf{\ ^{\shortmid }}h^{6}\{dp_{5}+[\ _{1}n_{k_{2}}+\ _{2}n_{k_{2}}\int
dp_{5}\frac{(\mathbf{\ ^{\shortmid }}\partial ^{5}\mathbf{\ ^{\shortmid }}%
h^{6})^{2}}{|\int dp_{5}\mathbf{\ ^{\shortmid }}\partial ^{5}[\ _{3}^{Q}J%
\mathbf{\ ^{\shortmid }}h^{6}]|\ (\mathbf{\ ^{\shortmid }}h^{6})^{5/2}}%
]dx^{k_{2}}\}+  \notag \\
&&\frac{(\mathbf{\ ^{\shortmid }}\partial ^{7}\mathbf{\ ^{\shortmid }}%
h^{8})^{2}}{|\int dp_{7}\mathbf{\ ^{\shortmid }}\partial ^{7}[\ _{4}^{Q}J%
\mathbf{\ ^{\shortmid }}h^{8}]|\ \mathbf{\ ^{\shortmid }}h^{8}}\{dp_{7}+%
\frac{\partial _{i_{3}}[\int dp_{7}(\ _{4}^{Q}J)\ \mathbf{\ ^{\shortmid }}%
\partial ^{7}\mathbf{\ ^{\shortmid }}h^{8}]}{\ \ _{4}^{Q}J\ \mathbf{\
^{\shortmid }}\partial ^{7}\mathbf{\ ^{\shortmid }}h^{8}}d\mathbf{\
^{\shortmid }}x^{i_{3}}\}^{2}+  \notag \\
&&\mathbf{\ ^{\shortmid }}h^{8}\{dE+[\ _{1}^{\shortmid }n_{k_{3}}+\
_{2}^{\shortmid }n_{k_{3}}\int dp_{7}\frac{(\mathbf{\ ^{\shortmid }}\partial
^{7}\mathbf{\ ^{\shortmid }}h^{8})^{2}}{|\int dp_{7}\mathbf{\ ^{\shortmid }}%
\partial ^{7}[\ \ _{4}^{Q}J\mathbf{\ ^{\shortmid }}h^{8}]|\ (\mathbf{\
^{\shortmid }}h^{8})^{5/2}}]d\mathbf{\ ^{\shortmid }}x^{k_{3}}\}.  \notag
\end{eqnarray}%
The procedure of generating of corresponding $\tau $-families of s-metrics
is described as follow: \newpage {\scriptsize 
\begin{eqnarray*}
&&%
\begin{tabular}{l}
\hline\hline
\begin{tabular}{lll}
& {\large \textsf{Table 12:\ Off-diagonal nonmetric cosmological pase space
configurations with fixed energy}} &  \\ 
& Exact solutions of $\ \ ^{\shortmid }\widehat{\mathbf{R}}_{\ \ \gamma
_{s}}^{\beta _{s}}(\tau )={\delta }_{\ \ \gamma _{s}}^{\beta _{s}}\ \
_{Q}^{s\shortmid }\mathbf{J}(\tau )\mathbf{\ }$(\ref{feq4afd}) on $%
T_{s}^{\ast }V$ transformed into a momentum version of nonlinear PDEs (\ref%
{eq1})-(\ref{e2c}) & 
\end{tabular}
\\ 
\end{tabular}
\\
&&%
\begin{tabular}{lll}
\hline\hline
&  &  \\ 
$%
\begin{array}{c}
\mbox{d-metric ansatz with} \\ 
\mbox{Killing symmetry }\partial _{3}=\partial _{t},\mathbf{\ ^{\shortmid }}%
\partial ^{8}%
\end{array}%
$ &  & $%
\begin{array}{c}
ds^{2}(\tau )\mathbf{\ }=g_{i_{1}}(x^{k_{1}})(dx^{i_{1}})^{2}+\underline{g}%
_{a_{2}}(x^{k_{1}},t)(dy^{a_{2}}+\underline{N}%
_{i_{1}}^{a_{2}}(x^{k_{1}},t)dx^{i_{1}})^{2} \\ 
+\mathbf{\ ^{\shortmid }}g^{a_{3}}(x^{k_{2}},p_{5})(dp_{a_{3}}+\mathbf{\
^{\shortmid }}N_{i_{2}a_{3}}(x^{k_{2}},p_{5})dx^{i_{2}})^{2} \\ 
+\mathbf{\ ^{\shortmid }}g^{a_{4}}(\mathbf{\ ^{\shortmid }}%
x^{k_{3}},p_{7})(dp_{a_{4}}+\mathbf{\ ^{\shortmid }}N_{i_{3}a_{4}}(\mathbf{\
^{\shortmid }}x^{k_{3}},p_{7})d\mathbf{\ ^{\shortmid }}x^{i_{3}})^{2},%
\mbox{
for }g_{i_{1}}=e^{\psi {(x}^{k_{1}}{)}}, \\ 
\underline{g}_{a_{2}}=\underline{h}_{a_{2}}(x^{k_{1}},t),\underline{N}%
_{i_{1}}^{3}=\ ^{2}\underline{n}_{i_{1}}=\underline{n}%
_{i_{1}}(x^{k_{1}},t),N_{i_{1}}^{4}=\ ^{2}\underline{w}_{i_{1}}=\underline{w}%
_{i_{1}}(x^{k_{1}},t), \\ 
\mathbf{\ ^{\shortmid }}g^{a_{3}}=\mathbf{\ ^{\shortmid }}%
h^{a_{3}}(x^{k_{2}},p_{5}),\mathbf{\ ^{\shortmid }}N_{i_{2}5}=\ _{\shortmid
}^{3}w_{i_{2}}=\mathbf{\ ^{\shortmid }}w_{i_{2}}(x^{k_{2}},p_{5}), \\ 
\mathbf{\ ^{\shortmid }}N_{i_{2}6}=\ _{\shortmid }^{3}n_{i_{2}}=\mathbf{\
^{\shortmid }}n_{i_{2}}(x^{k_{2}},p_{5}), \\ 
\mathbf{\ ^{\shortmid }}g^{a_{4}}=\mathbf{\ ^{\shortmid }}h^{a_{4}}(\mathbf{%
\ ^{\shortmid }}x^{k_{3}},p_{7}),\mathbf{\ ^{\shortmid }}N_{i_{3}7}=\ \
_{\shortmid }^{4}w_{i_{3}}=\mathbf{\ ^{\shortmid }}%
w_{i_{3}}(x^{k_{3}},p_{7}), \\ 
N_{i_{3}8}=\ _{\shortmid }^{4}n_{i_{3}}=\mathbf{\ ^{\shortmid }}%
n_{i_{3}}(x^{k_{3}},p_{7}),%
\end{array}%
$ \\ 
Effective matter sources &  & $\mathbf{\ _{Q}^{\shortmid }J}_{\ \nu
_{s}}^{\mu _{s}}(\tau )\mathbf{\ }=[\ _{1}^{Q}J({x}^{k_{1}})\delta
_{j_{1}}^{i_{1}},\ \ _{2}^{Q}\underline{J}({x}^{k_{1}},t)\delta
_{b_{2}}^{a_{2}},\ _{3}^{\shortmid Q}J({x}^{k_{2}},p_{5})\delta
_{b_{3}}^{a_{3}},\ _{4}^{\shortmid Q}J({x}^{k_{3}},p_{7})\delta
_{b_{4}}^{a_{4}}],$ \\ \hline
Nonlinear PDEs (\ref{eq1})-(\ref{e2c}) &  & $%
\begin{tabular}{lll}
$%
\begin{array}{c}
\psi ^{\bullet \bullet }+\psi ^{\prime \prime }=2\ \ _{1}^{Q}J; \\ 
\ ^{2}\underline{\varpi }^{\diamond _{2}}\ \underline{h}_{3}^{\diamond
_{2}}=2\underline{h}_{3}\underline{h}_{4}\ \ _{2}^{Q}\underline{J}; \\ 
\ ^{2}\underline{n}_{k_{1}}^{\diamond _{2}\diamond _{2}}+\ ^{2}\underline{%
\gamma }\ ^{2}\underline{n}_{k_{1}}^{\diamond _{2}}=0; \\ 
\ ^{2}\underline{\beta }\ ^{2}\underline{w}_{i_{1}}-\ ^{2}\underline{\alpha }%
_{i_{1}}=0;%
\end{array}%
$ &  & $%
\begin{array}{c}
\ ^{2}\underline{\varpi }{=\ln |\partial _{4}\underline{{h}}_{4}/\sqrt{|%
\underline{h}_{3}\underline{h}_{4}|}|,} \\ 
\ ^{2}\underline{\alpha }_{i_{1}}=(\partial _{4}\underline{h}_{3})\
(\partial _{i_{1}}\ ^{2}\underline{\varpi }), \\ 
\ ^{2}\underline{\beta }=(\partial _{4}\underline{h}_{4})\ (\partial _{3}\
^{2}\underline{\varpi }),\  \\ 
\ \ ^{2}\underline{\gamma }=\partial _{4}\left( \ln |\underline{h}%
_{3}|^{3/2}/|\underline{h}_{4}|\right) , \\ 
\partial _{1}q=q^{\bullet },\partial _{2}q=q^{\prime }, \\ 
\partial _{4}q=\partial _{t}q=q^{\diamond _{2}}%
\end{array}%
$ \\ 
$%
\begin{array}{c}
\mathbf{\ ^{\shortmid }}\partial ^{5}(\ _{\shortmid }^{3}\varpi )\ \mathbf{\
^{\shortmid }}\partial ^{5}\mathbf{\ ^{\shortmid }}h^{6}=2\mathbf{\
^{\shortmid }}h^{5}\mathbf{\ ^{\shortmid }}h^{6}\ \ _{3}^{\shortmid Q}J; \\ 
\ _{\shortmid }^{3}\beta \ _{\shortmid }^{3}w_{i_{2}}-\ _{\shortmid
}^{3}\alpha _{i_{2}}=0; \\ 
\mathbf{\ ^{\shortmid }}\partial ^{5}(\mathbf{\ ^{\shortmid }}\partial ^{5}\
_{\shortmid }^{3}n_{k_{2}})+\ _{\shortmid }^{3}\gamma \mathbf{\ ^{\shortmid }%
}\partial ^{5}(\ _{\shortmid }^{3}n_{k_{2}})=0;%
\end{array}%
$ &  & $%
\begin{array}{c}
\\ 
\ _{\shortmid }^{3}\varpi {=\ln |\mathbf{\ ^{\shortmid }}\partial ^{5}%
\mathbf{\ ^{\shortmid }}h^{6}/\sqrt{|\mathbf{\ ^{\shortmid }}h^{5}\mathbf{\
^{\shortmid }}h^{6}|}|,} \\ 
\ _{\shortmid }^{3}\alpha _{i_{2}}=(\mathbf{\ ^{\shortmid }}\partial ^{5}%
\mathbf{\ ^{\shortmid }}h^{6})\ (\partial _{i_{2}}\ _{\shortmid }^{3}\varpi
), \\ 
\ _{\shortmid }^{3}\beta =(\mathbf{\ ^{\shortmid }}\partial ^{5}\mathbf{\
^{\shortmid }}h^{6})\ (\mathbf{\ ^{\shortmid }}\partial ^{5}\ _{\shortmid
}^{3}\varpi ),\  \\ 
\ \ _{\shortmid }^{3}\gamma =\mathbf{\ ^{\shortmid }}\partial ^{5}\left( \ln
|\mathbf{\ ^{\shortmid }}h^{6}|^{3/2}/|\mathbf{\ ^{\shortmid }}h^{5}|\right)
,%
\end{array}%
$ \\ 
$%
\begin{array}{c}
\mathbf{\ ^{\shortmid }}\partial ^{7}(\ _{\shortmid }^{4}\varpi )\ \mathbf{\
^{\shortmid }}\partial ^{7}\ \mathbf{\ ^{\shortmid }}h^{8}=2\mathbf{\
^{\shortmid }}h^{7}\mathbf{\ ^{\shortmid }}h^{8}\ _{4}^{\shortmid Q}J; \\ 
\ _{\shortmid }^{4}\beta \ _{\shortmid }^{4}w_{i_{3}}-\ _{\shortmid
}^{4}\alpha _{i_{3}}=0; \\ 
\mathbf{\ ^{\shortmid }}\partial ^{7}(\mathbf{\ ^{\shortmid }}\partial ^{7}\
_{\shortmid }^{4}n_{k_{3}})+\ _{\shortmid }^{4}\gamma \mathbf{\ ^{\shortmid }%
}\partial ^{7}(\ _{\shortmid }^{4}n_{k_{3}})=0;%
\end{array}%
$ &  & $%
\begin{array}{c}
\\ 
\ _{\shortmid }^{4}\varpi {=\ln |\mathbf{\ ^{\shortmid }}\partial ^{7}%
\mathbf{\ ^{\shortmid }}h^{8}/\sqrt{|\mathbf{\ ^{\shortmid }}h^{7}\mathbf{\
^{\shortmid }}h^{8}|}|,} \\ 
\ _{\shortmid }^{4}\alpha _{i_{3}}=(\mathbf{\ ^{\shortmid }}\partial ^{7}%
\mathbf{\ ^{\shortmid }}h^{8})\ (\mathbf{\ ^{\shortmid }}\partial _{i_{3}}\
_{\shortmid }^{4}\varpi ), \\ 
\ _{\shortmid }^{4}\beta =(\mathbf{\ ^{\shortmid }}\partial ^{7}\mathbf{\
^{\shortmid }}h^{8})\ (\mathbf{\ ^{\shortmid }}\partial ^{7}\ _{\shortmid
}^{4}\varpi ),\  \\ 
\ \ _{\shortmid }^{4}\gamma =\mathbf{\ ^{\shortmid }}\partial ^{7}\left( \ln
|\mathbf{\ ^{\shortmid }}h^{8}|^{3/2}/|\mathbf{\ ^{\shortmid }}h^{7}|\right)
,%
\end{array}%
$%
\end{tabular}%
$ \\ \hline
$%
\begin{array}{c}
\mbox{ Gener.  functs:}\ \underline{h}_{4}(x^{k_{1}},t), \\ 
\ ^{2}\underline{\Psi }(x^{k_{1}},t)=e^{\ ^{2}\underline{\varpi }},\ ^{2}%
\underline{\Phi }(x^{k_{1}},t) \\ 
\mbox{integr. functs:}\ h_{4}^{[0]}(x^{k_{1}}),\  \\ 
_{1}n_{k_{1}}(x^{i_{1}}),\ _{2}n_{k_{1}}(x^{i_{1}}); \\ 
\mbox{ Gener.  functs:}\mathbf{\ ^{\shortmid }}h^{5}(x^{k_{2}},p_{5}), \\ 
\ \ _{\shortmid }^{3}\Psi (x^{k_{2}},p_{5})=e^{\ \ _{\shortmid }^{3}\varpi
},\ _{\shortmid }^{3}\Phi (x^{k_{2}},p_{5}) \\ 
\mbox{integr. functs:}\ h_{6}^{[0]}(x^{k_{2}}),\  \\ 
_{1}^{3}n_{k_{2}}(x^{i_{2}}),\ _{2}^{3}n_{k_{2}}(x^{i_{2}}); \\ 
\mbox{ Gener.  functs:}\mathbf{\ ^{\shortmid }}h^{7}(\mathbf{\ ^{\shortmid }}%
x^{k_{3}},p_{7}), \\ 
\ \ _{\shortmid }^{4}\Psi (x^{k_{2}},p_{7})=e^{\ \ \ _{\shortmid }^{4}\varpi
},\ \ \ _{\shortmid }^{4}\Phi (\mathbf{\ ^{\shortmid }}x^{k_{3}},p_{7}), \\ 
\mbox{integr. functs:}\ h_{8}^{[0]}(\mathbf{\ ^{\shortmid }}x^{k_{3}}),\  \\ 
_{1}^{4}n_{k_{3}}(\mathbf{\ ^{\shortmid }}x^{i_{3}}),\ _{2}^{4}n_{k_{3}}(%
\mathbf{\ ^{\shortmid }}x^{i_{3}}); \\ 
\mbox{\& nonlinear symmetries}%
\end{array}%
$ &  & $%
\begin{array}{c}
\ ((\ ^{2}\underline{\Psi })^{2})^{\diamond _{2}}=-\int dt\ \ _{2}^{Q}%
\underline{J}\underline{h}_{3}^{\ \diamond _{2}}, \\ 
(\ ^{2}\underline{\Phi })^{2}=-4\ _{2}\underline{\Lambda }\underline{h}_{3},
\\ 
h_{3}=h_{3}^{[0]}-(\ ^{2}\underline{\Phi })^{2}/4\ _{2}\underline{\Lambda },%
\underline{h}_{3}^{\diamond _{2}}\neq 0,\ _{2}\underline{\Lambda }\neq
0=const; \\ 
\\ 
\mathbf{\ ^{\shortmid }}\partial ^{5}((\ \ _{\shortmid }^{3}\Psi
)^{2})=-\int dp_{5}\ \ _{3}^{\shortmid Q}J\mathbf{\ ^{\shortmid }}\partial
^{5}\mathbf{\ ^{\shortmid }}h^{6}, \\ 
(\ \ _{\shortmid }^{3}\Phi )^{2}=-4\ _{3}^{\shortmid }\Lambda \mathbf{\
^{\shortmid }}h^{6}, \\ 
\mathbf{\ ^{\shortmid }}h^{6}=\mathbf{\ ^{\shortmid }}h_{[0]}^{6}-(\ \
_{\shortmid }^{3}\Phi )^{2}/4\ _{3}\Lambda ,\mathbf{\ ^{\shortmid }}\partial
^{5}\mathbf{\ ^{\shortmid }}h^{6}\neq 0,\ _{3}^{\shortmid }\Lambda \neq
0=const; \\ 
\\ 
\mathbf{\ ^{\shortmid }}\partial ^{7}((\ _{\shortmid }^{4}\Psi )^{2})=-\int
dp_{7}\ \ _{4}^{\shortmid Q}J\mathbf{\ ^{\shortmid }}\partial ^{7}\mathbf{\
^{\shortmid }}h^{8}, \\ 
(\ _{\shortmid }^{4}\Phi )^{2}=-4\ _{4}^{\shortmid }\Lambda \mathbf{\
^{\shortmid }}h^{8}, \\ 
\mathbf{\ ^{\shortmid }}h^{8}=\mathbf{\ ^{\shortmid }}h_{[0]}^{8}-(\
_{\shortmid }^{4}\Phi )^{2}/4\ _{4}^{\shortmid }\Lambda ,\mathbf{\
^{\shortmid }}\partial ^{7}\mathbf{\ ^{\shortmid }}h^{8}\neq 0,\
_{4}^{\shortmid }\Lambda \neq 0=const;%
\end{array}%
$ \\ \hline
Off-diag. solutions, $%
\begin{array}{c}
\mbox{d--metric} \\ 
\mbox{N-connec.}%
\end{array}%
$ &  & $%
\begin{tabular}{l}
$%
\begin{array}{c}
\ g_{i}=e^{\ \psi (x^{k})}\mbox{ as a solution of 2-d Poisson eqs. }\psi
^{\bullet \bullet }+\psi ^{\prime \prime }=2~\ _{1}^{Q}J; \\ 
\underline{h}_{4}=-(\ ^{2}\underline{\Psi }^{\diamond _{2}})^{2}/4\ \
_{2}^{Q}\underline{J}^{2}\underline{h}_{3}; \\ 
\underline{h}_{3}=\underline{h}_{3}^{[0]}-\int dt(\ ^{2}\underline{\Psi }%
^{2})^{\diamond _{2}}/4\ \ _{2}^{Q}\underline{J}=\underline{h}_{3}^{[0]}-\
^{2}\underline{\Phi }^{2}/4\ _{2}\underline{\Lambda }; \\ 
\underline{w}_{i_{1}}=\partial _{i_{1}}\ \ ^{2}\underline{\Psi }/\ \partial
\ ^{2}\underline{\Psi }^{\diamond _{2}}=\partial _{i_{1}}\ \ ^{2}\underline{%
\Psi }^{2}/\ \partial _{t}\ ^{2}\underline{\Psi }^{2}|; \\ 
\underline{n}_{k_{1}}=\ _{1}n_{k_{1}}+\ _{2}n_{k_{1}}\int dt(\ ^{2}%
\underline{\Psi }^{\diamond _{2}})^{2}/\ \ _{2}^{Q}\underline{J}^{2}|%
\underline{h}_{3}^{[0]}-\int dt(\ ^{2}\underline{\Psi }^{2})^{\diamond
_{2}}/4\ \ _{2}^{Q}\underline{J}^{2}|^{5/2};%
\end{array}%
$ \\ 
$%
\begin{array}{c}
\mathbf{\ ^{\shortmid }}h^{5}=-(\mathbf{\ ^{\shortmid }}\partial ^{5}\
_{\shortmid }^{3}\Psi )^{2}/4\ \ _{3}^{\shortmid Q}J^{2}\mathbf{\
^{\shortmid }}h^{6}; \\ 
\mathbf{\ ^{\shortmid }}h^{6}=\mathbf{\ ^{\shortmid }}h_{[0]}^{6}-\int dp_{5}%
\mathbf{\ ^{\shortmid }}\partial ^{5}((\ \ _{\shortmid }^{3}\Psi )^{2})/4\ \
_{3}^{\shortmid Q}J=\mathbf{\ ^{\shortmid }}h_{[0]}^{6}-(\ \ _{\shortmid
}^{3}\Phi )^{2}/4\ _{3}^{\shortmid }\Lambda ; \\ 
w_{i_{2}}=\partial _{i_{2}}(\ _{\shortmid }^{3}\Psi )/\mathbf{\ ^{\shortmid }%
}\partial ^{5}(\ _{\shortmid }^{3}\Psi )=\partial _{i_{2}}(\ _{\shortmid
}^{3}\Psi )^{2}/\ \mathbf{\ ^{\shortmid }}\partial ^{5}(\ _{\shortmid
}^{3}\Psi )^{2}|; \\ 
n_{k_{2}}=\ _{1}n_{k_{2}}+\ _{2}n_{k_{2}}\int dp_{5}(\mathbf{\ ^{\shortmid }}%
\partial ^{5}\ _{\shortmid }^{3}\Psi )^{2}/\ \ _{3}^{\shortmid Q}J^{2}|%
\mathbf{\ ^{\shortmid }}h_{[0]}^{6}- \\ 
\int dp_{5}\mathbf{\ ^{\shortmid }}\partial ^{5}((\ _{\shortmid }^{3}\Psi
)^{2})/4\ \ _{3}^{\shortmid Q}J^{2}|^{5/2};%
\end{array}%
$ \\ 
$%
\begin{array}{c}
\mathbf{\ ^{\shortmid }}h^{7}=-(\mathbf{\ ^{\shortmid }}\partial ^{7}\
_{\shortmid }^{4}\Psi )^{2}/4\ \ _{4}^{\shortmid Q}J^{2}\mathbf{\
^{\shortmid }}h^{8}; \\ 
\mathbf{\ ^{\shortmid }}h^{8}=\mathbf{\ ^{\shortmid }}h_{[0]}^{8}-\int dp_{7}%
\mathbf{\ ^{\shortmid }}\partial ^{7}((\ _{\shortmid }^{4}\Psi )^{2})/4\ \
_{4}^{\shortmid Q}J=h_{8}^{[0]}-(\ \ _{\shortmid }^{4}\Phi )^{2}/4\ \
_{4}^{\shortmid }\Lambda ; \\ 
\mathbf{\ ^{\shortmid }}w_{i_{3}}=\mathbf{\ ^{\shortmid }}\partial
_{i_{3}}(\ \ _{\shortmid }^{4}\Psi )/\ \mathbf{\ ^{\shortmid }}\partial
^{7}(\ _{\shortmid }^{4}\Psi )=\mathbf{\ ^{\shortmid }}\partial _{i_{3}}(\ \
_{\shortmid }^{4}\Psi )^{2}/\ \mathbf{\ ^{\shortmid }}\partial ^{7}(\
_{\shortmid }^{4}\Psi )^{2}|; \\ 
\mathbf{\ ^{\shortmid }}n_{k_{3}}=\ _{1}^{\shortmid }n_{k_{3}}+\
_{2}^{\shortmid }n_{k_{3}}\int dp_{7}(\ _{\shortmid }^{4}\Psi )^{2}/\ \
_{4}^{\shortmid Q}J^{2}|h_{8}^{[0]}- \\ 
\int dp_{7}\mathbf{\ ^{\shortmid }}\partial ^{7}((\ \ _{\shortmid }^{4}\Psi
)^{2})/4\ \ _{4}^{\shortmid Q}J^{2}|^{5/2}.%
\end{array}%
$%
\end{tabular}%
$ \\ \hline\hline
\end{tabular}%
\end{eqnarray*}%
}

\subsubsection{Locally anisotropic nonmetric cosmological solutions with
variable energy parameter}

Table 13 is a momentum phase version of Table 8. In this subsection, we
summarize the AFCDM for constructing $\tau $-families of locally anisotropic
nonmetric cosmological rainbow solutions. {\scriptsize 
\begin{eqnarray*}
&&%
\begin{tabular}{l}
\hline\hline
\begin{tabular}{lll}
& {\large \textsf{Table 13:\ Off-diagonal nonmetric cosmological pase space
configurations with variable energy}} &  \\ 
& Exact solutions of $\ \ \ ^{\shortmid }\widehat{\mathbf{R}}_{\ \ \gamma
_{s}}^{\beta _{s}}(\tau )={\delta }_{\ \ \gamma _{s}}^{\beta _{s}}\ \
_{Q}^{s\shortmid }\mathbf{J}(\tau )\mathbf{\ }$(\ref{feq4afd}) on $%
T_{s}^{\ast }V$ transformed into a momentum version of nonlinear PDEs (\ref%
{eq1})-(\ref{e2c}) & 
\end{tabular}
\\ 
\end{tabular}
\\
&&%
\begin{tabular}{lll}
\hline\hline
&  &  \\ 
$%
\begin{array}{c}
\mbox{d-metric ansatz with} \\ 
\mbox{Killing symmetry }\partial _{3}=\partial _{t},\mathbf{\ ^{\shortmid }}%
\partial ^{7}%
\end{array}%
$ &  & $%
\begin{array}{c}
ds^{2}(\tau )=g_{i_{1}}(x^{k_{1}})(dx^{i_{1}})^{2}+\underline{g}%
_{a_{2}}(x^{k_{1}},t)(dy^{a_{2}}+\underline{N}%
_{i_{1}}^{a_{2}}(x^{k_{1}},t)dx^{i_{1}})^{2} \\ 
+\mathbf{\ ^{\shortmid }}g^{a_{3}}(x^{k_{2}},p_{5})(dp_{a_{3}}+\mathbf{\
^{\shortmid }}N_{i_{2}a_{3}}(x^{k_{2}},p_{5})dx^{i_{2}})^{2} \\ 
+\mathbf{\ ^{\shortmid }}g^{a_{4}}(\mathbf{\ ^{\shortmid }}%
x^{k_{3}},p_{7})(dp_{a_{4}}+\mathbf{\ ^{\shortmid }}N_{i_{3}a_{4}}(\mathbf{\
^{\shortmid }}x^{k_{3}},p_{7})d\mathbf{\ ^{\shortmid }}x^{i_{3}})^{2},%
\mbox{
for }g_{i_{1}}=e^{\psi {(x}^{k_{1}}{)}}, \\ 
\underline{g}_{a_{2}}=\underline{h}_{a_{2}}(x^{k_{1}},t),\underline{N}%
_{i_{1}}^{3}=\ ^{2}\underline{n}_{i_{1}}=\underline{n}_{i_{1}}(x^{k_{1}},t),%
\underline{N}_{i_{1}}^{4}=\ ^{2}\underline{w}_{i_{1}}=\underline{w}%
_{i_{1}}(x^{k_{1}},t), \\ 
\mathbf{\ ^{\shortmid }}g^{a_{3}}=\mathbf{\ ^{\shortmid }}%
h^{a_{3}}(x^{k_{2}},p_{5}),\mathbf{\ ^{\shortmid }}N_{i_{2}5}=\ _{\shortmid
}^{3}w_{i_{2}}=\mathbf{\ ^{\shortmid }}w_{i_{2}}(x^{k_{2}},p_{5}), \\ 
\mathbf{\ ^{\shortmid }}N_{i_{2}6}=\ _{\shortmid }^{3}n_{i_{2}}=\mathbf{\
^{\shortmid }}n_{i_{2}}(x^{k_{2}},p_{5}), \\ 
\mathbf{\ ^{\shortmid }}\underline{g}^{a_{4}}=\mathbf{\ ^{\shortmid }}%
\underline{h}^{a_{4}}(\mathbf{\ ^{\shortmid }}x^{k_{3}},E),\mathbf{\
^{\shortmid }}\underline{N}_{i_{3}7}=\ \ _{\shortmid }^{4}\underline{n}%
_{i_{3}}=\mathbf{\ ^{\shortmid }}\underline{n}_{i_{3}}(x^{k_{3}},E), \\ 
\mathbf{\ ^{\shortmid }}\underline{N}_{i_{3}8}=\ _{\shortmid }^{4}\underline{%
w}_{i_{3}}=\mathbf{\ ^{\shortmid }}\underline{w}_{i_{3}}(x^{k_{3}},E),%
\end{array}%
$ \\ 
Effective matter sources &  & $\mathbf{\ ^{\shortmid Q}J}_{\ \nu _{s}}^{\mu
_{s}}(\tau )=[\ _{1}^{Q}J({x}^{k_{1}})\delta _{j_{1}}^{i_{1}},\ _{2}^{Q}%
\underline{J}({x}^{k_{1}},y^{3})\delta _{b_{2}}^{a_{2}},\ _{3}^{\shortmid
Q}J({x}^{k_{2}},p_{5})\delta _{b_{3}}^{a_{3}},\ _{4}^{\shortmid Q}\underline{%
J}({x}^{k_{3}},E)\delta _{b_{4}}^{a_{4}},],$ \\ \hline
Nonlinear PDEs (\ref{eq1})-(\ref{e2c}) &  & $%
\begin{tabular}{lll}
$%
\begin{array}{c}
\psi ^{\bullet \bullet }+\psi ^{\prime \prime }=2\ \ _{1}^{Q}J; \\ 
\ ^{2}\underline{\varpi }^{\diamond _{2}}\ \underline{h}_{3}^{\diamond
_{2}}=2\underline{h}_{3}\underline{h}_{4}\ _{2}^{Q}\underline{J}; \\ 
\ ^{2}\underline{n}_{k_{1}}^{\diamond _{2}\diamond _{2}}+\ ^{2}\underline{%
\gamma }\ ^{2}\underline{n}_{k_{1}}^{\diamond _{2}}=0; \\ 
\ ^{2}\underline{\beta }\ ^{2}\underline{w}_{i_{1}}-\ ^{2}\underline{\alpha }%
_{i_{1}}=0;%
\end{array}%
$ &  & $%
\begin{array}{c}
\ ^{2}\underline{\varpi }{=\ln |\partial _{4}\underline{{h}}_{4}/\sqrt{|%
\underline{h}_{3}\underline{h}_{4}|}|,} \\ 
\ ^{2}\underline{\alpha }_{i_{1}}=(\partial _{4}\underline{h}_{3})\
(\partial _{i_{1}}\ ^{2}\underline{\varpi }), \\ 
\ ^{2}\underline{\beta }=(\partial _{4}\underline{h}_{4})\ (\partial _{3}\
^{2}\underline{\varpi }),\  \\ 
\ \ ^{2}\underline{\gamma }=\partial _{4}\left( \ln |\underline{h}%
_{3}|^{3/2}/|\underline{h}_{4}|\right) , \\ 
\partial _{1}q=q^{\bullet },\partial _{2}q=q^{\prime }, \\ 
\partial _{4}q=\partial _{t}q=q^{\diamond _{2}}%
\end{array}%
$ \\ 
$%
\begin{array}{c}
\mathbf{\ ^{\shortmid }}\partial ^{5}(\ _{\shortmid }^{3}\varpi )\ \mathbf{\
^{\shortmid }}\partial ^{5}\mathbf{\ ^{\shortmid }}h^{6}=2\mathbf{\
^{\shortmid }}h^{5}\mathbf{\ ^{\shortmid }}h^{6}\ \ _{3}^{\shortmid Q}J; \\ 
\ _{\shortmid }^{3}\beta \ _{\shortmid }^{3}w_{i_{2}}-\ _{\shortmid
}^{3}\alpha _{i_{2}}=0; \\ 
\mathbf{\ ^{\shortmid }}\partial ^{5}(\mathbf{\ ^{\shortmid }}\partial ^{5}\
_{\shortmid }^{3}n_{k_{2}})+\ _{\shortmid }^{3}\gamma \mathbf{\ ^{\shortmid }%
}\partial ^{5}(\ _{\shortmid }^{3}n_{k_{2}})=0;%
\end{array}%
$ &  & $%
\begin{array}{c}
\\ 
\ _{\shortmid }^{3}\varpi {=\ln |\mathbf{\ ^{\shortmid }}\partial ^{5}%
\mathbf{\ ^{\shortmid }}h^{6}/\sqrt{|\mathbf{\ ^{\shortmid }}h^{5}\mathbf{\
^{\shortmid }}h^{6}|}|,} \\ 
\ _{\shortmid }^{3}\alpha _{i_{2}}=(\mathbf{\ ^{\shortmid }}\partial ^{5}%
\mathbf{\ ^{\shortmid }}h^{6})\ (\partial _{i_{2}}\ _{\shortmid }^{3}\varpi
), \\ 
\ _{\shortmid }^{3}\beta =(\mathbf{\ ^{\shortmid }}\partial ^{5}\mathbf{\
^{\shortmid }}h^{6})\ (\mathbf{\ ^{\shortmid }}\partial ^{5}\ _{\shortmid
}^{3}\varpi ),\  \\ 
\ \ _{\shortmid }^{3}\gamma =\mathbf{\ ^{\shortmid }}\partial ^{5}\left( \ln
|\mathbf{\ ^{\shortmid }}h^{6}|^{3/2}/|\mathbf{\ ^{\shortmid }}h^{5}|\right)
,%
\end{array}%
$ \\ 
$%
\begin{array}{c}
\mathbf{\ ^{\shortmid }}\underline{\partial }^{8}(\ _{\shortmid }^{4}%
\underline{\varpi })\ \mathbf{\ ^{\shortmid }}\underline{\partial }^{8}%
\mathbf{\ ^{\shortmid }}\underline{h}^{7}=2\ \mathbf{^{\shortmid }}%
\underline{h}^{7}\mathbf{\ ^{\shortmid }}\underline{h}^{8}\ _{4}^{\shortmid
Q}\underline{J}; \\ 
\mathbf{\ ^{\shortmid }}\underline{\partial }^{8}(\mathbf{\ ^{\shortmid }}%
\underline{\partial }^{8}\ _{\shortmid }^{4}\underline{n}_{k_{3}})+\
_{\shortmid }^{4}\underline{\gamma }\mathbf{\ ^{\shortmid }}\underline{%
\partial }^{8}(\ _{\shortmid }^{4}\underline{n}_{k_{3}})=0; \\ 
\ _{\shortmid }^{4}\underline{\beta }\ _{\shortmid }^{4}\underline{w}%
_{i_{3}}-\ _{\shortmid }^{4}\underline{\alpha }_{i_{3}}=0;%
\end{array}%
$ &  & $%
\begin{array}{c}
\\ 
\ _{\shortmid }^{4}\underline{\varpi }{=\ln |\mathbf{\ ^{\shortmid }}%
\underline{{\partial }}^{8}\mathbf{\ ^{\shortmid }}\underline{{h}}^{7}/\sqrt{%
|\mathbf{\ ^{\shortmid }}\underline{h}^{7}\mathbf{\ ^{\shortmid }}\underline{%
h}^{8}|}|,} \\ 
\ _{\shortmid }^{4}\underline{\alpha }_{i_{3}}=(\mathbf{\ ^{\shortmid }}%
\underline{\partial }^{8}\mathbf{\ ^{\shortmid }}\underline{h}^{7})\ (%
\mathbf{\ ^{\shortmid }}\partial _{i_{3}}\ _{\shortmid }^{4}\underline{%
\varpi }), \\ 
\ _{\shortmid }^{4}\underline{\beta }=(\mathbf{\ ^{\shortmid }}\underline{%
\partial }^{8}\mathbf{\ ^{\shortmid }}\underline{h}^{7})\ (\mathbf{\
^{\shortmid }}\underline{\partial }^{8}\ _{\shortmid }^{4}\underline{\varpi }%
),\  \\ 
\ \ _{\shortmid }^{4}\underline{\gamma }=\mathbf{\ ^{\shortmid }}\underline{%
\partial }^{8}\left( \ln |\mathbf{\ ^{\shortmid }}\underline{h}^{7}|^{3/2}/|%
\mathbf{\ ^{\shortmid }}\underline{h}^{8}|\right) ,%
\end{array}%
$%
\end{tabular}%
$ \\ \hline
$%
\begin{array}{c}
\mbox{ Gener.  functs:}\ \underline{h}_{4}(x^{k_{1}},t), \\ 
\ ^{2}\underline{\Psi }(x^{k_{1}},t)=e^{\ ^{2}\underline{\varpi }},\ ^{2}%
\underline{\Phi }(x^{k_{1}},t), \\ 
\mbox{integr. functs:}\ \underline{h}_{3}^{[0]}(x^{k_{1}}),\  \\ 
_{1}\underline{n}_{k_{1}}(x^{i_{1}}),\ _{2}\underline{n}_{k_{1}}(x^{i_{1}});
\\ 
\mbox{ Gener.  functs:}\mathbf{\ ^{\shortmid }}h^{5}(x^{k_{2}},p_{5}), \\ 
\ \ _{\shortmid }^{3}\Psi (x^{k_{2}},p_{5})=e^{\ _{\shortmid }^{3}\varpi },\
_{\shortmid }^{3}\Phi (x^{k_{2}},p_{5}) \\ 
\mbox{integr. functs:}\ h_{6}^{[0]}(x^{k_{2}}),\  \\ 
_{1}^{3}n_{k_{2}}(x^{i_{2}}),\ _{2}^{3}n_{k_{2}}(x^{i_{2}}); \\ 
\mbox{ Gener.  functs:}\mathbf{\ ^{\shortmid }}h^{7}(\mathbf{\ ^{\shortmid }}%
x^{k_{3}},p_{7}), \\ 
\ \ _{\shortmid }^{4}\underline{\Psi }(x^{k_{2}},E)=e^{\ _{\shortmid }^{4}%
\underline{\varpi }},\ _{\shortmid }^{4}\underline{\Phi }(\mathbf{\
^{\shortmid }}x^{k_{3}},E) \\ 
\mbox{integr. functs:}\ \underline{h}_{7}^{[0]}(\mathbf{\ ^{\shortmid }}%
x^{k_{3}}),\  \\ 
_{1}^{4}\underline{n}_{k_{3}}(\mathbf{\ ^{\shortmid }}x^{i_{3}}),\ _{2}^{4}%
\underline{n}_{k_{3}}(\mathbf{\ ^{\shortmid }}x^{i_{3}}); \\ 
\mbox{\& nonlinear symmetries}%
\end{array}%
$ &  & $%
\begin{array}{c}
\ ((\ ^{2}\underline{\Psi })^{2})^{\diamond _{2}}=-\int dt\ _{2}^{Q}%
\underline{J}\underline{h}_{3}^{\ \diamond _{2}}, \\ 
(\ ^{2}\underline{\Phi })^{2}=-4\ _{2}\underline{\Lambda }\underline{h}_{3},
\\ 
h_{3}=h_{3}^{[0]}-(\ ^{2}\underline{\Phi })^{2}/4\ _{2}\underline{\Lambda },%
\underline{h}_{3}^{\diamond _{2}}\neq 0,\ _{2}\underline{\Lambda }\neq
0=const; \\ 
\\ 
\mathbf{\ ^{\shortmid }}\partial ^{5}((\ \ _{\shortmid }^{3}\Psi
)^{2})=-\int dp_{5}\ _{3}^{\shortmid Q}J\mathbf{\ ^{\shortmid }}\partial ^{5}%
\mathbf{\ ^{\shortmid }}h^{6}, \\ 
(\ \ _{\shortmid }^{3}\Phi )^{2}=-4\ _{3}^{\shortmid }\Lambda \mathbf{\
^{\shortmid }}h^{6}, \\ 
\mathbf{\ ^{\shortmid }}h^{6}=\mathbf{\ ^{\shortmid }}h_{[0]}^{6}-(\
_{\shortmid }^{3}\Phi )^{2}/4\ \ _{3}^{\shortmid }\Lambda ,\mathbf{\
^{\shortmid }}\partial ^{5}\mathbf{\ ^{\shortmid }}h^{6}\neq 0,\
_{3}^{\shortmid }\Lambda \neq 0=const; \\ 
\\ 
\mathbf{\ ^{\shortmid }}\underline{\partial }^{8}((\ _{\shortmid }^{4}%
\underline{\Psi })^{2})=-\int dE\ _{4}^{\shortmid Q}\underline{J}\mathbf{\
^{\shortmid }}\underline{\partial }^{8}\mathbf{\ ^{\shortmid }}\underline{h}%
^{7}, \\ 
(\ _{\shortmid }^{4}\underline{\Phi })^{2}=-4\ _{4}^{\shortmid }\underline{%
\Lambda }\mathbf{\ ^{\shortmid }}\underline{h}^{7}, \\ 
\mathbf{\ ^{\shortmid }}\underline{h}^{7}=\mathbf{\ ^{\shortmid }}\underline{%
h}_{[0]}^{7}-(\ _{\shortmid }^{4}\underline{\Phi })^{2}/4\ _{4}^{\shortmid }%
\underline{\Lambda },\mathbf{\ ^{\shortmid }}\underline{\partial }^{8}%
\mathbf{\ ^{\shortmid }}\underline{h}^{7}\neq 0,\ _{4}^{\shortmid }%
\underline{\Lambda }\neq 0=const;%
\end{array}%
$ \\ \hline
Off-diag. solutions, $%
\begin{array}{c}
\mbox{d--metric} \\ 
\mbox{N-connec.}%
\end{array}%
$ &  & $%
\begin{tabular}{l}
$%
\begin{array}{c}
\ g_{i}=e^{\ \psi (x^{k})}\mbox{ as a solution of 2-d Poisson eqs. }\psi
^{\bullet \bullet }+\psi ^{\prime \prime }=2~\ _{1}^{Q}J; \\ 
\underline{h}_{4}=-(\ ^{2}\underline{\Psi }^{\diamond _{2}})^{2}/4\ \
_{2}^{Q}\underline{J}^{2}\underline{h}_{3}; \\ 
\underline{h}_{3}=\underline{h}_{3}^{[0]}-\int dt(\ ^{2}\underline{\Psi }%
^{2})^{\diamond _{2}}/4\ \ _{2}^{Q}\underline{J}=\underline{h}_{3}^{[0]}-\
^{2}\underline{\Phi }^{2}/4\ _{2}\underline{\Lambda }; \\ 
\underline{w}_{i_{1}}=\partial _{i_{1}}\ ^{2}\underline{\Psi }/\ \partial \
^{2}\underline{\Psi }^{\diamond _{2}}=\partial _{i_{1}}\ ^{2}\underline{\Psi 
}^{2}/\ \partial _{t}\ ^{2}\underline{\Psi }^{2}|; \\ 
\underline{n}_{k_{1}}=\ _{1}n_{k_{1}}+\ _{2}n_{k_{1}}\int dt(\ ^{2}%
\underline{\Psi }^{\diamond _{2}})^{2}/\ \ _{2}^{Q}\underline{J}^{2}|%
\underline{h}_{3}^{[0]}-\int dt(\ ^{2}\underline{\Psi }^{2})^{\diamond }/4\
\ _{2}^{Q}\underline{J}^{2}|^{5/2};%
\end{array}%
$ \\ 
$%
\begin{array}{c}
\mathbf{\ ^{\shortmid }}h^{5}=-(\mathbf{\ ^{\shortmid }}\partial ^{5}\
_{\shortmid }^{3}\Psi )^{2}/4\ \ _{3}^{\shortmid Q}J^{2}\mathbf{\
^{\shortmid }}h^{6}; \\ 
\mathbf{\ ^{\shortmid }}h^{6}=\mathbf{\ ^{\shortmid }}h_{[0]}^{6}-\int dp_{5}%
\mathbf{\ ^{\shortmid }}\partial ^{5}((\ \ _{\shortmid }^{3}\Psi )^{2})/4\ \
_{3}^{\shortmid Q}J=\mathbf{\ ^{\shortmid }}h_{[0]}^{6}-(\ \ _{\shortmid
}^{3}\Phi )^{2}/4\ _{3}^{\shortmid }\Lambda ; \\ 
w_{i_{2}}=\partial _{i_{2}}(\ _{\shortmid }^{3}\Psi )/\mathbf{\ ^{\shortmid }%
}\partial ^{5}(\ _{\shortmid }^{3}\Psi )=\partial _{i_{2}}(\ _{\shortmid
}^{3}\Psi )^{2}/\ \mathbf{\ ^{\shortmid }}\partial ^{5}(\ _{\shortmid
}^{3}\Psi )^{2}|; \\ 
n_{k_{2}}=\ _{1}n_{k_{2}}+\ _{2}n_{k_{2}}\int dp_{5}(\mathbf{\ ^{\shortmid }}%
\partial ^{5}\ _{\shortmid }^{3}\Psi )^{2}/\ \ _{3}^{\shortmid Q}J^{2}|%
\mathbf{\ ^{\shortmid }}h_{[0]}^{6}- \\ 
\int dp_{5}\mathbf{\ ^{\shortmid }}\partial ^{5}((\ _{\shortmid }^{3}\Psi
)^{2})/4\ \ _{3}^{\shortmid Q}J^{2}|^{5/2};%
\end{array}%
$ \\ 
$%
\begin{array}{c}
\mathbf{\ ^{\shortmid }}\underline{h}^{8}=-(\mathbf{\ ^{\shortmid }}%
\underline{\partial }^{8}\ _{\shortmid }^{4}\underline{\Psi })^{2}/4\ \
_{4}^{\shortmid Q}\underline{J}\mathbf{\ }^{2}\mathbf{\ ^{\shortmid }}%
\underline{h}^{7}; \\ 
\mathbf{\ ^{\shortmid }}\underline{h}^{7}=\mathbf{\ ^{\shortmid }}\underline{%
h}_{[0]}^{7}-\int dE\mathbf{\ ^{\shortmid }}\underline{\partial }^{8}((\
_{\shortmid }^{4}\underline{\Psi })^{2})/4\ \ _{4}^{\shortmid Q}\underline{J}%
\mathbf{\ }=\underline{h}_{[0]}^{7}-(\ _{\shortmid }^{4}\underline{\Phi }%
)^{2}/4\ \ _{4}^{\shortmid }\underline{\Lambda }; \\ 
\mathbf{\ ^{\shortmid }}\underline{n}_{k_{3}}=\ _{1}^{\shortmid }\underline{n%
}_{k_{3}}+\ _{2}^{\shortmid }\underline{n}_{k_{3}}\int dE(\ _{\shortmid }^{4}%
\underline{\Psi })^{2}/\ \ _{4}^{\shortmid Q}\underline{J}\mathbf{\ }^{2}|%
\underline{h}_{[0]}^{7}- \\ 
\int dE\mathbf{\ ^{\shortmid }}\underline{\partial }^{8}((\ _{\shortmid }^{4}%
\underline{\Psi })^{2})/4\ \ _{4}^{\shortmid Q}\underline{J}\mathbf{\ }%
^{2}|^{5/2}; \\ 
\mathbf{\ ^{\shortmid }}\underline{w}_{i_{3}}=\mathbf{\ ^{\shortmid }}%
\partial _{i_{3}}(\ \ _{\shortmid }^{4}\underline{\Psi })/\ \mathbf{\
^{\shortmid }}\partial ^{8}(\ _{\shortmid }^{4}\underline{\Psi })=\mathbf{\
^{\shortmid }}\partial _{i_{3}}(\ _{\shortmid }^{4}\underline{\Psi })^{2}/%
\mathbf{\ ^{\shortmid }}\partial ^{8}(\ _{\shortmid }^{4}\underline{\Psi }%
)^{2}|.%
\end{array}%
$%
\end{tabular}%
$ \\ \hline\hline
\end{tabular}%
\end{eqnarray*}%
}

Here, we provide an example of $\tau $-families of off-diagonal nonmetric
cosmological rainbow metrics: 
\begin{eqnarray}
&&d\widehat{s}_{[8d]}^{2}(\tau )=\widehat{g}_{\alpha _{s}\beta
_{s}}(x^{k},t,p_{5},E;\underline{h}_{3},\mathbf{\ ^{\shortmid }}h^{6},%
\mathbf{\ ^{\shortmid }}\underline{h}^{7};\ _{1}^{Q}J,\ _{2}^{Q}\underline{J}%
,\ _{3}^{\shortmid Q}J,\ _{4}^{\shortmid Q}\underline{J};\ _{1}\Lambda ,\
_{2}\underline{\Lambda },\ _{3}^{\shortmid }\Lambda ,\ _{4}^{\shortmid }%
\underline{\Lambda })d\mathbf{\ ^{\shortmid }}u^{\alpha _{s}}d\mathbf{\
^{\shortmid }}u^{\beta _{s}}=e^{\psi (x^{k},\ \ _{1}^{Q}J)}  \notag \\
&&\lbrack (dx^{1})^{2}+(dx^{2})^{2}]+\underline{h}_{3}[dy^{3}+(\
_{1}n_{k_{1}}+4\ _{2}n_{k_{1}}\int dt\frac{(\underline{h}_{3}{}^{\diamond
_{2}})^{2}}{|\int dt\ \ _{2}^{Q}\underline{J}\underline{h}_{3}{}^{\diamond
_{2}}|(\underline{h}_{3})^{5/2}})dx^{k_{1}}]+\frac{(\underline{h}%
_{3}{}^{\diamond _{2}})^{2}}{|\int dt\ \ _{2}^{Q}\underline{J}\underline{h}%
_{3}{}^{\diamond _{2}}|\ \underline{h}_{3}}  \label{lc8cstp8a} \\
&&\lbrack dt+\frac{\partial _{i_{1}}(\int dt\ \ _{2}^{Q}\underline{J}\ 
\underline{h}_{3}{}^{\diamond _{2}}])}{\ \ \ _{2}^{Q}\underline{J}\ 
\underline{h}_{3}{}^{\diamond _{2}}}dx^{i_{1}}]+\frac{(\mathbf{\ ^{\shortmid
}}\underline{\partial }^{5}\mathbf{\ ^{\shortmid }}\underline{h}^{6})^{2}}{{%
|\int dp_{5}\mathbf{\ ^{\shortmid }}\partial ^{5}[\ \ _{3}^{\shortmid Q}J%
\mathbf{\ ^{\shortmid }}h^{6}]|\ \mathbf{\ ^{\shortmid }}h^{6}}}\{dp_{5}+%
\frac{\partial _{i_{2}}[\int dp_{5}(\ _{3}^{\shortmid Q}J)\mathbf{\
^{\shortmid }}\partial ^{5}\mathbf{\ ^{\shortmid }}h^{6}]}{\ \
_{3}^{\shortmid Q}J\mathbf{\ ^{\shortmid }}\partial ^{5}\ \mathbf{%
^{\shortmid }}h^{6}}dx^{i_{2}}\}^{2}  \notag \\
&&+\ ^{\shortmid }h^{6}\{dp_{5}+[\ _{1}n_{k_{2}}+\ _{2}n_{k_{2}}\int dp_{5}%
\frac{(\mathbf{\ ^{\shortmid }}\partial ^{5}\mathbf{\ ^{\shortmid }}%
h^{6})^{2}}{|\int dp_{5}\mathbf{\ ^{\shortmid }}\partial ^{5}[\
_{3}^{\shortmid Q}J\mathbf{\ ^{\shortmid }}h^{6}]|\ (\mathbf{\ ^{\shortmid }}%
h^{6})^{5/2}}]dx^{k_{2}}\}+\mathbf{\ ^{\shortmid }}\underline{h}%
^{7}\{dp_{7}+[\ _{1}^{\shortmid }\underline{n}_{k_{3}}+\ _{2}^{\shortmid }%
\underline{n}_{k_{3}}\int dp_{7}  \notag \\
&&\frac{(\mathbf{\ ^{\shortmid }}\underline{\partial }^{8}\mathbf{\
^{\shortmid }}\underline{h}^{7})^{2}}{|\int dE\mathbf{\ ^{\shortmid }}%
\underline{\partial }^{8}[\ \ _{4}^{\shortmid Q}\underline{J}\mathbf{\
^{\shortmid }}\underline{h}^{7}]|\ (\mathbf{\ ^{\shortmid }}\underline{h}%
^{7})^{5/2}}]d\mathbf{\ ^{\shortmid }}x^{k_{3}}\}+\frac{(\mathbf{\
^{\shortmid }}\underline{\partial }^{8}\mathbf{\ ^{\shortmid }}\underline{h}%
^{7})^{2}}{|\int dE\mathbf{\ ^{\shortmid }}\underline{\partial }^{8}[\
_{4}^{\shortmid Q}\underline{J}\mathbf{\ ^{\shortmid }}\underline{h}^{7}]|\ 
\mathbf{\ ^{\shortmid }}\underline{h}^{7}}\{dE+\frac{\partial _{i_{3}}[\int
dE(\ \ _{4}^{\shortmid Q}\underline{J})\ \mathbf{\ ^{\shortmid }}\underline{%
\partial }^{8}\mathbf{\ ^{\shortmid }}\underline{h}^{7}]}{\ \
_{4}^{\shortmid Q}\underline{J}\ \mathbf{\ ^{\shortmid }}\underline{\partial 
}^{8}\mathbf{\ ^{\shortmid }}\underline{h}^{7}}d\mathbf{\ ^{\shortmid }}%
x^{i_{3}}\}^{2}.  \notag
\end{eqnarray}%
Finally, we note that a typical $\tau $-family of quasi-stationary rainbow
metrics on $T^{\ast }\mathbf{V}$, constructed for changing indices $%
7\longleftrightarrow 8$ and respective dependencies on coordinates and
Killing symmetry on $s=4$, is defined by s-metrics with explicit dependence
on $E$-variable. The locally anisotropic nonmetric cosmological s-metrics (%
\ref{lc8cstp8a}) consist of examples of phase space rainbow s-metrics
constructed on $T^{\ast }\mathbf{V}$ and defining FH geometric flow models.

\end{document}